\documentclass[11pt, a4paper, headsepline, footsepline, pointlessnumbers]{scrbook}

\pdfoutput=1

\usepackage{url}
\usepackage{color}
\usepackage{rotating}
\usepackage[ngerman,english]{babel}
\usepackage{setspace}

\usepackage[latin1]{inputenc}
\usepackage{longtable}
\usepackage[it, bf]{caption}
\usepackage{amsfonts}
\usepackage{amsmath}
\usepackage{mathrsfs}
\usepackage{epsfig}
\usepackage[clearempty]{titlesec}
\usepackage{booktabs}
\usepackage{hhline}
\usepackage{array}
\usepackage{subfigure}
\usepackage{floatflt}
\usepackage{graphicx}% Include figure files
\usepackage{dcolumn}% Align table columns on decimal point
\usepackage{bm}% bold math
\usepackage{mathrsfs} 
\usepackage{amssymb}

\newcommand{\ttbar}      {\mbox{$t\bar{t}$}}

\def\pt{\ensuremath{p_{\mathrm{T}}}} % Subscript roman not italic (EE)
\def\met{\mbox{\ensuremath{\not\!\!{E_{\mathrm{T}}}}}}
\def\mex{\mbox{\ensuremath{\not\!\!{E_x}}}}
\def\mexy{\mbox{\ensuremath{\not\!\!{E_{x,y}}}}}
\def\mey{\mbox{\ensuremath{\not\!\!{E_y}}}}
\def\et{\mbox{\ensuremath{E_{\mathrm{T}}}}}
\newcommand{\mtw}{m_{\mathrm T}^W}

\def\MeV{\; \mathrm{MeV}}
\def\GeV{\; \mathrm{GeV}}
\def\TeV{\; \mathrm{TeV}}
\def\pb{\; \mathrm{pb}}
\def\ifb{\; \mathrm{fb}^{-1}}
\def\ipb{\; \mathrm{pb}^{-1}}
\def\sec{\; \mathrm{s}}
\def\ns{\; \mathrm{ns}}
\def\T{\; \mathrm{T}}
\def\Tm{\; \mathrm{Tm}}
\def\mm{\; \mathrm{mm}}

\def\ms{\; \mathrm{ms}}
\def\min{\; \mathrm{min}}
\def\mrad{\; \mathrm{mrad}}
\def\mum{\; \mu\mathrm{m}}
\def\mus{\; \mu\mathrm{s}}
\def\invcms{\; \mathrm{cm}^{-2}\mathrm{s}^{-1}}

\newcommand{\rhadone}{R_{\mathrm{had}_1}}
\newcommand{\rhad}{R_{\mathrm{had}}}
\newcommand{\reta}{R_\eta}
\newcommand{\wtwo}{w_2}
\newcommand{\rphi}{R_\phi}
\newcommand{\wsthree}{w_{s \, 3}}
\newcommand{\wstot}{w_{s \, \mathrm{tot}}}
\newcommand{\fside}{F_{\mathrm{side}}}
\newcommand{\deltae}{\Delta E}
\newcommand{\eratio}{E_{\mathrm{ratio}}}

\newcommand{\ttg}{\ttbar\gamma}
\newcommand{\ptcone}{\mbox{$\pt^{\mathrm{cone20}}$}}

\newcommand{\feg}{f_{e \to \gamma}}
\newcommand{\Zee}{\mbox{$Z \to ee$} }
\newcommand{\Zeg}{\mbox{$Z \to e\gamma_{\mathrm{fake}}$} }

\newcommand{\emptynull}{\textcolor{white}{0}}
\newcommand{\emptyplus}{\textcolor{white}{+}}
\newcommand{\emptyminus}{\textcolor{white}{-}}

\newcommand{\emptypercent}{\textcolor{white}{\%}}

\usepackage{cite}
\usepackage{mcite}
\usepackage{multirow}
\usepackage{hyperref}

\renewcommand\arraystretch{1.25} % default : 1.0

\begin{document} 

\pagenumbering{roman} 

\begin{titlepage}
  \begin{center}
    \begin{onehalfspace} 
      \null
      \vspace{15mm}
      {\LARGE \bf
        Measurement of the inclusive \boldmath$\ttg$ cross section\\ at \mbox{$\sqrt{s} = 7 \TeV$} with the ATLAS detector\unboldmath \\
      }
      \vspace{0.15\textheight}
      {\Large
        \selectlanguage{ngerman}
        Dissertation\\
        \vspace{0.05\textheight}
        zur Erlangung des mathematisch-naturwissenschaftlichen Doktorgrades\\
        \glqq Doctor rerum naturalium\grqq\\
        der Georg-August-Universit\"at G\"ottingen\\
        \vspace{0.15\textheight}
        vorgelegt von\\
        Johannes Erdmann\\
        aus Bonn\\
        \vspace{0.15\textheight}
        G\"ottingen, 2012\\
        \selectlanguage{english}
      }
    \end{onehalfspace}
  \end{center}
\end{titlepage} 

\clearpage

\null
\vspace{0.85\textheight}
\begin{tabbing}
xxxxxxxxxxxxxx \=                   \kill
Referent:       \> Prof. Dr. Arnulf Quadt      \\
Korreferentin:  \> Jun.-Prof. Dr. Lucia Masetti \\
\end{tabbing}
\vspace{0.01\textheight}
Tag der m\"undlichen Pr\"ufung: 29.05.2012 \\

\clearpage

\begin{center}
  \null
  \vfil
  {\LARGE \bf
    \begin{onehalfspace}
      Measurement of the inclusive \boldmath$\ttg$ cross section\\
      at \mbox{$\sqrt{s} = 7 \TeV$} with the ATLAS detector\unboldmath\\
    \end{onehalfspace}
  }
  \vfil
  {\Large
    by\\
    Johannes Erdmann\\
  }
  \vfil
  {
    \parbox[t]{140mm}{A first measurement of the $\ttg$ cross section in $pp$ collisions at the LHC using \mbox{$1.04 \ifb$} of data taken with the ATLAS detector is presented.
A total of 122 candidate events were identified in the single electron and single muon channels.
The contributions from background processes with prompt photons, and with electrons or hadrons misidentified as photons were estimated reducing the
dependence on simulations by the use of data-driven techniques.
The resulting cross section times branching ratio into the single lepton and dilepton decay channels for photons with \mbox{$\pt > 8 \GeV$} reads
\begin{equation*}
  \sigma_{t\bar{t}\gamma} \cdot \mathrm{BR} = 1.9 \pm 0.5 \, \mathrm{(stat.)} \pm 0.8 \, \mathrm{(syst.)} \pm 0.1 \, \mathrm{(lumi.)} \pb \, ,
\end{equation*}
which is consistent with the Standard Model expectation from theoretical calculations.
The significance of the $\ttg$ signal was estimated to $2.5 \, \sigma$.
}
  }
  \vfil
\end{center}
\begin{figure}[b]
  \begin{minipage}{\textwidth}
    \begin{raggedright}
      \begin{tabular}{@{}l@{}}
        Post address:  \\
        Friedrich-Hund-Platz 1 \\
        37077 G\"ottingen  \\
        Germany      \\
      \end{tabular}
    \end{raggedright}
    \hfill        
    \begin{raggedleft}
      \begin{tabular}{@{}r@{}}
        II.Physik-UniG\"o-Diss-2012/03\\
        II.~Physikalisches Institut\\
        Georg-August-Universit\"at G\"ottingen\\
        April 2012\\
      \end{tabular}
    \end{raggedleft}
  \end{minipage}
  \vspace*{5mm}
\end{figure}

\cleardoublepage

\begin{flushright}
\textcolor{white}{ }\\
\vspace{0.25\textwidth}
And God saw that the light was good. \\
\vspace{0.75cm}
\textit{Genesis 1:4}
\end{flushright}

\cleardoublepage

\tableofcontents 
 
\clearpage 
 
\pagenumbering{arabic}
\setcounter{page}{1}

\addchap{Introduction}

When the electron was discovered in 1897, the first elementary particle had been found.
This can be considered the birth of particle physics, the science of elementary particles and their fundamental interactions.
The Standard Model of particle physics (SM) is in agreement with nearly all phenomena observed over the last decades, which have partly been measured with excellent precision.
However, there are also strong arguments that it might need to be embedded in a more general theory.

The Large Hadron Collider (LHC) at CERN, Geneva, is a proton-proton collider, which started operation in September 2008.
It was built to search for the still undiscovered Higgs boson and for physics beyond the Standard Model, but also to perform precision tests of SM processes.
In 2011, protons were brought to collision at a centre-of-mass energy of \mbox{$7 \TeV$} with instantaneous luminosities of up to several $10^{33} \invcms$.
The data used for the analysis presented in this thesis were taken with the ATLAS detector, a general purpose detector, which has been
designed to cope with the uniquely high collision rates, energies and instantaneous luminosities provided by the LHC.

The top quark is the heaviest of the known elementary particles.
It has been discovered in 1995~\cite{topobsD0,topobsCDF} at the Tevatron, Batavia (Illinois), and since then many of its properties have been measured at the Tevatron
and at the LHC.
Amongst others, the production, the mass, the decay and the spin properties of the top quark have been studied.
Another fundamental quantity is the electromagnetic charge of the top quark, which in the Standard Model is predicted to be $+\frac{2}{3}$ of the proton charge.
Recently, an alternative charge hypothesis of $-\frac{4}{3}$ the proton charge was ruled out~\cite{topchargeATLAS,topchargeCMS,topchargeD0,topchargeCDF}.

In the SM, the electromagnetic charge of a particle determines its electromagnetic coupling -- the interaction with other
electrically charged particles by the exchange of photons.
Hence, the amount of photon radiation from a particle is directly sensitive to its electromagnetic coupling.

Accordingly, top quark pair ($\ttbar$) events with additional photons in the final state, denoted $\ttg$ events in the following, are sensitive to the
electromagnetic coupling of the top quark.
The investigation of such events is an important test of the SM, because it provides a direct measure of the charge of the top quark
and of a possible anomalous structure of its electromagnetic interaction.

An important step towards precision tests of the coupling itself is the measurement of the production cross section of $\ttg$ events,
$\sigma_{t\bar{t}\gamma}$, at the LHC.
This thesis presents the first measurement of $\sigma_{t\bar{t}\gamma} \cdot \rm{BR}$ at a centre-of-mass energy of \mbox{$7 \TeV$}, where BR is the branching ratio
into $\ttg$ decays with one or two leptons in the final state (single lepton and dilepton channel, respectively).
\newline

The measurement was performed in the single lepton channel, which features a large variety of particles in the final state and, hence,
also of experimental signatures, which need to be identified and distinguished: photons, electrons, muons, jets from partons, and
missing transverse energy from the neutrino, which escapes the detector without interacting.
In particular, the presence of the photon added complexity to the analysis:
an analysis strategy which reduced the dependence of the measurement on simulations was set up and the contributions from the most important
background processes were estimated from data.

Dominant background contributions are processes in which a hadron or an electron is misidentified as a photon.
In addition, also processes with real photons in the final state were considered, such as the production of $W$ bosons with additional jets and a prompt photon.

A template fit to the sum of the transverse momenta in a cone around the photon candidate was performed, from which the number of $\ttg$ events in the selected
data sample and its statistical uncertainty were estimated.
This was translated into a measurement of $\sigma_{t\bar{t}\gamma} \cdot \rm{BR}$, and the statistical and systematic uncertainties of the measurement were evaluated.
Finally, the statistical significance of the result was estimated.

This thesis is divided into 14 chapters:
the theoretical background and the experimental setup are introduced in chapters~1 and~2.
In chapters~3~--~6, the modelling of the relevant processes and the selection of a data sample enhanced in $\ttg$ production are described.
In chapter~7, the analysis strategy is outlined, which is then detailed in chapters~8~--~11.
The derivation of the systematic uncertainties is described in chapter~12 and the final result is presented in chapter~13.
In chapter~14, the results are summarised and an outlook for further studies is given.

This analysis was first presented at the HCP conference 2012 in Paris~\cite{ttgATLAS}.
The results presented in this thesis include improved re-evaluations of several background contributions and of the systematic uncertainties, which
resulted in a lower measured $\ttg$ cross section and a lower statistical significance of the $\ttg$ signal.
\newline

Natural units are used throughout this thesis ($\hbar = c = 1$), and hence masses, energies and momenta are expressed in units of $[\mathrm{GeV}]$.
To avoid confusion, length and time are given in usual SI units ($[\mathrm{m}]$ and $[\mathrm{s}]$), because they do not refer to subnuclear but to macroscopic
detector quantities.

In many histograms, the last (first) bin contains also the sum of all entries above (below) the range of the histogram (\textit{overflow} or \textit{underflow} bin,
respectively).
This is noted for each histogram in the description.

\chapter{\boldmath$\ttg$ production in the Standard Model and beyond\unboldmath}
\label{sec:theory}

In the Standard Model of particle physics (SM), the top quark is the weak isospin partner of the bottom quark.
It is the most massive elementary particle known today.
The top quark is special in the SM, because its Yukawa coupling of the order of~1 results in a large mass, which
is much larger than the masses of all other particles.

The production of top quark pair ($\ttbar$) events with an additional photon in the final state ($\ttg$ events) is predicted by the SM,
and the amount of photons radiated from top quarks is given by the electromagnetic coupling of the top quark.
Discrepancies in $\ttg$ production with respect to the SM prediction would indicate an anomalous structure of the $t\gamma$-vertex.

In Sec.~\ref{sec:standardmodel}, a brief summary of the SM is given.
More detailed descriptions of the SM can be found in various text books, for example in Ref.~\cite{donoghue, griffiths}.
The properties of the top quark are discussed in more detail in Sec.~\ref{sec:topquark}, together with an overview of experimental results.
In Sec.~\ref{sec:ttgproduction}, the production of $\ttg$ events in the SM and beyond is summarised.

\section{Brief summary of the Standard Model}
\label{sec:standardmodel}

The SM is a theory of the interactions between elementary particles.
It is based on a quantum field theory in which interactions are introduced by local gauge symmetries.
The SM is very successful in describing a large variety of phenomena in particle physics.

\begin{figure}[h!]
\begin{center}
\includegraphics[width=0.4\textwidth]{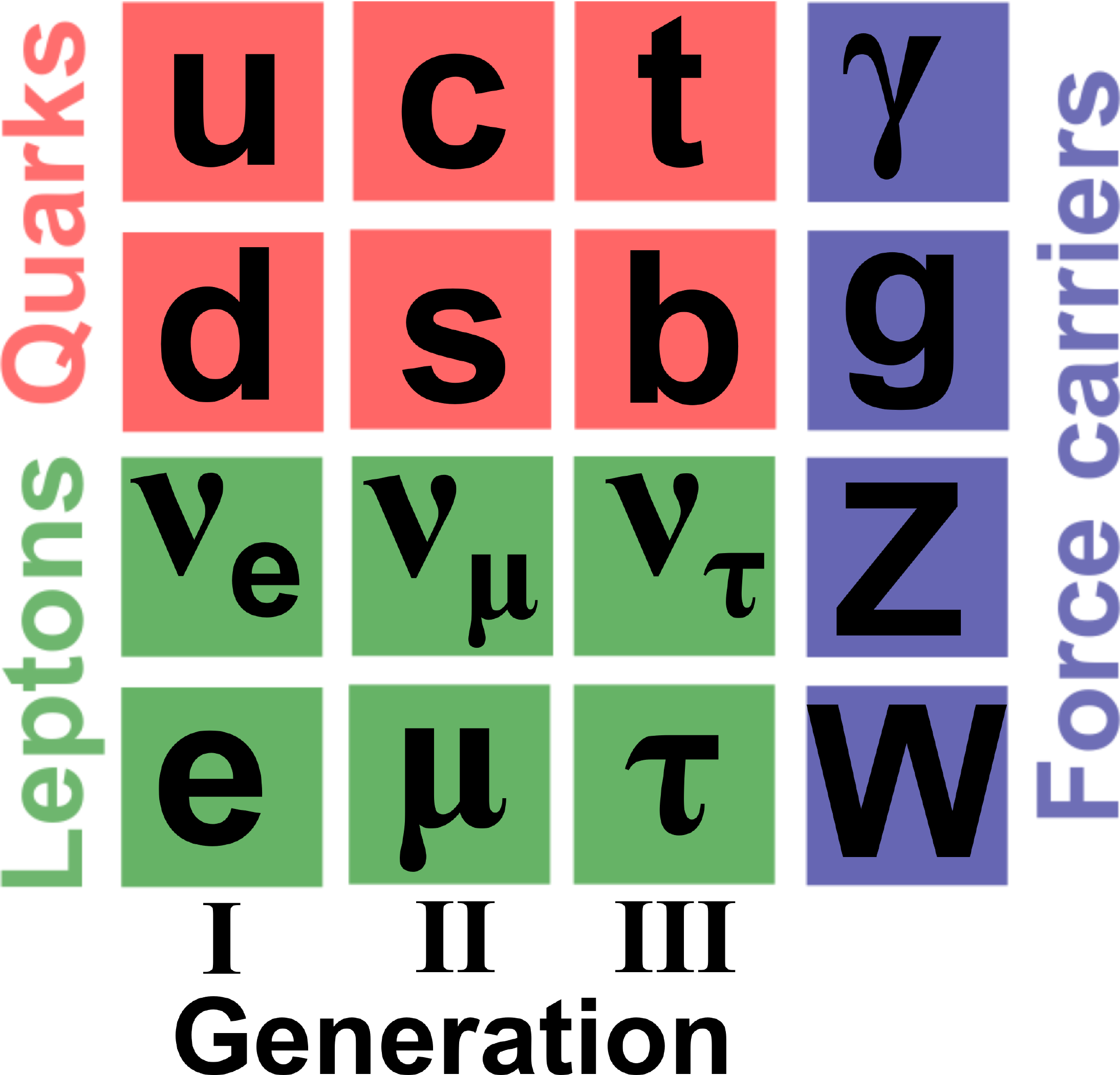}
\caption[Known elementary particles in the Standard Model]{
  Known elementary particles in the Standard Model:
  there are three generations of quarks and leptons.
  The interactions are mediated by four types of gauge bosons.
  The yet undiscovered Higgs boson is not depicted.}
\label{fig:standardmodel}
\end{center}
\end{figure}

Fig.~\ref{fig:standardmodel} shows a representation of the known elementary particles in the SM.
Several properties of the particles are shown in an overview in Tab.~\ref{tab:SMfermion} and~\ref{tab:SMboson}.
The fermions -- leptons and quarks~-- are spin--$\frac{1}{2}$ particles.
The interactions between fermions are described by the exchange of spin--$1$ gauge bosons:
the electromagnetic force is carried by photons ($\gamma$), the weak force by $W^\pm$ and $Z$ bosons and the strong force by gluons ($g$).
While photons and gluons are thought to be massless, $W^\pm$ and $Z$ bosons are massive.
Only particles which are electrically charged interact via the electromagnetic force.
Weak charge is carried by all fermions, which therefore interact via the weak force.
Colour charge, however, is only carried by quarks and gluons, which hence interact via the strong force.
The gravitational force is not described by the SM.

\begin{table}[h!]
  \center
  \begin{tabular}{|l|r@{}l c r@{}l l|r@{}l|c|c|}
    \hline
    Fermion & \multicolumn{6}{c|}{Mass} & \multicolumn{2}{c|}{Electric} & $I_3$ & Colour \\
    & \multicolumn{6}{c|}{} & \multicolumn{2}{c|}{charge} & & charge \\
    \hline
    up quark      & 0 & .0015 & -- & 0 & .0033 & GeV & $+\frac{2}{3}$ & $e$ & $+\frac{1}{2}$ & yes \\
    down quark    & 0 & .0035 & -- & 0 & .0060 & GeV & $-\frac{1}{3}$ & $e$ & $-\frac{1}{2}$ & yes \\
    charm quark   & 1 & .27& $^+_-$ & $^{\emptyplus 0}_{\emptyminus 0}$ & $^{.07\emptyplus}_{.11\emptyminus}$ & GeV & $+\frac{2}{3}$ & $e$ & $+\frac{1}{2}$ & yes \\
    strange quark & 0 & .104 & $^+_-$ & $^{\emptyplus 0}_{\emptyminus 0}$ & $^{.026 \emptyplus}_{.034\emptyminus}$ & GeV & $-\frac{1}{3}$ & $e$ & $-\frac{1}{2}$ & yes \\
    top quark     & 173 & .2 & $\pm$ & 0 & .9 & GeV & $+\frac{2}{3}$ & $e$ & $+\frac{1}{2}$ & yes \\
    bottom quark  & 4 & .20 & $^+_-$ & $^{\emptyplus 0}_{\emptyminus 0}$ & $^{.17 \emptyplus}_{.07\emptyminus}$ & GeV & $-\frac{1}{3}$ & $e$ & $-\frac{1}{2}$ & yes \\
    \hline
    electron neutrino    &  & & $<$ & 0 & .000002  & MeV & 0 & & $+\frac{1}{2}$ & no \\
    electron      & 0 & .510998910 & $\pm$ & 0 & .000000013 & MeV & $-$ & $e$ & $-\frac{1}{2}$ & no \\
%    electron      & \multicolumn{5}{r}{\textcolor{white}{$\pm$} 0.510998910} & & $-$ & $e$ & $-\frac{1}{2}$ & no \\
%                  & \multicolumn{5}{r}{$\pm$ 0.000000013} & MeV & & & & \\
    muon neutrino &  & & $<$ & 0 & .19 & MeV & 0 & & $+\frac{1}{2}$ & no \\
    muon          & 105 & .658367 & $\pm$ & 0 & .000004 & MeV & $-$ & $e$ & $-\frac{1}{2}$ & no \\
%    muon          & \multicolumn{5}{r}{\textcolor{white}{$\pm$} 105.658367} & & $-$ & $e$ & $-\frac{1}{2}$ & no \\
%                  & \multicolumn{5}{r}{$\pm$ 0.000004} & MeV & & & & \\
    tau neutrino  & & & $<$ & 18 & .2 & MeV & 0 & & $+\frac{1}{2}$ & no \\
    tau     & 1776 & .84 & $\pm$ & 0 & .17 & MeV & $-$ & $e$ & $-\frac{1}{2}$ & no \\
    \hline
  \end{tabular}
  \caption[Overview of fermion properties]{
    Overview of the masses~\cite{pdg,topmassTevatron}, electric charges, third components of the weak isospin, $I_3$, and colour charges of quarks (upper part) and leptons (lower part).
  }
  \label{tab:SMfermion}
\end{table}

\begin{table}[h!]
  \center
  \begin{tabular}{|l|r@{}l c r@{}l l|r@{}l|c|c|}
    \hline
    Gauge boson & \multicolumn{6}{c|}{Mass} \\
    \hline
    photon & & & $<$ & 1 & $\cdot 10^{-18}$ & eV \\
    gluon & & & & 0 & & \\
    $W^\pm$ boson & 80 &.385 & $\pm$ & 0 & .015 & GeV \\
    $Z$ boson & 91 &.1876 & $\pm$ & 0 & .0021 & GeV \\
    \hline
  \end{tabular}
  \caption[Overview of boson masses]{
    Overview of the masses~\cite{pdg,wmass} of the gauge bosons.
  }
  \label{tab:SMboson}
\end{table}

Leptons and quarks exist in three generations, where the masses of the particles increase from generation to generation.
Additionally, for every fermion, there is an antiparticle with the same properties like the particle, but with opposite values of the additive quantum
numbers, such as electric charge and the third component of the weak isospin, $I_3$.

Each quark generation consists of an up-type quark with $I_3 = +\frac{1}{2}$ and its down-type quark partner with $I_3 = -\frac{1}{2}$.
Due to the parity violating nature of the weak force, only left-handed quarks form doublets of up- and down-type quarks, while right-handed quarks
form singlets.
The quark doublets read:
up~($u$) and down~($d$) quark, charm~($c$) and strange~($s$) quark, and top~($t$) and bottom~($b$) quark.
Up-type quarks have an electric charge\footnote{$e$ is the absolute value of the charge of the electron.} of $+\frac{2}{3}e$,
the charge of the down-type quarks is $-\frac{1}{3}e$.

The lepton doublets consist of a lepton with electric charge $-e$ (electron $e$, muon $\mu$, tau $\tau$) and the corresponding neutrino
($\nu_e$, $\nu_\mu$, $\nu_\tau$), which is electrically neutral.
The charged leptons form right-handed singlets.
Since neutrinos are assumed to be massless in the SM, no right-handed neutrino singlets are foreseen.
In neutrino oscillation experiments, however, it was shown that neutrinos have non-vanishing masses~\cite{pdg}.
Although the neutrino masses have not yet been measured, the differences in the squares of their masses were measured in the oscillation experiments.
However, the neutrino masses must be very small compared to the scales present in high energy physics experiments and can hence be ignored in this
context.

Mathematically, the SM is formulated as a renormalisable, Lorentz invariant perturbative quantum field theory.
Interactions are introduced by local gauge symmetries.
The structure of the gauge groups is $\rm{SU}_C(3) \times \rm{SU}_L(2) \times \rm{U}_Y(1)$,
where $\rm{SU}_C(3)$ is the gauge group for Quantum Chromodynamics~(QCD) \cite{politzer1, politzer2, gross}, which describes the strong interaction, and
$\rm{SU}_L(2) \times \rm{U}_Y(1)$ is the gauge group for the unified electromagnetic and weak interactions~\cite{glashow, weinberg, salam}.
At the electroweak scale of \mbox{$246 \GeV$}, the symmetry between electromagnetic and weak interactions is spontaneously broken via the Higgs
mechanism~\cite{higgs, englert, kibble}.
This mechanism describes the generation of particle masses in the SM, but also requires the yet undiscovered Higgs
boson~\cite{atlashiggs,cmshiggs} to exist.

%The Lagrangian of the SM, $\mathcal{L}_{\rm SM}$, consists of a sector for QCD, an electroweak sector and a Higgs
%sector:
%\begin{equation}
%\mathcal{L}_{\rm SM} = \mathcal{L}_{\rm QCD} + \mathcal{L}_{\rm EW} + \mathcal{L}_{\rm H} \, .
%\end{equation}
After breaking of the electroweak symmetry, the Lagrangian of the electromagnetic interaction only is described by Quantum Electrodynamics (QED):
\begin{equation}
%\mathcal{L}_{\rm QED} = \bar{\psi} \left( i \gamma^\mu D_\mu - m \right) \psi - \frac{1}{4} F_{\mu \nu} F^{\mu \nu}
\mathcal{L}_{\rm QED} = \bar{\psi} \left( i \gamma^\mu \partial_\mu - m \right) \psi - \frac{1}{4} F_{\mu \nu} F^{\mu \nu} - q \bar{\psi} \gamma^\mu A_\mu \psi
\label{eq:tqvertex}
\end{equation}
The first two terms in Eq.~(\ref{eq:tqvertex}) are the kinetic terms for the fermion field $\psi$ and the photon field $A_\mu$, respectively.
The term $\left( - q \bar{\psi} \gamma^\mu A_\mu \psi \right)$ describes the interaction between fermions and photons, which is determined
by the electric charge $q$ of the fermion.

With $\ttg$ events, the interaction between top quarks and photons can be studied.
Hence, a measurement of $\ttg$ events can be interpreted as a measurement of the charge of the top quark in the SM, or as a measurement of
the structure of the interaction term.
This is discussed in more detail in Sec.~\ref{sec:ttgBSM}.

Although current data are in agreement with the SM, it is widely believed to be just an effective theory at low energies, because several
questions remain unanswered in the SM.
Examples are the hierarchy problem involving the fine-tuning of large corrections to the mass of the Higgs boson,
missing explanations for dark matter and dark energy, the large number of free parameters in the SM, the unification
of the electroweak and the strong interactions, and the inclusion of gravity in the model.

In case the SM is just an effective theory, it would need to be embedded in a more general theory and measurements would differ from SM predictions
starting at a certain energy scale.
Several models, such as supersymmetry~\cite{susy}, models with extra dimensions~\cite{largeExtraDim,warpedExtraDim} or technicolor~\cite{TC1,TC2,TC3,TC4}
provide solutions to some of the issues of the SM and predict new phenomena to appear at a scale of the order of \mbox{$1 \TeV$}.

\section{The top quark in the Standard Model}
\label{sec:topquark}

In this section, only some aspects of top quark physics can be highlighted.
Detailed information on top quark physics at hadron colliders can be found in Ref.~\cite{quadt}.
The current status of top quark physics is summarised in Ref.~\cite{pdg}.

\subsection[Top quark production in $pp$ collisions]{Top quark production in \boldmath$pp$ collisions\unboldmath}
\label{sec:topproduction}

\begin{figure}[h]
\begin{center}
\includegraphics[width=0.32\textwidth]{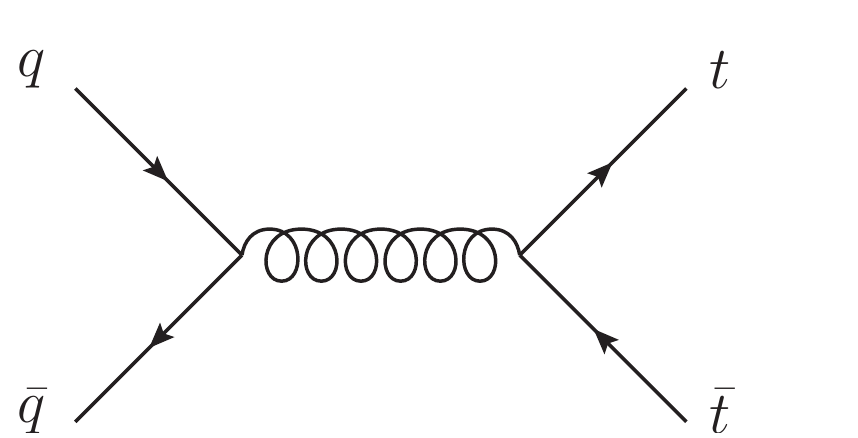} \\
\includegraphics[width=0.32\textwidth]{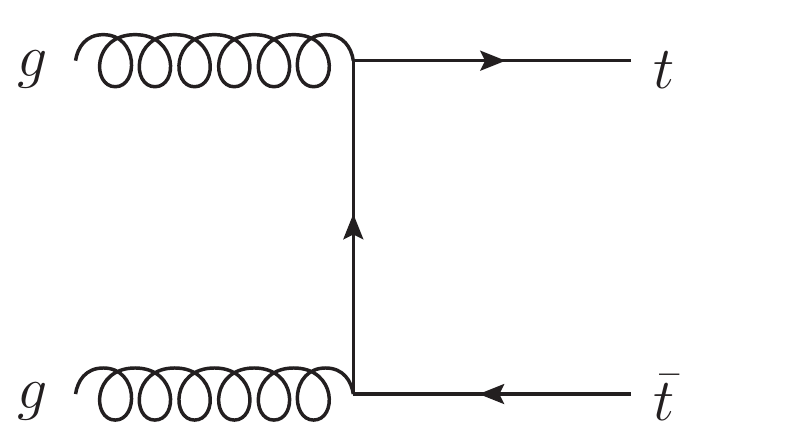}
\includegraphics[width=0.32\textwidth]{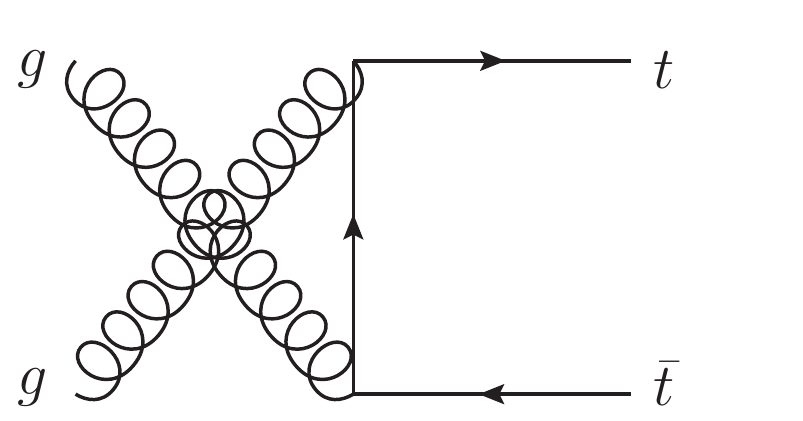}
\includegraphics[width=0.32\textwidth]{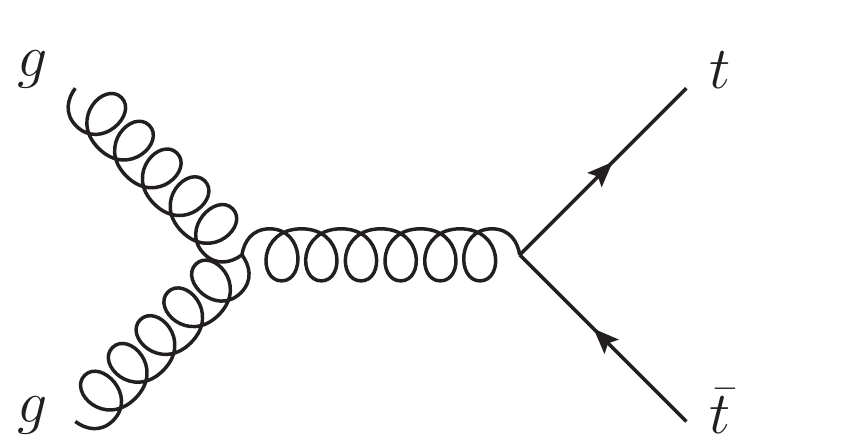}
\caption[Diagrams for $\ttbar$ production via the strong interaction]{
  Diagrams for $\ttbar$ production via the strong interaction in leading order:
  the upper plot shows the quark-antiquark annihilation, the lower plots show the different
  contributing diagrams from gluon-gluon fusion.}
\label{fig:topproduction}
\end{center}
\end{figure}

In proton-proton ($pp$) collisions, top quarks are dominantly produced in pairs via the strong interaction, but also the production of single
top quarks via the weak interaction is possible.
The contributing diagrams for $\ttbar$ production in leading order (LO) are shown in Fig.~\ref{fig:topproduction}.
They can be categorised into $s$-channel quark-antiquark annihilation (upper diagram) and gluon-gluon fusion in the $t$-, $u$- and $s$-channels (lower
diagrams).

In contrast to $e^+e^-$ colliders, at high-energy hadron colliders the interacting particles are partons, which are confined in the colliding hadrons.
Hence, the parton-parton cross section $\hat{\sigma}^{i j \to t\bar{t}}$ for partons $i$ and $j$ needs to be convolved with the parton distribution
functions (PDFs)~\cite{GribovLipatov,AltarelliParisi,Dokshitzer}
of the hadron, $f_i(x_i, \mu_F^2)$, which describe the probability to find a parton $i$ within the hadron with a fraction $x_i$ of its momentum.
The PDFs are evaluated at a factorisation scale $\mu_F$, which separates perturbative QCD from non-perturbative effects.
Hence, in $pp$ collisions, the $\ttbar$ production cross section reads~\cite{factorization,quadt}:
\begin{equation}
\sigma^{pp \to t\bar{t}}(\sqrt{s}, m_t) = \sum_{i,j = q,\bar{q},g} \int \mathrm{d}x_i \mathrm{d}x_j f_i(x_i, \mu_F^2) f_j(x_j, \mu_F^2) \cdot
\hat{\sigma}^{i j \to t\bar{t}} (m_t, \sqrt{\hat{s}}, x_i, x_j, \alpha_s(\mu_R^2), \mu_R^2) \; \mathrm{,}
\label{eq:factorisation}
\end{equation}
where $m_t$ is the top quark mass, $\sqrt{s}$ is the centre-of-mass energy of the $pp$ collision, $\sqrt{\hat{s}}$ is the centre-of-mass energy of
the parton-parton system and $\mu_R$ is the renormalisation scale, which is introduced to allow for finite-order calculations in QCD perturbation theory.
A typical choice for $\mu_R$ as well as for $\mu_F$ is the energy scale for the process under study.
In processes with top quarks, the top quark mass is often chosen as this energy scale.
While $\hat{\sigma}^{i j \to t\bar{t}}$ can be calculated in perturbative QCD, the PDFs can not and need to be extracted
from data measured in $e p$, $p\bar{p}$ and $pp$ collisions and from other sources~\cite{CTEQ, MSTW, NNPDF, HERAPDF}.

An approximate next-to-next-to-leading order (NNLO) calculation for $\sigma^{pp \to t\bar{t}}$ at \mbox{$\sqrt{s} = 7 \TeV$} \cite{topxsec_th1,topxsec_th2}
yielded a prediction of \mbox{$165 \; ^{+11}_{-16} \pb$} using HATHOR~\cite{hathor} and CTEQ PDFs~\cite{CTEQ} at a top mass of \mbox{$172.5 \GeV$}, where
the uncertainty is due to the uncertainties on the PDFs, and on the factorisation and renormalisation scales.

\subsection{Top quark decay}
\label{sec:topdecay}

In the SM, about 99.8\% of the top quarks decay into a $W^+$ boson and a $b$-quark with a lifetime of about \mbox{$0.5 \cdot 10^{-24} \sec$}~\cite{pdg}.
Decays to a $W^+$ boson and a $d$- or $s$-quark are strongly suppressed by the CKM matrix elements
$V_{td} = 0.00359 \pm 0.00016$ and $V_{ts} = 0.0415 \, ^{+0.0010}_{-0.0011}$~\cite{pdg}.
Since the lifetime of top quarks is shorter than the typical time for hadronisation, they do not form bound states before they decay.
Hence, the top quark is the only known quark for which the quark properties are accessible, such as for example its electromagnetic coupling
(cf.~Sec.~\ref{sec:topproperties} and~\ref{sec:ttgproduction}).

\begin{figure}[h]
\begin{center}
\includegraphics[width=0.48\textwidth]{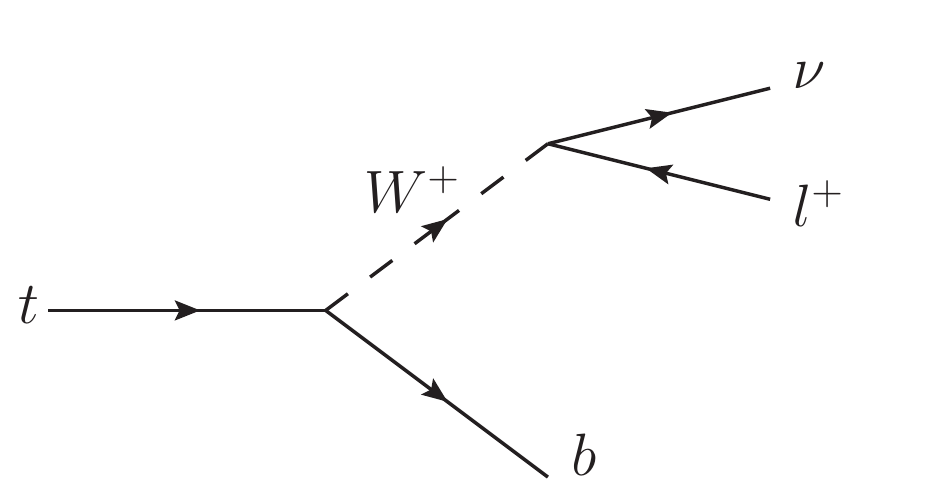}
\includegraphics[width=0.48\textwidth]{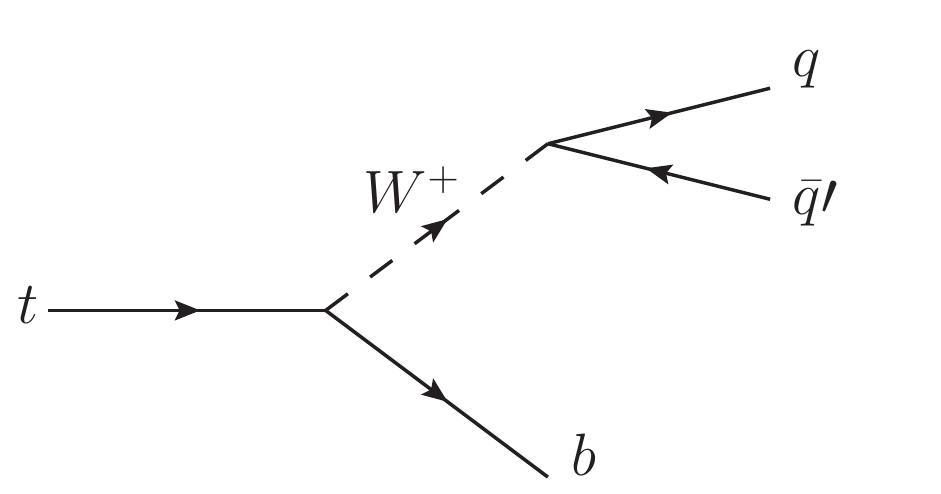}
\caption[Top quark decay]{
  Top quark decay: almost all top quarks decay into a bottom quark and a $W^+$ boson.
  The $W^+$ boson can then either decay into a charged lepton and the corresponding neutrino (left),
  or into an up-type quark and a down-type antiquark (right).}
\label{fig:topdecay}
\end{center}
\end{figure}

The $W^+$ boson from the decay of the top quark can either decay into a pair of up-type quark ($u$ or $c$) and down-type antiquark
($\bar{d}$, $\bar{s}$ or $\bar{b}$), or into a charged lepton and the corresponding neutrino. 
This is illustrated in Fig.~\ref{fig:topdecay}.
The dominant decay of the antitop quark analogously reads \mbox{$\bar{t} \rightarrow W^- \bar{b}$}.

\subsection[Experimental signatures of $\ttbar$ production]{Experimental signatures of \boldmath$\ttbar$ production\unboldmath}
\label{sec:topsignatures}

\begin{figure}[h]
\begin{center}
\includegraphics[width=0.6\textwidth]{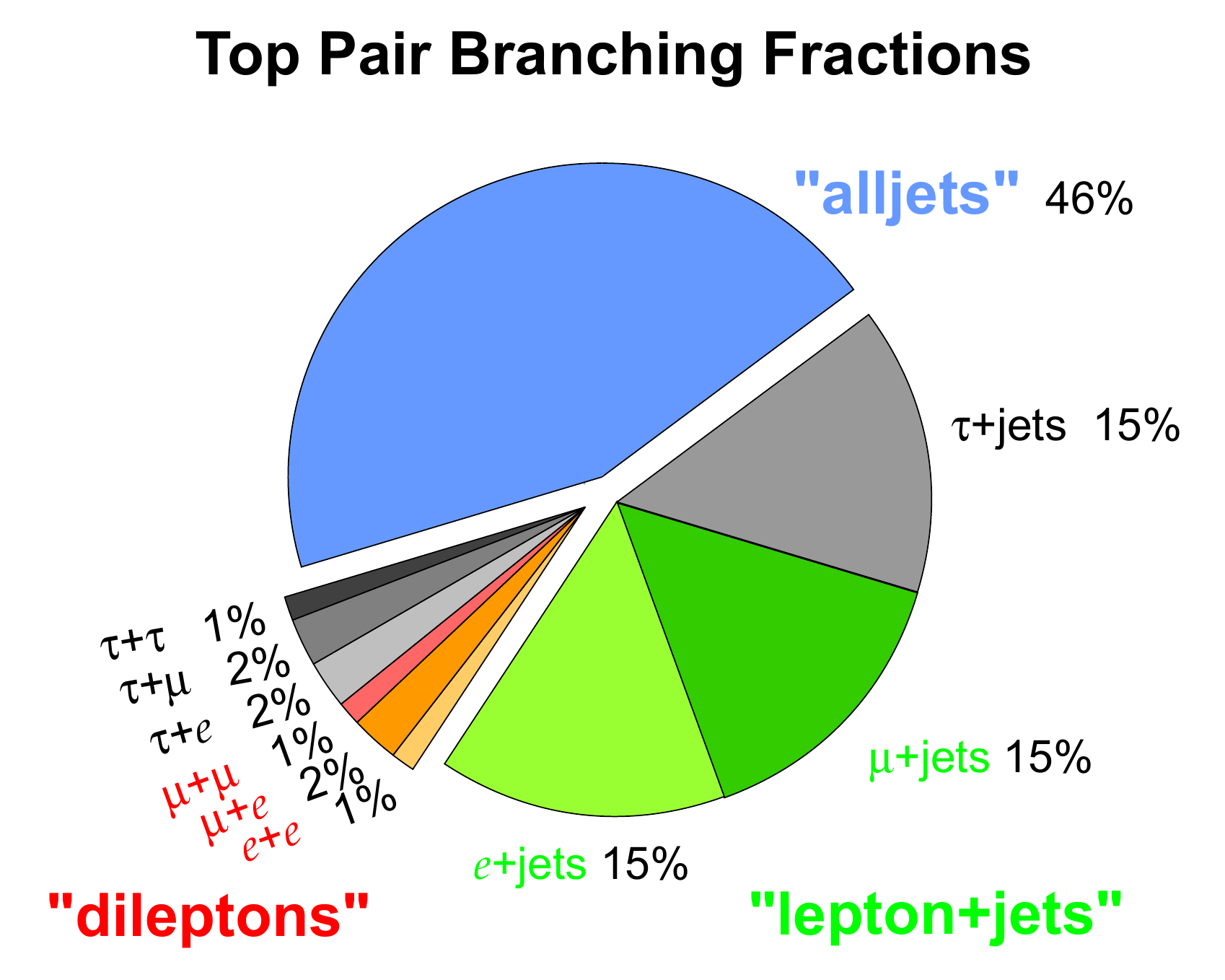}
\caption[Experimental signatures of $\ttbar$ decays]{
  Experimental signatures of top quark pair decays and their branching ratios~\cite{piechart}:
  in about 30\% of the cases, top quark pairs decay into either the single electron or the single muon channel.}
\label{fig:piechart}
\end{center}
\end{figure}

Since $W$ bosons can decay into quarks (hadronic decay) or leptons (leptonic decay), the decay
of top quark pairs can be categorised into three channels, depending on the number of leptons and quarks in the final state.
Since quarks hadronise and form jets, bundles of particles, the categories are labelled accordingly:
the all-hadronic (or alljets) channel, where both $W$ bosons decay hadronically,
the dilepton channel, where both $W$ bosons decay leptonically, and the single lepton (or lepton+jets) channel with one $W$ boson decaying leptonically
and the other one decaying hadronically.
The branching ratios of the respective channels are illustrated in Fig.~\ref{fig:piechart}.

While the all-hadronic channel has the largest branching ratio, at hadron colliders it suffers from large background contributions
from multijet production.
From an experimental point of view, electrons and muons provide a clear signature, while the identification of $\tau$-leptons is more involved, because
of the different leptonic and hadronic decay modes of the $\tau$-lepton.

The single lepton channel is sometimes called the \textit{golden channel}, because of its clear signature and relatively high branching ratio
(about 30\%).
It is characterised by a high-energetic electron or muon, four high-energetic jets, out of which two are originating from $b$-quarks,
and a large imbalance of the momentum in the transverse plane due to the neutrino, which escapes detection.
Depending on the lepton type, the final state is called \textit{single electron} or \textit{single muon channel}.
Since electrons and muons from $\tau$-decays are experimentally indistinguishable from electrons and muons from the direct decay of the $W$ boson,
in the analysis presented in this thesis,
\mbox{$\tau \rightarrow e \bar{\nu}_e \nu_\tau$} and \mbox{$\tau \rightarrow \mu \bar{\nu}_\mu \nu_\tau$} decays were included in the electron and the muon
channel, respectively.

\subsection[Background processes for single lepton decays of $\ttbar$ events]{Background processes for single lepton decays of \boldmath$\ttbar$ events\unboldmath}
\label{sec:topbackgrounds}

The background processes for the single electron and single muon channels can be divided into reducible and irreducible background processes.
Irreducible background processes feature the same final state as the signal.
This is the case for the production of leptonically decaying $W$ bosons in association with jets ($W$+jets).
These jets arise from QCD corrections to the $W$ boson production and can originate from gluons and light quarks, but also from $b$-quarks.
Moreover, events with electroweakly produced single top quarks may feature additional jets from QCD corrections and may be misreconstructed as $\ttbar$
events.

Reducible background processes have a different experimental final state, but one or more particles are not correctly reconstructed or are
just outside of the detector acceptance.
Momentum imbalance may always originate from miscalibrations of jet energies.
Hence, $Z$+jets events with $Z \rightarrow ee$ or $Z \rightarrow \mu\mu$, where one lepton is not reconstructed correctly, may be misreconstructed as
$\ttbar$ events.
The same holds true for $WW$, $WZ$ and $ZZ$ production with additional jets from initial or final state radiation (ISR/FSR).
However, $WW \to l\nu q\bar{q}^\prime$ and $WZ \to l\nu q\bar{q}$ decays are part of the irreducible background, strictly speaking.
Finally, jets in multijet events may give rise to signatures similar to electrons or muons from $W$ boson decays:
hadrons from jet fragmentation may be misidentified as electrons, and electrons and muons within jets coming from $B$-meson decays may be misidentified
as isolated electrons or muons from a $W$ decay, respectively.
Due to the large cross section for multijet events at hadron colliders, such events are expected to have a non-negligible background contribution,
even if the misidentification rates are small.

\subsection{Top quark properties and experimental results}
\label{sec:topproperties}

In addition to the production cross section for $\ttbar$ and single top events in
$pp$~\cite{topXsecATLAS, topXsecCMS, singletopATLAS_t, singletopATLAS_s, singletopATLAS_Wt, singletopCMS_t, singletopCMS_Wt}
and $p\bar{p}$~\cite{topXsecD0dilepton, topXsecCDFcombination, singletopD0, singletopCDF} collisions,
various properties of the top quark have been studied.
In particular, the mass of the top quark has been measured with high precision to \mbox{$173.2 \pm 0.9 \GeV$} at the Tevatron~\cite{topmassTevatron}.
Measurements at the LHC are in agreement with this result~\cite{topmassATLAS, topmassCMS}.

The knowledge about the couplings of the top quark is still limited, although the SM provides clear predictions for them.
The structure of the $Wtb$-vertex has been studied in the parity violating weak decay of the top quark~\cite{whelicityTevatron, whelicityATLAS, whelicityCMS}
and the strength of the coupling has been tested in the production of single top quarks~\cite{singletopCMS_t,singletopD0,singletopCDF}.
To date, results are consistent with SM expectations.

%Its large mass makes the top quark a special particle in the SM.
%Hence, in many models beyond the SM, alternative processes for the production or decay of top quarks are predicted.
%In the following some related measurements in production and decay of $\ttbar$ events are highlighted.
%
Since the top quark decays before it can form bound states, the spin of the two top quarks from $\ttbar$ production is transferred to the decay products.
A certain correlation between the top quark spins is expected in the SM, which may be altered by additional production processes.
To date, measurements of the spin correlation at \mbox{$\sqrt{s} = 7 \TeV$} are consistent with SM
expectations~\cite{spincorrD0, spincorrCDF, spincorrATLAS}.

Also the measurement of the charge asymmetry in $\ttbar$ production is a test of the production process.
The asymmetry originates from interference effects in NLO between initial and final state gluon radiation as well as from interferences between
Born and box diagrams.
It is small in the SM, but significantly enhanced in several alternative models~\cite{chargeasym}.
Measurements performed by the CDF and D\O\ collaborations show discrepancies of up to more than $3 \sigma$ from SM expectations~\cite{asymCDF,asymCDFupdate,asymD0}.
Since in $pp$ collisions the asymmetry is less pronounced than in $p\bar{p}$ collisions, ATLAS and CMS have not yet gained the sensitivity of the
Tevatron experiments.
To date, measurements of the asymmetry in $pp$ collisions are consistent with SM expectations~\cite{asymATLAS, asymCMS}.

Alternative top quark decays have been searched for and exclusion limits have been set for example on flavour changing neutral current decays, such
as \mbox{$t \rightarrow q Z$}~\cite{tZqD0, tZqCDF, tZqATLAS, tZqCMS} and \mbox{$t \rightarrow q \gamma$}~\cite{tgammaqCDF}.
Also, the branching ratio $t \rightarrow W b$~\cite{RbD0, RbCDF, RbCMS} has been measured.
To date, all results are consistent with the expectations from the SM.

The measurement of the number of jets produced in association with $\ttbar$ events~\cite{topnjetATLAS} is a first measurement towards
tests of the strong coupling of the top quark.

In the SM, the electromagnetic coupling is given by the electric charge of the top quark, which is predicted to be \mbox{$+\frac{2}{3}e$}.
The pair production of \textit{exotic top quarks} with an electric charge of \mbox{$-\frac{4}{3}e$}~\cite{exotictop1, exotictop2} results in the
same final state as SM $\ttbar$ production.
Analyses in which the top quark charge is measured from the charge of the lepton and the associated $b$-quark could rule out the exotic top quark
scenario with up to 99\% confidence level~\cite{topchargeATLAS,topchargeCMS,topchargeD0,topchargeCDF}.

$\ttg$ events directly probe the electromagnetic coupling of the top quark without necessarily assuming the structure of the $t\gamma$-vertex predicted in the SM.
Hence, they provide a test of the SM which is complementary to analyses considering only the exotic top quark scenario.
In $p\bar{p}$ collisions, a first measurement of the $\ttg$ cross section and the ratio of the cross sections for $\ttg$ and $\ttbar$ production
at \mbox{$\sqrt{s} = 1.96  \TeV$} has been performed by the CDF collaboration~\cite{ttgammaCDF}.
The $\ttg$ cross section was measured to \mbox{$0.18 \pm 0.08 \pb$}, and the signal significance was estimated to 3.0 standard deviations.
The ratio of the $\ttg$ and $\ttbar$ cross sections has been measured to \mbox{$0.024 \pm 0.009$}, consistent with SM expectations.
$\ttg$ production is discussed in more detail in Sec.~\ref{sec:ttgproduction}.

A variety of additional measurements of the properties of the top quark itself, of its decay and of processes with top quarks in the final state
has been performed.
These measurements are not listed in this thesis, but an overview can be found in Ref.~\cite{pdg}.

Due to its large mass and its large Yukawa coupling, the top quark may play a special role in electroweak symmetry breaking
and for the discovery of phenomena beyond the SM.
For instance, new heavy particles could manifest themselves in modifications of the SM predictions for top quark properties, for example in the
production process, which would then be observed in measurements of the charge asymmetry.
Moreover, since top quarks decay before they form bound hadronic states, the top quark is a unique opportunity to study the properties of a bare
quark, such as the couplings or the spin properties.
In particular, top quarks are the only quarks for which the electromagnetic coupling is directly accessible.

\section[$\ttg$ production in $pp$ collisions]{\boldmath$\ttg$ production in $pp$ collisions\unboldmath}
\label{sec:ttgproduction}

\subsection[Production of $\ttg$ events in the Standard Model]{Production of \boldmath$\ttg$ events in the Standard Model\unboldmath}
\label{sec:ttgSM}

As outlined in Sec.~\ref{sec:topproperties}, the production of $\ttg$ events is sensitive to the electromagnetic coupling of the top quark.
In the SM, it is solely given by the electric charge of the top quark --~cf. Eq.~(\ref{eq:tqvertex}).
Hence, within the SM, a measurement of the $\ttg$ cross section can be interpreted as a measurement of the top quark charge.

%The production of $\ttg$ events can be categorised into events with photons radiated in the production of the $\ttbar$ system
%(\textit{radiative production}) and events with photons radiated in the decay of one of the top quarks
%(\textit{radiative decay})~\cite{baur}.
%
In $\ttbar$ events, photons can be radiated from all charged particles, including the top quark, but also from incoming quarks and the charged
decay products of the top quarks.
Artificially, the contributing diagrams can be divided into diagrams from \textit{radiative production} and \textit{radiative decay}, but the
$\ttg$ final state is only well-defined when all interference terms between the different processes are taken into account.

\begin{figure}[p]
\begin{center}
\includegraphics[width=0.35\textwidth]{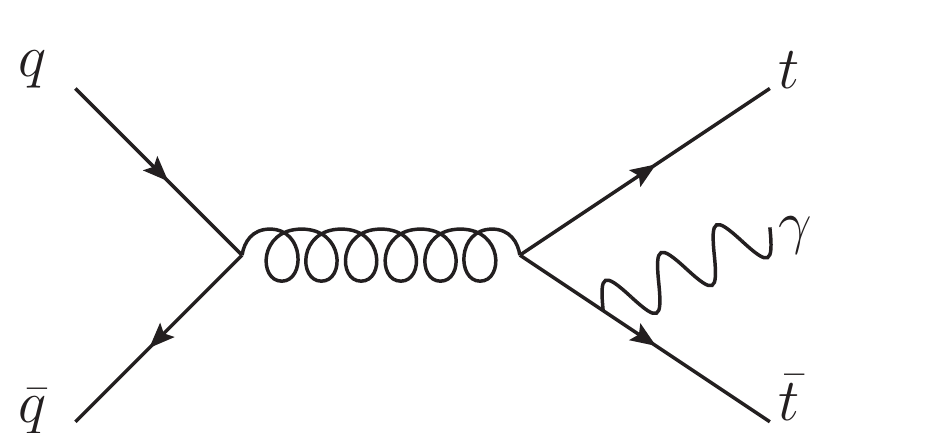}
\includegraphics[width=0.35\textwidth]{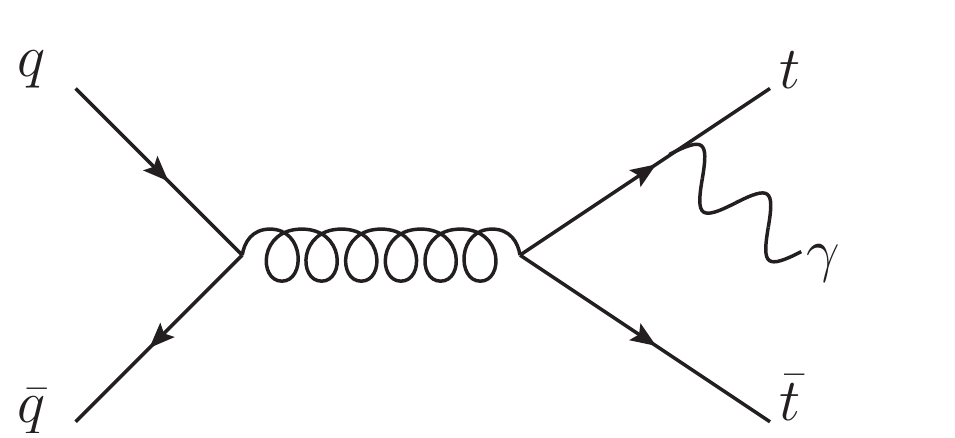} \\
\includegraphics[width=0.35\textwidth]{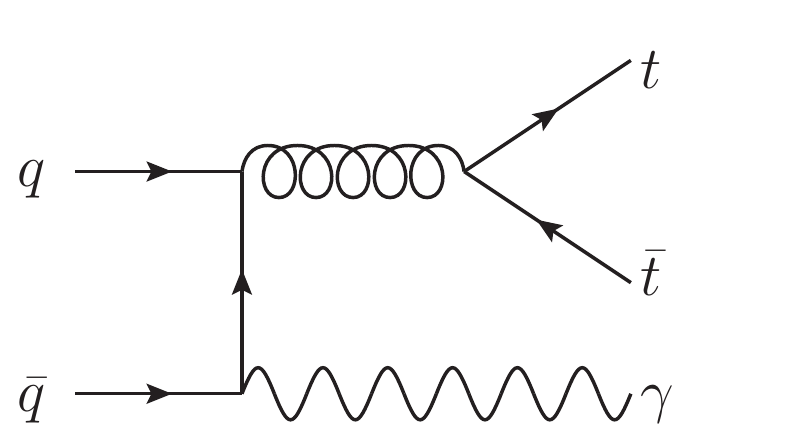}
\includegraphics[width=0.35\textwidth]{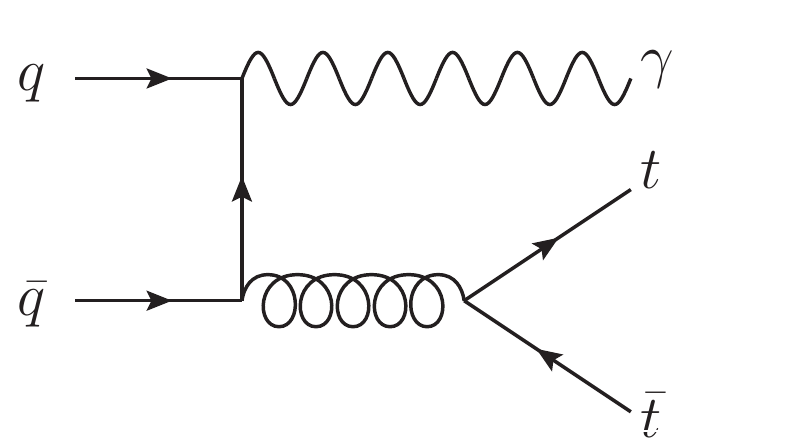}
\caption[LO diagrams for $\ttg$ production in quark-antiquark annihilation]{
  Leading order diagrams for $\ttg$ production in quark-antiquark annihilation.}
\label{fig:ttgproduction_qq}
\vspace{0.025\textwidth}
\includegraphics[width=0.35\textwidth]{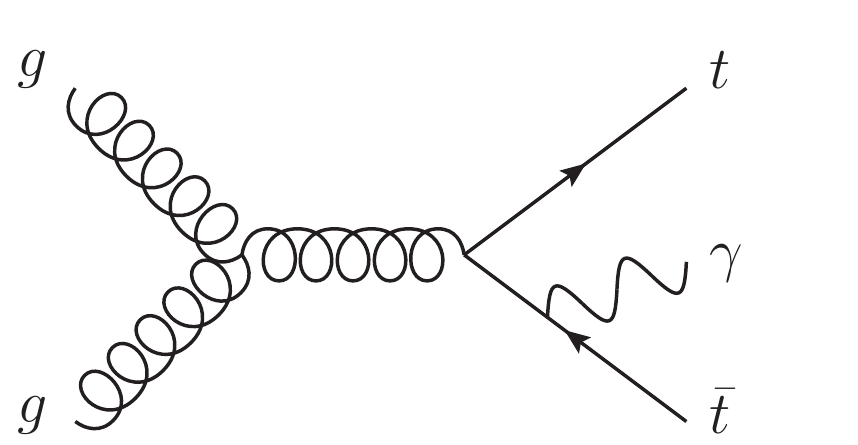}
\includegraphics[width=0.35\textwidth]{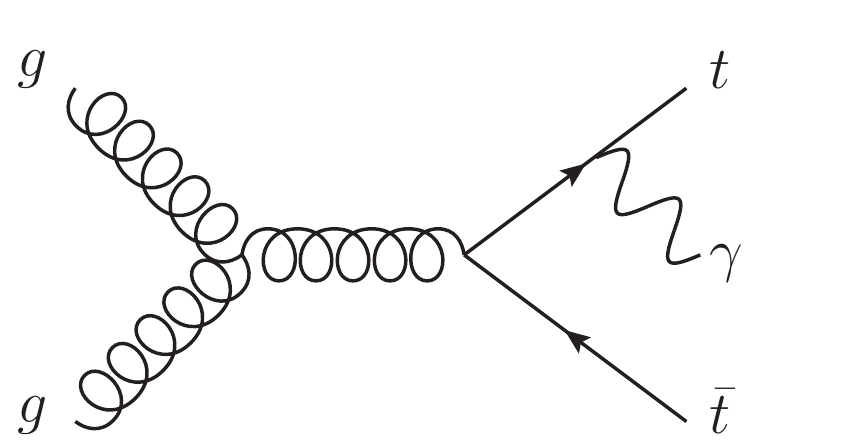} \\
\includegraphics[width=0.35\textwidth]{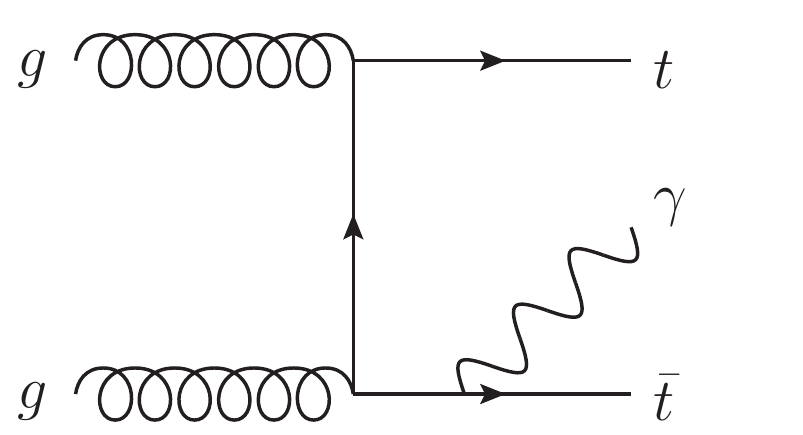}
\includegraphics[width=0.35\textwidth]{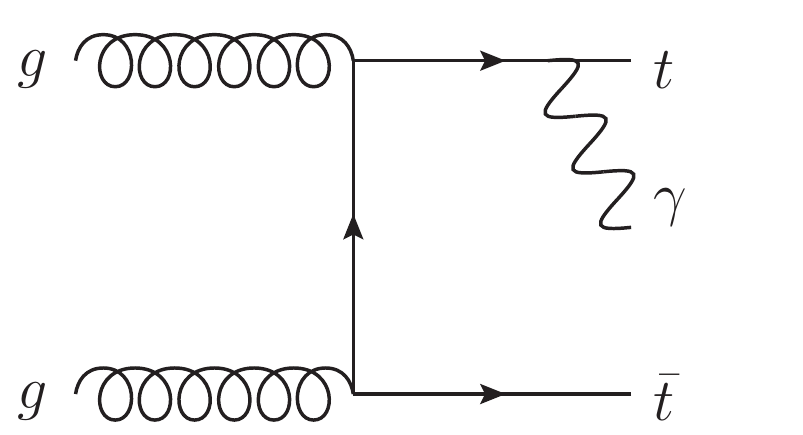} \\
\includegraphics[width=0.35\textwidth]{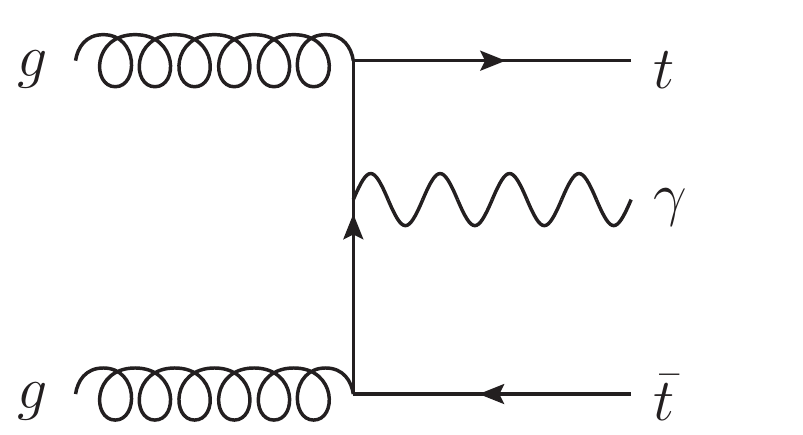}
\includegraphics[width=0.35\textwidth]{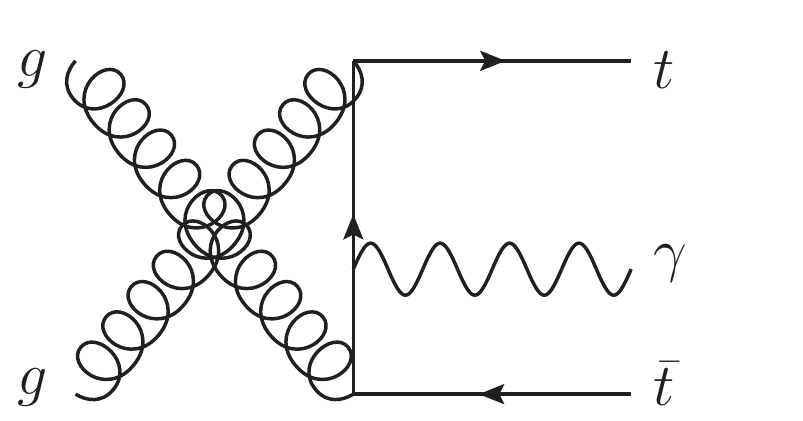} \\
\includegraphics[width=0.35\textwidth]{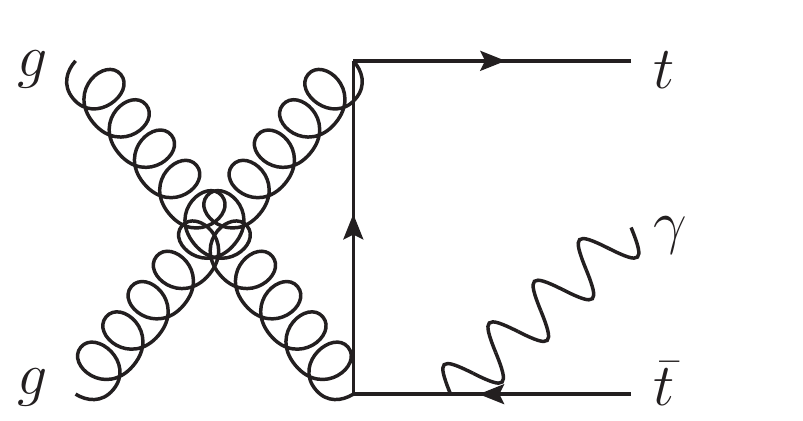}
\includegraphics[width=0.35\textwidth]{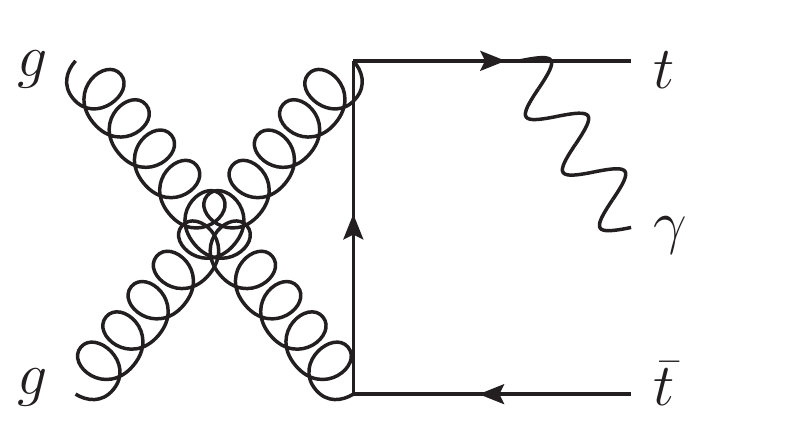}
\caption[LO diagrams for $\ttg$ production in gluon-gluon fusion]{
  Leading order diagrams for $\ttg$ production in gluon-gluon fusion.}
\label{fig:ttgproduction_gg}
\end{center}
\end{figure}

\begin{figure}[h]
\begin{center}
\includegraphics[width=0.32\textwidth]{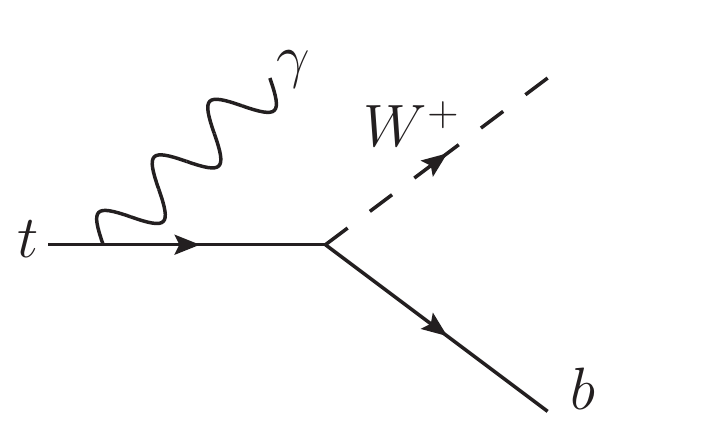}
\includegraphics[width=0.32\textwidth]{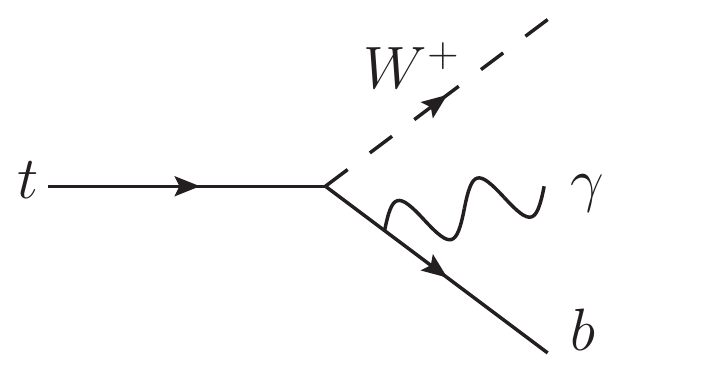}
\includegraphics[width=0.32\textwidth]{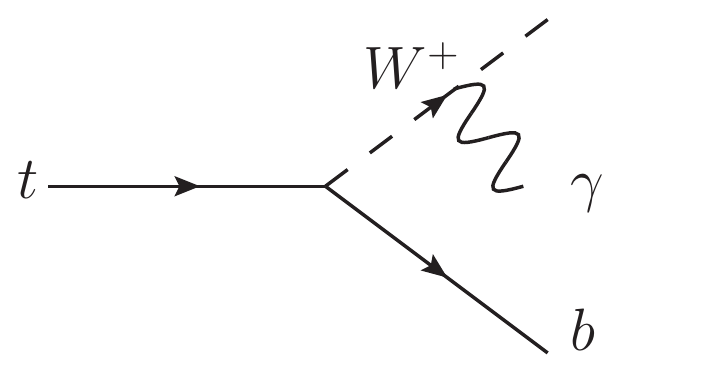}
\caption[Top quark decay with additional photon radiation]{
  Top quark decay with additional photon radiation:
  photons can be radiated from the top quark, the bottom quark or the $W$ boson.}
\label{fig:ttgdecay}
\end{center}
\end{figure}

In LO, radiative $\ttg$ production can occur in quark-antiquark annihilation (Fig.~\ref{fig:ttgproduction_qq})
or gluon-gluon fusion (Fig.~\ref{fig:ttgproduction_gg})~\cite{ttgSimATLAS} -- similarly to $\ttbar$ production (Sec.~\ref{sec:topproduction}).
Photons can be radiated from the incoming quarks or from the top quarks.
%For comparison with the radiation from top quarks in the radiative decay, it is worth pointing out that top quarks in radiative
%production are only on their mass shell after the photon radiation.
The LO processes for the radiative decay are sketched in Fig.~\ref{fig:ttgdecay}.
After a top quark pair is produced, a photon can be radiated either from the decaying top quark, the $W$ boson or the $b$-quark.
Photons can also be radiated from the charged decay products of the $W$ boson, the charged lepton or the quarks, respectively, which is not
illustrated in Fig.~\ref{fig:ttgdecay}.

It has been shown that interferences between the different diagrams are not negligible~\cite{baur} and,
hence, a separation between processes with photons radiated from top quarks and processes with photons radiated from other particles is not well defined.
From Eq.~(\ref{eq:tqvertex}) it could be concluded that the cross section for $\ttg$ production is proportional to the square of the top
quark charge.
Since also diagrams with photons radiated from other particles need to be added to the definition of the \textit{inclusive $\ttg$ cross section},
the dependence on the top quark charge is not just simply quadratic.

A calculation of the inclusive cross section in $pp$ collisions at \mbox{$\sqrt{s} = 14 \TeV$} in next-to-leading order (NLO) QCD is
available~\cite{kfactor}.
In this calculation, infrared divergencies were avoided by a minimum photon transverse momentum of \mbox{$20 \GeV$}.
In order to get rid of collinear divergencies, photons were required to fulfil a Frixione-type parton isolation~\cite{frixioneisolation}
with a width parameter of~0.4.
The resulting cross section in NLO was found to be larger than the calculation in LO and, hence,
the ratio of the two calculations, the so-called NLO \textit{$k$-factor}, differs from unity.

For \mbox{$\sqrt{s} = 7 \TeV$}, this $k$-factor was estimated to $2.6 \pm 0.5$~\cite{kfactorPrivate}.
For this calculation, the cut on the photon transverse momentum was adjusted to \mbox{$8 \GeV$} in order to be applicable to the
simulations used in the measurement presented in this thesis (Sec.~\ref{sec:signalmodelling}).
The effect of different isolation criteria used in the calculation and the simulation was expected to be well covered by the
uncertainty on the $k$-factor, which was estimated by changing the renormalisation and factorisation scales from the nominal value of $2 m_t$
to $m_t$.

The $k$-factor was applied to the $\ttg$ LO calculation obtained with the WHIZARD Monte Carlo generator~\cite{whizard, omega}, which is described
in more detail in Sec.~\ref{sec:signalmodelling}.
For \mbox{$\sqrt{s} = 7 \TeV$} an estimate of \mbox{$\sigma_{t\bar{t}\gamma} \cdot \mathrm{BR} = 2.1 \pm 0.4 \pb$} was obtained, where BR is the branching
ratio into the single lepton and dilepton decay modes (see Sec.~\ref{sec:signalmodelling}).

\subsection[Production of $\ttg$ events beyond the Standard Model]{Production of \boldmath$\ttg$ events beyond the Standard Model\unboldmath}
\label{sec:ttgBSM}

In Sec.~\ref{sec:topproperties}, it has been mentioned that the top quark may be a window to physics beyond the SM.
In particular, it may be of interest for an understanding of electroweak symmetry breaking, because its mass is of the same order as the electroweak scale.
In order to observe deviations from SM predictions, precision tests of the properties of the top quark are necessary, where
tests in the electroweak sector are especially interesting.
In $\ttg$ events, deviations from the SM electromagnetic $t\gamma$-vertex $\Gamma^\mu = - i Q_t e \gamma^\mu$ can be searched for.

The following generalised form of the $t\gamma$-vertex is a natural extension of the SM~\cite{formfactor1, formfactor2}:
\begin{equation*}
\Gamma^\mu \left( q^2 \right) = - i Q_t e \left[ \gamma^\mu \left( F_1^V \left( q^2 \right)+ F_1^A \left( q^2 \right) \gamma_5 \right)
+ i \frac{\sigma^{\mu \nu}}{2 m_t} q_\nu \left( F_2^V \left( q^2 \right) + F_2^A \left( q^2 \right) \gamma_5 \right) \right] \; ,
\end{equation*}
where $q$ is the photon four-momentum, $Q_t$ is the value of the top quark charge in units of $e$, and $m_t$ is the mass of the top quark.
$F_1^V$ and $F_1^A$ are the form factors for vector and axial-vector couplings, respectively.
The form factors $F_2^V$ and $F_2^A$ represent magnetic and electric dipole moments of the top quark.
In the SM, all form factors vanish at tree level except for $F_1^V$, which is equal to unity.
$F_2^V$ and $F_2^A$ receive non-zero contributions only when higher loop corrections are considered.
Deviations from these predictions, for example enhanced dipole moments, would indicate the presence of phenomena beyond the SM.

At hadron colliders, a precise determination of the form factors will be challenging due to the limited precisions in the measurement
of the four-momenta and in the identification of the particles involved.
However, with growing statistics at the LHC, the sensitivity to the electromagnetic form factors will increase.
Previous studies in the $\ttg$ topology~\cite{ttgSimATLAS,baur} have focused on the discrimination of two different scenarios
for the electromagnetic charge of the top quark ($+ \frac{2}{3}e$ and $- \frac{4}{3}e$, cf. Sec.~\ref{sec:topproperties}).
However, in Ref.~\cite{baur} it has been estimated that the charge of the top quark
can be measured with a precision of about 10\% with an integrated luminosity of \mbox{$10 \ifb$} at \mbox{$\sqrt{s} = 14 \TeV$}.

Electron-positron colliders would provide a cleaner environment for studying the electromagnetic coupling of the top quark.
For an integrated luminosity of \mbox{$10 \ifb$} at \mbox{$\sqrt{s} = 500 \GeV$}, a precision of 5~--~10\% on the axial form factors has been
predicted~\cite{topatee}.
Even better limits would be achieved at future photon-photon colliders~\cite{topatgammagamma}.

\chapter{The ATLAS experiment at the LHC}
\label{sec:experiment}

The ATLAS detector is one of the main experiments located at the Large Hadron Collider (LHC) at CERN.
It has been built to cover a large physics program with collisions of protons and heavy ions at unprecedented energies and extremely high rates
and instantaneous luminosities.

In Sec.~\ref{sec:LHC}, the accelerator is introduced.
In Sec.~\ref{sec:ATLAS}, the detector and its different subcomponents as well as the detector readout are described.

\section{The Large Hadron Collider and its experiments}
\label{sec:LHC}

\begin{figure}[h]
\begin{center}
\includegraphics[width=0.9\textwidth]{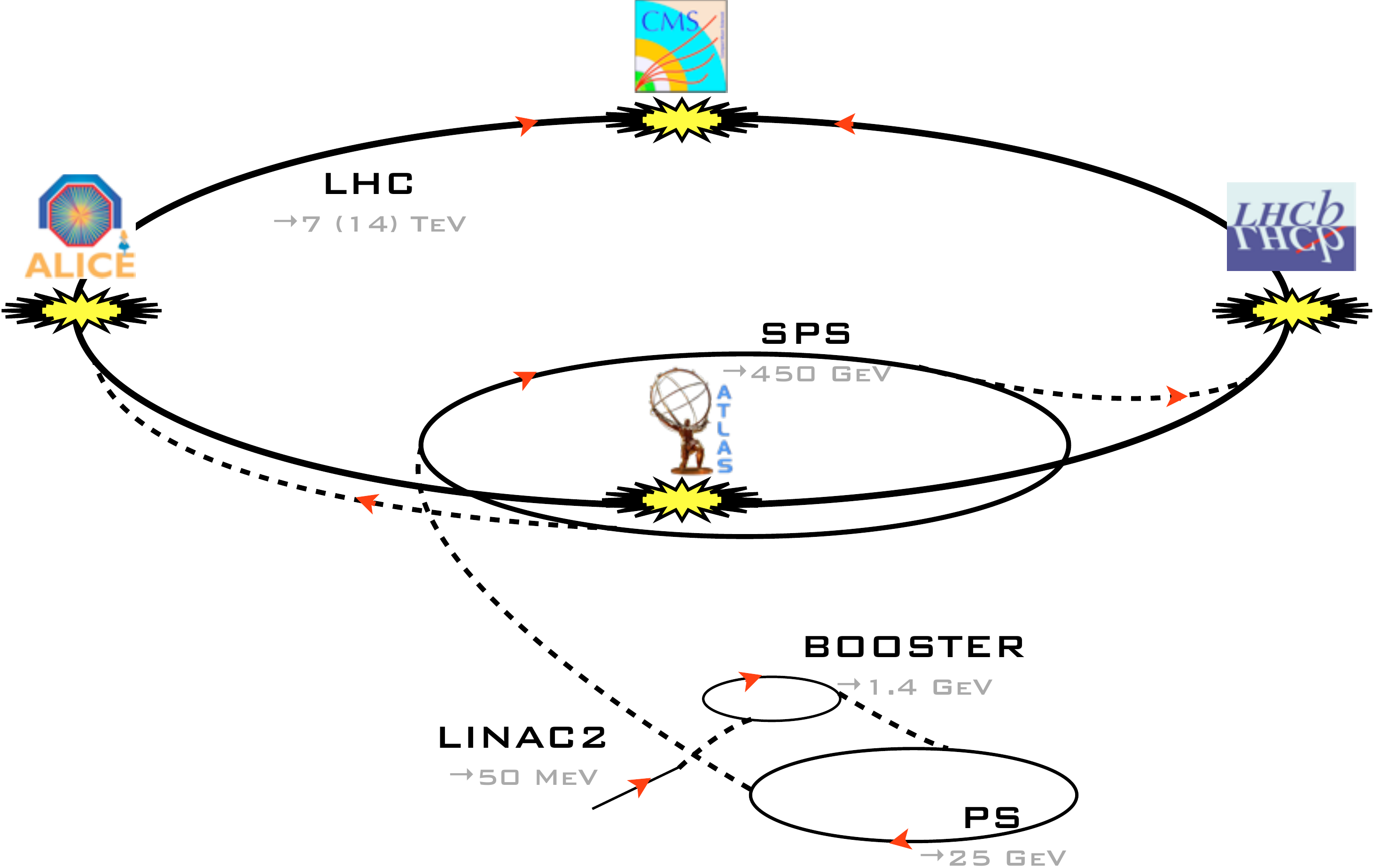}
\caption[Sketch of the Large Hadron Collider]{
  Sketch of the Large Hadron Collider, its preaccelerators and the four main experiments at the interaction points (not to scale)~\cite{anna}.}
\label{fig:lhc}
\end{center}
\end{figure}

The LHC~\cite{LHCreport12, *LHCreport3} is a circular $pp$ collider at CERN near Geneva, Switzerland.
In addition to protons, also heavy ions can be brought to collision.
The accelerator is located in the tunnel of the dismounted Large Electron-Positron Collider, which has a circumference of 27~km.
Fig.~\ref{fig:lhc} shows a sketch of the LHC and its preaccelerators.
Protons from the ionisation of hydrogen atoms are accelerated to \mbox{$50 \MeV$} in a linear collider~(LINAC2) before entering the Proton Synchrotron
Booster~(BOOSTER).
In the Proton Synchrotron~(PS), the protons gain a total energy of \mbox{$25 \GeV$}.
Finally, they enter the Super Proton Synchrotron~(SPS), in which they gain an energy of \mbox{$450 \GeV$}, which is the nominal injection energy for the LHC.

Bunches of protons are injected into separate beam pipes in both directions around the LHC ring and are accelerated
with a system of radio cavities to an energy of \mbox{$3.5 \TeV$}, corresponding to \mbox{$\sqrt{s} = 7 \TeV$} (2010~--~2011).
This is roughly 3.5 times more than the highest energies achieved at any other particle collider before.
Superconducting NbTi magnets with a field of up to \mbox{$8.33 \T$} keep the protons on their circular track.
In 2012, the proton energy has been increased to \mbox{$4 \TeV$}, corresponding to \mbox{$\sqrt{s} = 8 \TeV$}.
In the following years, the design centre-of-mass energy of \mbox{$\sqrt{s} = 14 \TeV$} is aimed for.

Each proton bunch consists of about $10^{11}$ particles.
With a minimal bunch spacing of \mbox{$25 \ns$} and a maximum of 2808 bunches, a $pp$ design luminosity of \mbox{$10^{34}\invcms$} can be achieved, 
which results in an average number of more than 20 inelastic collisions per bunch crossing depending on the beam focusing.

In 2011, the largest part of the data was taken with a bunch spacing of \mbox{$50 \ns$}~\cite{fournier}.
With this setting, up to 1380 bunches were circulated.
Luminosities of up to \mbox{$3.65 \cdot 10^{33} \invcms$} were achieved with up to $1.5 \cdot 10^{11}$ protons per bunch.

There are four main experiments at the LHC, which are also depicted in Fig.~\ref{fig:lhc}:
ATLAS, which is discussed in more detail in Sec.~\ref{sec:ATLAS}, CMS~\cite{cms}, ALICE~\cite{alice} and LHCb~\cite{lhcb} 
are located at the four interaction points of the two beams.
ATLAS and CMS have been designed to cover a wide variety of physics in $pp$ collisions, ranging from precision tests of the SM to
searches for the Higgs boson and searches for new phenomena.
The design choices for the ALICE detector were guided by the needs of a detector for ion-ion collisions with strong emphasis on the tracking
detectors rather than on calorimetry.
LHCb is a detector optimised for the study of $b$-physics, in particular measurements of CP violation in the $b$-sector.
Since $b\bar{b}$ systems are highly boosted at the LHC, LHCb has an asymmetric design to focus
on the forward region on one side of the $pp$ interaction point.

\begin{figure}[h]
\begin{center}
\includegraphics[width=0.58\textwidth, angle=270]{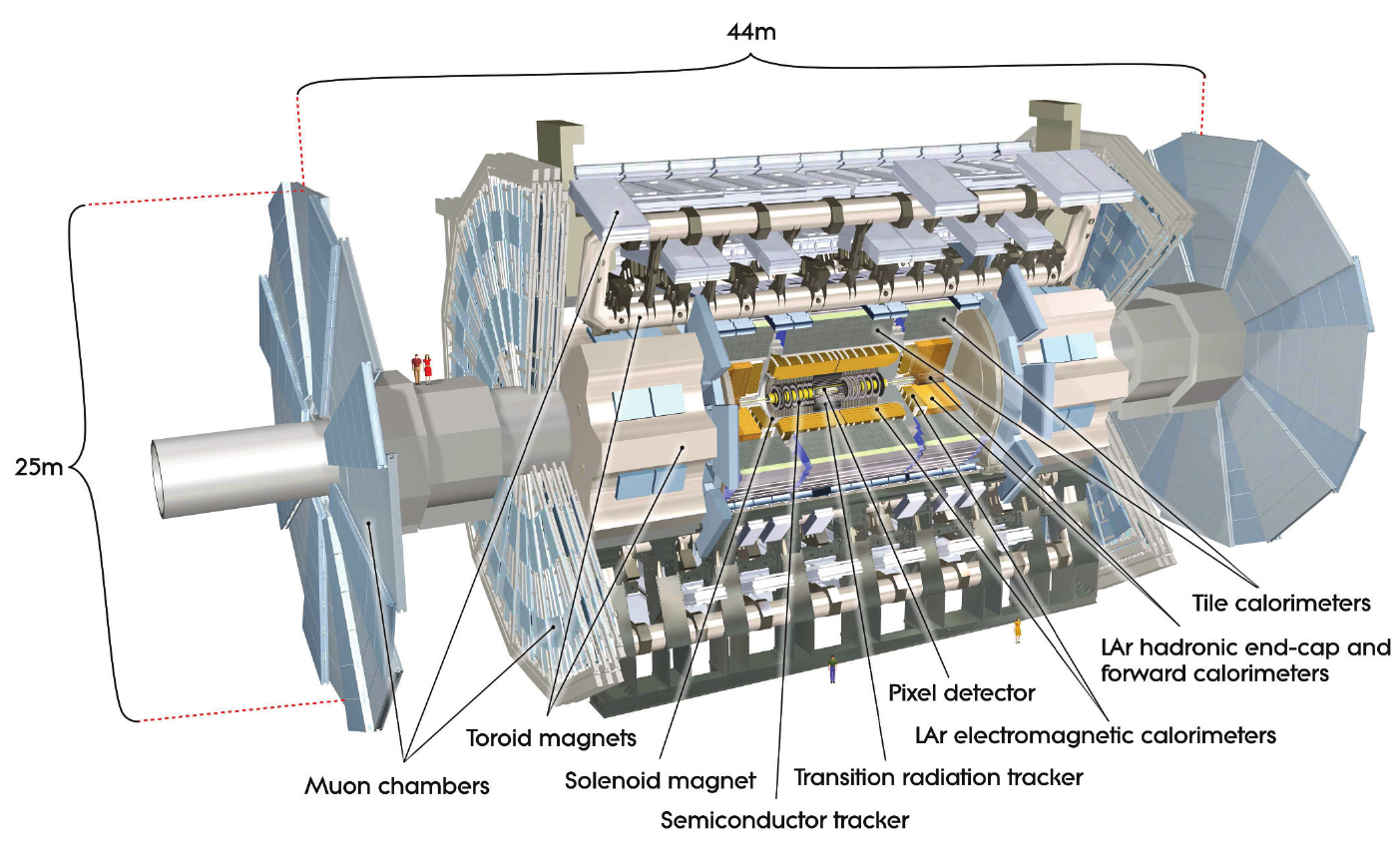}
\caption[Drawing of the ATLAS detector]{
  Drawing of the ATLAS detector showing the different detectors and magnet systems~\cite{detectorpaper}.}
\label{fig:ATLAS}
\end{center}
\end{figure}

\section{The ATLAS experiment}
\label{sec:ATLAS}

Fig.~\ref{fig:ATLAS} shows a drawing of the ATLAS detector~\cite{detectorpaper, ATLASTDR} with its solenoid and toroid magnets and the different
subdetector systems.
The subdetectors~--~inner tracking detector, calorimeters and muon spectrometer~--~are briefly described in
Sec.~\ref{sec:ID}~--~\ref{sec:muonspectrometer}.
The magnet, trigger and data acquisition systems are introduced in Sec.~\ref{sec:magnets} and~\ref{sec:triggerDAQ}, respectively.

All detector systems have been designed to cope with two main challenges set by the LHC:
on the one hand, the high event rate puts special requirements to the detector.
Fast and radiation-hard electronics and sensor elements are necessary as well as an efficient trigger and data acquisition system.
On the other hand, very good particle identification is the key to an efficient suppression of the large background contribution from multijet
production at a hadron collider, as well as from additional inelastic interactions from the same bunch crossing~(pile-up).

\begin{figure}[h]
\begin{center}
\includegraphics[width=0.6\textwidth]{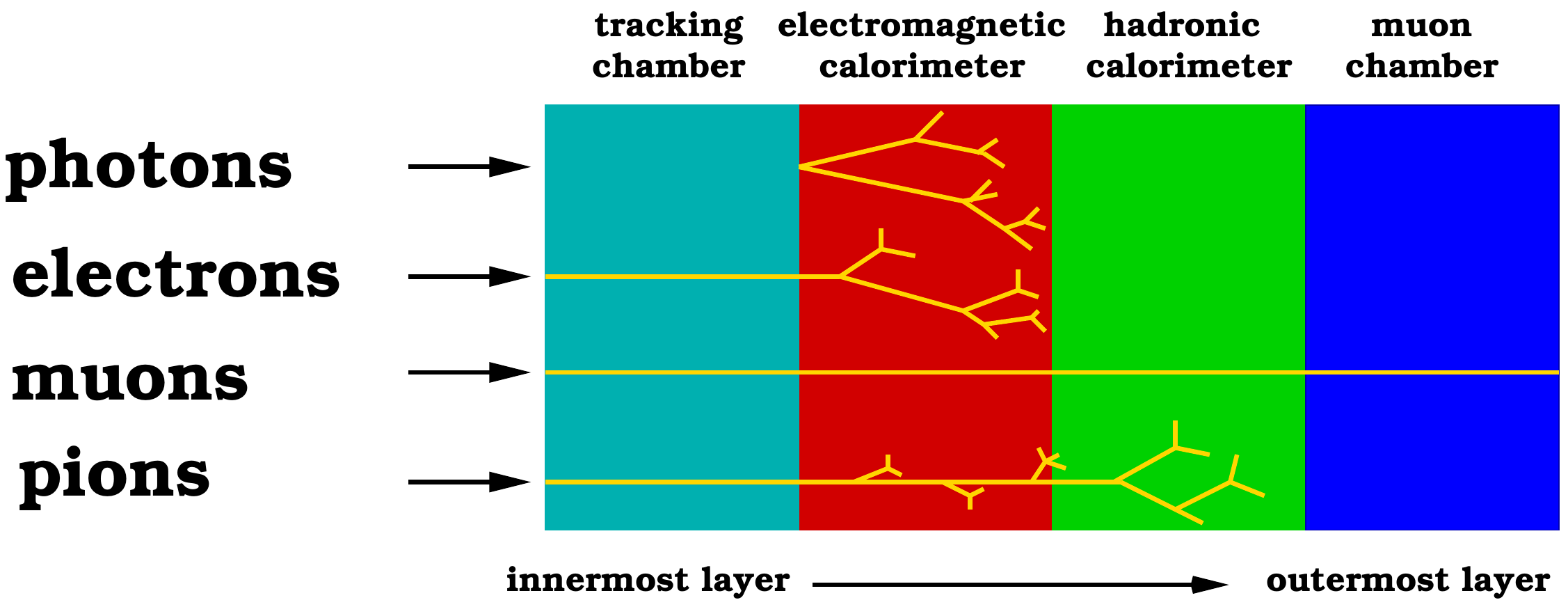}
\caption[Schematic view of the interactions of particles in a general purpose detector]{
  Schematic view of the interactions of different types of particles in a general purpose detector at a modern high energy particle collider.}
\label{fig:particlesInDet}
\end{center}
\end{figure}

The design of the ATLAS detector follows the structure illustrated in Fig.~\ref{fig:particlesInDet}, which shows schematically
the interactions of different types of particles with the detector material:
the innermost detector layer is a tracking detector, in which electrically charged particles are traced for example by creating electron-hole pairs
in semiconductors or by ionising gas.
Typically, the tracking detector is embedded in a magnetic field so that the momenta of the particles can be measured from the curvature
of the tracks.

The next detector layers consist of electromagnetic~(EM) and hadronic calorimeters.
The calorimeters are massive and therefore induce electromagnetic showers
from electrons and photons by pair production and bremsstrahlung, and hadronic showers by various processes between hadrons and matter.
Electron and photon showers are typically contained
in the electromagnetic calorimeter while hadronic showers range into the hadronic calorimeter.

Since almost all muons produced at the LHC are minimum ionising particles, they are the only electrically charged particles which may pass the
calorimeters and reach the outermost layer, which is made of tracking chambers.
Hence, these detectors are called muon chambers.

In principle, it is desirable to cover the full solid angle with sensitive detector material.
However, in the very forward region some space must be kept open for the beam pipes.
Moreover, support structures, cables, cooling systems etc. need to be included in the detector design and reduce the sensitive volume.

\subsection{Inner detectors}
\label{sec:ID}

\begin{figure}[h]
\centering
\includegraphics[width=0.36\textwidth, angle=270]{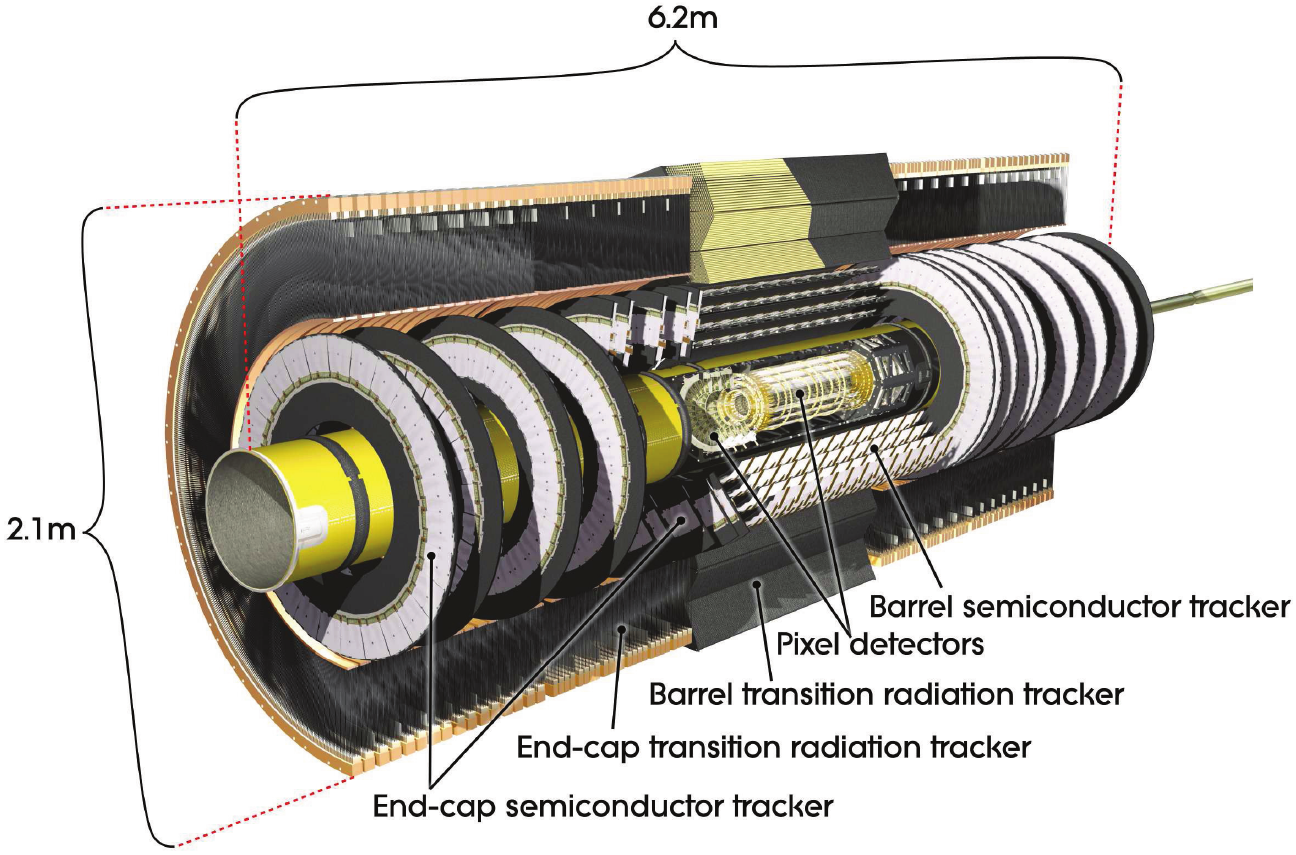}
\includegraphics[width=0.34\textwidth, angle=270]{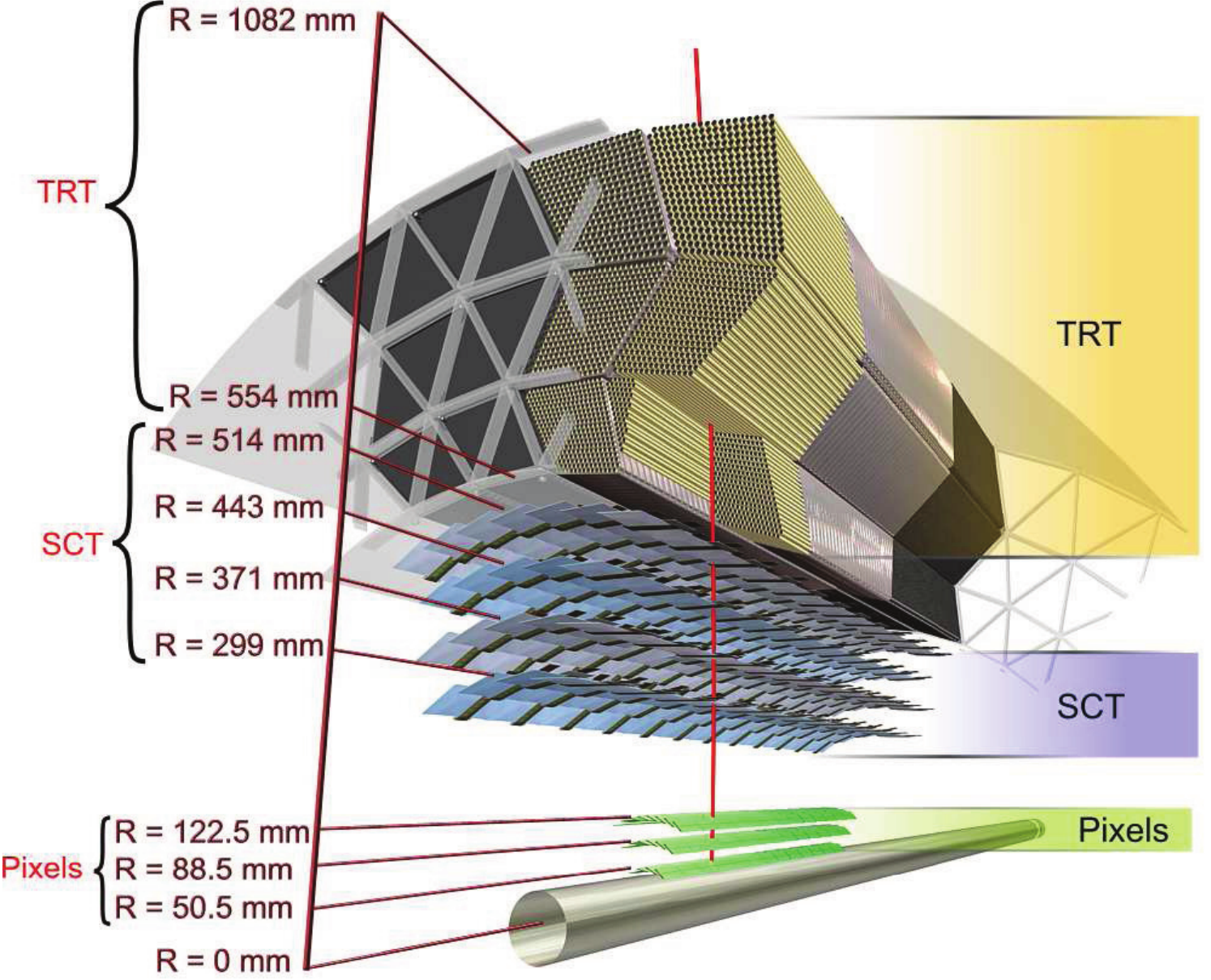}
\caption[Overview of the Inner Detector]{
  Overview of the Inner Detector~\cite{detectorpaper}:
  the left figure shows a longitudinal section of the Inner Detector with the different subdetectors.
  The right figure shows a transverse section and illustrates the distances of the different detector layers from the beam line.}
\label{fig:ID}
\end{figure}

The ATLAS Inner Detector (ID) consists of three subdetector systems:
the Pixel detector and the Semiconductor Tracker~(SCT), which use silicon semiconductor
technology, and the Transition Radiation Tracker~(TRT), which exploits the transition radiation produced in a gas mixture of Xe, CO$_2$ and O$_2$.
Fig.~\ref{fig:ID} shows a longitudinal and a transverse section of the ID.
In particular, the distances of the different subdetector layers from the beam line are illustrated.
The whole ID is embedded in a \mbox{$2 \T$} solenoidal field (Sec.~\ref{sec:magnets}).

With three concentric cylinders~(barrel part) and  three endcap disks, perpendicular to the beam axis, the Pixel detector
covers a range\footnote{ATLAS uses a right-handed coordinate system with its origin at the nominal interaction point (IP) in the centre
  of the detector and the $z$-axis along the beam pipe. The $x$-axis points from the IP to the centre of the LHC ring, and the $y$-axis points
  upwards. Cylindrical coordinates ($r$, $\phi$) are used in the transverse plane, where $\phi$ is the azimuthal angle around the beam pipe. The
  pseudorapidity is defined in terms of the polar angle $\theta$ as \mbox{$\eta = - \ln \left[ \tan \left( \frac{\theta}{2} \right) \right]$}.} of
\mbox{$|\eta| < 2.5$}.
Each of the 1744 sensors consists of a segmented silicon wafer with pixels of minimum area \mbox{$50 \times 400 \mum^2$} and 46080 readout channels.
The innermost pixel layer, the so-called $b$-layer, is as close to the beam line as \mbox{$50.5 \mm$} and allows for a precise extrapolation of tracks
to the vertices.
This is crucial for any $b$-tagging strategy based on impact parameters and the identification of secondary vertices.

The SCT consists of four layers in the barrel and nine endcap disks.
It covers the range \mbox{$|\eta| < 2.5$}.
The sensors use silicon microstrip technology with a strip pitch of \mbox{$80 \mum$}.
In the barrel, the strips are arranged parallel to the beam line, while in the disks, the strips are oriented radially.
Modules are arranged back-to-back with a small stereo angle of \mbox{$40 \mrad$} to allow for a measurement of the azimuth angle in each layer.
A typical track yields three space-points in the Pixel detector and eight in the SCT.
Together, the silicon trackers ensure the measurement of the track momenta and the identification of primary and secondary vertices.

In the barrel part of the TRT, there are 73 planes of straw tubes filled with gas, which are arranged parallel to the beam axis.
In the endcap, there are 160 straw planes, oriented radially.
The TRT covers a range of \mbox{$|\eta| < 2.0$}, in which the separation of electrons from charged pions is improved by exploiting transition radiation.
Although the TRT does not provide track information in the direction along the beam line, pattern recognition and the measurement of the track momenta
become more robust by using the signals from the TRT.

The total amount of material of the ID is as large as roughly 0.5 electromagnetic radiation
lengths\footnote{The radiation length is defined as the typical amount of material traversed by an electron after which it has lost
$\frac{1}{e}$ of its original energy by bremsstrahlung.}
$X_0$ for \mbox{$|\eta| < 0.6$}.
For \mbox{$0.6 < |\eta| < 1.37$} and \mbox{$1.52 < |\eta| < 2.5$}, the amount of material reaches up to \mbox{$1.5 X_0$}.
In the barrel-to-endcap transition region at \mbox{$1.37 < |\eta| < 1.52$}, the amount of material is even larger.
Electrons and photons in this region were not taken into account in this analysis.
A particular consequence of the sizable amount of material in front of the calorimeters is that a large fraction of photons convert into \mbox{$e^+e^-$}
pairs in the ID volume.

\subsection{Calorimeters}
\label{sec:calorimeter}

\begin{figure}[h]
\begin{center}
\includegraphics[width=0.4\textwidth, angle=270]{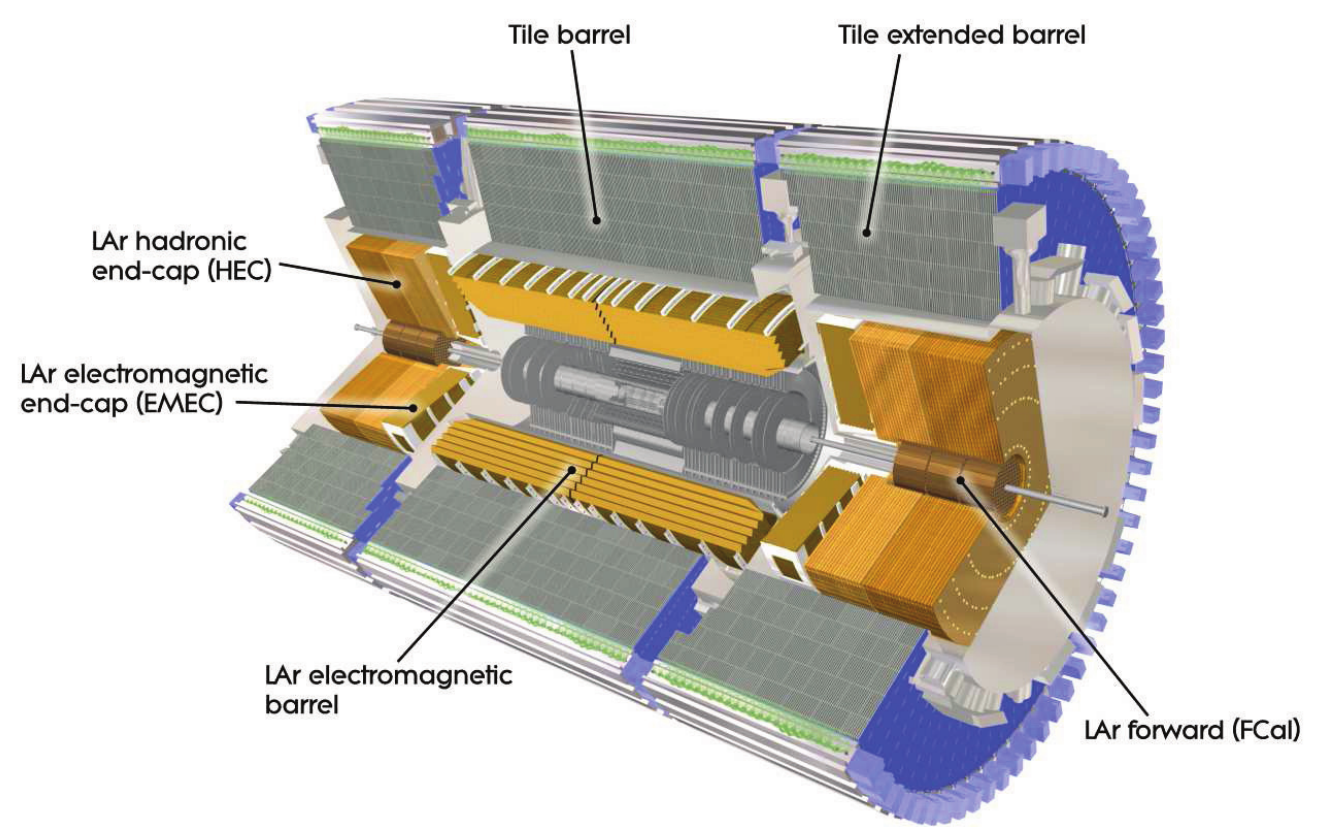}
\caption[Overview of the calorimeter system]{
  Overview of the calorimeter system~\cite{detectorpaper}:
  the different subdetectors of the electromagnetic and hadronic calorimeter are shown.}
\label{fig:calorimeters}
\end{center}
\end{figure}

Fig.~\ref{fig:calorimeters} shows an overview of the different electromagnetic and hadronic calorimeters of the ATLAS detector.
All calorimeters are sampling calorimeters consisting of alternating layers of dense absorber material and active material, where only the active
material is used for the energy measurement.
This design allows for a compact size of the calorimeter system.

The hadronic calorimeter in the barrel~(Tile) uses steel as absorber and scintillators as active material.
All other calorimeters use liquid argon~(LAr) technology with different types of absorbers:
lead in the electromagnetic barrel~(EMB) and the electromagnetic endcap calorimeter (EMEC),
copper in the hadronic endcap calorimeter~(HEC) and the electromagnetic part of the Forward Calorimeter~(FCal),
and tungsten in the hadronic part of the FCal.
The LAr calorimeters are placed in three cryostats: one for the barrel and one for each endcap.

The technologies have been chosen to provide fast readout, radiation hardness and high containment of electromagnetic and hadronic showers to ensure
a precise measurement of their energies.
The energy flux is varying in the different detector regions.
Especially in the very forward region, which is covered by the FCal (\mbox{$3.1 < |\eta| < 4.9$}), extremely high fluxes from minimum bias events
drove the design towards dense absorber material and small LAr gaps.

All calorimeters are finely granulated and also segmented longitudinally to allow for a precise determination of the position of the showers and to
distinguish different shower types by the use of shower shapes.
This is particularly important for the central region, which is devoted to precision measurements of electrons and photons:
the EMB (\mbox{$|\eta| < 1.475$}) is segmented into three longitudinal layers, where the first layer, the so-called LAr strips, provide
a very fine granularity in $\eta$ of 0.0031.
To ensure continuous coverage in azimuth and to enable fast readout, the lead absorbers are folded into an accordion shaped structure.

A similar design as for the EMB has been used for the EMEC, which is divided into two wheels covering the ranges \mbox{$1.375 < |\eta| < 2.5$}
and \mbox{$2.5 < |\eta| < 3.2$}.
The inner wheel has a coarser granularity in $\eta$ and $\phi$, limiting the region devoted to precision physics to \mbox{$|\eta| < 2.5$}.
A thin LAr layer (presampler) is placed in front of the EMB and the EMEC for \mbox{$|\eta| < 1.8$} to correct for energy lost in front of the calorimeter.

The Tile calorimeter is located behind the EMB and the EMEC and is divided into three longitudinal layers.
It consists of a central barrel (\mbox{$|\eta| < 1.0$}) and an extended-barrel part (\mbox{$0.8 < |\eta| < 1.7$}).
The radial depth is about 7.4 nuclear interaction
lengths\footnote{The nuclear interaction length $\lambda_I$ for hadrons is defined in analogy to the electromagnetic radiation length $X_0$
for electrons and photons.} ($\lambda_I$).

The HEC is a traditional LAr sampling calorimeter covering the region (\mbox{$1.5 < |\eta| < 3.2$}), which is placed behind the EMEC in the same
cryostat.
It consists of two independent wheels, each of which is divided longitudinally into two parts.

%The design energy resolution for electrons and jets are summarised in Tab.~\ref{tab:designresolution}.
Altogether, the calorimeters cover the range \mbox{$|\eta| < 4.9$} and, thus, provide good hermiticity to ensure also a precise measurement
of the imbalance of the transverse momentum.
Over the whole range in $\eta$, the total thickness of the calorimeter system is approximately \mbox{$10 \lambda_I$}, ensuring good containment of
electromagnetic and hadronic showers and limiting punch-through effects to the muon spectrometer.

\subsection{Muon detectors}
\label{sec:muonspectrometer}

The ATLAS muon system covers the range \mbox{$|\eta| < 2.7$} and is designed to measure the momenta of muons exiting the calorimeter system starting
at energies above \mbox{$\sim 3 \GeV$}.
The tracks of the muons are bent by the magnetic field of the air-core toroid system in the barrel and in the endcaps (Sec.~\ref{sec:magnets}).
The fields in the barrel and in the endcaps are oriented such that muon tracks in both regions are mostly orthogonal to the field lines.

\begin{figure}[h]
\begin{center}
\includegraphics[width=0.4\textwidth, angle=270]{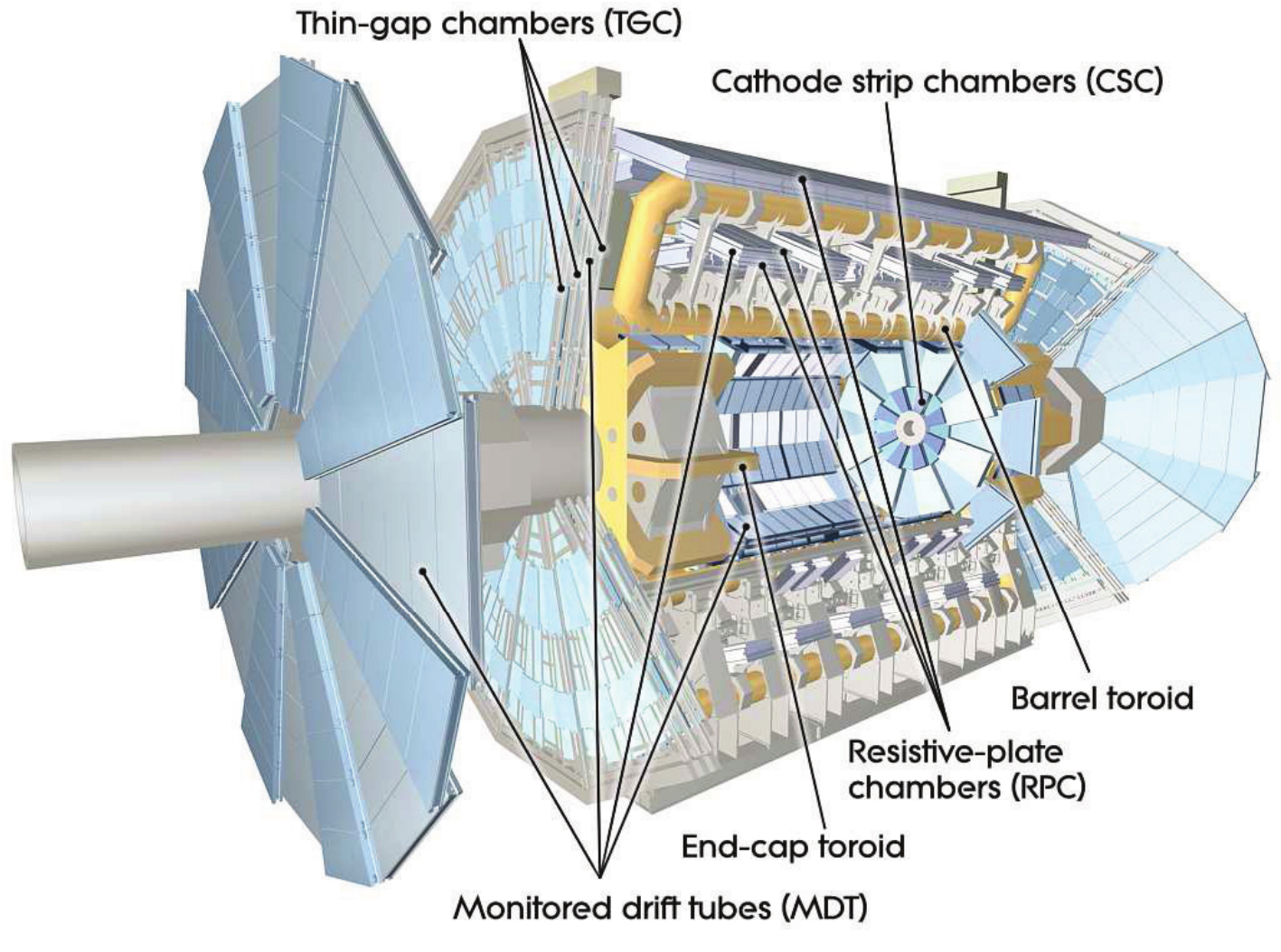}
\caption[Overview of the muon system]{
  Overview of the muon system~\cite{detectorpaper}:
  the different types of tracking and trigger chambers are shown.}
\label{fig:muons}
\end{center}
\end{figure}

An overview of the different subsystems is shown in Fig.~\ref{fig:muons}:
the muon system consists of high-precision tracking chambers as well as trigger systems.
In the barrel part, Monitored Drift Tubes~(MDTs) are used for tracking and Resistive Plate Chambers~(RPCs) for triggering (Sec.~\ref{sec:triggerDAQ}).
In the endcaps, tracking information is provided by Cathode Strip Chambers~(CSCs) and Thin Gap Chambers~(TGCs) are used for triggering.
In the barrel as well as in the endcaps, muons typically cross three longitudinal layers of the muon spectrometer.
The muon system is divided into eight octants with overlaps in $\phi$ to avoid gaps in the detector coverage.

The technologies for the tracking systems have been chosen such that high precision can be achieved given the level of the particle flux.
The MDTs in the barrel part follow a robust and reliable detector design.
Since each tube contains only one sense wire, the simple geometry allows for the prediction of deformations as well as for a precise reconstruction.
As the particle flux is increasing with $|\eta|$, the CSCs are more suited for the endcap region:
the higher granularity of the multiwire proportional chambers facilitates to cope with the increasing rates.

The choice of the technologies for the trigger chambers was driven by the requirement for fast and highly efficient trigger capabilities
given the different conditions present in the barrel and endcap regions during data taking.
Additionally, an adequate resolution of the transverse momentum of the tracks was required.
In the barrel, RPCs provide good spatial and time resolution.
In the endcap regions, however, higher particle fluxes as well as the need for a higher granularity required a different
technology: TGCs are used for the region \mbox{$1.05 < |\eta| < 2.4$}.
They are based on the same principle as multiwire proportional chambers and fulfil the needs in terms of rate capability and granularity.
With RPCs and TGCs, a time resolution of \mbox{$15 - 25 \ns$} can be achieved, which is sufficient for fast trigger decisions and a good association of
tracks to bunch crossings.

The benchmark for the tracking performance of the muon spectrometer is set by a 10\% resolution on the transverse momentum of \mbox{$1 \TeV$}
muons~\cite{detectorpaper}.
To achieve this goal, the position of the MDT wires and the CSC strips must be known with a precision better than \mbox{$30 \mum$}.
Therefore, a high-precision optical alignment system was set up to monitor the relative position of the MDT chambers and their internal deformations.

\subsection{Magnet system}
\label{sec:magnets}

\begin{figure}[h]
\centering
  \begin{minipage}[t][0.55\textwidth][b]{0.45\textwidth}
  \includegraphics[width=0.9\textwidth, angle=270]{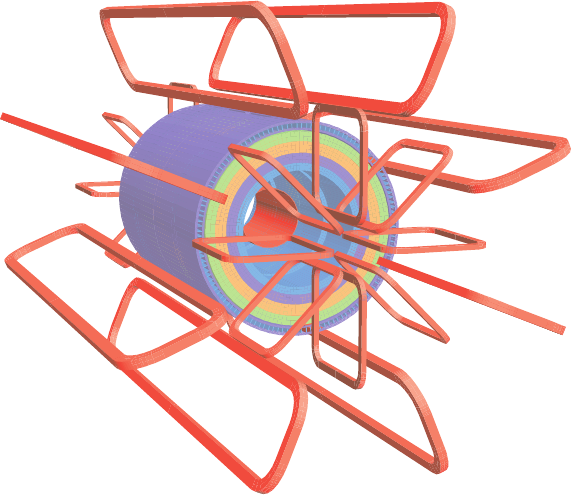}
  \caption[Sketch of the magnet system]{
    Sketch of the magnet system~\cite{detectorpaper}:
    the solenoid and the toroids are shown, as well as the tile calorimeter.}
  \label{fig:magnets}
  \end{minipage}
  \hspace{1cm}
  \begin{minipage}[t][0.55\textwidth][b]{0.45\textwidth}
  \includegraphics[width=0.9\textwidth]{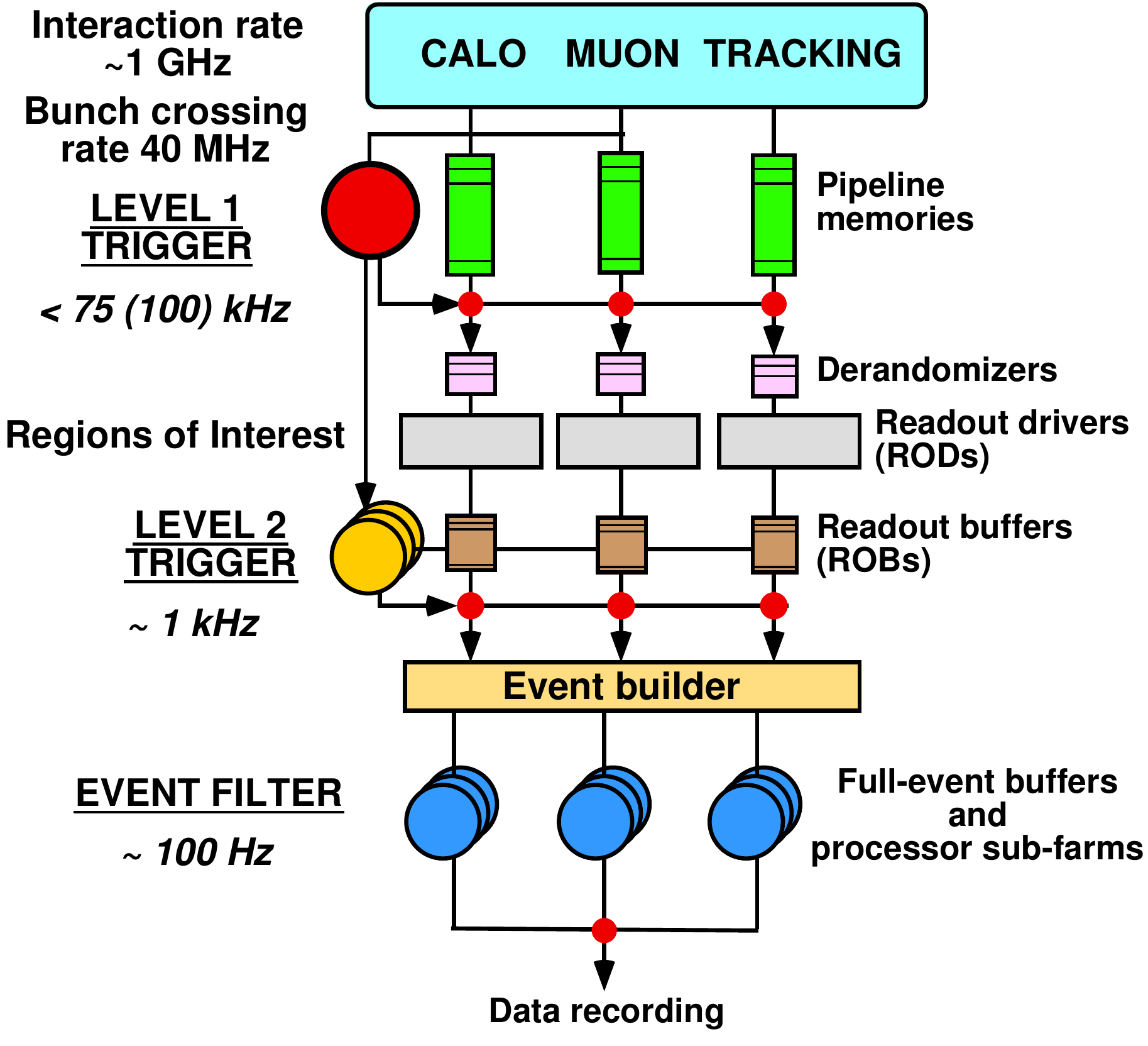}
  \caption[Sketch of the trigger chain]{
    Sketch of the trigger chain and the different trigger levels~\cite{ATLASTDR}.
    The indicated rates are orders of magnitude only.
  }
  \label{fig:triggerdaq}
  \end{minipage}
\end{figure}

The ATLAS magnetic system consists of four large superconducting magnets: a central solenoid and three toroid magnets in the barrel and the
two endcaps.
A sketch of the magnet system is shown in Fig.~\ref{fig:magnets}.
The solenoid and the toroids are shown, as well as the tile calorimeter.

The central solenoid provides an axial field with a strength of \mbox{$2 \T$}.
The magnetic flux is returned by the tile calorimeter and its girder structure.
The solenoid was designed to be particularly lightweight and to minimise the amount of material in front of the calorimeter system
to which it contributes only a total of 0.66 electromagnetic radiation lengths.

The toroid systems provide magnetic fields with a bending power of \mbox{$1.5 - 5.5 \Tm$} in the barrel and \mbox{$1 - 7.5 \Tm$} in the endcap regions.
Each system consists of eight air-core coils placed in aluminium housings.
The toroidal fields contain non-uniformities which need to be known to high precision to allow for an accurate measurement of muon momenta.
Hence, 1800 Hall sensors were installed in the muon spectrometer volume to enable the monitoring of the magnetic field.

\subsection{Trigger and data acquisition}
\label{sec:triggerDAQ}

Assuming a bunch spacing of \mbox{$25 \ns$} and approximately 20 inelastic interactions per bunch crossing\footnote{As mentioned in Sec.~\ref{sec:LHC}, in 2011 most of the data was taken with a bunch-spacing of \mbox{$50 \ns$}.}, the event rate at the ATLAS detector is
of the order of 1~GHz.
A three-level trigger system was set up to reduce this rate to about 200~Hz.
In 2011 data taking, the real trigger rate was indeed of the order of 300~Hz~\cite{fournier}.
The triggers need to suppress minimum bias events very strongly while efficiently selecting rare physics events.
The data acquisition system~(DAQ) gathers the data from the different detector subsystems and buffers them until a trigger decision is received.
When the event is not rejected by one of the trigger levels, the data are recorded permanently.
Fig.~\ref{fig:triggerdaq} shows a sketch of the ATLAS trigger chain indicating the order of magnitude of the trigger rates at the different trigger
levels.

The first trigger level~(L1) is a hardware-based trigger, which reduces the event rate to approximately 75~kHz.
Muons, electrons, photons, jets and hadronically decaying $\tau$-leptons with high transverse momenta are searched for as well as a large momentum
imbalance in the transverse plane and a large total transverse energy.
The muon trigger chambers are used as well as the calorimeter system with reduced granularity.
Within less than \mbox{$2.5 \mus$}, Regions-of-Interest~(RoI) are identified in $\eta$-$\phi$-space, which serve as seeds for the decision at the second
trigger level~(L2).

The high level trigger is composed of the L2 and the Event Filter (EF), both of which are software-based trigger systems.
At L2, the energy and direction of the RoIs are further investigated and also the types of the trigger objects are analysed.
Within \mbox{$40 \ms$} a decision is made, and the event rate is reduced to below 3.5~kHz.
The EF further decreases the rate down to roughly 200~Hz.
Events passing the EF are stored permanently.
The full event information is available at the EF level and, hence, energies and directions of the trigger objects are estimated with higher precision
than at L1 and L2.
In particular, the discrimination between the different particle types is enhanced by the use of the ID tracking system and calorimeter
shower shapes.

Selections of different trigger signatures are collected in so-called trigger menus.
For triggers with very high rates, only a fraction of the triggered events can be selected on a random basis
in order to perform cross-checks and studies of less rare physics processes.
The trigger menus are adjusted to the data taking conditions, in particular to the instantaneous luminosity, in order to make optimal use of the
band width available for storage.

\subsection[Performance in $pp$ collisions]{Performance in \boldmath$pp$ collisions\unboldmath}
\label{sec:performance}

The performance of the different subsystems of the ATLAS detector was studied with the data taken in 2010 and 2011.
Of particular importance is the agreement with Monte Carlo~(MC) simulations (Ch.~\ref{sec:modelling}).

\begin{figure}[h!]
\begin{center}
\includegraphics[height=0.325\textwidth]{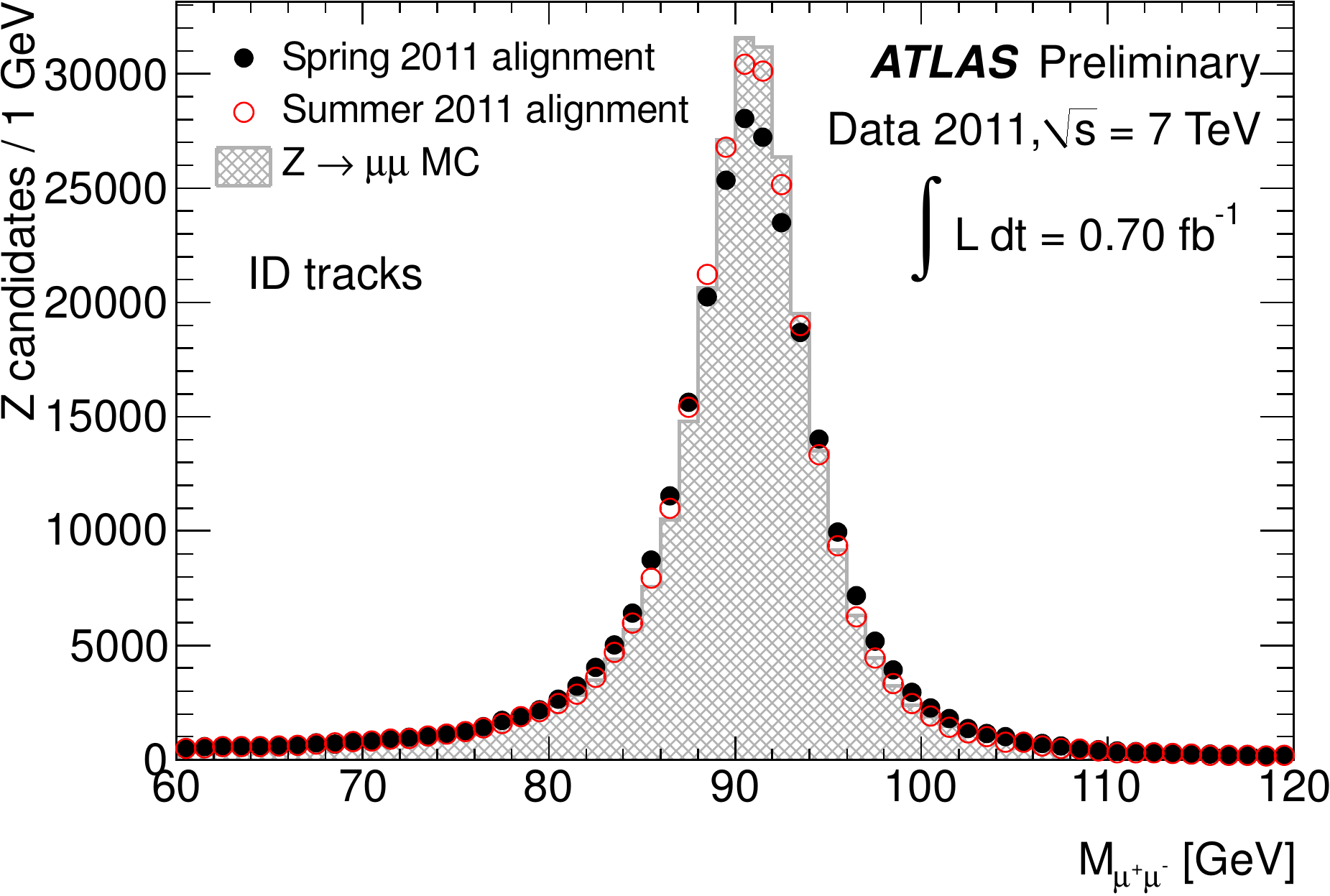}
\includegraphics[height=0.325\textwidth]{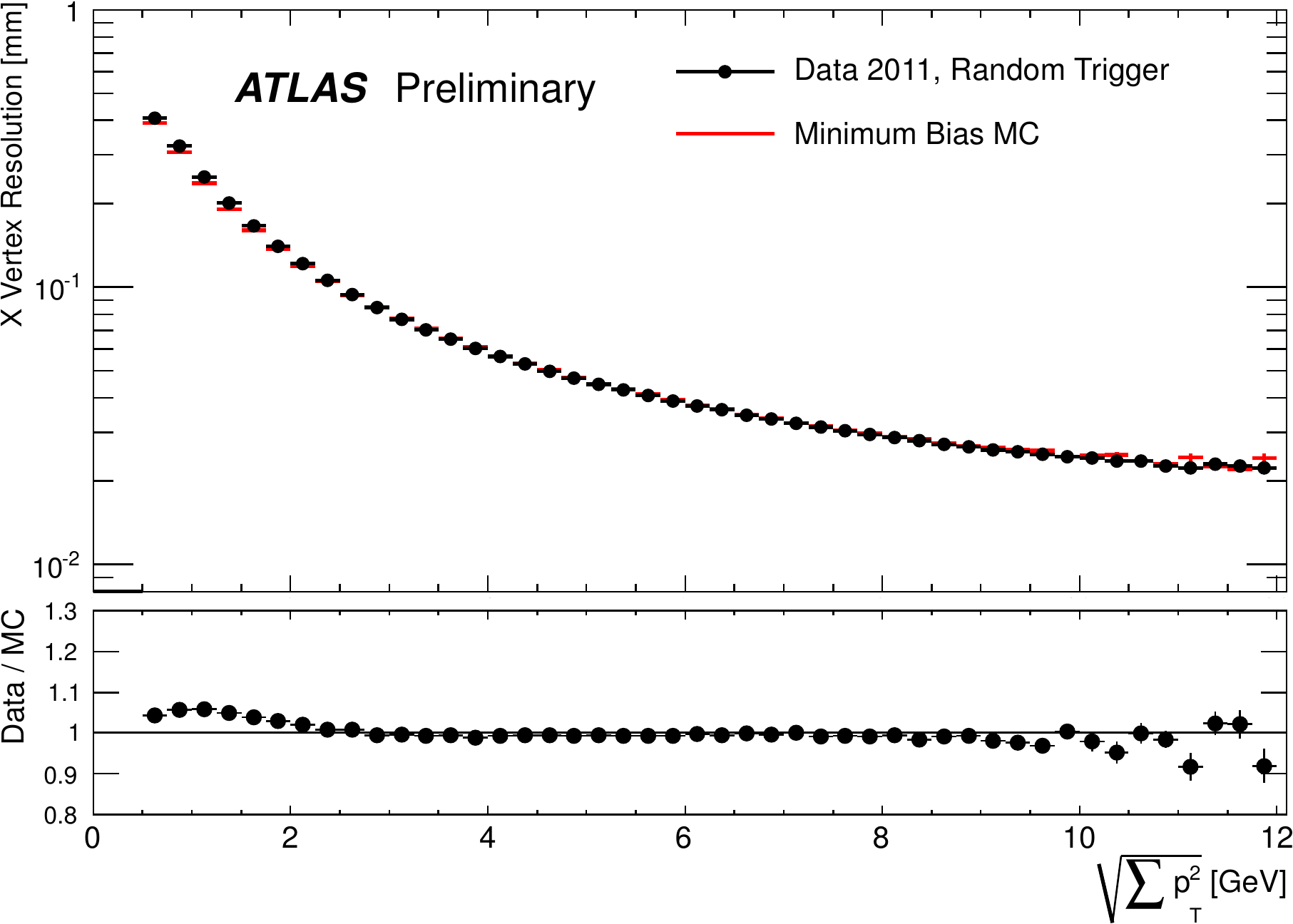}
\caption[Performance of the Inner Detector]{
  Performance of the Inner Detector in $pp$ collisions~\cite{idperformance}:
  the left plot shows the invariant mass of $\mu^+\mu^-$ pairs around the mass of the $Z$ boson from ID tracks only in \mbox{$0.70 \ifb$} of data.
  In shaded grey, the expectation from MC simulations is shown. The full and open circles show different versions of the relative alignment
  of the ID subsystems.
  The right plot shows the estimated resolution of the vertex position in $x$-direction as a function of the squared sum of the $\pt$ of the tracks
  associated to the vertex in 1.5 million minimum bias events.
  Below the plot, the ratio of data and MC simulations is shown.
}
\label{fig:IDperformance}
\end{center}
\end{figure}

The left plot of Fig.~\ref{fig:IDperformance} shows the invariant mass of $\mu^+\mu^-$ pairs around the mass of the $Z$ boson in \mbox{$0.70 \ifb$} of data,
where the muon momentum was measured using ID tracks only.
An early and an improved version of the alignment of the ID subdetectors were used to measure the distribution (full and open circles,
respectively).
In shaded grey, the expectation from MC simulations is shown.
The width of the reconstructed $Z$ mass distribution is a measure for the ID track momentum resolution.
The resolution in data with improved alignment is generally well reproduced by the MC simulations.
However, the resolution is slightly worse in data, so that the distribution in the MC simulations needs to be broadened to match the
data distribution.

The right plot of Fig.~\ref{fig:IDperformance} shows the estimated resolution of the vertex position in $x$-direction as a function of the squared sum
of the transverse momentum\footnote{The transverse momentum is defined as \mbox{$\pt = p \sin \theta = p / \cosh \eta$}.}
$\pt$ of the tracks associated to the vertex $\sqrt{\sum \pt^2}$ in 1.5 million minimum bias events~\cite{ATLASminbias}.
As expected, the vertex resolution improves with increasing $\sqrt{\sum \pt^2}$.
The general trend of the data is well described by the MC simulations although there are discrepancies in particular for low $\sqrt{\sum \pt^2}$.
The distributions for the resolution in $y$- and $z$-direction show a similar behaviour~\cite{idperformance}.

\begin{figure}[h!]
\begin{center}
%\vspace{0.025\textwidth}
\includegraphics[height=0.35\textwidth]{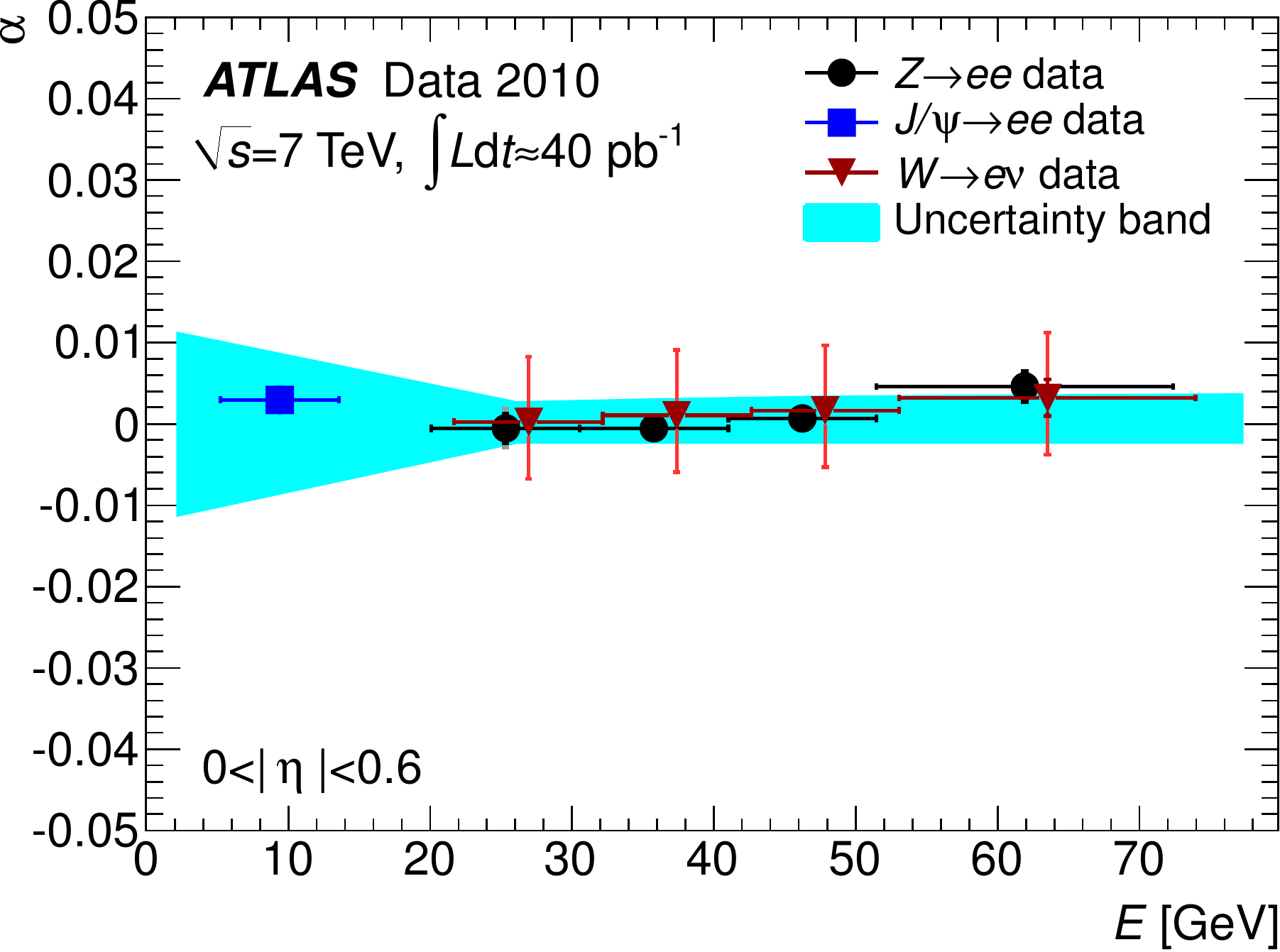}
\includegraphics[height=0.35\textwidth]{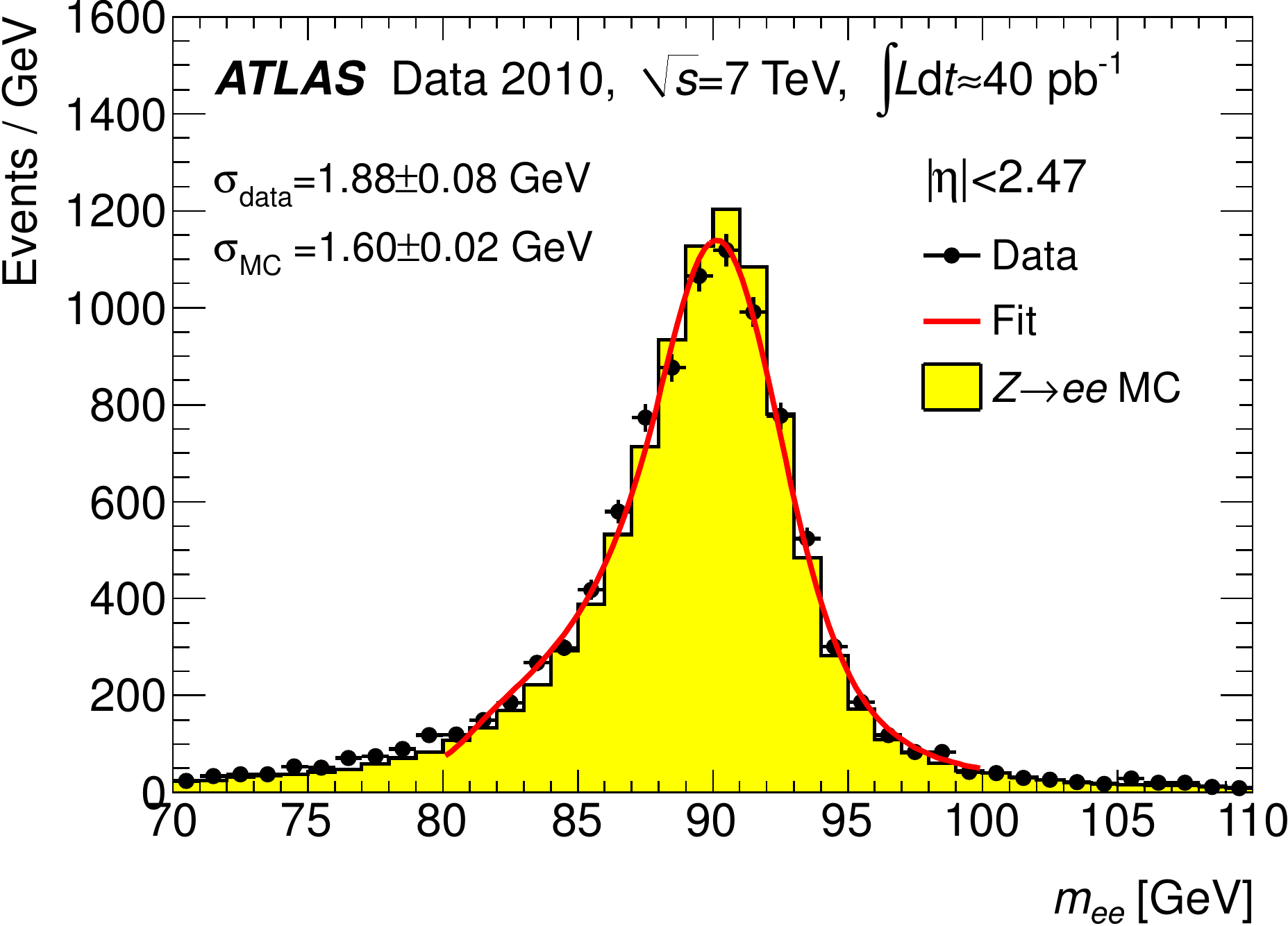}
\caption[Electron energy performance]{
  Performance of the electron energy measurement in $pp$ collisions~\cite{electronperformance}:
  the left plot shows the correction to the electron energy scale derived from \Zee, \mbox{$J/\psi \to ee$} and \mbox{$W \to e\nu$}
  events with the associated uncertainties for \mbox{$0 < |\eta| < 0.6$}.
  The right plot shows the di-electron invariant mass distribution around the mass of the $Z$ boson for data and MC simulations.
  For both plots, \mbox{$40 \ipb$} of data were used.
}
\label{fig:electronperformance}
\end{center}
\end{figure}

The left plot of Fig.~\ref{fig:electronperformance} shows the correction to the electron energy scale derived from \mbox{$Z \to ee$},
\mbox{$J/\psi \to ee$} and \mbox{$W \to e\nu$} events with the associated uncertainties for \mbox{$0 < |\eta| < 0.6$} in \mbox{$40 \ipb$} of data.
The measurements for the different processes are consistent within uncertainties and the electron energy scale is known to a level better than 1\%.
For higher $|\eta|$, the uncertainties are slightly larger~\cite{electronperformance}.

The right plot of Fig.~\ref{fig:electronperformance} shows the di-electron invariant mass distribution around the mass of the $Z$ boson in \mbox{$40 \ipb$}
of data and in MC simulations.
The width of the distribution is a measure of the electron resolution.
The MC simulations describe the data well, although the $Z$ mass resolution is slightly narrower in data, which can be corrected for by enlarging
the electron resolution in the simulations.

\begin{figure}[h!]
\begin{center}
\includegraphics[width=0.325\textwidth, angle=270]{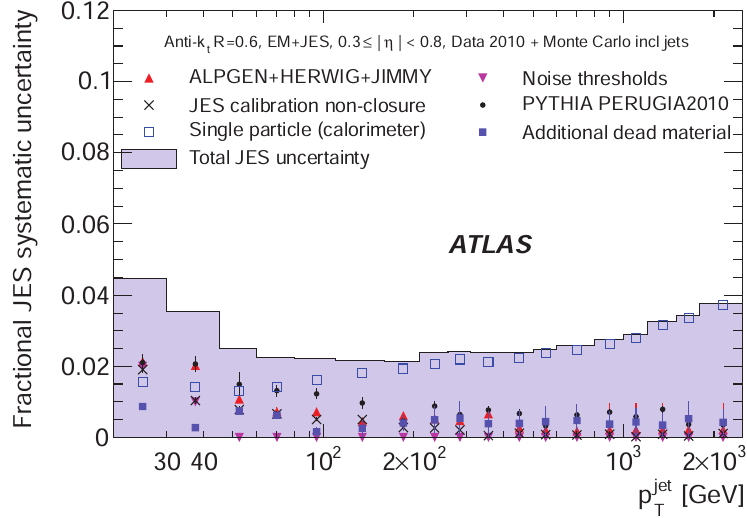}
\includegraphics[width=0.325\textwidth, angle=270]{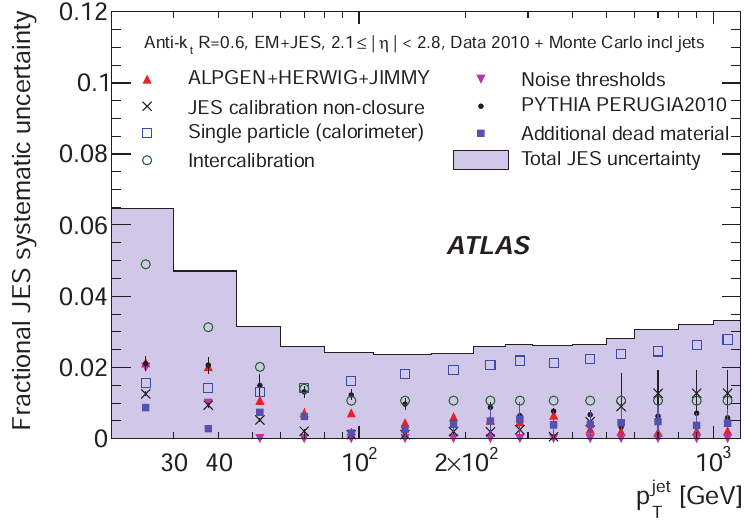}
\caption[Jet energy performance]{
  Performance of the jet energy measurement in $pp$ collisions~\cite{jetperf}:
  the total uncertainty on the jet energy scale as well as different contributions to the uncertainty are shown as a function of the $\pt$ of the jet.
  The left plot shows jets in the region \mbox{$0.3 \leq |\eta| < 0.8$}, the right plot shows jets in the region \mbox{$2.1 \leq |\eta| < 2.8$}.
  For the derivation of the uncertainties the data taken in 2010 as well as MC simulations were used.
}
\label{fig:jetperformance}
\vspace{0.025\textwidth}
\includegraphics[height=0.35\textwidth]{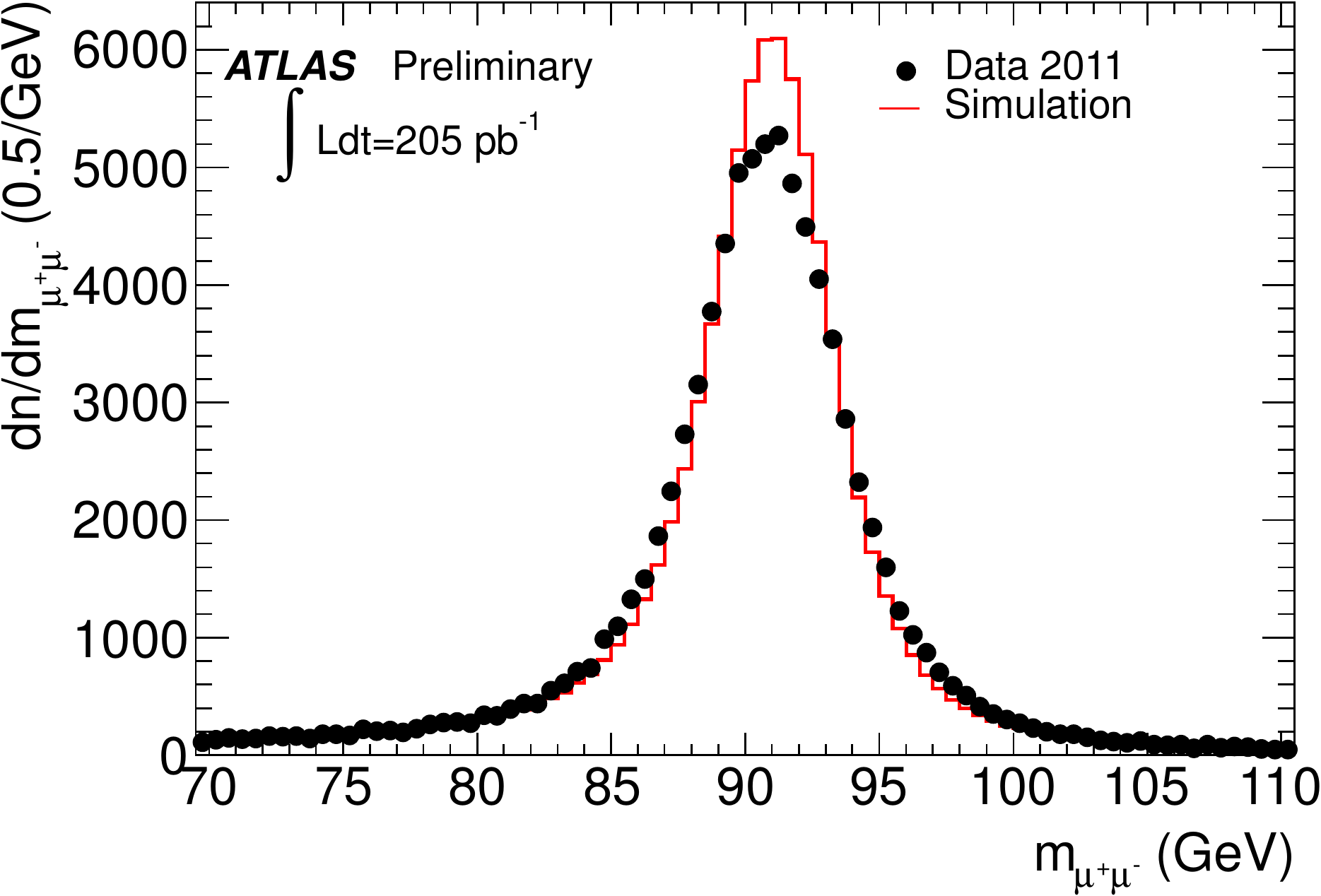}
\quad
\includegraphics[height=0.35\textwidth]{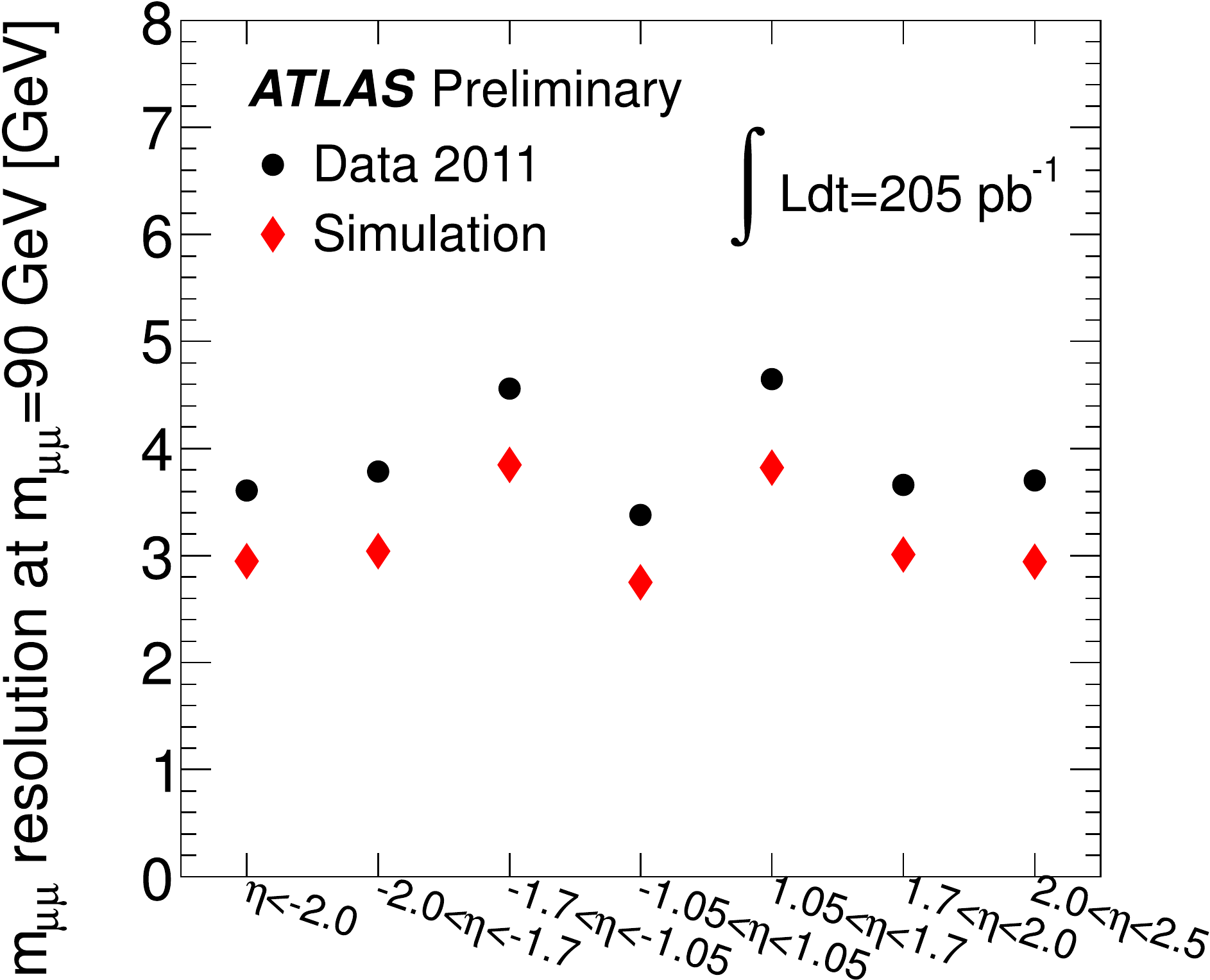}
\caption[Muon momentum performance]{
  Performance of the muon momentum measurement in $pp$ collisions~\cite{muonperformanceplots}:
  the left plot shows the invariant di-muon mass around the mass of the $Z$ boson in \mbox{$205 \ipb$} of data together with the expectation
  from MC simulations.
  The right plot shows the width of the distribution around an invariant mass of \mbox{$90 \GeV$} for different regions of the muon pseudorapidity.
  The resolutions measured in data are compared to the resolutions from MC simulations.
}
\label{fig:muonperformance}
\end{center}
\end{figure}

Fig.~\ref{fig:jetperformance} shows the total uncertainty on the jet energy scale~(JES) as well as different contributions to the uncertainty as a
function of the $\pt$ of the jet.
The left plot shows jets in the region \mbox{$0.3 \leq |\eta| < 0.8$}, the right plot shows jets in the region \mbox{$2.1 \leq |\eta| < 2.8$}.
For the derivation of the uncertainties the data taken in 2010 as well as MC simulations were used.

Contributions to the total JES uncertainty are shown for
the underlying MC model (up-pointing triangles and solid circles),
the noise threshold of the clustering algorithm (down-pointing triangles),
the in-situ calibration using the balance in $\pt$ of dijet events, called $\eta$-intercalibration (crosses and open circles),
the response to single hadrons (open squares), 
and the knowledge of the detector material (full squares).

The total JES uncertainty is of the order of 4\% (6\%) for low-$\pt$ jets in the range \mbox{$0.3 \leq |\eta| < 0.8$} (\mbox{$2.1 \leq |\eta| < 2.8$}) and
decreases to roughly 2\% (2.5\%) for jets with \mbox{$60 \GeV < \pt < 800 \GeV$}.
The uncertainty in the low-$\pt$-region is dominated by MC modelling uncertainties and uncertainties due to the $\eta$-intercalibration.
For higher $\pt$, the JES uncertainty increases again to 4\% (3\%) due to the limited knowledge of the response to high-energetic single
hadrons.

The left plot in Fig.~\ref{fig:muonperformance} shows the invariant di-muon mass around the mass of the $Z$ boson in \mbox{$205 \ipb$} of data.
The expectation from MC simulations is also depicted and shows a slightly narrower distribution than observed in data.

The right plot in Fig.~\ref{fig:muonperformance} shows the width of the distribution around an invariant mass of \mbox{$90 \GeV$} for different regions of
the muon $\eta$.
The resolution is similar for all $\eta$-regions.
Only in the barrel-to-endcap transition regions at \mbox{$1.05 < |\eta| < 1.7$}, the resolution is slightly worse.
The resolution is broader in data than in MC simulations consistently for all $\eta$-regions, which is taken into account by
correcting the muon momentum resolution in the simulations.

\chapter{Description of the analysed data set}
\label{sec:data}

\begin{figure}[h]
\begin{center}
\includegraphics[width=0.46\textwidth]{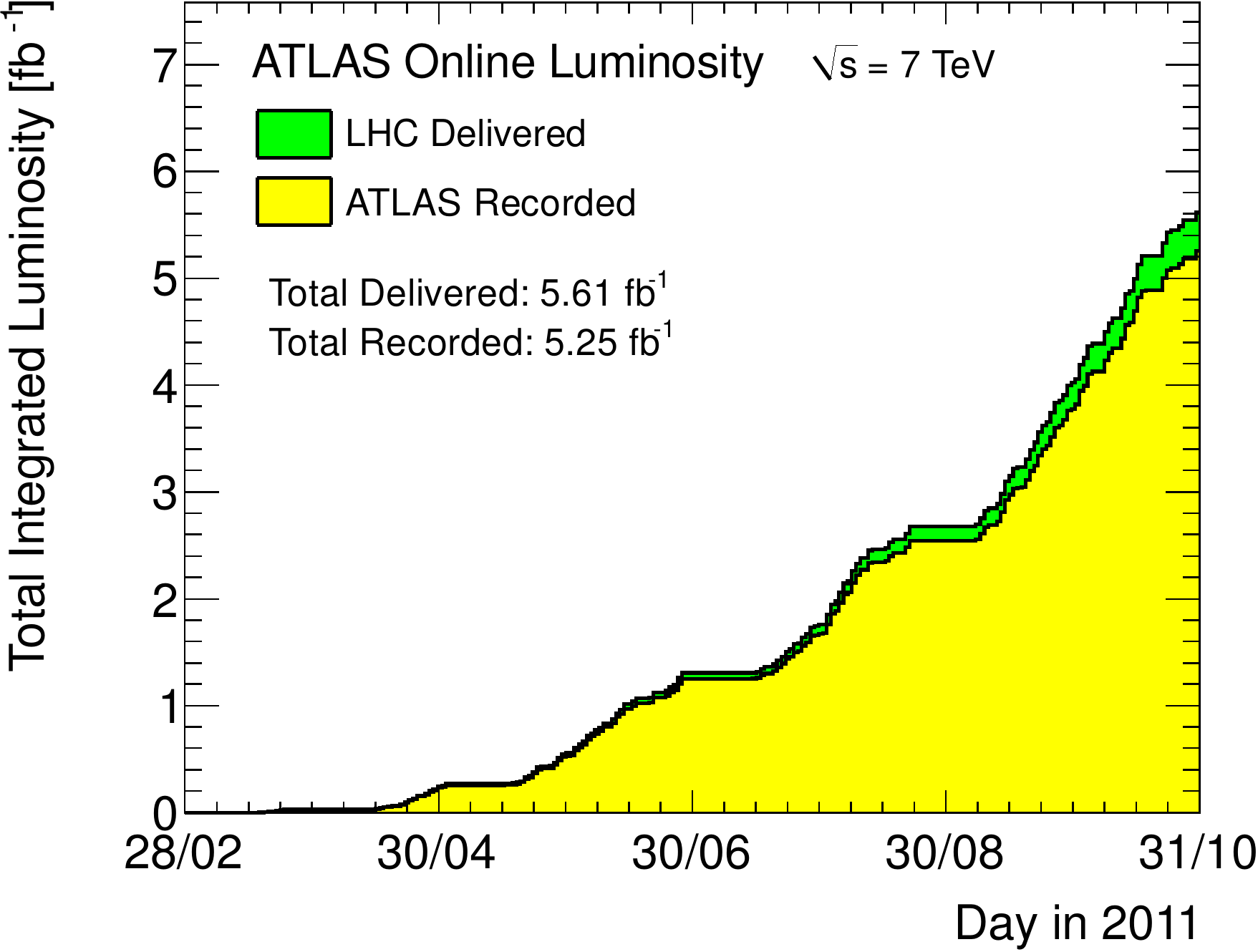}
\hspace{0.5cm}
\includegraphics[width=0.46\textwidth]{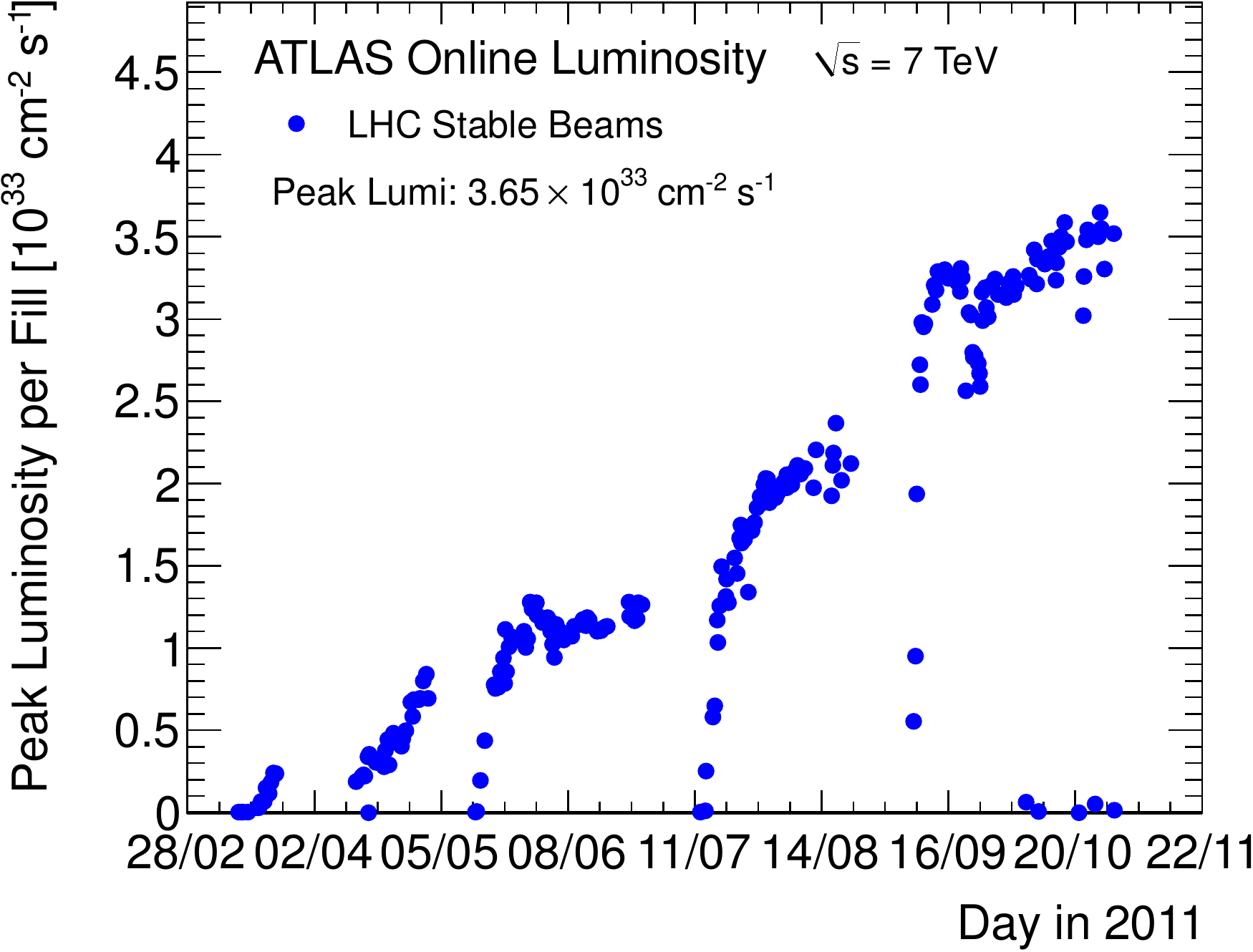}
\caption[Integrated luminosity in 2011]{
  Left: integrated luminosity in 2011~\cite{lumi} --
  the luminosity delivered by the LHC is shown as well as the luminosity recorded by the ATLAS experiment.
  Right: peak luminosity in 2011 delivered by the LHC~\cite{lumi}.}
\label{fig:lumi}
\end{center}
\end{figure}

The ATLAS experiment started data taking at \mbox{$\sqrt{s} = 7 \TeV$} in 2010 and recorded roughly \mbox{$45 \ipb$} of data in the first
year~\cite{lumi}.
In 2011, the instantaneous luminosity of the LHC was very much increased and a total of \mbox{$5.61 \ifb$} of data were provided by the LHC throughout the
year.
The data taking efficiency of the ATLAS experiment was very good: \mbox{$5.25 \pm 0.19 \ifb$} of data were recorded (left plot in Fig.~\ref{fig:lumi}).

For the analysis presented in this thesis, \mbox{$1.04 \ifb$} of data were analysed.
The data were taken between March 2011 (data taking period~B) and June 2011 (period~H).
The instantaneous luminosity was increased in several steps during this time (right plot in Fig.~\ref{fig:lumi}):
the number of protons per bunch and the number of bunches present in the LHC were increased,
the focusing of the proton beams at the interaction point was improved and the bunch spacing time was reduced.

The increasing instantaneous luminosity lead to a larger average number of interactions per bunch crossing (pile-up).
The increase in the number of protons and the improvements in focusing enhanced the contribution from \textit{in-time pile-up}, i.e. additional
interactions in the same bunch crossing.
Shorter time differences between the bunches and longer bunch trains created additional \textit{out-of-time pile-up}, which consists
of overlaid interactions from different bunch crossings.
Out-of-time pile-up effects originate from the finite readout time of the subdetector systems, which may lead to a wrong assignment of detector
signals to bunch crossings and to an intrinsic integration of signals from different bunch crossings.

\begin{figure}[h]
\begin{center}
\includegraphics[width=0.5\textwidth]{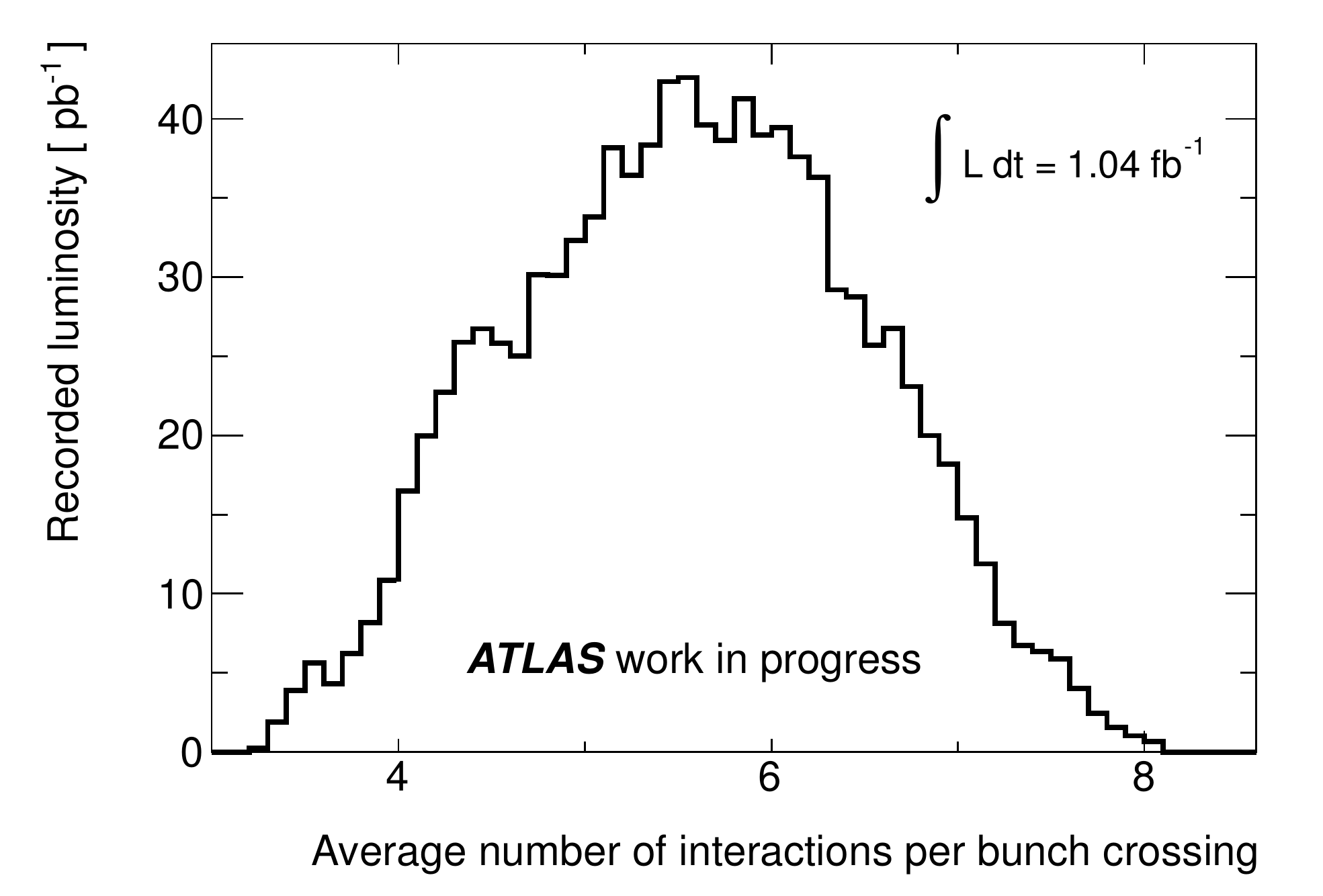}
\caption[Average number of interactions per bunch crossing]{
  Average number of interactions per bunch crossing for the \mbox{$1.04 \ifb$} of data which were analysed in this thesis.}
\label{fig:mu}
\end{center}
\end{figure}

The average number of interactions per bunch crossing for the data analysed in this thesis is shown in Fig.~\ref{fig:mu}.
It varies between three and eight interactions depending on the data taking period.

Not only the configuration of the LHC proton beams was changing during data taking, but also the conditions of the ATLAS detector and readout itself.
Although the ATLAS detector was operated in a stable mode in general, it was unavoidable that certain failures in the subdetector systems
temporarily deteriorated the quality of the data taken.
Examples are short periods of increased noise in the LAr calorimeters and modules in the pixel detector which were temporarily unreachable for
readout.
The effect on the quality of the data caused by each of these defects was studied.
Based on the defects present in each short time period of data taking~(luminosity blocks of the order of \mbox{$1 \min$}), a list of good luminosity blocks in each
run of data taking was compiled~(good runs list).
Only data fulfilling the good runs list criteria were taken into account for the analysis as well as for the calculation of the integrated luminosity.

In about 84\% of the data considered in this analysis, six front-end boards in the barrel of the EMB were not operational due to broken optical links.
This issue was permanently present until the end of period H and required a special treatment, which is described in Sec.~\ref{sec:MET}.
It was also corrected for in the MC simulations and was considered as a source of systematic
uncertainty~(Sec.~\ref{sec:syst_detectormodelling}).

The data were reconstructed with release \texttt{AtlasPhysics-16.6.5.5.1} of the ATLAS reconstruction framework ATHENA~\cite{athena}.
The analysis of the data was then performed using C++ and the ROOT framework~\cite{root}.

Data were bundled in three different data streams according to the triggers that fired: \texttt{Egamma}, \texttt{Muons} and \texttt{JetTauEtmiss}.
In order to reduce the amount of data to be processed in each analysis, certain preselections (skims) of the different data streams were provided.
The following skims were used:
in the \texttt{Egamma} stream at least one electron candidate with loose shower shape criteria and a
transverse energy\footnote{The transverse energy is defined in analogy to the transverse momentum: \mbox{$\et = E \sin \theta = E / \cosh \eta$}.}
of \mbox{$\et > 20 \GeV$} was required, and
in the \texttt{Muons} stream at least one muon candidate with \mbox{$\pt > 18 \GeV$} was required.
In the \texttt{JetTauEtmiss} stream, at least four jets with \mbox{$\pt > 20 \GeV$} and two jets with \mbox{$\pt > 40 \GeV$} were required, or at least
five jets with a \mbox{$\pt > 20 \GeV$}.

The object definitions and requirements on the transverse momenta in the skims of the \texttt{Egamma} and \texttt{Muons} streams are looser than the
requirements which were actually used for the analysis~(Ch.~\ref{sec:objects} and~\ref{sec:selection}).
The skim of the \texttt{JetTauEtmiss} stream was intended for $\ttbar$ analyses in the all-hadronic channel, but was also used as a control region
in this analysis~(Ch.~\ref{sec:strategy} and~\ref{sec:faketemplate}).

\chapter{Signal and background modelling}
\label{sec:modelling}

Simulations of physics processes and detector responses are crucial for the modelling of signal and background events in modern high energy physics.
As the processes involved can be described by probability density functions, Monte Carlo~(MC) simulation techniques are a natural choice, but also
methods to extract contributions from certain processes from data are established.

According to the factorisation theorem, Eq.~(\ref{eq:factorisation}), the hard scattering process can be separated from non-perturbative QCD effects.
There are specific programs for the generation of hard processes, which can be interfaced with programs which provide models for the non-perturbative
evolution of the final state including parton showering and hadronisation.
There are two main models for hadronisation: the Lund string model, which is implemented in the PYTHIA generator~\cite{pythia}, and the HERWIG cluster
model, which is implemented in the HERWIG generator~\cite{herwig}, which is commonly used together with the JIMMY generator for multiple parton
scattering~\cite{jimmy}.
These generators also provide models for the underlying event, which adds contributions from ISR and FSR, multiple parton
interactions, beam remnants and pile-up contributions to the final state.

All particles which are (meta-)stable on time scales of the order of \mbox{$\Delta t = l / c$}, with $l$ the
distance of the first detector layer from the interaction point and $c$ the speed of light,
%or metastable, such as $B$-mesons or $\tau$-leptons,
were passed to a detector simulation using Geant4~\cite{geant1,geant2}.
Geant4 is a general framework for the simulation of the interaction of particles with matter based on MC techniques.
A detailed description of the ATLAS detector was used as an input for the Geant4 simulation~\cite{detectorsimulation}.
The output of the detector simulation was then passed through the same reconstruction software which was used for the reconstruction of data.

The signal process (Sec.~\ref{sec:signalmodelling}) and a part of the background processes (Sec.~\ref{sec:backgroundmodelling}) were modelled using MC
simulations.
However, the MC modelling of certain background processes is not sufficiently reliable and methods for the estimation of these contributions from data
needed to be applied, as discussed in Sec.~\ref{sec:backgroundphotons}.

\section[Simulation of $\ttg$ events]{Simulation of \boldmath$\ttg$ events\unboldmath}
\label{sec:signalmodelling}

The simulation of the $\ttg$ signal process was performed with the WHIZARD MC generator~\cite{whizard, omega}.
The full seven-particle final state was calculated in the single lepton and dilepton $\ttg$ decay channels:
$l \nu_l q \bar{q}\prime b\bar{b} \gamma$ and $l \nu_l \tilde{l} \nu_{\tilde{l}} b\bar{b} \gamma$, with
\mbox{$l / \tilde{l} = e^\pm, \, \mu^\pm, \, \tau^\pm$} and \mbox{$\nu_l / \nu_{\tilde{l}}$} the corresponding antineutrino.

The WHIZARD MC generator was developed especially for automated calculations of matrix elements (MEs) in LO.
For a given initial and final state it calculates the full ME, taking into account all possible contributing diagrams.
Hence, all diagrams shown in Fig.~\ref{fig:ttgproduction_qq} and~\ref{fig:ttgproduction_gg} were included together with the different
decay modes of the $W$ boson, and interference effects were properly taken into account.

The masses of the $u$-,~$d$-,~$s$- and~$c$-quarks, and the masses of electrons
are much smaller than the typical scale of the hard $\ttg$ process, which is of the order of the top quark mass.
In order to simplify the calculation of the MEs, the masses of these particles were approximated by zero.
The masses of muons and $\tau$-leptons were set to \mbox{$m_\mu = 105 \MeV$} and \mbox{$m_\tau = 1776 \MeV$}, respectively.
The masses of $b$- and top quarks were set to \mbox{$m_b = 4.2 \GeV$} and \mbox{$m_t = 172.5 \GeV$}, respectively.

Since the photon is massless, the radiation of photons from charged particles is collinear and infrared divergent.
Hence, the phase space for photon radiation needs to be reduced in finite order calculations in order to obtain a finite cross section for the
$\ttg$ process:
infrared divergencies were avoided by requiring a minimal $\pt$ of \mbox{$8 \GeV$} for photons from the hard process.
To avoid collinear divergencies, the invariant masses of pairs of massless particles were required to be larger than \mbox{$5 \GeV$}.
For consistency, the invariant mass cuts were not only applied for electrons, but also for muons and $\tau$-leptons.

The following invariant masses were considered in the single lepton decay mode:
$m(q_1, q_2)$,
$m(q_1, \gamma)$,
$m(q_2, \gamma)$,
$m(l, \gamma)$,
$m(Q_1, \gamma)$,
$m(Q_2, \gamma)$,
$m(g_1, q_1)$,
$m(g_1, q_2)$,
$m(g_2, q_1)$, and
$m(g_2, q_2)$,
where $q_1$ and $q_2$ are the quarks from the decay of the hadronic $W$ boson,
$l$ is the charged lepton from the decay of the leptonic $W$ boson,
$Q_1$ and $Q_2$ are the incoming quarks in the case of quark-antiquark annihilation,
and $g_1$ and $g_2$ are the incoming gluons in the case of gluon-gluon fusion.
In addition, for each incoming quark~$i$, the invariant mass $m(Q_i, q_j)$ was considered if $q_j$ is the antiparticle of $Q_i$.
The requirements on $Q_1$ and $Q_2$ were dropped for $b$-quarks, because the latter were not assumed to be massless in the event generation.
In the dilepton decay mode the invariant mass criterion on $m(l, \gamma)$ was required to hold for both leptons.

The renormalisation and factorisation scales for the $\ttg$ process were set to $2 m_t$ and
a LO cross section of \mbox{$0.84 \pb$} was calculated with WHIZARD.
As discussed in Sec.~\ref{sec:ttgSM}, the $k$-factor for the $\ttg$ process was estimated to \mbox{$2.6 \pm 0.5$}, which leads to a prediction for the
$\ttg$ cross section times branching ratio (BR) into the single lepton and dilepton channels of
\mbox{$\sigma_{t\bar{t}\gamma} \cdot \mathrm{BR} = (2.1 \pm 0.4) \pb$}.

For the ME calculation, CTEQ6L1 PDFs were used.
Modified LO PDFs based on the MRST~\cite{mrst} set, called \texttt{MRST2007lomod}, available from the
LHAPDF package~\cite{lhapdf}, were used for the parton shower generation.
Parton shower and underlying event were added to the $\ttg$ events using HERWIG (version~6.510) and JIMMY, respectively.

\section{Background modelling}
\label{sec:backgroundmodelling}

Traditionally, many of the background processes to $\ttbar$ events are estimated from MC simulations, because they provide a good description of the data.
Backgrounds from multijet events, however, are known to be poorly modelled by MC generators and need to be estimated from data.
A region which is strongly enhanced in multijet production is defined, and the contribution in the signal region is obtained with
extrapolation methods.
Details on the treatment of the multijet background in this analysis are given in Sec.~\ref{sec:QCDgamma}.

At hadron colliders, additional jets are frequently produced by ISR and FSR of quarks and gluons.
Hadrons in these jets can be misidentified as photons (hadron fakes) and therefore analyses with photons in the final state
typically feature sizable background contributions from this source.

For two reasons, the use of MC simulations for the hadron fake contribution is disfavoured with respect to estimates from data:
firstly, the exact simulation of geometric shapes of the electromagnetic clusters in the calorimeter (shower shapes) requires very detailed
detector understanding and is hence challenging~\cite{electronperformance}.
Secondly, the description of jet fragmentations with a leading neutral hadron ($\pi^0$, $\eta$, \dots) by MC simulations is known to be difficult.
Such neutral hadrons are likely to give rise to photon-like signals via the decay into two photons.

Hence, the strategy of this analysis was set up in a way to minimise the dependence on MC simulations (Ch.~\ref{sec:strategy}).
In particular, the background from hadrons misidentified as photons was estimated from data.

The estimates from MC simulations only make up a small part of the final background estimate, but MC simulations were also used for cross-check
studies.
The MC modelling of the different background contributions is described in the following.
%\newline

\subsubsection{\bf Top quark pair production}
$\ttbar$ events were produced with the MC@NLO~\cite{mcatnlo} generator (version~3.41) using CTEQ6.6~\cite{cteq66} PDFs.
For the simulation of the parton shower and the underlying event, MC@NLO was interfaced to the HERWIG (version~6.510) and JIMMY generators, for which
the AUET1 tune to ATLAS data~\cite{herwigjimmytunes} was used.
The $\ttbar$ cross section was calculated to \mbox{$165 \, ^{+11}_{-16} \pb$} in approximate NNLO with HATHOR~\cite{hathor}.

Photons are also produced in the $\ttbar$ simulation, which may lead to a $\ttg$ signature:
two kinds of processes can occur:
HERWIG produces real photons in the fragmentation processes and also
allows for QED corrections in the production and decay of the top quark pairs using the PHOTOS package~\cite{photos}.
In order to avoid double-counting of $\ttg$ events in the samples generated with WHIZARD and MC@NLO,
contributions which fulfil the requirements for the signal phase space, as defined in Sec.~\ref{sec:signalmodelling}, were removed from the MC@NLO sample.

\begin{figure}[h]
\begin{center}
\includegraphics[width=0.4\textwidth]{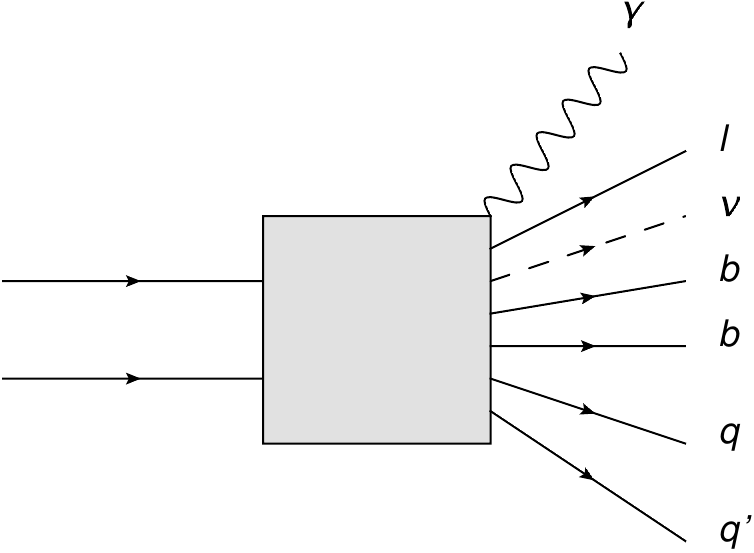} \qquad
\includegraphics[width=0.4\textwidth]{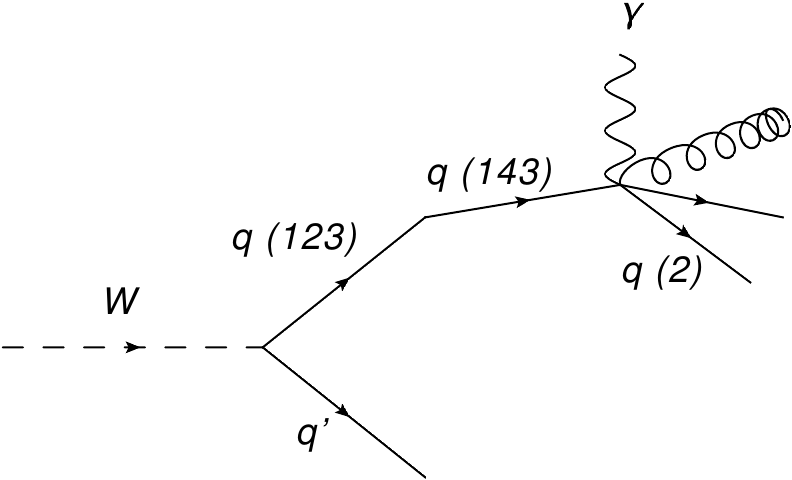}
\caption[Illustration of $\gamma$ radiation from quarks in $\ttbar$ events in WHIZARD and MC@NLO]{
  Illustration of the different treatment of photon radiation from quarks in $\ttbar$ events in the WHIZARD~(left) and
  MC@NLO~(right) MC generators.}
\label{fig:MCoverlap}
\end{center}
\end{figure}

The definition of the signal phase space involves the invariant mass cuts described in Sec.~\ref{sec:signalmodelling}.
Due to the different handling of photon radiation in WHIZARD and HERWIG+PHOTOS, the application of these cuts is not trivial as illustrated
in Fig.~\ref{fig:MCoverlap}:
in WHIZARD, photon radiation is handled as part of the ME calculation and the invariant mass cuts are applied to the seven-particle final
state (left plot).
In the MC@NLO $\ttbar$ sample, photon radiation is added a posteriori.
Photons can be radiated from the incoming quarks, the top quarks, the $W$ boson or the decay products of the $W$ boson (not illustrated in the figure).

Particularly different with respect to WHIZARD is the treatment of the radiation from quarks, which is part of the parton shower process in
HERWIG (right plot):
a quark from the $W$ decay (HERWIG status code~123 or~124) is translated to a jet four-vector (status code~143 or~144)~\cite{herwig}.
%Energy and momentum do not need to be conserved in this step.
The constituents of the jet, which may comprise photons, are listed as the decay products of the jet four-vector.
Thus, photon radiation is not treated in a single step, but as part of the parton shower process, and the
definition of the quark to be considered for \mbox{$m(q, \gamma)$} is ambiguous.

The following approach is considered closest to the treatment in WHIZARD:
the four-momenta of all particles in the parton shower are added up except for the photon.
The combined four-momentum is then used as an estimate for
the four-momentum of the quark \textit{after} photon radiation in the evaluation of \mbox{$m(q, \gamma)$}.
However, a systematic uncertainty is assigned to this ambiguity as described in Sec.~\ref{sec:QCDgamma}.

For the study of systematic uncertainties related to MC generators, additional samples for the $\ttbar$ process were available.
POWHEG~\cite{powheg} was used as an alternative NLO generator.
It was interfaced with HERWIG and JIMMY as well as with PYTHIA (version 6).
For studies of the amount of ISR and FSR, different samples with modified values for the ISR and FSR parameters
were generated with AcerMC~\cite{acermc} interfaced to PYTHIA.
The parameters were varied in ranges currently not excluded by experimental data~\cite{CSCbook}, comparable to
those used in the Perugia Soft/Hard tune variations~\cite{perugia}.
%\newline

\subsubsection{\bf Single top production}
%{\bf Single top production:}
Single top processes were simulated with the MC@NLO generator interfaced to HERWIG and JIMMY.
In order to avoid an overlap of the $Wt$-channel with $\ttbar$ production, the diagram-removal scheme described in Ref.~\cite{WtchannelDiagramRemoval} was
applied.

The single top cross sections were taken from the approximate NNLO calculations in the $t$-, $s$- and $Wt$-production channels, which
yielded \mbox{$64.6 \, ^{+2.7}_{-2.0} \pb$}, \mbox{$4.6 \pm 0.2 \pb$} and
\mbox{$15.7 \pm 1.1 \pb$}~\cite{singletop_t_xsec,singletop_s_xsec,singletop_Wt_xsec}.
%\newline

\subsubsection{\bf \boldmath$W$ boson production in association with jets ($W$+jets)\unboldmath}
$W$+jets events were simulated with the ALPGEN~\cite{alpgen} generator using the CTEQ6L1~\cite{CTEQ} PDFs.
For the simulation of the parton shower and the underlying event, ALPGEN was interfaced to the HERWIG and JIMMY generators, for which
the AUET1 tune to ATLAS data for CTEQ6.1~\cite{herwigjimmytunes} was used.

The process was simulated for different parton multiplicities in the ME, and the matching to the parton shower was applied exclusively for
parton multiplicities smaller than five and inclusively for the \mbox{$2 \to 7$} process with five partons in the final state.
In order to enhance statistics for $W$ boson production processes with additional heavy flavour jets ($c$- or $b$-jets), additional samples for the
processes $W$+$c$+jets, $W$+$c\bar{c}$+jets, and $W$+$b\bar{b}$+jets with up to five partons in the final state were produced.
The overlap with the flavour inclusive samples was removed in order to avoid double-counting~\cite{shermanThesis}.

The cross sections of the different parton multiplicity samples were calculated in NNLO using the FEWZ~\cite{fewz} and ZWPROD~\cite{zwprod} codes.
One inclusive $W$+jets sample was constructed by weighting the individual samples corresponding to their integrated luminosities.
%\newline

The uncertainty on the cross section was evaluated using Berends-Giele scaling~\cite{berendsgiele1, berendsgiele2}:
the uncertainty increases with the number of jets and was estimated to 24\% for each jet in the final state, to be treated uncorrelated.
Hence, for events with at least four jets, an uncertainty of 48\% was obtained.
An additional uncertainty on the amount of the fraction of $b$- and $c$-quarks produced in association with the $W$ boson was added in quadrature:
the fraction of events with a $b\bar{b}$ or $c\bar{c}$ pair was scaled by a factor of \mbox{$1.63 \pm 0.76$} with respect to the generation with ALPGEN.
The fraction of events with only one $c$-quark was scaled by a factor of \mbox{$1.11 \pm 0.35$}.
The fraction of events without $c$- or $b$-quarks was corrected correspondingly in order to preserve the total predicted number of $W$+jets events.

\subsubsection{\bf \boldmath$Z$ boson production in association with jets ($Z$+jets)\unboldmath}
$Z$+jets production was treated similarly to $W$+jets production:
ALPGEN, HERWIG and JIMMY were used with the same PDFs and generator tunes.
Also, the same scheme for the generation of samples with the different parton multiplicities was applied.
The statistics for $Z$+jets production with heavy flavour jets was enhanced by the use of $Z$+$b\bar{b}$+jets samples with up to five partons in the final
state.
The overlap with the flavour inclusive samples was removed.

As for $W$+jets production, the cross sections were calculated in NNLO with the FEWZ and ZWPROD codes, and one inclusive
$Z$+jets sample was constructed by a reweighting of the individual samples.
%\newline

The uncertainty on the cross section was estimated to be the same as for $W$+jets production, that is 48\% for events with at least four jets in the final
state.

\subsubsection{\bf \boldmath$W$ boson production in association with jets and photons ($W$+jets+$\gamma$)\unboldmath}
$W$+$\gamma$ events with additional jets in the final state were generated with ALPGEN.
Parton shower and underlying event were added using HERWIG and JIMMY, and the CTEQ6L1 PDFs were used~--~as for the $W$+jets
sample.
As for the $W$+jets generation, the matching of the parton shower to the ME was performed exclusively for the zero to four parton
samples, and inclusively for the five parton sample (\mbox{$2 \to 8$} process).
To avoid infrared divergencies, a minimum $\pt$ of \mbox{$10 \GeV$} was required for the photon.

The relative weights of the different parton multiplicities were taken from the ALPGEN event generation.
The absolute cross section for the $W$+jets+$\gamma$ sample was used for the measurement presented in this thesis.
%\newline

\subsubsection{\bf Diboson production (\boldmath$WW$, $WZ$, $ZZ$)\unboldmath}
%{\bf \boldmath Diboson production ($WW$, $WZ$, $ZZ$):\unboldmath}
$WW$, $WZ$ and $ZZ$ events were generated with HERWIG.
The cross sections calculated with HERWIG were corrected by $k$-factors obtained with the MCFM code~\cite{mcfm}, which read
1.48 for $WW$, 1.30 for $WZ$, and 1.60 for $ZZ$ production.
The uncertainties on the diboson cross sections were estimated to 5\% following the approach in Ref.~\cite{ATLAStopXsec_3pb}.
\newline

For the simulation of pile-up effects, minimum bias events generated with PYTHIA were overlayed to the hard processes.
The pile-up configuration corresponded to a bunch spacing of \mbox{$50 \ns$}, which represents most of the data taken in 2011~\cite{fournier}.
The pile-up rate was kept variable in the simulation of the minimum bias events, and the MC samples were reweighted so that the distribution of
the number of interactions per bunch crossing was the same in MC simulations and in the data analysed (Fig.~\ref{fig:mu}).

\chapter{Object definitions}
\label{sec:objects}

Particles traversing the detector produce characteristic signatures in the different subdetectors (Sec.~\ref{sec:ATLAS}),
which are used to identify the particle type.
Combined objects were constructed using information from the Inner Detector~(ID), the calorimeter system and the muon spectrometer~(MS), which were then
compared to the object definitions for electrons (Sec.~\ref{sec:electron}), muons (Sec.~\ref{sec:muon}), jets (Sec.~\ref{sec:jet}) and photons
(Sec.~\ref{sec:photon}).
These definitions combine a high probability for the identification of real electrons, muons, jets and photons
with a low probability of misidentifying another object as one of these particles.
Moreover, $b$-tagging was used to identify jets from $b$-quarks (Sec.~\ref{sec:btagging}).

Since the initial momentum of the colliding partons in the transverse plane is small, an imbalance of the transverse momentum of the whole event
(missing transverse energy $\met$) indicates the presence of high-energetic, undetected particles, such as neutrinos.
The energy and momentum measurements in the whole detector were used to measure $\met$ (Sec.~\ref{sec:MET}).

\section{Electron definition}
\label{sec:electron}

Electron candidate objects~\cite{electronperformance} were selected by searching for large energy deposits in the electromagnetic (EM) calorimeter using
a fixed-size window in $\eta$-$\phi$-space with an ID track pointing in its direction.
The energy deposited in the EM cluster was corrected for energy losses in front of the calorimeter, lateral leakage outside of the
cluster window and energy deposited behind the calorimeter.
The electron four-vector was built from the cluster energy and the direction measurements from the ID track.%,

Only the central part of the calorimeter was used and $|\eta_{\mathrm{cluster}}|$ was required to be smaller than 2.47.
The transition region from the barrel to the endcap calorimeter, \mbox{$1.37 < |\eta_{\mathrm{cluster}}| < 1.52$}, was not considered.
The $\et$ of the electron was required to be larger than \mbox{$25 \GeV$} in the analysis, but also electrons with an $\et$ down
to \mbox{$15 \GeV$} were used for the estimation of the photon isolation properties (Ch.~\ref{sec:photontemplate}).

In order to suppress backgrounds from other particles misidentified as electrons, cut-based sets of quality criteria (menus)
provided increasing background rejection: the so-called \texttt{loose}, \texttt{medium} and \texttt{tight} menus.
In the following, the \texttt{tight} menu is described, which yields an overall efficiency of roughly 75\%.

Electron clusters tend to be smaller in size than clusters from hadrons within jets.
Hence, several observables constructed from the geometrical shape of EM clusters (shower shapes), such as their lateral width or
the energy in the highest-energetic calorimeter cells, were used in the \texttt{tight} menu.
Also, the fraction of the energy deposited in the hadronic calorimeter, which is typically very small for electrons, was exploited to suppress
backgrounds from jets.
The transition radiation in the TRT was used in addition to discriminate against charged hadrons.
In order to assure that tracks are not accidentally associated to clusters, quality criteria on the number of hits in the silicon trackers,
a good geometrical matching of the track direction and the cluster position as well as of the track momentum and the cluster energy were required.
The track also had to point back closely to the primary vertex.
Backgrounds from converted photons (\mbox{$\gamma \to e^+e^-$}) were suppressed by requiring a hit in the Pixel $b$-layer, because most photons
do not convert before they reach this first detector layer.

Electrons which were also reconstructed as photon candidates were not considered, but treated as photons
(cf. \textit{photon recovery procedure} in Sec.~\ref{sec:photon}).
To further suppress backgrounds from jets, the energy of the electron candidate in a cone of
\mbox{$\Delta R = 0.2$} around the electron
direction\footnote{$\Delta R = \sqrt{\Delta \eta^2 + \Delta \phi^2}$, with
\mbox{$\Delta \eta = \eta_{\rm cell} - \eta$} and
\mbox{$\Delta \phi = \phi_{\rm cell} - \phi$}.} (isolation energy) was required to be less than \mbox{$3.5 \GeV$}.
The isolation energy was corrected for average energy deposits from pile-up events.
Since the electron identification and jet finding algorithms (Sec.~\ref{sec:jet}) are independent of each other, most electrons were also reconstructed
as jets.
In order to avoid double-counting, jets which were closer than 0.2 in $\eta$-$\phi$-space to an electron were disregarded.
Electrons close to a region in the EM calorimeter which was known to feature a broken optical link (Ch.~\ref{sec:data})
or a dead high-voltage channel were ignored.
MC simulations were corrected for this effect.

Electron signatures were used in the trigger menus at L1, L2 and EF (Sec.~\ref{sec:triggerDAQ}).
At L1, trigger towers
%of $0.1 \times 0.1$ in $\eta$-$\phi$-space
above certain $\et$ thresholds were searched for.
% with a coarse granularity.
At L2, a simplified version of the offline reconstruction algorithm
% described above
was used.
The final trigger decision at EF level used the full offline algorithm with slightly looser requirements. % than applied offline.
For the data analysed, the \texttt{EF\_e20\_medium} trigger was used for the selection of candidate events in the electron channel.
This trigger required a minimum $\et$ of \mbox{$20 \GeV$} at EF level and the \texttt{medium} trigger shower shape menu.
The thresholds at L1 and L2 had to be adjusted during data taking to maintain manageable event rates given the increasing instantaneous luminosity.

%An electron object was considered \textit{matched} to the trigger object, if the distance in $\eta$-$\phi$-space to the trigger object was
%smaller than 0.15.
%
Trigger, reconstruction and identification efficiencies were measured in data using tag-and-probe methods in \Zee
and \mbox{$W \to e\nu$} events~\cite{electronperformance}:
while one electron was used to define the sample (tag),
the efficiencies of the other electron (probe) were measured.
In case of $W$ events, $\met$ was used as a tag.
%To illustrate this, a comparison of data and MC efficiencies for a looser electron identification (\texttt{loose++} menu) than used in this analysis
%is shown in Fig.~\ref{fig:electron_efficiency}.
Discrepancies between data and MC were accounted for by scale factors (SFs):
\mbox{$\mathrm{SF}(\et, \eta) = \varepsilon_{\mathrm{data}}(\et, \eta) / \varepsilon_{\mathrm{MC}}(\et, \eta)$},
where $\varepsilon_{\mathrm{data}}$ and $\varepsilon_{\mathrm{MC}}$ are the efficiencies in data and MC, respectively.
MC events were weighted with the SFs, so that the efficiencies in the MC simulations yielded those in data.

The SFs for the \texttt{EF\_e20\_medium} trigger were derived in 18 bins in $\eta$ and vary between 0.97 and~1.00 with uncertainties smaller than 0.01.
The SFs for the electron reconstruction were derived in three bins in $|\eta|$ and vary between 0.98 and~1.01 with uncertainties smaller than 0.02.
For the electron identification, the SFs were divided into 18 bins in $\eta$ and five bins in $\et$.
The range for these SFs is 0.95~--~1.12 with uncertainties smaller than 0.04.

The energy scale and resolution of electrons was studied in data using \Zee events~\cite{electronperformance}.
In order to match the distribution of the di-electron invariant mass in data, the electron energy resolution was corrected in MC simulations.
Additionally, small corrections to the energy scale in data were applied.

\section{Muon definition}
\label{sec:muon}

Muons were reconstructed using tracks in the MS and in the ID using the \texttt{MuId} algorithm\footnote{The
\texttt{MuId} algorithm is also called ``chain 2'' reconstruction in Ref.~\cite{muonMBperf} and~\cite{muonPerf}.}~\cite{muonMBperf,muonPerf}.
The pattern recognition started from track segments in the MS, which were extrapolated to the ID.
If a matching ID track for a MS track segment was found, a combined fit to the ID and MS muon hits was performed and a combined muon
object was built.
The muon momentum was measured from the combined fit.
Only muons with a $\pt$ of at least \mbox{$20 \GeV$} fulfilling \mbox{$|\eta| < 2.5$} were considered in this analysis.

Several isolation criteria were applied in order to suppress backgrounds from muons produced in jets, in particular muons from leptonically decaying
$B$-mesons in $b$-jets.
The energy deposited in the calorimeter in a cone of \mbox{$\Delta R = 0.3$} around the muon candidate was required to be smaller than \mbox{$4 \GeV$}.
Also the sum of the transverse momenta of the tracks in the ID in a cone of \mbox{$\Delta R = 0.3$} around the muon track had to be smaller than \mbox{$4 \GeV$}.
Additionally, if a jet with minimum $\pt$ of \mbox{$20 \GeV$} was closer than 0.4 in $\eta$-$\phi$-space to the muon, the muon was not considered isolated
and was disregarded.

Muon signatures were used in the trigger menu at L1, L2 and EF (Sec.~\ref{sec:triggerDAQ}).
The triggers at L1 were based on coincidences in $\eta$ and $\phi$ of hits in the different layers of the RPC and TGC systems requiring a minimum
$\pt$ of the trigger object candidate.
At L2 and EF level, track finding algorithms using information from the MS and the ID were used.
In this analysis, the \texttt{EF\_mu18} trigger chain was used which required a combined muon with \mbox{$\pt > 18 \GeV$} at EF level.

Trigger, reconstruction and identification efficiencies were measured in data using tag-and-probe methods in \mbox{$Z \to \mu^+\mu^-$} events.
Discrepancies between data and MC were accounted for by SFs~\cite{muonPerf}.
MC events were weighted with the SFs, so that the efficiencies in the MC simulations yielded those in data.
Events in which the muon object had a $\pt$ larger than \mbox{$150 \GeV$} were disregarded, because muon trigger SFs were not available for larger
$\pt$ due to an issue with the trigger modelling in MC simulations in this region.

The SFs for the \texttt{EF\_mu18} trigger were derived in three bins in $\pt$, seven bins in $\eta$ and three bins in $\phi$.
For \mbox{$\pt < 120 \GeV$}, they vary between 0.88 and 1.21 with uncertainties up to 0.09 in the barrel, and between 0.98 and 1.04
with uncertainties smaller than 0.02 in the endcap region.
For larger transverse momenta, the measurement from $Z$ boson decays was limited by statistics and SFs down to 0.66 and up to 1.48
were derived in certain regions of the phase space with correspondingly large uncertainties.
The SFs for the muon reconstruction were derived in 20 bins in $\eta$ and vary between 0.98 and~1.01 with uncertainties which do not exceed 0.005.
For the muon  identification, one inclusive SF was derived, which reads:
\mbox{$1.0008 \pm 0.0003 \, \mathrm{(stat.)} \pm 0.0003 \, \mathrm{(syst.)}$}.

The momentum scale and resolution of muons were measured in data using \mbox{$Z \to \mu^+\mu^-$} events \cite{muonResolution}.
In order to match the distribution of the di-muon invariant mass in data, the muon momentum scale and resolution were
corrected in MC simulations.

\section{Jet definition}
\label{sec:jet}

Jets were reconstructed with the anti-$k_t$ algorithm~\cite{antikt,fastjet} with a distance parameter of \mbox{$R = 0.4$}.
Jets were built from topological calorimeter clusters based on the significance of the energy deposit in the calorimeter cells with respect
to their noise level~\cite{2004TB}.
The clusters, and hence the jets, were calibrated to the EM scale, which is smaller than the hadronic energy scale due to the
non-compensating nature of the ATLAS calorimeters.
The hadronic energy scale was restored using MC based correction factors depending on the jet $\pt$ and $\eta$~\cite{jetperf}.
In this analysis, jets were required to have a minimum $\pt$ of \mbox{$25 \GeV$}.
This requirement was lowered to \mbox{$20 \GeV$} for the muon isolation criterion.

In order to ensure that the selected jets originated from the hard scattering process,
quality criteria~\cite{jetperf} were applied to all jets with a minimum $\pt$ of \mbox{$20 \GeV$}.
These criteria strongly reduce beam-gas and beam-halo events arising from the interactions of the proton
beam with gas in the beam pipe and with the collimators upstream the detector, respectively.
They also suppress backgrounds from cosmic ray muons and from calorimeter signals which were due to large electronics noise.
Since all of these background contributions tend to create artificial high-$\pt$ signals in the detector, the rejection of events with jets failing the
criteria improves the quality of the $\met$ measurement.

The uncertainty on the jet energy scale (JES) was derived from the response of single hadrons from test beam measurements, in-situ techniques applied in
collision data, and input from MC simulations~\cite{jetperf}.
Dijet events were used to test the relative energy scale of the different regions of the detector given the varying design of the calorimeter and material
in front of the calorimeter.
The JES was validated in-situ comparing the jet energy to the momentum carried by the tracks associated to the jet, and by exploring
the balance in $\pt$ of $\gamma$+jet events, photon events with hadronic recoil, and multijet events with one high-$\pt$ jet~\cite{jetperf}.
The total JES uncertainties in the ranges \mbox{$0.3 \leq |\eta| < 0.8$} and \mbox{$2.1 \leq |\eta| < 2.8$} were presented in
Fig.~\ref{fig:jetperformance}.

The jet reconstruction efficiency was measured comparing jets built from calorimeter clusters to jets built from tracks~\cite{jetperf}.
The jet energy resolution was estimated using the $\pt$-balance in dijet events~\cite{JER}.

\section{Definition of the missing transverse energy}
\label{sec:MET}

The measurement of $\met$~\cite{METperf} was based on the energy measurement in the calorimeters and the muon momentum measurements in the MS.
In order to calibrate the different energy depositions in the calorimeter correctly, different objects were identified and their energy was taken into
account at their correct energy scale.
Firstly, electrons and photons with a minimum $\et$ of \mbox{$10 \GeV$} were identified requiring the respective \texttt{tight} shower shape menus
$\left(E_{x,y}^{\rm electrons} \; {\rm and} \; E_{x,y}^{\rm photons}\right)$.
Secondly, anti-$k_t$ jets with a $\pt$ of at least \mbox{$20 \GeV$} were taken into account at the hadronic energy scale $\left(E_{x,y}^{\rm jets} \right)$.
Jets with transverse momenta between~$7$ and \mbox{$20 \GeV$} (soft jets) were not calibrated to the hadronic scale, but were included in the $\met$ definition
at the EM scale $\left(E_{x,y}^{\rm soft \, jets}\right)$.
The energy in topological clusters not used in any of the reconstructed objects was also included at the EM scale $\left( E_{x,y}^{\rm cell \, out} \right)$.

Muons reconstructed with the \texttt{MuId} algorithm were taken into account up to \mbox{$|\eta| < 2.7$}.
They were required to feature a combined ID and MS track for \mbox{$|\eta| < 2.5$}.
The energy deposition in the calorimeter around the muon track was included in case the muon was separated from any anti-$k_t$ jet by a minimum distance
of 0.3 in $\eta$-$\phi$-space
$\left(E_{x,y}^{\rm muons}\right)$.
Hence, the full definition of the $\met$ reads\footnote{$E_x$ and $E_y$ are defined in analogy to the transverse energy $\et$.}:
\begin{eqnarray}
  \mexy & = & - \left( E_{x,y}^{\rm electrons} + E_{x,y}^{\rm photons} + E_{x,y}^{\rm jets} + E_{x,y}^{\rm soft \, jets} + p_{x,y}^{\rm muons}+  E_{x,y}^{\rm cell \, out} \right)
  \; , \nonumber \\
%  \met & = & \sqrt{\left(\mex\right)^2+\left(\mey\right)^{2}}
  \met & = & \sqrt{\mex^2+\mey^{2}} \; .
  \label{eq:met}
\end{eqnarray}

As mentioned in Sec.~\ref{sec:jet}, the quality of the $\met$ measurement was assured by rejecting events where a jet with \mbox{$\pt > 20 \GeV$}
did not fulfil the jet quality criteria.
Also events with jets with \mbox{$\pt > 20 \GeV$} closer than 0.1 in $\eta$-$\phi$-space to a region which was known to feature
a broken optical link (Ch.~\ref{sec:data}) or a dead high-voltage channel in the LAr calorimeter were disregarded.
MC simulations were corrected for this effect.
Events with large noise in the LAr calorimeter were also rejected.

The object definitions used for the $\met$ calculation were chosen consistent with those described in Sec.~\ref{sec:electron}, \ref{sec:muon},
\ref{sec:jet} and~\ref{sec:photon}, and the uncertainties of the energy and momentum scales of the different $\met$ contributions were propagated
consistently to the $\met$.

\section[$b$-tagging]{\boldmath$b$-tagging\unboldmath}
\label{sec:btagging}

$b$-tagging was used to identify jets originating from the $b$-quarks of the top quark decays.
The relatively long lifetime of $B$-mesons leads to displaced secondary vertices and impact parameters of their decay products.
Algorithms were constructed to discriminate $b$- from light jets exploiting these properties.
Two $b$-tagging algorithms were combined in this analysis~\cite{btaggingperf,jetfitter}: \texttt{I3PD} and \texttt{JetFitter}.

\texttt{JetFitter} makes use of the topology of weak $b$- and $c$-hadron decays inside jets to discriminate between $b$-, $c$- and light jets.
Different properties of the reconstructed decay vertices, such as their masses, momenta, flight-length significances and track
multiplicities, were used to build a likelihood discriminant.
The second algorithm, \texttt{I3PD}, uses the transverse and longitudinal impact parameter significances $d_0 / \sigma_{d_0}$ and $z_0 / \sigma_{z_0}$
of the tracks associated to the jet.
Since both algorithms use different observables sensitive to the jet flavour, their combination is advantageous.
A single discriminating observable for $b$-tagging was constructed using a neural network.

For this analysis, a working point for the neural network output was chosen such that approximately 70\% of the $b$-jets in simulated $\ttbar$ events
were correctly $b$-tagged.
This corresponds to misidentification rates of roughly 20\% for $c$- and 1\% for light flavour jets.

The $b$-tagging efficiency was measured in data using different techniques~\cite{btagcalibration, system8}.
These exploit in particular the relative $\pt$ of the muon from semileptonic $b$-decays and the reconstruction of $D^{*+}$ mesons from
\mbox{$b \to X \mu D^{*+}$} decays.
Mistagging rates were measured using the invariant mass of the tracks associated to a secondary vertex and events with negative impact parameters.
From these measurements, SFs were derived to reproduce the data tagging and mistagging efficiencies in MC simulations.

The SFs for the $b$-tagging efficiency were derived in five bins in $\pt$, and vary between 0.94 and 0.99 with uncertainties between 0.06 and 0.15.
The mistagging SFs were estimated in eight bins in $\pt$ for jets with \mbox{$|\eta| < 1.2$} and \mbox{$1.2 < |\eta| < 2.5$}, and they vary between
0.98 and 1.27 with uncertainties between 0.11 and 0.22.

\section{Photon definition}
\label{sec:photon}

The reconstruction of photon and electron candidate objects is done by one single \texttt{egamma} algorithm, so that objects are unambiguously
identified as either photons or electrons~\cite{expPhotonPerf}.
The \texttt{egamma} algorithm takes into account that photons may convert into $e^+e^-$ pairs by the interaction with the material in front of the
calorimeter, and different reconstruction paths are foreseen for \textit{unconverted} and \textit{converted} photons.
While unconverted photons do not feature a track pointing to their EM clusters, converted photons typically have two tracks pointing to the cluster
(two-track conversions).
There are also asymmetric conversions into an $e^+e^-$ pair with two very different transverse momenta, so that the low-$\pt$ track is
likely not to be reconstructed, which leads to \textit{one-track conversions}.
%One-track conversion candidates are difficult to discriminate against electron candidates, since both feature exactly one track pointing to a
%cluster in the EM calorimeter.
%However,
The tracks of one-track conversions often miss hits in the first ID layers, depending on where the conversion into the $e^+e^-$ pair took place,
and also energy and momentum measurements tend to be not consistent with an electron object.

Unconverted photons were reconstructed by finding fixed-size tower clusters with significant energy in the EM calorimeter and no
ID track matched to it.
Clusters with tracks associated to them were treated as electrons/conversion candidates.
The next step was the \textit{photon recovery procedure}~\cite{expPhotonPerf}, which resolves the ambiguity between electrons and photons.
This is particularly important for converted photons, but also unconverted photons with erroneously associated tracks need to be recovered.

Conversion vertices were reconstructed by either fitting two tracks under the assumption that they originated from a massless particle or by identifying
single tracks that did not feature hits in the inner layers of the ID.
EM clusters were then checked for conversion vertices matching the cluster centre in $\eta$-$\phi$-space when extrapolated to the
calorimeter surface.

Converted photons were identified by comparing the tracks from the associated conversion vertices to the track which matches the cluster best.
Additionally, clusters with the best matching high-$\pt$ track without hits in the silicon trackers (TRT-only track) were considered as conversions
candidates.
Refined requirements were applied using the ratio of calorimeter energy and track momentum ($E/p$), and the presence of hits in the Pixel $b$-layer.
In addition, unconverted photons with low-$\pt$ TRT-only tracks and low-$\pt$ tracks with a large $E/p$ ratio were identified.

A special treatment was applied to converted photon candidates with a track missing a hit in the Pixel $b$-layer, which would have been identified
as an electron candidate if the hit had been present: if the track passed through a module in the $b$-layer which was known to feature readout
problems during data taking, the candidate was not considered a photon.

The photon energy was calibrated using the energy deposited in the EM calorimeter including the presampler.
Cluster sizes of \mbox{$3 \times 5$} and \mbox{$3 \times 7$} cells were used in the second layer of the barrel calorimeter for unconverted and converted
photon candidates, respectively~\cite{CSCbook}.
A larger cluster size was used for converted photons to account for broader showers in $\phi$ due to bremsstrahlung from the $e^+e^-$ pair.
The cluster energy was corrected for energy losses in front of the calorimeter, lateral leakage outside of the cluster and energy deposited behind
the EM calorimeter.
These corrections were parametrised as a function of the energy depositions in the presampler, the three calorimeter layers, and the $\eta$
of the photon object.

Photon four-vectors were constructed from calorimeter information only.
Assuming a photon mass of zero, the cluster energy was used together with the $\eta$- and $\phi$-position of the cluster in the second calorimeter layer,
in which the bulk of the photon energy is typically deposited.

Only photons with \mbox{$|\eta| < 2.37$} were considered, thus limiting the acceptance to the pseudorapidity range with particularly fine granularity
in the first calorimeter layer (LAr strips).
The transition region from the barrel to the endcap calorimeter, \mbox{$1.37 < |\eta| < 1.52$}, was not considered.
Photons were required to have a minimum $\et$ of \mbox{$15 \GeV$}.

A \texttt{loose} and a \texttt{tight} cut-based menu using photon shower shapes and hadronic leakage
were used to purify the sample of selected photon candidates.
In the following, only the \texttt{tight} menu is described.
Since the clusters of unconverted and converted photons had slightly different properties, the cuts on the various observables used in the \texttt{tight}
menu were tuned separately for these two kinds of photons.
The cuts are binned in seven $|\eta|$-regions accounting for the varying amount of material in front of the calorimeter and the different granularities:
\mbox{$[0.0, 0.6)$},
\mbox{$[0.6, 0.8)$},
\mbox{$[0.8, 1.15)$},
\mbox{$[1.15, 1.37)$},
\mbox{$[1.52, 1.81)$},
\mbox{$[1.81, 2.01)$}, and
\mbox{$[2.01, 2.37)$}.
An overview of the different observables used in the \texttt{tight} menu including the definitions and symbols used further on is given in
Tab.~\ref{tab:showershapes}.

\begin{table}[htp]
\centering
\begin{tabular}[h] {|p{0.26\textwidth}|l|p{0.57\textwidth}|}
  \hline
  Category & Symbol & Description \\
  \hline
  Hadronic leakage & $\rhadone$
                   & Ratio of the $\et$ in the first layer of the hadronic calorimeter to the $\et$ of the EM cluster (used for
                     \mbox{$|\eta| < 0.8$} and \mbox{$|\eta| > 1.37$}) \\
                   & $\rhad$
                   & Ratio of the $\et$ in the whole hadronic calorimeter to the $\et$ of the EM cluster (used for \mbox{$0.8 < |\eta| < 1.37$}) \\
  Second calorimeter layer & $\reta$
                   & Ratio of the cell energies in \mbox{$3 \times 7$} and \mbox{$7 \times 7$} cells in $\eta$-$\phi$ \\
                   & $\wtwo$
                   & Lateral width of the shower \\
                   & $\rphi$
                   & Ratio of the cell energies in \mbox{$3 \times 3$} and \mbox{$3 \times 7$} cells in $\eta$-$\phi$ \\
  First calorimeter layer (LAr~strips) & $\wsthree$
                   & Shower width for three cells around the maximum cell in the first layer \\
                   & $\wstot$
                   & Total shower width in $\eta$ in the first layer \\
                   & $\fside$
                   & Fraction of the energy outside of the core of the three central cells but within seven cells in the first layer (in~$\eta$) \\
                   & $\deltae$
                   & Difference between the energy in the second maximum cell and the energy reconstructed in the cell
                     with the minimal value found between the first and second maximum cells in the first layer \\
                   & $\eratio$
                   & Ratio of the energy difference of the two largest energy deposits along $\eta$ in the first layer over the sum of these energies \\
  \hline
\end{tabular}\\
\caption[Overview of the \texttt{tight} menu shower shapes] {
  Overview of the shower shapes used in the \texttt{tight} menu~\cite{expPhotonPerf}.}
\label{tab:showershapes}
\end{table}

\begin{figure}[htp]
\begin{center}
\includegraphics[width=0.6\textwidth]{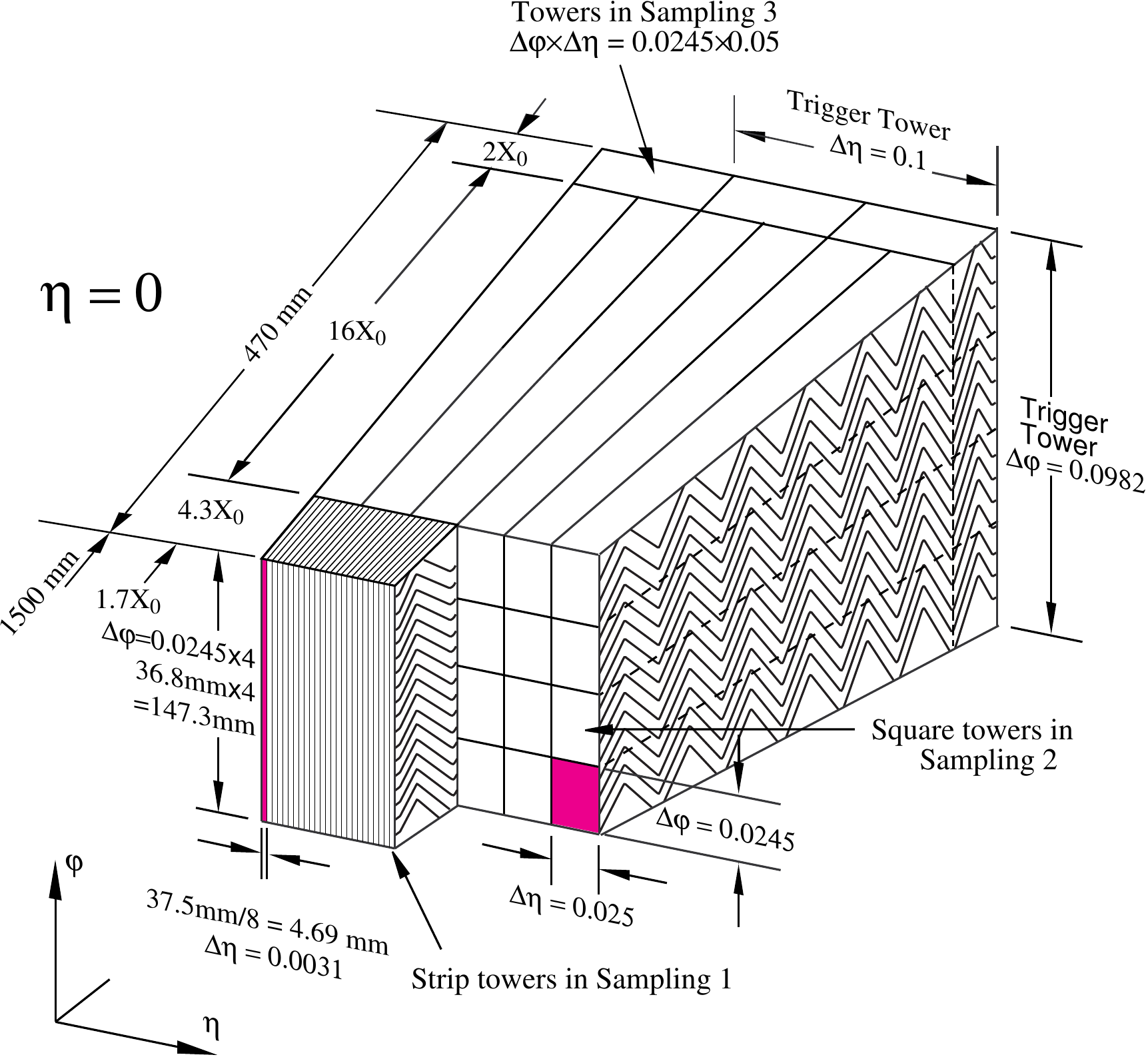}
\caption[Sketch illustrating the LAr strips in the EMB]{
  Sketch of a module of the EMB: in particular, the LAr strips (first longitudinal layer) are depicted~\cite{LArTDR}.}
\label{fig:accordion}
\end{center}
\end{figure}

As for electrons, the energy in the hadronic calorimeter behind the EM cluster (hadronic leakage) and the cluster widths in the second
layer of the EM calorimeter were used to discriminate real photons from hadrons produced in jet fragmentation.
The latter tend to deposit a significant amount of energy in the hadronic calorimeter and produce broader clusters.

In order to suppress backgrounds from jets which fragment with a high-energetic hadron decaying to two photons ($\pi^0$ mesons, $\eta$
and $\eta^\prime$ mesons \dots), the first calorimeter layer (LAr \textit{strips}) was used.
The LAr strips are finely granulated in $\eta$, as illustrated in Fig.~\ref{fig:accordion}.
%, which shows a sketch of the three longitudinal layers of the LAr barrel calorimeter.
The photons from the meson decays are typically very close to each other and therefore give rise to one single cluster in the EM
calorimeter.
However, this cluster originating from two photons tends to be slightly broader than single photon clusters, which is exploited by the
observables $\wsthree$, $\wstot$ and~$\fside$, and also tends to have a second maximum within the cluster, to which $\deltae$ and~$\eratio$
are sensitive.

No isolation criterion was included in the photon definition, although real photons are expected to feature significantly less activity in a small
cone around the photon candidate than fake photons from hadrons inside jets.
However, the track isolation $\ptcone$, as defined in Ch.~\ref{sec:strategy}, was used in a template fit to estimate the amount of
hadrons misidentified as photons directly from data.

Photons close to a region in the EM calorimeter which was known to feature a broken optical link (Ch.~\ref{sec:data})
or a dead high-voltage channel in the LAr calorimeter were ignored.
MC simulations were corrected for this effect.
Moreover, EM clusters with very narrow energy deposits and a large contribution from untypical electronics pulse shapes in the LAr calorimeter
were disregarded, because they were most probable to originate from large electronics noise.

As mentioned already in Sec.~\ref{sec:electron}, the \texttt{egamma} algorithm and the jet finding algorithm are independent of each other.
Hence, not only electrons are double-counted as jets, but also photons.
In order to avoid this effect, the corresponding jets needed to be removed, which was done based on a geometrical matching in $\eta$-$\phi$-space of jets
and photons:
jets which were closer than \mbox{$\Delta R = 0.1$} to photons were disregarded.

\begin{figure}[h]
\centering
\includegraphics[width=0.49\textwidth]{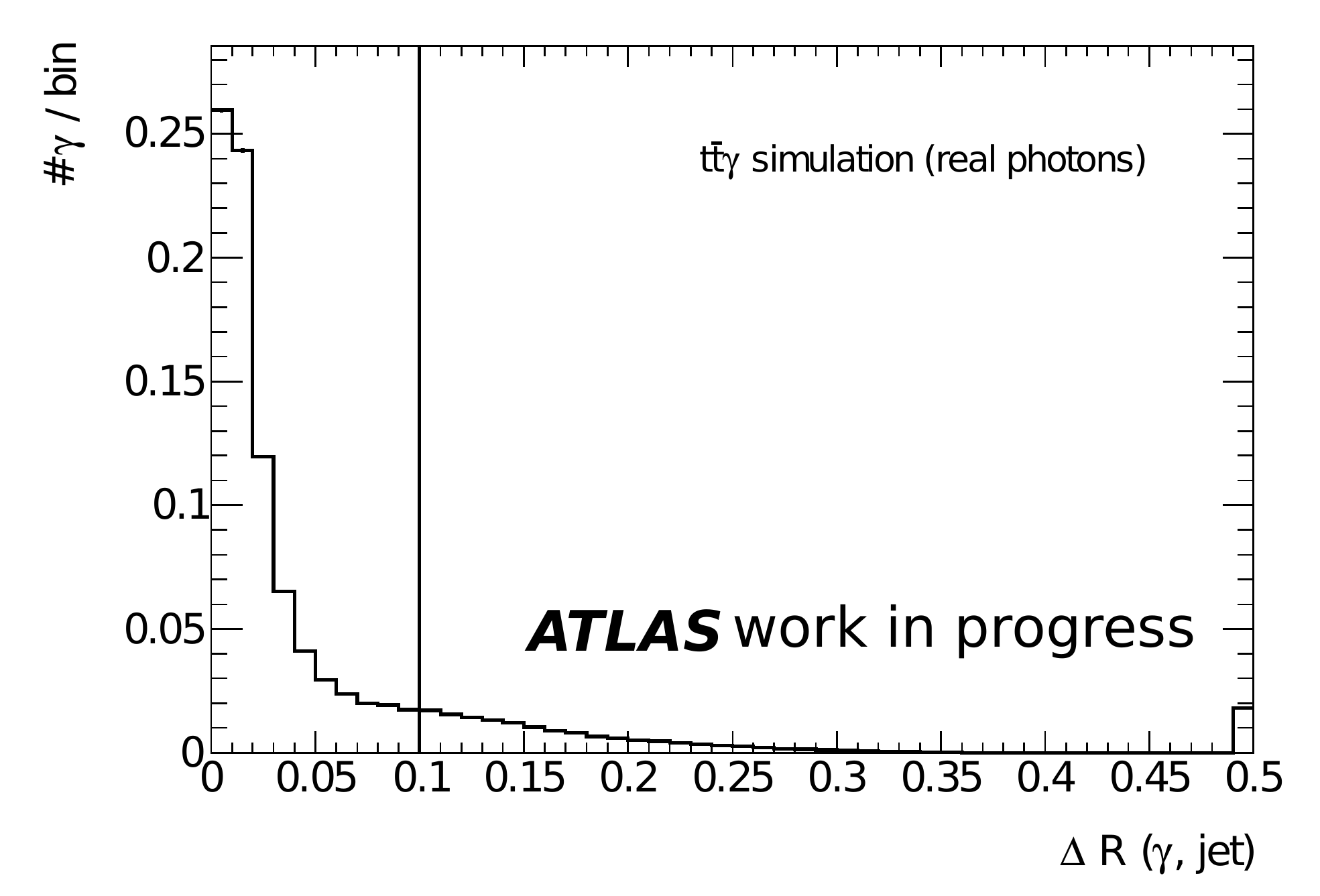}
\includegraphics[width=0.49\textwidth]{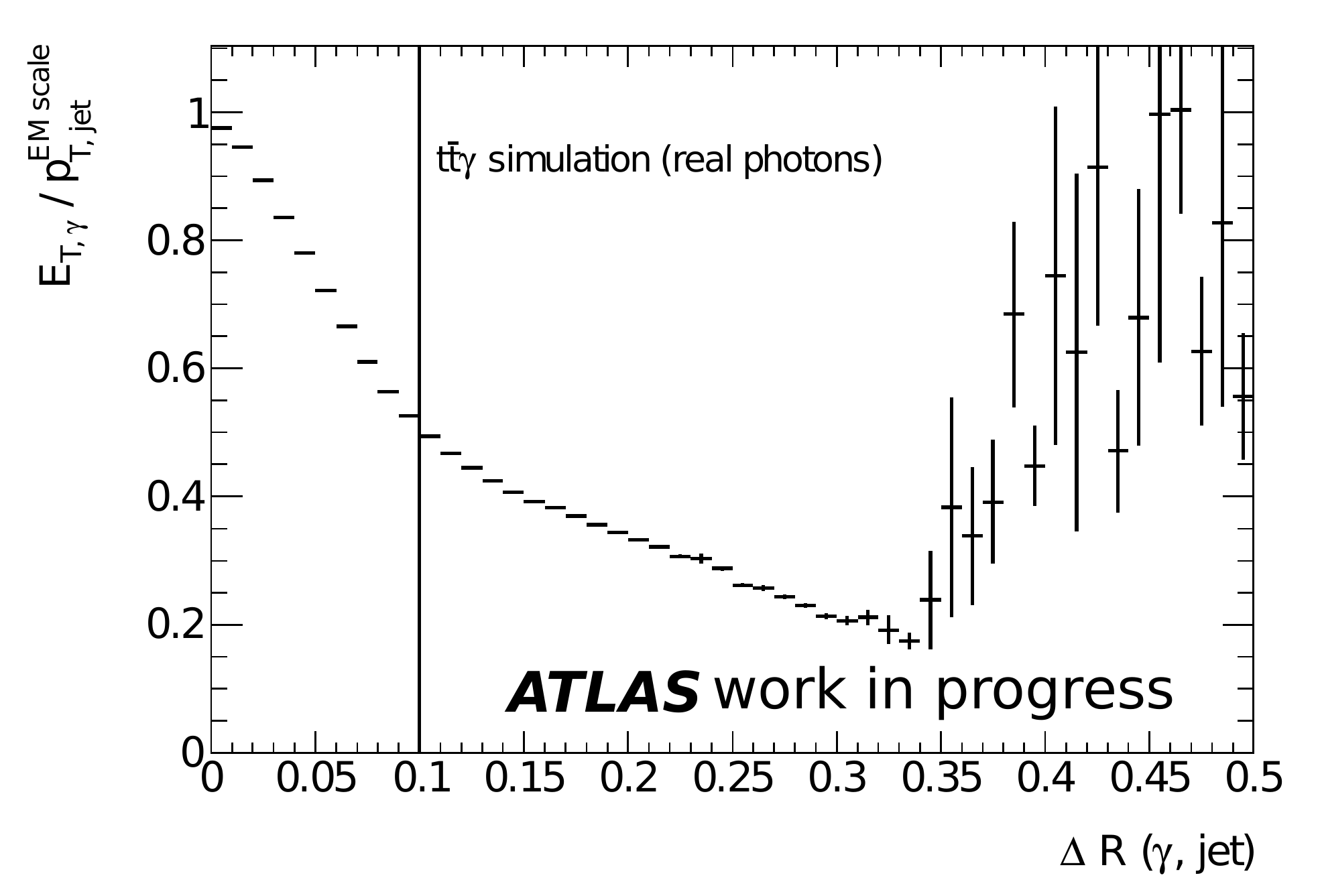}
\caption[Distance between photons and closest jet]{
  Left: the distance in $\eta$-$\phi$-space between real photons and the closest reconstructed jet in $\ttg$ simulation.
  The effect of double-counting real photons by the jet reconstruction is clearly visible.
  The last bin includes the overflow bin.
  Right: the average ratio of real photon $\et$ and closest jet $\pt$ on electromagnetic scale as a function of the distance between photon and
  jet.
  Jets with centres that differ from the photon direction include energy deposits from additional particles.
  The cut applied at \mbox{$\Delta R = 0.1$} is indicated by a vertical line.
}
\label{fig:photonjetoverlap}
\end{figure}

The left plot in Fig.~\ref{fig:photonjetoverlap} shows the distance in $\eta$-$\phi$-space between photons and the closest reconstructed jet in simulated
$\ttg$ events.
The photon objects were required to be real photons, that means originating from true simulated photons.
For roughly 85\% of the photons, the closest jet centre differs by less than 0.1 in $\eta$-$\phi$-space, which indicates that the photon was also
reconstructed as a jet and was hence double-counted.

The right plot in Fig.~\ref{fig:photonjetoverlap} shows the average ratio of the transverse energy of real photons and the transverse momentum
of the jet closest to the photon as a function of the distance between the photon and the jet.
The jet $\pt$ was taken on the EM scale to be comparable to the photon $\et$.
In cases where the jet axis was very close to the photon direction, the jet $\pt$ was found to be similar to the photon $\et$, which means that the
jet does not contain additional particles and mainly consists of the photon object.
If the jet algorithm picks up additional particles close to the real photon, the $\pt$ of the jet exceeds the photon $\et$ and also the jet axis
differs from the photon direction.
Larger distances between photons and jets indicate that more additional energy was included in the jet on average.

Hence, there are two effects which needed to be addressed:
the double-counting of photons as jets was avoided by removing jets closer to good photons than \mbox{$\Delta R = 0.1$}.
%In these cases, the jet $\pt$ did not differ by much from the photon $\et$, and the jet basically consisted of the
In these cases, the photon was found to be still the main component of the jet
comparing\footnote{It is emphasised that both, jet $\pt$ and photon $\et$, are based on energy measurements in the calorimeters.}
photon $\et$ and jet $\pt$.
The second effect was the presence of particles in the proximity of the photon, which happened when a real jet was so close to the photon that both
particles were reconstructed as one single jet.
This was indicated by larger \mbox{$\Delta R$} values for the closest jet.
Since such events featured two overlapping objects, just removing the jet would have biased the jet reconstruction.
Hence, the whole event was removed if it featured a \mbox{$\Delta R$} value between~0.1 and~0.5.

\begin{figure}[h!]
\centering
\includegraphics[width=0.49\textwidth]{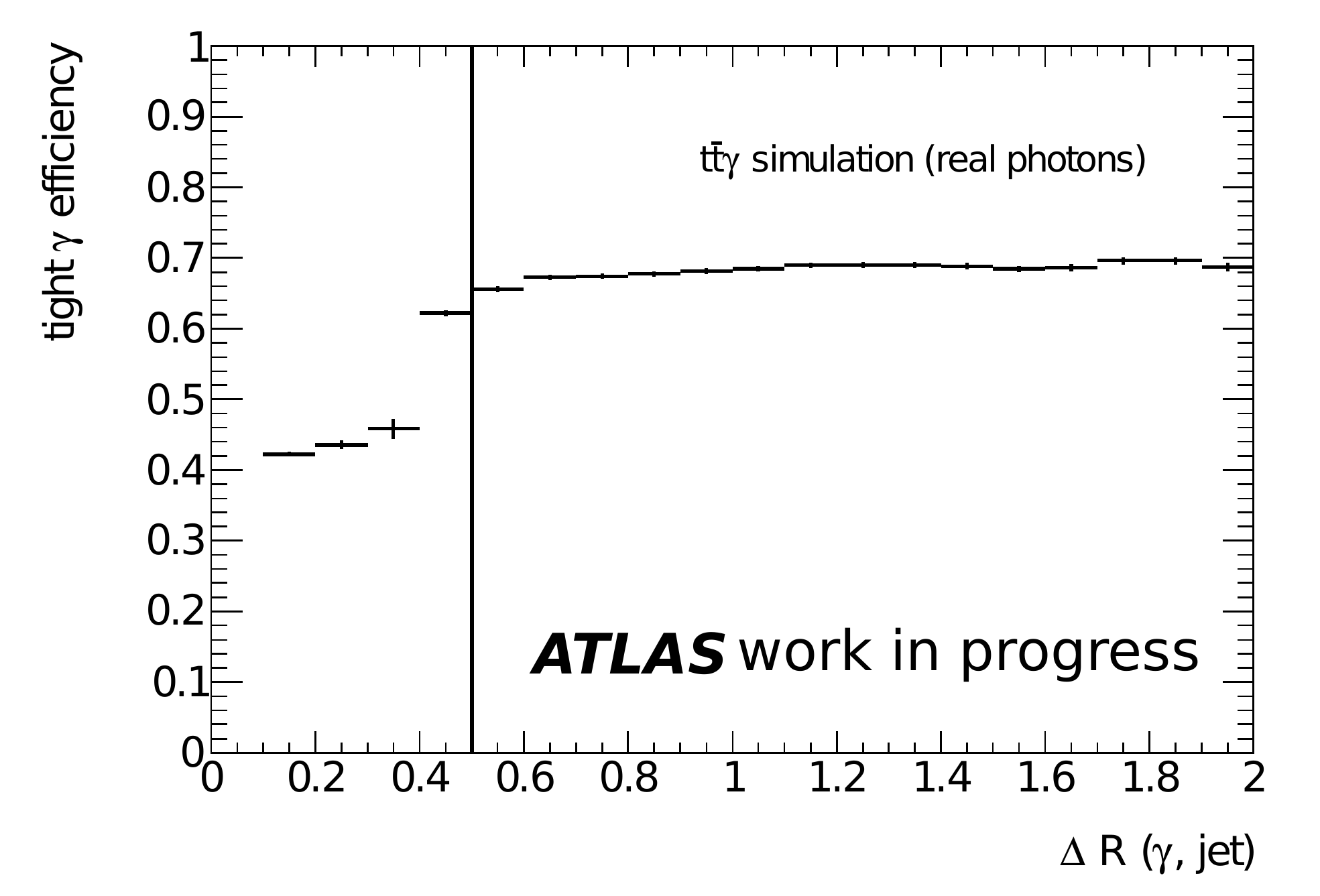}
\includegraphics[width=0.49\textwidth]{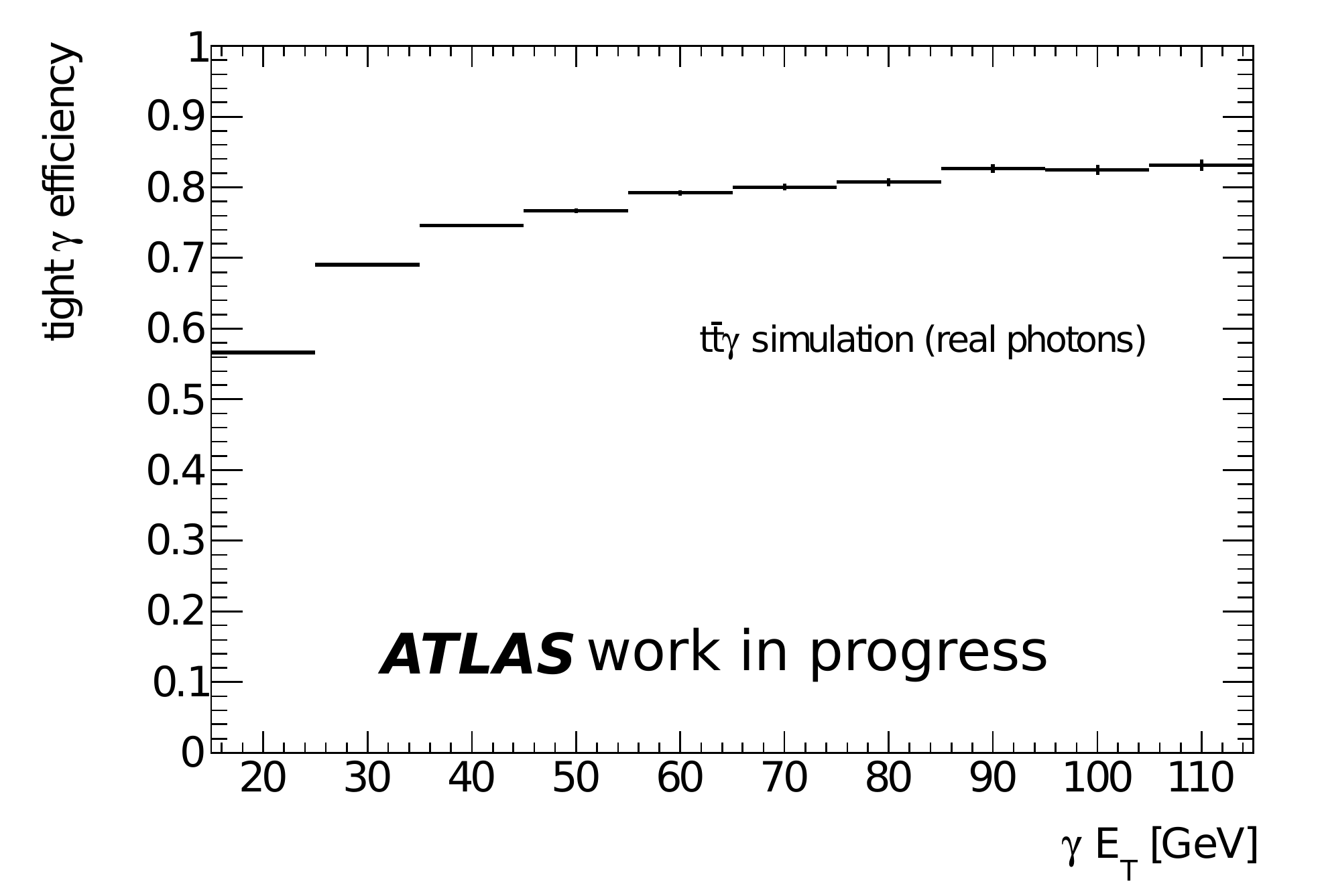}
\caption[Photon identification efficiency]{
  Left: photon identification (\texttt{tight} menu) efficiency for real photons in $\ttg$ simulation as a function of the
  \mbox{$\Delta R$} between photon and closest jet with a minimal distance of \mbox{$\Delta R = 0.1$}.
  The efficiency drops significantly for small distances in $\eta$-$\phi$-space.
  The cut at \mbox{$\Delta R = 0.5$} is indicated by a vertical line.
  Right: photon identification efficiency for good real photons in $\ttg$ simulation as a function of the photon $\et$.
  Only photons with a minimal distance of \mbox{$\Delta R = 0.5$} to the closest jet were taken into account.
}
\label{fig:photonefficiency}
\includegraphics[width=0.49\textwidth]{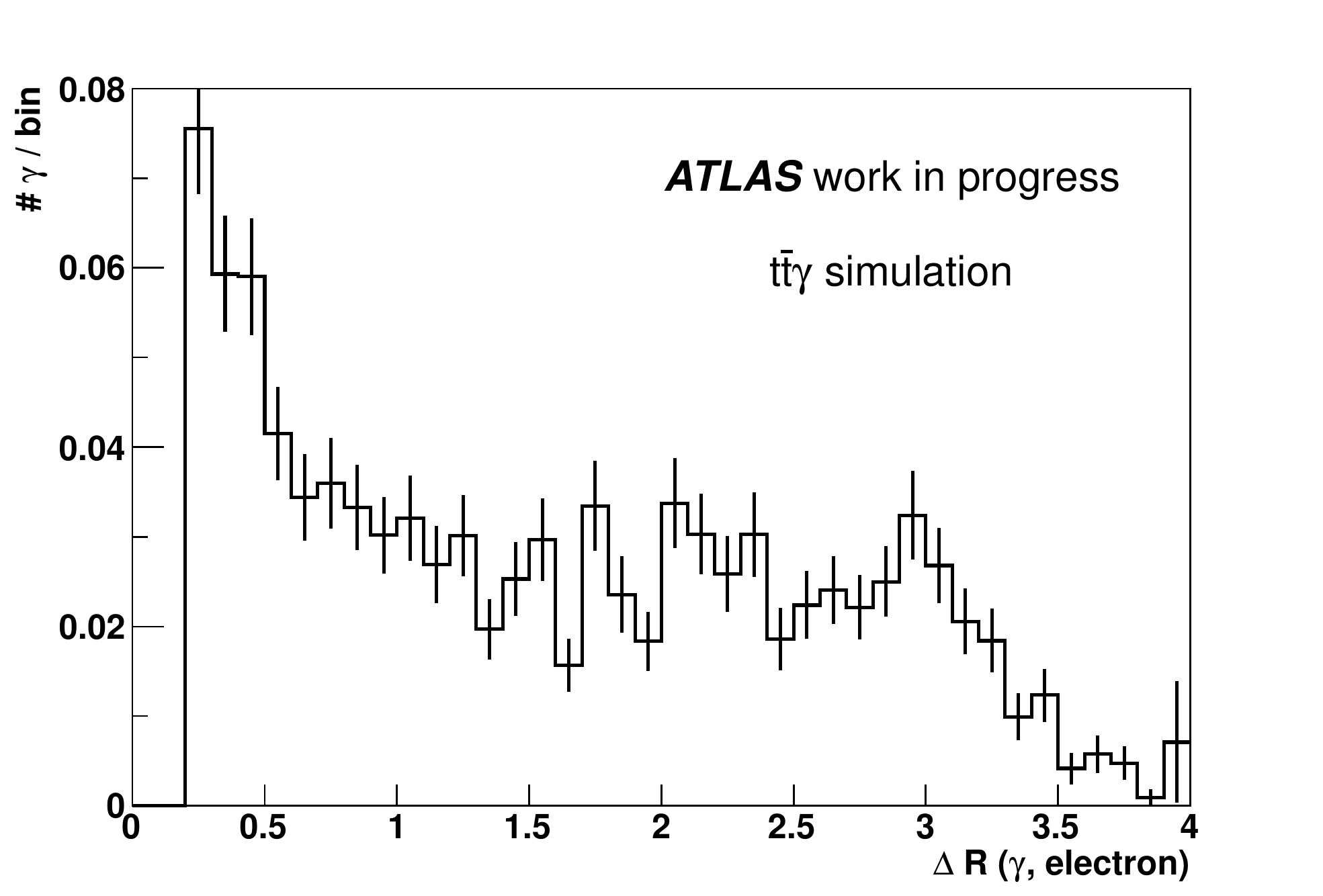}
\includegraphics[width=0.49\textwidth]{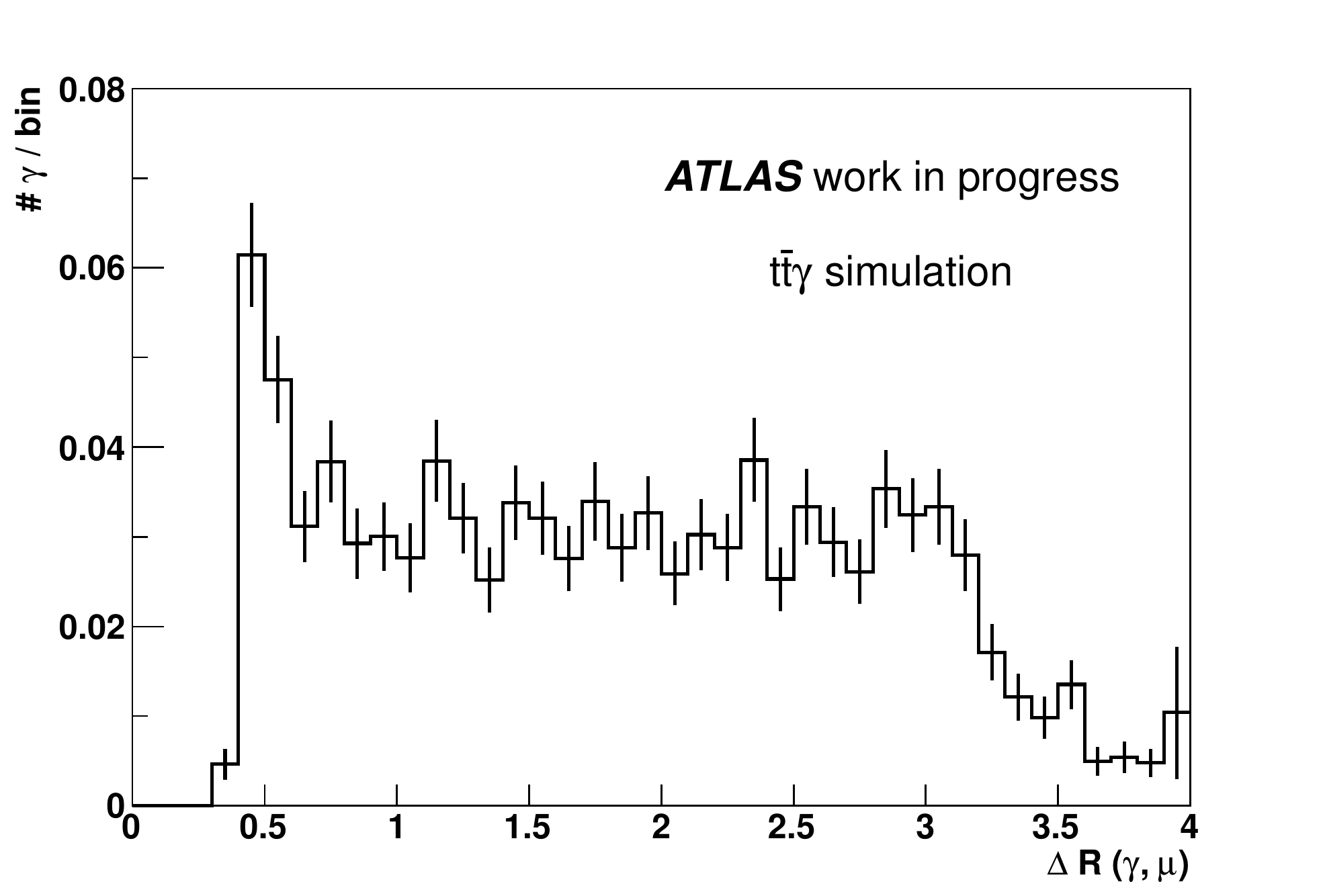}
\caption[Distance between photons and leptons]{
  The left plots shows the normalised distribution of the distance in $\eta$-$\phi$-space between photons and electrons after a full event
  selection (Ch.~\ref{sec:selection}) in simulated $\ttg$ events.
  The right plots shows the distance between photons and muons.
  Photon and electron objects are separated by a $\Delta R$ of at least 0.2.
  The minimal distance between photons and muons is of the order of 0.4.
  In both plots, the last bin includes the overflow bin.
}
\label{fig:photonleptondeltaR}
\end{figure}

The upper limit of \mbox{$\Delta R = 0.5$} was motivated by the left plot in Fig.~\ref{fig:photonefficiency}, which shows the photon identification
efficiency for real photons in $\ttg$ simulation as a function of the \mbox{$\Delta R$} between the photon and the closest jet after jets with
\mbox{$\Delta R < 0.1$} were already removed.
The identification efficiency is defined with respect to reconstructed photons.
It drops significantly for small distances in $\eta$-$\phi$-space, which indicates a distortion of the photon shower shape variables
used in the \texttt{tight} identification menu by the energy deposits of surrounding particles from a close-by jet.
The efficiency was found to be close to constant for \mbox{$\Delta R > 0.5$}, which was hence chosen as the upper limit for events to be disregarded due to the
proximity of jets to photons.

Photons and electrons as well as photons and muons were implicitly separated in $\eta$-$\phi$-space by their respective object definitions:
the double-counting of electromagnetic clusters as electrons and photons was avoided by the photon recovery procedure described above.
Muons were not considered if they were closer than 0.4 in $\eta$-$\phi$-space to a jet (Sec.~\ref{sec:muon}).
Since photons were also reconstructed as jets, this translated to a minimal $\Delta R$ requirement between muons and photons.
Fig.~\ref{fig:photonleptondeltaR} shows the normalised distributions of the distance in $\eta$-$\phi$-space between photons and electrons (left),
and between photons and muons (right), respectively, in simulated $\ttg$ events.
A full event selection as described in Ch.~\ref{sec:selection} was applied.
It can be seen that photons and electrons are separated by a $\Delta R$ of at least 0.2.
The minimal distance between photons and muons is of the order of 0.4.

The right plot in Fig.~\ref{fig:photonefficiency} shows the identification efficiency as a function of the photon $\et$ after the requirements
on the photon-jet distance were applied.
The efficiency rises with increasing photon $\et$ and approaches a constant value of roughly 80\%.
For low-$\et$ photons, the efficiency was found to be significantly lower and only as large as roughly 55\%.
The reason is the decreasing discriminating power between showers from real photons and from fake photons from \mbox{$\pi^0 \to \gamma \gamma$} decays,
because the shapes of low-energetic showers are subject to stronger statistical fluctuations than high-energetic showers.

\begin{figure}[p]
\begin{center}
\includegraphics[width=0.44\textwidth]{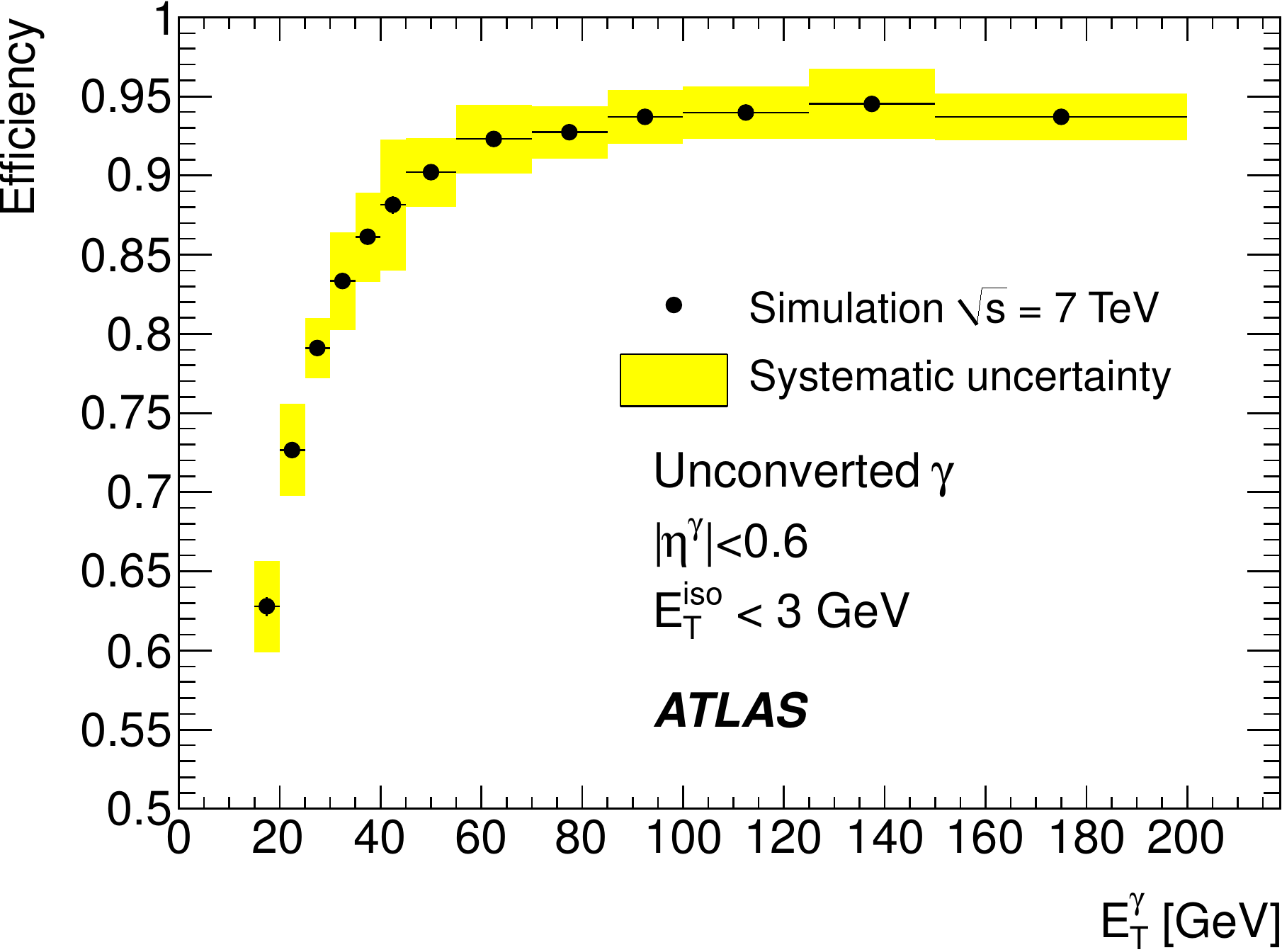}
\includegraphics[width=0.44\textwidth]{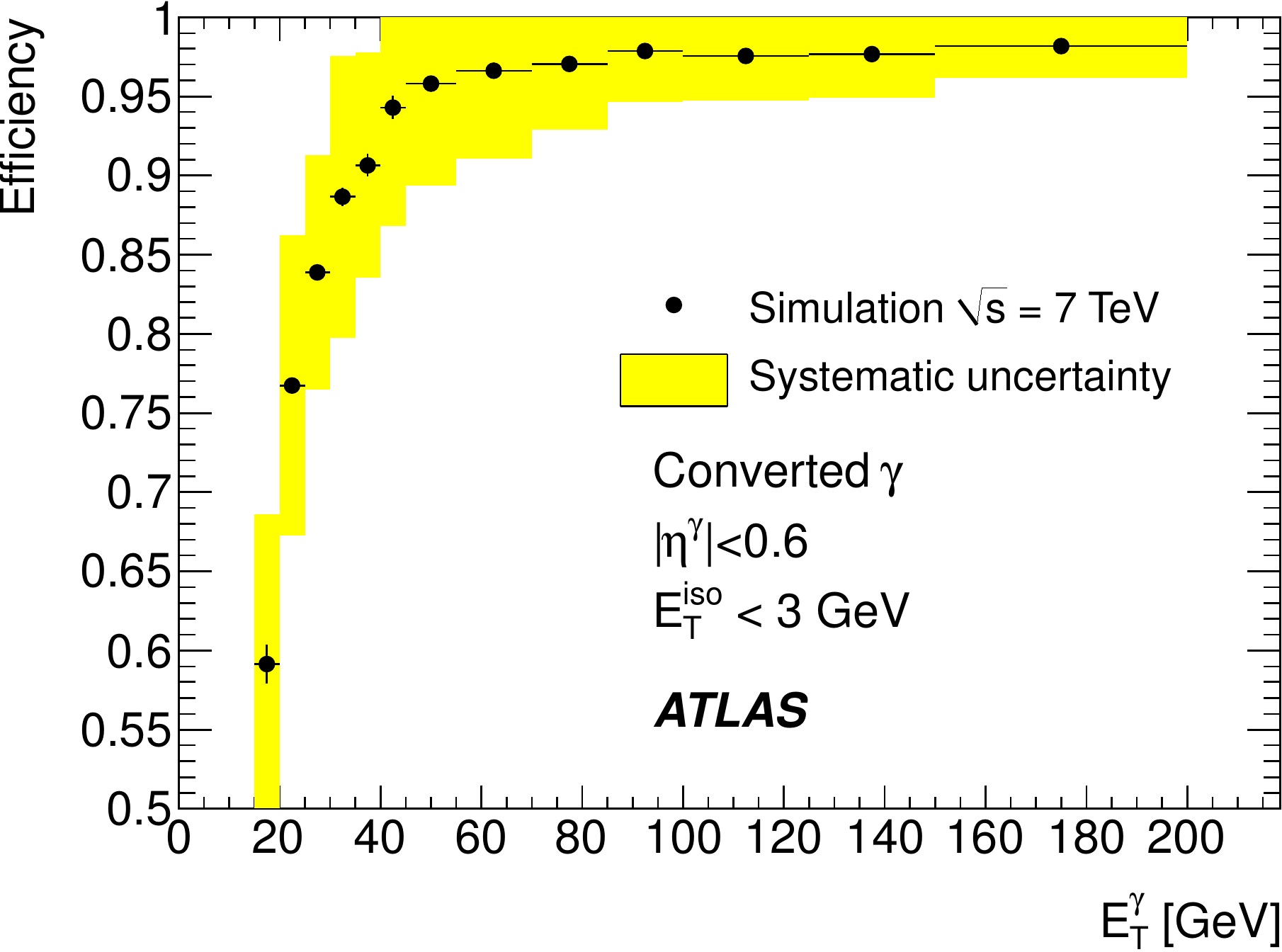} \\
\includegraphics[width=0.44\textwidth]{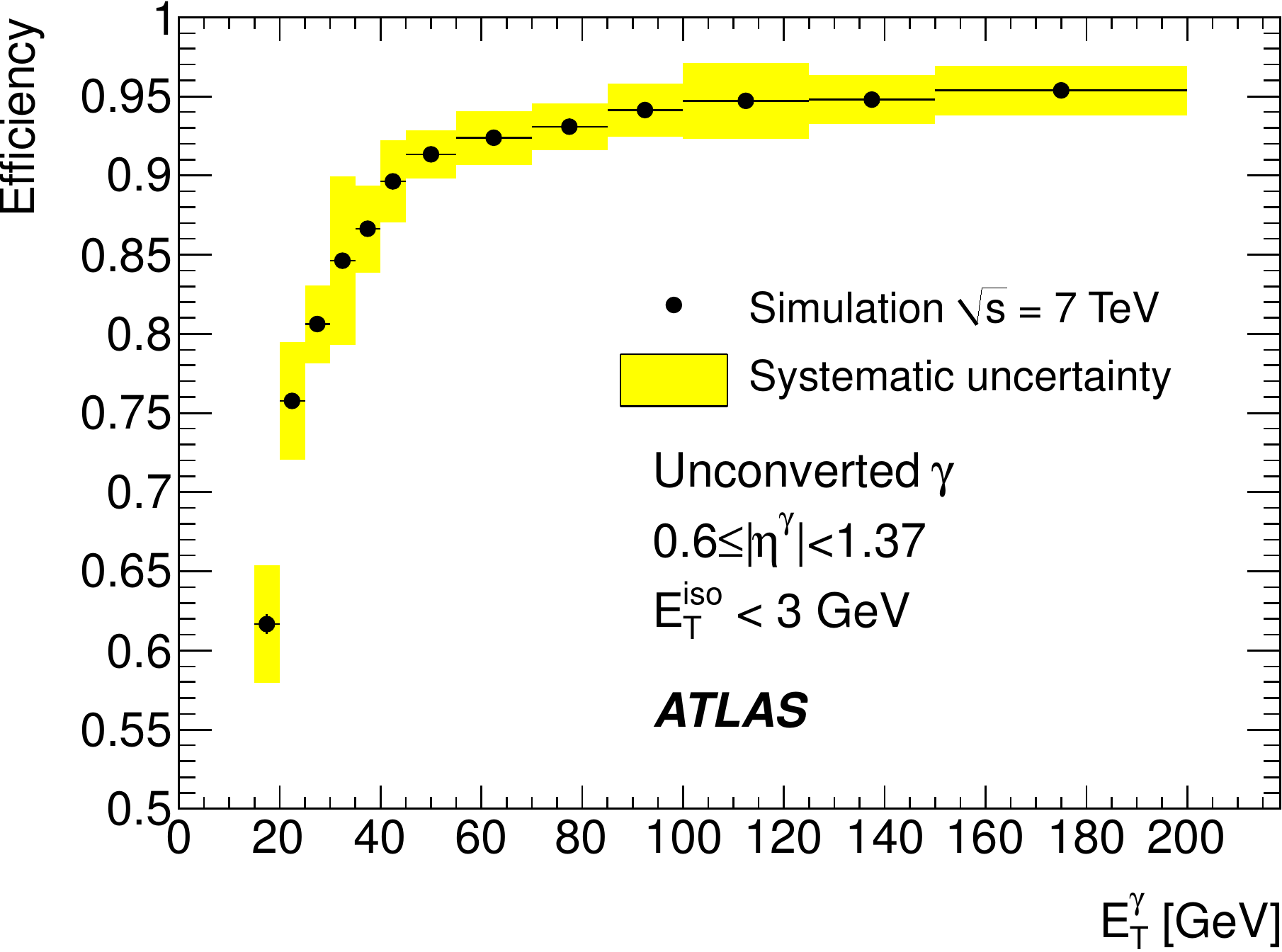}
\includegraphics[width=0.44\textwidth]{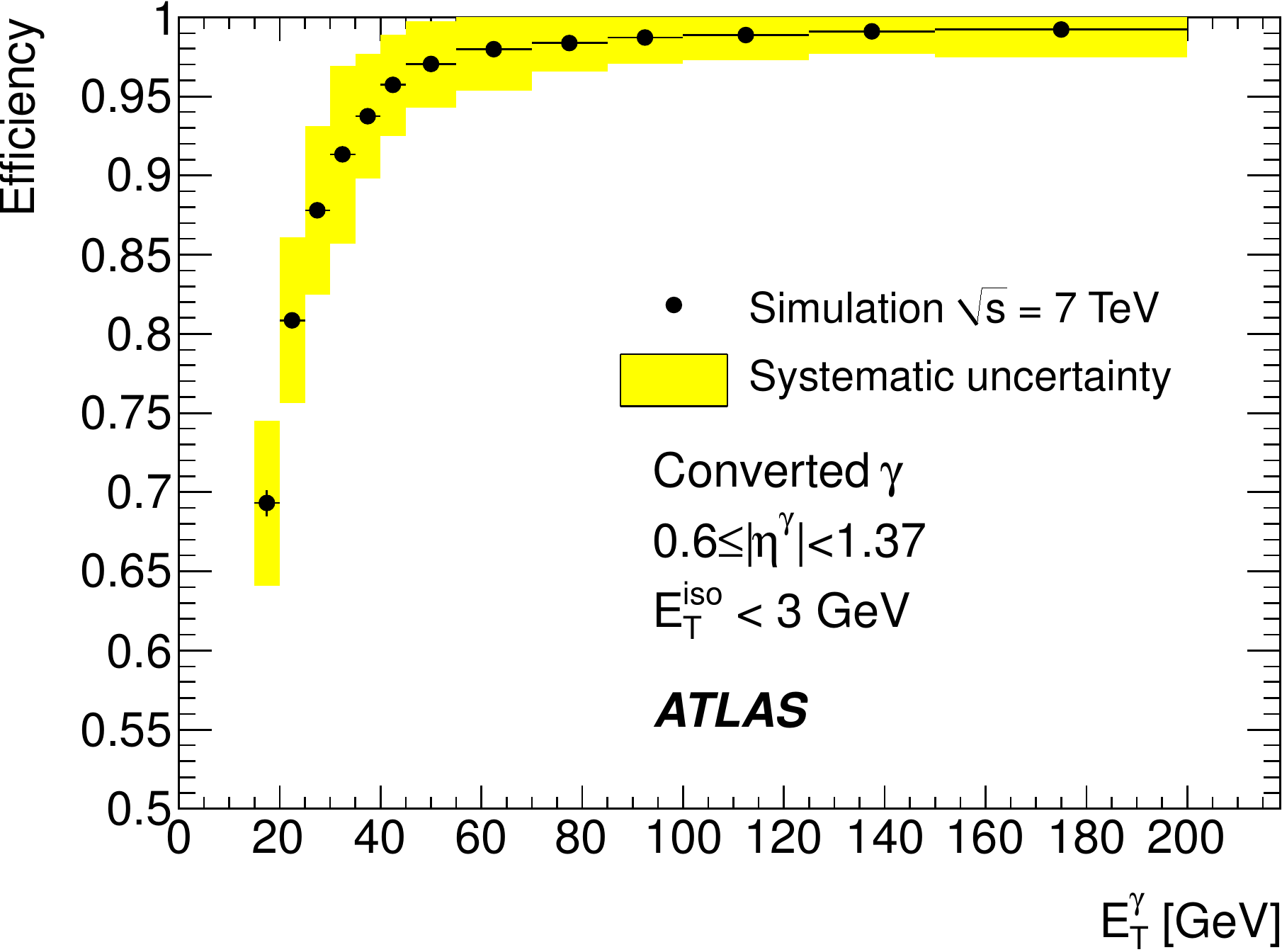} \\
\includegraphics[width=0.44\textwidth]{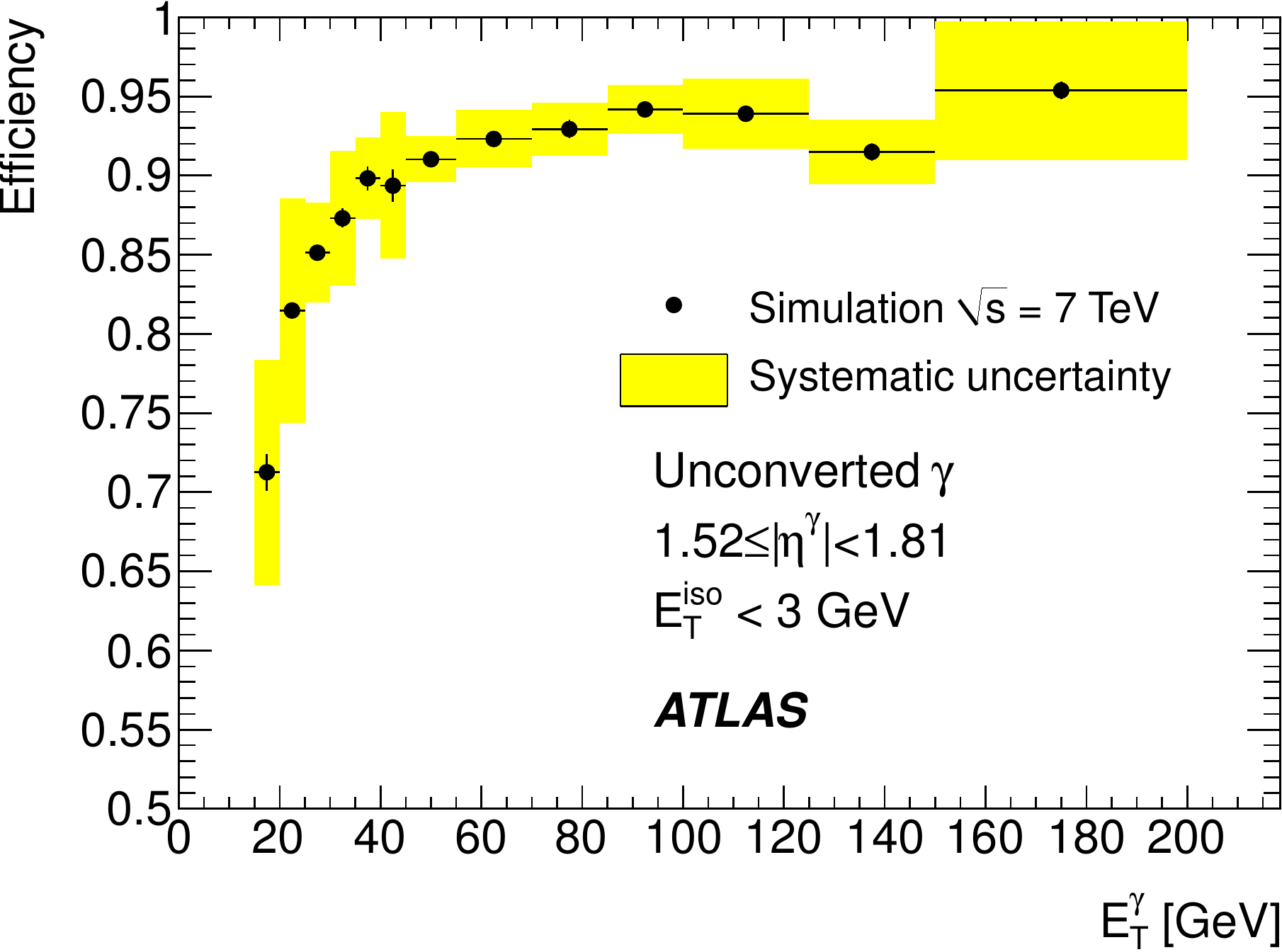}
\includegraphics[width=0.44\textwidth]{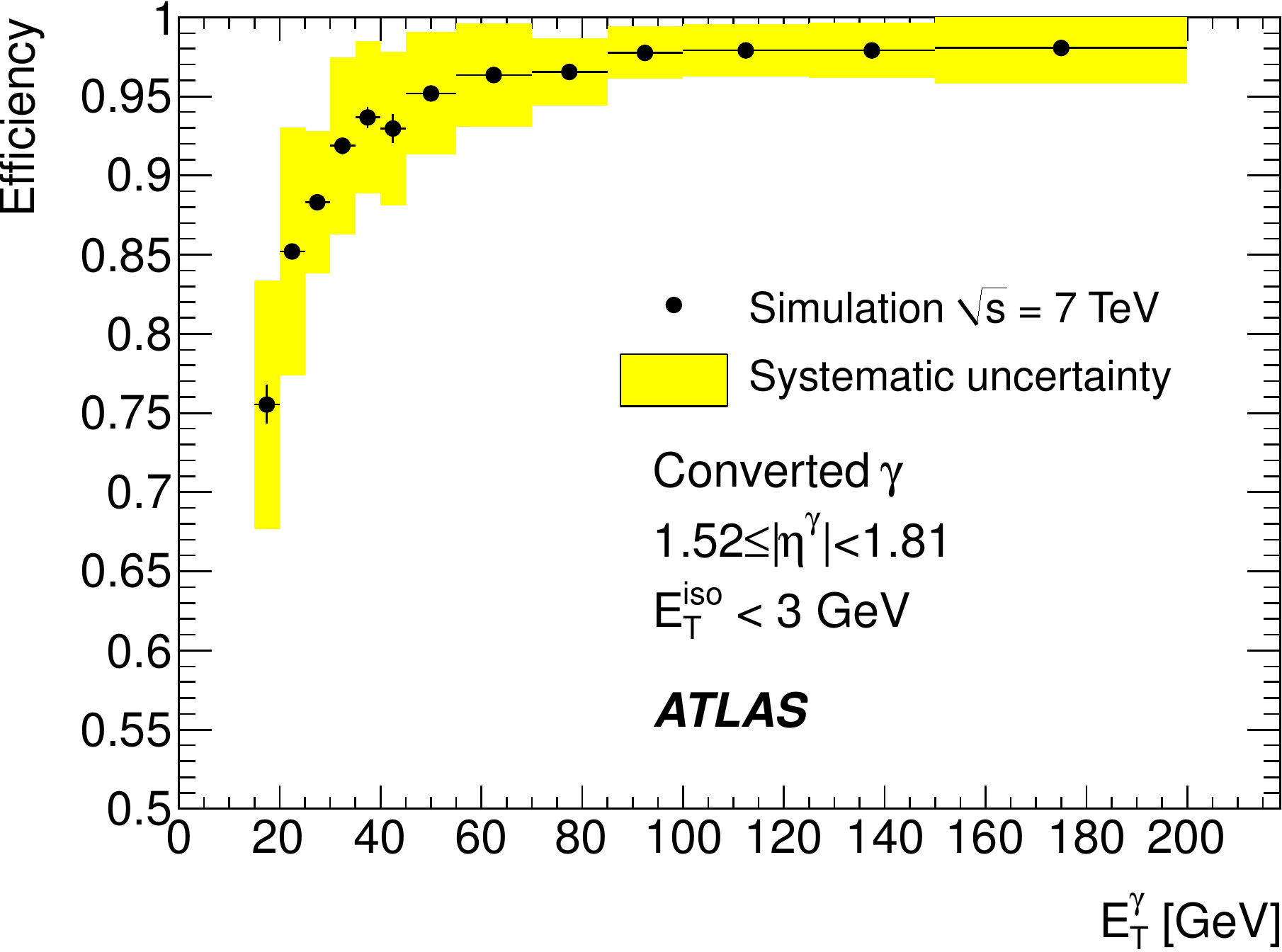} \\
\includegraphics[width=0.44\textwidth]{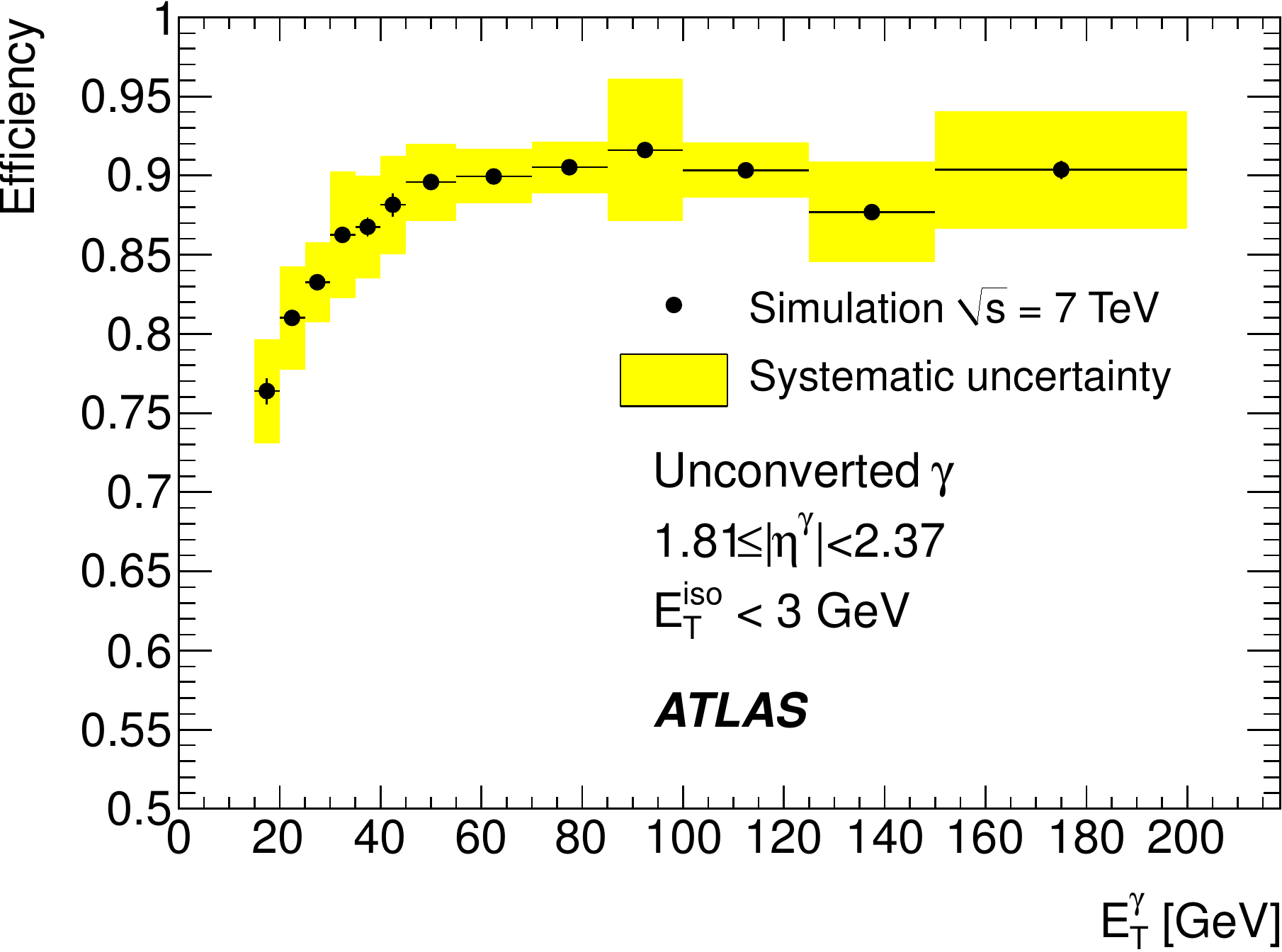}
\includegraphics[width=0.44\textwidth]{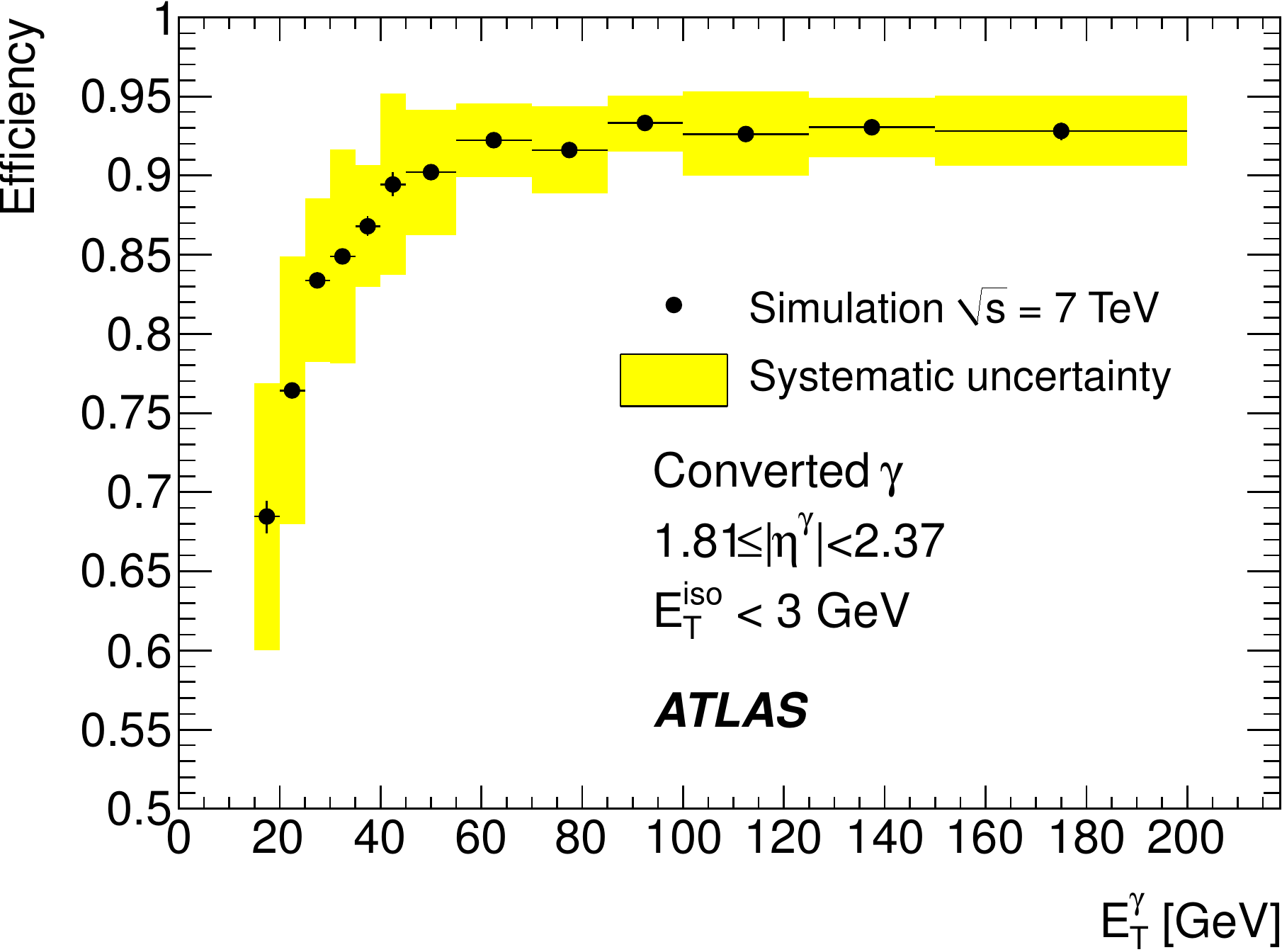}
\caption[Identification efficiency for unconverted and converted photons]{
  Identification efficiency for unconverted and converted photons in bins of $\et$ and $\eta$ derived for the
  isolated diphoton cross section measurement~\cite{diphoton, *diphoton_note}.
  The isolation energy in the calorimeter in a cone of size \mbox{$\Delta R = 0.4$} around the photon was required to be less than \mbox{$3 \GeV$}.
  The efficiencies themselves were not used in this analysis but the relative systematic uncertainty was used.}
\label{fig:photonefficiency_diphoton}
\end{center}
\end{figure}

Since there is no easily accessible reference process for photons in the energy range used in this analysis, the measurement of the photon
identification efficiencies as well as the energy scale and resolution is not trivial.
For the correct simulation of the identification efficiencies, the shower shapes used in the \texttt{tight} menu were shifted in the MC simulations
so that their efficiencies matched the ones in data.
This is a procedure which was adopted before in the isolated prompt photon~\cite{promptphoton} and the isolated diphoton cross section
measurements~\cite{diphoton, *diphoton_note}.
It accounts for the average mismodelling of the shower shapes in the simulation.

The shifts were obtained by comparing the shower shape distributions for photons in data to the distributions obtained
from true photons in MC simulations, where the photons originated from jet fragmentation or $\gamma$+jet processes.
For this comparison, the energy in a cone of 0.4 in $\eta$-$\phi$-space around the photon candidate was required to be less than \mbox{$3 \GeV$},
so that a purer sample of photon candidates was selected.
Fig.~\ref{fig:photonefficiency_diphoton} shows the identification efficiencies as used in the diphoton measurement~\cite{diphoton, *diphoton_note}
for the \texttt{tight} menu in the isolated photon sample for unconverted and converted photons in bins of $\et$ and $\eta$.
The values are larger than those given in the right plot in Fig.~\ref{fig:photonefficiency}, because of the correlation between calorimeter
isolation and the shape of the cluster in the calorimeter.

In this analysis, the same shower shape shifts as in Ref.~\cite{diphoton, *diphoton_note} were used to correct the efficiencies in MC simulations, but
the actual efficiencies shown in Fig.~\ref{fig:photonefficiency_diphoton} were not used.
Fig.~\ref{fig:photonefficiency_diphoton} also shows the systematic uncertainty on the efficiency measurement.
The systematic uncertainties used in this analysis were taken as the relative uncertainties shown in the plots.

The photon energy scale and energy resolution were taken from the measurements of \Zee events~\cite{electronperformance} mentioned in Sec.~\ref{sec:electron}.
Slight modifications were applied to the energy scale correction to account for the differences in the interaction of electrons and photons with the
material in front of the calorimeter and with the presampler.

\chapter{Event selection}
\label{sec:selection}

The signature of $\ttg$ events in the single lepton decay channel is defined by a high-$\et$ electron or a high-$\pt$ muon, large $\met$, four high-$\pt$ jets,
and a high-$\et$ photon.
Two of these jets are $b$-jets originating from $b$-quarks.

The background processes for $\ttg$ production are the same as for $\ttbar$ production, but with one or more additional photons in the final state, which can
be radiated from all charged particles.
Moreover, events may feature hadrons and electrons misidentified as photons.

Some of the background processes to $\ttbar$ production (Sec.~\ref{sec:topbackgrounds}) have cross sections which largely exceed the
$\ttbar$ production cross section at hadron colliders.
This holds in particular for multijet production and for $W$ and $Z$ production in association with jets.

In a first step, the \textit{preselection}, a pure sample of events containing top quark pairs was obtained by applying selection criteria
which suppress different background components to $\ttbar$ production while maintaining a reasonable efficiency for $\ttbar$ events
(Sec.~\ref{sec:preselection}).
These requirements were motivated by previous analyses in the single lepton $\ttbar$ decay channel at the ATLAS experiment (cf. for example
Ref.~\cite{ATLAStopXsec_3pb}).
In particular, the use of $b$-tagging information enhanced the signal-over-background ratio.

In a second step, the \textit{final event selection}, photon-specific criteria were applied (Sec.~\ref{sec:finalselection}).
Sec.~\ref{sec:yields} presents the event yields for data, which are compared to the expectations for the signal and the different background
contributions.

\section{Preselection}
\label{sec:preselection}

Events in the electron and muon channels were required to be triggered by the \texttt{EF\_e20\_medium} and \texttt{EF\_mu18} trigger chain,
respectively (Sec.~\ref{sec:electron} and Sec.~\ref{sec:muon}).
Non-collision background was rejected by selecting only events with a good primary vertex with at least five tracks associated to it.

Exactly one good lepton was required in each event:
in the electron channel, an electron with an $\et$ of at least \mbox{$25 \GeV$} was required.
In the muon channel, exactly one muon with a minimum $\pt$ of \mbox{$20 \GeV$} had to be present.
Events in which the muon $\pt$ exceeded \mbox{$150 \GeV$} were disregarded (Sec.~\ref{sec:muon}).
In both channels, no lepton of the other type beyond the respective $\et$ or $\pt$ threshold was allowed to be present.
Electrons were required to match the trigger object that fired the \texttt{EF\_e20\_medium} trigger:
the distance in $\eta$-$\phi$-space between these objects had to be smaller than 0.15.
Muons were required to match the trigger object that fired the \texttt{EF\_mu18} trigger at L2 as well as at EF level in MC simulations:
a requirement of 0.15 on the distance between the two objects was imposed.
In data, trigger matching was not applied in the muon channel due to a software issue at the L2 level.
A systematic uncertainty was assigned to this feature (Sec.~\ref{sec:syst_detectormodelling}).

The offline requirements on the electron $\et$ and the muon $\pt$ of \mbox{$25 \GeV$} and \mbox{$20 \GeV$}, respectively, were sufficiently high to ensure
that effects due to the trigger threshold were avoided:
in this region, the trigger efficiencies were constant as a function of the lepton $\et$ or $\pt$, respectively.
Backgrounds from multijet production with jets misidentified as electrons or with muons from the decay of heavy flavour hadrons were strongly suppressed
by the electron and muon definitions presented in Ch.~\ref{sec:objects}.

In order to avoid the rare cases where a muon was also reconstructed as an electron, events in which a good electron and a muon shared a track were discarded.
For this check, muons were required to fulfil all identification criteria apart from the distance requirement between muons and the closest jet.

Events with jets that failed certain quality criteria (Sec.~\ref{sec:jet}) were disregarded, as well as events with potentially incorrect data in
the LAr calorimeters (Sec.~\ref{sec:MET}).
By these means, biases of the $\met$ measurement and the jet reconstruction efficiency were avoided.

Since multijet events are typically balanced in the transverse plane, the presence of large $\met$ provides strong discrimination
between multijet events and processes with $\met$ from neutrinos in the final state, as in $W$+jets and $\ttbar$ production.
However, $\met$ may originate from the miscalibrations of jet energies.
Moreover, the transverse mass of the $W$ boson, as defined by
\begin{equation*}
\mtw = \sqrt{2 \cdot \left( \pt^l \cdot \met - p_x^l \cdot \mex - p_y^l \cdot \mey \right)} \, , \quad l = e, \mu \, ,
\end{equation*}
discriminates between events with true $W$ bosons and multijet or $Z$+jets production.

A $\met$ of at least \mbox{$35 \GeV$} and \mbox{$20 \GeV$} was required in the electron and muon channel, respectively.
The requirement was stronger in the electron channel, because multijet events feature more electron than muon candidate objects.
Additionally, in the electron channel a transverse $W$ mass of \mbox{$25 \GeV$} was required.
In the muon channel the sum of $\met$ and $\mtw$ was required to be larger than \mbox{$60 \GeV$}, hence exploiting the correlation between $\met$ and $\mtw$.

All background processes feature strongly decreasing cross sections with an increasing number of jets in the final state.
Since $\ttbar$ events typically possess many jets with high $\pt$, a minimum of four jets with $\pt$ larger than \mbox{$25 \GeV$} was required, which
enhanced the total signal-over-background ratio.

Finally, at least one of the jets had to be $b$-tagged using the algorithm described in Sec.~\ref{sec:btagging}.
This requirement provided additional discrimination power between $\ttbar$ events and the background processes, which feature only a small amount of
$b$- and $c$-jets.

\begin{figure}[p]
\begin{center}
\includegraphics[width=0.32\textwidth]{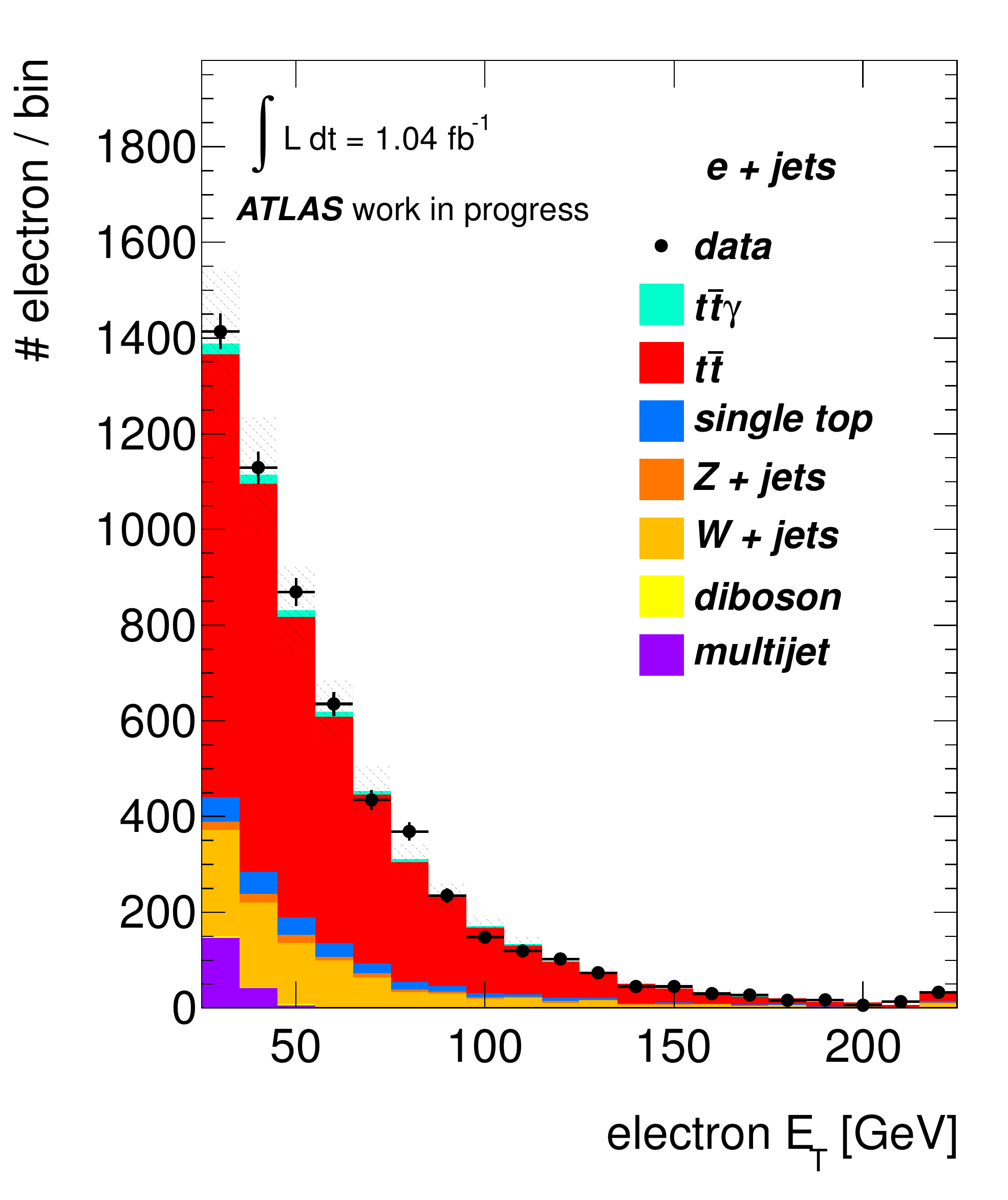}
\includegraphics[width=0.32\textwidth]{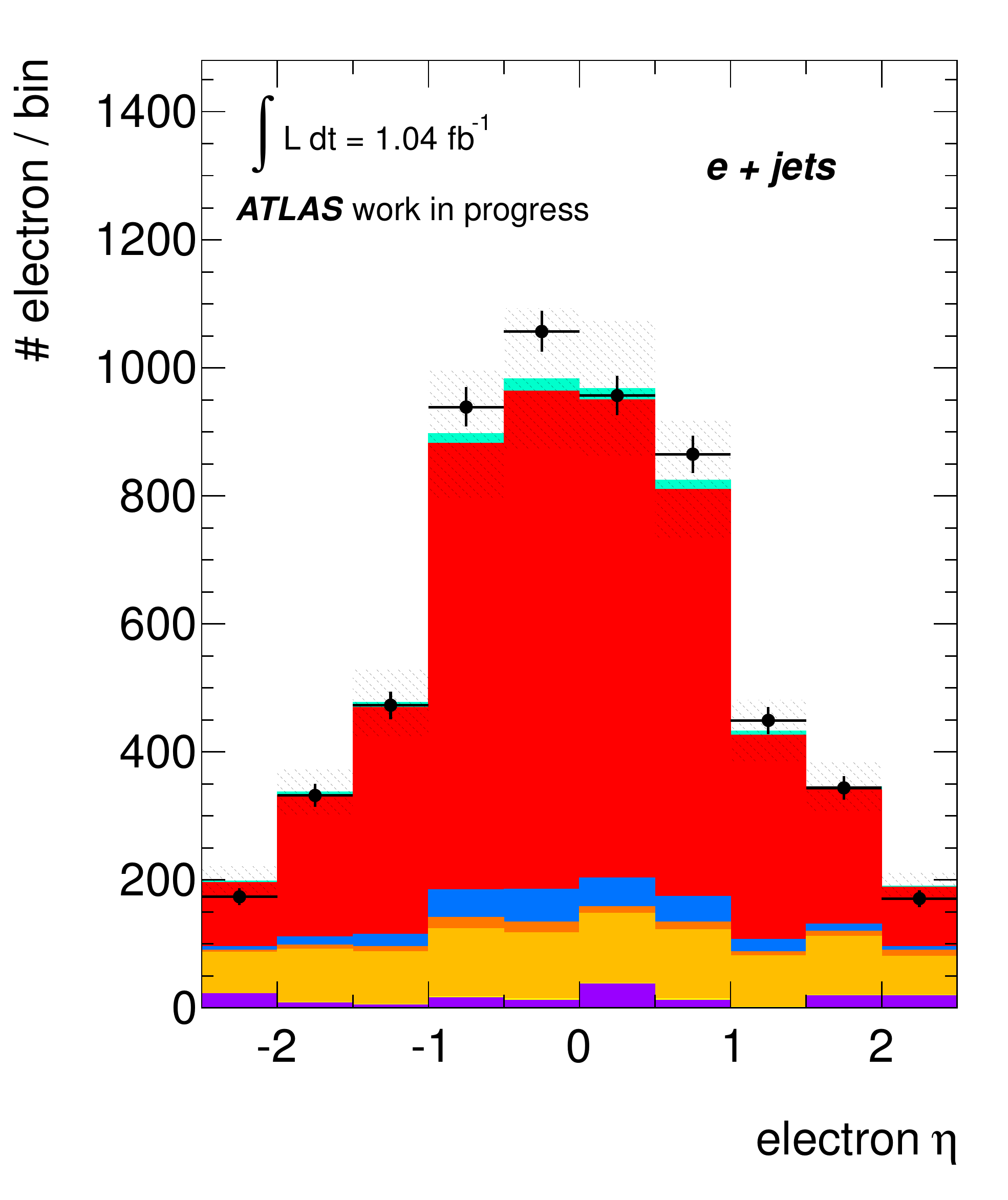}
\includegraphics[width=0.32\textwidth]{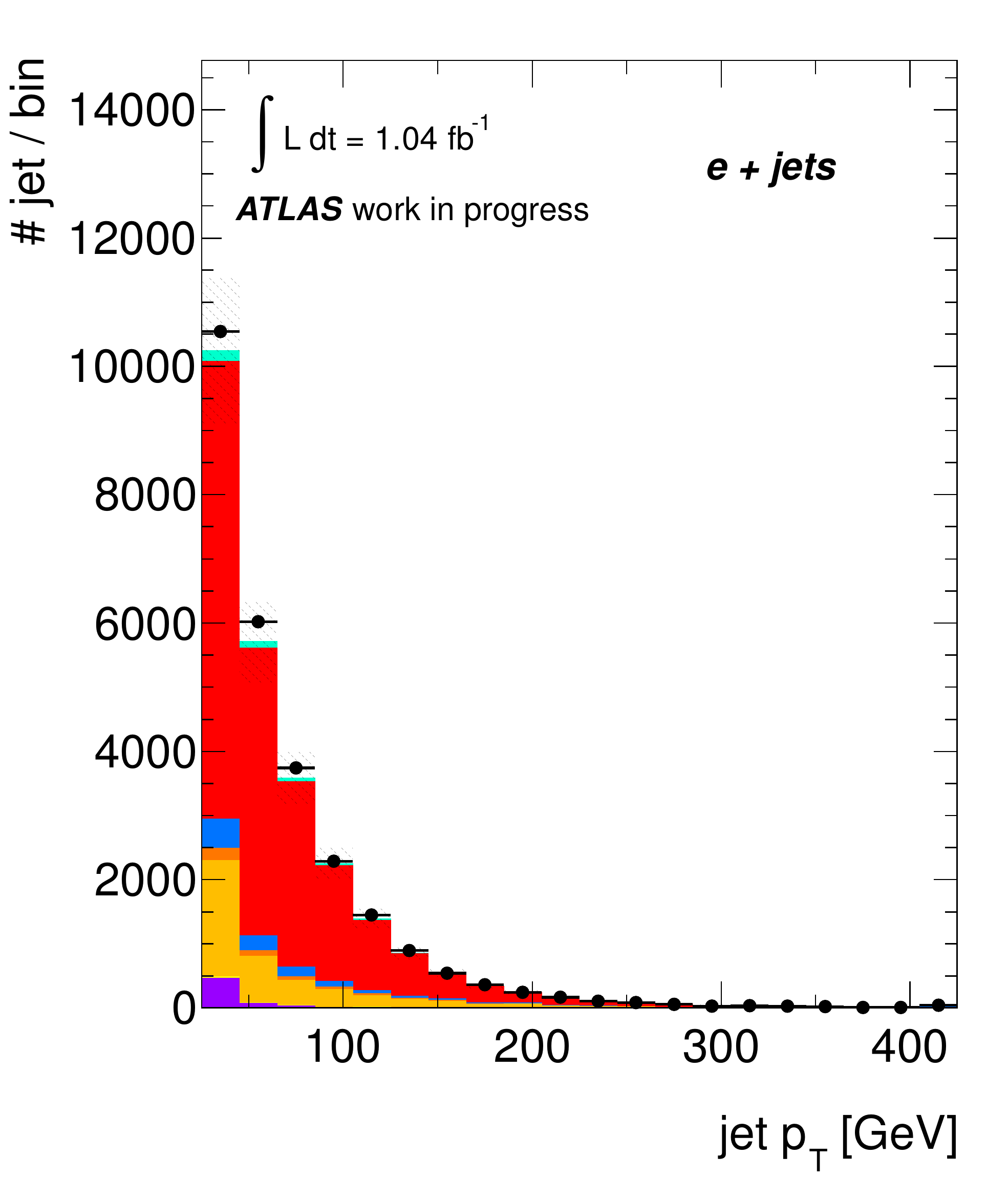} \\
\includegraphics[width=0.32\textwidth]{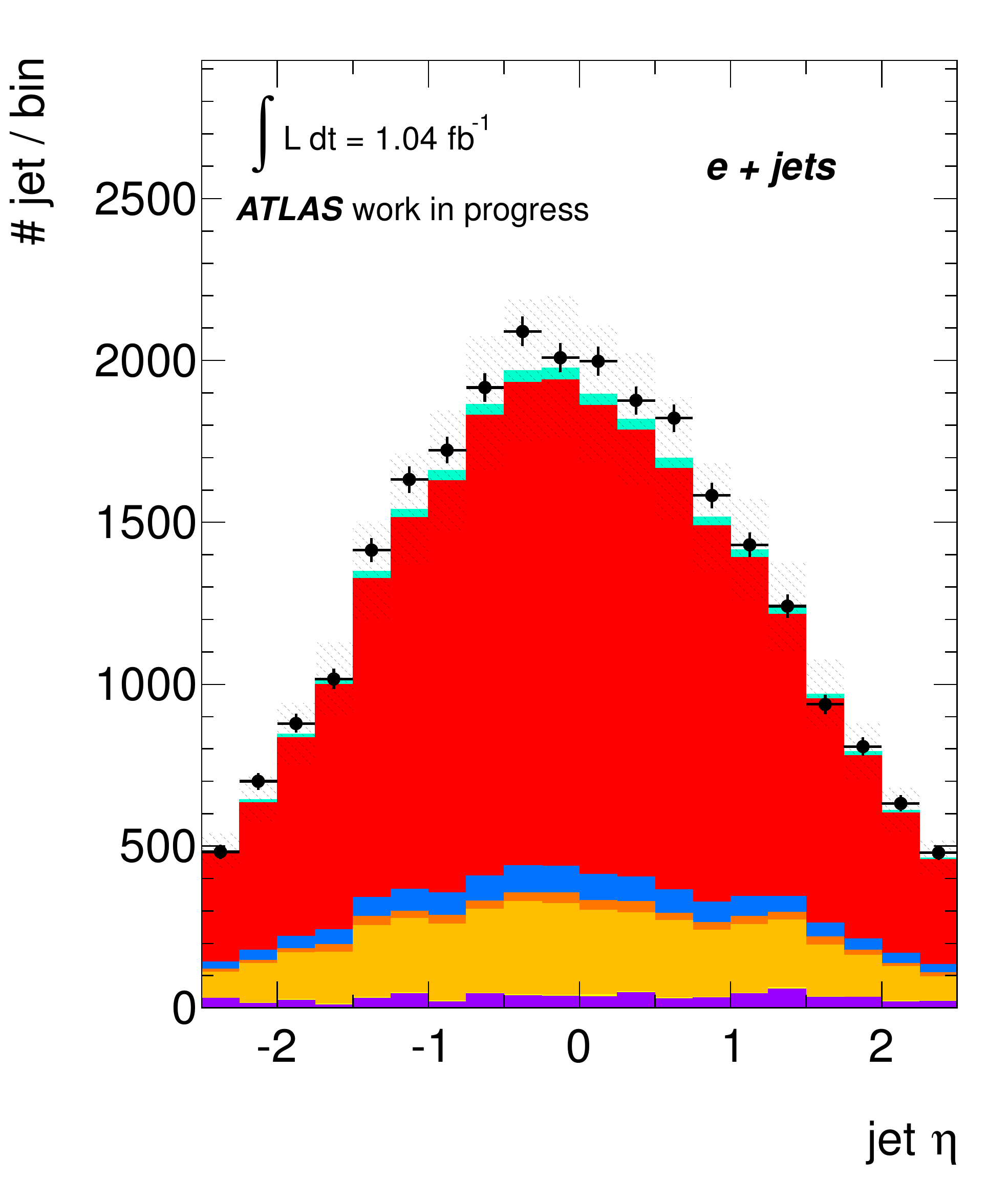}
\includegraphics[width=0.32\textwidth]{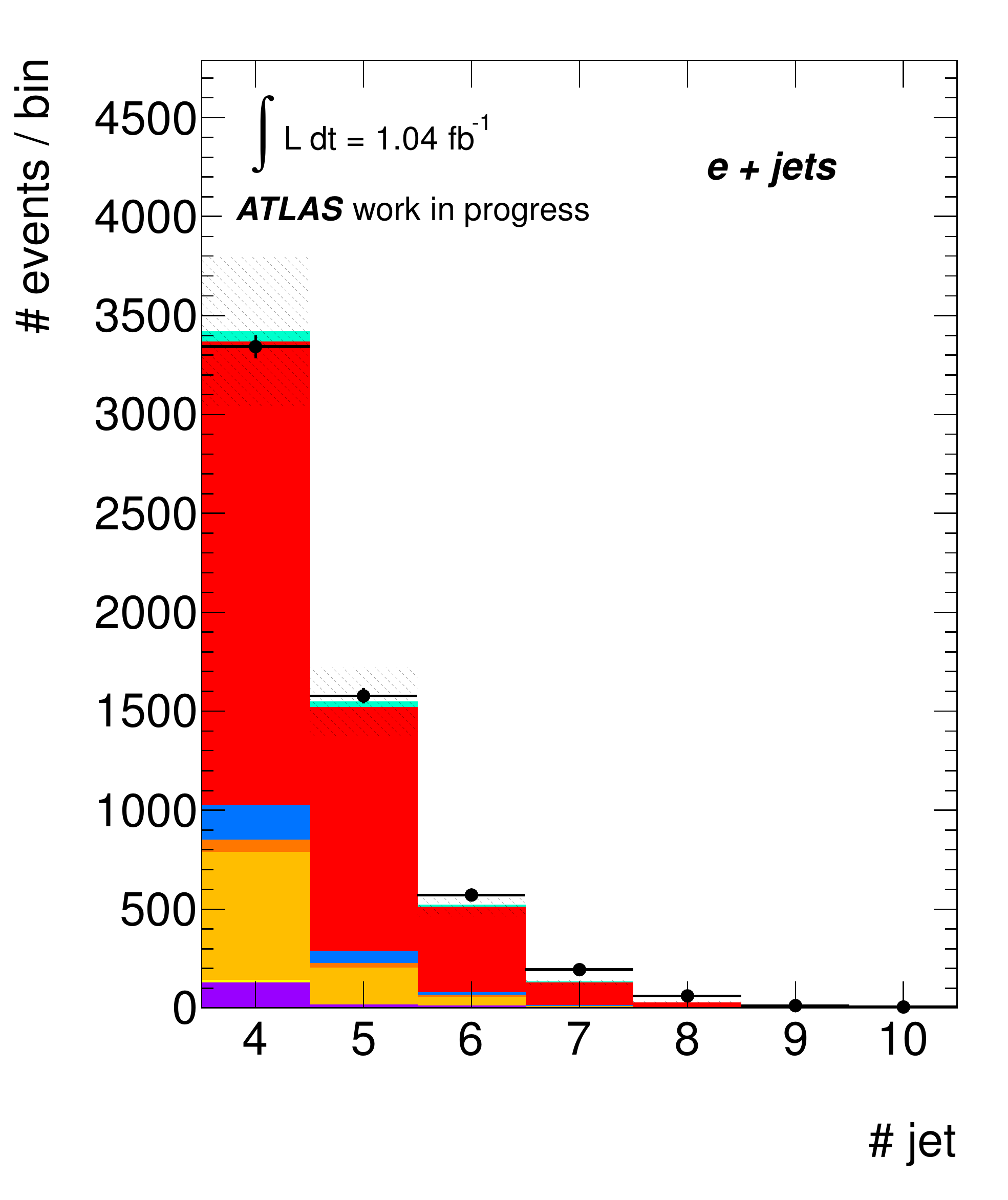}
\includegraphics[width=0.32\textwidth]{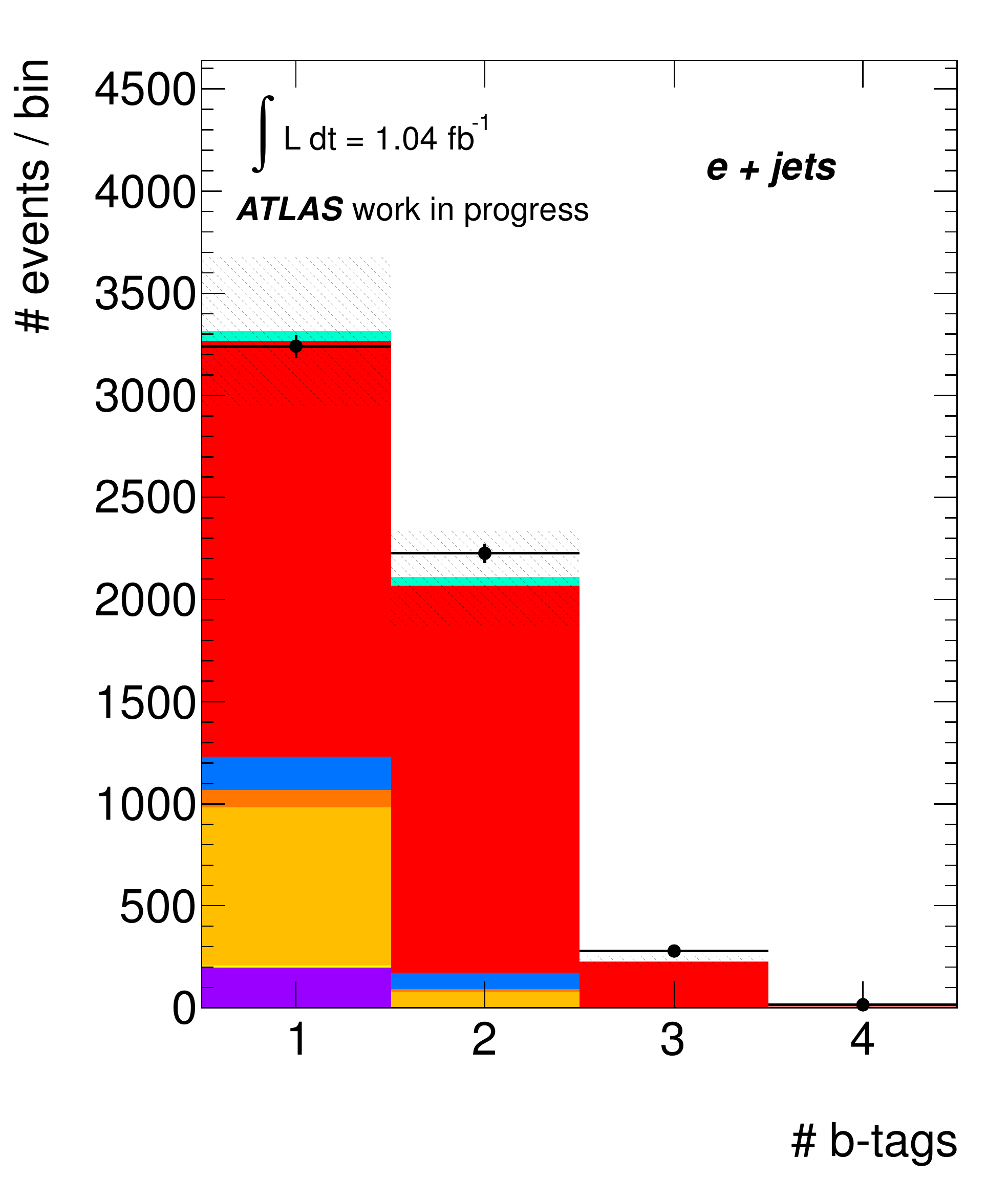} \\
\includegraphics[width=0.32\textwidth]{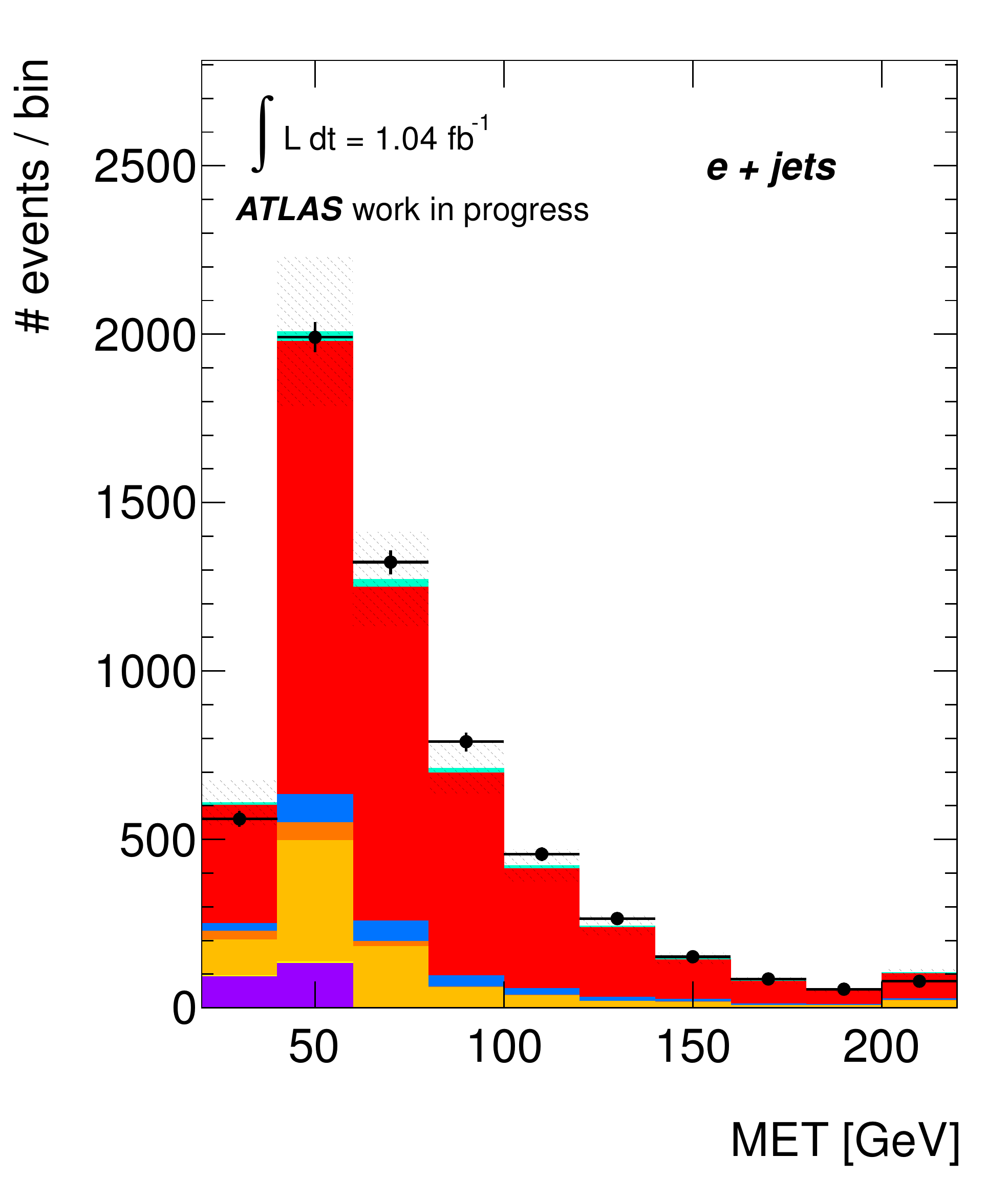}
\includegraphics[width=0.32\textwidth]{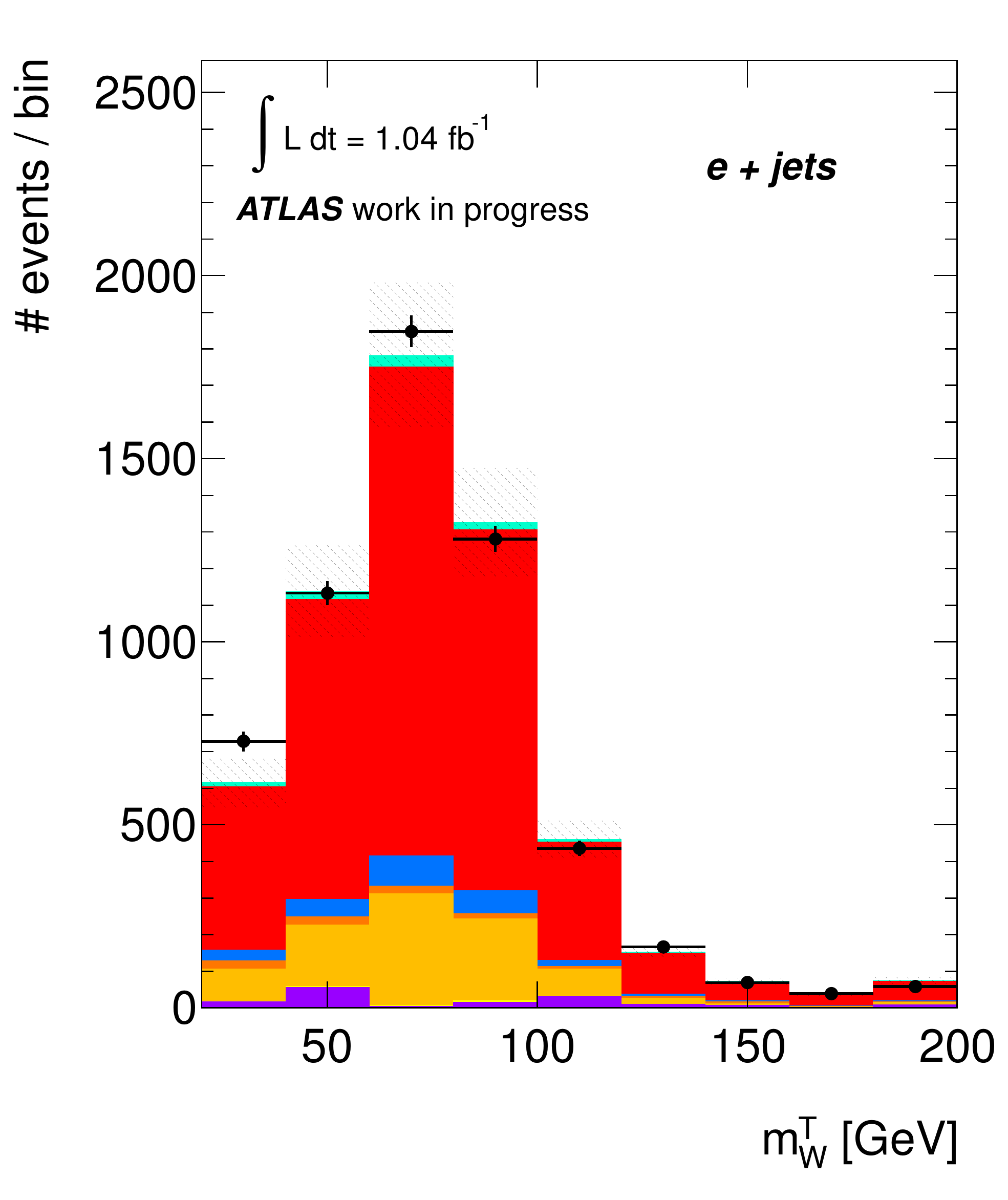}
\caption[Kinematic distributions after preselection (e+jets)]{
  Kinematic distributions after the preselection in the electron channel.
  Shown are the distributions from data, and the expectations from MC simulations and a data-driven estimate of the multijet
  background: electron $\et$ and $\eta$, jet $\pt$ and $\eta$, the number of jets, the number of $b$-tagged jets, $\met$ and $\mtw$.
  The hashed uncertainty band represents the uncertainty on the expectation.
  In all plots, except for the $\eta$ distributions, the last bin includes the overflow bin.
}
\label{fig:controlplots_e}
\end{center}
\end{figure}

\begin{figure}[p]
\begin{center}
\includegraphics[width=0.32\textwidth]{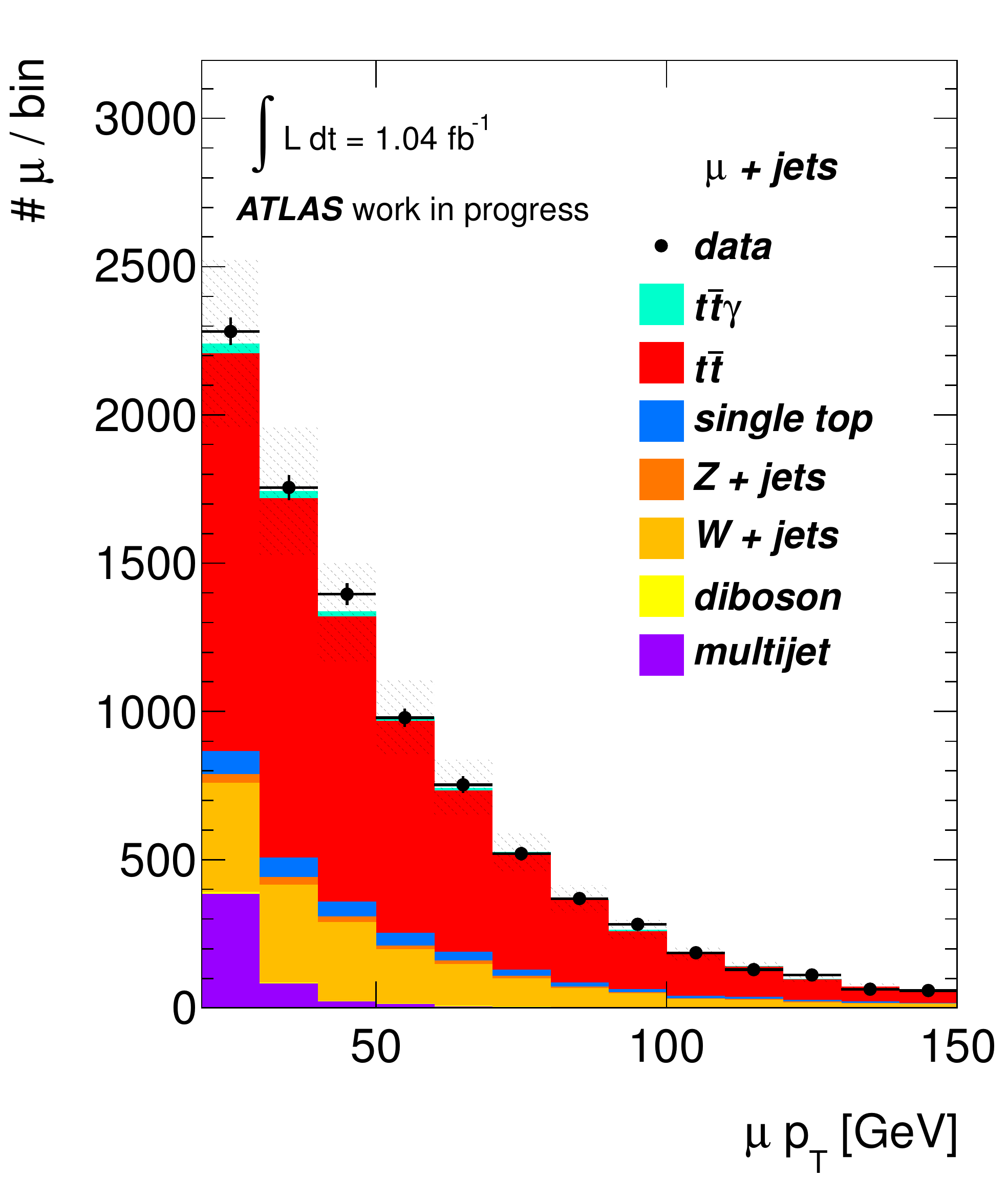}
\includegraphics[width=0.32\textwidth]{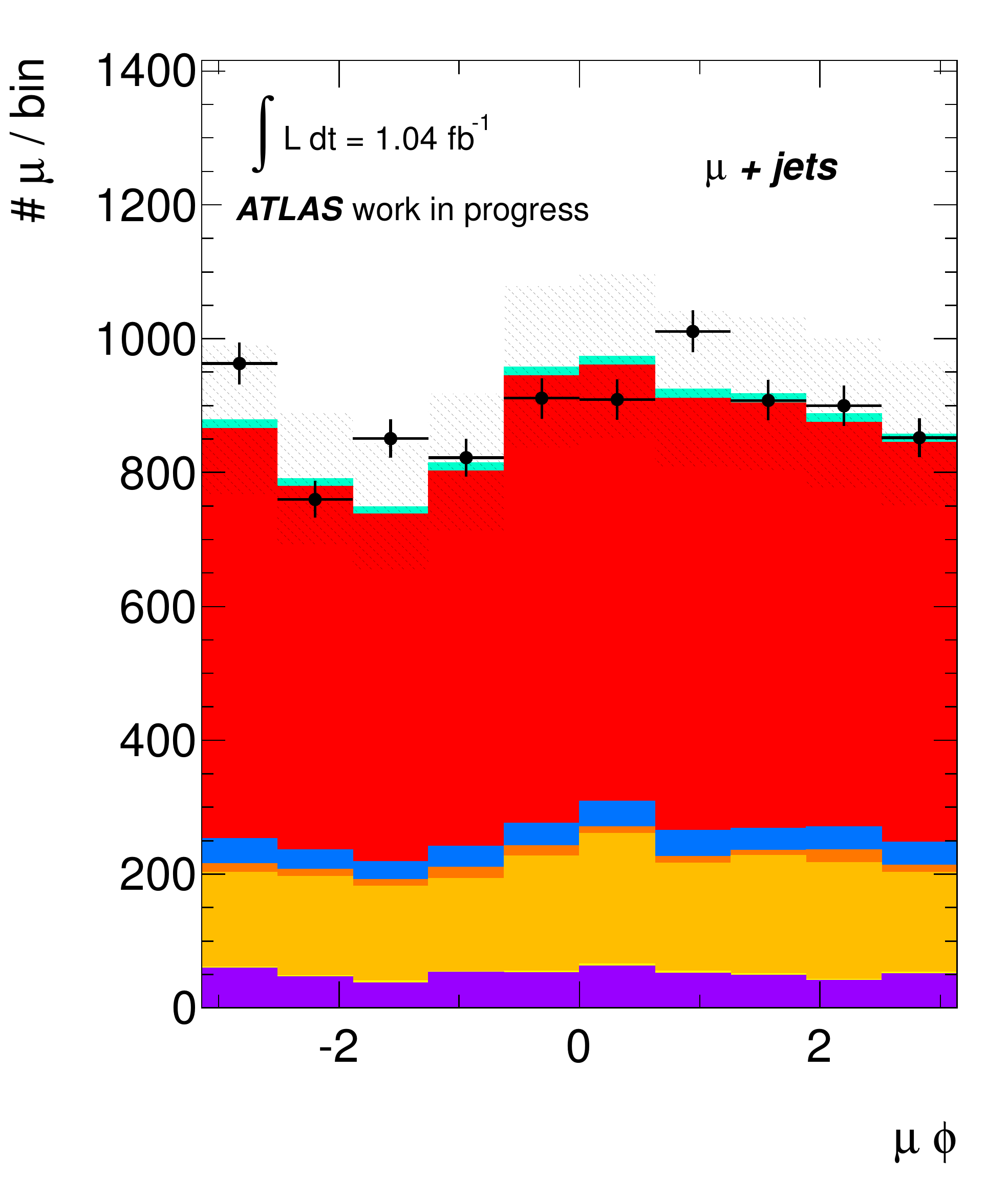}
\includegraphics[width=0.32\textwidth]{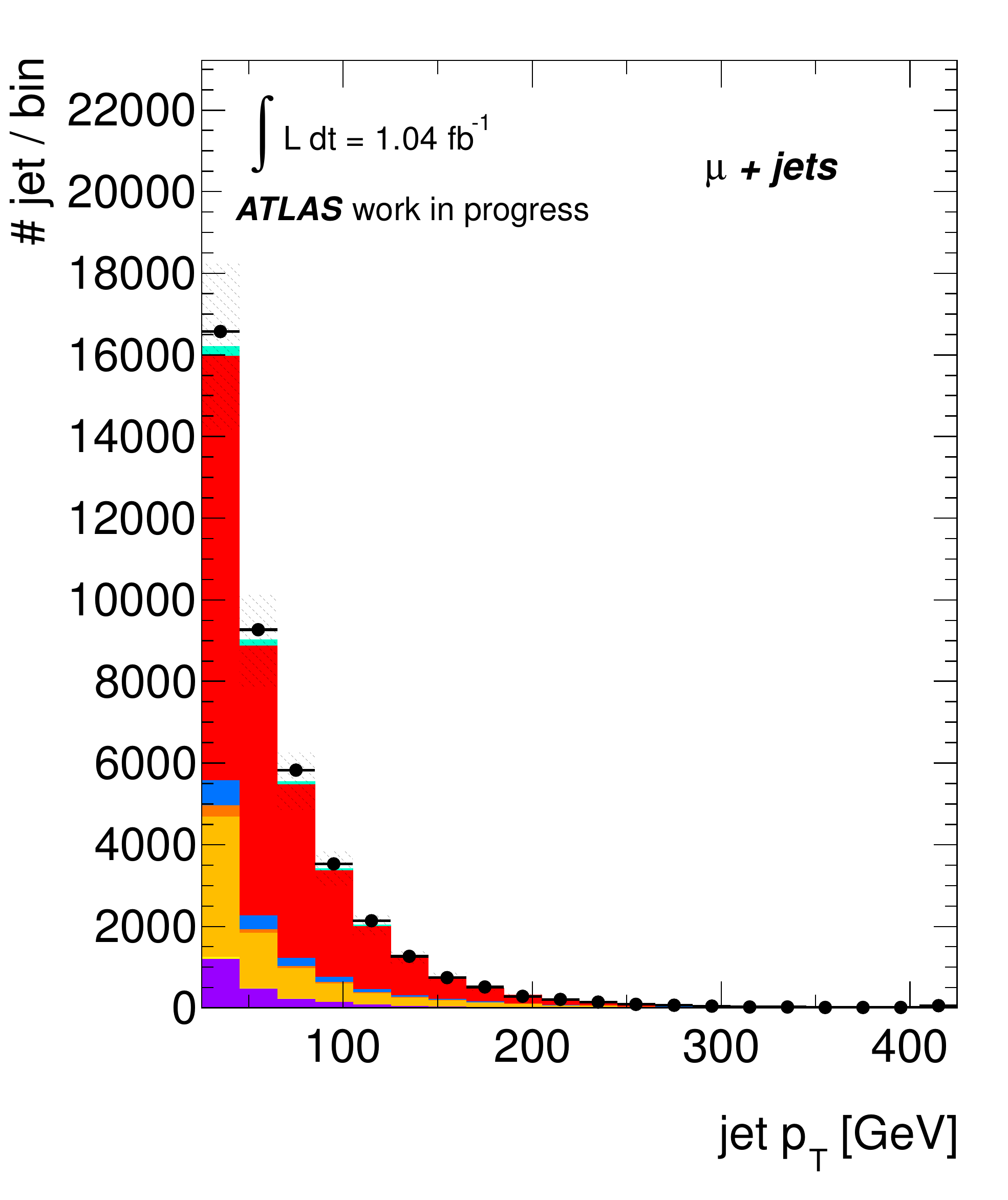} \\
\includegraphics[width=0.32\textwidth]{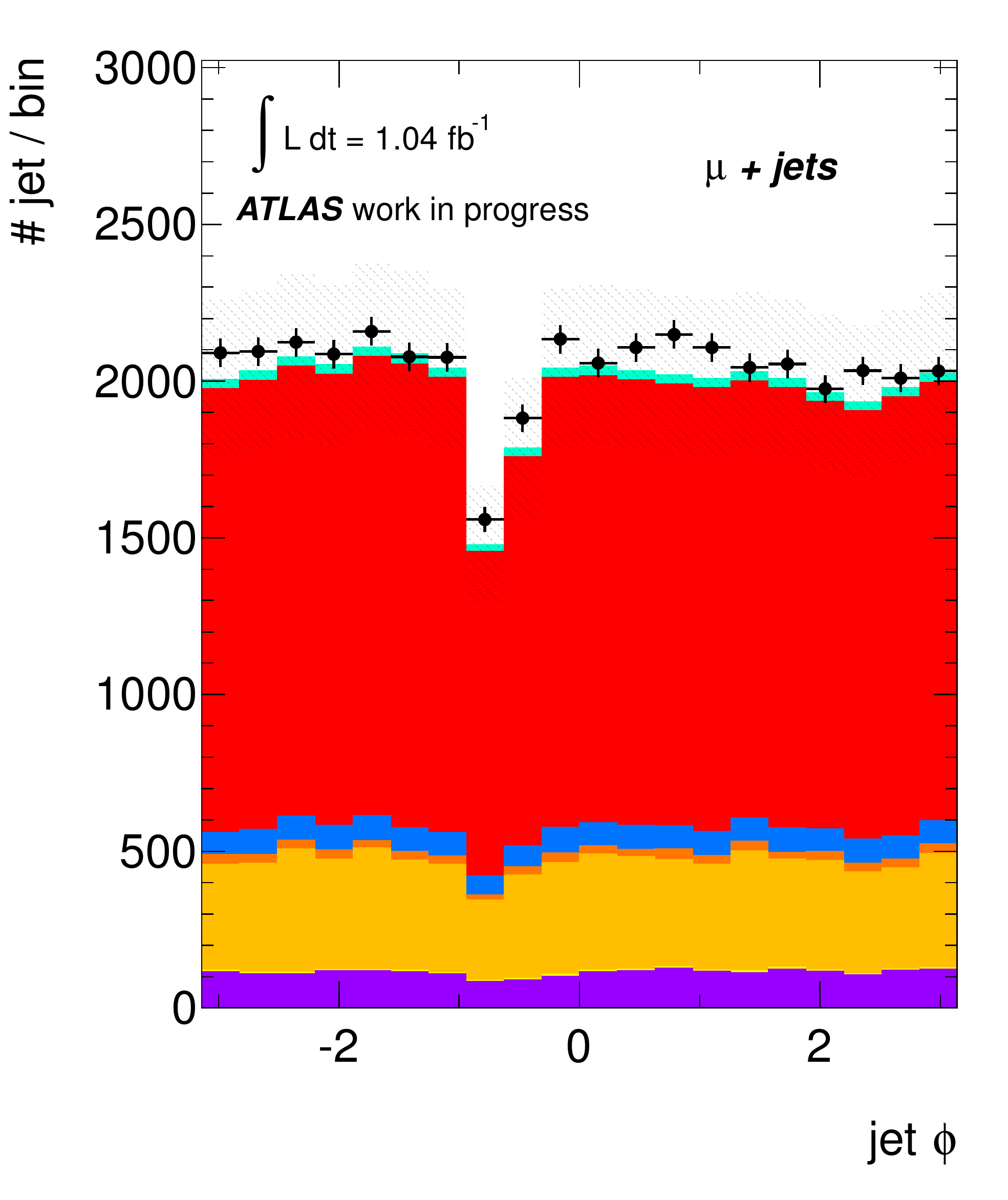}
\includegraphics[width=0.32\textwidth]{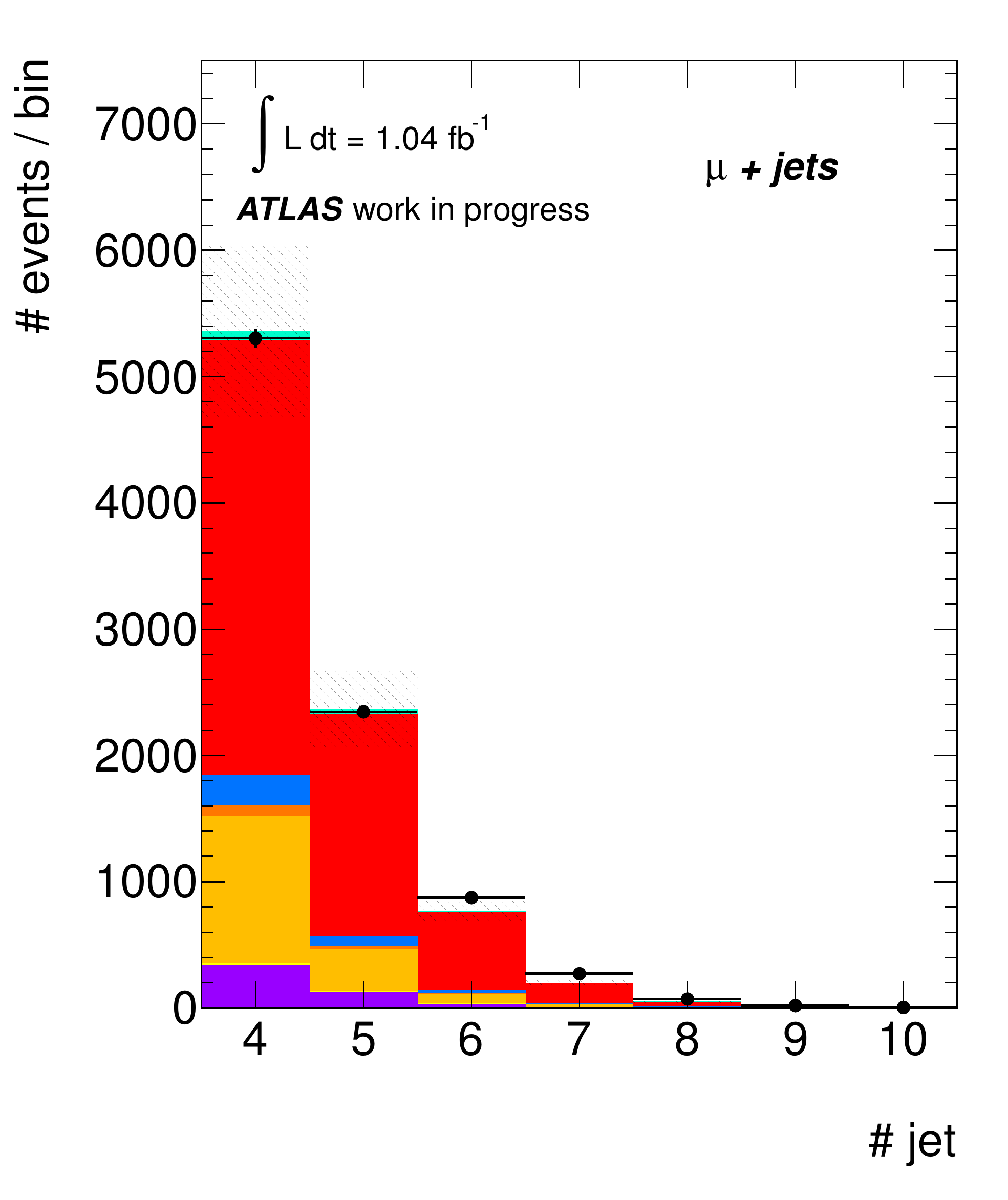}
\includegraphics[width=0.32\textwidth]{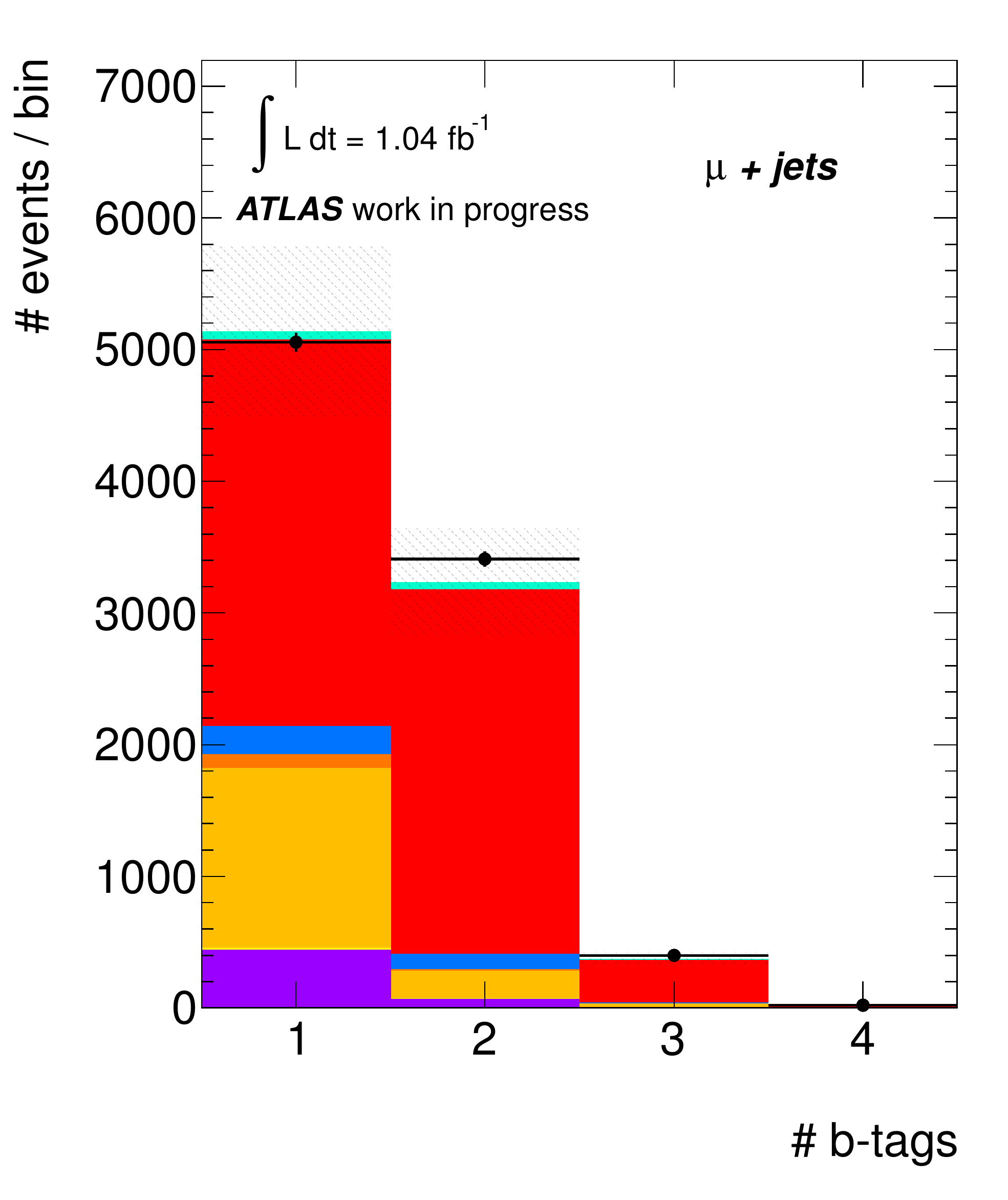} \\
\includegraphics[width=0.32\textwidth]{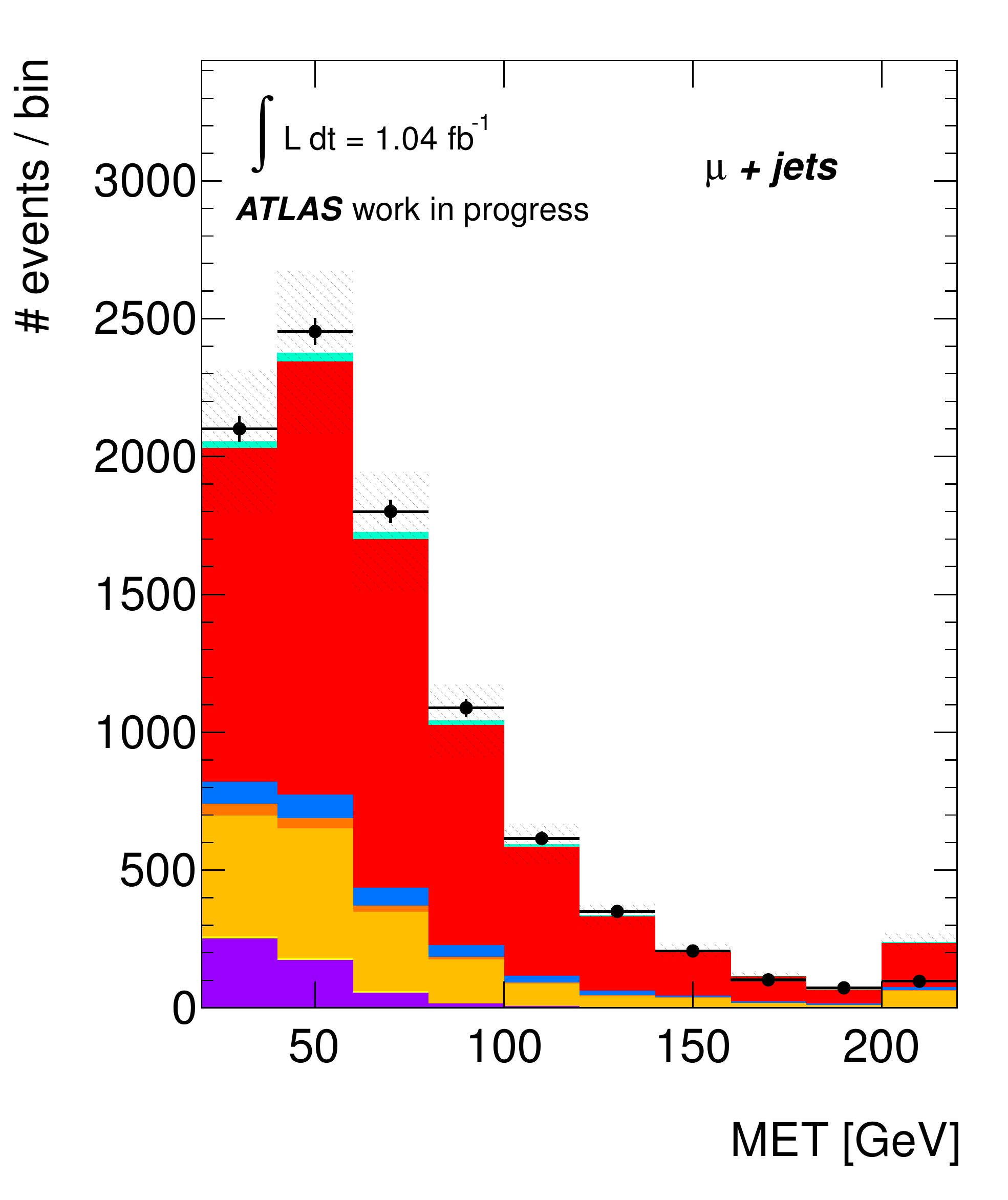}
\includegraphics[width=0.32\textwidth]{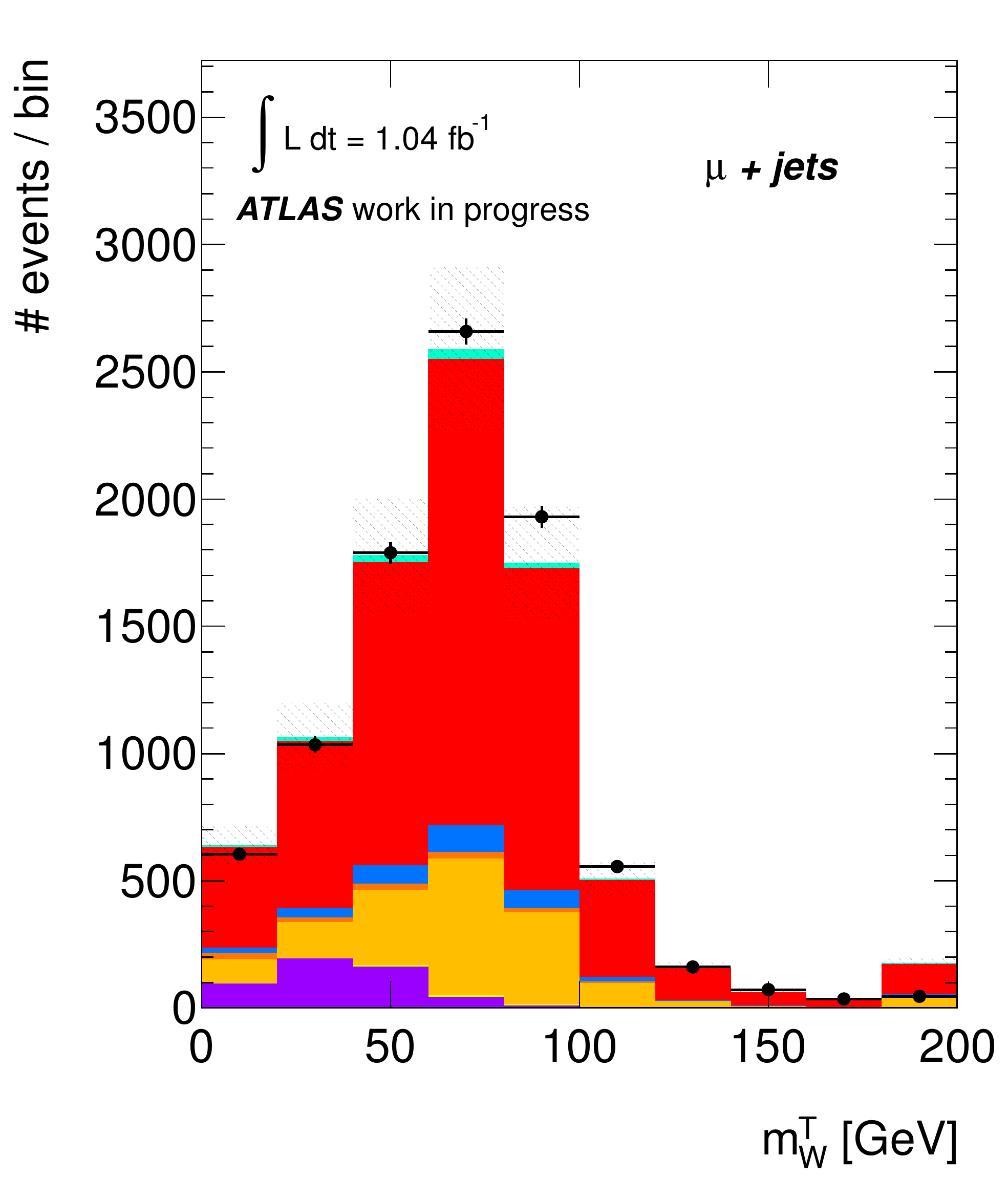}
\caption[Kinematic distributions after preselection ($\mu$+jets)]{
  Kinematic distributions after the preselection in the muon channel.
  Shown are the distributions from data, and the expectations from MC simulations and a data-driven estimate of the multijet
  background: muon $\pt$ and $\phi$, jet $\pt$ and $\phi$, the number of jets, the number of $b$-tagged jets, $\met$ and $\mtw$.
  The hashed uncertainty band represents the uncertainty on the expectation.
  In all plots, except for the $\phi$ distributions, the last bin includes the overflow bin.
}
\label{fig:controlplots_mu}
\end{center}
\end{figure}

Fig.~\ref{fig:controlplots_e} and~\ref{fig:controlplots_mu} show a selection of kinematic distributions after the preselection in the electron and muon
channels, respectively.
The data distributions are compared to the expectations from MC simulations and a data-driven estimate of the multijet
background (see Sec.~\ref{sec:QCDgamma}).
The yields and their uncertainties for the different contributing processes are described in Sec.~\ref{sec:yields}.

In the electron (muon) channel, the electron (muon) $\et$ ($\pt$) and $\eta$ ($\phi$) as well as the $\pt$ and $\eta$ ($\phi$) of all selected jets are shown.
Moreover, the total number of jets, the total number of $b$-tagged jets, the $\met$ and $\mtw$ are depicted.
Despite a slightly overall lower expectation in the number of events than observed in data, well covered by the uncertainty on the expectation, the agreement with the observed
data distributions is good in both lepton channels.
This holds also for the slightly asymmetric distribution of the muon $\phi$ and the significantly lower number of jets at the value of $\phi$ which
corresponds to the position of the broken optical link in the LAr calorimeter around \mbox{$\phi = -1$} (Ch.~\ref{sec:data}).

\section{Final event selection}
\label{sec:finalselection}

\begin{figure}[h]
\centering
\includegraphics[width=0.49\textwidth]{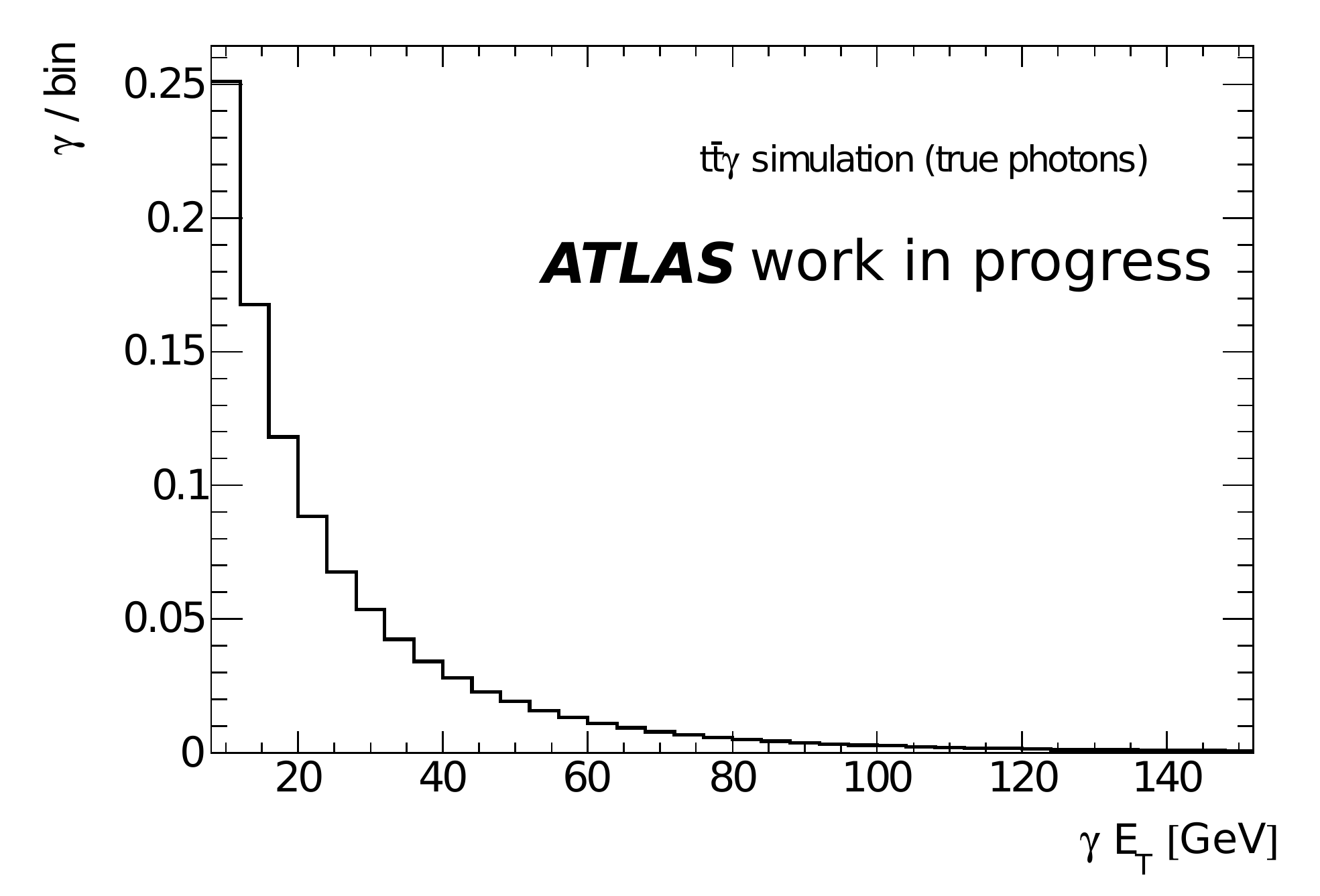}
\caption[True photon $\et$ in $\ttg$ simulation]{
  Distribution of the true photon $\et$ in $\ttg$ simulations.
  The last bin includes the overflow bin.
}
\label{fig:photonpt}
\end{figure}

The preselection, as detailed in the previous section, improves the $\ttbar$-over-background ratio and defined a sample dominated by
$\ttbar$ production.
$\ttg$ candidate events were identified in this sample by requiring the presence of a good photon with \mbox{$\et > 15 \GeV$}.
The photon $\et$ requirement was chosen as low as possible, because the distribution of the photon $\et$ in $\ttg$ events is dropping strongly
towards high values of $\et$, as shown in Fig.~\ref{fig:photonpt}.

In order to suppress $Z$+jets events with one of the electrons misidentified as a photon, the invariant mass of the electron and the photon in the
electron channel was required to be outside a window of \mbox{$5 \GeV$} around the $Z$ boson mass: \mbox{$m(e\gamma) \notin [86 \GeV, 96 \GeV]$}.
As discussed in Sec.~\ref{sec:photon}, the distance in $\eta$-$\phi$-space between good photons and the closest jet had to be larger than 0.5.

\begin{figure}[h!]
\begin{center}
\includegraphics[width=0.32\textwidth]{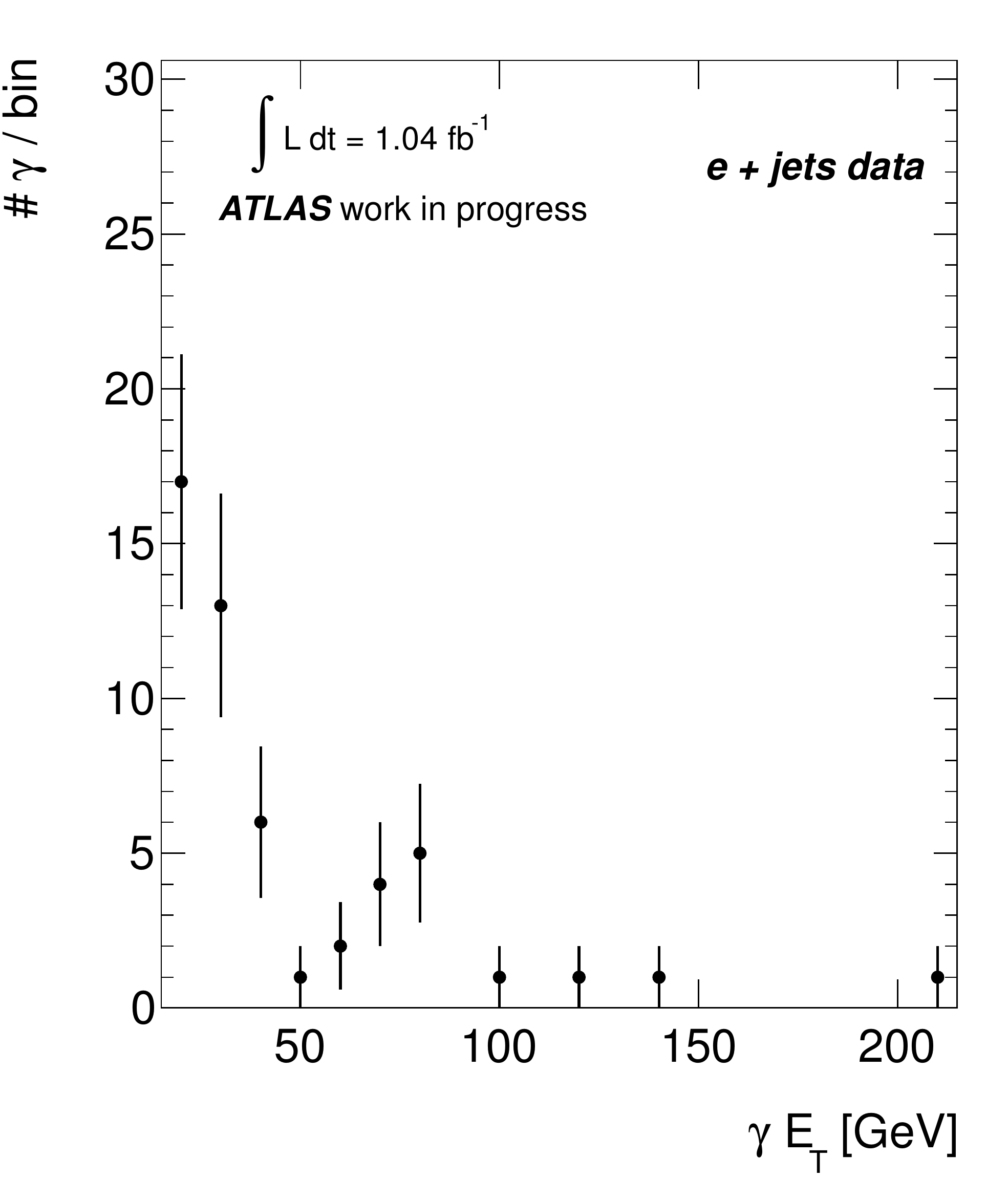}
\includegraphics[width=0.32\textwidth]{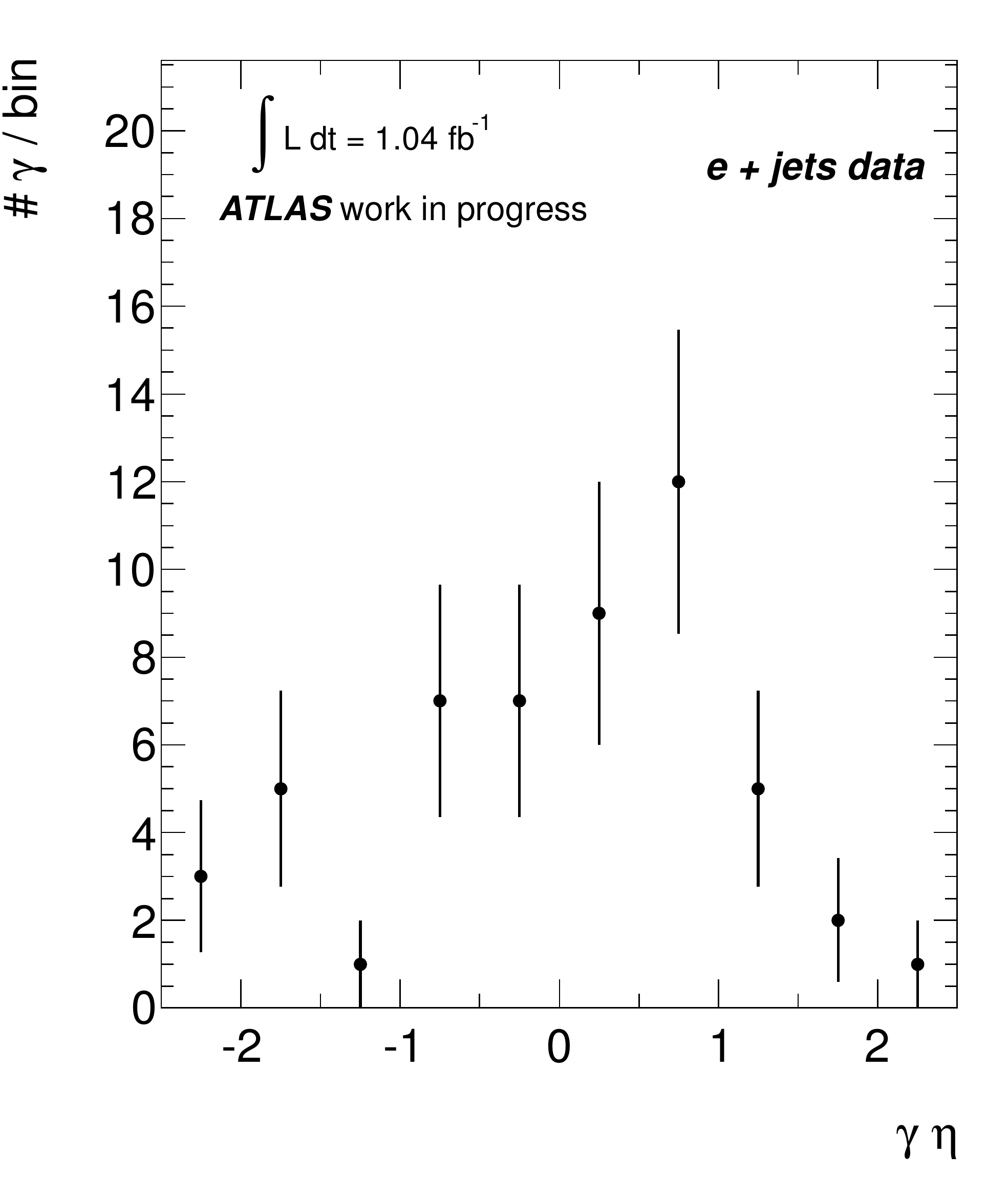}
\includegraphics[width=0.32\textwidth]{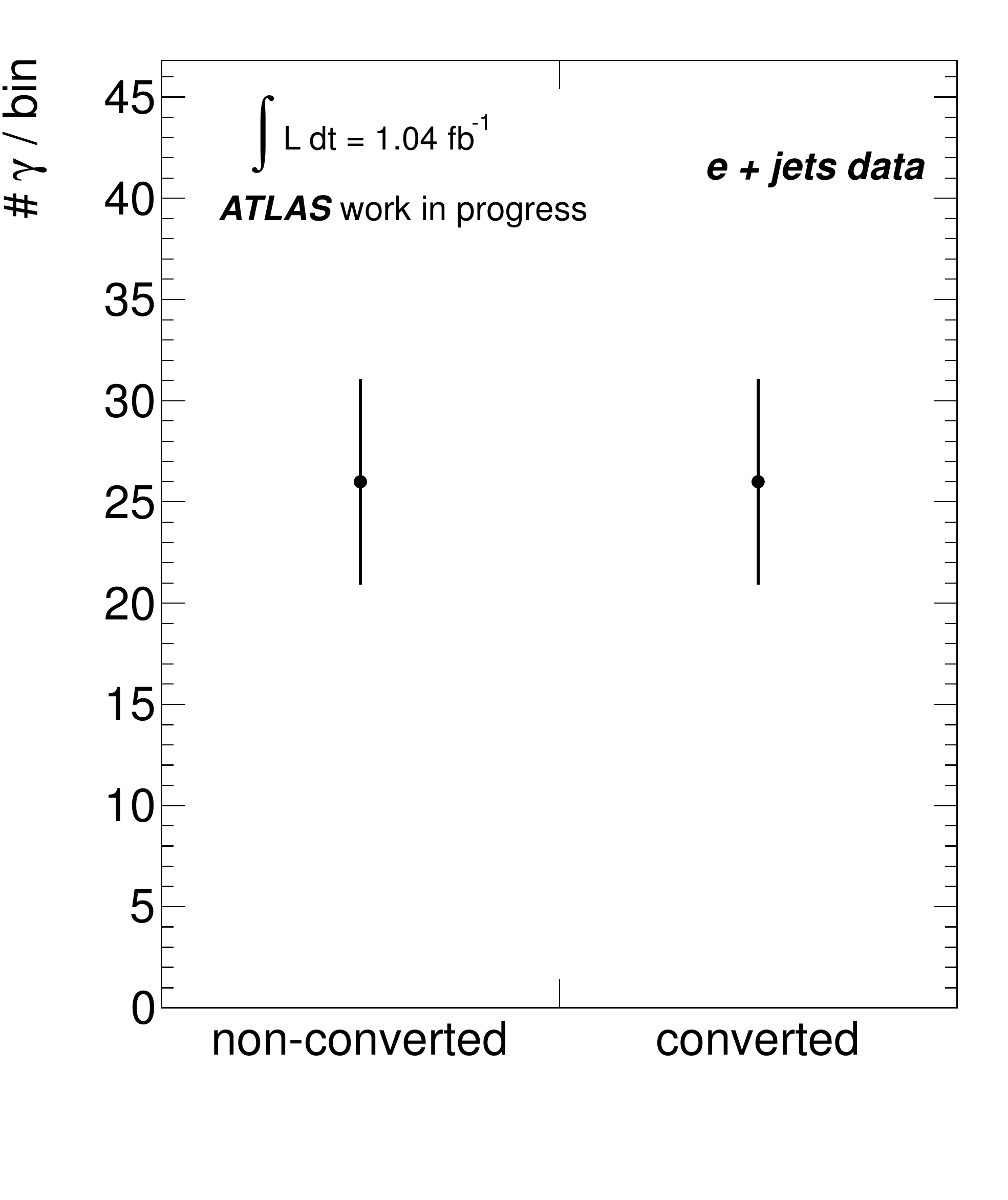} \\
\caption[Photon distributions (e+jets)]{
  Photon distributions for the $\ttg$ event candidates in data in the single electron channel.
  Shown are the photon $\et$ and $\eta$, and the fraction of non-converted and converted photons.
  In the $\et$ distribution, the last bin includes the overflow bin.
}
\label{fig:photonkinematics_el}
\includegraphics[width=0.32\textwidth]{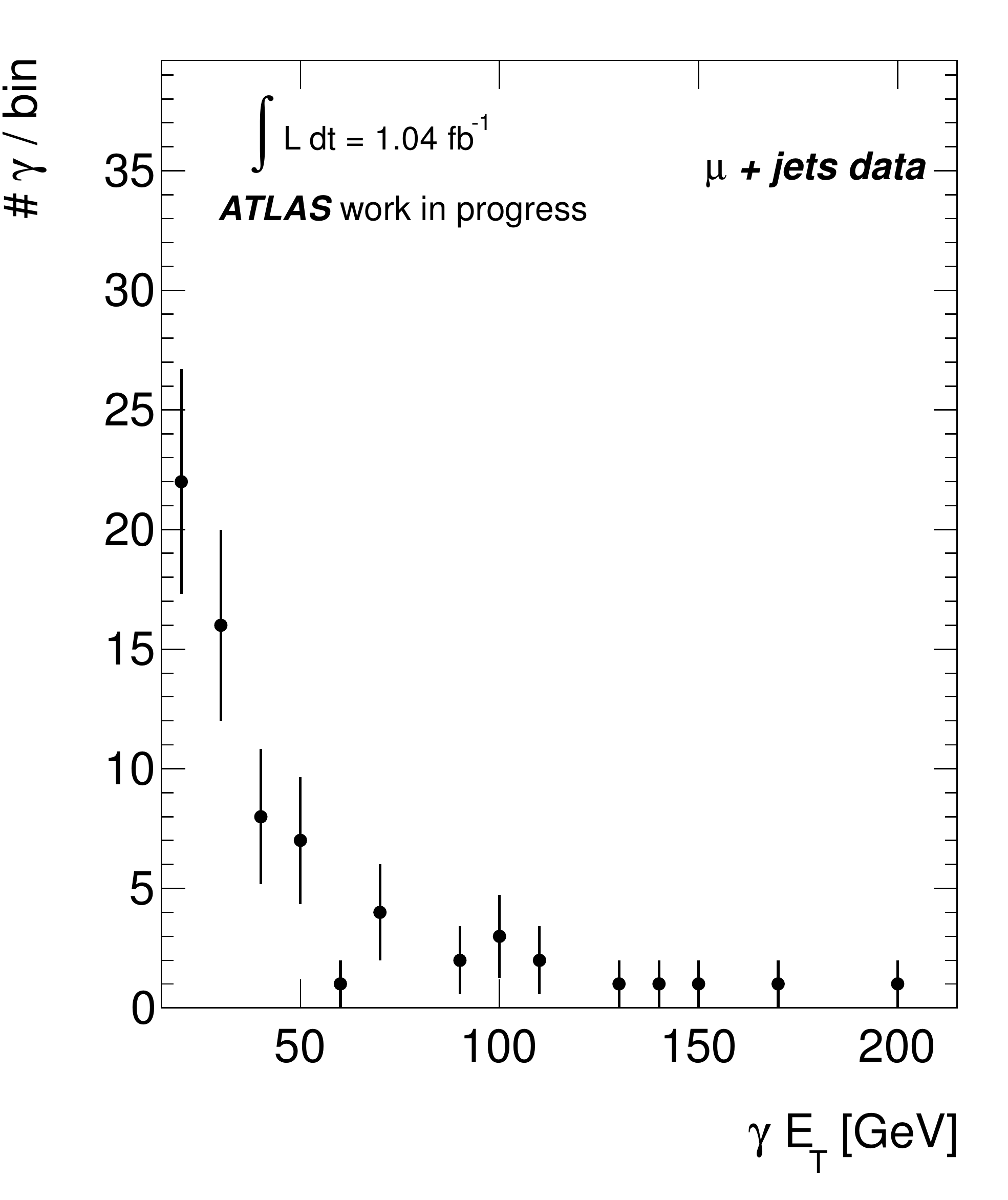}
\includegraphics[width=0.32\textwidth]{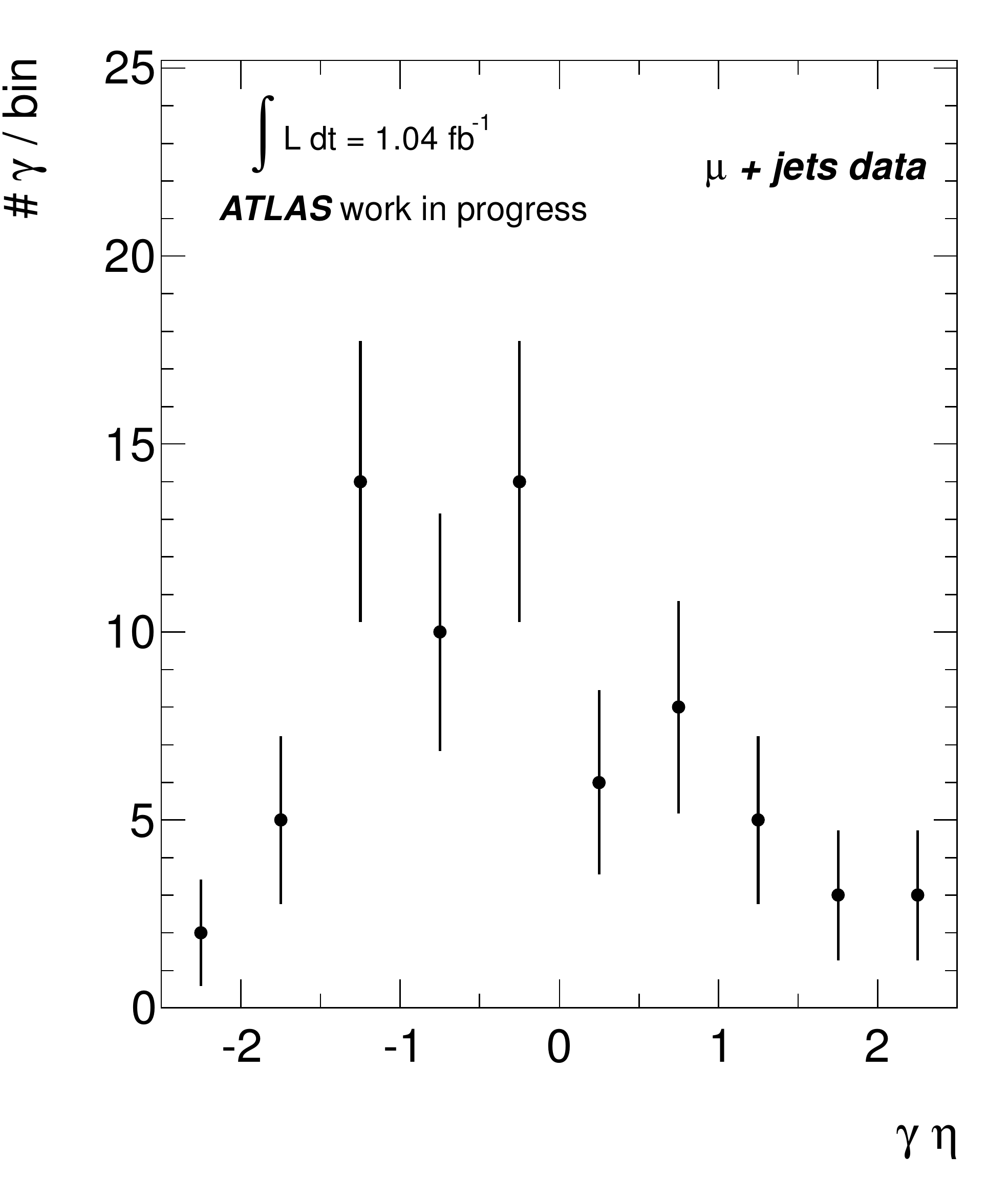}
\includegraphics[width=0.32\textwidth]{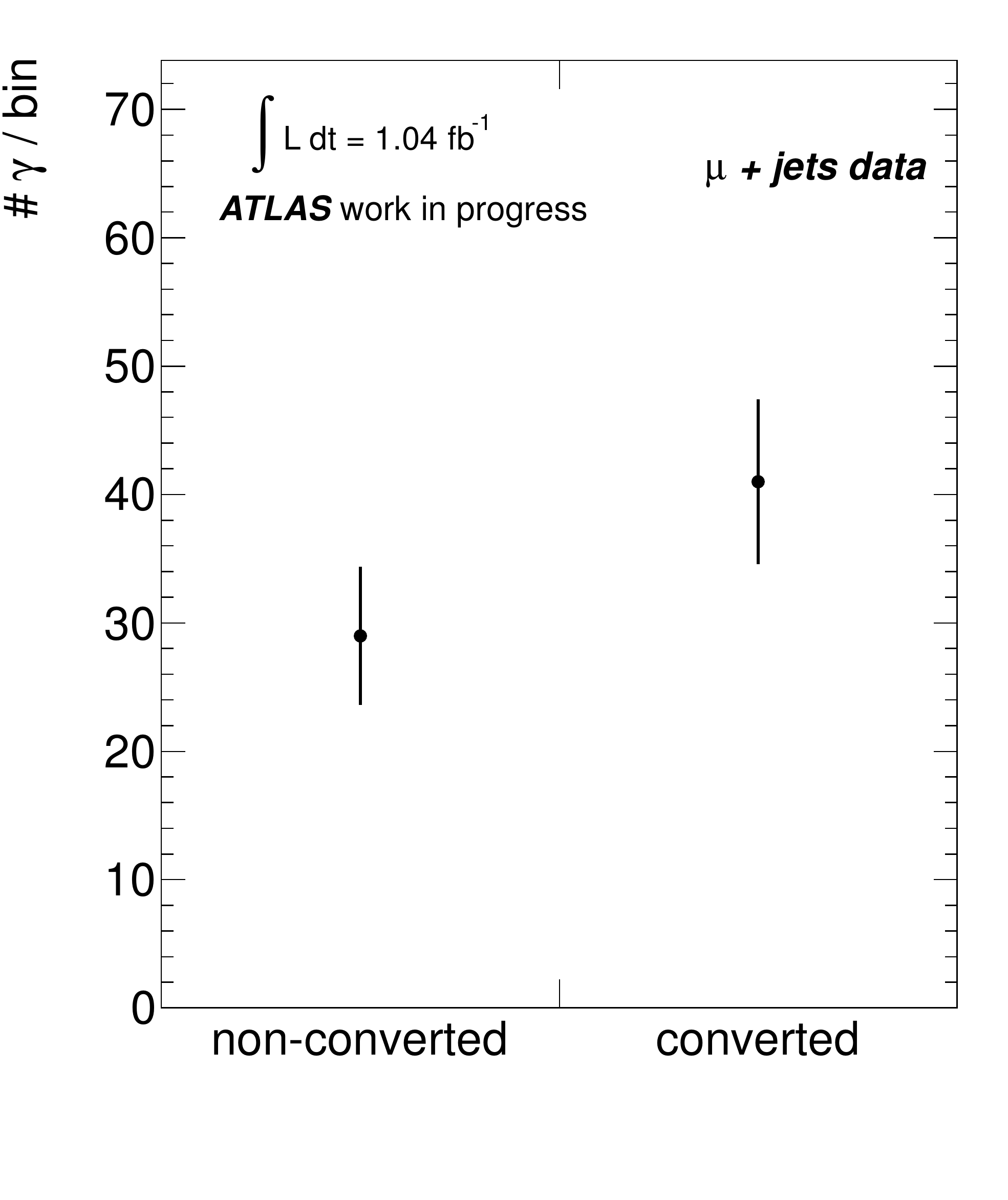} \\
\caption[Photon distributions ($\mu$+jets)]{
  Photon distributions for the $\ttg$ event candidates in data in the single muon channel.
  Shown are the photon $\et$ and $\eta$, and the fraction of non-converted and converted photons.
  In the $\et$ distribution, the last bin includes the overflow bin.
}
\label{fig:photonkinematics_mu}
\end{center}
\end{figure}

Fig.~\ref{fig:photonkinematics_el} and~\ref{fig:photonkinematics_mu} show the photon $\et$, $\eta$ and the fraction of non-converted
and converted photons of the $\ttg$ event candidates in the electron and muon channel, respectively.
The amount of real $\ttg$ events within these candidates was estimated from a template fit to the $\ptcone$ distribution as described in
Sec.~\ref{sec:strategy}.
This estimate is data-driven, because simulations are not trusted for the description of the background rates.
Consequently, no expectations from simulations are shown in these plots.

It is worth pointing out that the $\ttg$ candidates contain a sizable fraction from hadrons misidentified as photons.
However, the $\et$ spectrum of the photon candidates drops towards large transverse momenta, as expected for true photons as well as for fake photons.
More photon candidates are found in the very central part of the detector than at larger values of $|\eta|$, and
photon candidates are non-converted and converted to approximately equal parts.

\section{Event yields}
\label{sec:yields}

Tab.~\ref{tab:cutflowpreselection} shows the event yields for the preselection for data and the expectations
from MC simulations as well as the estimation of the multijet background from data in the electron and muon channels, respectively.
The yields before and after requiring at least one $b$-tagged jet in the event selection are shown.

For the uncertainties on the expectation of the different processes, only the uncertainty on the cross section calculations were taken into account
for $\ttg$, $\ttbar$, single top and diboson production as mentioned in Ch.~\ref{sec:modelling}.
The procedure for the estimation of the multijet background is described in Sec.~\ref{sec:QCDgamma}.
The uncertainty on the multijet background was estimated conservatively to 100\% if at least one $b$-tagged jet was required in the events, and to 50\%
otherwise.

Additional systematic uncertainties on the expectations were not taken into account, because this table is only intended to illustrate that the yields
observed in data are well under control for the preselection.
The yields observed in data are roughly 100 events larger than expected in both lepton channels after the full preselection,
which is well covered by the uncertainty on the expectation.
Overall, the yields observed in data are consistent with the sum of the expectations within the considered uncertainties before and after the
$b$-tagging requirement.

Tab.~\ref{tab:cutflowselection} shows the event yields for data and the expectations for signal and
the different background sources for the final event selection.
The $\ttg$ signal expectation from the WHIZARD MC simulation yields \mbox{$22 \pm 4$} and \mbox{$28 \pm 6$} events in the electron
and muon channels, respectively.
This corresponds to a combined efficiency and acceptance of 0.97\% and 1.27\% with respect to the total generated signal.
The efficiency of the final event selection for $\ttg$ events was found to be roughly 20\% with respect to the
preselection (cf. Tab.~\ref{tab:cutflowpreselection} and~\ref{tab:cutflowselection}).

The yields for $\ttg$ events outside of the signal phase space, \textit{$\ttg$ background} as described in Sec.~\ref{sec:backgroundmodelling}, are
discussed in Sec.~\ref{sec:ttgbkg}.
The contributions from $W$+jets+$\gamma$, multijet+$\gamma$, $Z$+jets+$\gamma$, single top+$\gamma$ and diboson+$\gamma$ production are discussed
in Sec.~\ref{sec:QCDgamma}~--~\ref{sec:restgamma}.
The contributions from electrons misidentified as photons in dileptonic $\ttbar$ decays, $Z$+jets, single top and diboson
events was estimated by measuring the misidentification rate in data (Ch.~\ref{sec:electronfake}).

Hadrons misidentified as photons can occur in all of the background processes by the presence of additional jets.
However, a prediction for the yield cannot be obtained from simulations as discussed in Sec.~\ref{sec:backgroundmodelling}.
In order to estimate the number of hadrons misidentified as photons in the selected data events, a template fit to the $\ptcone$
distribution was performed (Ch.~\ref{sec:strategy}).
The final result of the fit including systematic uncertainties is presented in Ch.~\ref{sec:results}.

As a cross-check for the final fit result, the amount of fake photons from hadrons was estimated from photon candidates with
\mbox{$\ptcone > 3 \GeV$}.
While only 2\% of true photons have $\ptcone$ values larger than \mbox{$3 \GeV$}, this holds true for 42\% of the hadron fakes.
%(Fig.~\ref{fig:templates_overlay}).
Hence, \mbox{$\ptcone > 3 \GeV$} defines a control region (CR) largely dominated by hadron fakes.

\begin{figure}[h]
\begin{center}
\includegraphics[width=0.49\textwidth]{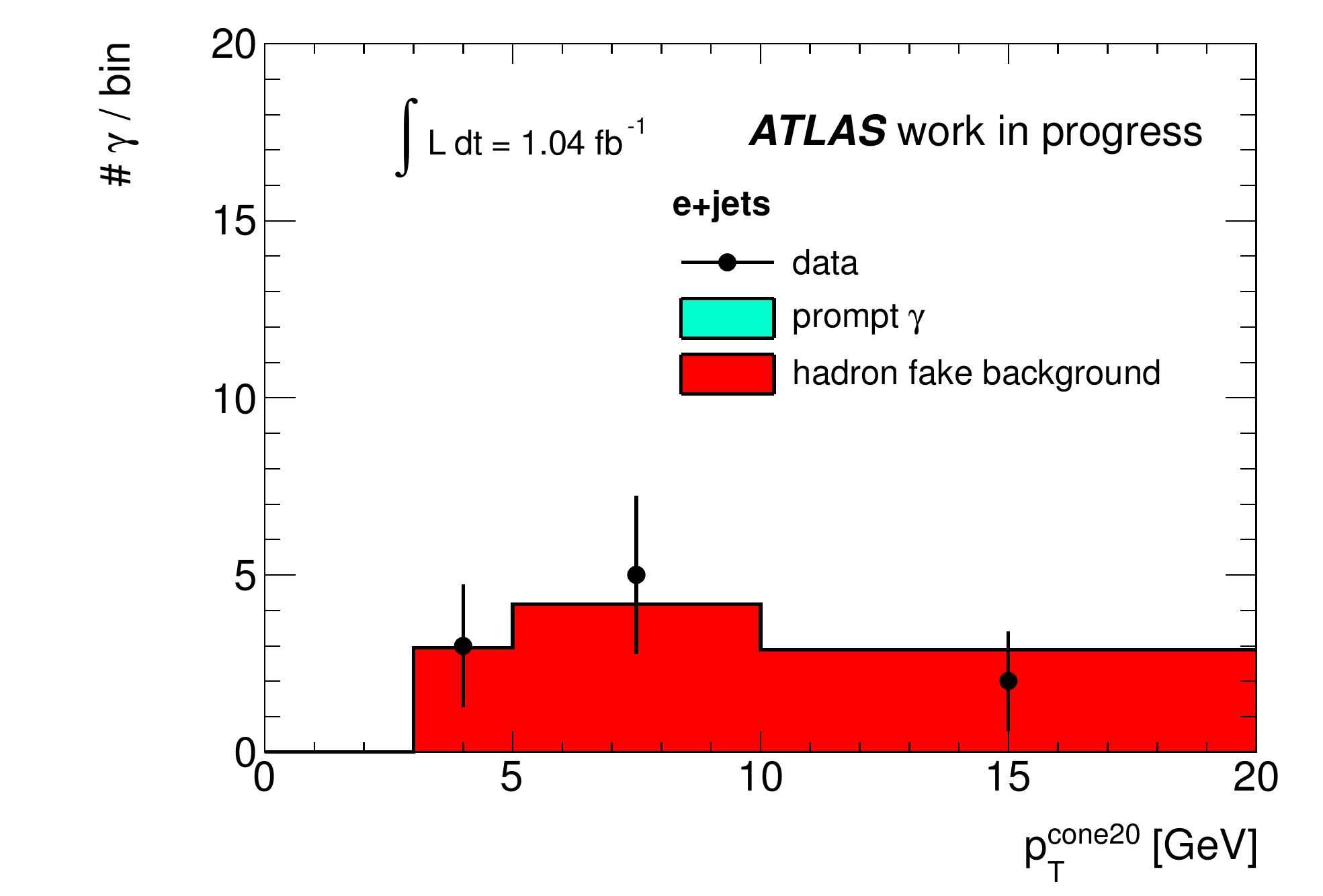}
\includegraphics[width=0.49\textwidth]{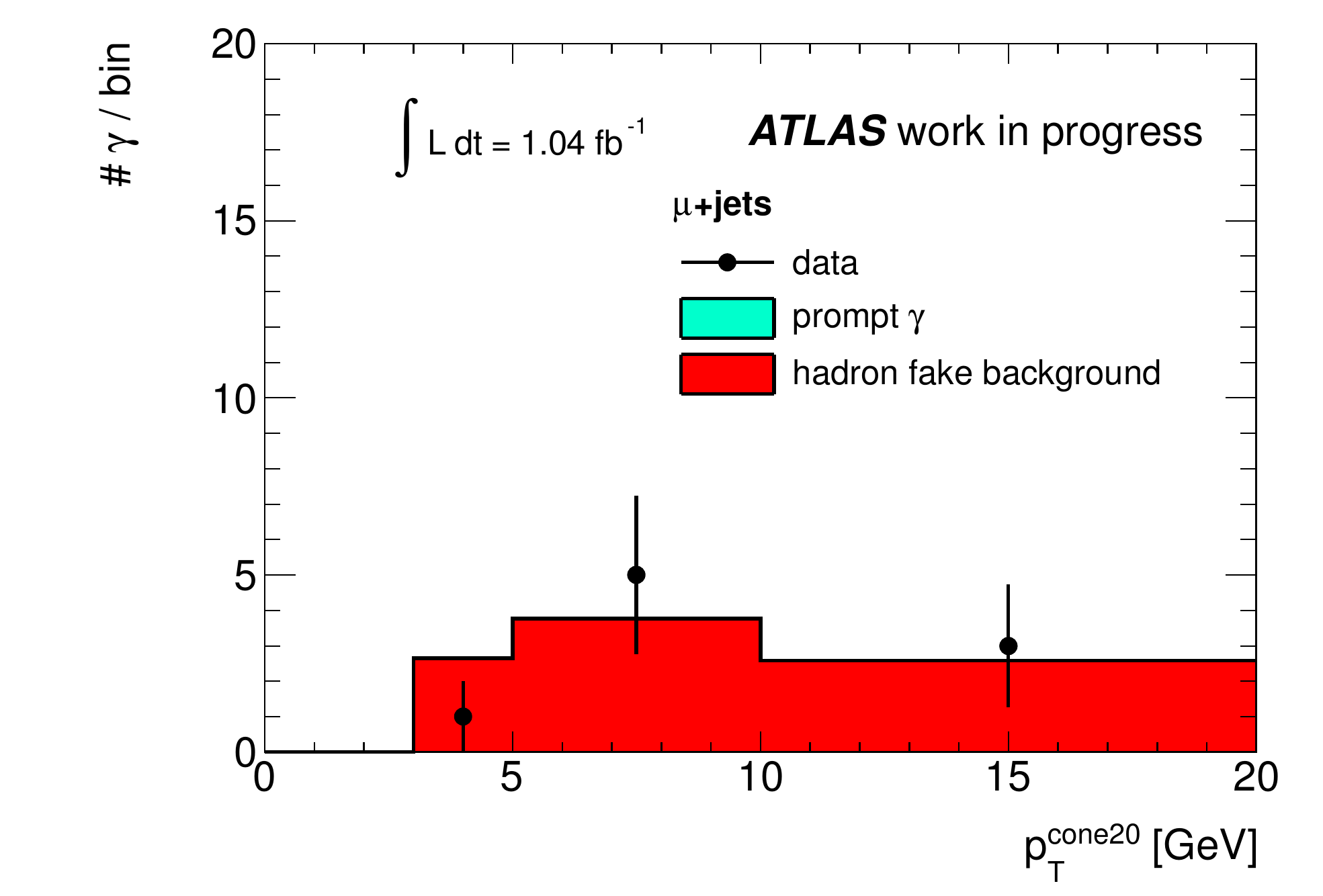}
\caption[Template fits to the $\ptcone$ distribution for \mbox{$\ptcone > 3 \GeV$}]{
  Template fits to the $\ptcone$ distribution for photon candidates with \mbox{$\ptcone > 3 \GeV$} in the electron (left) and the
  muon channel (right) with templates for true photons and for hadrons misidentified as photons.
  The data distribution is compatible with being only due to hadron fakes.
  In both plots, the last bin includes the overflow bin.
}
\label{fig:hadrononlyfits}
\end{center}
\end{figure}

Fig.~\ref{fig:hadrononlyfits} shows template fits, as described in Ch.~\ref{sec:strategy}, to the $\ptcone$ distribution for photon candidates in the
CR in the electron (left) and the muon channel (right).
Templates for true photons and for hadrons misidentified as photons were used.
The data distribution is compatible with being only due to hadron fakes.

The fits yield \mbox{$10 \pm 3$} and \mbox{$9 \pm 3$} hadron fake events in the CR in the electron and muon channel, respectively, which
translates into expectations of \mbox{$24 \pm 8$} and \mbox{$22 \pm 7$} events for the whole $\ptcone$ spectrum.
The uncertainty quoted is the statistical uncertainty from the fit only.
The numbers from the final template fit (cf. Ch.~\ref{sec:results}) read
\mbox{$20 \, ^{+7}_{-6} \, \mathrm{(stat.)} \pm 3 \, \mathrm{(syst.)}$} and
\mbox{$27 \, ^{+8}_{-7} \, \mathrm{(stat.)} \pm 4 \, \mathrm{(syst.)}$} in the electron and muon channel, respectively.
The final estimates in both channels are consistent with the cross-check in the region \mbox{$\ptcone > 3 \GeV$} within the statistical uncertainties.
However, it needs to be pointed out that the statistical uncertainties of the final and the cross-check estimates are partly correlated.
The final estimate is believed to be closer to the real hadron fake contribution, because it takes into account more information from the whole
$\ptcone$ spectrum.

The uncertainties on the other background contributions quoted in Tab.~\ref{tab:cutflowselection} are the total uncertainties including statistical
and systematic uncertainties as derived in Ch.~\ref{sec:electronfake} and Ch.~\ref{sec:backgroundphotons}.
In data, 52 and 70 events in the electron and muon channels were identified, respectively.
The sum of the signal and background expectations yield \mbox{$60 \pm 10$} and \mbox{$67 \pm 10$}.
Hence, the number of observed events in data is compatible with the expectation within the uncertainties in both lepton channels.

\begin{table}[h]
\centering
\begin{tabular}[h]{|l|rcr|rcr|rcr|rcr|}
\hline
           & \multicolumn{6}{c|}{e+jets} & \multicolumn{6}{c|}{$\mu$+jets} \\
\hline
           & \multicolumn{3}{c|}{before $b$-tag} & \multicolumn{3}{c|}{after $b$-tag} & \multicolumn{3}{c|}{before $b$-tag} & \multicolumn{3}{c|}{after $b$-tag} \\
\hline
$\ttg$     &   108 & $\pm$ &   21 &   95 & $\pm$ &  19 &   144 & $\pm$ &   28 &  126 & $\pm$ &   25 \\
$\ttbar$   &  4710 & $\pm$ &  460 & 4160 & $\pm$ & 400 &  6840 & $\pm$ &  660 & 6040 & $\pm$ &  590 \\
$W$+jets   &  5600 & $\pm$ & 2700 &  890 & $\pm$ & 460 & 10000 & $\pm$ & 4900 & 1600 & $\pm$ &  800 \\
$Z$+jets   &   630 & $\pm$ &  300 &   99 & $\pm$ &  48 &   860 & $\pm$ &  410 &  122 & $\pm$ &   59 \\
Single top &   394 & $\pm$ &   17 &  265 & $\pm$ &  13 &   552 & $\pm$ &   22 &  360 & $\pm$ &   17 \\
Diboson    &    79 & $\pm$ &    4 &   13 & $\pm$ &   1 &   125 & $\pm$ &    6 &   22 & $\pm$ &    1 \\
Multijet   &   790 & $\pm$ &  390 &  150 & $\pm$ & 150 &  1600 & $\pm$ &  800 &  490 & $\pm$ &  490 \\
\hline
Sum        & 12300 & $\pm$ & 2800 & 5670 & $\pm$ & 630 & 20200 & $\pm$ & 5000 & 8760 & $\pm$ & 1100 \\
\hline
Data       & 11856 &       &      & 5761 &       &     & 18978 &       &      & 8863 &       &      \\
\hline
\end{tabular}
\caption[Event yields for data and expectations (preselection)] {
  Event yields for data and expectations for signal and the different background contributions for \mbox{$1.04 \ifb$} for the preselection in both lepton
  channels.
}
\label{tab:cutflowpreselection}
\end{table}

\begin{table}[h]
\centering
\begin{tabular}[h]{|l|r@{}l c r@{}l|r@{}l c r@{}l|}
\hline
                      & \multicolumn{5}{c|}{e+jets} & \multicolumn{5}{c|}{$\mu$+jets} \\
\hline
$\ttg$                & 22 &    & $\pm$ & 4 &                          & 28 &    & $\pm$ & 6 &   \\
Background $\ttg$   &  0 &.8  & $^{+}_{-}$ & $^{\emptyplus 1}_{\emptyminus 0}$ & $^{.1\emptyplus}_{.8\emptyminus}$   &  1 &.3  & $^{+}_{-}$ & $^{\emptyplus 1}_{\emptyminus 1}$ & $^{.9\emptyplus}_{.3\emptyminus}$ \\
$W$+jets+$\gamma$     &  1 &.8  & $\pm$ & 0 &.7                        &  3 &.7  & $\pm$ & 1 &.4 \\
$Z$+jets+$\gamma$     &  1 &.3  & $^+_-$ & $^{\emptyplus 2}_{\emptyminus 1}$ & $^{.4\emptyplus}_{.3\emptyminus}$ &  1 &.6  & $^+_-$ & $^{\emptyplus 2}_{\emptyminus 1}$ & $^{.3\emptyplus}_{.6\emptyminus}$ \\
Single top+$\gamma$ &  0 &.6  & $^+_-$ & $^{\emptyplus 0}_{\emptyminus 0}$ & $^{.7\emptyplus}_{.6\emptyminus}$ &  0 &.2  & $^+_-$ & $^{\emptyplus 0}_{\emptyminus 0}$ & $^{.3\emptyplus}_{.2\emptyminus}$ \\
Diboson+$\gamma$    &  0 &.16 & $^{+}_{-}$ & $^{\emptyplus 0}_{\emptyminus 0}$ & $^{.34\emptyplus}_{.16\emptyminus}$ &  0 &.04 & $^{+}_{-}$ & $^{\emptyplus 0}_{\emptyminus 0}$ & {$^{.18\emptyplus}_{.04\emptyminus}$} \\
Multijet+$\gamma$   &  1 &.2 & $^{+}_{-}$ & $^{\emptyplus 1}_{\emptyminus 1}$ & $^{.6\emptyplus}_{.2\emptyminus}$ &  0 &.3 & $^{+}_{-}$ & $^{\emptyplus 1}_{\emptyminus 0}$ & $^{.0\emptyplus}_{.3\emptyminus}$ \\
Dileptonic $\ttbar$ ($e \to \gamma$) &  6 &.8  & $\pm$ & 2 &.3 &  9 &.6  & $\pm$ & 2 &.7 \\
$Z$+jets ($e \to \gamma$) &  1 &.7  & $^+_-$ & $^{\emptyplus 3}_{\emptyminus 1}$ & $^{.1\emptyplus}_{.7\emptyminus}$ &  0 &.7  & $^+_-$ & $^{\emptyplus 1}_{\emptyminus 0}$ & $^{.8\emptyplus}_{.7\emptyminus}$ \\
Single top $Wt$-channel ($e \to \gamma$) &  0 &.22  & $^+_-$ & $^{\emptyplus 0}_{\emptyminus 0}$ & $^{.25\emptyplus}_{.22\emptyminus}$ & -0 &.10  & $\pm$ & 0 & .10 \\
Diboson ($e \to \gamma$) &  0 &.04  & $^+_-$ & $^{\emptyplus 0}_{\emptyminus 0}$ & $^{.14\emptyplus}_{.04\emptyminus}$ &  0 &.00  & $^+_-$ & $^{\emptyplus 0}_{\emptyminus 0}$ & $^{.14\emptyplus}_{.00\emptyminus}$ \\
Hadrons   misidentified as photons & 24 &    & $\pm$ & 8 &                          & 22 &    & $\pm$ & 7 &  \\
\hline
Sum                   & 60 &    & $\pm$ & 10&                          & 67 &    & $\pm$ & 10& \\
\hline
Data                  & 52 &    &       &   &                          & 70 &    &       &   & \\
\hline
\end{tabular}
\caption[Event yields for data and expectations (final selection)] {
  Event yields for data and expectations for signal and the different background contributions for \mbox{$1.04 \ifb$} for the final event selection
  in both lepton channels.
}
\label{tab:cutflowselection}
\end{table}

\chapter{Analysis strategy}
\label{sec:strategy}

Background processes to $\ttg$ production feature either real photons, such as $W$+jets+$\gamma$ production, electrons misidentified
as photons, or hadrons misidentified as photons (hadron fakes).
The strategy for the analysis was set up to cope in particular with the sizable background from hadron fakes.
$\tau$-leptons may be misidentified as photons if they decay into an electron or into hadrons.
Both cases are covered by the treatments of electrons and hadrons misidentified as photons.

The treatment of the background from events with hadrons misidentified as photons needs particular attention, because it is not well modelled by MC
simulations as discussed in Sec.~\ref{sec:backgroundmodelling}.
The analysis strategy takes this into account and the amount of signal $\ttg$ events as well as the amount of hadron fakes were estimated from a
template fit to the photon isolation distribution of the selected $\ttg$ candidates.
While prompt photons are generally isolated, hadron fakes are typically surrounded by other particles from the fragmentation process and hence
a good discrimination between prompt photons and hadron fakes can be achieved by the use of isolation observables.

Backgrounds with real photons or electrons misidentified as photons cannot be distinguished from $\ttg$ events by considering the photon isolation,
and they were estimated separately:
the probability for electrons to be misidentified as photons was measured in data using a sample largely enriched in \Zee events (Ch.~\ref{sec:electronfake}).
Background contributions with real photons were estimated partly from control regions in data and partly from MC simulations
(Ch.~\ref{sec:backgroundphotons}).

\subsubsection{\boldmath Photon isolation $\left( \ptcone \right)$\unboldmath}

Isolation observables are constructed from the surrounding energy in the calorimeter or from the tracks close to the photon candidate.
The transverse isolation energy in the calorimeter in a narrow cone around the photon depends on the photon $\eta$, because
of the varying amount of material in front of the calorimeter.
Given the limited amount of $\ttg$ candidate events, calorimeter isolation was hence disfavoured as a discriminating variable
with respect to the $\ptcone$ observable, which is independent of the photon $\eta$ to first order (Ch.~\ref{sec:photontemplate}).
$\ptcone$ is defined as the sum of the transverse momenta of the tracks in a cone of \mbox{$\Delta R = 0.2$} around the photon
candidate~\cite{expPhotonPerf}.
A small cone size of 0.2 was chosen in order to avoid a bias to $\ptcone$ due to the presence of nearby particles from the $\ttg$ final state.

Tracks were required to have a $\pt$ of at least \mbox{$1 \GeV$}, at least seven hits in the Pixel and SCT detectors, and a hit in the
Pixel $b$-layer.
Tracks associated to conversion vertices closer than 0.1 to the photon in $\eta$-$\phi$-space were not considered.
The transverse and longitudinal impact parameters of the tracks with respect to the primary vertex had to be smaller than \mbox{$1 \mm$} to reduce biases from
tracks that originated from pile-up interactions.
The longitudinal impact parameter requirement was not included in the definition of $\ptcone$ for photons~\cite{expPhotonPerf}, although
it was used for electrons.
This requirement was added for photons for a consistent treatment of electron and photon isolation,
given that the photon distribution was estimated using electrons from \Zee decays (Ch.~\ref{sec:photontemplate}).
In order to add the requirement on the longitudinal impact parameter, the photon $\ptcone$ observable was recalculated.
%
%In order to add the requirement on the longitudinal impact parameter, the photon $\ptcone$ observable had to be recalculated\footnote{The
%recalculation was not straightforward, because the conversion vertex information was not available in the format used for analysis.}, which was done
%with an algorithm described in App.~\ref{sec:app_ptcone20}.

\begin{figure}[h]
  \begin{center}
    \includegraphics[width=0.49\textwidth]{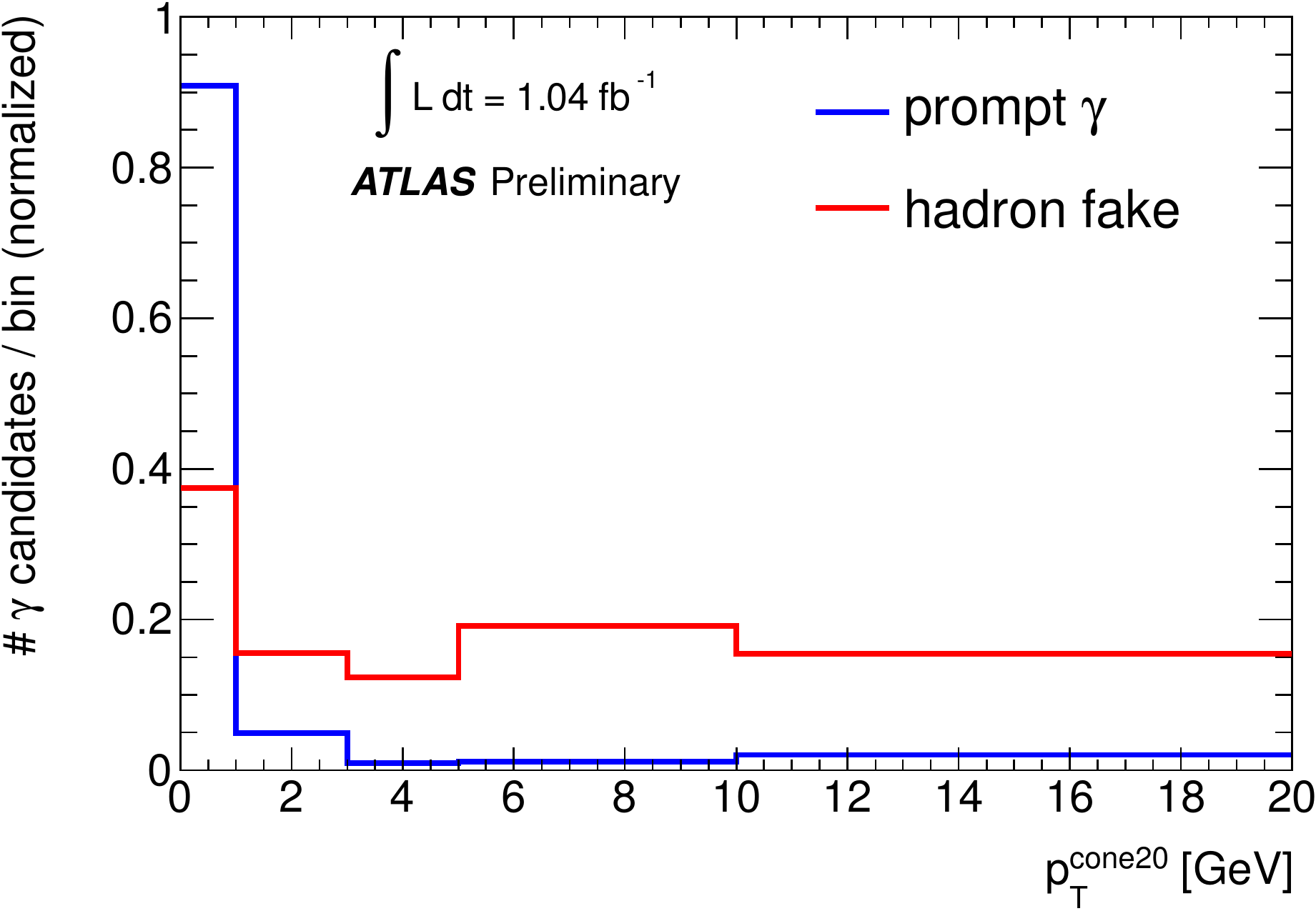}
    \caption[$\ptcone$ templates for prompt and fake photons]{
      $\ptcone$ templates for prompt photons and fake photons from hadrons, normalised to unity.
      The last bin includes the overflow bin.
      Details about the derivation of the templates can be found in Ch.~\ref{sec:photontemplate} and~\ref{sec:faketemplate}.
    }
    \label{fig:templates_overlay}
  \end{center}
\end{figure}

Fig.~\ref{fig:templates_overlay} shows the $\ptcone$ distributions normalised to unity (templates) for prompt photons and for hadrons misidentified as photons.
The derivation of the templates is described in detail in Ch.~\ref{sec:photontemplate} and~\ref{sec:faketemplate}.
The bin sizes were chosen in studies using simulations such that the expected bin contents for the hadron fake distribution were similar.

\subsubsection{Description of the template fit}

The $\ptcone$ distribution of the photons from the selected $\ttg$ candidate events was fitted using template
distributions of real photons and hadron fakes in order to estimate the expected number of signal events ($s$).
The template distributions are those presented in Fig.~\ref{fig:templates_overlay}.
A binned likelihood fit was performed in five bins:
$[0 \GeV, 1 \GeV)$,
$[1 \GeV, 3 \GeV)$,
$[3 \GeV, 5 \GeV)$,
$[5 \GeV, 10 \GeV)$
and~$[10 \GeV, \infty)$.

The number of events in each bin $i$ of the signal template distribution relates to $s$ by
\begin{equation}
  s_i = \varepsilon_i \cdot s \, ,
  \label{eq:efficiency}
\end{equation}
where $\varepsilon_i$ describes the acceptance and selection efficiency for $\ttg$ events, and the probability to end up in bin $i$.
For each background~$j$, templates were built to describe their respective contribution $b_i^j$ in the $i$-th bin of the isolation distribution.
Hence, the sum of all contributions in each bin reads:
\begin{equation}
  \lambda_i = s_i + \sum_{j=1}^{N_{\rm bkg}} b_i^j \, .
  \label{eq:expectation}
\end{equation}

The following likelihood was then maximised in the fit using Markov Chain Monte Carlo implemented in the Bayesian Analysis Toolkit~\cite{bat}
\begin{equation}
  L = \prod_{i=1}^{N_{\rm bins}} P(N_i | \lambda_i) \cdot \prod_{j=1}^{N_{\rm bkg}} P(b^j) \cdot P(s) \, ,
  \label{eq:likelihood}
\end{equation}
where $N_i$ is the number of observed events in bin $i$ of the isolation distribution.
$P(N_i | \lambda_i)$ is the Poissonian probability to observe $N_i$ data events given an expectation of $\lambda_i$.
$P(b^j)$ is the probability for the $j$-th background contribution, and $P(s)$ is the probability for the signal contribution.

The background probabilities were either chosen to be constant in a range $[b^j_{\rm min}, b^j_{\rm max}]$, if the background yield was
treated as a free parameter
\begin{equation*}
  P(b^j) = \left\{ {\begin{matrix} \frac{1}{b^j_{\rm max} - b^j_{\rm min}} \, , & b^j_{\rm min} \leq b^j \leq b^j_{\rm max} \\ 0, & {\rm else} \end{matrix}} \right. \, ,
\end{equation*}
or fixed to a background estimate $\bar{b}^j$:
\begin{equation*}
  P(b^j) = \delta \left( \bar{b}^j - b^j \right) \, ,
\end{equation*}
where $\delta(x)$ is the delta distribution.
The uncertainty on the background estimate $\bar{b}^j$ was then treated as a source of systematic uncertainty (Ch.~\ref{sec:syst_backgroundmodelling}).

Tab.~\ref{tab:templateparameters} gives an overview of the different parameters of the template fit and their respective probabilities:
as already mentioned, the hadron fake contribution was treated as a free parameter, and
a constant background probability was assigned to it covering the whole range of hadron fake contributions between 0\% and 100\%.

The $\ttg$ signal contribution was also treated as a free parameter:
\begin{equation*}
  P(s) = \left\{ {\begin{matrix} \frac{1}{s_{\rm max} - s_{\rm min}} \, , & s_{\rm min} \leq s \leq s_{\rm max} \\ 0, & {\rm else} \end{matrix}} \right. \, ,
\end{equation*}
with $s$ covering a range of signal fractions between 0\% and 100\%.
It is worth stressing that the parameter describing the number of signal events accounts for the acceptance and selection efficiency and therefore
represents the total number of $\ttg$ events extrapolated to the whole signal phase space.
Acceptance and efficiency were estimated in MC simulations, and systematic uncertainties were evaluated accordingly
(Sec.~\ref{sec:syst_signalmodelling} and~\ref{sec:syst_detectormodelling}).

\begin{table}[b]
\centering
\begin{tabular}[h] {|p{0.18\textwidth}|p{0.175\textwidth}|p{0.125\textwidth}|p{0.35\textwidth}|}
  \hline
  Process & Parameter & Parameter probability & Note \\
  \hline
  $\ttg$ signal & free & constant & accounts for acceptance and selection efficiency \\
  True photon bkg. & fixed to estimate & delta \mbox{function} & systematic uncertainties by up- and down-variation \\
  Electron fakes & fixed to estimate & delta \mbox{function} & systematic uncertainties by up- and down-variation \\
  Hadron fakes & free & constant & - \\
  \hline
\end{tabular}\\
\caption[Signal and background parameters in the template fit]{
  Parameters and parameter probabilities for the signal and background contributions used in the template fit.
}
\label{tab:templateparameters}
\end{table}

The contributions from background processes with electrons misidentified as photons and from processes with true photons were fixed to the
estimates derived in Sec.~\ref{sec:egammaapplication} and Ch.~\ref{sec:backgroundphotons}, respectively.
Systematic uncertainties were evaluated by up- and down-variations of the different contributing processes (Sec.~\ref{sec:syst_backgroundmodelling}).

The template fit was performed in both lepton channels simultaneously.
One combined likelihood was constructed in order to estimate the expected number of signal events $s$, which,
combining Eq.~(\ref{eq:efficiency}),~(\ref{eq:expectation}) and~(\ref{eq:likelihood}), explicitly reads:
\begin{eqnarray*}
  L & = &
  \prod_{i=1}^{N_{\rm bins}} P\left(N_{i, \rm{e+jets}} \left| \lambda_{i, \rm{e+jets}} = \varepsilon_{i, \rm{e+jets}} \cdot s + \sum_{j=1}^{N_{\rm bkg}} b_{i, \rm{e+jets}}^j \right. \right) \cdot
  \prod_{j=1}^{N_{\rm bkg}} P\left(b^j_{\rm{e+jets}}\right) \cdot \\
  & & \prod_{i=1}^{N_{\rm bins}} P\left(N_{i, \mu\rm{+jets}} \left| \lambda_{i, \mu\rm{+jets}} = \varepsilon_{i, \mu\rm{+jets}} \cdot s + \sum_{j=1}^{N_{\rm bkg}} b_{i, \mu\rm{+jets}}^j \right. \right) \cdot
  \prod_{j=1}^{N_{\rm bkg}} P\left(b^j_{\mu\rm{+jets}}\right) \cdot
  P(s) \, .
\end{eqnarray*}
As a cross-check, also separate fits in the electron and muon channels were performed.

\chapter{Derivation of the prompt photon template}
\label{sec:photontemplate}

In order to reduce the dependence on MC simulations, the template distributions for prompt photons were derived using a data-driven approach.
Since the definition of a data sample which contains mainly prompt photons is not straightforward, the prompt photon template was derived from
electrons from \mbox{$Z \to e^+e^-$} data,
because the isolation properties of electrons and photons are very similar given the similar detector signature.
Small differences in the $\ptcone$ distribution between photons and electrons were corrected for using simulated \mbox{$Z \to e^+e^-$}
and $\ttg$ events.

A data sample dominated by \mbox{$Z \to e^+e^-$} events was selected by the following criteria:
two oppositely charged electrons had to be present in the event, and their invariant mass had to be within a \mbox{$50 \GeV$} window around the mass of
the $Z$ boson.
Since the background from multijet events is small close to the $Z$ mass peak, the number of selected events was increased by loosening the requirements
on the electron objects with respect to the definitions presented in Sec.~\ref{sec:electron}:
only \texttt{medium} shower shape criteria were applied, which still lead to a high signal-over-background ratio.
In order to avoid a bias on the $\ptcone$ distribution from the calorimeter isolation criterion, the latter was disregarded.

Events were triggered by the \texttt{EF\_e20\_medium} trigger, which is the same trigger as used for the selection of the $\ttg$ event
candidates.
Consequently, the first electron was required to have an $\et$ of at least \mbox{$25 \GeV$}.
The $\et$ threshold for the second electron, however, was lowered to \mbox{$15 \GeV$}, which corresponds to the lower threshold for the photon $\et$
in this analysis.
%\footnote{Since track information is not used for the estimate of the photon four-vector,
%$\et$ and $\et$ are interchangeably for photons.}.

Selected events were required to have a good primary vertex with at least five tracks associated to it.
Events with large noise in the LAr calorimeter were rejected.

\begin{figure}[h]
  \begin{center}
    \includegraphics[width=0.49\textwidth]{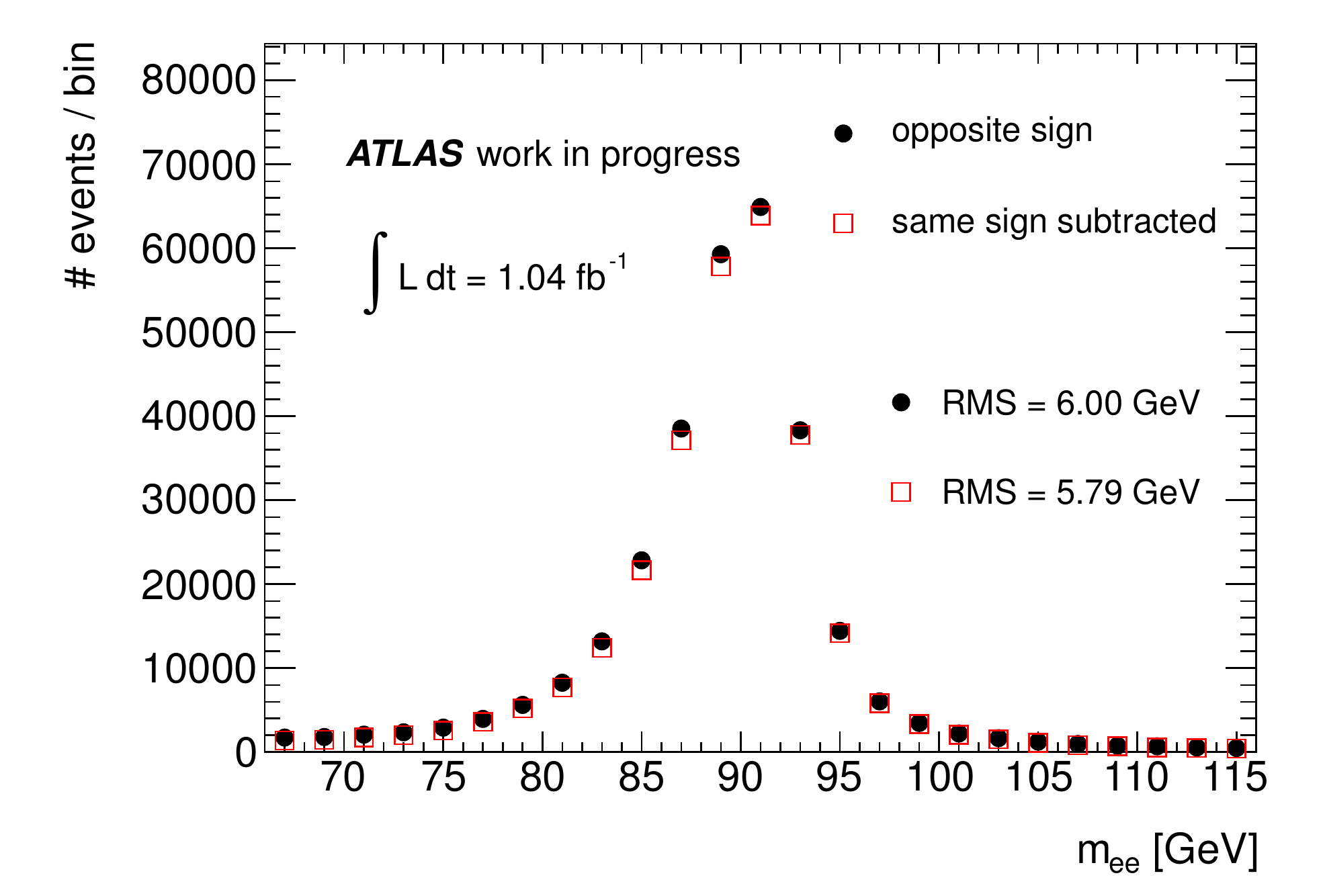}
    \caption[Distribution of the di-electron invariant mass in data]{
      Distribution of the di-electron invariant mass in data.
      Electrons fulfil the \texttt{medium} shower shape criteria.
      The distribution for opposite-sign pairs is shown with the solid circles.
      When the contribution from same-sign electron pairs was subtracted, the distribution with the open squares was obtained.
      The first bin includes the underflow bin, and the last bin includes the overflow bin.
    }
    \label{fig:zmass}
  \end{center}
\end{figure}

Fig.~\ref{fig:zmass} shows the distribution of the di-electron invariant mass in the whole data set of \mbox{$1.04 \ifb$}.
The solid circles show the distribution for electron pairs with opposite charge.
A small contribution from multijet events with two jets misidentified as electrons was accounted for
by subtracting the contribution from electron pairs with the same electric charge, as shown by the distribution with the open squares.
Since no real electron is present in multijet events, the measurement of the electric charge is arbitrary and pairs with same and opposite charge occur
equally often.
Charge misidentification in \mbox{$Z \to e^+e^-$} events also contributes to the opposite-sign distribution, but does not introduce a bias to the
$\ptcone$ distribution of the electrons.
Hence, subtracting the opposite-sign contribution removes the background from multijet events from the same-sign $\ptcone$ distribution.

When the contribution from same-sign electron pairs is subtracted, the width of the invariant mass distribution decreases (Fig.~\ref{fig:zmass}):
the root mean square is reduced from \mbox{$6.00 \GeV$} to \mbox{$5.79 \GeV$}.
The fact that the resolution of the $Z$ boson mass peak was found to be improved after the subtraction of the same-sign contribution indicates that
the background from multijet events was indeed suppressed.

\begin{figure}[h]
  \begin{center}
    \includegraphics[width=0.49\textwidth]{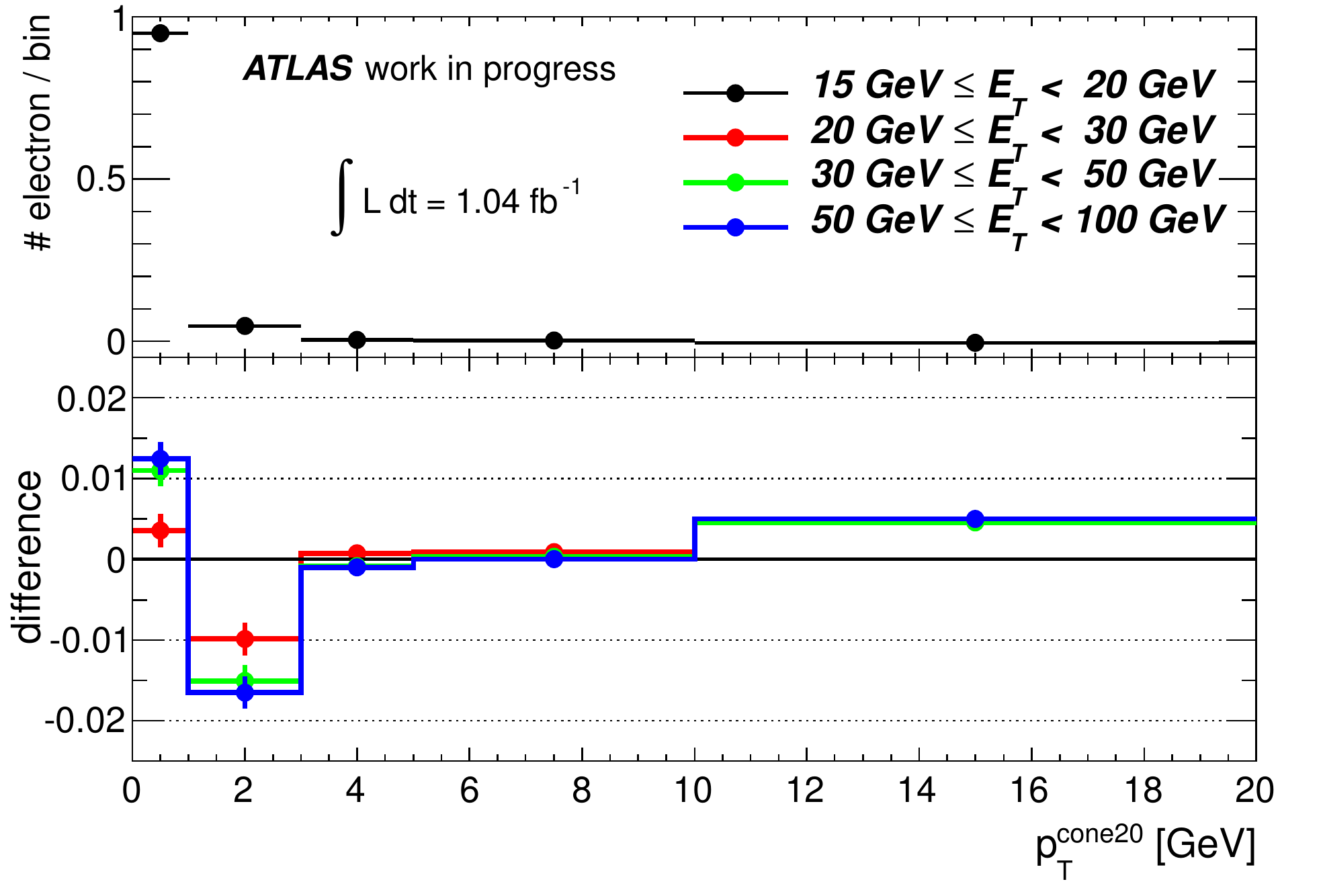}
    \includegraphics[width=0.49\textwidth]{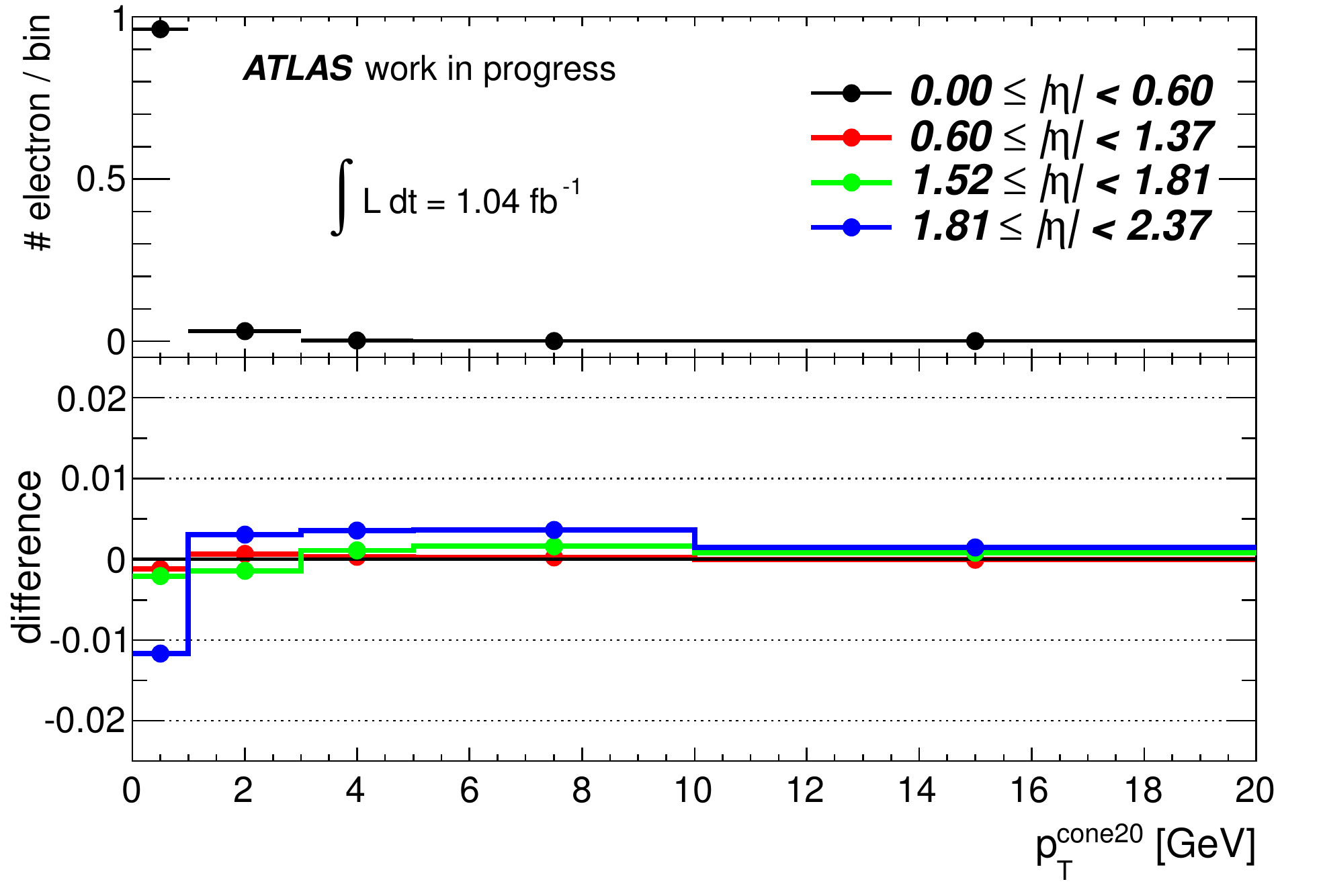}
    \caption[$\ptcone$ distributions for electrons]{
      $\ptcone$ distributions for electrons from \mbox{$Z \to e^+e^-$} decays for different regions of $\et$ (left) and $|\eta|$ (right).
      The upper part of the plots show the regions \mbox{$15 \GeV \leq \et < 20 \GeV$} and \mbox{$0 \leq \eta < 0.6$}, respectively.
      The lower part of the plots show the difference with respect to this distribution for the other regions in $\et$ and $\eta$.
      In both plots, the last bin includes the overflow bin.
    }
    \label{fig:electrontemplate_pt_eta}
  \end{center}
\end{figure}

The $\ptcone$ distributions for the electrons from this data sample are shown in Fig.~\ref{fig:electrontemplate_pt_eta} for different values of $\et$
(left plot) and different regions in $|\eta|$ (right plot):
$[15 \GeV, 20 \GeV)$,
$[20 \GeV, 30 \GeV)$,
$[30 \GeV, 50 \GeV)$,
$[50 \GeV, 100 \GeV)$, and
$[0, 0.60)$,
$[0.60, 1.37)$,
$[1.52, 1.81)$,
$[1.81, 2.37)$, respectively.
The upper part of the plots show the regions \mbox{$15 \GeV \leq \et < 20 \GeV$} and \mbox{$0 \leq \eta < 0.6$}, respectively.
The lower part of the plots show the difference with respect to this distribution for the other regions in $\et$ and $\eta$.
The $\ptcone$ observable is stable with respect to the electron $\et$ and $\eta$.
Discrepancies between different $\et$ and $|\eta|$ bins are smaller than 2\% and were ignored given the size of the systematic uncertainties
discussed in Ch.~\ref{sec:systematics}.

Differences between the $\ptcone$ distributions of electrons and photons were studied in simulated events.
The same event selection as for the selection of the \mbox{$Z \to e^+e^-$} data events was applied to \mbox{$Z \to e^+e^-$} simulations in order
to obtain electron distributions.
These distributions were then compared to those from real photons from simulated $\ttg$ events.

\begin{figure}[p]
  \begin{center}
    \includegraphics[width=0.435\textwidth]{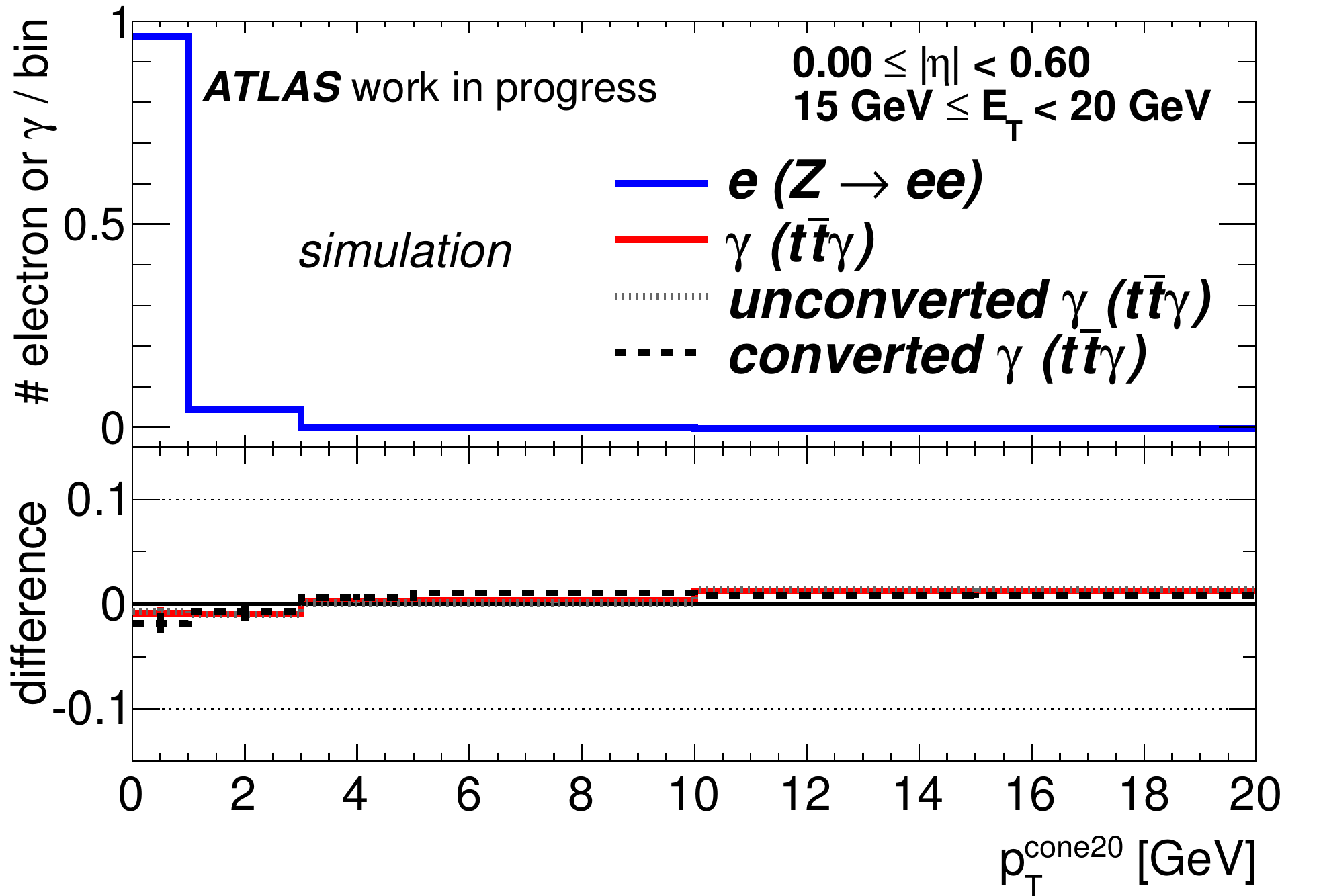}
    \includegraphics[width=0.435\textwidth]{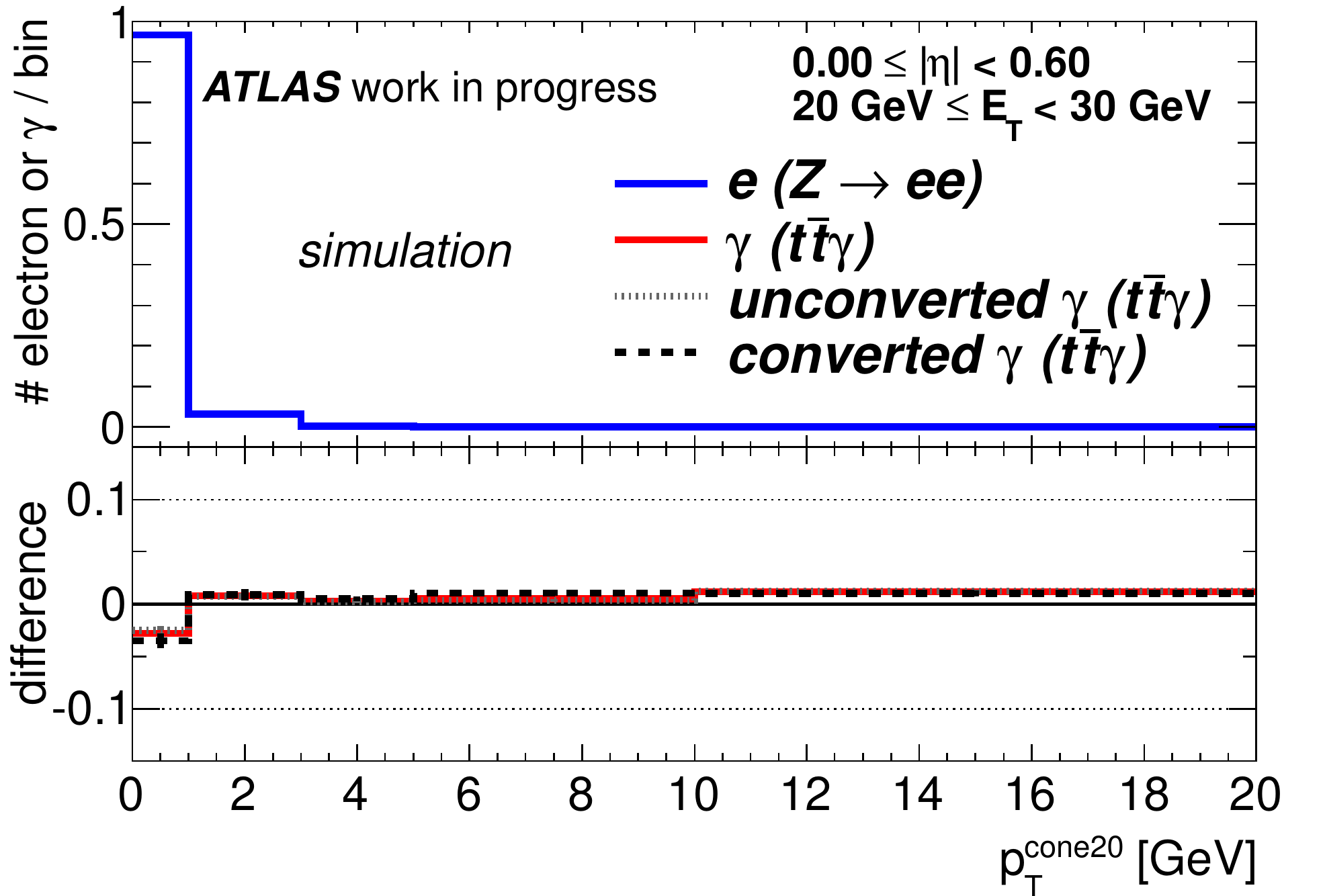}
    \includegraphics[width=0.435\textwidth]{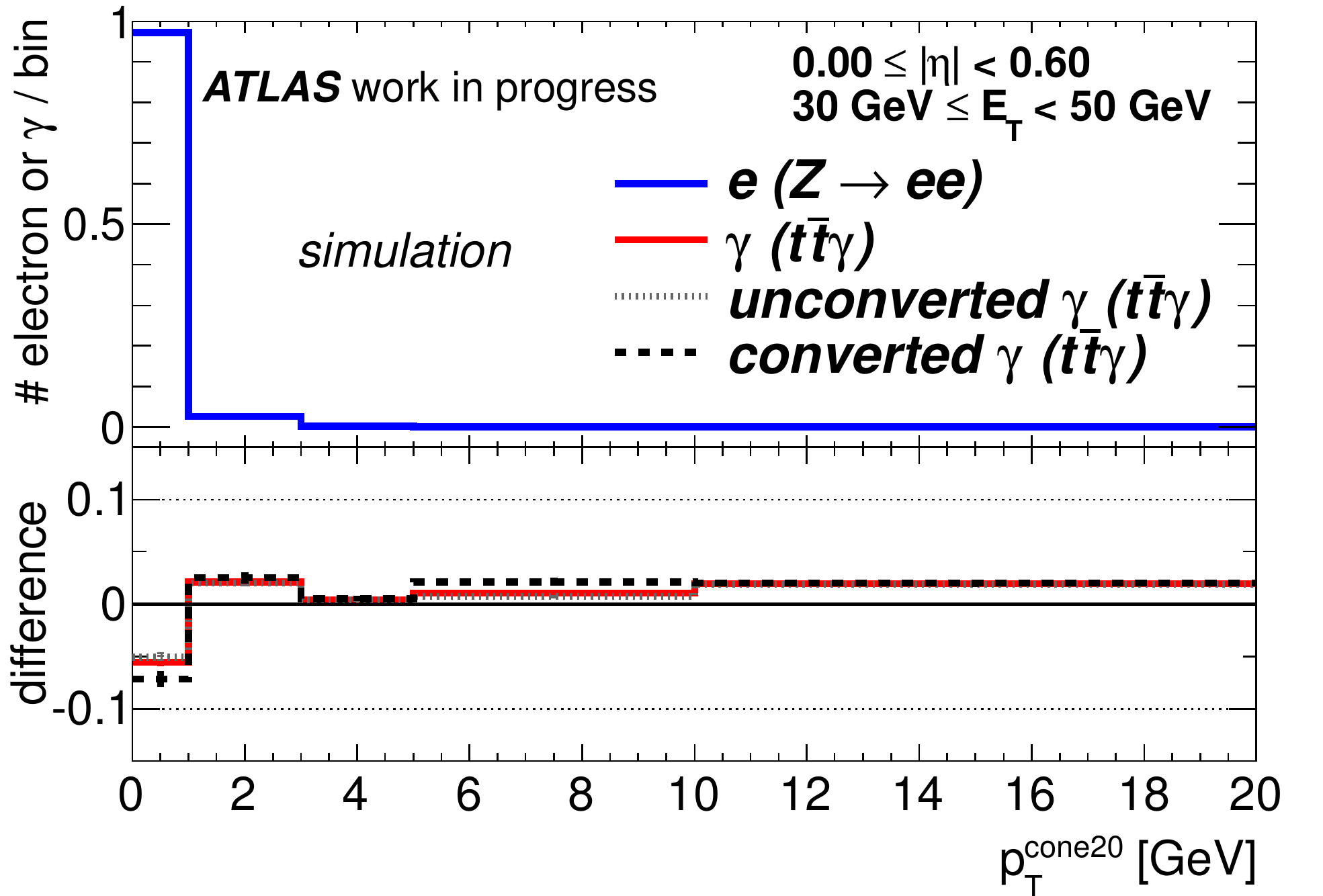}
    \includegraphics[width=0.435\textwidth]{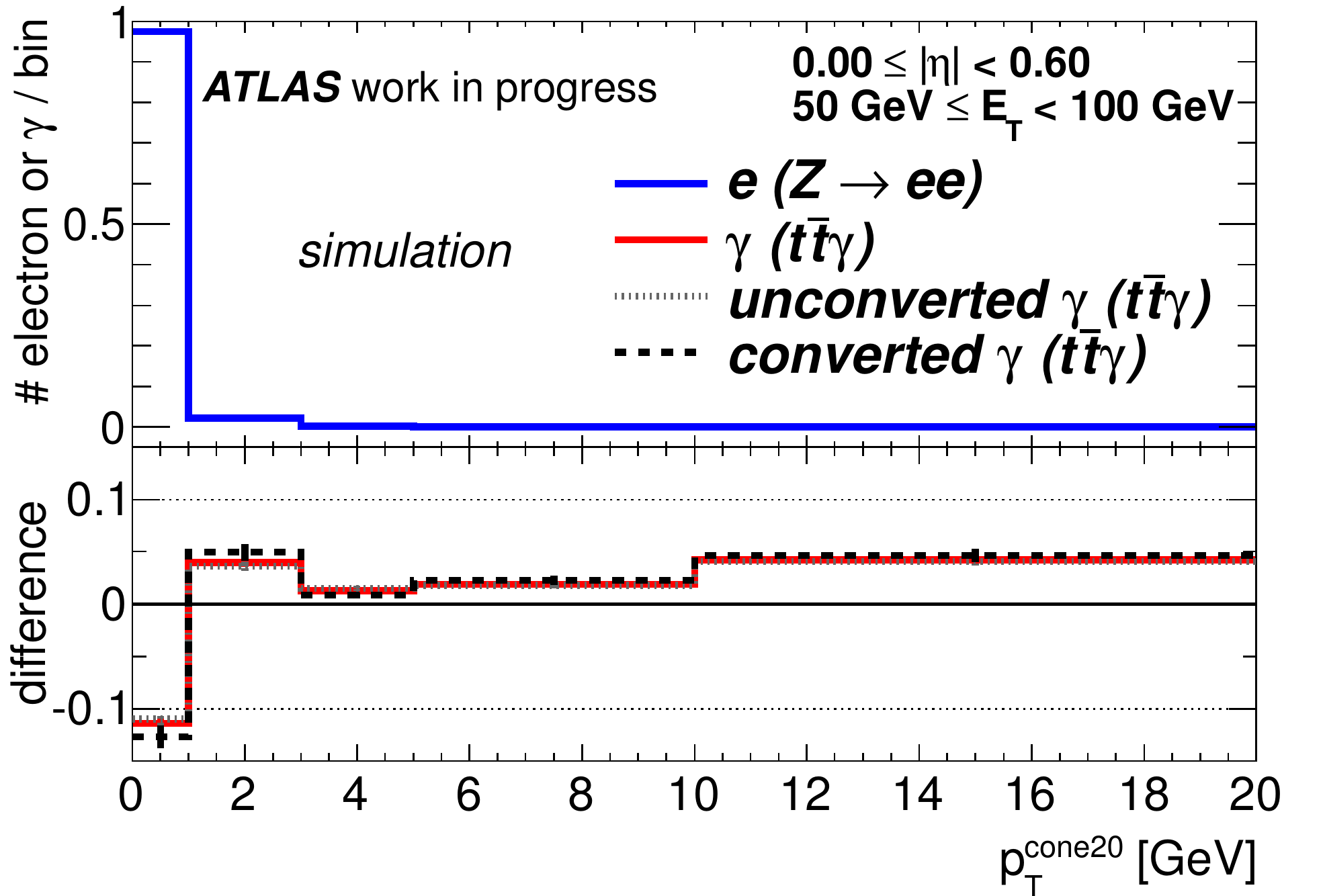}
    \includegraphics[width=0.435\textwidth]{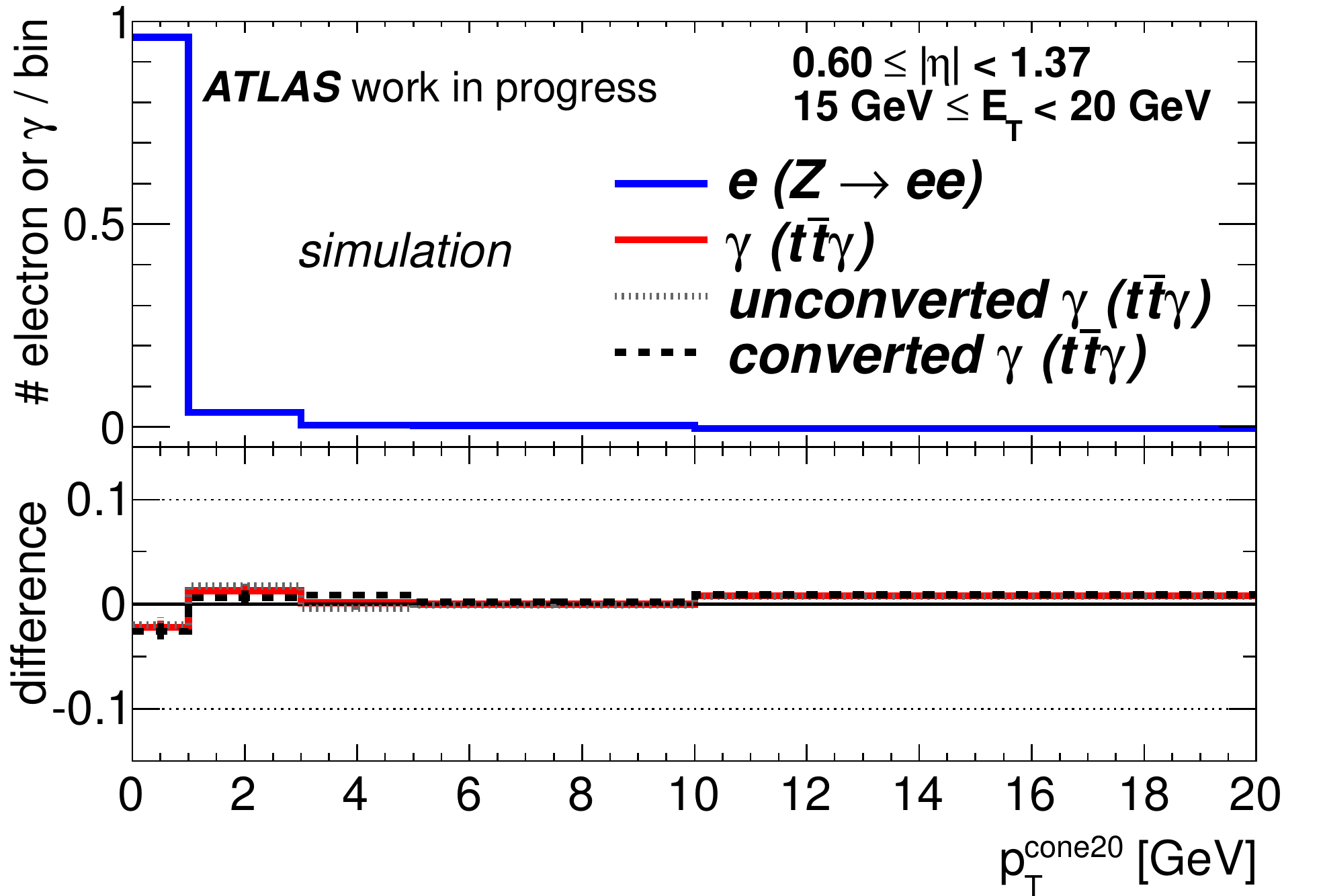}
    \includegraphics[width=0.435\textwidth]{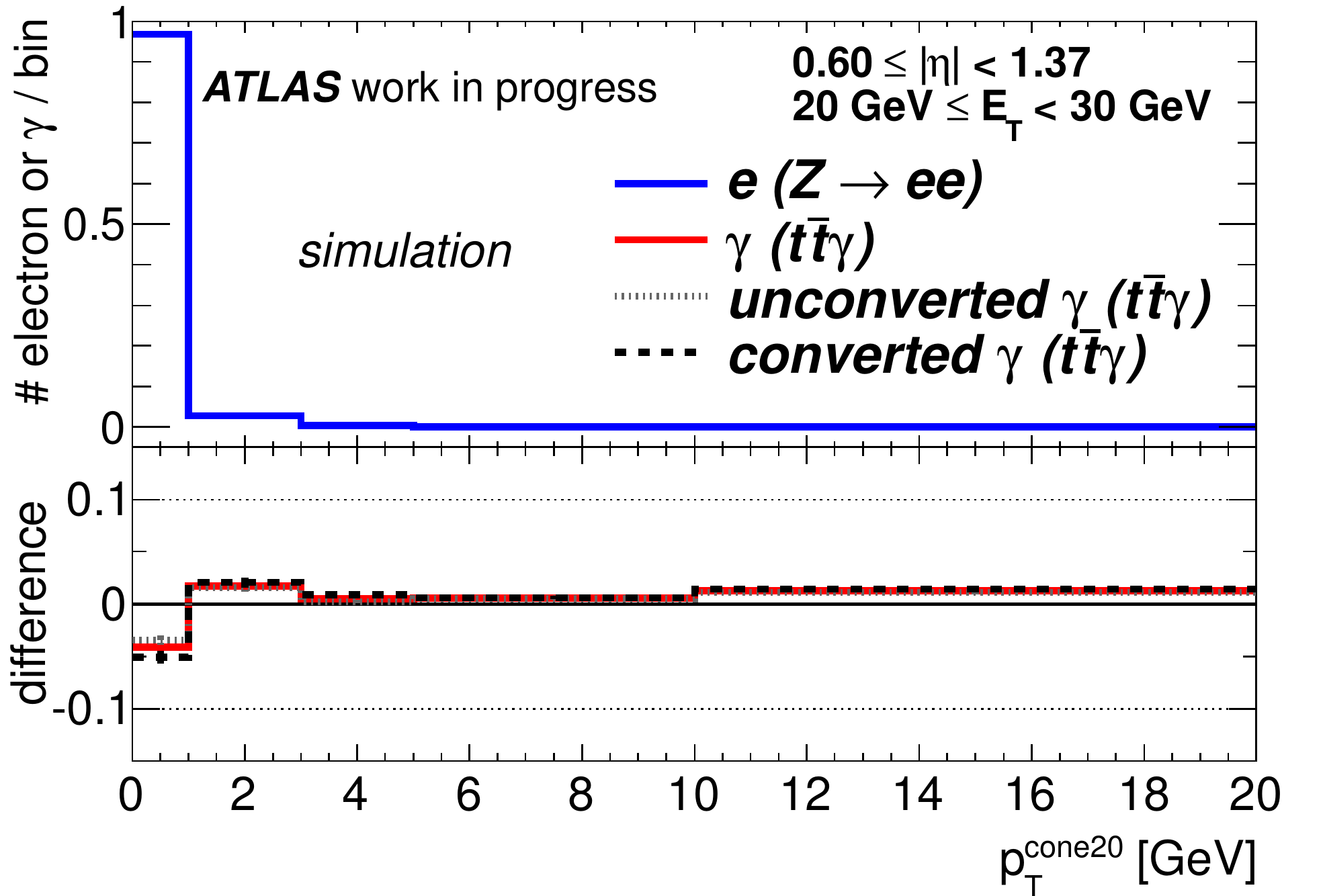}
    \includegraphics[width=0.435\textwidth]{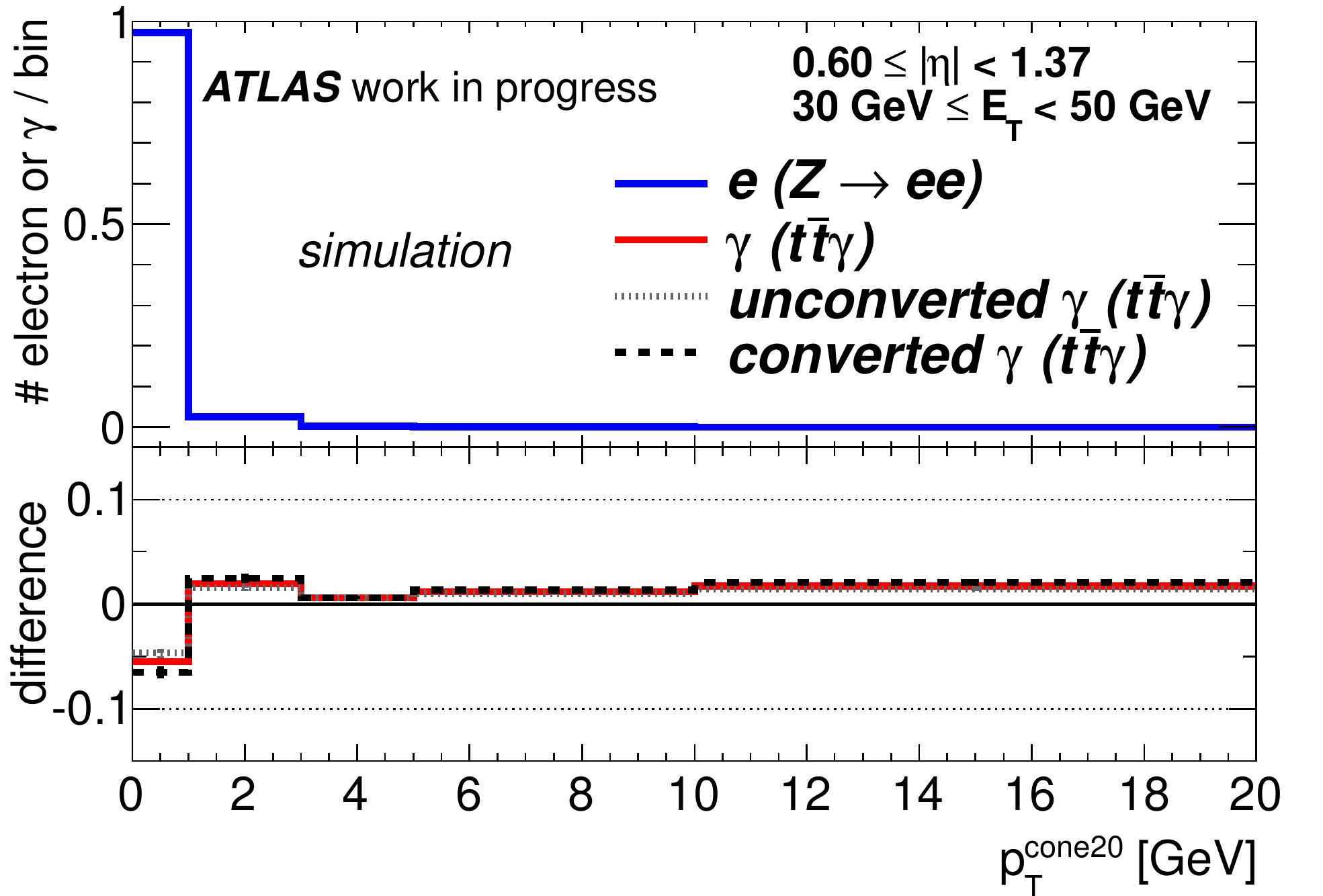}
    \includegraphics[width=0.435\textwidth]{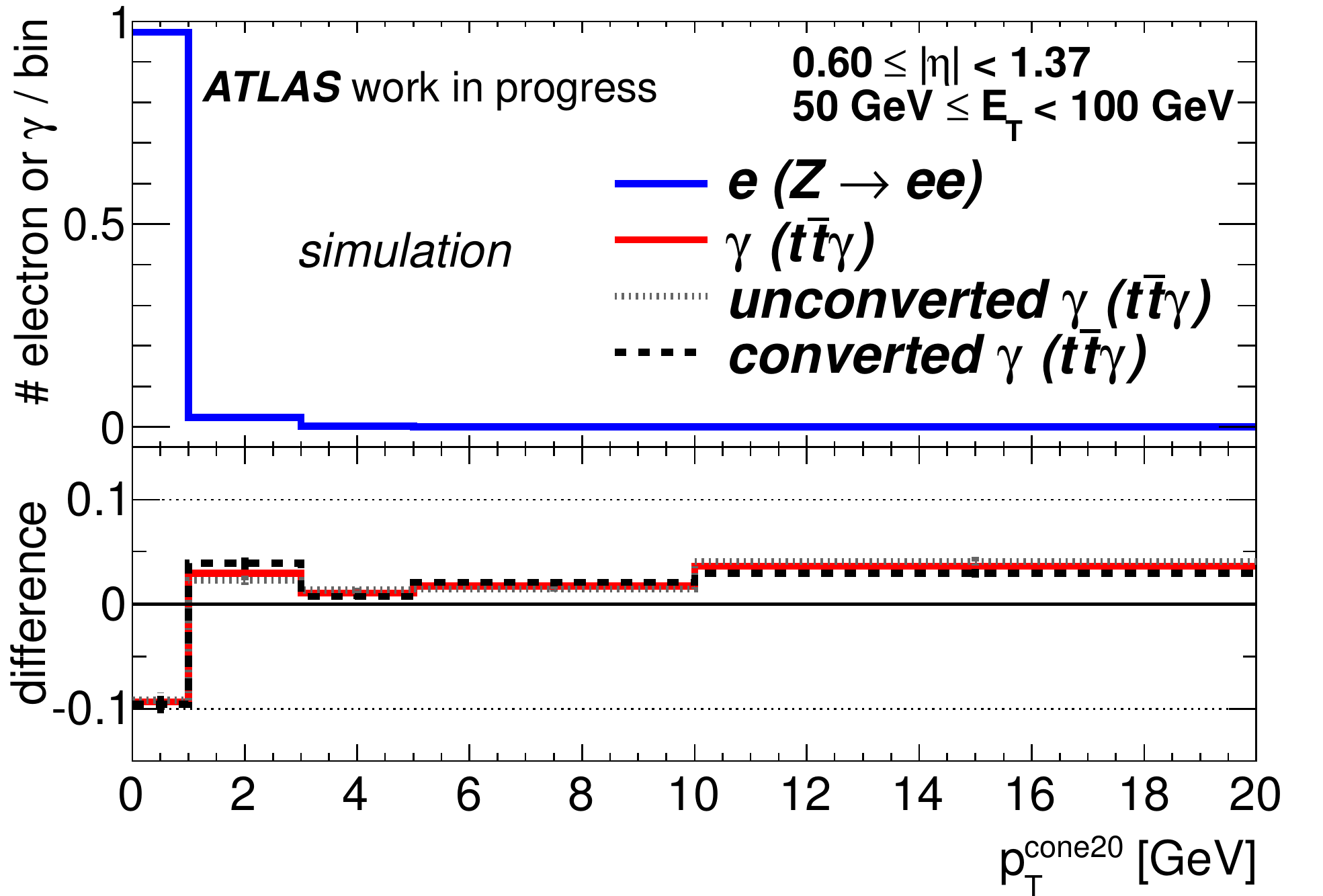}
    \caption[$\ptcone$ distributions for electrons and photons from simulation (1)]{
      $\ptcone$ distributions for electrons from simulated \mbox{$Z \to e^+e^-$} decays (upper part of each plot)
      in different bins of $\et$ for \mbox{$0 \leq |\eta| < 0.60$} (four upper plots) and \mbox{$0.60 \leq |\eta| < 1.37$} (four lower plots)
      normalised to unity.
      The lower part of each plot shows the difference of the distribution of photons from simulated $\ttg$ events (solid line) with respect
      to the electron distribution.
      Additionally, the distributions for unconverted (dotted grey line) and converted photons (dashed black line) from $\ttg$ simulations are depicted.
      In all plots, the last bin includes the overflow bin.
    }
    \label{fig:extrapolation_1}
  \end{center}
\end{figure}

\begin{figure}[p]
  \begin{center}
    \includegraphics[width=0.435\textwidth]{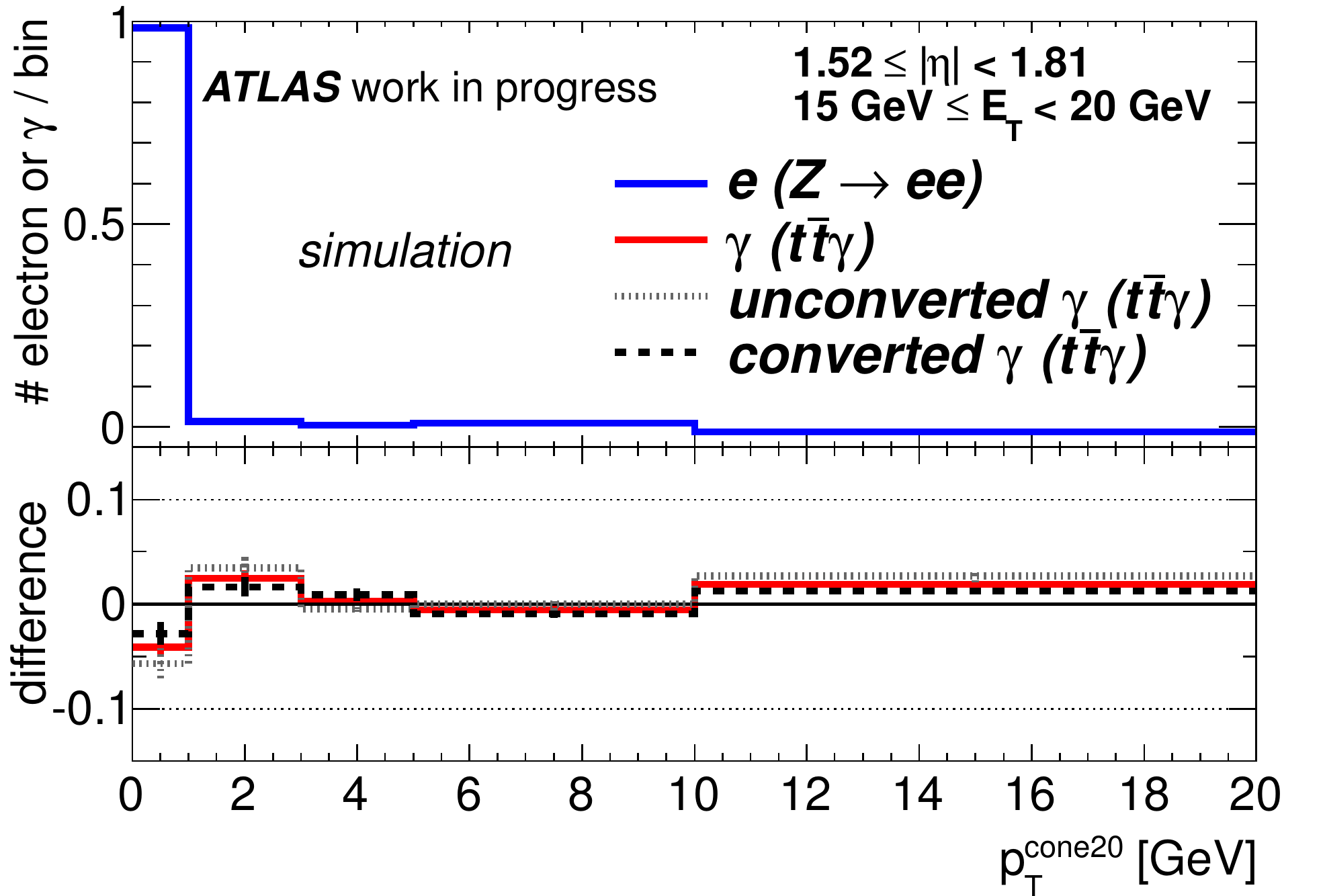}
    \includegraphics[width=0.435\textwidth]{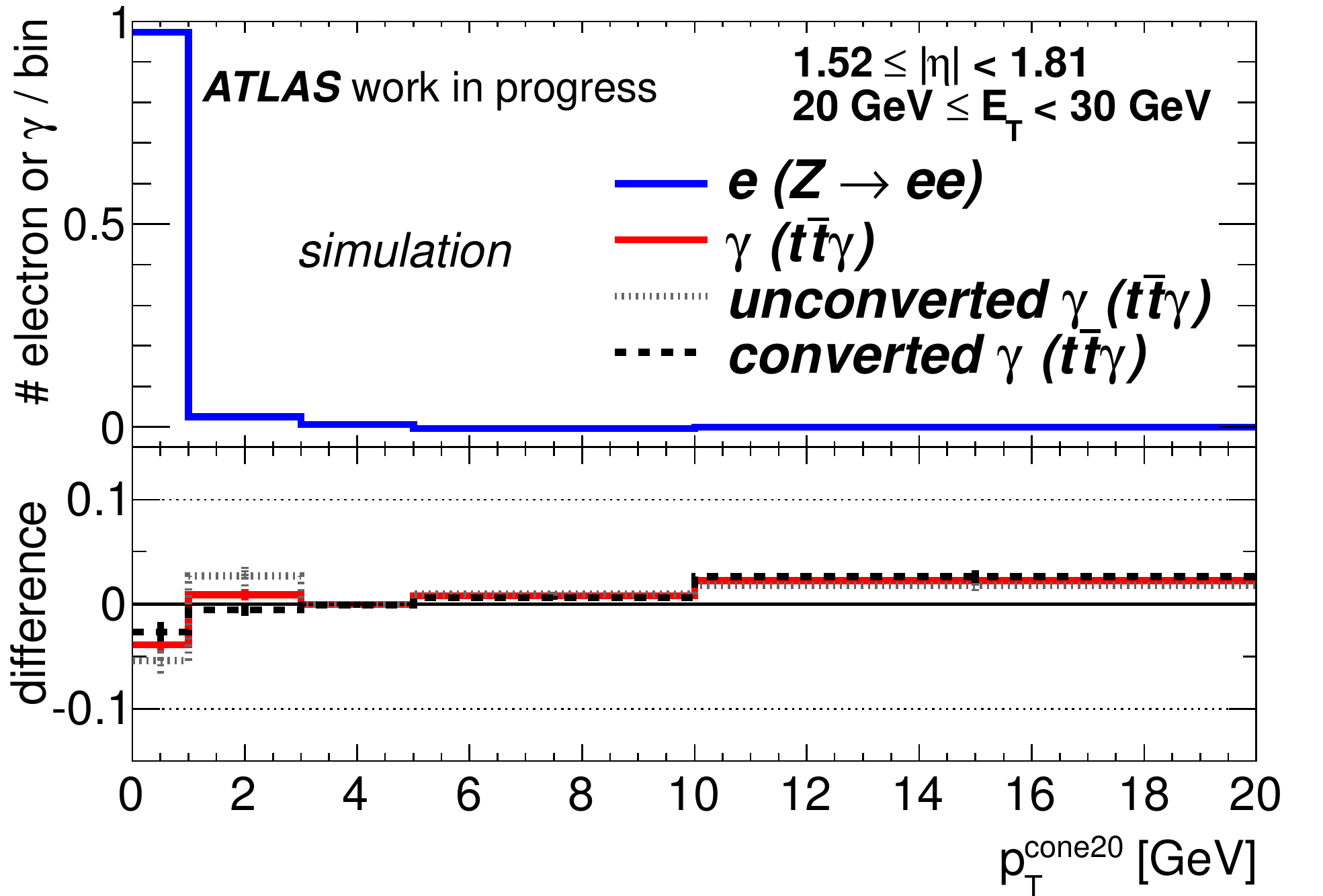}
    \includegraphics[width=0.435\textwidth]{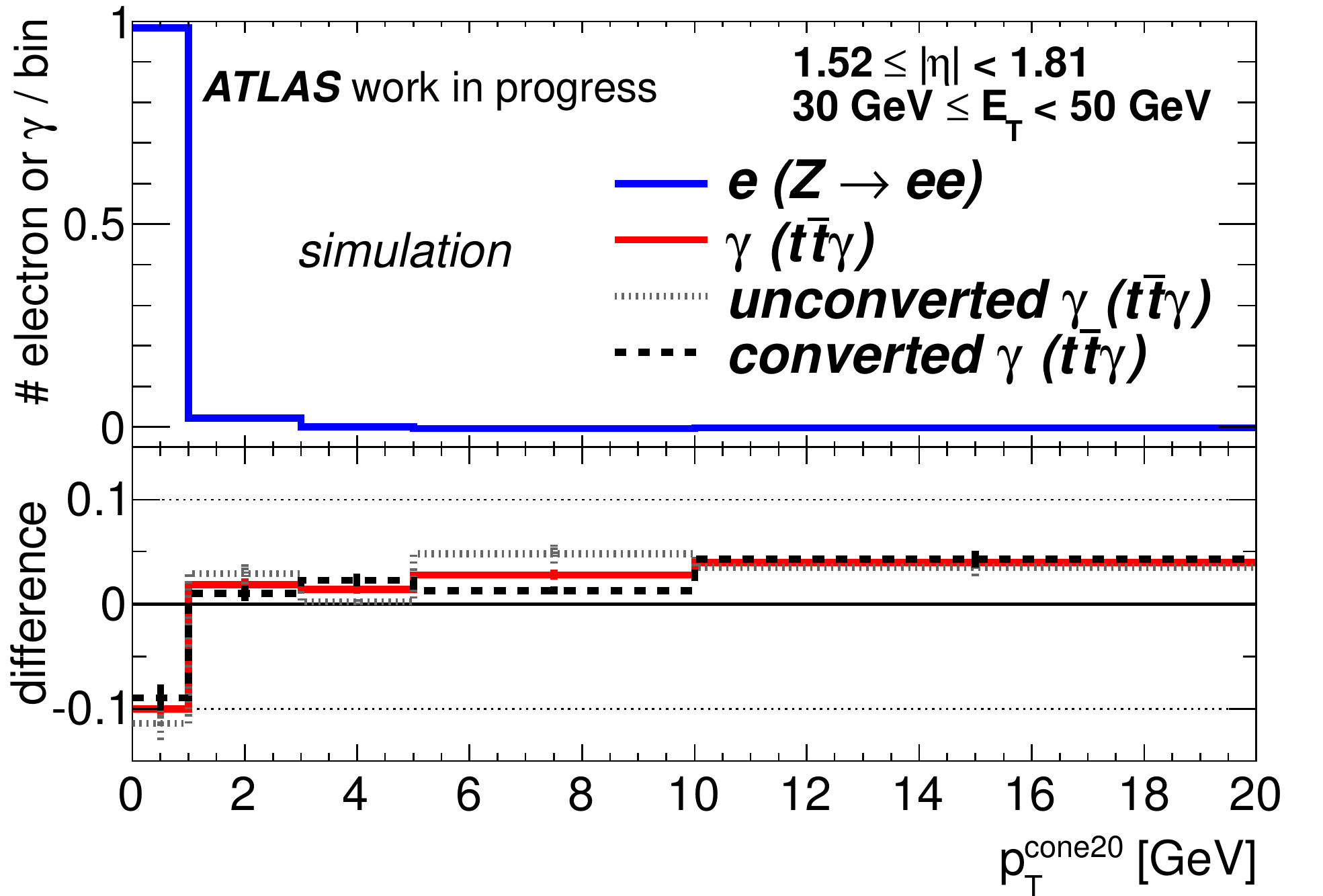}
    \includegraphics[width=0.435\textwidth]{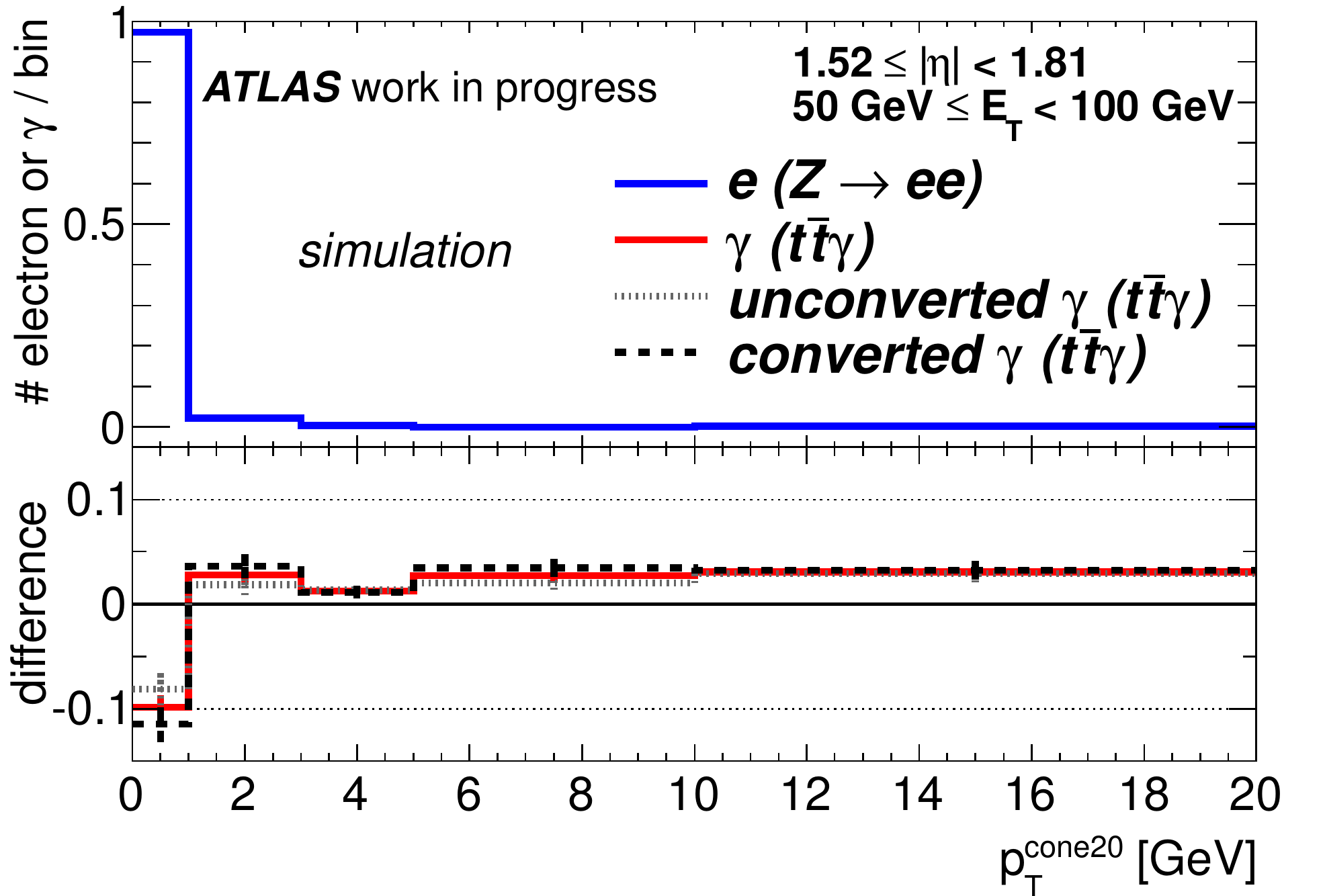}
    \includegraphics[width=0.435\textwidth]{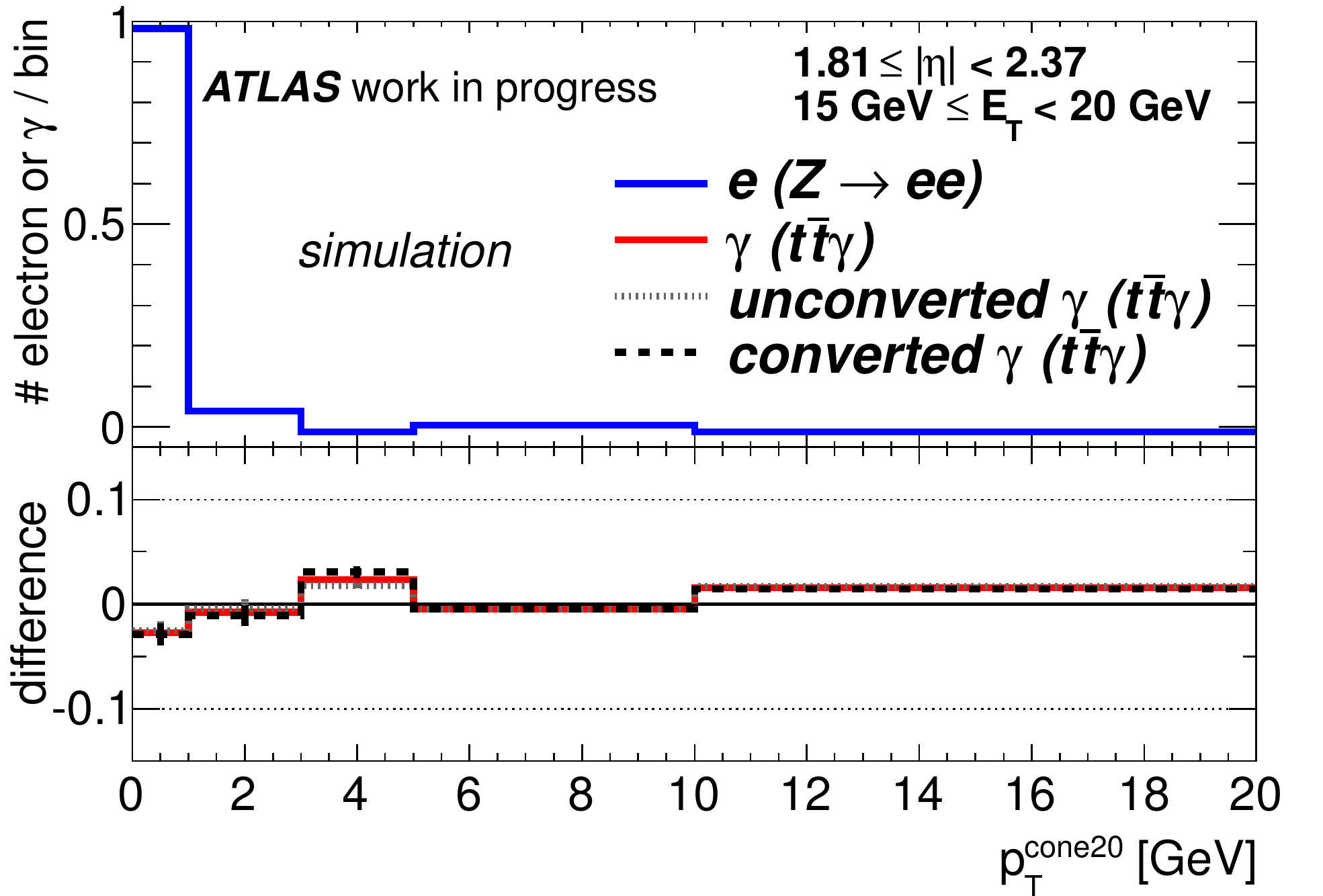}
    \includegraphics[width=0.435\textwidth]{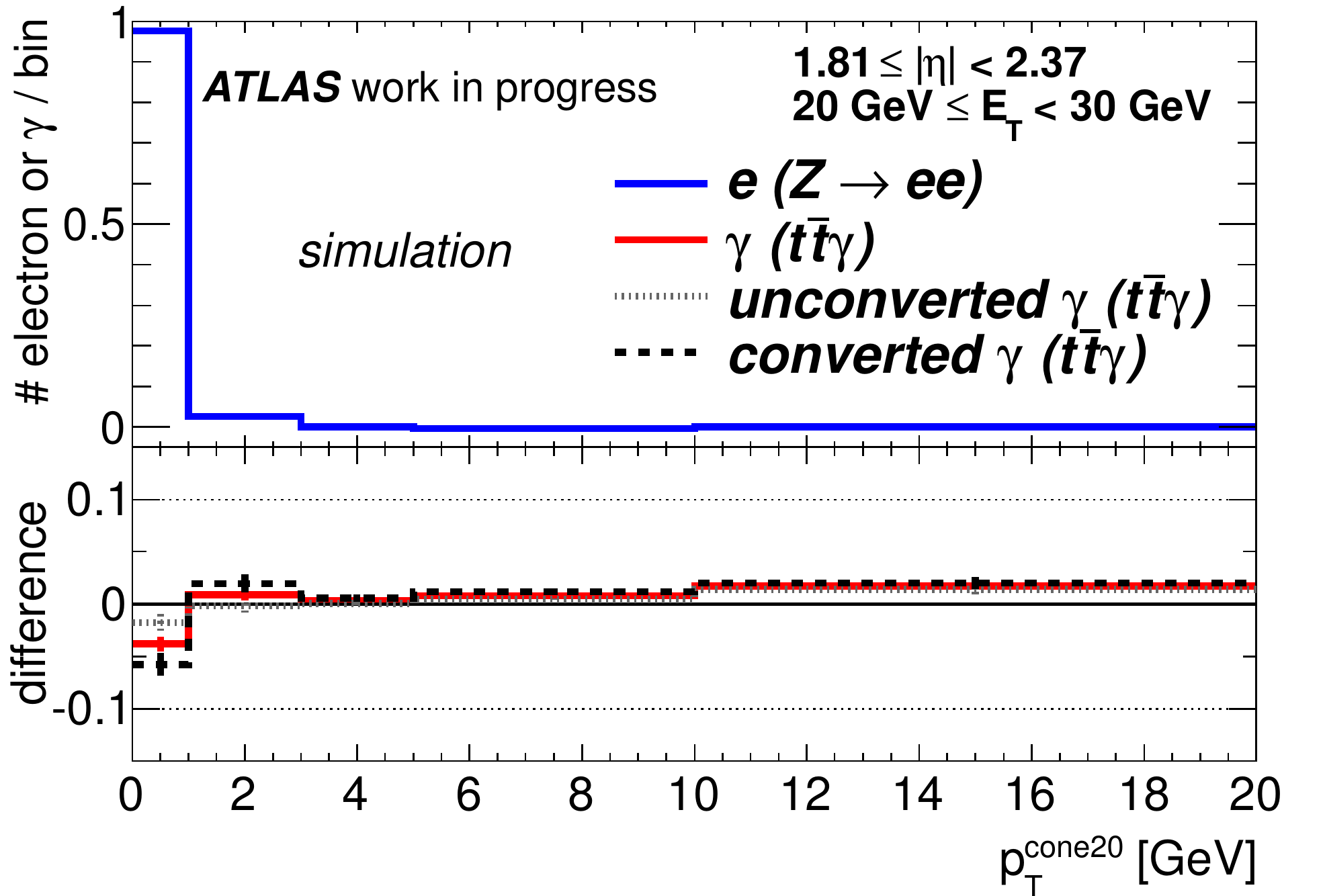}
    \includegraphics[width=0.435\textwidth]{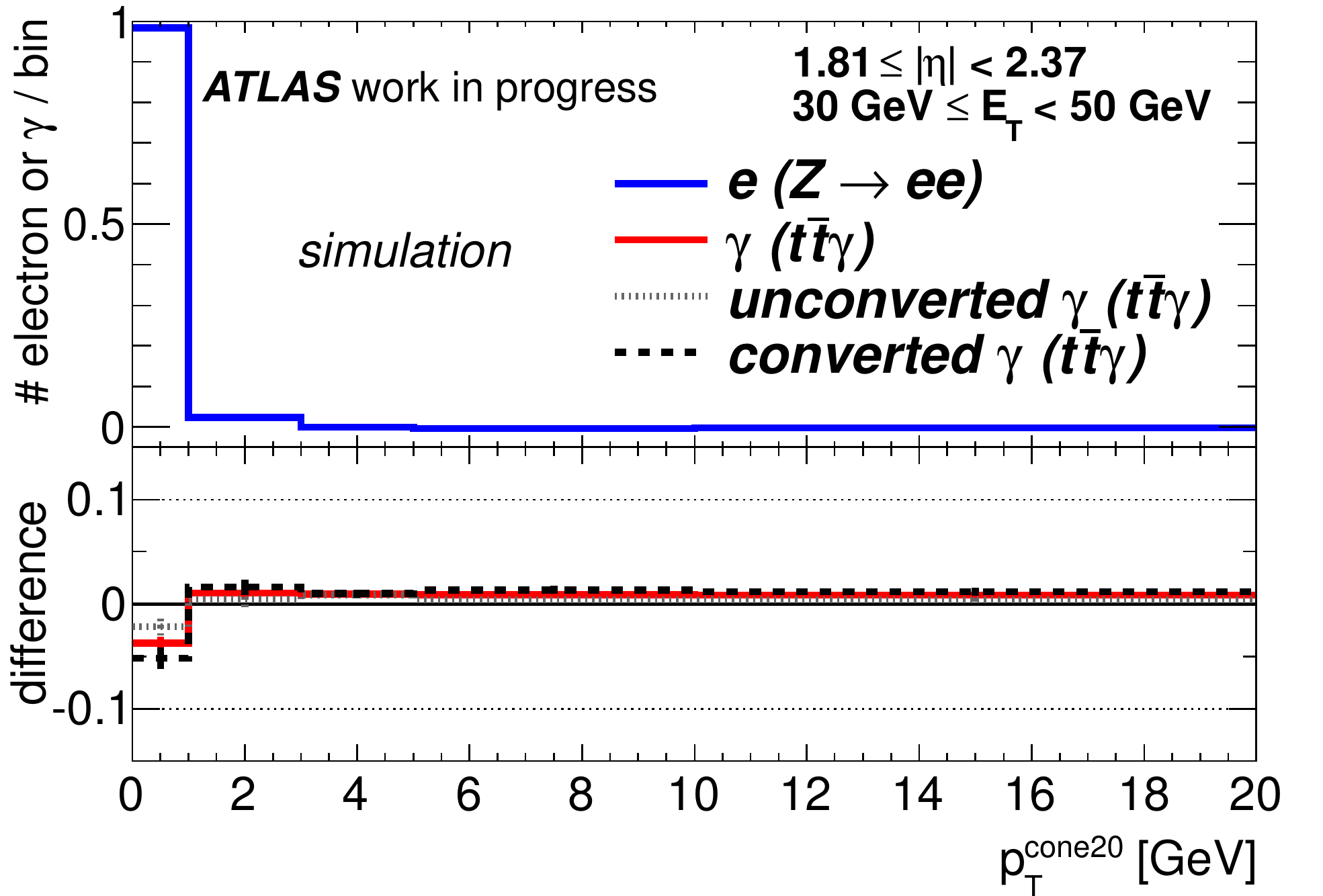}
    \includegraphics[width=0.435\textwidth]{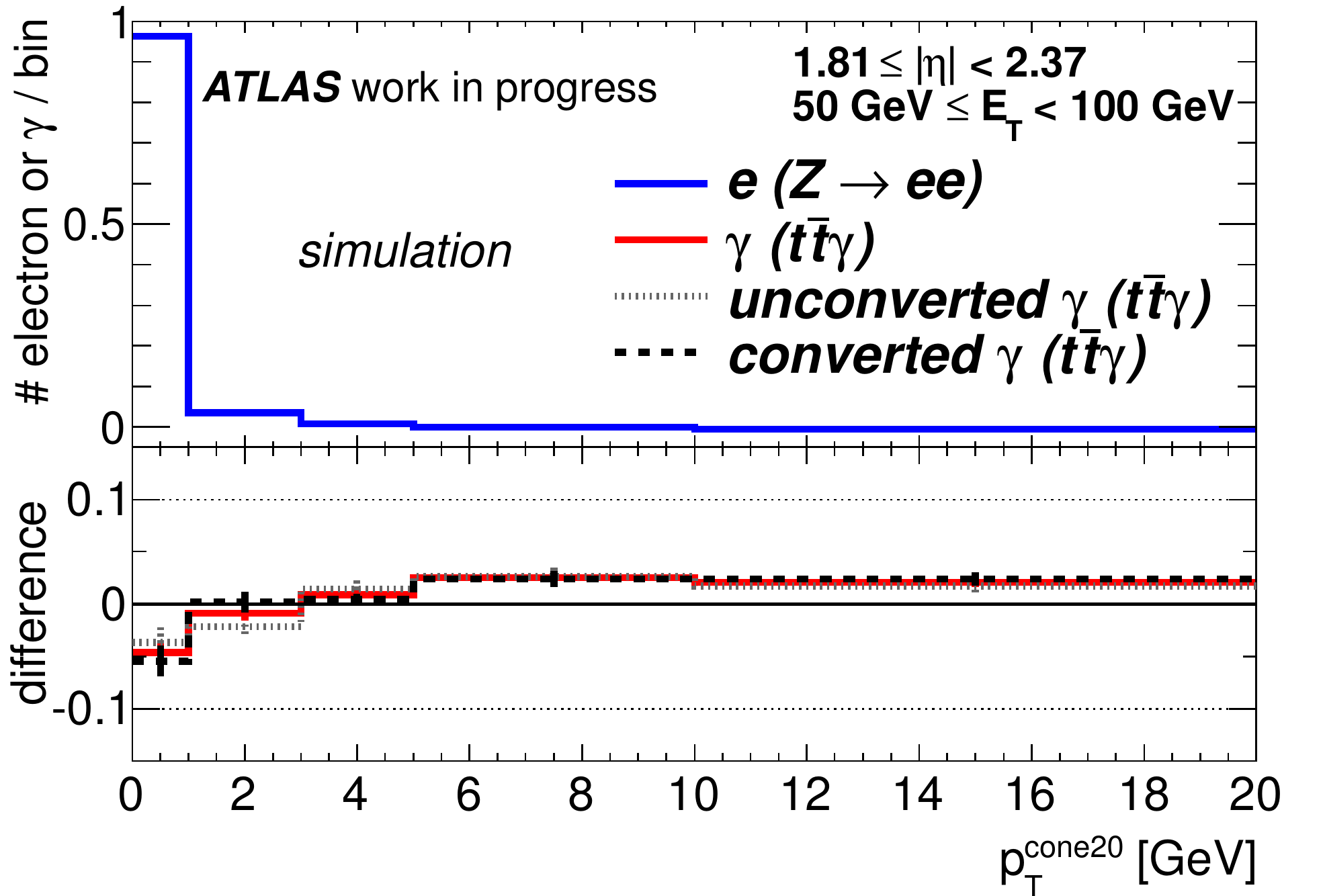}
    \caption[$\ptcone$ distributions for electrons and photons from simulation (2)]{
      $\ptcone$ distributions for electrons from simulated \mbox{$Z \to e^+e^-$} decays (upper part of each plot)
      in different bins of $\et$ for \mbox{$1.52 \leq |\eta| < 1.81$} (four upper plots) and \mbox{$1.81 \leq |\eta| < 2.37$} (four lower plots)
      normalised to unity.
      The lower part of each plot shows the difference of the distribution of photons from simulated $\ttg$ events (solid line) with respect
      to the electron distribution.
      Additionally, the distributions for unconverted (dotted grey line) and converted photons (dashed black line) from $\ttg$ simulations are depicted.
      In all plots, the last bin includes the overflow bin.
    }
    \label{fig:extrapolation_2}
  \end{center}
\end{figure}

Fig.~\ref{fig:extrapolation_1} and~\ref{fig:extrapolation_2} show the electron distribution in the upper part of each plot.
The lower part of each plot shows the difference of the distribution of photons from simulated $\ttg$ events (solid line) with respect
to the electron distribution.
Additionally, distributions for unconverted (dotted grey line) and converted photons (dashed black line) are shown
in four different regions of $\et$ and $|\eta|$.

The differences between the shapes of the electron and photon $\ptcone$ distributions are small for low $\et$ and increase 
up to 0.1 in the first two bins with increasing electron and photon $\et$ in the different $|\eta|$-regions.
The $|\eta|$-region \mbox{$[1.81, 2.37)$} is special, because the acceptance of the TRT ends at \mbox{$|\eta| < 2.0$}, and
the differences between electron and photon $\ptcone$ are smaller than in the other $|\eta|$-regions.
No significant differences between unconverted and converted photons were observed, so that both photon types were treated together.

A priori no differences between electrons and photons are expected for the $\ptcone$ distributions.
If there were inherent differences between electrons and photons, they would unlikely contribute to the differences observed in the comparison
of \mbox{$Z \to e^+e^-$} and $\ttg$ simulations:
one-track conversion photons are by construction very similar to electrons, and the differences between electrons and photons are significantly
larger than those between the $\ptcone$ distributions of unconverted and converted photons.
Hence, the difference in the topologies of \mbox{$Z \to e^+e^-$} and $\ttg$ events is expected to cause the observed electron-photon differences.

\begin{figure}[h!]
  \begin{center}
    \includegraphics[width=0.49\textwidth]{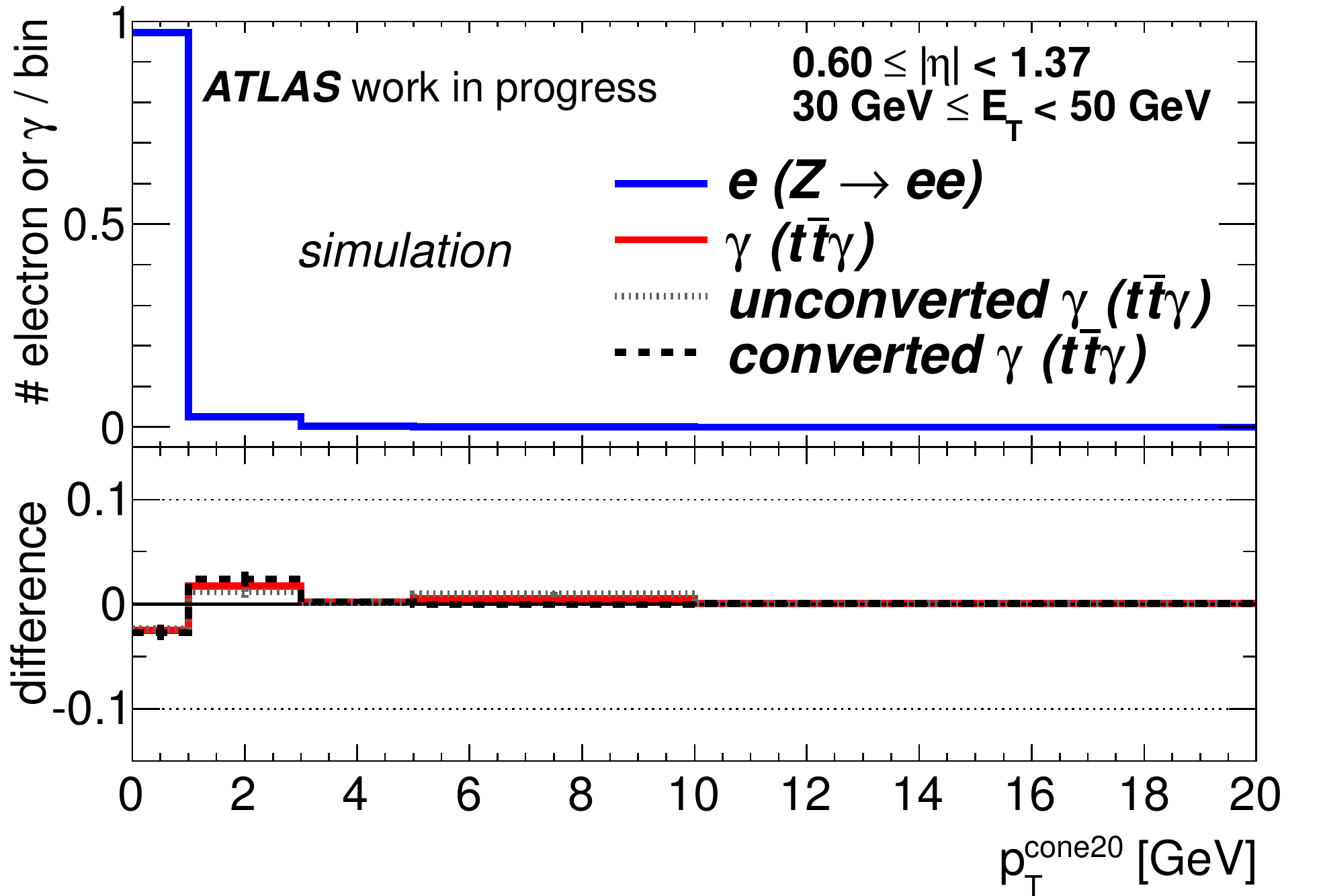}
    \includegraphics[width=0.49\textwidth]{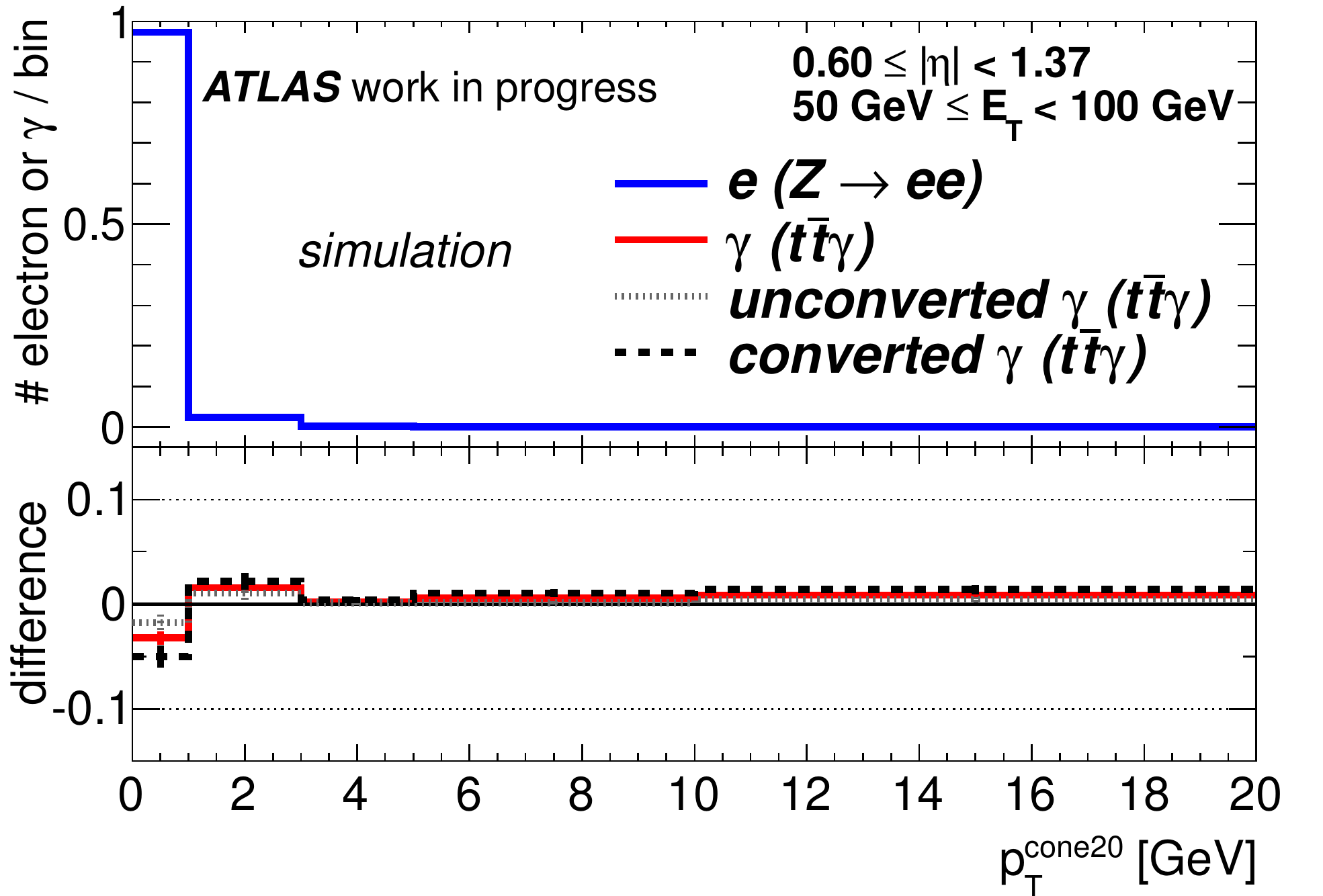}
    \caption[$\ptcone$ distributions for electrons and photons from simulation (3)]{
      $\ptcone$ distributions for electrons from simulated \mbox{$Z \to e^+e^-$} decays (upper part of the plots)
      normalised to unity for \mbox{$0.60 \leq |\eta| < 1.37$} and \mbox{$30 \GeV \leq \et < 50 \GeV$}
      (left plot) as well as for \mbox{$50 \GeV \leq \et < 100 \GeV$} (right plots).
      The lower part of each plot shows the difference of the distribution of photons from simulated $\ttg$ events (solid line) with respect
      to the electron distribution.
      Additionally, the distributions for unconverted (dotted grey line) and converted photons (dashed black line) from $\ttg$ simulations are depicted.
      In both plots, the last bin includes the overflow bin.\\
      Photons closer than 0.2 in $\eta$-$\phi$-space to a true lepton were not considered and the agreement with the electron distribution
      is improved with respect to the plots in Fig.~\ref{fig:extrapolation_1} and~\ref{fig:extrapolation_2}.
    }
    \label{fig:extrapolation_3}
  \end{center}
\end{figure}

Fig.~\ref{fig:extrapolation_3} shows the $\ptcone$ distributions for simulated electrons and photons,
as already shown in Fig.~\ref{fig:extrapolation_1} and~\ref{fig:extrapolation_2}.
However, only photons that had a minimum distance of 0.2 in $\eta$-$\phi$-space from a true electron or muon were considered.
Such leptons only needed to be present among the generated particles before detector simulation, but did not need to fulfil all quality criteria of
the lepton definitions (Sec.~\ref{sec:electron} and~\ref{sec:muon}) and did not even need to be reconstructed at all.
The track of such leptons, however, is likely to have a large contribution to the $\ptcone$ value of the close-by photon.
This effect is particularly relevant for dileptonic $\ttbar$ decays.

Only two example plots for the region \mbox{$0.60 \leq |\eta| < 1.37$} for transverse momenta between $30$ and \mbox{$50 \GeV$} (left plot) and $50$
and \mbox{$100 \GeV$}
(right plot) are shown.
For reference, the remaining plots are presented in App.~\ref{sec:app_extrapolation}.
Compared to Fig.~\ref{fig:extrapolation_1}, the agreement between the electron and photon distributions is significantly improved.
In particular, the increased tail in the photon distributions, as seen in the higher $\et$ bins in Fig.~\ref{fig:extrapolation_1}
and Fig.~\ref{fig:extrapolation_2} is suppressed.
The differences between the electron and photon distributions were hence concluded to be mainly due to the different topologies of \mbox{$Z \to e^+e^-$}
and $\ttg$ events.

\begin{figure}[h!]
  \begin{center}
    \includegraphics[width=0.49\textwidth]{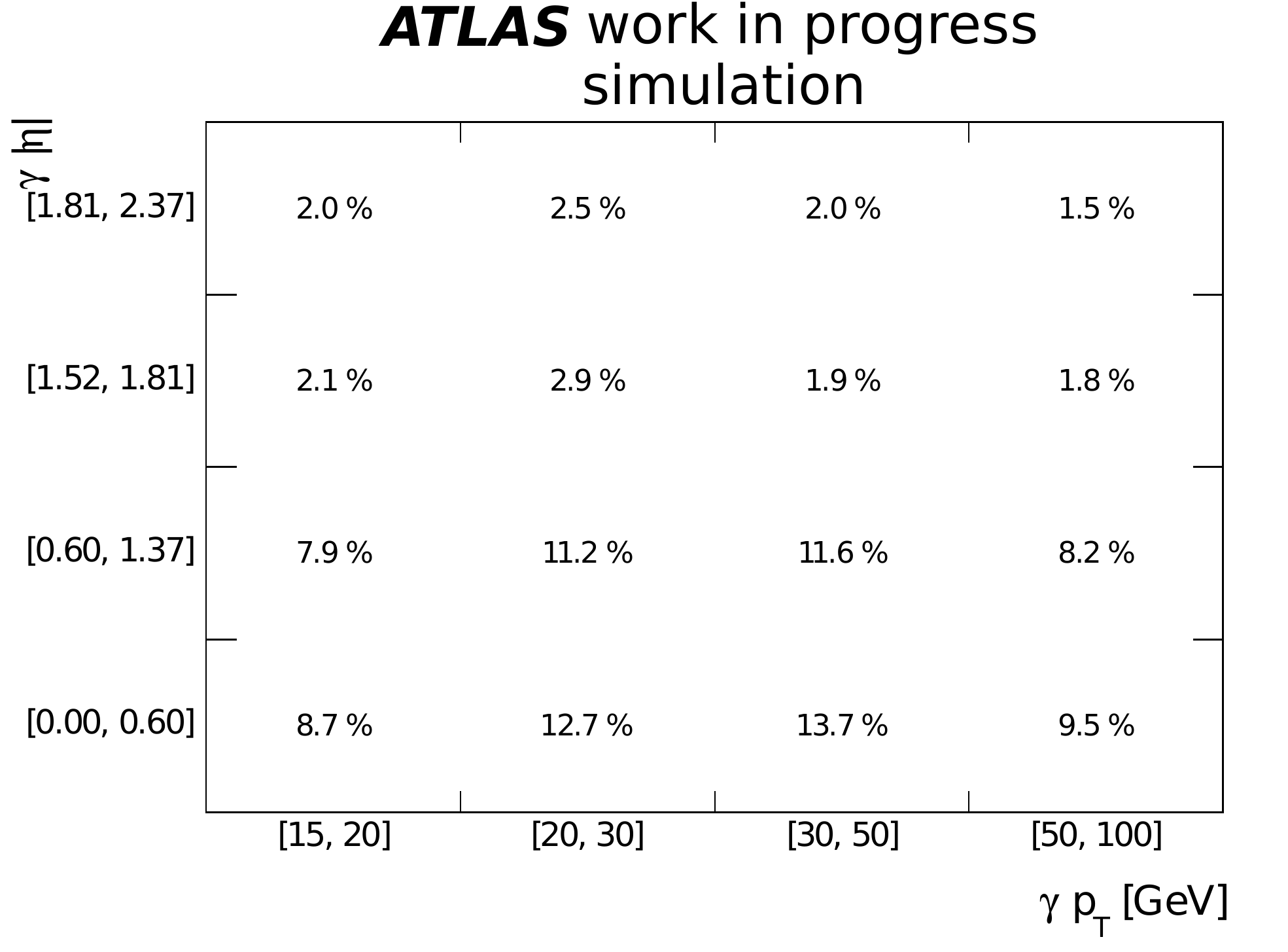}
    \caption[Distribution of $\et$ and $|\eta|$ for real photons in $\ttg$ simulations]{
      Distribution of the $\et$ and $|\eta|$ of real photons in $\ttg$ MC simulation in 16 bins,
      as used for the derivation of the photon $\ptcone$ templates from the electron templates extracted from \mbox{$Z \to e^+e^-$} data.
    }
    \label{fig:extrapolation_etapt}
  \end{center}
\end{figure}

The differences between the electron and photon distributions in the MC simulations
were used to derive a prompt photon template $s_\gamma$ from the electron template $s_e$ measured in data:
\begin{eqnarray*}
  s_\gamma = s_e + \Delta s^{\rm{MC}}(\et, |\eta|) \quad {\rm with} \quad \Delta s^{\rm MC} = s_\gamma^{\rm{MC}} - s_e^{\rm{MC}} \, .
\end{eqnarray*}
The derivation was done in 16 bins, four in $\et$ and four in $|\eta|$, as detailed above.
The relative weights of the different extrapolation bins were taken from the $\et$ and $|\eta|$ distributions of real photons in $\ttg$ MC simulation,
presented in Fig.~\ref{fig:extrapolation_etapt}.
The resulting prompt photon template was presented in Fig.~\ref{fig:templates_overlay}.
The systematic uncertainty assigned to this electron-to-photon extrapolation is discussed in Sec.~\ref{sec:syst_backgroundmodelling}.

\chapter{Background events with hadrons misidentified as photons}
\label{sec:faketemplate}

As discussed in Sec.~\ref{sec:backgroundmodelling} and Ch.~\ref{sec:strategy}, simulations cannot be trusted for the modelling
of the properties of hadrons misidentified as photons, and the contribution from hadron fakes was hence estimated using a template
fit to the $\ptcone$ distribution of the photon candidates in data.
The $\ptcone$ template distribution for the hadron fakes was derived from a control region (CR) in data largely enriched in hadron fakes,
as described in this chapter.

The CR was defined by requiring that at least one out of the following four shower shape variables from the LAr strips
failed the cut from the \texttt{tight} photon menu: $\deltae$, $\fside$, $\wsthree$ and $\eratio$ (cf. Tab.~\ref{tab:showershapes}).
All other criteria of the good photon definition detailed in Sec.~\ref{sec:photon} needed to be satisfied.
Such photon objects are called \textit{hadron fake candidates} in the following.

Variables constructed from the LAr strips were chosen, because they are sensitive to the core of the photon cluster and are hence expected not to be
correlated to photon isolation variables, such as $\ptcone$.
This procedure was previously used in the prompt photon~\cite{promptphoton} and diphoton cross section measurements~\cite{diphoton}, where
the CR was shown to model the properties of hadron fakes well, although these analyses used calorimeter isolation instead of $\ptcone$.

\begin{figure}[h]
  \begin{center}
    \includegraphics[width=0.49\textwidth]{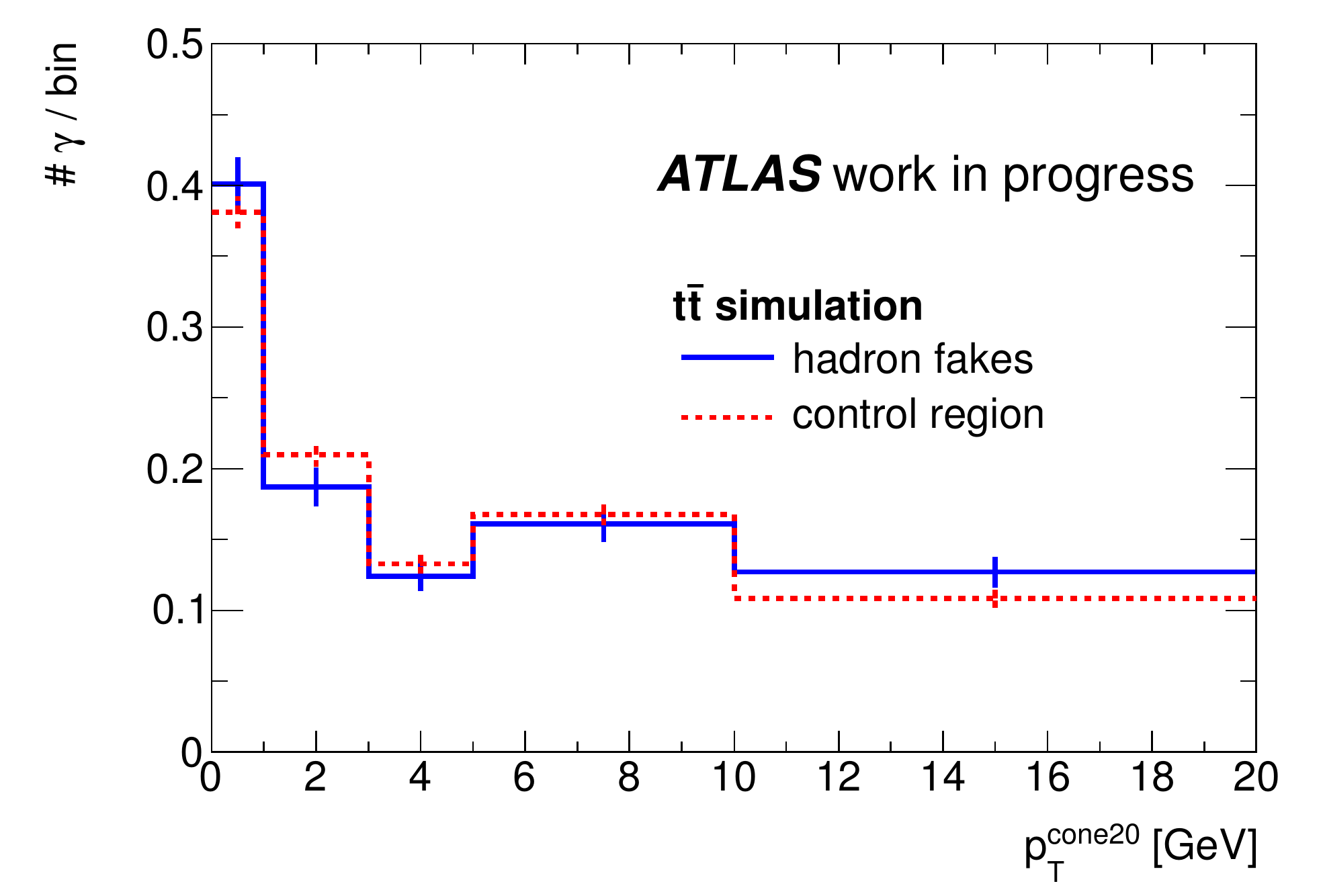}
    \caption[$\ptcone$ distributions of true hadron fakes and hadron fake candidates]{
      Normalised $\ptcone$ distributions for true hadron fakes (solid line) and hadron fake candidates (dashed line) from $\ttbar$ MC simulation.
      The last bin includes the overflow bin.
    }
    \label{fig:ptconecheck}
  \end{center}
\end{figure}

It was therefore checked that also the $\ptcone$ distribution in the $\ttbar$ environment is well modelled by the choice of this CR.
Fig.~\ref{fig:ptconecheck} shows the $\ptcone$ distribution for true hadron fakes (solid line) and hadron fake candidates (dashed line)
from $\ttbar$ MC simulation.
The distributions for true hadron fakes and the background model agree within uncertainties.
The four shower shapes used for the definition of the CR have hence been shown to be sufficiently uncorrelated with the $\ptcone$ variable, and
the hadron fake candidate selection therefore provides a good estimator for the $\ptcone$ distribution of hadron fakes in the $\ttbar$ environment.

\begin{figure}[h]
  \begin{center}
    \includegraphics[width=0.49\textwidth]{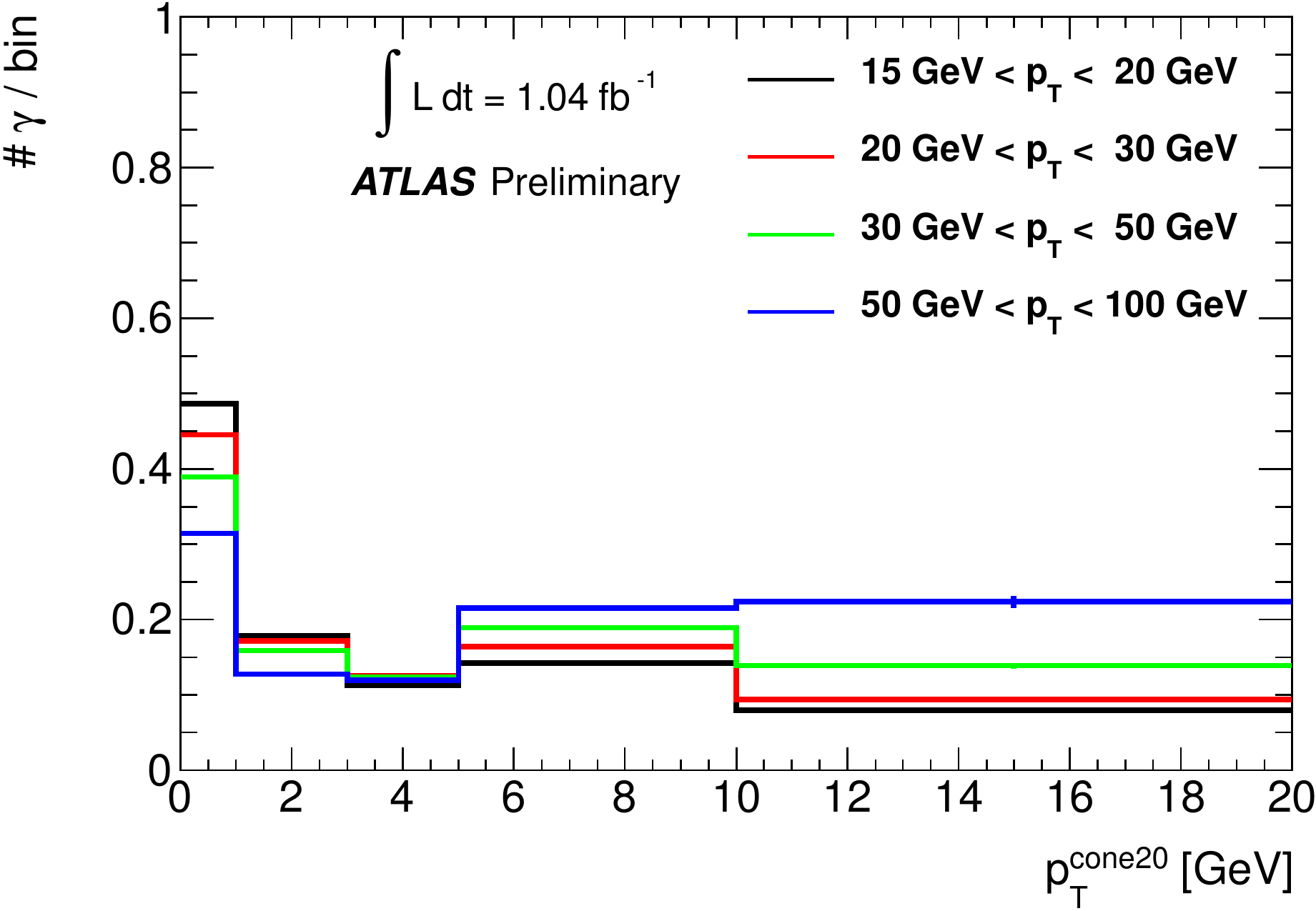}
    \includegraphics[width=0.49\textwidth]{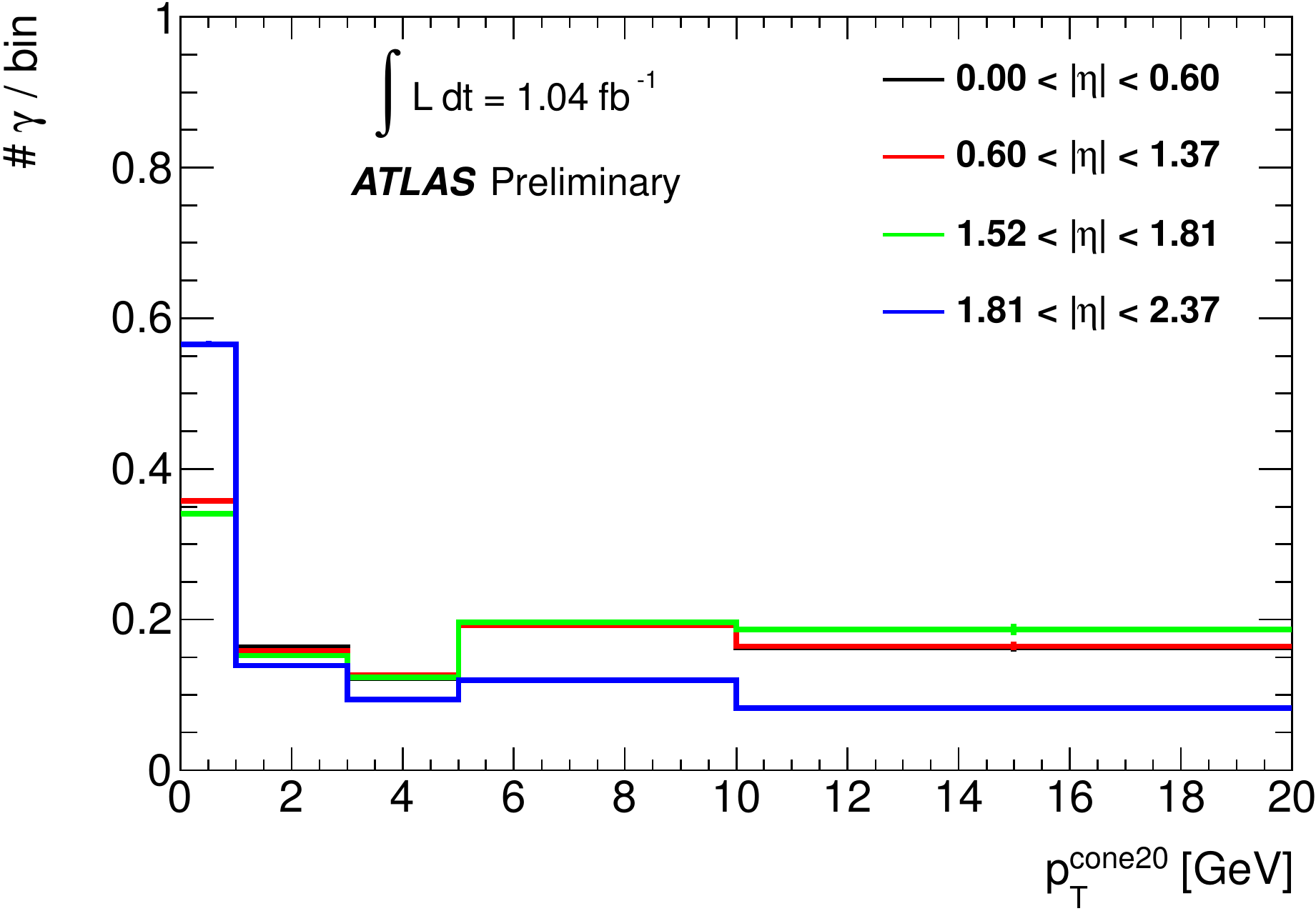}
    \caption[$\ptcone$ distributions for hadron fake candidates from data]{
      Normalised $\ptcone$ distributions for hadron fake candidates from the data control region for different regions of $\et$ (left) and $|\eta|$
      (right).
      In both plots, the last bin includes the overflow bin.
    }
    \label{fig:backgroundtemplate_pt_eta}
  \end{center}
\end{figure}

In order to derive the $\ptcone$ distributions from data, events from the skim of the \texttt{JetTauEtmiss} stream (Ch.~\ref{sec:data}) were used.
Fig.~\ref{fig:backgroundtemplate_pt_eta} shows the $\ptcone$ distributions for hadron fake candidates in different regions of $\et$ and $|\eta|$
as derived from data.
In contrast to the electron distributions, the distributions for hadron fake candidates show a clear dependence on the $\et$ of the object.
This behaviour is expected, because hadrons with large $\et$ are likely to originate from the jet fragmentation of high-$\pt$ partons, which features more
track activity than the fragmentation of lower-$\pt$ partons.

Two regions in $|\eta|$ can be distinguished: very similar $\ptcone$ distributions are observed for photons in the region
\mbox{$0 \leq |\eta| < 1.81$}, while they differ significantly from the $\ptcone$ distribution in the region \mbox{$1.81 \leq |\eta| < 2.37$}.
This is explained by the acceptance limit of the TRT at \mbox{$|\eta| = 2.0$}.

In order to correctly construct the fake hadron template, the templates for the different $\et$ and $\eta$-regions needed to be reweighted
according to the $\et$ and $\eta$ spectra of the hadron fakes within the selected $\ttg$ candidates.
In the following, the procedure for the estimation of the $\et$ spectrum as well as of the fraction of photons in the high-$|\eta|$-region
\mbox{$1.81 \leq |\eta| < 2.37$} is described.

\begin{figure}[h]
  \begin{center}
    \includegraphics[width=0.49\textwidth]{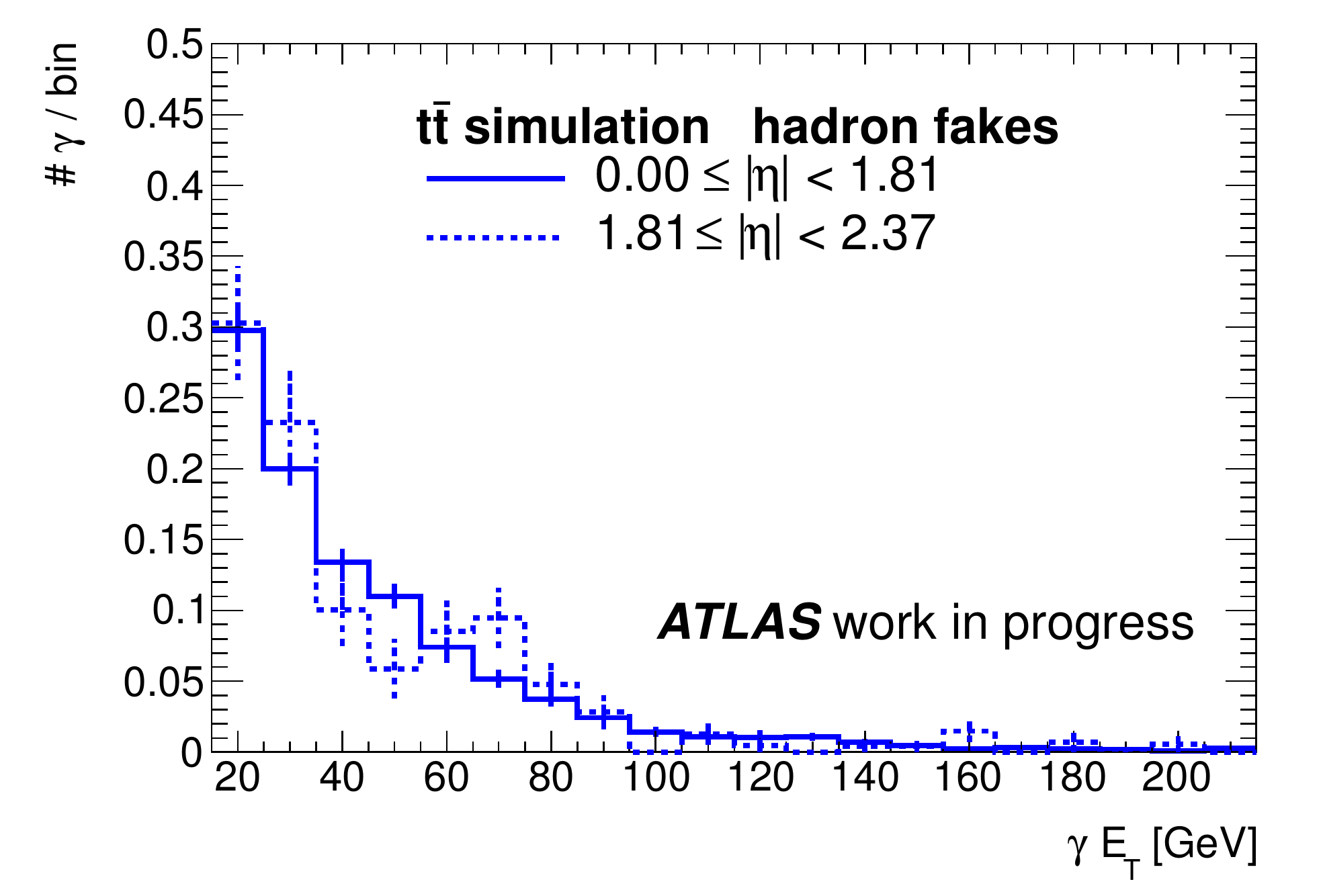}
    \includegraphics[width=0.49\textwidth]{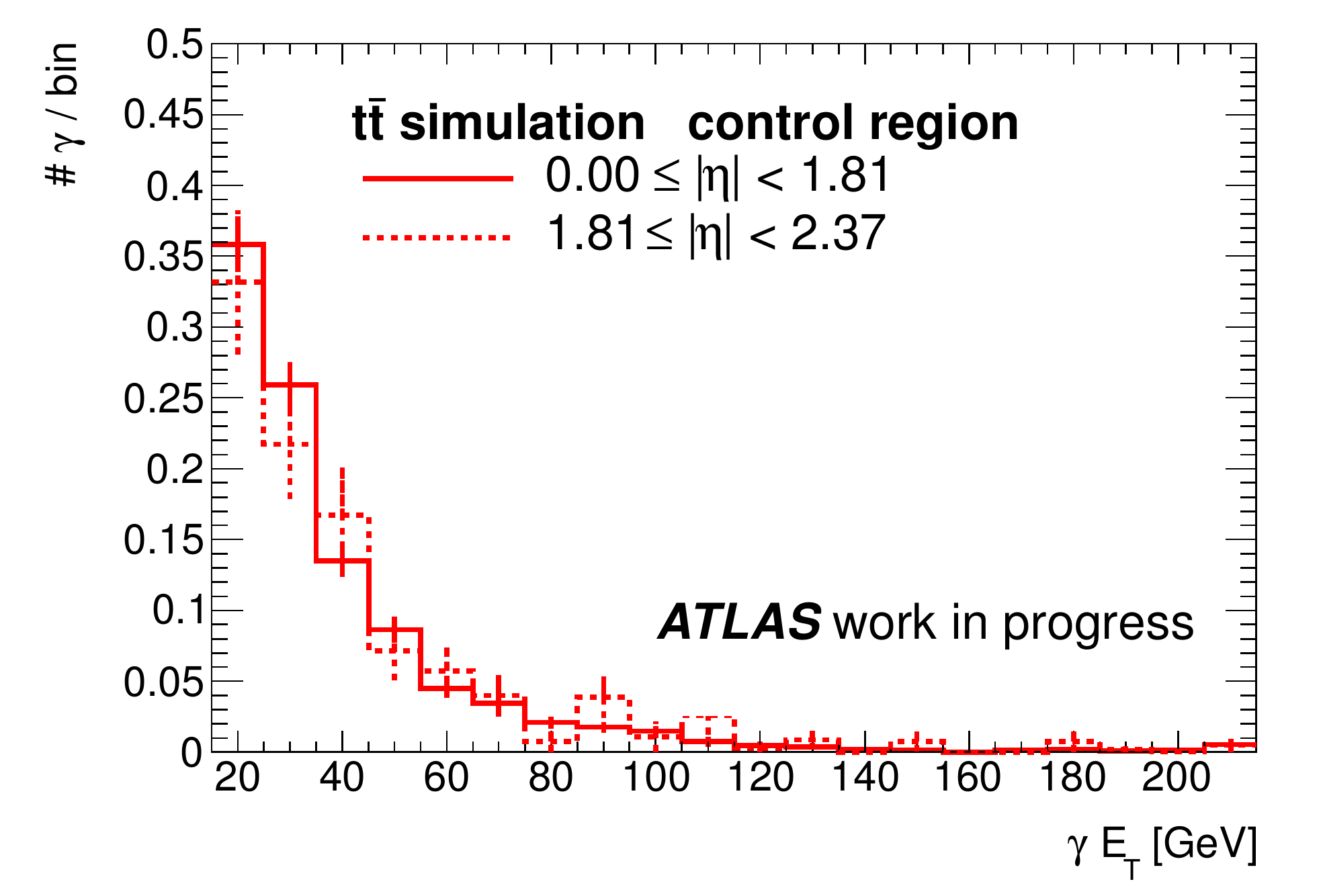}
    \caption[$\et$ spectra of true hadron fakes and hadron fake candidates]{
      $\et$ spectra of true hadron fakes (left) and hadron fake candidates (right) in $\ttbar$ simulations
      for \mbox{$0 \leq |\eta| < 1.81$} (solid line) and \mbox{$1.81 \leq |\eta| < 2.37$} (dashed line).
      In both plots, the last bin includes the overflow bin.
    }
    \label{fig:backgroundphotons_2}
    \vspace{0.025\textwidth}
    \includegraphics[width=0.49\textwidth]{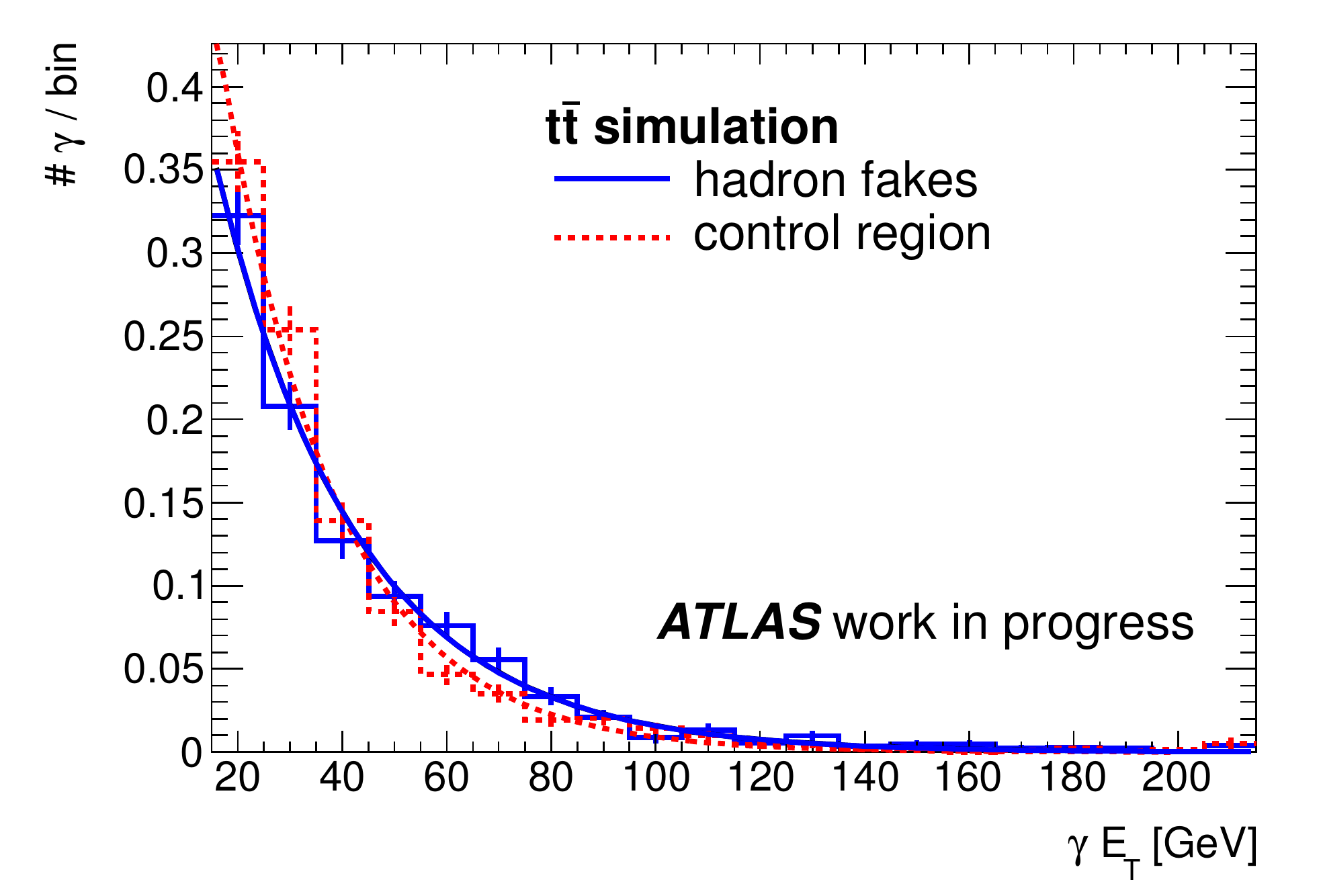}
    \includegraphics[width=0.49\textwidth]{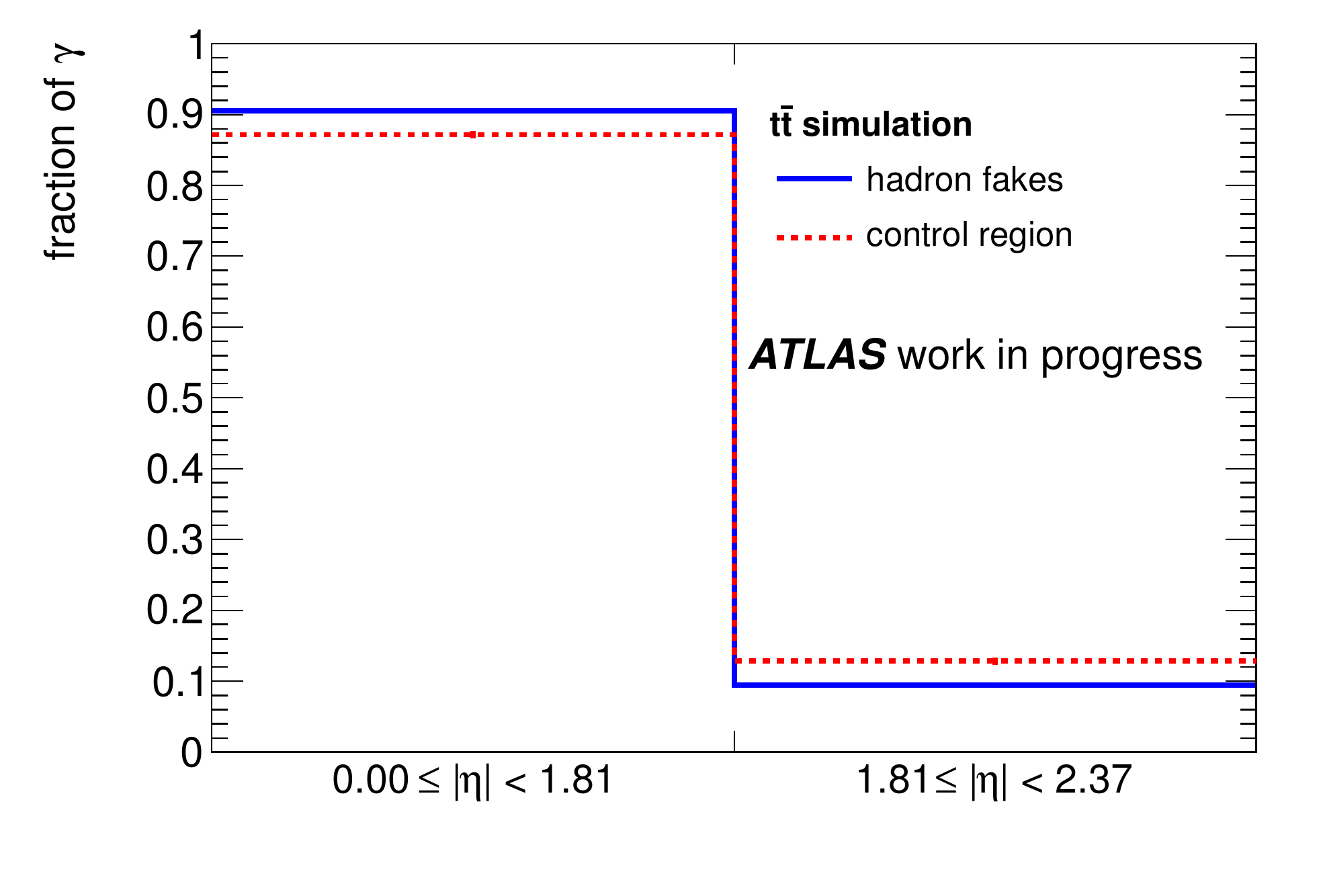}
    \caption[$\et$ spectrum and high-$|\eta|$ fraction for true hadron fakes and hadron fake candidates]{
      The left plot shows the $\et$ spectra of true hadron fakes (solid line) and hadron fake candidates (dashed line) in $\ttbar$ simulations.
      An exponential fit to both spectra is also depicted.
      The right plot shows the fraction of photons with $|\eta|$ smaller than 1.81 and larger than 1.81.
      In the $\et$ distribution, the last bin includes the overflow bin.
    }
    \label{fig:backgroundphotons_1}
  \end{center}
\end{figure}

First, it is shown that the $\et$ spectra of fake photons are very similar in the low- and the high-$|\eta|$-regions.
Fig.~\ref{fig:backgroundphotons_2} shows the photon $\et$ distributions from a full event selection in $\ttbar$ simulations for true hadron fakes
and hadron fake candidates.
In both cases, the $\et$ distributions are similar in the two $|\eta|$-regions.
Hence, the reweighting of the $\ptcone$ templates can be performed independently in $\et$ and $\eta$.

\begin{figure}[h]
  \begin{center}
    \includegraphics[width=0.49\textwidth]{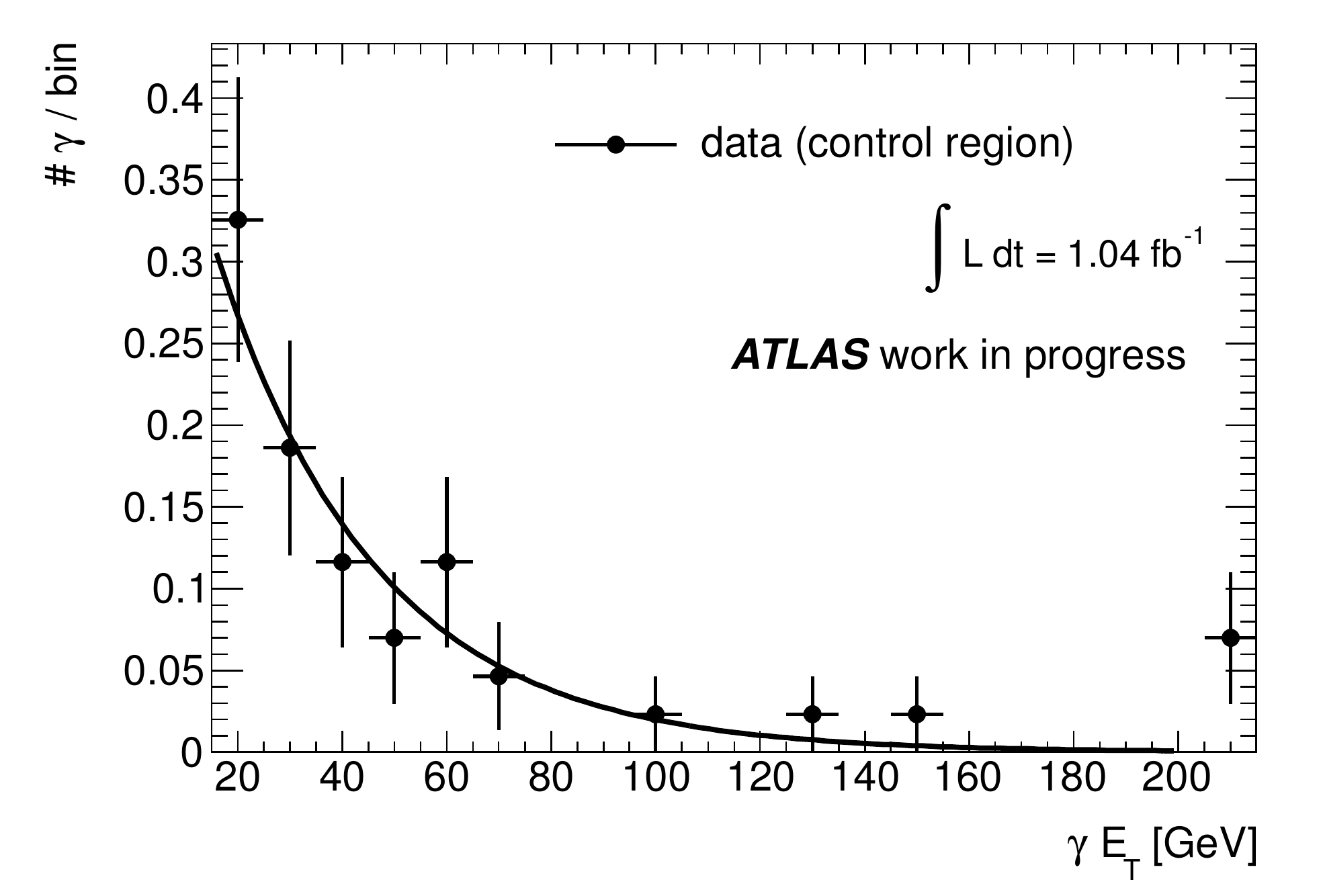}
    \includegraphics[width=0.49\textwidth]{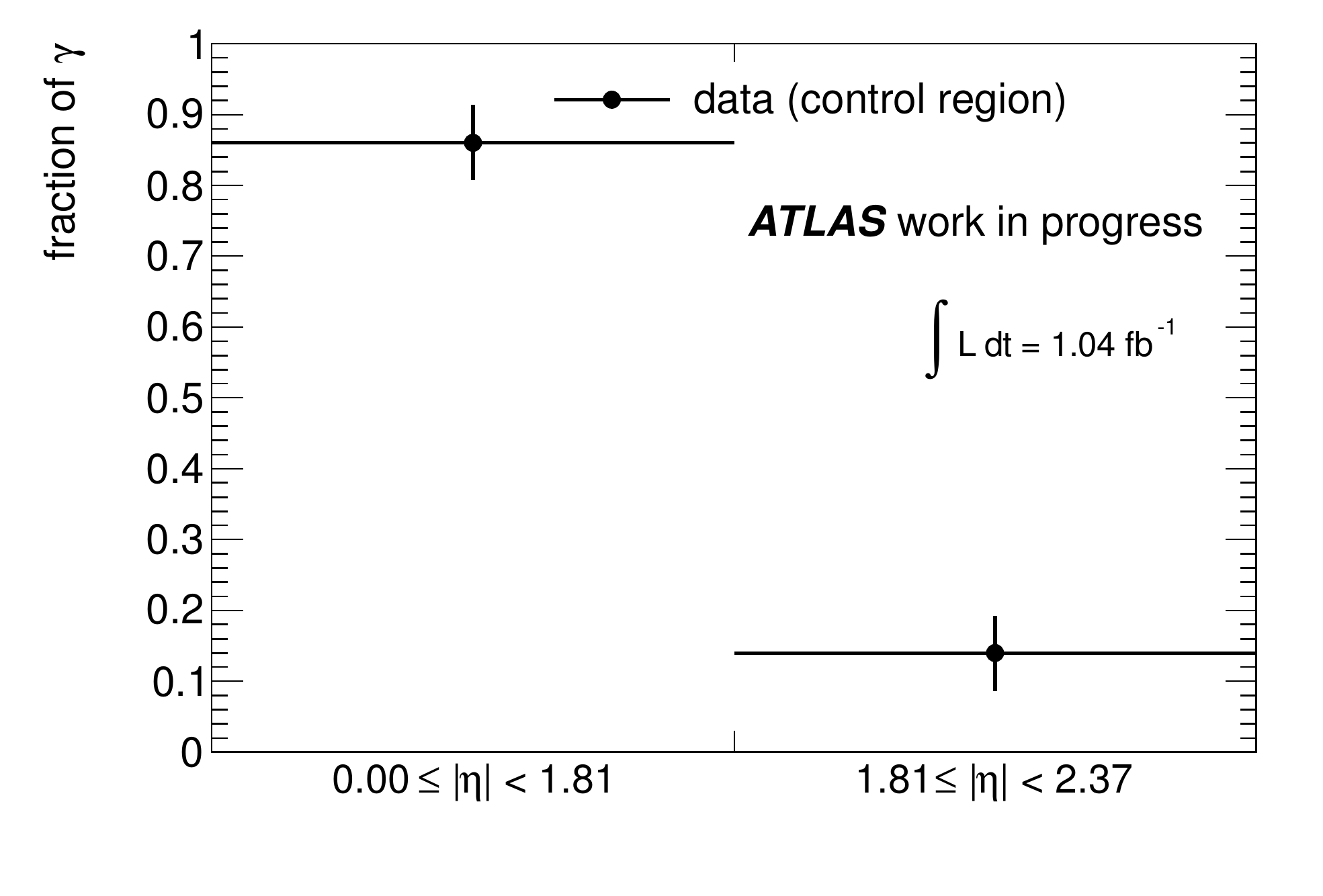}
    \caption[$\et$ spectrum and high-$|\eta|$ fraction for hadron fake candidates from data]{
      The left plot shows the estimated $\et$ spectrum of hadron fake candidates in data in both lepton channels together.
      An exponential fit to the spectrum is also shown.
      The right plot shows the fraction of photons with $|\eta|$ smaller than 1.81 and larger than 1.81.
      In the $\et$ distribution, the last bin includes the overflow bin.
    }
    \label{fig:backgroundselection}
  \end{center}
\end{figure}

The idea for the estimation of the fake photon $\et$ spectrum and the high-$|\eta|$ fraction is the following:
the good photon definition is replaced by the hadron fake candidate definition in the event selection.
The resulting photon $\et$ and $\eta$ spectra in data are then used to estimate the distributions for hadron fakes passing the actual good photon
definition.

This approach was validated using $\ttbar$ MC simulations:
the left plot of Fig.~\ref{fig:backgroundphotons_1} shows the $\et$ spectra for hadron fakes and hadron fake candidates from a full event selection.
The right plot shows the fraction of photons with \mbox{$0 \leq |\eta| < 1.81$} and \mbox{$1.81 \leq |\eta| < 2.37$}, respectively.

The $\et$ spectrum of the hadron fakes is modelled by an exponential function and the left plot of Fig.~\ref{fig:backgroundphotons_1} shows
the exponential fits to the photon $\et$ spectra:
the exponential model is fitting well the MC distributions.
The mean lifetimes extracted from the fits read \mbox{$27.1 \pm 0.9 \GeV$} and \mbox{$21.7 \pm 0.7 \GeV$} for true hadron fakes and hadron fake
candidates, respectively.
The difference of 25\% is taken into account as a systematic uncertainty on the exponential estimated from data.

The event selection with the hadron fake candidate definition yields 17 events in data in the single electron channel and 26 events in the single muon channel.
Since no differences between the photon spectra in the two lepton channels are expected, the events in both channels were combined in order to increase
the available statistics.
The left plot of Fig.~\ref{fig:backgroundselection} shows the $\et$ spectrum of the hadron fake candidates from this selection.
The exponential fit is also shown.
The entries in the last bin, which contains the overflow bin, are neglected.
The fit yields a mean lifetime of \mbox{$31 \pm 11 \GeV$}, which is compatible within uncertainties with the mean lifetime observed in MC simulations.
Considering an additional 25\% uncertainty as discussed above, results in an estimate of \mbox{$31 \pm 13 \GeV$}.
Due to the limited statistics, the uncertainty on the mean lifetime of the exponential is large.
This uncertainty is considered in the evaluation of the systematic uncertainties in Sec.~\ref{sec:syst_backgroundmodelling}.

The effect of the last bin, which contains the overflow bin, on the estimation of the $\et$ spectrum is negligible:
taking the last bin into account did not change the result of the exponential fit significantly.
Even when the weight for the high-$\et$-region was enhanced by 0.07, which corresponds to the three events in the last bin,
this was found to be covered by the systematic variations of the $\et$ spectrum, for which the weight of the high-$\et$-region was varied
by $+0.12$ and $-0.19$, respectively.

The right plot of Fig.~\ref{fig:backgroundselection} shows the fraction of hadron fake candidates with \mbox{$0 \leq |\eta| < 1.81$} and with
\mbox{$1.81 \leq |\eta| < 2.37$}.
The estimated fractions are in agreement with the MC estimates shown in Fig.~\ref{fig:backgroundphotons_1} within the statistical uncertainties.

The model for the hadron fake $\et$ spectrum and high-$|\eta|$ fraction was used to reweight the eight data templates
\mbox{$b^\mathrm{had} \left( \ptcone \, \left| \, \et, |\eta| \right. \right)$}
from the four different regions in $\et$ and the two regions in $|\eta|$ to one single template for the background from hadrons
$b^\mathrm{had} \left( \ptcone \right)$:
\begin{equation*}
%  b^\mathrm{had} \left( \ptcone \right)
%  = \mathrm{w}_\mathrm{{exp}}\left(\et\right) \cdot \mathrm{w}_\mathrm{frac} \left(|\eta|\right) \cdot 
%  b^\mathrm{had} \left( \ptcone \, \left| \, \et, |\eta| \right. \right) \, ,
  b^\mathrm{had} \left( \ptcone \right)
  = \sum_{\et \, {\rm bins}} \; \sum_{|\eta| \, {\rm bins}} \; \mathrm{w}_\mathrm{{exp}}\left(\et\right) \cdot \mathrm{w}_\mathrm{frac} \left(|\eta|\right) \cdot 
  b^\mathrm{had} \left( \ptcone \, \left| \, \et, |\eta| \right. \right) \, ,
\end{equation*}
where $\mathrm{w}_\mathrm{{exp}}\left(\et\right)$ and $\mathrm{w}_\mathrm{frac} \left(|\eta|\right)$ are the weights derived from the exponential fit to the $\et$
spectrum and the high-$|\eta|$ fraction for the given bin in $\et$ and $|\eta|$, respectively.
The resulting template $b^\mathrm{had}$ was presented in Fig.~\ref{fig:templates_overlay}.

\chapter{Background events with electrons misidentified as photons}
\label{sec:electronfake}

Electron and photon objects are very similar: since their electromagnetic clusters have nearly identical properties, electrons and photons
are only distinguished by the tracks associated to the cluster.
Unconverted photons do not feature associated tracks, and converted photons are discriminated from electrons by requiring that the associated
tracks belong to a conversion vertex (Sec.~\ref{sec:photon}).

However, electrons may be misidentified as photon objects when either the electron track was only reconstructed with poor quality or not at all.
Electromagnetic clusters from electrons may also feature additional tracks pointing to them, which may originate from close-by jet activity, from the
underlying event, or from another collision in the same bunch crossing (pile-up).
Such an electron may be misidentified as a converted photon.

The electron-to-photon misidentification rate $\feg$ is defined as the probability for a true electron to be misidentified as a photon object.
$\feg$ was measured in data using \Zee events using a tag-and-probe method as described in Sec.~\ref{sec:egammafakerate}.
Scale factors (SFs) with respect to the rate observed in MC simulations were derived in 16 bins of the photon $\et$ and $\eta$.
The main backgrounds with misidentified electrons are dileptonic $\ttbar$ decays and $Z$+jets production, but also single top
and diboson production contribute.
MC simulations of these backgrounds were corrected using the SFs as described in Sec.~\ref{sec:egammaapplication}.

\section{Estimate of the misidentification rate}
\label{sec:egammafakerate}

\Zee decays provide a clean environment for the study of electron properties.
The measurement of $\feg$ was based on the ratio of \Zee and \Zeg events in data, where in \Zeg one electron was misidentified as a photon.
In \Zeg events, the electron was required to have a larger $\et$ than the photon.
Therefore, the electron with the larger $\et$ in \Zee events and the electron in \mbox{$Z \to e\gamma_{\mathrm{fake}}$},
called \textit{tag electron} in the following, originated in most cases from the higher-$\et$ electron in \Zee events.

Let $\varepsilon_1$ denote the combined trigger, reconstruction and identification efficiency for the tag electron.
Let $\varepsilon_2$ denote the combined reconstruction and identification efficiency for the second electron in the \Zee selection.
Then, the numbers of events in the \Zee sample ($N_{ee}$) and in the \Zeg sample ($N_{e\gamma}$) are given by
\begin{eqnarray*}
  N_{ee} & = & \varepsilon_1 \cdot \varepsilon_2 \cdot N \, , \; \rm{and} \\
  N_{e\gamma} & = & \varepsilon_1 \cdot \feg \cdot N \, ,
\end{eqnarray*}
where $N$ is the total number of true \Zee events in the acceptance region.

\begin{figure}[h]
  \begin{center}
    \includegraphics[width=0.49\textwidth]{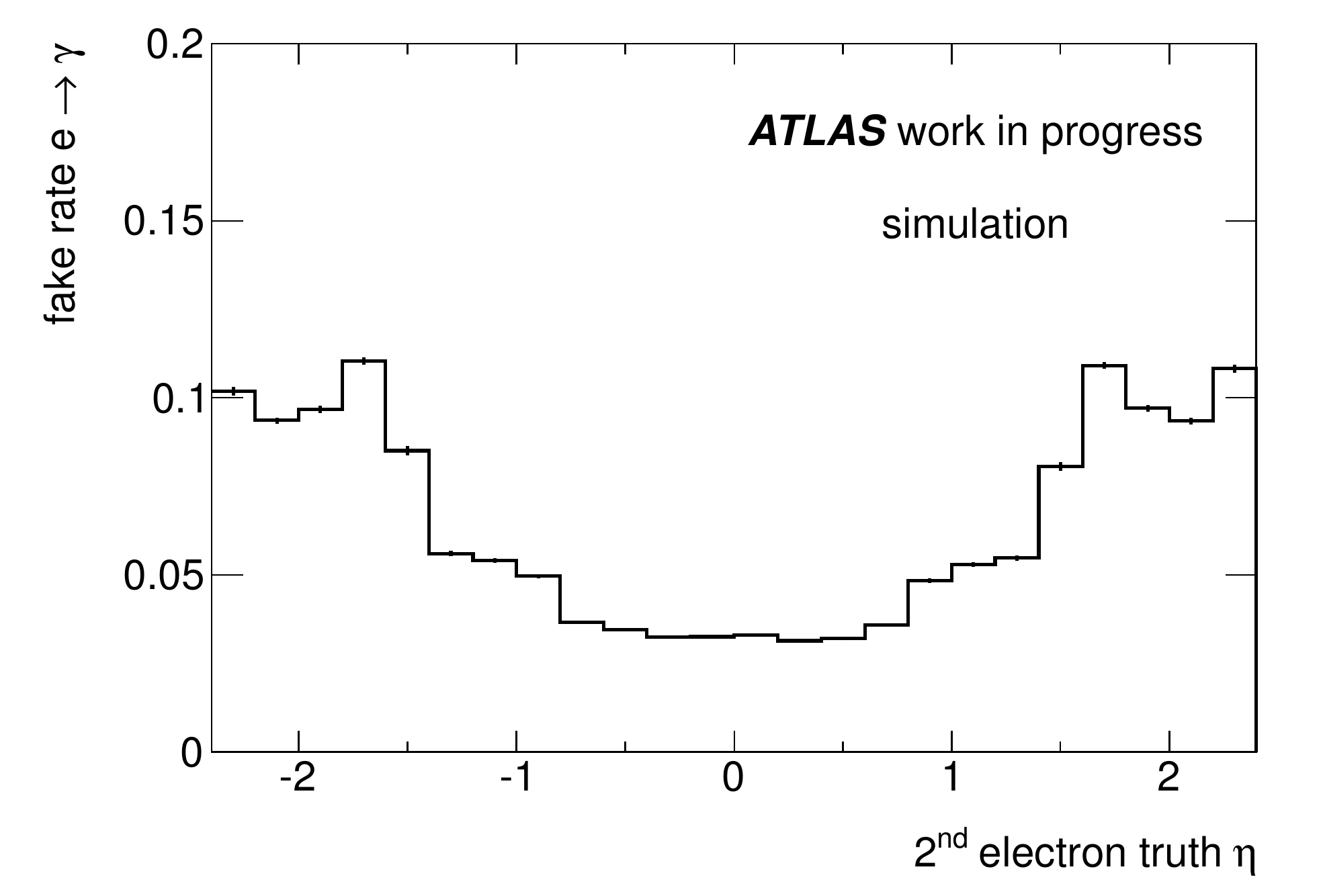}
    \includegraphics[width=0.49\textwidth]{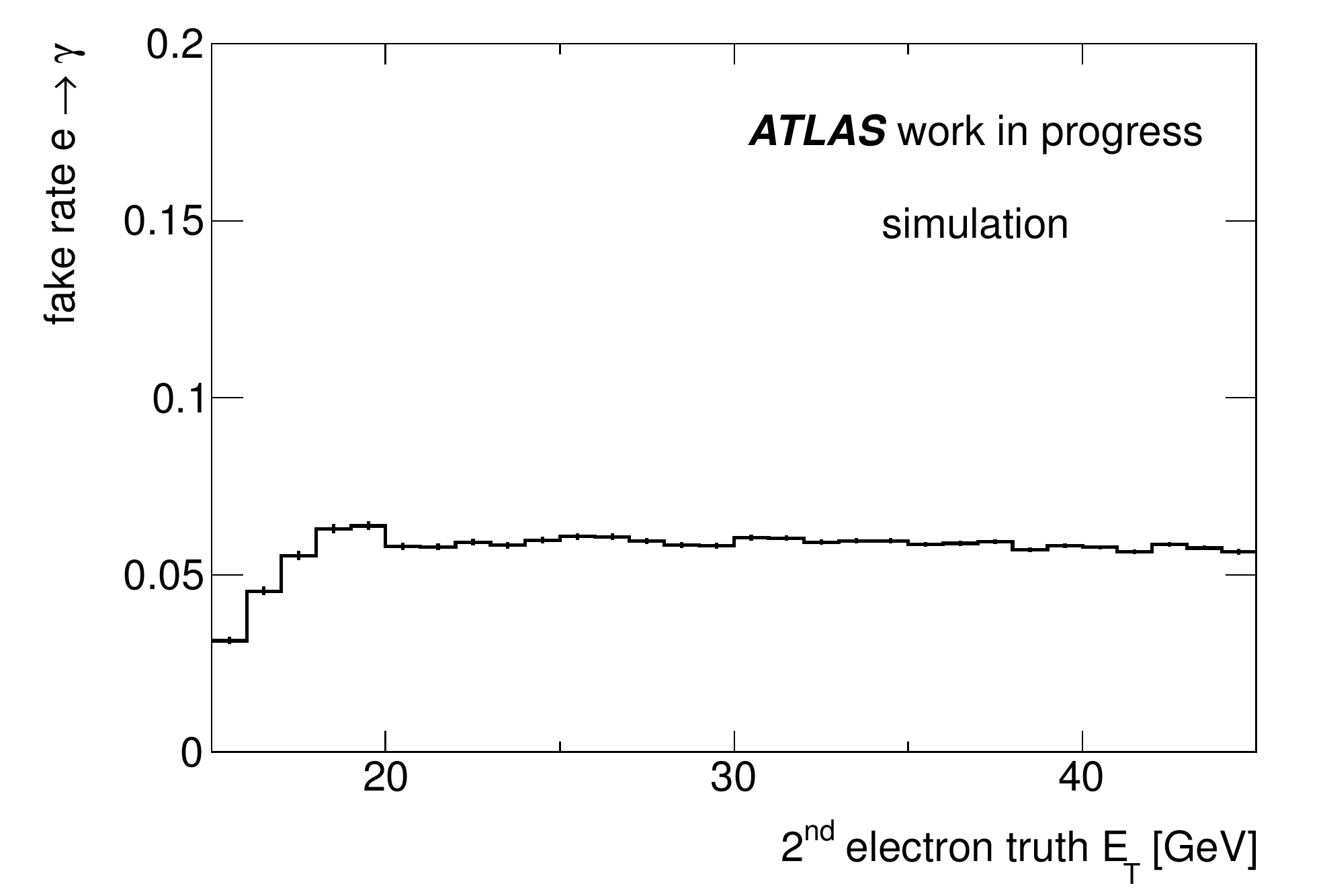}
    \caption[True electron-to-photon misidentification from MC]{
      Misidentification rate $\feg$ from MC simulations as a function of the true electron $\eta$ (left) and $\et$ (right).
    }
    \label{fig:truefakerate}
  \end{center}
\end{figure}

Hence, $\feg$ is given by
\begin{equation*}
  \feg = \varepsilon_2 \cdot \frac{N_{e\gamma}}{N_{ee}} \, .
\end{equation*}
Fig.~\ref{fig:truefakerate} shows $\feg$ as a function of the true electron $\eta$ (left) and $\et$ (right) in simulations.
The dependence on $\eta$ is explained by the varying granularity of the calorimeter and the changing amount of material in front of it.
$\feg$ may be affected by the altering shape of calorimeter showers and the varying performance of the track reconstruction.
While for high electron $\et$ the misidentification rate is constant at roughly 6\%,
for low $\et$, a turn-on is observed, because in this analysis photons were required to have a minimum $\et$ of \mbox{$15 \GeV$}.

A scale factor was derived in order to correct simulations so that they describe the misidentification rate observed in data:
\begin{equation}
  \mathrm{SF} = \frac{\left. \frac{N_{e\gamma}}{N_{ee}} \right|_{\mathrm{data}}}{\left. \frac{N_{e\gamma}}{N_{ee}} \right|_{\mathrm{MC}}} \quad .
  \label{eq:SF}
\end{equation}
$\varepsilon_2$ cancels out, because the simulations were corrected for discrepancies with respect to data using SFs for electron
reconstruction and identification (Sec.~\ref{sec:electron}).

For the measurement of the misidentification rate, events triggered by the \texttt{EF\_e20\_medium} chain were selected which featured also
a good vertex with at least five associated tracks.
At least one good electron (Sec.~\ref{sec:electron}) with \mbox{$\et > 25 \GeV$} was required, which was closer than 0.15 in $\eta$-$\phi$-space
to a trigger object (tag electron).
Only events with good data taking conditions of the LAr calorimeters were considered.

For the selection of \Zee events, a second good electron with \mbox{$\et > 15 \GeV$} was required (\textit{probe electron}).
In order to avoid a bias from the trigger, the first electron had to fire the trigger, and the second electron was used for the measurement of the
misidentification rate, where the second electron was chosen to be the electron with the lower $\et$.
The invariant mass of the two electrons was required to lie within a \mbox{$\pm 50 \GeV$} window around the $Z$ boson mass.
This wide mass range was chosen in order to include side bands with large background contributions from multijet production.

For the selection of \Zeg events, a photon with \mbox{$\et > 15 \GeV$} was required (\textit{probe photon}).
The $\et$ of the photon had to be lower than the $\et$ of the tag electron.
The invariant mass of the electron and the photon also had to lie within a \mbox{$\pm 50 \GeV$} window around the $Z$ boson mass.

\begin{figure}[h]
  \begin{center}
    \includegraphics[width=0.49\textwidth]{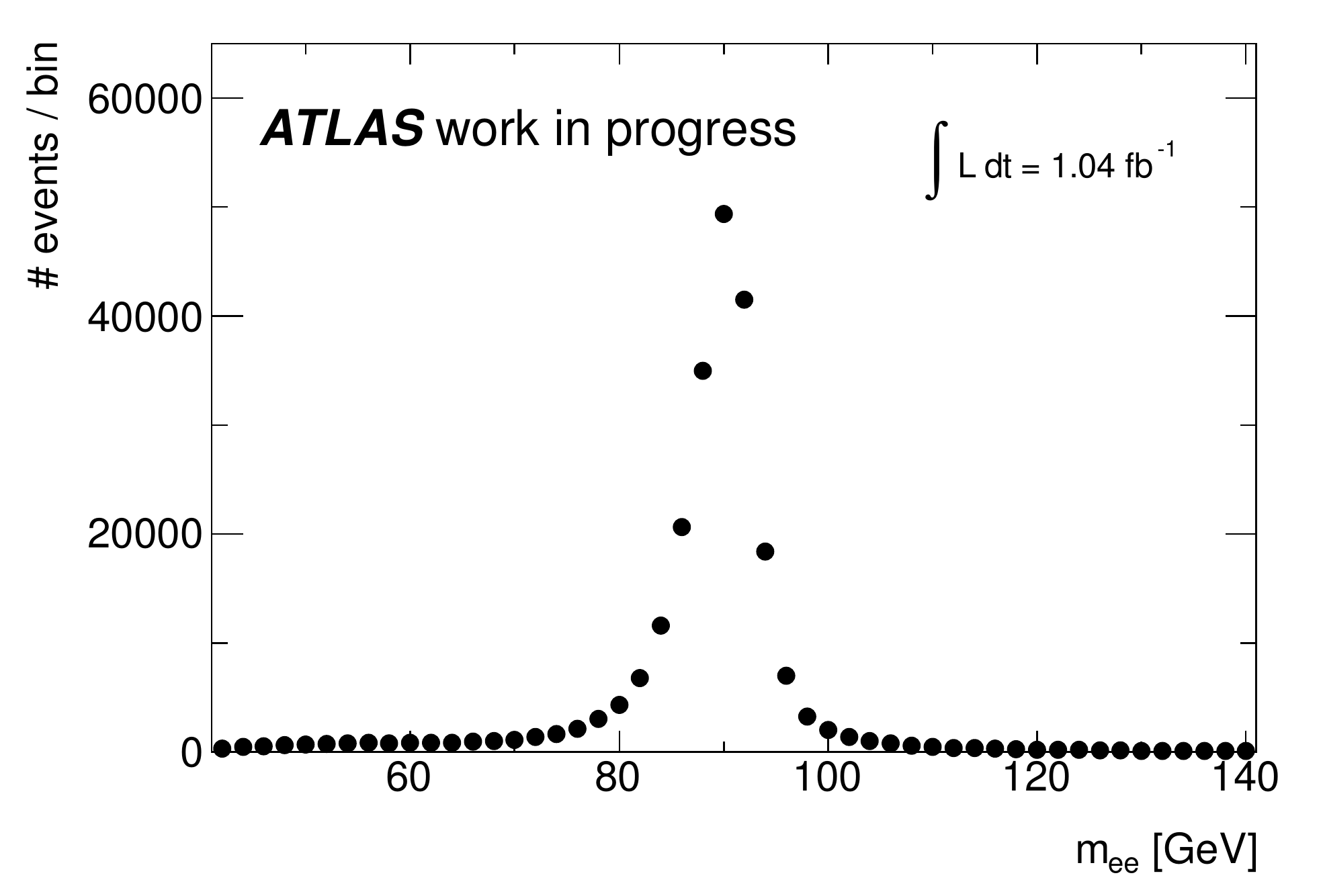}
    \includegraphics[width=0.49\textwidth]{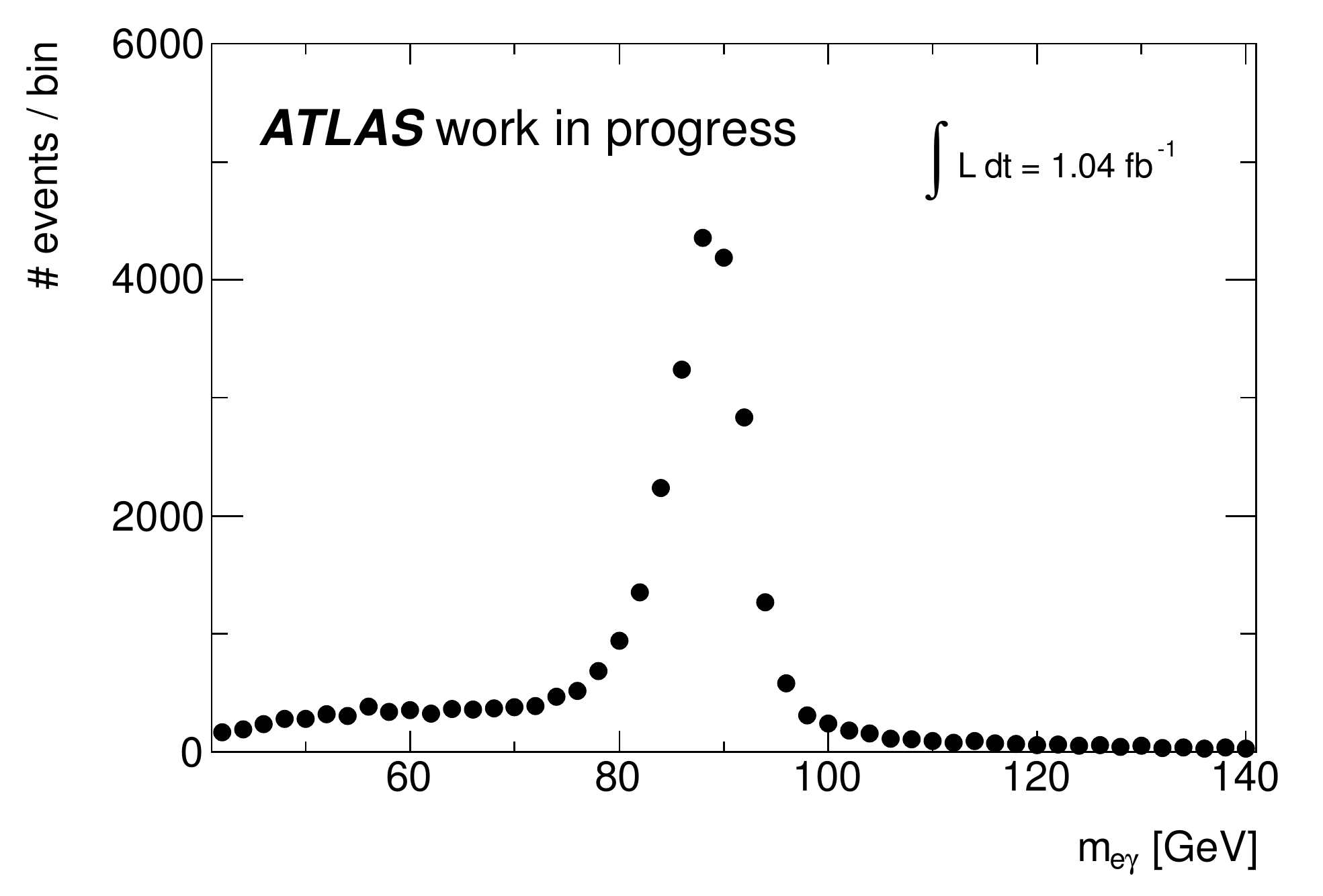}
    \caption[Electron-electron and electron-photon invariant mass distributions in data]{
      Invariant mass distributions for the two electrons from the \Zee selection (left), and the electron and the photon from the \Zeg selection (right)
      in data.
    }
    \label{fig:ZeeZegamma}
  \end{center}
\end{figure}

The left plot in Fig.~\ref{fig:ZeeZegamma} shows the invariant mass distributions from \mbox{$1.04 \ifb$} of data for the two electrons from the \Zee
selection.
The right plot shows the invariant mass distribution for the electron and the photon from the \Zeg selection.
Both distributions feature a distinct $Z$ boson mass peak, which indicates that the photon candidates from the \Zeg selection were indeed dominated by
misidentified electrons from the $Z$ boson decay.

\subsubsection{Estimation of the background contribution from multijet events}

Hadrons from jet fragmentation produced in multijet events may be misidentified as electrons or photons.
Since the cross section for multijet production is much larger than the cross section for $Z$ boson production, multijet events with two
misidentified hadrons may contribute sizeably to the \Zee and \Zeg samples.

A broad invariant mass window around the $Z$ boson mass was chosen in order to include regions of very high and low invariant masses, which
feature a larger fraction of multijet events than the central peak.
These side bands were used to estimate the contribution from multijet events in a narrow window of \mbox{$10 \GeV$} around the $Z$ mass.
The purer \Zee and \Zeg samples in the narrow mass window were then used for the measurement of the misidentification rate $\feg$.

In order to estimate the multijet background, the invariant mass distributions presented in Fig.~\ref{fig:ZeeZegamma} were fitted with a signal
and a background model:
while the multijet background was modelled with an exponential, the signal $Z$ peak was modelled with the convolution
\mbox{$\mathrm{CB} \mathord{*} \mathrm{BW}$} of a Breit-Wigner distribution
($\mathrm{BW}$) for the $Z$ boson mass and a Crystal-Ball function $\mathrm{CB}$ in order to describe the mass resolution:
\begin{equation*}
  \mathrm{CB}\left( m ; \alpha, n, \bar{m}, \sigma \right) = \rm{norm.} \cdot
\left\{ {
    \begin{matrix} \left( \frac{n}{|\alpha|} \right)^n \cdot \exp\left(-\frac{|\alpha|^2}{2}\right)
      \cdot \left( \frac{n}{|\alpha|}  - |\alpha| - \frac{m - \bar{m}}{\sigma} \right)^{-n}
      & , & \frac{m - \bar{m}}{\sigma} \leq -\alpha \\
      & & \\
      \exp \left( - \frac{\left( m - \bar{m} \right)^2}{2 \cdot \sigma^2} \right)
      & , & \rm{else}
    \end{matrix}} \quad .
\right.
\end{equation*}

\begin{figure}[h]
  \begin{center}
    \includegraphics[width=0.49\textwidth]{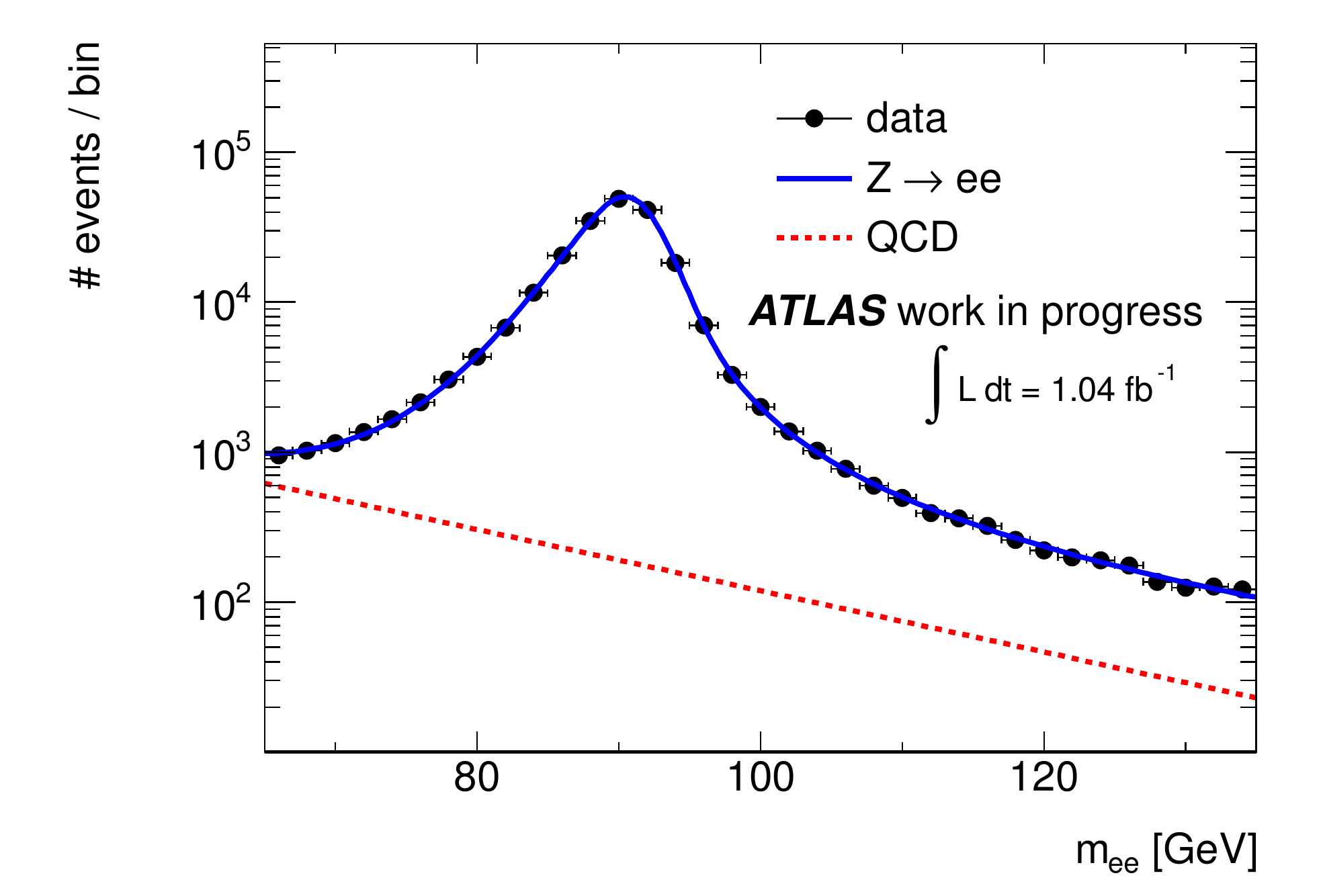}
    \includegraphics[width=0.49\textwidth]{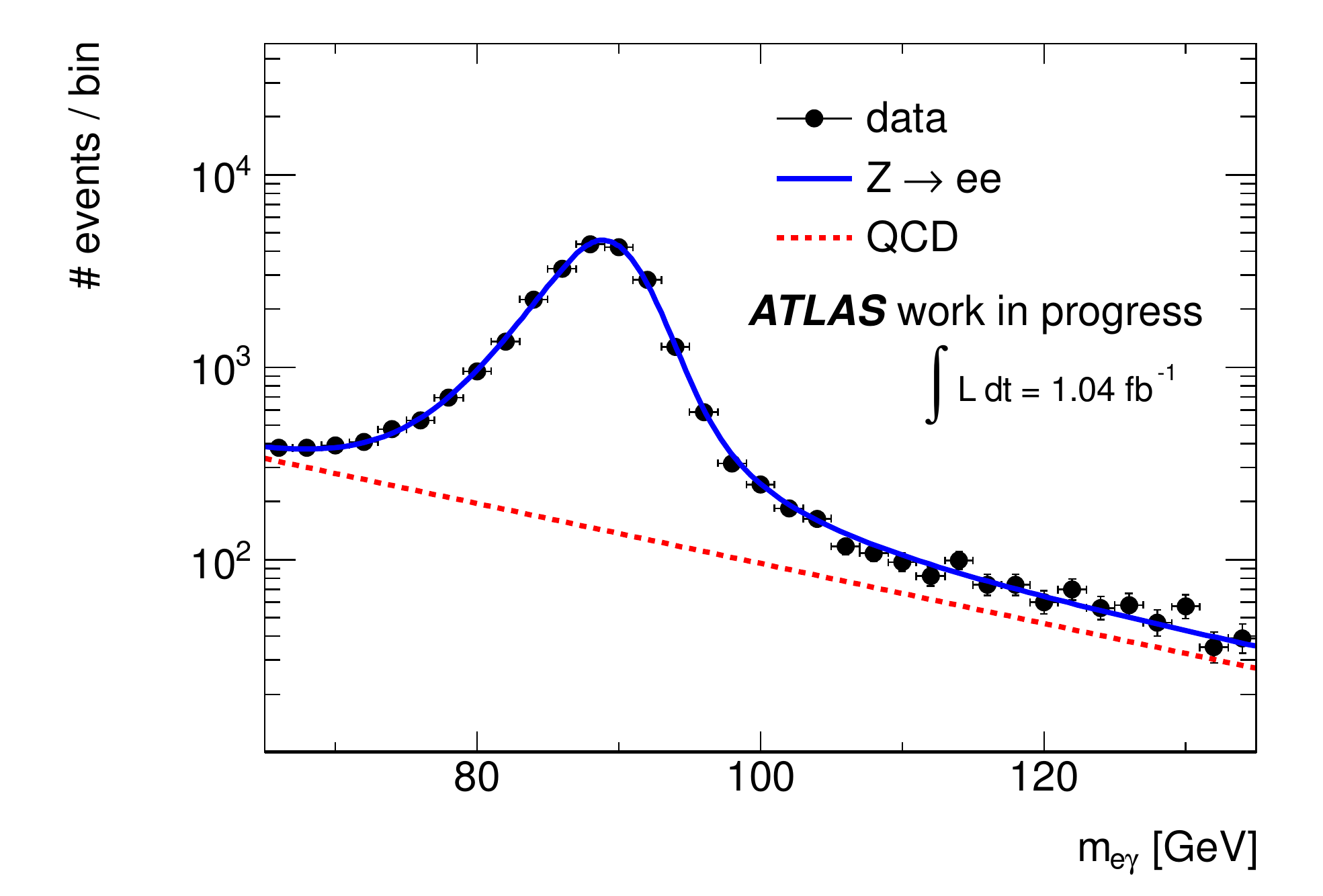}
    \caption[$m(ee)$ and $m(e\gamma)$ in data with signal and background fit]{
      Invariant mass distributions for the two electrons from the \Zee selection (left) and for the electron and the photon from the \Zeg
      selection (right) in data.
      The distributions were fitted with a convolution of a Breit-Wigner distribution and a Crystal-Ball function for the signal contribution
      (solid line), and an exponential for the background modelling (dashed line).
    }
    \label{fig:CB_globalfit}
  \end{center}
\end{figure}

Fig.~\ref{fig:CB_globalfit} shows the fits to the invariant mass distributions for the two electrons from the \Zee selection (left) and for the electron
and the photon from the \Zeg selection (right) in data.
The combination of the background fit (dashed line) and the signal fit (solid line) fits the data very well.
The fit range was chosen to \mbox{$[65 \GeV, 135 \GeV]$}.
In order to allow for a variation of the limits for the evaluation of the systematic uncertainty on the fit stability, the upper limit was chosen to
be slightly lower than the maximum value used in the selection (\mbox{$141 \GeV$}).
In the low-mass region (\mbox{$< 60 \GeV$}),
however, the signal distribution does not follow the \mbox{$\mathrm{CB} \mathord{*} \mathrm{BW}$} shape, because of the minimum
requirements on the electron and photon $\et$ of \mbox{$25 \GeV$} and \mbox{$15 \GeV$}, respectively.
In order to avoid a bias due to this effect, the lower bound was chosen at \mbox{$65 \GeV$}.

While the fit in the \Zee sample yields a contribution of \mbox{$0.97 \pm 0.04 \, \%$} from multijet events, the fit in the \Zeg sample yields a multijet
contribution of \mbox{$6.5 \pm 0.3 \, \%$}.
The uncertainties are the statistical ones from the fit only.
As expected, the multijet contribution is larger in the \Zeg sample, because no isolation criterion was applied to photons.

The fraction of multijet events is expected to decrease with the object's $\et$, because the $\pt$ spectrum for jets drops exponentially, while
the $\et$ spectrum for electrons from $Z$ boson decays drops less strongly due to the underlying Breit-Wigner distribution.
In addition, a dependence on $\eta$ can be expected, because the varying detector geometry changes the probability to misidentify objects from
jet fragmentation as electrons.
Hence, the multijet background in the \Zeg sample was estimated in four bins of $\et$ and four bins of $|\eta|$ of the probe photon:
the bins in $|\eta|$ were the same as used in Ch.~\ref{sec:photontemplate} and~\ref{sec:faketemplate}, which read
\mbox{$[0, 0.6)$}, \mbox{$[0.6, 1.37)$}, \mbox{$[1.52, 1.81)$}, and \mbox{$[1.81, 2.37)$}.
The following $\et$ bins were chosen: \mbox{$[15 \GeV, 20 \GeV)$}, \mbox{$[20 \GeV, 30 \GeV)$}, \mbox{$[30 \GeV, 40 \GeV)$}, and \mbox{$[40 \GeV, 45 \GeV)$}.
Events where the photon had an $\et$ of more than \mbox{$45 \GeV$} were not taken into account, because in these events also the tag electron had an $\et$
of at least \mbox{$45 \GeV$}.
$Z$ decays in which both decay products have a very large $\et$ are rare and also feature an invariant mass distribution which differs from the
form of the \mbox{$\mathrm{CB} \mathord{*} \mathrm{BW}$} signal model.

\begin{figure}[h!]
  \begin{center}
    \includegraphics[width=0.49\textwidth]{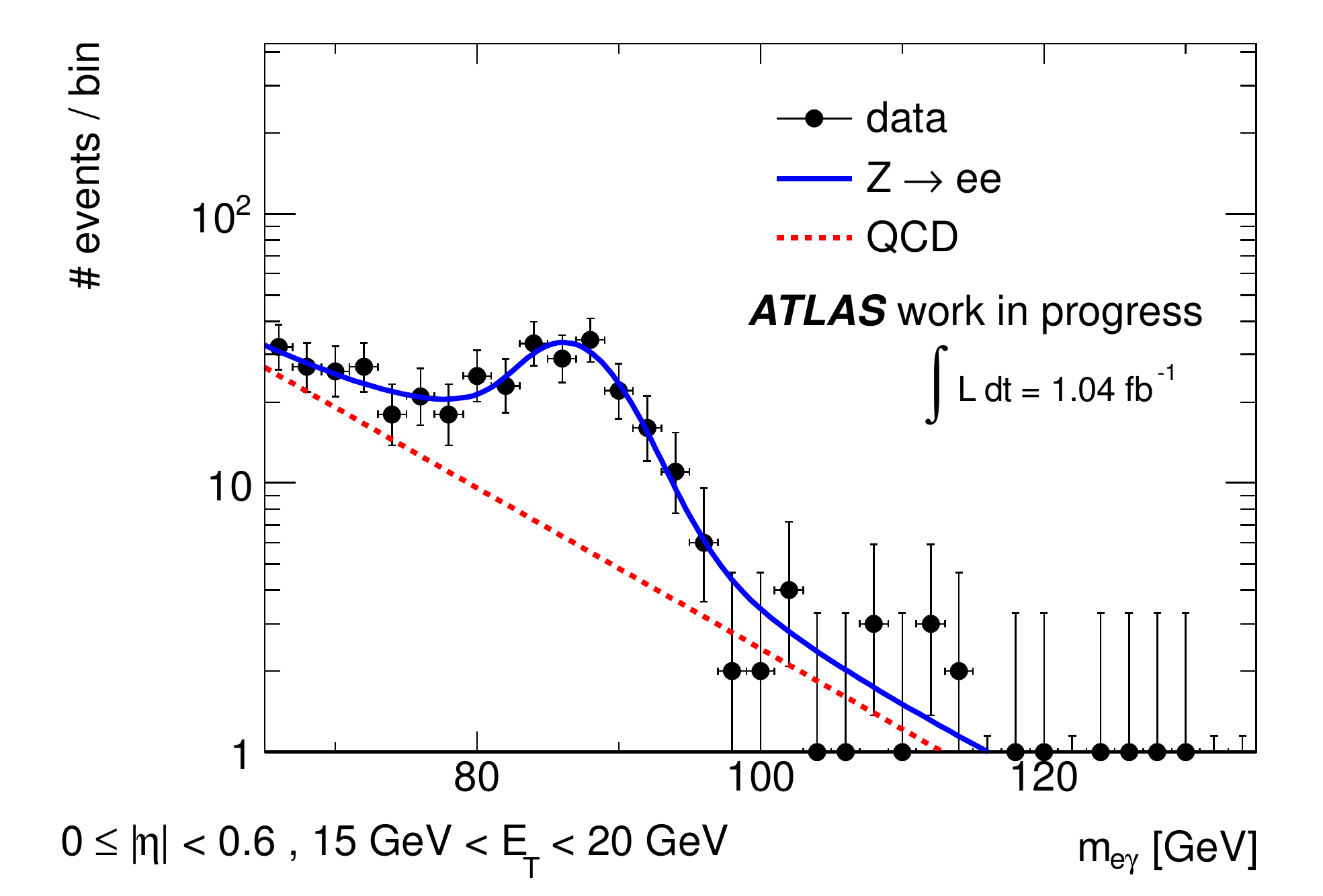}
    \includegraphics[width=0.49\textwidth]{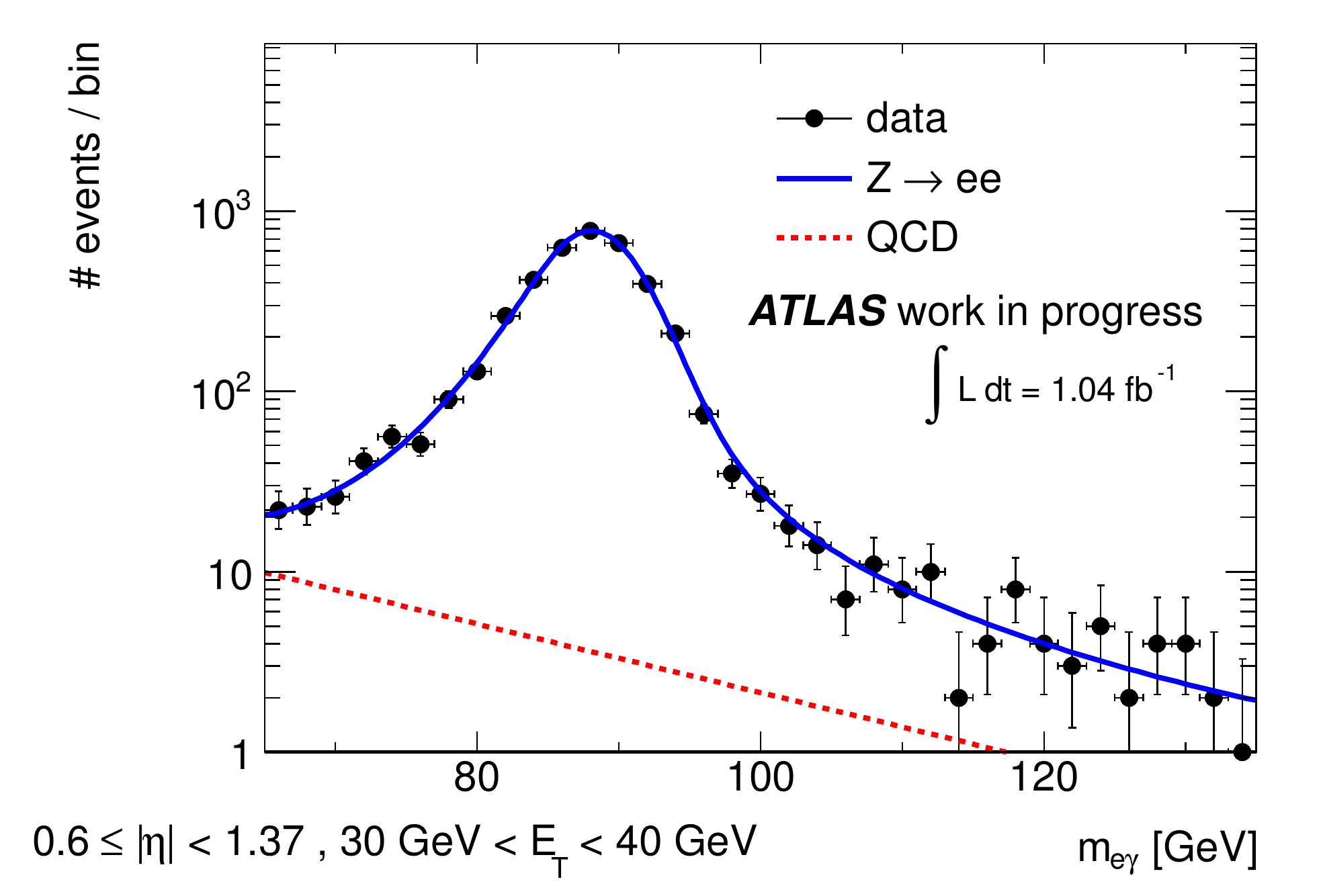}
    \caption[$m(e\gamma)$ in data with signal and background fit in two $\eta$- and $\et$-regions]{
      Invariant mass distributions for the electron and the photon from the \Zeg selection in data
      for photons in the region \mbox{$0 \leq |\eta| < 0.6$} and \mbox{$15 \GeV \leq \et < 20 \GeV$} (left),
      and in the region \mbox{$0.6 \leq |\eta| < 1.37$} and \mbox{$30 \GeV \leq \et < 40 \GeV$} (right).
      The distributions were fitted with a convolution of a Breit-Wigner distribution and a Crystal-Ball function for the signal contribution
      (solid line), and an exponential for the background modelling (dashed line).
    }
    \label{fig:CB_egamma_regions}
    \vspace{0.025\textwidth}
    \includegraphics[width=0.49\textwidth]{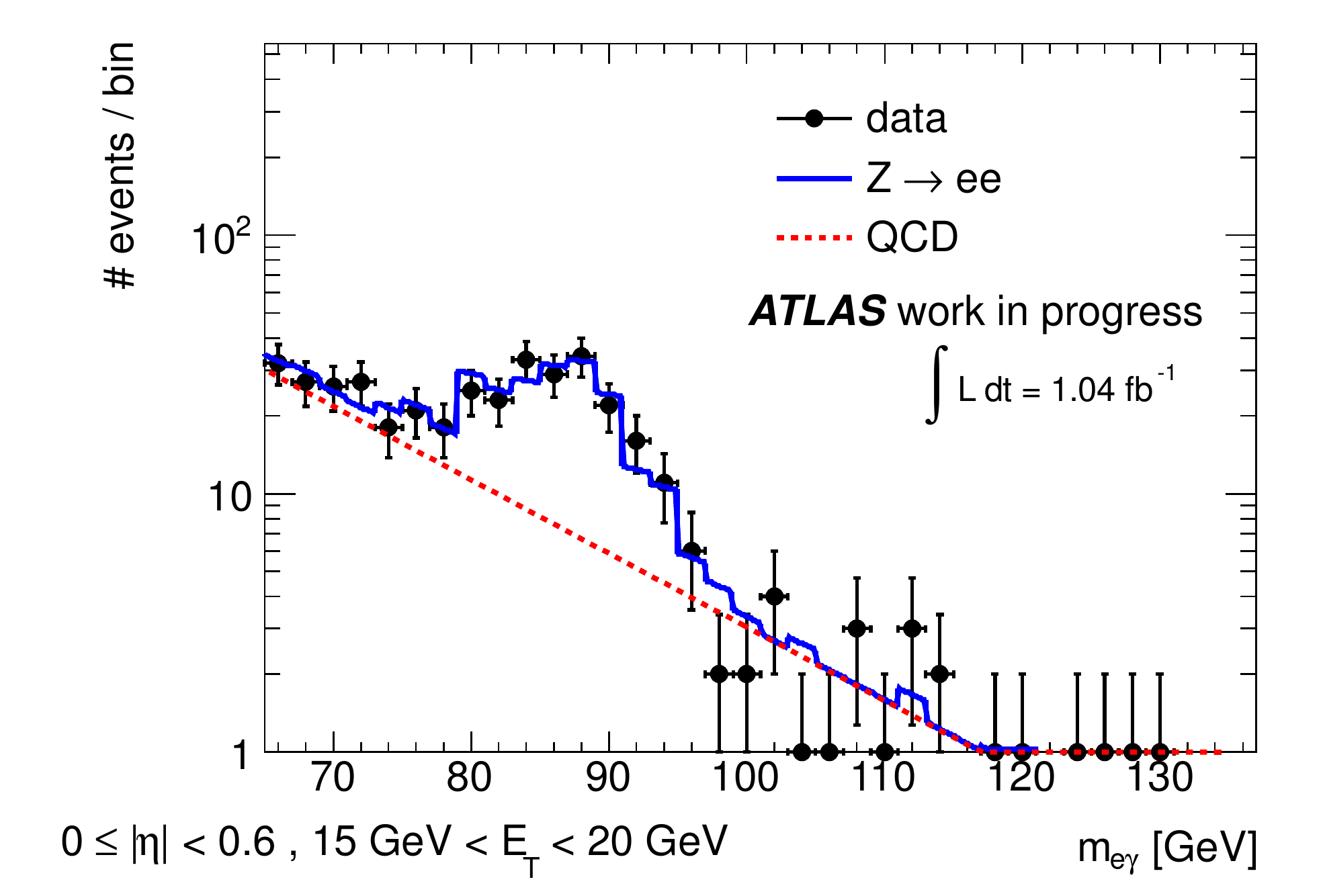}
    \includegraphics[width=0.49\textwidth]{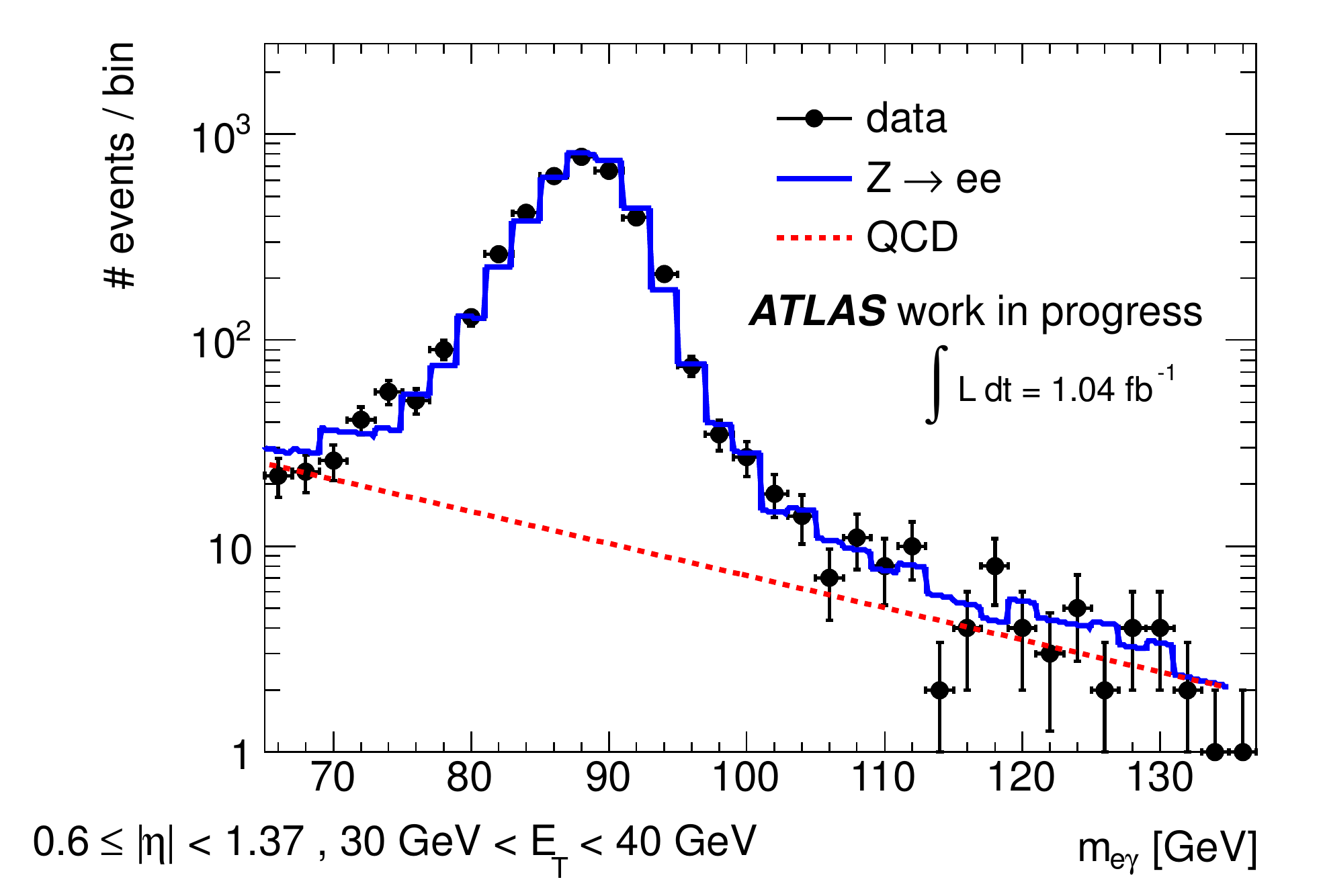}
    \caption[$m(e\gamma)$ in data with signal and background fit in two $\eta$- and $\et$-regions]{
      Invariant mass distributions for the electron and the photon from the \Zeg selection in data
      for photons in the region \mbox{$0 \leq |\eta| < 0.6$} and \mbox{$15 \GeV \leq \et < 20 \GeV$} (left),
      and in the region \mbox{$0.6 \leq |\eta| < 1.37$} and \mbox{$30 \GeV \leq \et < 40 \GeV$} (right).
      The distributions were fitted with templates derived from MC simulations (solid line), and an exponential for the background modelling
      (dashed line).
    }
    \label{fig:Template_egamma_regions}
  \end{center}
\end{figure}

Fig.~\ref{fig:CB_egamma_regions} shows the fits for the \Zeg selection for photons in the region \mbox{$0 \leq |\eta| < 0.6$} and
\mbox{$15 \GeV \leq \et < 20 \GeV$} (left), and in the region \mbox{$0.6 \leq |\eta| < 1.37$} and
\mbox{$30 \GeV \leq \et < 40 \GeV$} (right).
The resulting fraction of multijet events in the \Zeg sample are summarised in Tab.~\ref{tab:egammaFakeRate_QCD} including the statistical
uncertainty from the fit.

\begin{table}[h]
\centering
\footnotesize
\begin{tabular}[h] {|c|c|r@{}l c r@{}l l c r@{}l l|r@{}l c r@{}l l|}
  \hline
  $|\eta|$-region & $\et$-region \mbox{$[\mathrm{GeV}]$} & \multicolumn{10}{c|}{\Zeg sample [\%]} & \multicolumn{6}{c|}{\Zee sample [\%]} \\
  \hline
  \mbox{$[0, 0.6)$} & \mbox{$[15, 20)$} & 27 &   & $\pm$ & 11 &   & (stat.) & $\pm$ & 7 &   & (syst.) & 2 &.3 & $\pm$ & 2&.3 & (syst.) \\
           & \mbox{$[20, 30)$} & 16 &   & $\pm$ &  1 &   & (stat.) & $\pm$ & 2 &   & (syst.) & 0 &.9 & $\pm$ & 0&.9 & (syst.) \\
           & \mbox{$[30, 40)$} &  1 &.0 & $\pm$ &  0 &.3 & (stat.) & $^+_-$ & $^{\emptyplus 2}_{\emptyminus 1}$ & & (syst.) & 0&.4 & $\pm$ & 0&.4 & (syst.) \\
           & \mbox{$[40, 45)$} &  0 &.5 & $\pm$ &  0 &.3 & (stat.) & $^+_-$ & $^{\emptyplus 1}_{\emptyminus 0}$ & $^{.1\emptyplus}_{.5\emptyminus}$ & (syst.) & 0&.1 & $\pm$ & 0&.1 & (syst.) \\
  \hline
  \mbox{$[0.6, 1.37)$} & \mbox{$[15, 20)$} &    33 &   & $\pm$ & 9 &   & (stat.) & $\pm$ & 19 &   & (syst.) & \multicolumn{6}{c|}{same as in} \\
              & \mbox{$[20, 30)$} &     6 &   & $\pm$ & 3 &   & (stat.) & $^+_-$ & $^{\emptyplus 10}_{\emptyminus \emptynull 6}$ & & (syst.) & \multicolumn{6}{c|}{first $|\eta|$-region} \\
              & \mbox{$[30, 40)$} &     0 &.9 & $\pm$ & 0 &.5 & (stat.) & $^+_-$ & $^{\emptyplus 2}_{\emptyminus 0}$ & $^{\emptyplus}_{.9\emptyminus}$ & (syst.) & & & & & & \\
              & \mbox{$[40, 45)$} &     0 &.2 & $\pm$ & 0 &.2 & (stat.) & $^+_-$ & $^{\emptyplus 0}_{\emptyminus 0}$ & $^{.8\emptyplus}_{.2\emptyminus}$ & (syst.) & & & & & & \\
  \hline
  \mbox{$[1.52, 1.81)$} & \mbox{$[15, 20)$} &   16 &   & $\pm$ & 6 &   & (stat.) & $\pm$ & 8 &   & (syst.) & \multicolumn{6}{c|}{same as in} \\
               & \mbox{$[20, 30)$} &    1 &.4 & $\pm$ & 1 &.2 & (stat.) & $^+_-$ & $^{\emptyplus 7}_{\emptyminus 1}$ & $^{\emptyplus}_{.4\emptyminus}$ & (syst.) & \multicolumn{6}{c|}{first $|\eta|$-region} \\
               & \mbox{$[30, 40)$} & $<$0 &.1 & $^+_-$ & $^{\emptyplus 0}_{\emptyminus 0}$ & $^{.9\emptyplus}_{.0\emptyminus}$ & (stat.) & $^+_-$ & $^{\emptyplus 2}_{\emptyminus 0}$ & & (syst.) & & & & & & \\
               & \mbox{$[40, 45)$} & $<$0 &.1 & $^+_-$ & $^{\emptyplus 6}_{\emptyminus 0}$ & & (stat.) & $^+_-$ & $^{\emptyplus 1}_{\emptyminus 0}$ & $^{.7\emptyplus}_{.0\emptyminus}$ & (syst.) & & & & & & \\
  \hline
  \mbox{$[1.81, 2.37)$} & \mbox{$[15, 20)$} &   15 &   & $\pm$ & 3 &   & (stat.) & $\pm$ & 3 &   & (syst.) & \multicolumn{6}{c|}{same as in} \\
               & \mbox{$[20, 30)$} &    1 &.5 & $\pm$ & 1 &.2 & (stat.) & $^+_-$ & $^{\emptyplus 6}_{\emptyminus 1}$ & $^{\emptyplus}_{.5\emptyminus}$ & (syst.) & \multicolumn{6}{c|}{first $|\eta|$-region} \\
               & \mbox{$[30, 40)$} &    0 &.1 & $^+_-$ & $^{\emptyplus 0}_{\emptyminus 0}$ & $^{.2\emptyplus}_{.1\emptyminus}$ & (stat.) & $^+_-$ & $^{\emptyplus 2}_{\emptyminus 0}$ & $^{\emptyplus}_{.1\emptyminus}$ & (syst.) & & & & & & \\
               & \mbox{$[40, 45)$} & $<$0 &.1 & $\pm$ & 6 &   & (stat.) & $^+_-$ & $^{\emptyplus 1}_{\emptyminus 0}$ & $^{.2\emptyplus}_{.0\emptyminus}$ & (syst.) & & & & & & \\
  \hline
\end{tabular}\\
\normalsize
\caption[Multijet estimation for $\feg$] {
  Overview of the estimation of the multijet background in the \Zee and \Zeg samples.
  The estimated fraction of the multijet background is given.
  In the case of the \Zeg sample, the statistical uncertainty from the fit as well as the estimated systematic uncertainty are presented.
  For the \Zee sample, the systematic uncertainty was estimated conservatively to be as large as 100\%.
}
\label{tab:egammaFakeRate_QCD}
\end{table}

Tab.~\ref{tab:egammaFakeRate_QCD} also shows systematic uncertainties on the multijet fraction.
They were estimated by comparing the result from the fit as described above with a fit where the
signal model was replaced by template distributions retrieved from $\Zee$ simulations.
Both, the tag electron and the probe photon, were required to be matched to one of the generated electrons from the $Z$ decay
within a cone of 0.2 in $\eta$-$\phi$-space.
Fig.~\ref{fig:Template_egamma_regions} shows two example fits in the same regions as shown in Fig.~\ref{fig:CB_egamma_regions}.
Generally, the estimated multijet fractions are of the same order in both approaches, but sizable differences of up to 12\% were observed in the
low-$\et$-region.
The difference was symmetrised and taken as a systematic uncertainty.

Additionally, the upper and lower limits of the fit range were varied by \mbox{$\pm \, 2 \GeV$} in order to evaluate the stability of the
fit, and a systematic uncertainty was estimated from the maximum deviation from the central value observed.
The differences were found to be smaller than the statistical uncertainties from the fit in 15 out of the 16 $\et$-$\eta$-regions.
Only for \Zeg events with photons in the region \mbox{$0.6 \leq |\eta| < 1.37$} and \mbox{$15 \GeV \leq \et < 20 \GeV$},
a systematic uncertainty of 15\% was found given a statistical fit uncertainty of 9\%.
The systematic uncertainties from the limit variation were added in quadrature to the uncertainties retrieved from the comparison with the
alternative modelling.

\begin{figure}[p]
  \begin{center}
    \includegraphics[width=0.49\textwidth]{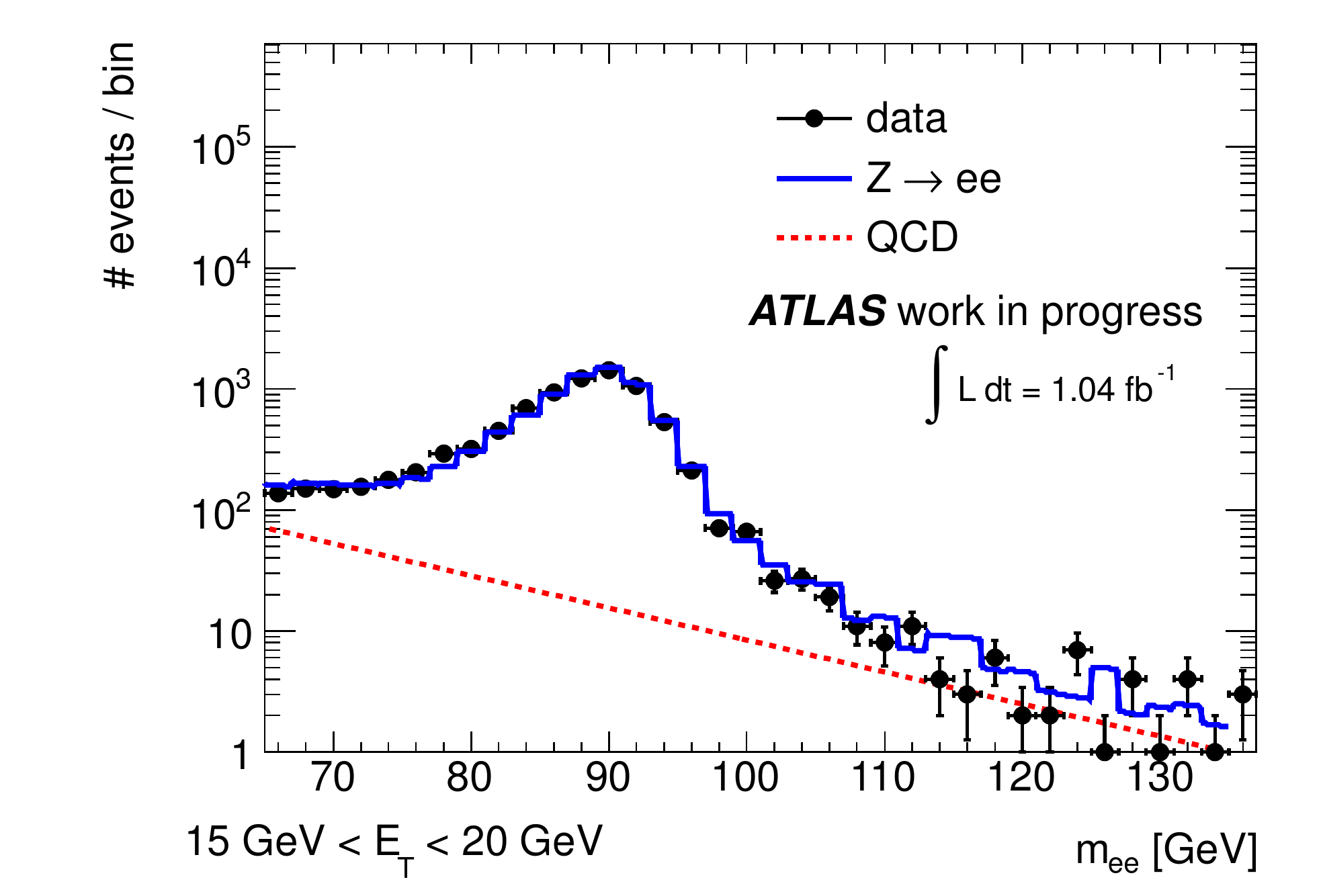}
    \includegraphics[width=0.49\textwidth]{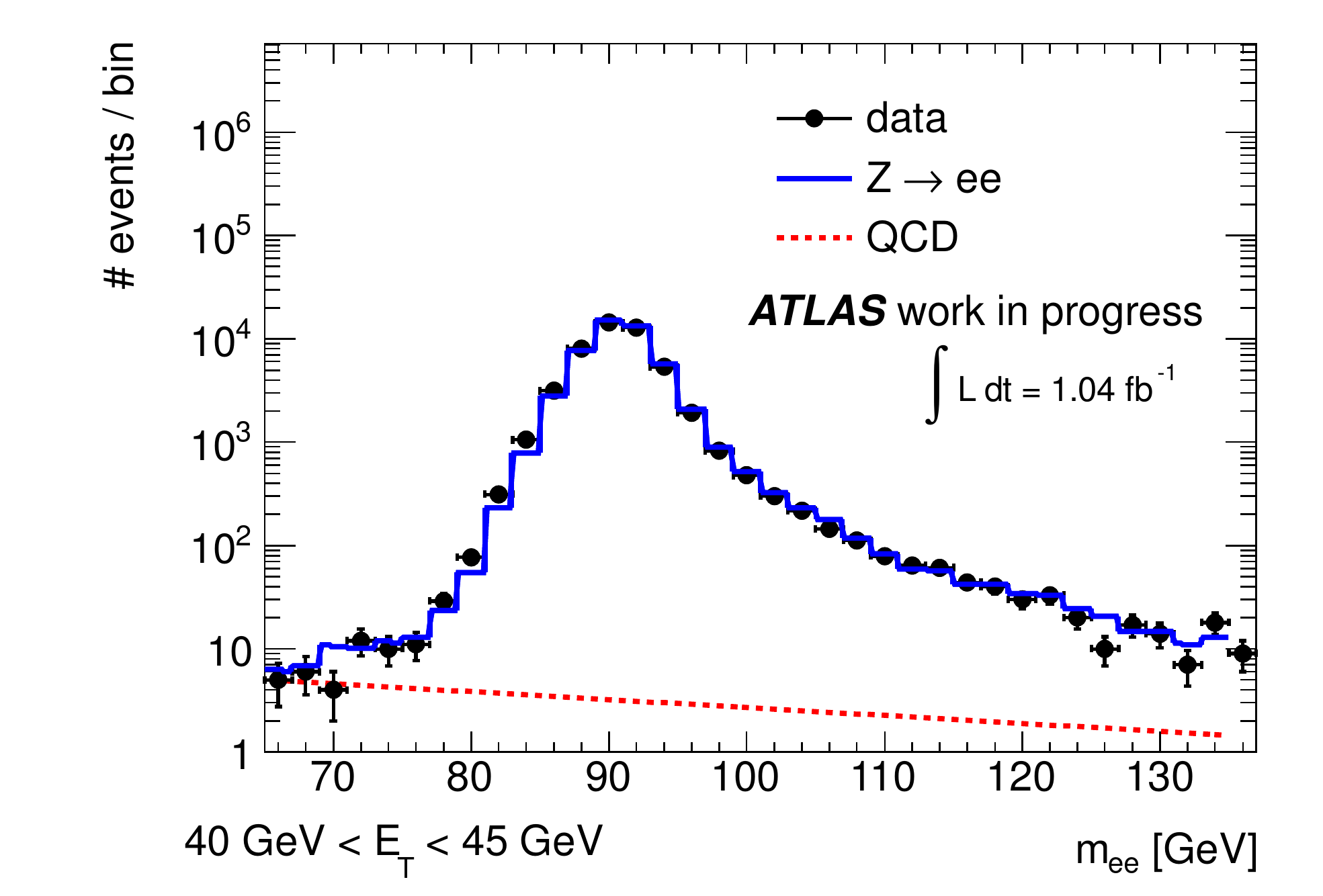}
    \caption[$m(ee)$ in data with signal and background fit in two $\et$-regions]{
      Invariant mass distributions for the two electrons from the \Zee selection in data
      for the probe electron in the region \mbox{$15 \GeV \leq \et < 20 \GeV$} (left),
      and in the region \mbox{$40 \GeV \leq \et < 45 \GeV$} (right).
      The distributions were fitted with a template derived from MC simulations (solid line), and an exponential for the background modelling
      (dashed line).
    }
    \label{fig:Template_ee_regions}
    \vspace{0.025\textwidth}
    \includegraphics[width=0.52\textwidth]{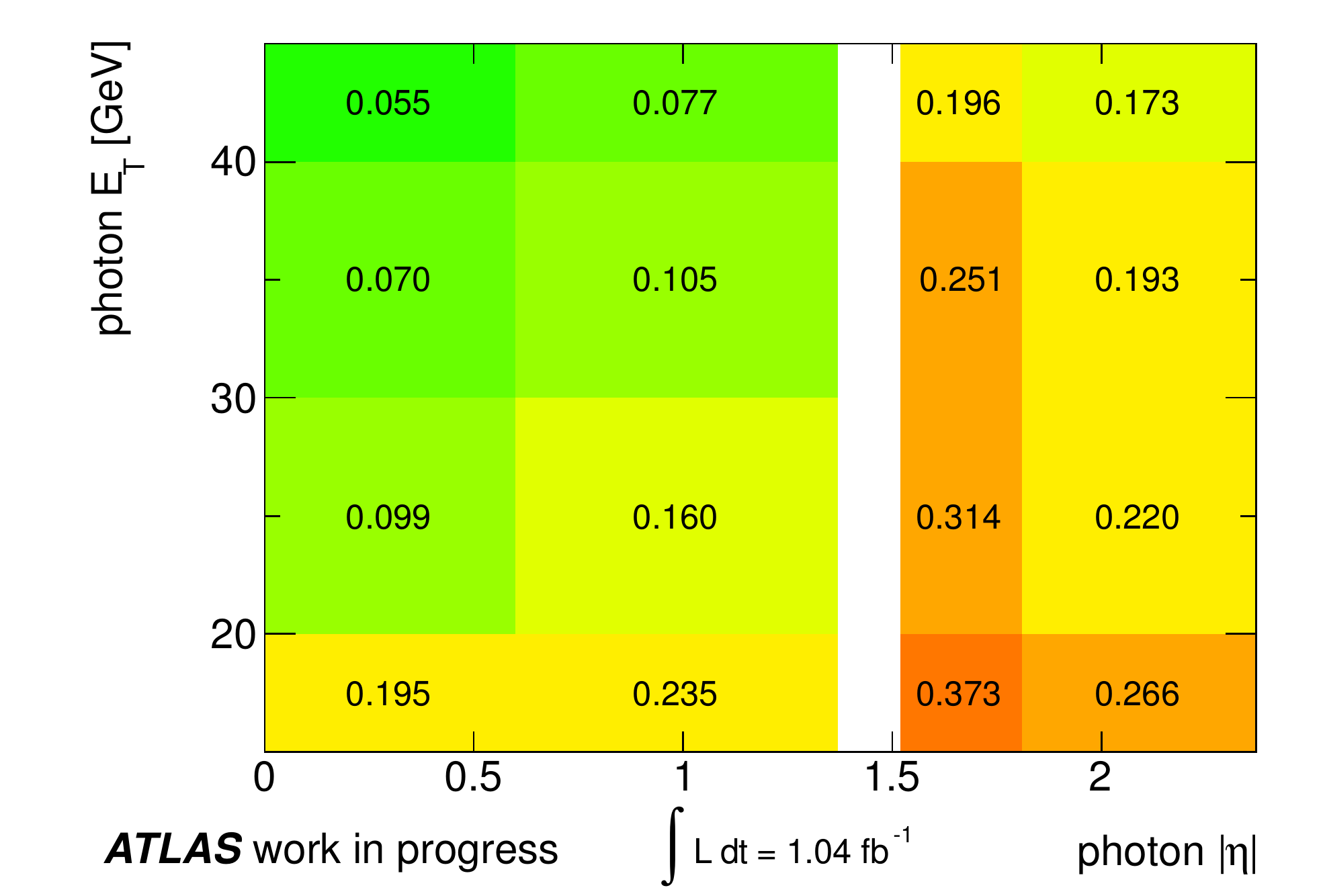}
    \caption[Ratio of events in the \Zeg and \Zee samples in data]{
      Ratio of the number of events in the \Zeg and in the \Zee sample in four bins of $\eta$ and four bins of $\et$ in data.
      The yields were corrected using the estimate for the multijet contribution.
    }
    \label{fig:neg_over_nee}
    \vspace{0.025\textwidth}
    \includegraphics[width=0.52\textwidth]{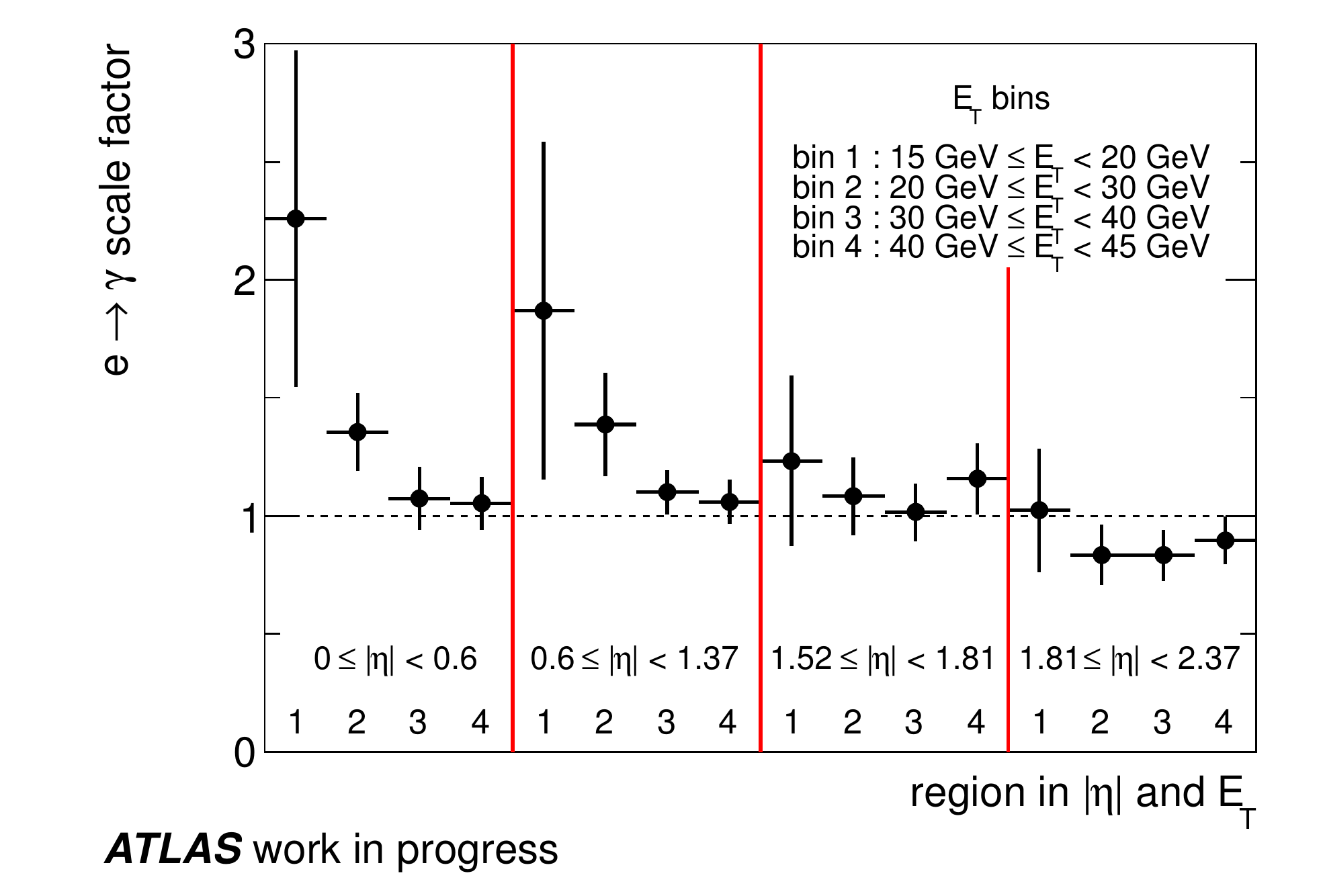}
    \caption[Scale factors for the electron misidentification rate]{
      Scale factors for the misidentification rate $\feg$ in 16 bins in $\et$ and $|\eta|$.
      The error bars show the sum of statistical and systematic uncertainties as presented in Tab.~\ref{tab:egammafakerate_uncertainties}.
    }
    \label{fig:SFs}
  \end{center}
\end{figure}

In the \Zee sample, the multijet background was estimated using the template fit method only.
The fraction of multijet events is low, even in the high- and low-mass regions, so that the kinematic bias of the $Z$ peak due to the minimal
requirements on the $\et$ of the electrons dominates and does not allow for a fit with the 
\mbox{$\mathrm{CB} \mathord{*} \mathrm{BW}$} signal model.
This is evident from the right plot in Fig.~\ref{fig:Template_ee_regions}, which shows the invariant mass distribution for the events from \Zee
sample with an $\et$ of the probe electron between \mbox{$40 \GeV$} and \mbox{$45 \GeV$}.
Clearly, the lower side of the spectrum cannot be described by a \mbox{$\mathrm{CB} \mathord{*} \mathrm{BW}$} signal model.
The template fit, however, describes the distribution reasonably well.
The fraction of multijet events was estimated in four regions of $\et$, similar to the estimate in the \Zeg sample.
An $\eta$-dependence was not accounted for, however, but a possible dependence is believed to be covered by the large systematic uncertainty assigned.
The left plot in Fig.~\ref{fig:Template_ee_regions} shows another example fit in the region \mbox{$[15 \GeV, 20 \GeV)$}.

The results of the multijet estimate in the \Zee sample are presented in Tab.~\ref{tab:egammaFakeRate_QCD}:
for low-$\et$ probe electrons, a fraction of 2.3\% was estimated.
It decreases with increasing electron $\et$.
The fraction of multijet background was found to be small in the \Zee sample.
However, also the effect on the overall estimation of $\feg$ was evaluated to be small (see below), and a conservative systematic uncertainty of 100\%
was assigned, which turned out to be negligible with respect to other systematic uncertainties affecting $\feg$.

\begin{sidewaystable}[p]
\centering
\footnotesize
\begin{tabular}[h] {|l|r@{}l|r@{}l|r@{}l|r@{}l|r@{}l|r@{}l|r@{}l|r@{}l|r@{}l|r@{}l|r@{}l|r@{}l|r@{}l|r@{}l|r@{}l|r@{}l|}
  \hline
  Source of uncertainty & \multicolumn{8}{c|}{\mbox{$0 \leq |\eta| < 0.6$}} & \multicolumn{8}{c|}{\mbox{$0.6 \leq |\eta| < 1.37$}} & \multicolumn{8}{c|}{\mbox{$1.52 \leq |\eta| < 1.81$}} & \multicolumn{8}{c|}{\mbox{$1.81 \leq |\eta| < 2.37$}} \\
  & \multicolumn{2}{c|}{1} & \multicolumn{2}{c|}{2} & \multicolumn{2}{c|}{3} & \multicolumn{2}{c|}{4} &
    \multicolumn{2}{c|}{1} & \multicolumn{2}{c|}{2} & \multicolumn{2}{c|}{3} & \multicolumn{2}{c|}{4} &
    \multicolumn{2}{c|}{1} & \multicolumn{2}{c|}{2} & \multicolumn{2}{c|}{3} & \multicolumn{2}{c|}{4} &
    \multicolumn{2}{c|}{1} & \multicolumn{2}{c|}{2} & \multicolumn{2}{c|}{3} & \multicolumn{2}{c|}{4} \\
  \hline
  Statistical      & 0&.10 &    0&.04 &    0&.03 &    0&.04
                   & 0&.07 &    0&.03 &    0&.02 &    0&.03
                   & 0&.07 &    0&.03 &    0&.02 &    0&.05
                   & 0&.05 &    0&.02 &    0&.02 &    0&.03 \\
  Multijet in \Zee & 0&.05 &    0&.01 & $<$0&.01 & $<$0&.01
                   & 0&.05 &    0&.01 & $<$0&.01 & $<$0&.01
                   & 0&.03 &    0&.01 & $<$0&.01 & $<$0&.01
                   & 0&.02 &    0&.01 & $<$0&.01 & $<$0&.01 \\
  Multijet in \Zeg & 0&.4  &    0&.04 &    0&.02 &    0&.01
                   & 0&.6  &    0&.15 &    0&.02 &    0&.01
                   & 0&.15 &    0&.08 &    0&.02 &    0&.08
                   & 0&.05 &    0&.05 &    0&.02 &    0&.05 \\
  $e$ SF (trigger) & 0&.02 &    0&.01 &    0&.01 &    0&.01
                   & 0&.02 &    0&.01 &    0&.01 &    0&.01
                   & 0&.01 &    0&.01 &    0&.01 &    0&.01
                   & 0&.01 &    0&.01 &    0&.01 &    0&.01 \\
$e$ SF (reconstruction)& 0&.03& 0&.02 &    0&.01 &    0&.01
                   & 0&.03 &    0&.02 &    0&.02 &    0&.01
                   & 0&.02 &    0&.01 &    0&.01 &    0&.02
                   & 0&.01 &    0&.01 &    0&.01 &    0&.01 \\
  $e$ SF (ID)      & 0&.4  &    0&.10 &    0&.05 &    0&.05
                   & 0&.4  &    0&.10 &    0&.05 &    0&.05
                   & 0&.2  &    0&.08 &    0&.05 &    0&.06
                   & 0&.2  &    0&.07 &    0&.04 &    0&.05 \\
  $e$ scale        & 0&.03 &    0&.02 &    0&.02 &    0&.01
                   & 0&.02 &    0&.03 &    0&.02 &    0&.02
                   & 0&.05 &    0&.05 &    0&.05 & $<$0&.01
                   & 0&.01 &    0&.02 &    0&.02 &    0&.01 \\
  $e$ resolution   & 0&.01 &    0&.01 &    0&.01 &    0&.01
                   & 0&.02 & $<$0&.01 &    0&.01 &    0&.01
                   & 0&.01 &    0&.01 &    0&.01 &    0&.02
                   & 0&.02 & $<$0&.01 &    0&.01 &    0&.02 \\
  $\gamma$ SF (ID) & 0&.3  &    0&.11 &    0&.09 &    0&.08
                   & 0&.15 &    0&.09 &    0&.06 &    0&.04
                   & 0&.14 &    0&.08 &    0&.06 &    0&.06
                   & 0&.13 &    0&.07 &    0&.05 &    0&.06 \\
  $\gamma$ scale   & 0&.05 &    0&.01 &    0&.01 &    0&.01
                   & 0&.01 &    0&.03 &    0&.03 & $<$0&.01
                   & 0&.01 &    0&.05 &    0&.06 &    0&.07
                   & 0&.01 &    0&.04 &    0&.04 &    0&.01 \\
$\gamma$ resolution& 0&.03 &    0&.02 & $<$0&.01 &    0&.02
                   & 0&.02 &    0&.01 & $<$0&.01 &    0&.02
                   & 0&.02 &    0&.02 &    0&.01 &    0&.02
                   & 0&.01 & $<$0&.01 &    0&.01 &    0&.02 \\
Pile-up conditions & 0&.2  &    0&.02 &    0&.08 &    0&.03
                   & 0&.01 &    0&.07 &    0&.01 &    0&.05
                   & 0&.16 &    0&.06 &    0&.03 &    0&.05
                   & 0&.03 &    0&.04 &    0&.07 &    0&.01 \\
  \hline
  Total            & 0&.7  &    0&.2  &    0&.1  &    0&.1
                   & 0&.7  &    0&.2  &    0&.1  &    0&.1
                   & 0&.4  &    0&.2  &    0&.1  &    0&.2
                   & 0&.3  &    0&.1  &    0&.1  &    0&.1 \\
  \hline
\end{tabular}\\
\normalsize
\caption[Uncertainties on the electron-to-photon misidentification rate] {
  Overview of the statistical and systematic uncertainties on the electron-to-photon misidentification rate for all regions in $\eta$ and $\et$.
  The $\et$ bins are labelled from 1 to 4, where the ranges are \mbox{$[15 \GeV, 20 \GeV)$}, 
  \mbox{$[20 \GeV, 30 \GeV)$},  \mbox{$[30 \GeV, 40 \GeV)$}, and \mbox{$[40 \GeV, 45 \GeV)$}.
}
\label{tab:egammafakerate_uncertainties}
\end{sidewaystable}

\subsubsection{Derivation of the scale factors}

The scale factors were calculated according to Eq.~(\ref{eq:SF}):
the number of events in the \Zeg and \Zee samples in data were corrected using the multijet background estimate derived above.
Fig.~\ref{fig:neg_over_nee} shows the ratio of the number of events in the \Zeg and the \Zee samples
\mbox{$N_{e\gamma} / N_{ee}$} in four bins of $\eta$ and four bins of $\et$ in data.

For the MC simulations, this ratio was calculated using only events in which the tag electron and the probe electron or photon,
respectively, were matched to the generated electrons from the $Z$ decay within a cone of size 0.2 in $\eta$-$\phi$-space.
Fig.~\ref{fig:SFs} shows the SFs in the 16 bins in $\et$ and $|\eta|$ together with the combined statistical and systematic uncertainty as
evaluated in the next section.

The SFs vary with the $\et$ and the $\eta$ of the probe object, and it was hence necessary to derive them as a function of the probe kinematics.
In general, the SFs are larger for lower $\et$ and close to unity for transverse energies between \mbox{$30 \GeV$} and \mbox{$45 \GeV$}.
The SFs in the lowest $\et$-bin for \mbox{$0 \leq |\eta| < 0.6$} and \mbox{$0.6 \leq |\eta| < 1.37$} are much larger than 1, but
they are also subject to large uncertainties.

The kinematic range of the probes was limited to \mbox{$45 \GeV$} in $\et$.
However, the SFs in the $\et$-bins \mbox{$[30 \GeV, 40 \GeV)$} and \mbox{$[40 \GeV, 45 \GeV)$} are compatible within the uncertainties in all $\eta$-regions,
and, hence, it is well-motivated to use the SFs for the bin \mbox{$[40 \GeV, 45 \GeV)$} also for higher transverse energies.

\subsubsection{Evaluation of systematic uncertainties}

Different sources of systematic uncertainties were considered for the $\feg$ SFs, an overview of which is given in
Tab.~\ref{tab:egammafakerate_uncertainties} together with the statistical uncertainty and the quadratic sum of all uncertainties.

The uncertainty on the multijet background in the \Zee and \Zeg were considered as well as several electron- and photon-specific uncertainties, the
evaluation of which is described in Ch.~\ref{sec:systematics}:
uncertainties on the electron SFs for triggering, reconstruction and identification, uncertainties on the electron energy scale and resolution,
and uncertainties on the photon identification, energy scale and resolution were considered.

The SFs for electrons detailed in Sec.~\ref{sec:electron} were only derived for electron transverse energies larger than \mbox{$25 \GeV$}.
However, SFs for the range \mbox{$15 \GeV$} to \mbox{$25 \GeV$} were derived for electrons without isolation requirement.
Since the efficiency of the isolation cut is as large as roughly 97\% for electrons, an additional uncertainty of 3\% was added to the SFs below
\mbox{$25 \GeV$} in order to conservatively account for the missing isolation cut.

The dependence on the pile-up conditions was evaluated by comparing $N_{ee}$ and $N_{e\gamma}$ in a low pile-up region with \mbox{1 -- 5} primary vertices
in the event to a high pile-up region with \mbox{6 -- 10} primary vertices.
The differences with respect to the nominal SFs are of the same order of magnitude as the statistical uncertainties, so the observed effect could
well originate from statistical fluctuations.
The largest difference between the low pile-up and high pile-up regime with respect to the nominal SFs was taken as a symmetric systematic uncertainty,
which is believed to be very conservative, because the MC simulations were reweighted to the pile-up conditions present during data taking.

Apart from the uncertainty on the pile-up conditions and the statistical uncertainty, the following systematic uncertainties dominate the
total SF uncertainty:
the estimate of the multijet background in the \Zeg sample, and the SFs for electron and photon identification, which feature large uncertainties
in particular for electrons and photons with low $\et$.

\section{Application to processes with two leptons in the final state}
\label{sec:egammaapplication}

The SFs for $\feg$ were applied to simulations of background events with a final state electron which was misidentified as a photon
and an additional real lepton.
Processes of interest are dileptonic decays of top quark pairs, \Zee and $Z \to \tau\tau$ events with additional jets,
single top production in association with a leptonically decaying $W$ boson, and diboson production with at least two final state leptons.

Tab.~\ref{tab:electronfake_yields} shows the resulting expectations for \mbox{$1.04 \ifb$}.
The uncertainties on the expectations include all systematic uncertainties as discussed in Ch.~\ref{sec:systematics}, in particular the uncertainty
on $\feg$ and uncertainties due to detector modelling.
Uncertainties on the MC normalisation from the uncertainties on the cross sections and the luminosity as well as uncertainties due to
limited MC statistics.
Uncertainties which affect $\feg$ as well as the selection efficiencies in the MC simulations were treated correlated.

The negative estimate in the muon channel from single top $Wt$ production is due to negative weights in the MC@NLO generation and the value
of $-0.1$ is compatible with zero within the uncertainties.
The largest contribution is from dileptonic $\ttbar$ decays, which features a larger yield in the muon channel due to the looser event selection
with respect to the electron channel (Sec.~\ref{sec:preselection}).

\begin{table}[h]
\centering
\begin{tabular}[h]{|l|r@{}l c r@{}l|r@{}l c r@{}l|}
  \hline
  Process & \multicolumn{5}{c|}{e+jets} & \multicolumn{5}{c|}{$\mu$+jets} \\
  \hline
  Dileptonic $\ttbar$ &  6 &.8  & $\pm$ & 2 &.3 &  9 &.6  & $\pm$ & 2 &.7 \\
  $Z$+jets &  1 &.7  & $^+_-$ & $^{\emptyplus 3}_{\emptyminus 1}$ & $^{.1\emptyplus}_{.7\emptyminus}$ &  0 &.7  & $^+_-$ & $^{\emptyplus 1}_{\emptyminus 0}$ & $^{.8\emptyplus}_{.7\emptyminus}$ \\
  Single top $Wt$-channel &  0 &.22  & $^+_-$ & $^{\emptyplus 0}_{\emptyminus 0}$ & $^{.25\emptyplus}_{.22\emptyminus}$ & -0 &.10  & $\pm$ & 0 & .10 \\
  Diboson &  0 &.04  & $^+_-$ & $^{\emptyplus 0}_{\emptyminus 0}$ & $^{.14\emptyplus}_{.04\emptyminus}$ &  0 &.00  & $^+_-$ & $^{\emptyplus 0}_{\emptyminus 0}$ & $^{.14\emptyplus}_{.00\emptyminus}$ \\
  \hline
\end{tabular}\\
\caption[Expected yields for processes with an electron misidentified as a photon]{
  Expected yields for processes with an electron misidentified as a photon for \mbox{$1.04 \ifb$} from MC simulations using the
  SFs for the electron-to-photon misidentification rate.
  The uncertainties quoted are the total systematic uncertainties.
}
\label{tab:electronfake_yields}
\end{table}

\chapter{Background events with prompt photons in the final state}
\label{sec:backgroundphotons}

The treatment of hadrons and electrons misidentified as photons has been discussed in the previous chapters.
The remaining backgrounds are processes with prompt photons in the final state.

Sec.~\ref{sec:ttgbkg} discusses the small background contribution from $\ttg$ production which does not fulfil the signal phase space cuts.
Background estimates of multijet and $W$+jets production with an additional prompt photon in the final state were derived with data-driven
approaches as described in Sec.~\ref{sec:QCDgamma} and~\ref{sec:Wjetsgamma}, respectively.
The remaining background contributions from $Z$+jets, single top and diboson production in association with a prompt photon were estimated
using MC simulations (Sec.~\ref{sec:restgamma}).

\section[$\ttg$ production outside of the signal phase space]{\boldmath$\ttg$ production outside of the signal phase space\unboldmath}
\label{sec:ttgbkg}

The $\ttg$ signal is only well-defined within a certain phase space, as for example defined by the invariant mass cuts required in
the signal modelling in this analysis (Sec.~\ref{sec:signalmodelling}).
In rare cases, however, $\ttg$ events outside of this signal phase space may fulfil the event selection.

\begin{figure}[h!]
\begin{center}
\includegraphics[width=0.6\textwidth]{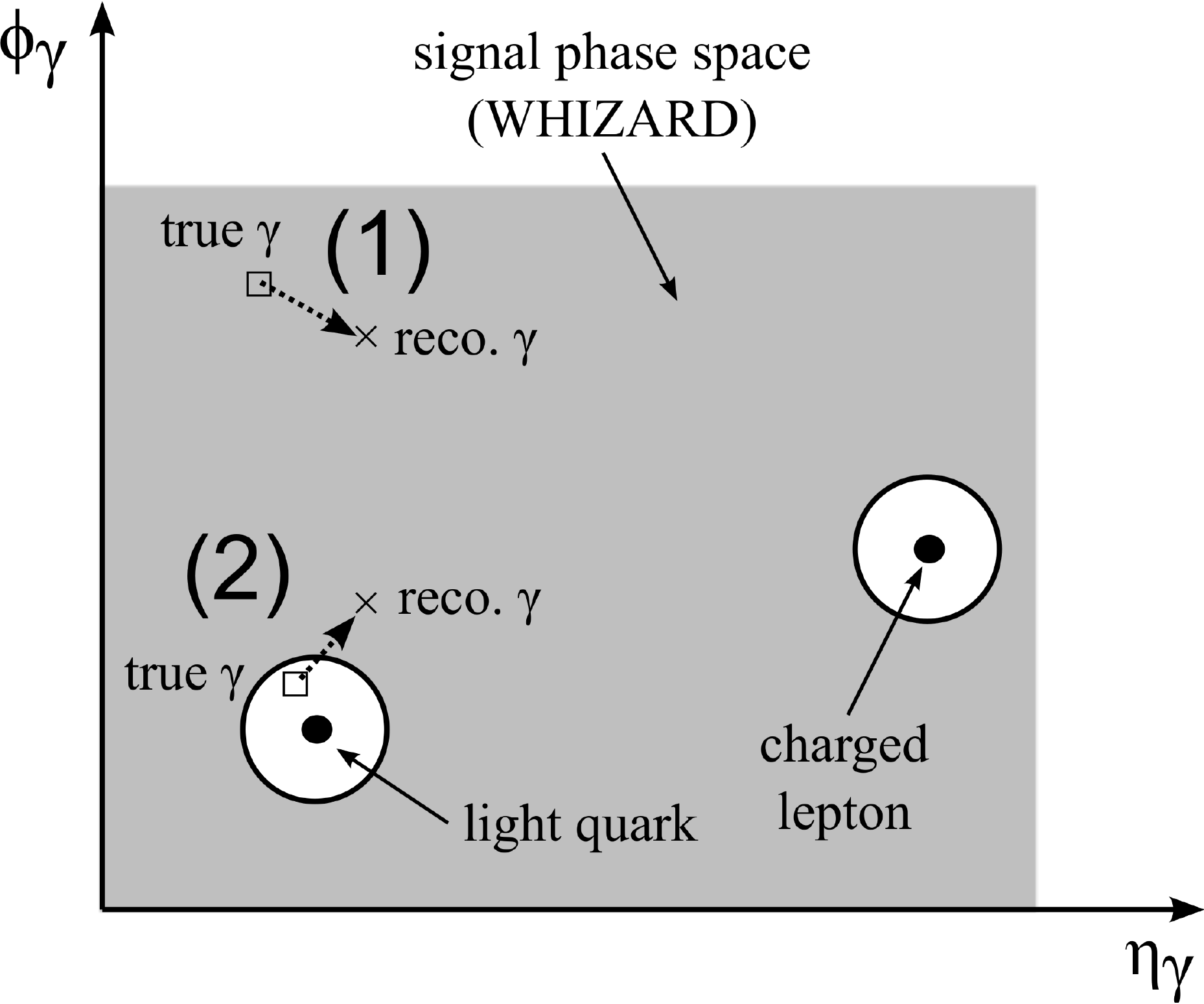}
\caption[Illustration of photons within and outside of the $\ttg$ signal phase space]{
  Illustration of photons within and outside of the $\ttg$ signal phase space:
  the plot shows the angular phase space in photon $\eta$ and $\phi$.
  Regions around charged massless particles, such as light quarks and charged leptons, are forbidden for true photons.
  True and the corresponding reconstructed photons are indicated by open squares and crosses, respectively.
  Case (1) shows a photon from the $\ttg$ signal phase space, while case (2) shows a photon from a background $\ttg$ event.
}
\label{fig:MCatNLO_WHIZARD_illustration}
\end{center}
\end{figure}

This is illustrated in Fig.~\ref{fig:MCatNLO_WHIZARD_illustration}, which shows a schematic representation of the angular phase space in $\eta$
and $\phi$ for photons in $\ttbar$ events.
Regions around charged massless fermions, such as light quarks and charged leptons, are forbidden for true photons, because
the invariant mass cuts applied in the WHIZARD generation effectively translate into minimal angular distances:
\begin{eqnarray*}
  m^2 & = & \left( p_f + p_\gamma \right)^2 = 2 \cdot E_f \cdot E_\gamma \cdot \left( 1 - \cos \alpha \right) >
  m_{\mathrm{cut}}^2 \\
  \Rightarrow \alpha & > & \arccos \left( 1 - \frac{m_{\mathrm{cut}}^2}{2 \cdot E_f \cdot E_\gamma} \right) \, ,
\end{eqnarray*}
with $\alpha$ the angle between the charged massless fermion and the photon.
The minimal value for $\alpha$ depends on the fermion and photon energies.

True photons and the corresponding reconstructed photons are indicated by open squares and crosses in Fig.~\ref{fig:MCatNLO_WHIZARD_illustration},
respectively.
Two cases are depicted:
case (1) shows a photon from the $\ttg$ signal phase space, while case (2) shows a photon from a \textit{background $\ttg$} event:
the true photon lies outside of the signal phase space, because it is too close to a light quark.
However, the position in $\eta$-$\phi$-space of the reconstructed photon
differs from the position of the true photon and lies actually within the measured signal phase space.

In order to estimate the contribution from background $\ttg$ events, $\ttbar$ events simulated with MC@NLO were used:
the $\ttbar$ sample also contains photon radiation in the phase space not generated with WHIZARD through QED corrections provided
by the PHOTOS package~\cite{photos}.
However, interference effects are not correctly taken into account and it is hence indicated to assign a conservative systematic uncertainty to the
$\ttg$ estimates obtained from this sample.

The MC@NLO sample does not only contain $\ttg$ events outside of the signal phase space, but also events within the latter.
In order to avoid double-counting, the overlap with the WHIZARD $\ttg$ sample was removed from the MC@NLO
$\ttbar$ simulation, as discussed in Sec.~\ref{sec:backgroundmodelling}.
After the overlap removal, an estimate of 0.8 and 1.3 background $\ttg$ events was derived for \mbox{$1.04 \ifb$} of data in the single electron and single muon
channels, respectively\footnote{Only reconstructed photons were considered which were matched to a true photon with a minimal $\pt$ of \mbox{$10 \GeV$}.}.

The origin of these events was found to be due to collinear radiation from leptons and quarks.
In the following, it is first shown that the event selection applied in this analysis is well-suited to select only events from the phase space
generated in the $\ttg$ simulation, and that the invariant mass cuts applied in the event generation were hence not chosen too loosely.
Then, it is shown that background $\ttg$ events are indeed due to collinear radiation from leptons and quarks.

\begin{figure}[h!]
\begin{center}
\includegraphics[width=0.49\textwidth]{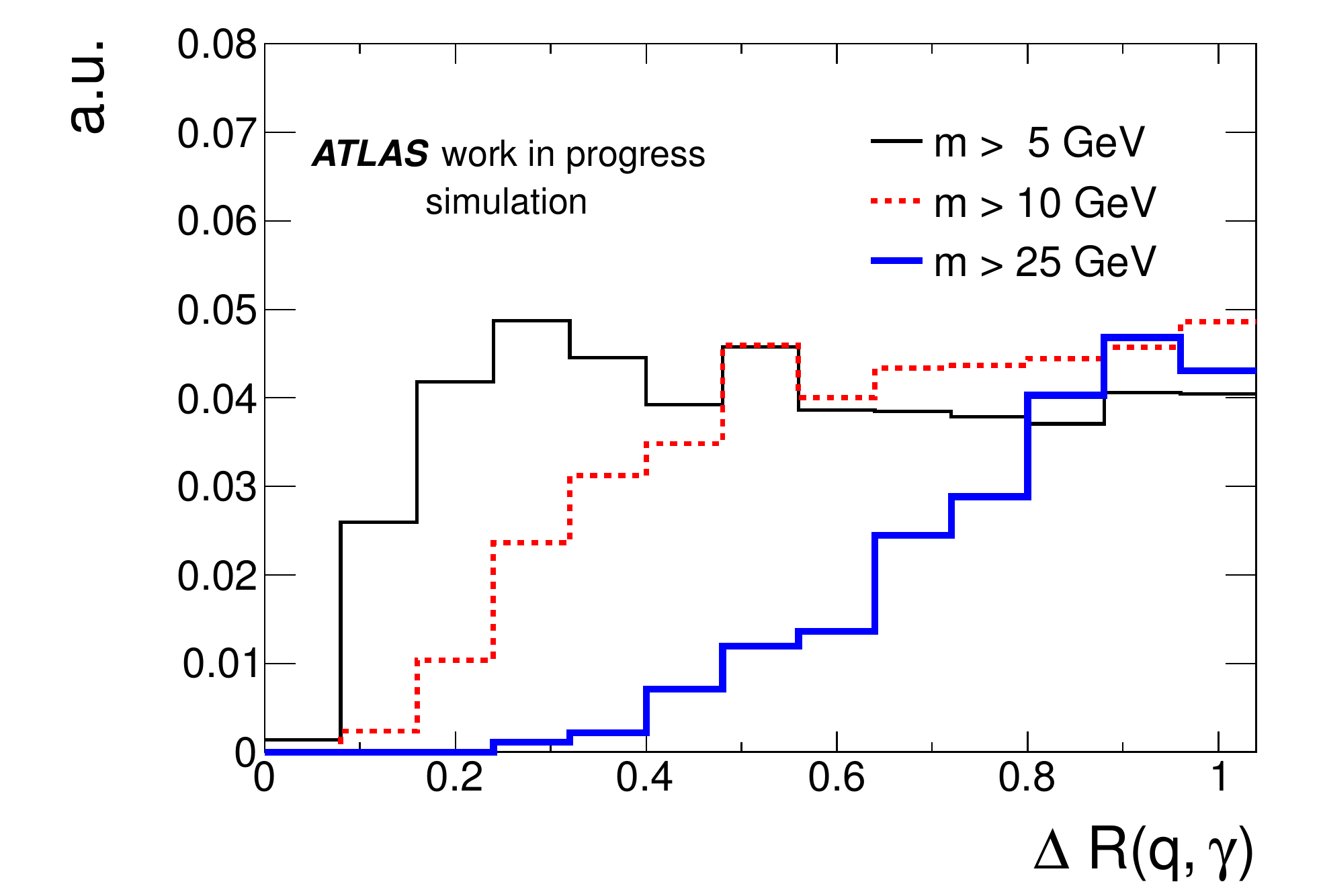}
\caption[$\Delta R$ between quarks and true photons in WHIZARD $\ttg$ simulations]{
  Smallest distance in $\eta$-$\phi$-space between light quarks and true photons for different invariant mass cuts
  $m(q, \gamma)$ in WHIZARD $\ttg$ simulations: \mbox{$5 \GeV$} (thin solid line), \mbox{$10 \GeV$} (dashed line), and \mbox{$25 \GeV$} (thick solid line).
}
\label{fig:whizard_dRqgamma}
\vspace{0.025\textwidth}
\includegraphics[width=0.49\textwidth]{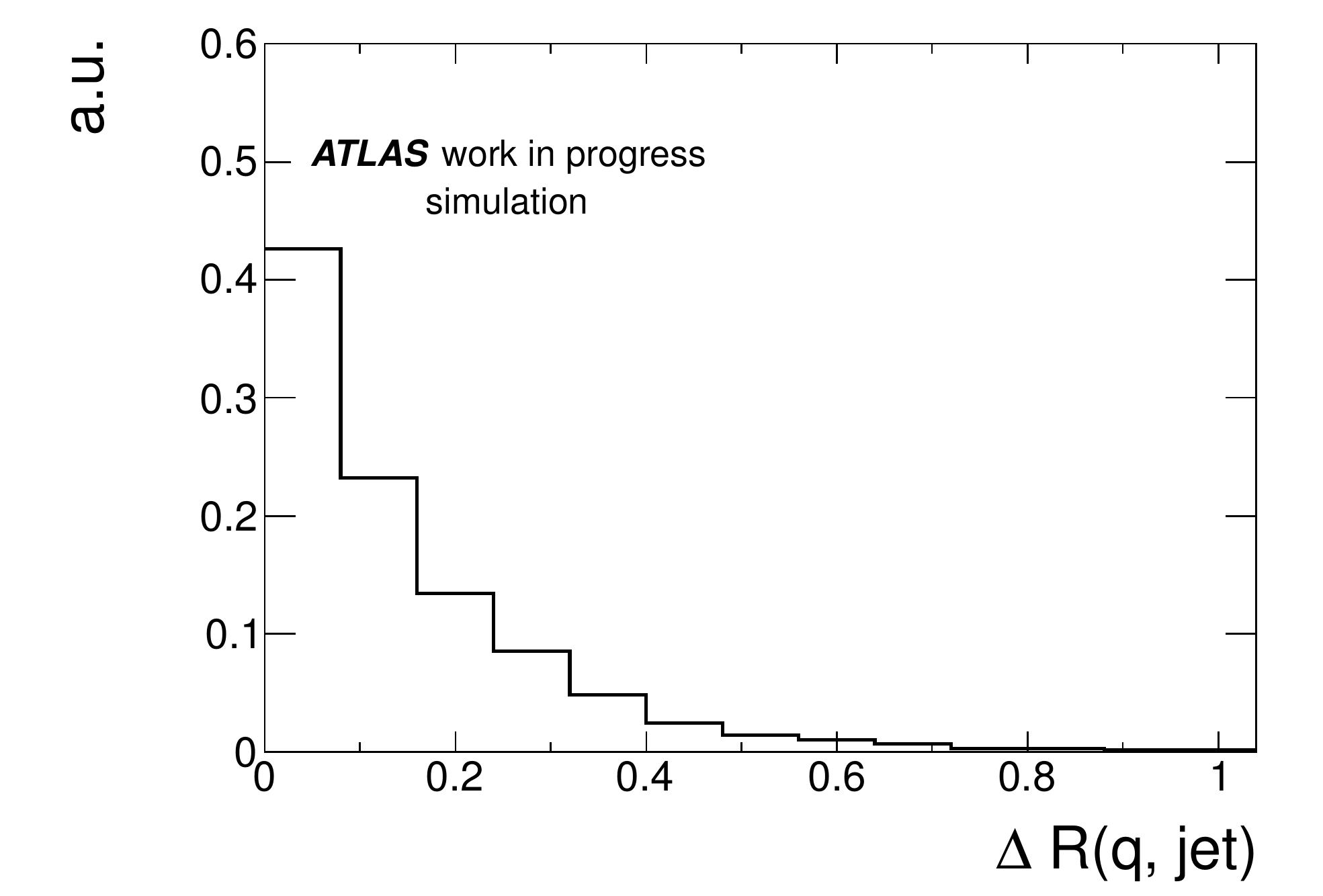}
\includegraphics[width=0.49\textwidth]{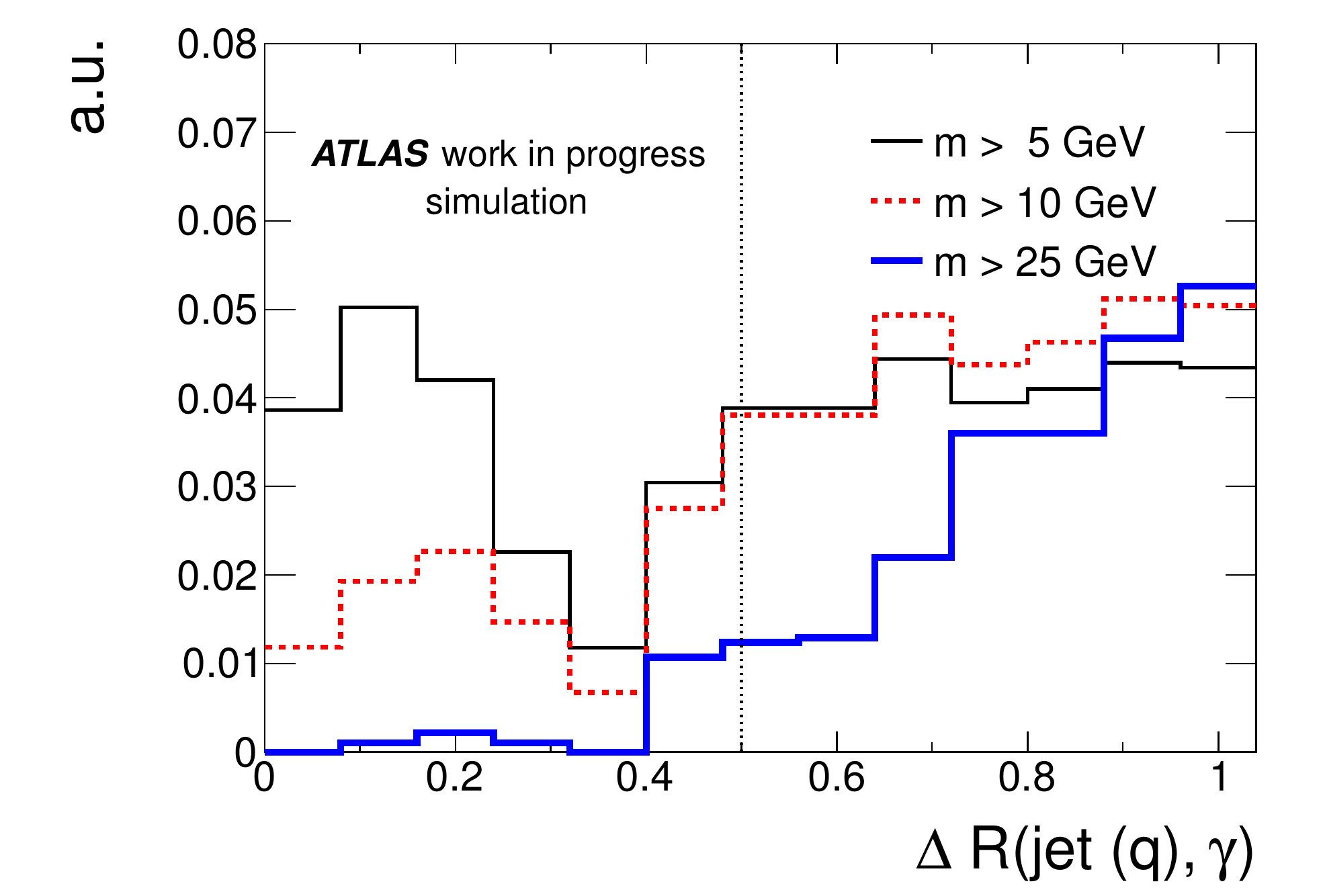}
\caption[$\Delta R$ between jets and true photons in WHIZARD $\ttg$ simulations]{
  The left plot illustrates the angular resolution of jets: the smallest distance in $\eta$-$\phi$-space between a light quark and the closest
  jet is shown.
  The right plot shows the smallest distance in $\eta$-$\phi$-space between jets matched to a light quark and true photons for different
  invariant mass cuts   $m(q, \gamma)$ in WHIZARD $\ttg$ simulations: \mbox{$5 \GeV$} (thin solid line), \mbox{$10 \GeV$} (dashed line), and
  \mbox{$25 \GeV$}
  (thick solid line).
  The dotted vertical line shows the minimal $\Delta R$ applied between jets and reconstructed photons in the event selection.
}
\label{fig:whizard_dRjetgamma}
\end{center}
\end{figure}

\subsubsection{Check of the invariant mass criteria}

The contribution of background $\ttg$ events was found to be as large as roughly 4\% of the yields predicted by the WHIZARD $\ttg$
simulation (Sec.~\ref{sec:yields}).
This could indicate that the signal phase space generated with WHIZARD is slightly too small for the considered event selection.

Fig.~\ref{fig:whizard_dRqgamma} shows the smallest distance in $\eta$-$\phi$-space between light quarks and true photons for different invariant mass
cuts $m(q, \gamma)$ of \mbox{$5 \GeV$}, \mbox{$10 \GeV$}, and \mbox{$25 \GeV$}.
For all three invariant mass cuts, a turn-on at a certain typical $\Delta R$ is observed.
The turn-on has a finite width because of the energy spectra of quarks and photons.
As expected, the turn-on value increases for larger invariant mass cuts.

In order to study the effect of the generation level cuts on the angular distances of reconstructed objects, the angular resolution needs to be
considered.
Jets are expected to feature the worst angular resolution of the considered objects:
the left plot in Fig.~\ref{fig:whizard_dRjetgamma} shows the distance between a light quark and the closest jet.
The angular resolution is of the order of only 0.15 in $\Delta R$.

The right plot in Fig.~\ref{fig:whizard_dRjetgamma} shows the smallest distance in $\eta$-$\phi$-space between jets and true photons, again for
invariant mass cuts of \mbox{$5 \GeV$}, \mbox{$10 \GeV$} and \mbox{$25 \GeV$}.
The jets were required to be matched to a light quark within \mbox{$\Delta R < 0.3$}.
For large values of $\Delta R$, the same trend as in Fig.~\ref{fig:whizard_dRqgamma} is observed with slightly broadened turn-on curves
due to the angular resolution.
At values of \mbox{$\Delta R < 0.4$}, a peak is observed which is more pronounced for the lower invariant mass cuts than for the cut at \mbox{$25 \GeV$}.
This peak is due to true photons which are close to a quark, which causes the photon and the jet to merge and form one single jet.
At values of \mbox{$\Delta R > 0.4$}, photons and jets are well separated, which is consistent with the distance parameter of 0.4 used for the
anti-$k_t$ jet algorithm (Sec.~\ref{sec:jet}).

Generally, the direction of reconstructed photons does not differ significantly from the direction of the corresponding true photon.
Hence, in order to avoid biases due to jet-photon merging, reconstructed photons were required to be separated by more than 0.5 in $\eta$-$\phi$-space
(Sec.~\ref{sec:photon}), as indicated by a dotted vertical line in the right plot of Fig.~\ref{fig:whizard_dRjetgamma}.
As seen in the plot, the cut value of 0.5 is well above the turn-on for an invariant mass cut of \mbox{$5 \GeV$}.

Hence, the angular cuts applied in the event selection appear to be sufficient to select only events from the signal phase space.
This does not only hold for the distances between jets and photons, but also for electrons and muons, which are separated from photons by
minimal distances in $\eta$-$\phi$-space (Sec.~\ref{sec:photon}).
It is therefore concluded that the origin of the background $\ttg$ events is not due to inappropriately high invariant mass cuts in the signal
definition.

\subsubsection{Photon radiation from charged leptons}

\begin{figure}[h]
\begin{center}
\includegraphics[width=0.49\textwidth]{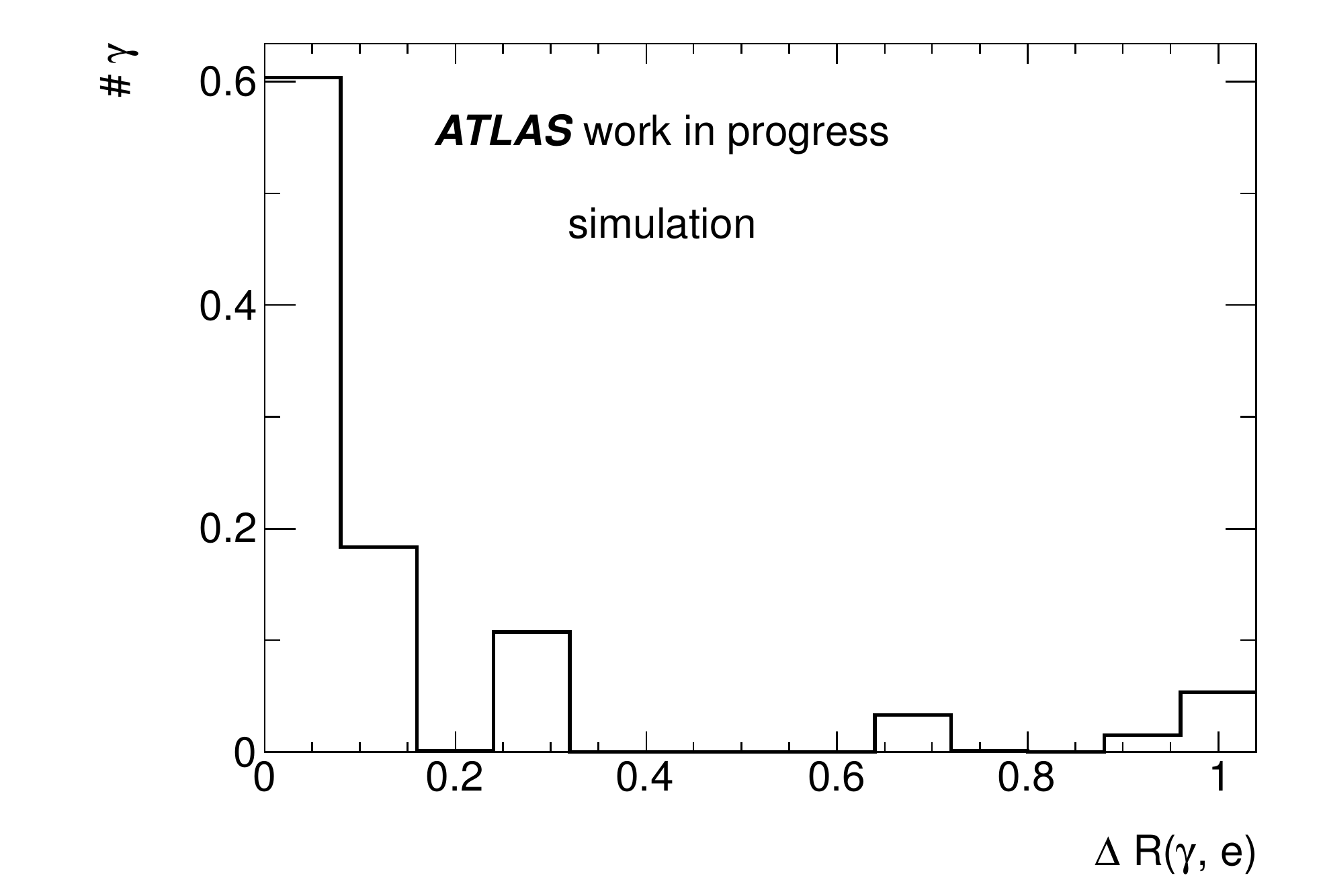}
\includegraphics[width=0.49\textwidth]{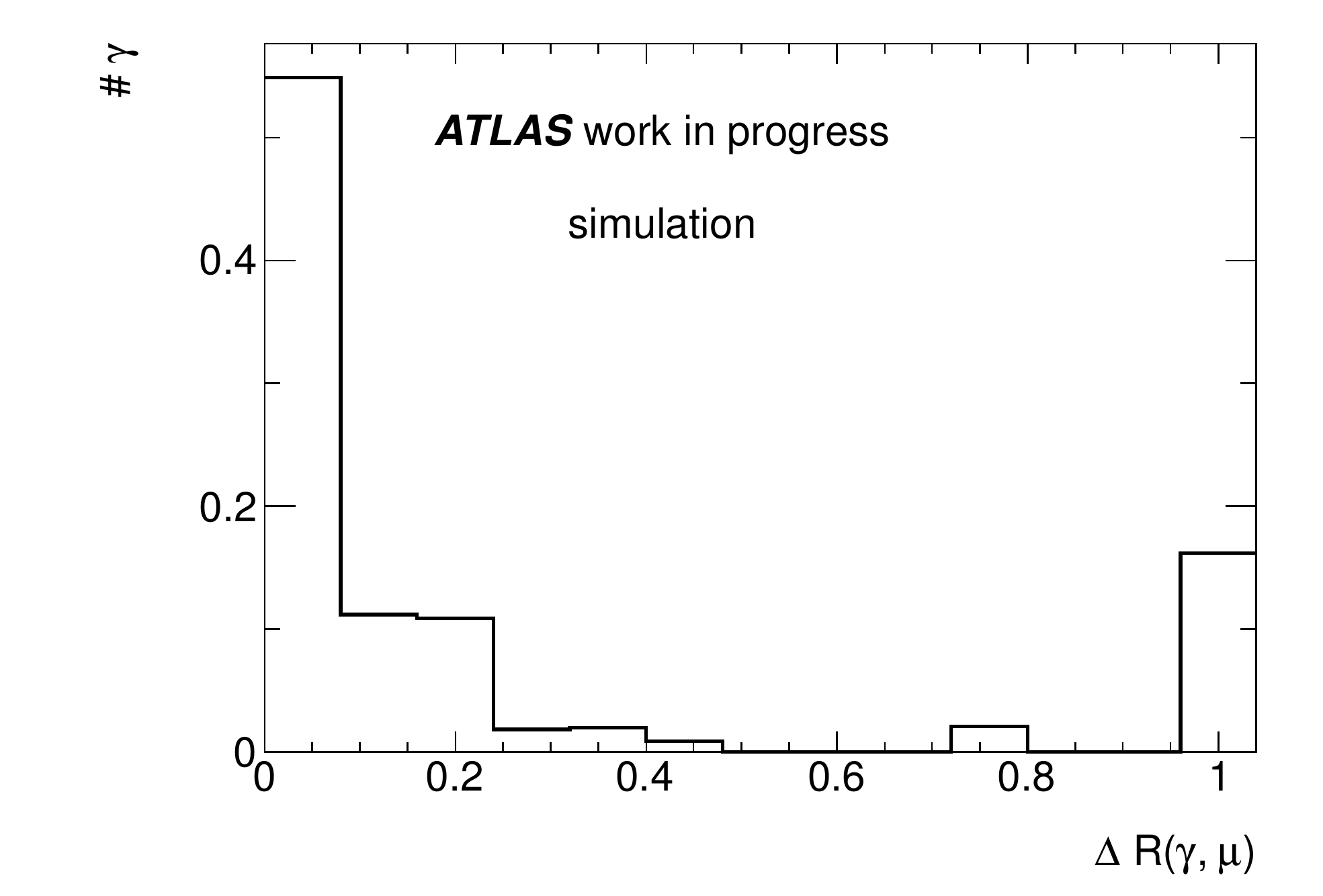}
\caption[Distance between true photons and the closest lepton for the $\ttg$ background]{
  Distance in $\eta$-$\phi$-space between true photons and the closest lepton in
  $\ttg$ background events from MC simulations with MC@NLO in the electron channel (left) and in the muon channel (right).
  In both plots, the last bin includes the overflow bin.
}
\label{fig:fsr}
\end{center}
\end{figure}

Fig.~\ref{fig:fsr} shows the distance in $\eta$-$\phi$-space between true photons and the closest true lepton in $\ttg$ background events
in the electron (left plot) and in the muon channel (right plot), respectively.
Photons at low $\Delta R$ values were radiated from the corresponding lepton, as checked with MC truth information.

Experimentally, photons and leptons have a minimal distance in $\eta$-$\phi$-space (Sec.~\ref{sec:photon}).
Hence, events with collinearly radiated photons can only pass the $\ttg$ event selection if the corresponding lepton was not identified as a lepton
object according to Sec.~\ref{sec:electron} or~\ref{sec:muon}.
These events are mostly dileptonic $\ttbar$ decays with one well-identified lepton and one lepton with a collinear photon radiation.
They are a natural part of the $\ttg$ background, because such events were not simulated with WHIZARD due to collinear divergences in the ME
calculation.

A summary of the contributions from radiation from electrons, muons and $\tau$-leptons is presented in Tab.~\ref{tab:backgroundttbar}:
most photons are radiated from electrons.
The radiation from muons has a smaller contribution.
Radiation from $\tau$-leptons contributes less than 0.1 events in both channels.
Altogether, collinear radiation from charged leptons represents 0.7 and 1.1 background $\ttg$ events in the electron and muon channel, respectively.
The systematic uncertainty on these contributions was conservatively estimated to be as large as 100\%, i.e. 0.7 and 1.1 events.

\subsubsection{Photons from jet fragmentation}

Background $\ttg$ events in single lepton $\ttbar$ decays are mostly due to photon production in jet fragmentation processes, called
\textit{bremsstrahlung} in the following.
Bremsstrahlung is partly covered by the WHIZARD simulation in cases where the final state photon fulfils the signal phase space criteria.
Contributions from outside of the signal phase space were estimated using photon radiation in the MC@NLO $\ttbar$ simulation.
For \mbox{$1.04 \ifb$}, a contribution of about 0.3 events in both lepton channels together was estimated (Tab.~\ref{tab:backgroundttbar}).

In Sec.~\ref{sec:backgroundmodelling}, ambiguities in the removal of the signal phase space from the MC@NLO simulation due to bremsstrahlung processes
were discussed:
there are several possibilities for the choice of the light quark to be used in the requirement on the invariant mass of light quarks and final
state photons.
In order to assign a systematic uncertainty arising from these ambiguities, different possibilities for the choice of the light quark were considered
(see Fig.~\ref{fig:MCoverlap} for an illustration of the particles with the different status codes):
\begin{itemize}
\item Variation (a) : light quark with status code 123/124
\item Variation (b) : light quark with status code 143/144
\item Variation (c) : light quark with status code 2
\end{itemize}
The yields for variations (a)~--~(c) are shown in Tab.~\ref{tab:bremsstrahlung} together with the default case, for which
the four-momenta of all particles in the parton shower were added up except for the photon and the combined four-momentum was used as an estimate for
the quark four-momentum after photon radiation.
The expected yields for \mbox{$1.04 \ifb$} vary, and the largest yield was considered as a conservative estimate of the systematic uncertainty.

\begin{table}[h]
  \center
  \begin{tabular}{|l|r@{}l c r@{}l|r@{}l c r@{}l|}
    \hline
    & \multicolumn{5}{c|}{e+jets} & \multicolumn{5}{c|}{$\mu$+jets} \\
    \hline
    Radiation from electrons & 0&.50 & $\pm$ & 0&.50 & 0&.74 & $\pm$ & 0&.74 \\
    Radiation from muons & 0&.16 & $\pm$ & 0&.16 & 0&.23 & $\pm$ & 0&.23 \\
    Radiation from $\tau$-leptons & 0&.04 & $\pm$ & 0&.04 & 0&.09 & $\pm$ & 0&.09 \\
    Bremsstrahlung & 0&.08 & $^+_-$ & $^{\emptyplus 0}_{\emptyminus 0}$ & $^{.37\emptyplus}_{.08\emptyminus}$ & 0&.20 & $^+_-$ & $^{\emptyplus 0}_{\emptyminus 0}$ & $^{.69\emptyplus}_{.20\emptyminus}$ \\
    \hline
    Sum & 0&.8 & $^+_-$ & $^{\emptyplus 1}_{\emptyminus 0}$ & $^{.1\emptyplus}_{.8\emptyminus}$ & 1&.3 & $^+_-$ & $^{\emptyplus 1}_{\emptyminus 1}$ & $^{.9\emptyplus}_{.3\emptyminus}$ \\
%    \hline
%    Sum without radiation from electrons & 0&.3 & $^+_-$ & $^{\emptyplus 0}_{\emptyminus 0}$ & $^{.5\emptyplus}_{.3\emptyminus}$ & 0&.5 & $^+_-$ & $^{\emptyplus 1}_{\emptyminus 0}$ & $^{.0\emptyplus}_{.5\emptyminus}$ \\
    \hline
  \end{tabular}
  \caption[Overview of the different $\ttg$ background contributions]{
    Overview of the different contributions from background $\ttg$ events estimated in $\ttbar$ events simulated with MC@NLO for \mbox{$1.04 \ifb$}
    together with the associated systematic uncertainties.
  }
  \label{tab:backgroundttbar}
  \vspace{0.025\textwidth}
  \begin{tabular}{|l|r@{}l|r@{}l|}
    \hline
    Bremsstrahlung definition & \multicolumn{2}{c|}{e+jets} & \multicolumn{2}{c|}{$\mu$+jets} \\
    \hline
    Variation (a) & 0&.37 & 0&.69 \\
    Variation (b) & 0&.05 & 0&.16 \\
    Variation (c) & 0&.15 & 0&.38 \\
    \hline
    Default & 0&.08 $^{+0.37}_{-0.08}$ & 0&.20 $^{+0.69}_{-0.20}$ \\
    \hline
  \end{tabular}
  \caption[Overview of systematic variations for background $\ttg$ contributions from bremsstrahlung]{
    Overview of the systematic variations for the background $\ttg$ contribution from bremsstrahlung
    for \mbox{$1.04 \ifb$} estimated in $\ttbar$ events simulated with MC@NLO.
    For the default estimate, the resulting systematic uncertainty is shown.
  }
  \label{tab:bremsstrahlung}
\end{table}

\subsubsection{Estimate and systematic uncertainty}

The total prediction of background $\ttg$ events reads 0.8 events in the electron and 1.3 events in the muon channel.
%not considering photons radiated from electrons as discussed above.
The systematic uncertainties on the different contributions are added linearly to conservatively account for correlations between the different
sources of background $\ttg$ events.
The uncertainties add up to 1.1 events in the electron and 1.9 events in the muon channel, respectively, leading to background $\ttg$ estimates of
\mbox{$0.8 ^{+1.1}_{-0.8}$} and \mbox{$1.3 ^{+1.9}_{-1.3}$}.

\section[Multijet production with a prompt photon]{Multijet production with a prompt photon}
\label{sec:QCDgamma}

Multijet events do not feature prompt electrons or muons, but may involve the production of prompt photons such as in $\gamma$+jet production,
or may feature real photons originating from jet fragmentation.
The background from multijet production with prompt photons was hence estimated in two steps:
in a first step, the \textit{matrix method} was used to estimate the contribution from processes which lead to good electron and good muon objects after
the preselection (Sec.~\ref{sec:preselection}), so-called \textit{fake leptons}.
Examples for such processes are leptonic decays of heavy flavour mesons, and hadrons from jet fragmentation, which may be misidentified as electrons
if a large fraction of the energy is deposited in the electromagnetic calorimeter.

The matrix method, discussed in detail in the following,
provided a sample of events which was used to estimate the yield as well as differential distributions of the multijet background.
Hence, in a second step, the final event selection (Sec.~\ref{sec:finalselection}) was applied to this sample, which yielded a sample of events
with a fake lepton and a photon candidate.
The fraction of real photons within these candidate photons was estimated using a template fit to the $\ptcone$ distribution as described in
Ch.~\ref{sec:strategy}.
The templates for prompt photons and hadrons misidentified as photons as derived in Ch.~\ref{sec:photontemplate} and~\ref{sec:faketemplate}
were used for this fit.

\subsubsection{Fake lepton contribution estimated with the matrix method}

The matrix method is a data-driven technique to estimate the amount of fake leptons with a minimised dependence on MC simulations.
It was successfully used for top quark analyses in the single lepton decay channel at ATLAS, cf. for example
Ref.~\cite{topchargeATLAS, topXsecATLAS, topmassATLAS, whelicityATLAS, spincorrATLAS, asymATLAS}.
The method is based on an additional \textit{loose} lepton definition compared to the \textit{tight} lepton definition used for the actual analysis.
By replacing the tight by the loose definition, a \textit{loose event selection} is obtained, which defines a \textit{loose data sample}.

The tight definitions for electrons and muons were those presented in Sec.~\ref{sec:electron} and~\ref{sec:muon}.
For the loose electron definition, the \texttt{tight} electron menu was replaced by the \texttt{medium} menu with looser shower shape requirements.
Additionally, a hit in the Pixel $b$-layer was required to reject backgrounds from converted photons.
The electron isolation cut was loosened to \mbox{$6 \GeV$} instead of \mbox{$3.5 \GeV$}.
For the loose muon definition, the requirements on the track and calorimeter isolations were disregarded.

The number of events in the loose sample $N_L$ consists of $N_L^{\mathrm{real}}$ events with real leptons and $N_L^{\mathrm{fake}}$ events with fake leptons:
\begin{equation}
  N_L = N_L^{\mathrm{real}} + N_L^{\mathrm{fake}} \, .
  \label{eq:mm1}
\end{equation}
The same holds for the number of events in the tight sample $N_T$:
\begin{equation}
  N_T = N_T^{\mathrm{real}} + N_T^{\mathrm{fake}} = \varepsilon \cdot N_L^{\mathrm{real}} + f \cdot N_L^{\mathrm{fake}} \, ,
  \label{eq:mm2}
\end{equation}
where $\varepsilon$ and $f$ are the probabilities for real and fake leptons in the loose sample to also fulfil the tight lepton definition.
If $\varepsilon$ and $f$ are known in addition to $N_L$ and $N_T$, the number of events with fake leptons in the tight sample $N_T^{\mathrm{fake}}$
can be calculated by solving Eq.~(\ref{eq:mm1}) and Eq.~(\ref{eq:mm2}).
The solution reads:
\begin{eqnarray*}
N_T^{\mathrm{fake}} & = & f \cdot N_L^{\mathrm{fake}} = f \cdot \frac{\varepsilon \cdot N_L - N_T}{\varepsilon - f}
= f \cdot \frac{\varepsilon \cdot \left( N_T + N_L^{\mathrm{non-tight}} \right) - N_T}{\varepsilon - f} \\
& = & \mathrm{w}_L^{\mathrm{non-tight}} \cdot N_L^{\mathrm{non-tight}} + \mathrm{w}_T \cdot N_T \; ,
\end{eqnarray*}
with
\begin{equation*}
\mathrm{w}_L^{\mathrm{non-tight}} = \frac{f \varepsilon}{\varepsilon - f} \; , \; {\rm and} \quad
\mathrm{w}_T = - f \cdot \frac{1 - \varepsilon}{\varepsilon - f} \; .
\end{equation*}

Not only the total yield of multijet events was estimated, but also kinematic distributions of the multijet background
were extracted by weighting events in the loose sample:
the weight $\mathrm{w}_T$ was applied to events which fulfilled the tight event selection;
the weight $\mathrm{w}_L^{\mathrm{non-tight}}$ was applied to events which fulfilled only the loose criteria, but not the tight criteria.
Typically the following relations hold:
\begin{equation*}
  \mathrm{w}_T < 0 \, , \quad \mathrm{w}_L^{\mathrm{non-tight}} > 0 \, , \quad \left| \mathrm{w}_T \right| \ll \left| \mathrm{w}_L^{\mathrm{non-tight}} \right| \, .
\end{equation*}

The efficiency for real leptons $\varepsilon$ was measured in \mbox{$Z \to l^+l^-$} events using the tag-and-probe method for electrons and muons.
The fake efficiency $f$ was estimated in a CR dominated by multijet production:
\begin{eqnarray*}
5 \GeV < \met < 20 \GeV \quad & {\rm (e+jets)} \; , \\
\mtw < 20 \GeV \quad {\rm and} \quad \met + \mtw < 60 \GeV \quad & {\rm (}\mu{\rm +jets)} \; .
\end{eqnarray*}
Contributions from processes with real leptons, as for example from $W$+jets and $Z$+jets production, were subtracted from the yields in the CR.

In the electron channel, real and fake efficiencies were parametrised as a function of the electron $\eta$ in order to account for the
varying detector geometry.
In the muon channel, the efficiencies were parametrised as a function of the muon $\eta$, and as a function of the highest jet $\pt$
present in the event:
a high $\pt$ indicates a lot of hadronic activity, which -- in turn -- is correlated to the muon isolation.
%Consequently, also the weights $\mathrm{w}_T$ and $\mathrm{w}_L^{\mathrm{non-tight}}$ depend on these kinematic variables.

\begin{figure}[p]
\begin{center}
\includegraphics[width=0.355\textwidth]{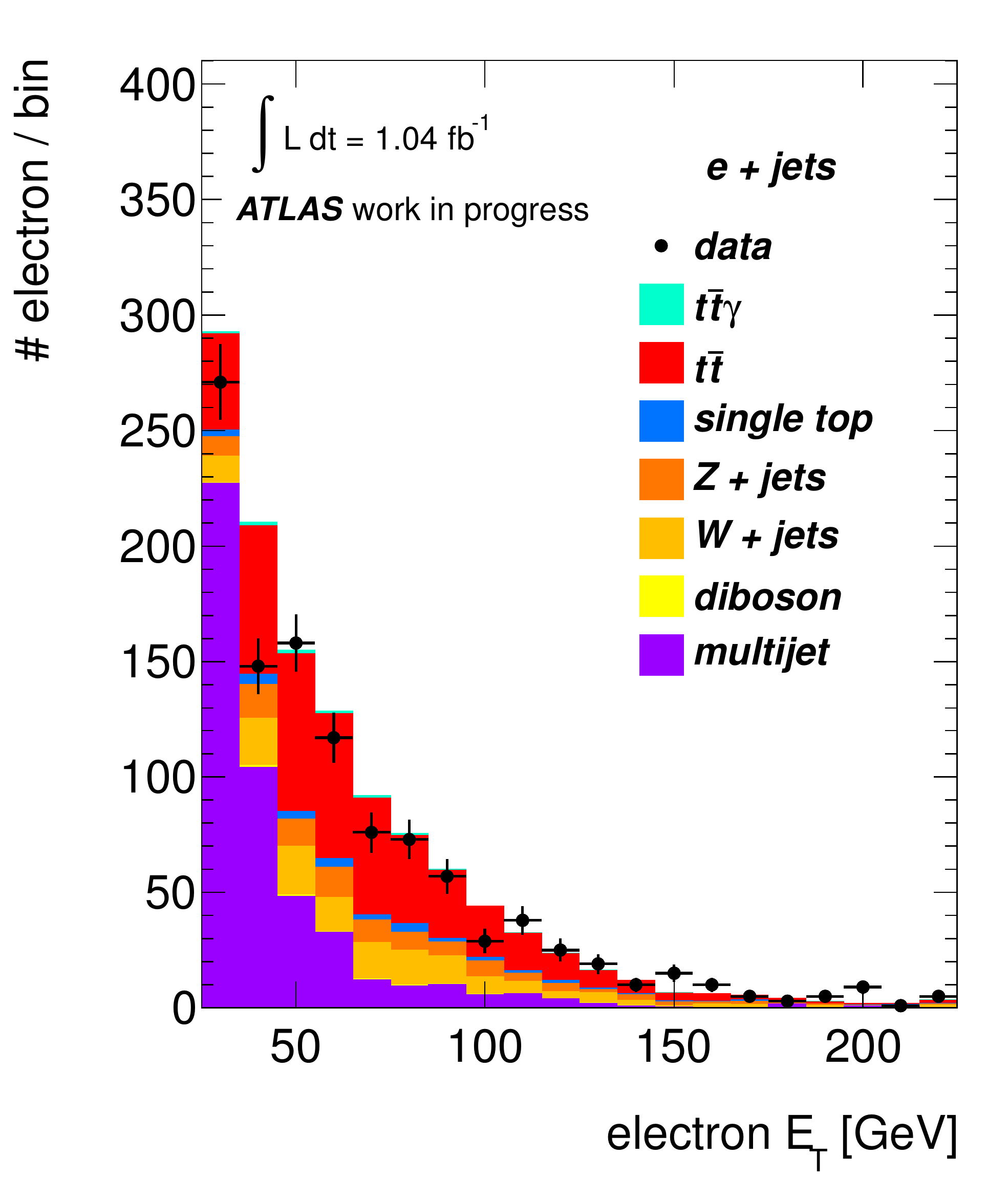}
\includegraphics[width=0.355\textwidth]{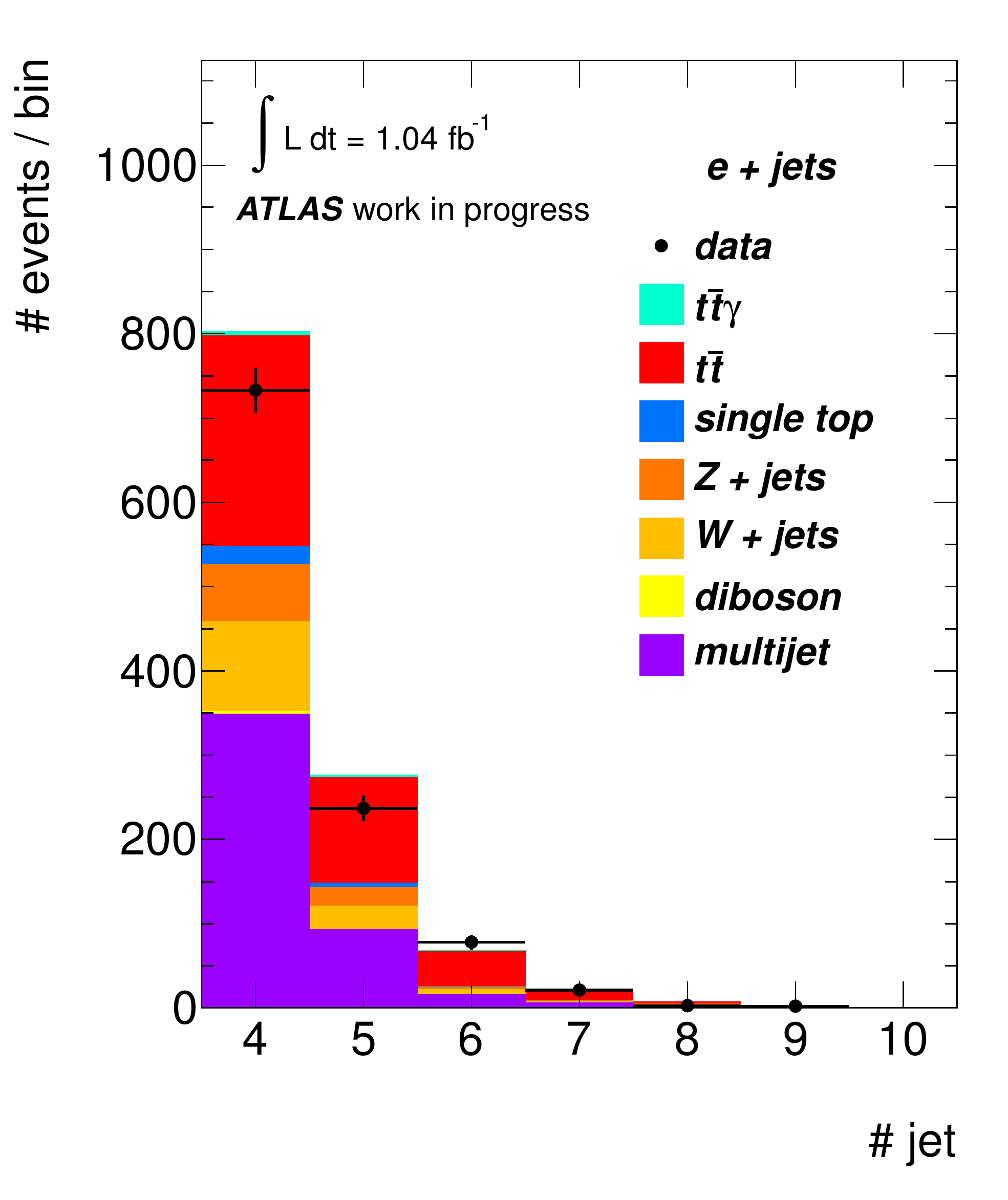} \\
\includegraphics[width=0.355\textwidth]{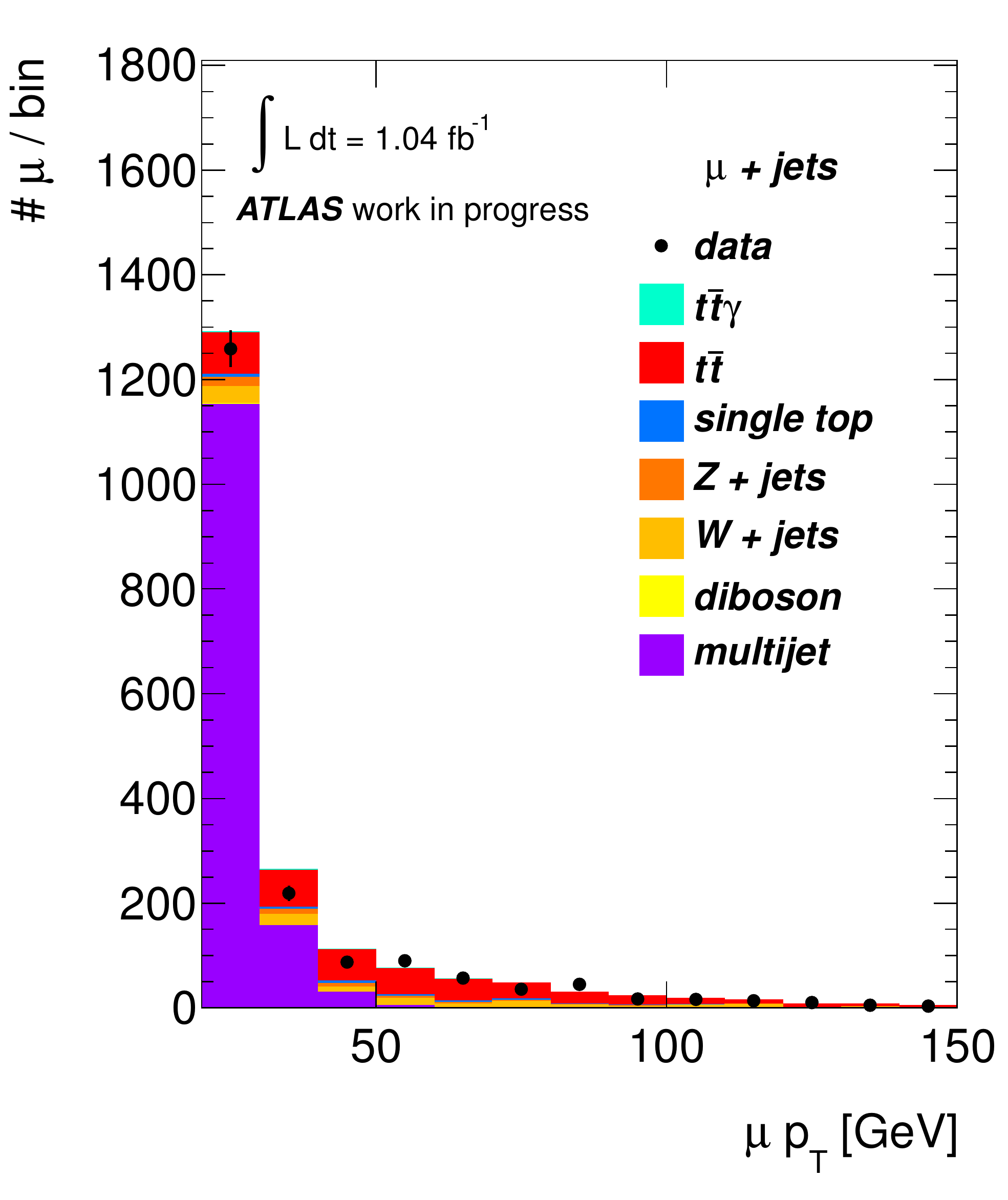}
\includegraphics[width=0.355\textwidth]{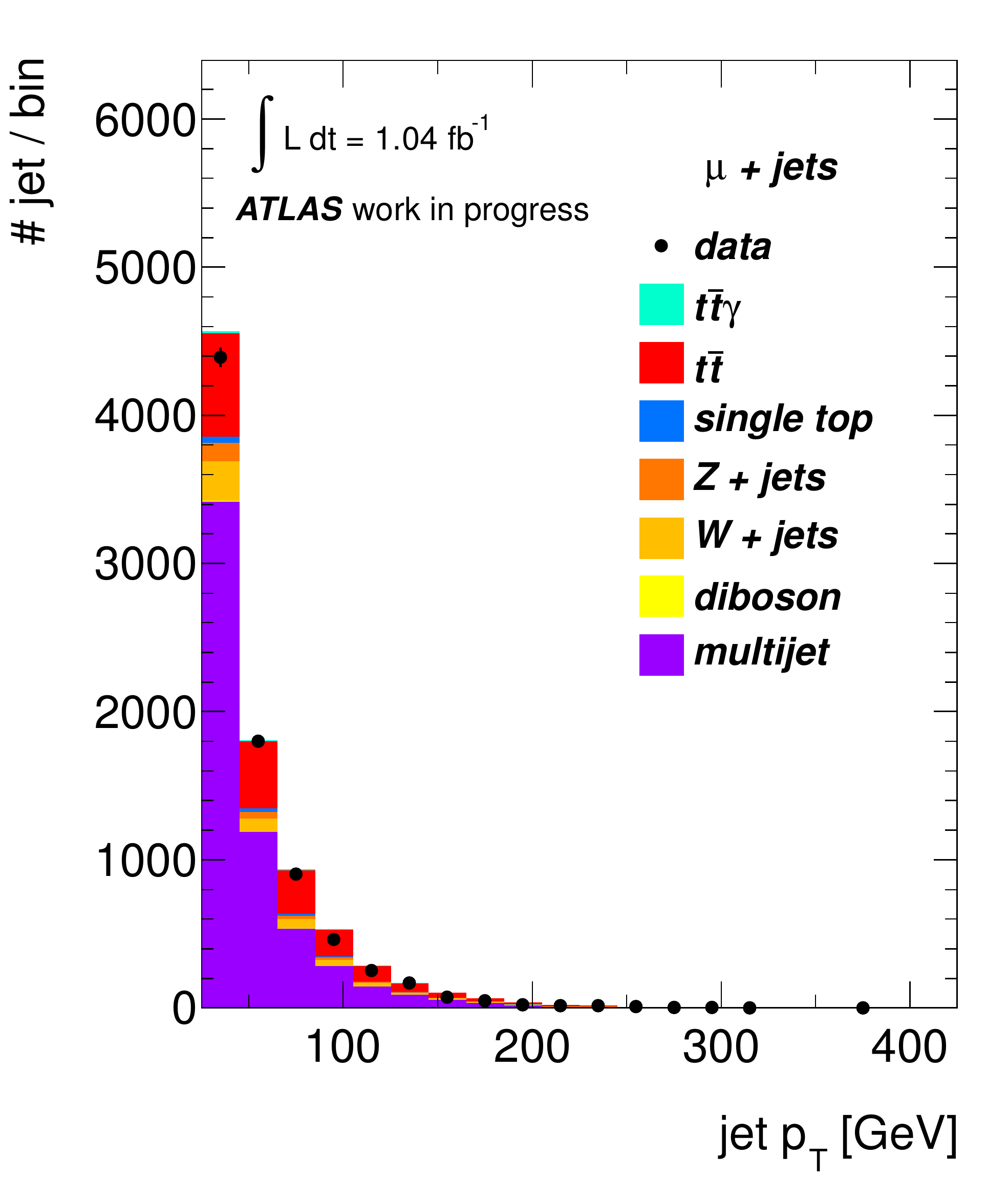}
\caption[Distributions in control regions largely enhanced in multijet background]{
  Distributions in control regions largely enhanced in multijet background for data and the expectations from MC simulations as well as the
  multijet estimate from the matrix method:
  the two plots in the upper row show the electron $\et$ distribution (left) and the number of jets (right) in the electron channel.
  The two plots in the lower row show the $\pt$ distributions of the muon (left) and of the jets (right) in the muon channel.
  In all plots, the last bin includes the overflow bin.
}
\label{fig:QCDCR}
\vspace{0.025\textwidth}
\includegraphics[width=0.45\textwidth]{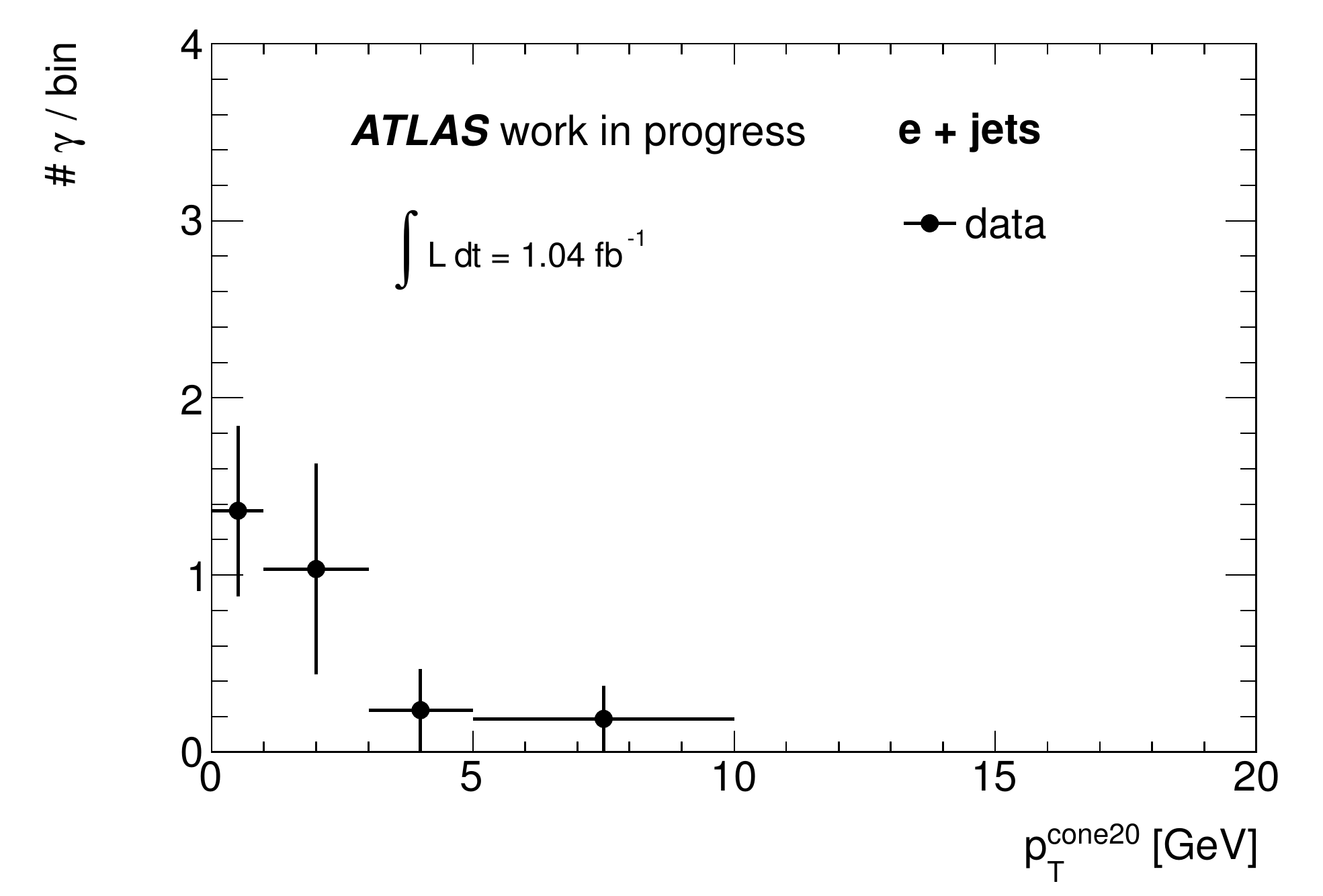}
\includegraphics[width=0.45\textwidth]{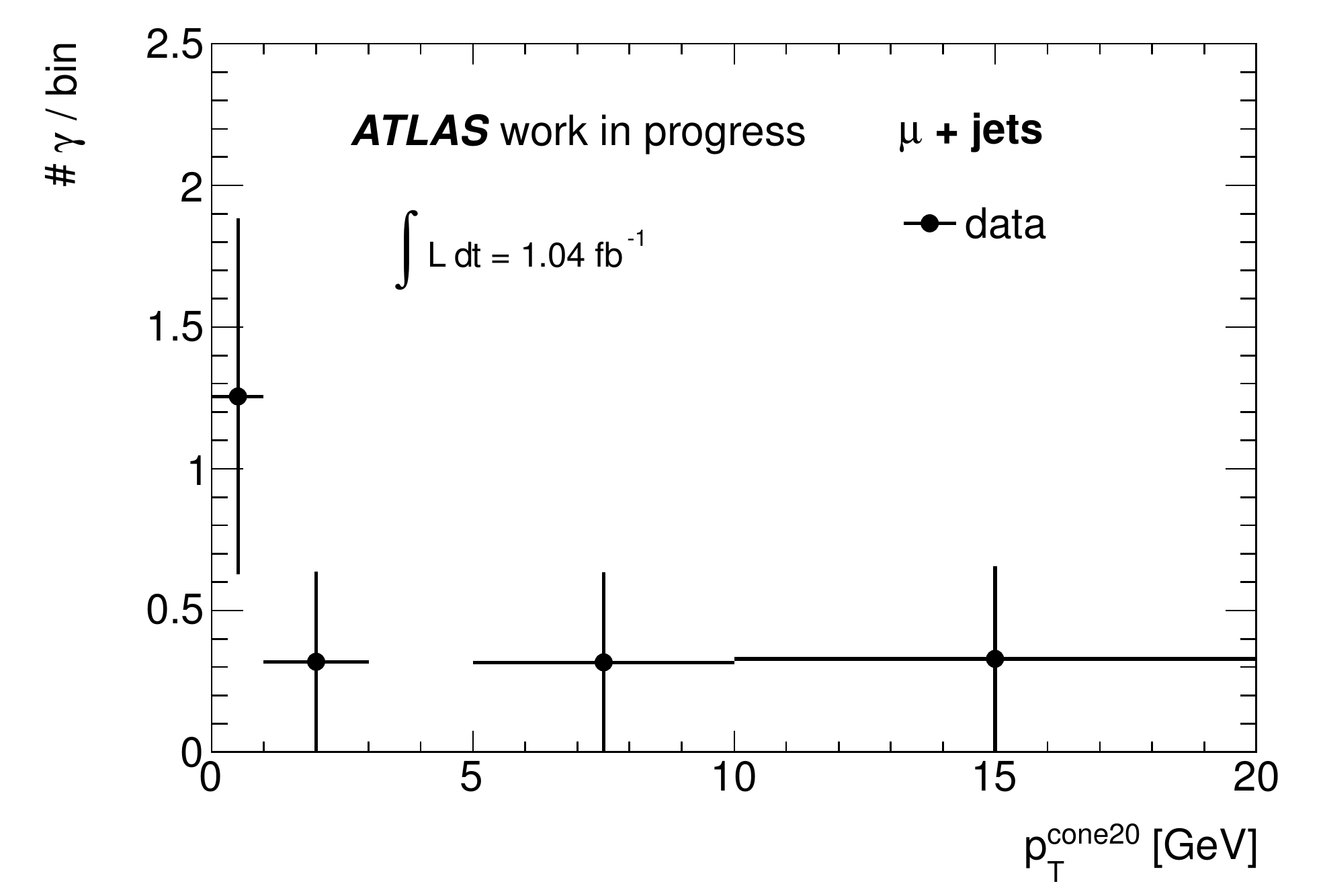}
\caption[$\ptcone$ distribution for photon candidates in events with a fake lepton]{
  $\ptcone$ distribution for photon candidates in events with a fake lepton in the electron channel (left) and the muon channel (right),
  respectively.
  In both plots, the last bin includes the overflow bin.
}
\label{fig:QCD_ptcone}
\end{center}
\end{figure}

Fig.~\ref{fig:QCDCR} shows distributions in the respective CRs for the electron and the muon channel after the $b$-tagging requirement.
The CRs are largely enhanced in multijet production and the data distributions as well as the expectations from MC simulations for various processes
are shown together with the multijet estimate from the matrix method.
The two plots in the upper row show the electron $\et$ distribution (left) and the number of jets (right) in the electron channel.
The two plots in the lower row show the $\pt$ distributions of the muon (left) and of the jets (right) in the muon channel.

The agreement of the sum of the expectations with data is reasonable.
Small disagreements can be explained by the inclusion of the photon in the $\met$ definition for this analysis, which was not included in the definition
of the CRs used in the derivation of the lepton misidentification rates.
In principle, this small change in the CR is expected to have a small effect on the overall multijet estimate, which is confirmed by the reasonable
agreement observed in Fig.~\ref{fig:QCDCR}.
These small differences between data and the sum of the expectations are covered by the systematic uncertainty of 100\% assigned to the multijet
background estimate in the signal region.

The yields obtained after the preselection with and without the $b$-tagging requirement were shown in Tab.~\ref{tab:cutflowpreselection}.
The uncertainty on the yield was estimated to 50\% for events without the $b$-tagging requirement and 100\% in events with at least one $b$-tagged jet.

\subsubsection{Prompt photon fraction estimated with the template fit}

In order to identify multijet events with an additional prompt photon, only events with a photon object according to the definition
in Sec.~\ref{sec:photon} were selected from the weighted loose sample.
Since loose electron objects may also be identified as photons, a minimal distance in $\eta$-$\phi$-space was required between loose electrons
and photons.

This yielded a sample correctly normalised for the presence of a fake lepton and a photon candidate.
The photon candidate could either be a prompt photon or a hadron misidentified as a photon, while the aim was to estimate only the contribution
featuring prompt photons.
As discussed in Ch.~\ref{sec:strategy}, $\ptcone$ is a good discriminating variable between prompt photons and hadrons misidentified as photons.
The $\ptcone$ distributions for the photon candidates in the weighted loose sample are shown in Fig.~\ref{fig:QCD_ptcone} for the electron channel
(left) and for the muon channel (right), respectively.
Both distributions feature a low number of expected events and show contributions in the low-$\ptcone$-region, which is dominated by prompt photons,
as well as in the high-$\ptcone$-region, which is largely dominated by hadrons misidentified as photons.
The total number of events reads 2.7 events in the electron and 2.2 events in the muon channel.

In order to isolate the contribution from prompt photons, a template fit to the $\ptcone$ distribution, as introduced in Ch.~\ref{sec:strategy},
was performed in both lepton channels separately.
However, a template fit cannot be performed to data distributions which feature fractional numbers of events for which the Poissonian probability
density function (pdf) is not well-defined.
This issue was solved by noting that the estimate presented in Fig.~\ref{fig:QCD_ptcone} was dominated by the events with positive weights, while
events with negative weights constitute only a small correction.
%Moreover, positive as well as negative weights were of the same order of magnitude, respectively.
Hence, the following procedure was applied:
the purity of prompt photons within the photon candidates was estimated using only the events with a positive weight, and the result was then
rescaled to the weighted yields (2.7 and 2.2 events, respectively).
This procedure is valid, because the photon isolation
and the loose/tight classification of the electron object are uncorrelated.

For the template fit, the prompt photon template was taken as derived in Ch.~\ref{sec:photontemplate}, because it depends only marginally on the
photon kinematics.
The hadron fake template, however, needed to be rederived with respect to Ch.~\ref{sec:faketemplate} by estimating the $\et$ spectrum of
hadron fakes as well as the fraction of hadron fakes with \mbox{$|\eta| > 1.81$}.
For the muon channel, the same procedure for the reweighting of the hadron fake template was applied as presented in Ch.~\ref{sec:faketemplate}.

In the electron channel, the statistics in the CR was too low to estimate the $\et$ spectrum and the fraction of photons with \mbox{$|\eta| > 1.81$}
for the hadron fakes.
The template estimated for the $\ttg$ topology was used instead (Ch.~\ref{sec:faketemplate}) and for the evaluation of systematic uncertainties,
the extreme templates in $\et$ and $\eta$ were used, as presented in Fig.~\ref{fig:backgroundtemplate_pt_eta}.

\begin{figure}[h!]
\begin{center}
\includegraphics[width=0.49\textwidth]{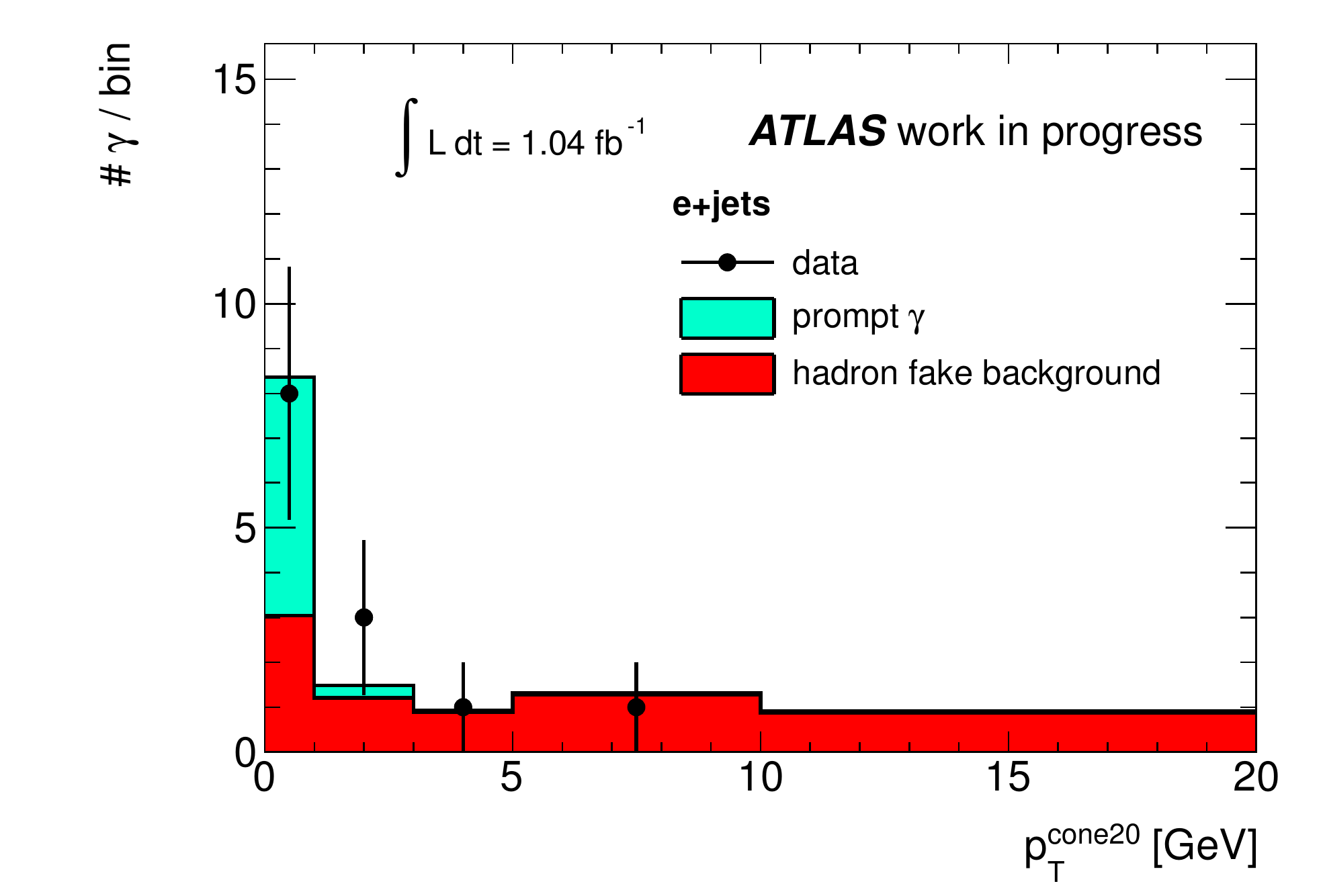}
\includegraphics[width=0.49\textwidth]{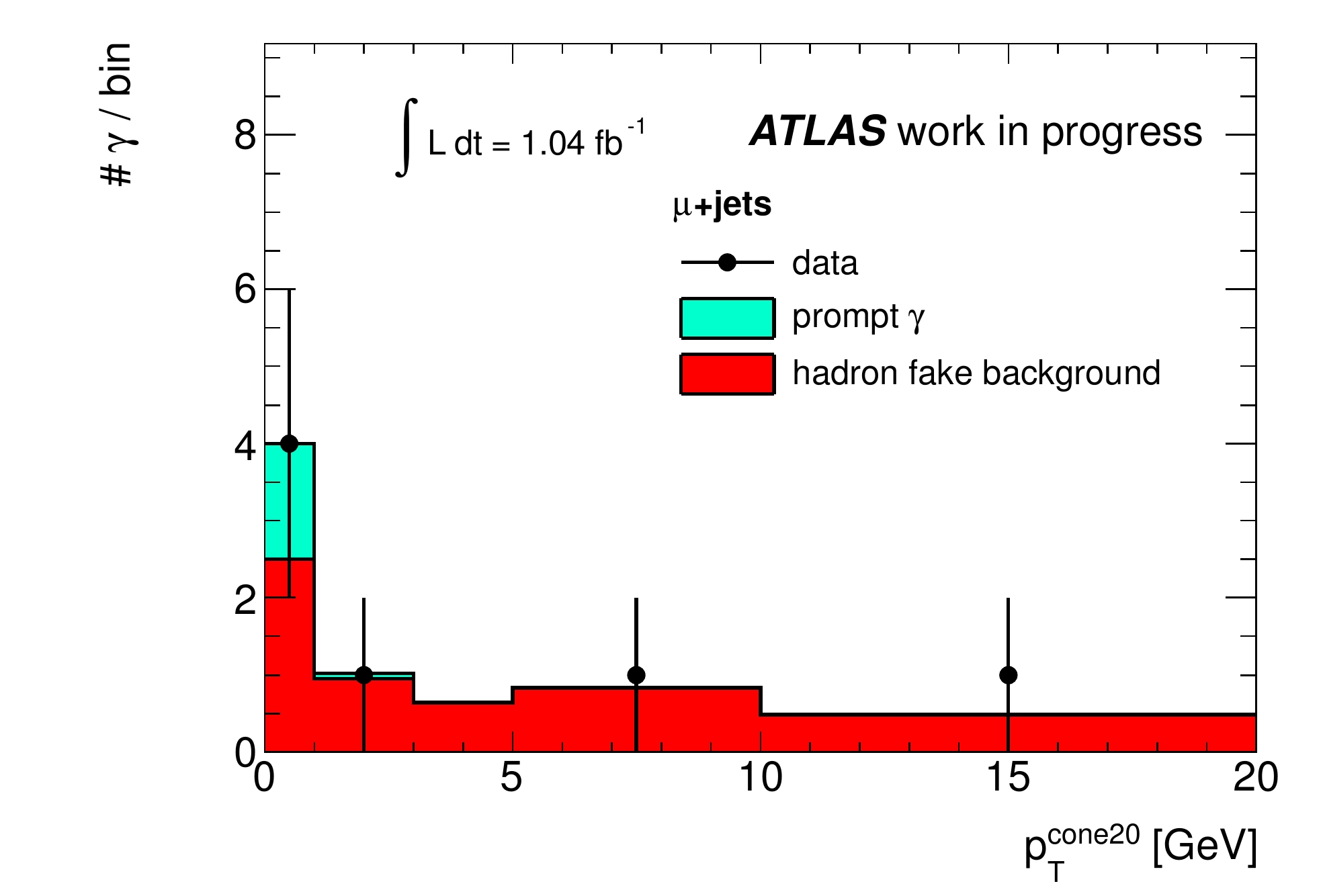}
\caption[Template fit to the photon $\ptcone$ distribution in events with a fake lepton]{
  Template fit to the photon $\ptcone$ distribution in events with a fake lepton in the electron channel (left) and the muon channel (right),
  respectively.
  Only events with a positive weight from the matrix method are shown and their respective weights were not used in the distribution.
  The results of the fit yielded the contributions from real photons and hadrons misidentified as photons.
  In both plots, the last bin includes the overflow bin.
}
\label{fig:QCD_templatefit}
\vspace{0.025\textwidth}
\includegraphics[width=0.49\textwidth]{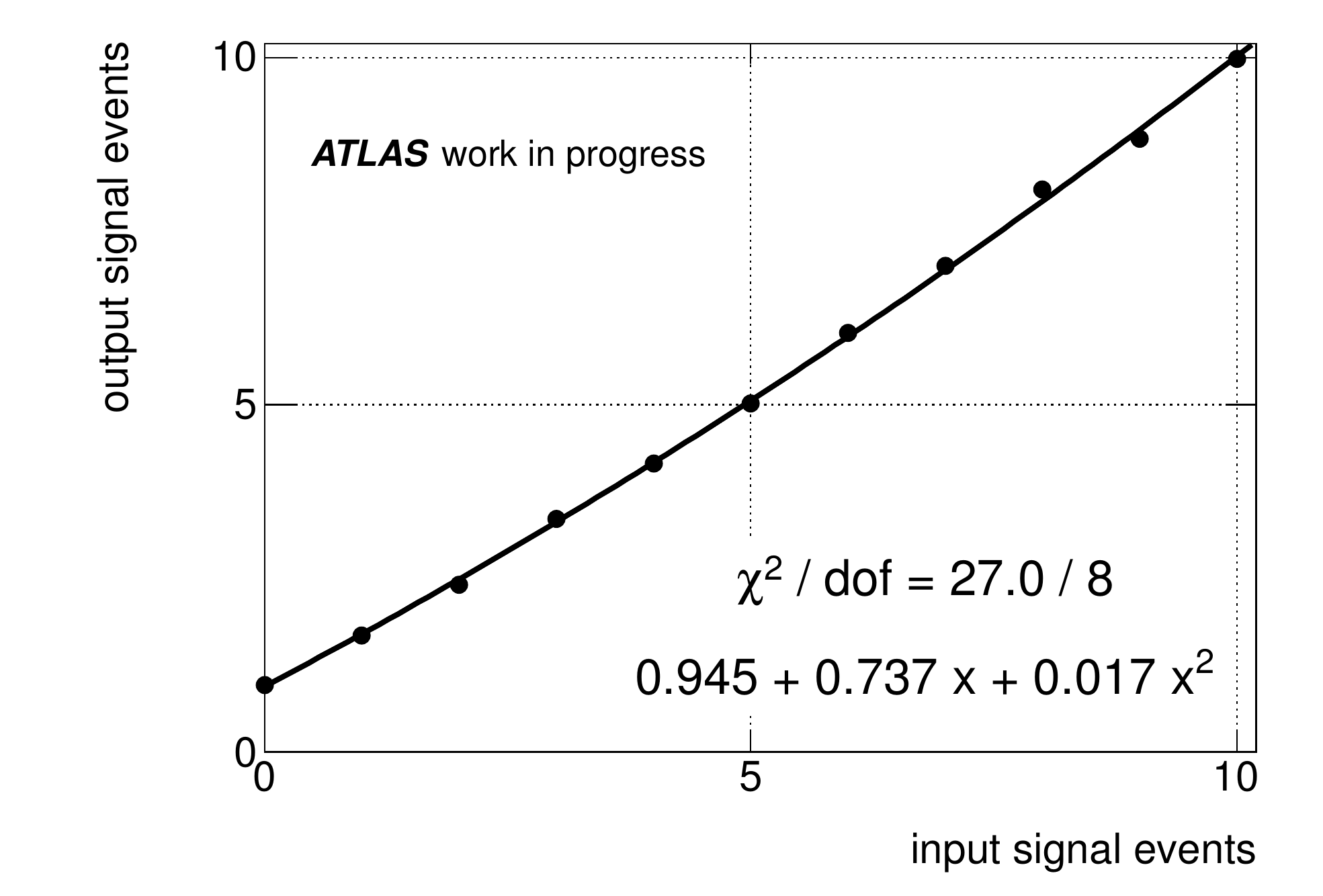}
\includegraphics[width=0.49\textwidth]{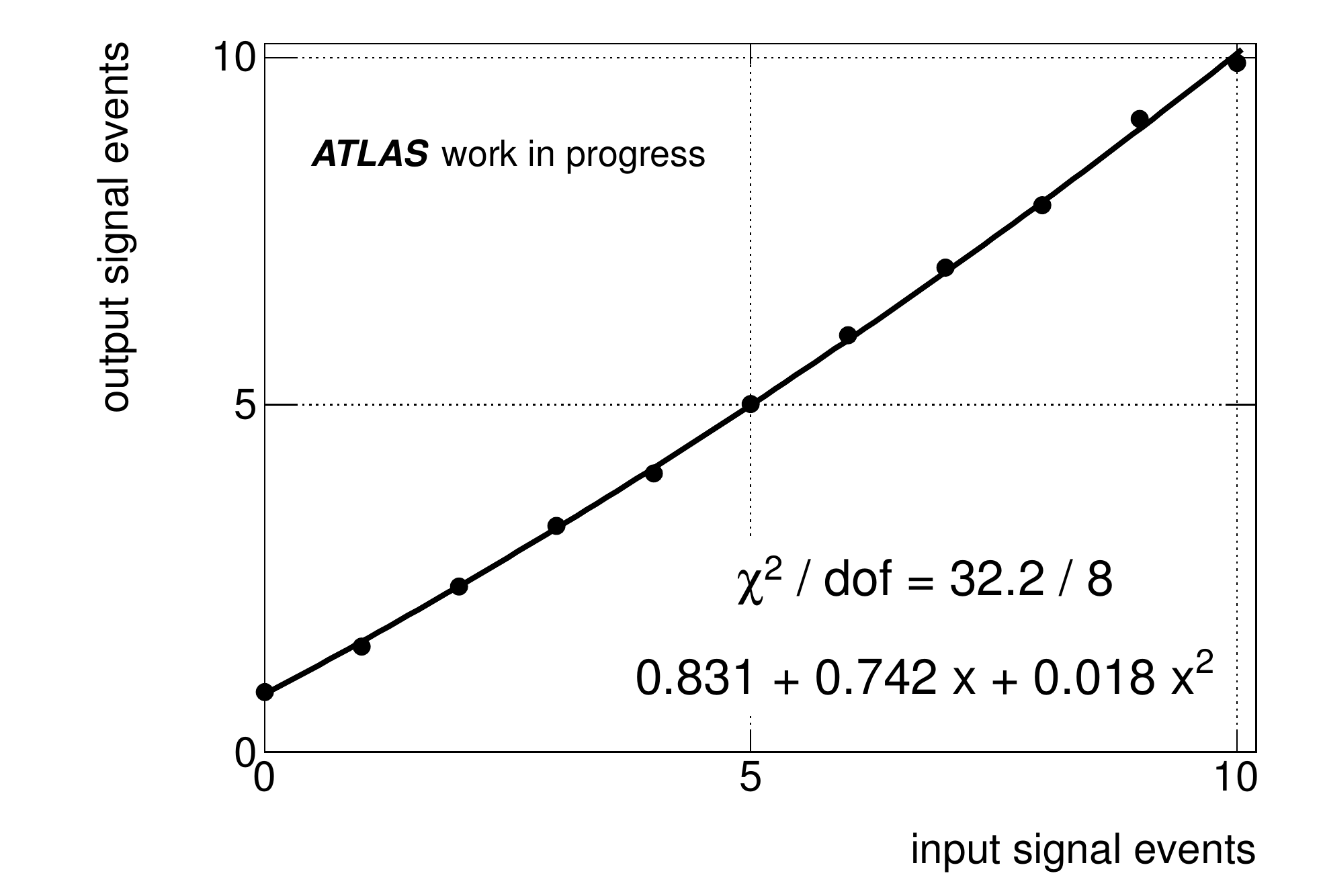}
\caption[Calibration curves for the multijet+$\gamma$ template fit]{
  Calibration curves in the electron (left) and the muon channel (right):
  pseudo-data were created with a variable amount of real photons ($x$-axis).
  The amount of hadrons misidentified as photons was set to the value from the fit to data (Fig.~\ref{fig:QCD_templatefit}).
  The mean output of real photon events from 5000 sets of pseudo-data is shown in the $y$-axis.
  A parabola was used to parametrise the resulting curves.
}
\label{fig:QCD_linearity}
\end{center}
\end{figure}

Fig.~\ref{fig:QCD_templatefit} shows the resulting template fits in the electron channel (left) and the muon channel (right), respectively.
In the electron channel, the fit yields \mbox{$5.7 \, ^{+3.5}_{-3.7}$} out of 13 events with a real photon.
In the muon channel, the result reads \mbox{$1.6 \, ^{+2.3}_{-1.6}$} real photon events out of 7 events.
The mode of the marginalised pdf was used as estimator and the statistical uncertainty was estimated using the smallest interval containing
68\% of the marginalised pdf.

In order to check if the fit result correctly estimated the real underlying number of multijet events with prompt photons, the following test
was performed:
pseudo-data were created by fluctuating each bin of the $\ptcone$ distribution individually around the expectation value according to the
Poissonian pdf.
While the expectation for the hadron fake contribution was fixed to the global mode of the actual fit, the expectation for the prompt photon
contribution was varied between 0 and 10 events.

For each input value for the number of prompt photons events,
Fig.~\ref{fig:QCD_linearity} shows the corresponding mean output of the fit from 5000 pseudo-data distributions.
For larger input values, the fit output corresponds to the input value.
For lower input values, however, the fit was found to be biased, because the Poissonian pdf is asymmetric around its mode for small expected values,
and the mean of the pdf is larger than the expected value.
A parabola was used to parametrise these calibration curves.

\begin{table}[h]
  \center
  \begin{tabular}{|l|c r@{}l|c r@{}l|}
    \hline
    Source & \multicolumn{6}{c|}{Uncertainty} \\
    & \multicolumn{3}{c|}{e+jets} & \multicolumn{3}{c|}{$\mu$+jets} \\
    \hline
    Yield       & & 1 &.2 & & 0 & .3 \\
    \hline
    Statistical & $^+_-$ & $^{\emptyplus 0}_{\emptyminus 0}$ & $^{.7\emptyplus}_{.9\emptyminus}$ & $^+_-$ & $^{\emptyplus 0}_{\emptyminus 0}$ & $^{.9\emptyplus}_{.3\emptyminus}$ \\
    \hline
    Fake lepton normalisation & $\pm$ & 1&.2 & $\pm$ & 0&.3 \\
    \hline
    Electron to photon extrapolation & $\pm$ & 0&.04 & $\pm$ & 0&.02 \\
    Fraction of converted prompt photons & $\pm$ & 0&.04 & $\pm$ & 0&.02 \\
    Pile-up dependence of the signal template & $\pm$ & 0&.03 & $\pm$ & 0&.02 \\
    Reweighting of the background templates ($\et$) & $\pm$ & 0&.35 & $\pm$ & 0&.05 \\
    Reweighting of the background templates ($\eta$) & $\pm$ & 0&.68 & $\pm$ & 0&.12 \\
    Fraction of converted hadron fakes & $\pm$ & 0&.07 & $\pm$ & 0&.05 \\
    Pile-up dependence of the background template & $\pm$ & 0&.03 & $\pm$ & 0&.03 \\
    \hline
    Sum (stat. + syst.) & $^+_-$ & $^{\emptyplus 1}_{\emptyminus 1}$ & $^{.6\emptyplus}_{.2\emptyminus}$ & $^+_-$ & $^{\emptyplus 1}_{\emptyminus 0}$ & $^{.0\emptyplus}_{.3\emptyminus}$ \\
    \hline
  \end{tabular}
  \caption[Uncertainties on the estimate of the multijet+$\gamma$ background]{
    Different sources of uncertainties on the estimate of multijet events with an additional prompt photon:
    the statistical uncertainty from the template fit, the uncertainty on the amount of fake leptons, and
    the systematic uncertainties associated to the template fit are presented.
  }
  \label{tab:QCD_uncertainties}
\end{table}

The results retrieved from the fit were corrected using the parabolas from Fig.~\ref{fig:QCD_linearity} and rescaled to the actual weighted
multijet estimate.
The final estimates read \mbox{$1.2 \, ^{+0.7}_{-0.9}$} and \mbox{$0.3 \, ^{+0.9}_{-0.3}$} in the electron and muon channel, respectively.
The uncertainty is the statistical uncertainty only as retrieved from the template fit.

Systematic uncertainties were evaluated for the fake lepton estimate as well as for the amount of prompt photons within the photon candidates.
An overview of all uncertainties considered is shown in Tab.~\ref{tab:QCD_uncertainties}.

The systematic uncertainty on the amount of fake leptons was estimated to 100\% of the final yield.
The systematic uncertainties on the amount of prompt photons were estimated by replacing the nominal prompt photon and hadron fake templates
in the template fit by systematically varied templates.
Pseudo-data were constructed and 5000 $\ptcone$ distributions were fitted to evaluate the impact of each systematic effect.
The procedure is described in detail in Ch.~\ref{sec:systematics}, where also the different templates from the systematic variations are presented.

The following sources of systematic uncertainties were considered for the prompt photon template:
the uncertainty on the extrapolation from electron to photon templates;
the uncertainty on the fraction of converted photons used for the derivation of the template;
the uncertainty on a possible dependence of the template on the pile-up conditions.

For the hadron fake template, the following sources of systematic uncertainties were considered:
uncertainties on the reweighting of the template in $\et$ and $\eta$;
the uncertainty on the fraction of converted photons used for the derivation of the template;
the uncertainty on a possible dependence of the template on the pile-up conditions.
The variations for the reweighting of the hadron fake templates were increased with respect to the description in Ch.~\ref{sec:systematics}
as stated above.

Dominant uncertainties were found to be the statistical uncertainty from the template fit as well as the uncertainty on the amount of fake leptons.
In the electron channel, also the uncertainty due to the reweighting of the fake hadron template was found to be sizable, which is due
to the low statistics in the CR used to estimate the reweighting parameters.
The final estimate in the electron channel reads: \mbox{$1.2 \, ^{+1.6}_{-1.2} \, \rm{(stat.+syst.)}$}.
The estimate in the muon channel reads: \mbox{$0.3 \, ^{+1.0}_{-0.3} \, \rm{(stat.+syst.)}$}.

\section[$W$+jets production with a prompt photon]{
\boldmath$W$+jets production with a prompt photon\unboldmath}
\label{sec:Wjetsgamma}

The background contribution from $W$+jets production with an additional photon in the final state was estimated using a data-driven approach
in order to reduce the dependence on MC simulations.
The event selection from Ch.~\ref{sec:selection} was modified in order to define a CR with an enhanced $W$+jets+$\gamma$ contribution:
one to three jets were required instead of four jets and none of the jets needed to be $b$-tagged.
The yield in the signal region was then extrapolated from the CR using MC simulated $W$+jets+$\gamma$ events.

The event selection in the CR yielded 1446 events in the electron and 2468 events in the muon channel.
However, background contributions were due to processes with prompt photons in the final state, but also from events with electrons and hadrons
misidentified as photons.

The upper part of Tab.~\ref{tab:W_bkg} shows the expected background contributions for events with a prompt photon in the final state:
$\ttbar$, $Z$+jets, single top, and diboson production with an additional photon in the final state were estimated using MC simulations.
Systematic uncertainties include uncertainties on the expected cross sections for the individual processes,
as well as uncertainties due to detector modelling effects as discussed in Sec.~\ref{sec:syst_detectormodelling} and uncertainties
due to limited MC statistics (cf. Sec.~\ref{sec:syst_backgroundmodelling}).
For the contributions from $Z$+jets+$\gamma$, single top+$\gamma$ and diboson+$\gamma$ production, an additional uncertainty of 100\% was added
as discussed in Sec.~\ref{sec:restgamma}

\begin{table}[h]
  \center
  \begin{tabular}{|l|r@{}l c r@{}l|r@{}l c r@{}l|}
    \hline
    Background contribution & \multicolumn{5}{c|}{e+jets} & \multicolumn{5}{c|}{$\mu$+jets} \\
    \hline
    $\ttg$                      &   36 &    & $\pm$ &  8 &    & 49 &    & $\pm$ & 11 &    \\
    $Z$+jets+$\gamma$           &   80 &    & $^+_-$ & $^{\emptyplus 100}_{\emptyminus \emptynull 80}$ & $^{\emptyplus}_{\emptyminus}$
                                &  260 &    & $^+_-$ & $^{\emptyplus 290}_{\emptyminus 260}$ & $^{\emptyplus}_{\emptyminus}$ \\
    single top+$\gamma$       &    6 & .6 & $\pm$ &  6 & .6
                                &    8 & .1 & $^+_-$ & $^{\emptyplus \emptynull \emptynull 8}_{\emptyminus \emptynull \emptynull 8}$ & $^{.2\emptyplus }_{.1\emptyminus}$ \\
    diboson+$\gamma$          &    6 & .8 & $^+_-$ & $^{\emptyplus \emptynull \emptynull 7}_{\emptyminus \emptynull \emptynull 6}$ & $^{.2\emptyplus }_{.8\emptyminus}$
                                &   12 & .4 & $^+_-$ & $^{\emptyplus \emptynull 12}_{\emptyminus \emptynull 12}$ & $^{.6\emptyplus }_{.4\emptyminus}$ \\
    multijet+$\gamma$         &   80 &    & $\pm$ & 40 &    & 75 &    & $\pm$ & 38 &    \\
    \hline
    $\ttg$ ($e \to \gamma$)     &    1 & .4 & $\pm$ &  0 & .4 &  1 & .2 & $\pm$ &  0 & .3 \\
    $\ttbar$ ($e \to \gamma$)   &   48 &    & $\pm$ &  9 &    & 70 &    & $\pm$ & 13 &    \\
    $Z$+jets ($e \to \gamma$)   &  220 &    & $\pm$ &130 &    & 14 &    & $\pm$ &  9 &    \\
    single top ($e \to \gamma$) &    5 & .8 & $\pm$ &  1 & .3 &  6 & .8 & $\pm$ &  1 & .3 \\
    diboson ($e \to \gamma$)    &    9 & .1 & $\pm$ &  2 & .4 &  8 & .4 & $\pm$ &  2 & .2 \\
    \hline
  \end{tabular}
  \caption[Expected yields for different background contributions in the $W$+jets CR]{
    Expected yields for different background contributions in the $W$+jets CR in the electron (left) and in the muon channel (right).
    The upper part shows the expectations from events with prompt photons in the final state.
    The lower part shows the expectations from events with electrons misidentified as photons.
  }
  \label{tab:W_bkg}
\end{table}

The contribution from multijet events with prompt photons was estimated using the approach discussed in Sec.~\ref{sec:QCDgamma}
with the matrix method and a template fit to the $\ptcone$ distribution.
Estimates of \mbox{$80 \pm 40$} and \mbox{$75 \pm 38$} events were obtained in the electron and muon channel, respectively.
In contrast to the estimate derived in Sec.~\ref{sec:QCDgamma} for the signal region, the reweighting of the hadron fake templates in $\et$ and $\eta$
was successfully performed in both lepton channels, because of increased statistics in the CR enhanced in hadron fakes.
Moreover, the result of the template fit turned out to be unbiased given the larger expected yields, and therefore a correction as depicted in
Fig.~\ref{fig:QCD_linearity} was found to be not necessary.

The lower part of Tab.~\ref{tab:W_bkg} shows the expected contributions from background events with an electron which was misidentified as a photon.
The yields were estimated using the scale factors for the electron-to-photon misidentification rate $\feg$ as described in Ch.~\ref{sec:electronfake}.
The quoted uncertainties include the systematic uncertainties on the expected cross sections for the individual processes,
the uncertainties due to detector modelling effects, uncertainties due to limited MC statistics and the uncertainties on the $\feg$ scale factors.

The largest background contributions in the CR with either prompt photons or electrons misidentified as photons were found to be from $Z$+jets and
multijet production.
These contributions also feature the largest absolute uncertainties on the background predictions.
As expected, the $Z$+jets production with a misidentified electron contributes significantly less in the muon compared to the electron channel:
the only contribution in the muon channel is from \mbox{$Z \to \tau \tau$} events with one $\tau$-lepton decaying to a muon and one $\tau$-lepton
decaying to an electron, which is then misidentified as a photon, while in the electron channel \Zee events contribute.
Additional sizable background contributions are due to $\ttg$ production and $\ttbar$ production with a misidentified electron.

The amount of events in the CR with hadrons misidentified as photons was estimated using a similar approach as used for the multijet+$\gamma$ estimate:
a template fit to the photon $\ptcone$ distribution was used to distinguish prompt photons and misidentified electrons from hadron fakes as described
in Ch.~\ref{sec:strategy}.
While the signal template was used as derived in Ch.~\ref{sec:photontemplate}, the template for misidentified hadrons needed to be rederived following
the approach in Ch.~\ref{sec:faketemplate}:
due to the $\et$ and $\eta$ dependence of the $\ptcone$ distribution from misidentified hadrons, the templates needed to be reweighted in
$\et$ and $\eta$ using the CR dominated by hadron fakes as introduced in Ch.~\ref{sec:faketemplate}.
The resulting hadron fake template represents the $\ptcone$ distribution for hadron fakes in the $W$+jets+$\gamma$ CR.

\begin{figure}[h]
\begin{center}
\includegraphics[width=0.49\textwidth]{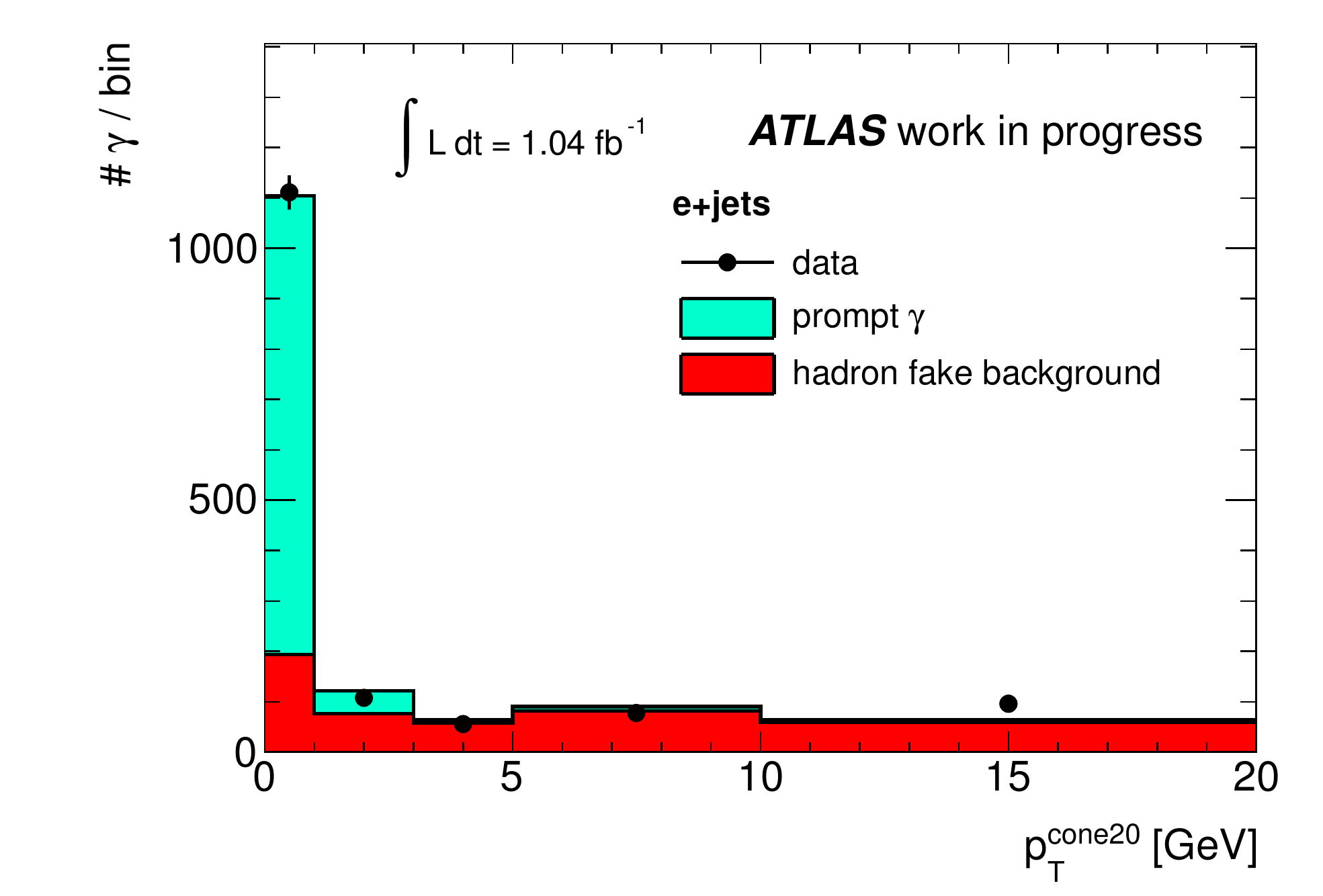}
\includegraphics[width=0.49\textwidth]{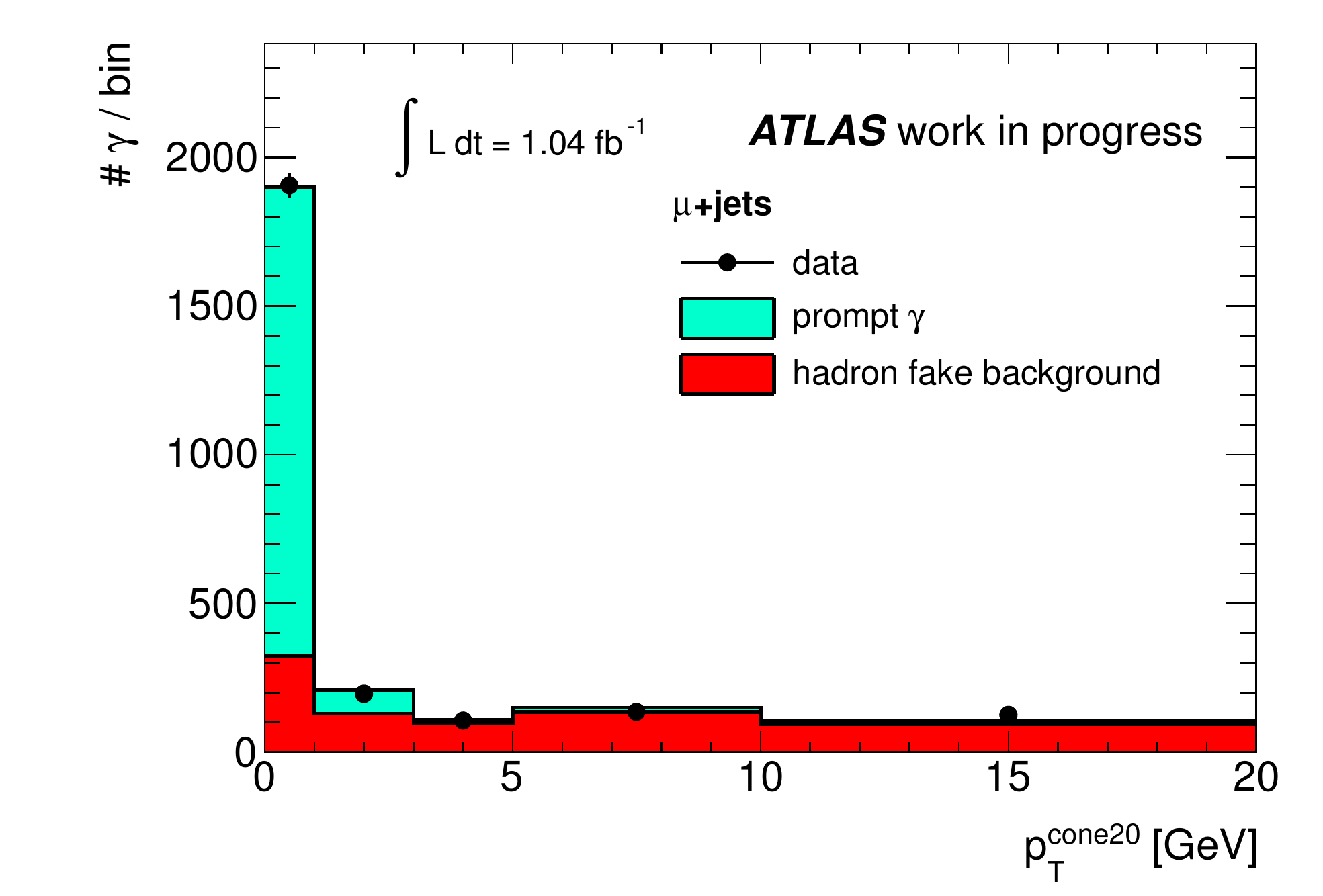}
\caption[Template fit to the photon $\ptcone$ distribution in events in the $W$+jets+$\gamma$ CR]{
  Template fit to the photon $\ptcone$ distribution in events in the $W$+jets+$\gamma$ control region in the electron (left) and the muon channel (right),
  respectively, using the templates for true photons and for hadrons misidentified as photons.
  In both plots, the last bin includes the overflow bin.
}
\label{fig:Wjetsgammafits}
\end{center}
\end{figure}

Fig.~\ref{fig:Wjetsgammafits} shows the resulting template fits in the $W$+jets+$\gamma$ CR in the electron (left) and the muon channel (right).
The prompt photon template represents the $W$+jets+$\gamma$ process as well as all background processes with prompt photons and misidentified electrons
as listed in Tab.~\ref{tab:W_bkg}.
After subtraction of the background expectations, the resulting $W$+jets+$\gamma$ yields in the CR read \mbox{$480 \, ^{+40}_{-30}$} and
\mbox{$1190 \, ^{+60}_{-50}$} in the electron and the muon channel, respectively, where the uncertainties represent just the statistical uncertainty of the
template fit.

These yields were extrapolated using $W$+jets+$\gamma$ simulations
to the signal region with four jets out of which at least one had to be $b$-tagged.
The yields in the signal region read \mbox{$1.84 \, ^{+0.17}_{-0.13}$} and \mbox{$3.72 \, ^{+0.18}_{-0.15}$}
in the electron and the muon channel, respectively, where, again, the uncertainties are only the statistical uncertainties from the template fit.

\begin{table}[h]
  \center
  \begin{tabular}{|l|r@{}l c r@{}l|r@{}l c r@{}l|}
    \hline
    Control region & \multicolumn{5}{c|}{e+jets} & \multicolumn{5}{c|}{$\mu$+jets} \\
    \hline
    Default (1 -- 3 jets) & 1 & .84 & $^+_-$ & $^{\emptyplus 0}_{\emptyminus 0}$ & $^{.17 \, \emptyplus}_{.13 \, \emptyminus}$
                          & 3 & .72 & $^+_-$ & $^{\emptyplus 0}_{\emptyminus 0}$ & $^{.18 \, \emptyplus}_{.15 \, \emptyminus}$ \\
    1 jet                 & 1 & .73 & $^+_-$ & $^{\emptyplus 0}_{\emptyminus 0}$ & $^{.18 \, \emptyplus}_{.18 \, \emptyminus}$
                          & 3 & .87 & $^+_-$ & $^{\emptyplus 0}_{\emptyminus 0}$ & $^{.22 \, \emptyplus}_{.18 \, \emptyminus}$ \\
    2 jets                & 2 & .06 & $^+_-$ & $^{\emptyplus 0}_{\emptyminus 0}$ & $^{.27 \, \emptyplus}_{.32 \, \emptyminus}$
                          & 3 & .58 & $^+_-$ & $^{\emptyplus 0}_{\emptyminus 0}$ & $^{.32 \, \emptyplus}_{.36 \, \emptyminus}$ \\
    3 jets                & 1 & .78 & $^+_-$ & $^{\emptyplus 0}_{\emptyminus 0}$ & $^{.59 \, \emptyplus}_{.57 \, \emptyminus}$
                          & 2 & .51 & $^+_-$ & $^{\emptyplus 0}_{\emptyminus 0}$ & $^{.60 \, \emptyplus}_{.70 \, \emptyminus}$ \\
    \hline
  \end{tabular}
  \caption[$W$+jets+$\gamma$ yields in the signal region using different control regions (CRs)]{
    Resulting $W$+jets+$\gamma$ yields in the signal region using different control regions (CRs) in the electron (left) and the muon channel (right):
    while the default CR features one to three jets, CRs with only one, two or three jets were investigated.
    The uncertainties represent the statistical uncertainties from the fit only.
  }
  \label{tab:W_CRs}
\end{table}

Tab.~\ref{tab:W_CRs} shows the resulting $W$+jets+$\gamma$ yields in the signal region using different control regions (CRs) in both lepton channels.
While the default CR features one to three jets, CRs with only one, two or three jets were investigated.
The uncertainties represent the statistical uncertainties from the fit only and the results in the different CR regions were found to be consistent
within the fit uncertainties.
Only the estimate in the muon channel using the three-jet-CR differs from the estimates from the other CRs, 
but the discrepancy is still only of the order of two standard deviations\footnote{The probability that at least one out of six measurements
deviates by at least two standard deviations from the expectation value is roughly 24\%.}.
The result is hence considered to be stable with respect to the choice of the CR.

\begin{table}[h]
  \center
  \begin{tabular}{|l|c r@{}l|c r@{}l|}
    \hline
    Source of uncertainty & \multicolumn{3}{c|}{e+jets} & \multicolumn{3}{c|}{$\mu$+jets} \\
    \hline
    Statistical & $^+_-$ & $^{\emptyplus\emptypercent 9}_{\emptyminus\emptypercent 7}$ & $^{.3 \, \%\emptyplus}_{.1 \, \%\emptyminus}$ &
                  $^+_-$ & $^{\emptyplus\emptypercent 4}_{\emptyminus\emptypercent 4}$ & $^{.9 \, \%\emptyplus}_{.0 \, \%\emptyminus}$ \\
    Template shapes                 & $\pm$ & 12 & .6 \% & $\pm$ &  8 & .8 \% \\
    Background MC cross sections and statistics     & $\pm$ & 15 & .0 \% & $\pm$ &  5 & .6 \% \\
    Prompt $\gamma$ background      & $\pm$ & 18 & .7 \% & $\pm$ & 22 & .0 \% \\
    $\feg$                          & $\pm$ &  4 & .6 \% & $\pm$ &  0 & .8 \% \\
    Luminosity                      & $\pm$ &  3 & .2 \% & $\pm$ &  1 & .3 \% \\
    Electron modelling              & $\pm$ &  8 & .0 \% & $\pm$ &  0 & .7 \% \\
    Muon modelling                  & $\pm$ &  0 & .1 \% & $\pm$ &  1 & .0 \% \\
    Jet and $\met$ modelling        & $\pm$ & 15 & .0 \% & $\pm$ & 21 & .2 \% \\
    Photon modelling                & $\pm$ &  4 & .1 \% & $\pm$ &  1 & .4 \% \\
    $b$-tagging modelling           & $\pm$ & 11 & .3 \% & $\pm$ & 17 & .8 \% \\
    Modelling of LAr readout issues & $\pm$ &  5 & .1 \% & $\pm$ &  4 & .2 \% \\
    \hline
    Total systematics               & $\pm$ & 36 & .3 \% & $\pm$ & 37 & .2 \% \\
    \hline
    Total                           & $\pm$ & 36 & \textcolor{white}{.0} \% & $\pm$ & 37 & \textcolor{white}{.0} \% \\
    \hline
  \end{tabular}
  \caption[Sources of uncertainties on the $W$+jets+$\gamma$ yield in the signal region]{
    Sources of uncertainties on the $W$+jets+$\gamma$ yield in the signal region in the electron (left) and in the muon channel (right).
  }
  \label{tab:W_systematics}
\end{table}

Tab.~\ref{tab:W_systematics} shows the contributions from different sources of uncertainties on the $W$+jets+$\gamma$ yields in the signal region in
both lepton channels.
Uncertainties are due to the shape of the prompt photon and the hadron fake template, as well as due to various detector modelling effects as described
in Ch.~\ref{sec:systematics}.
Additional sources of systematic uncertainties are the background estimates:
for estimates derived from MC simulations, uncertainties originate from the uncertainties on the production cross sections, from the uncertainty on the
luminosity, and from the uncertainties on the $\feg$ scale factors.
Moreover, the amount of background events with prompt photons in the final state was found to be subject to large uncertainties as described in
Sec.~\ref{sec:QCDgamma} for multijet production and in Sec.~\ref{sec:restgamma} for $Z$+jets+$\gamma$, single top+$\gamma$ and diboson+$\gamma$
production.

The largest sources of systematic uncertainties were found to be due to the shapes of the templates, the cross sections for the processes
estimated from MC simulations,
the background estimates from processes with prompt photons in the final state, and the modelling of jets, $\met$ and $b$-tagging.
The uncertainty on the MC normalisation is much larger in the electron than in the muon channel.
This can be understood since the electron channel features larger background contributions (Tab.~\ref{tab:W_bkg})
with respect to the $W$+jets+$\gamma$ expectation and is hence subject to larger uncertainties.
The uncertainty on the jet and $\met$ modelling is largely dominated by the uncertainty on the jet energy scale (JES).

Most systematic uncertainties due to the detector modelling affect the expected yields in the CR as well as the yields in the signal
region and are therefore expected to have a reduced impact on the extrapolation to the signal region using the $W$+jets+$\gamma$ simulation.
However, this does not hold for the uncertainty due to the modelling of the $b$-tagging performance, because a $b$-tag was not required in the CR, which
explains that the systematic uncertainty due to $b$-tagging is of the same order of magnitude as the uncertainties due to the JES modelling,
while for the final cross section (Ch.~\ref{sec:results}), the JES uncertainties dominate.

The final estimates for $W$+jets+$\gamma$ in the signal region including statistical and systematic uncertainties read \mbox{$1.8 \pm 0.7$} and \mbox{$3.7 \pm 1.4$}
in the electron and muon channel, respectively.
From MC simulations \mbox{$2.5 \pm 1.2$} and \mbox{$4.1 \pm 2.0$} events were expected, where the uncertainty is just the 48\% uncertainty on the $W$+jets
normalisation (Sec.~\ref{sec:backgroundmodelling}) and hence does not include additional systematic uncertainties.
The data-driven estimate is hence consistent with the expectation from MC simulation within the uncertainties.
Moreover, the uncertainty was reduced with respect to the uncertainty associated to the MC expectation.

\section{Additional backgrounds with a prompt photon}
\label{sec:restgamma}

The remaining background sources from $Z$+jets, single top and diboson production with an additional prompt photon in the final state were estimated
using MC simulations as described in Sec.~\ref{sec:backgroundmodelling}.
In Tab.~\ref{tab:Z_st_diboson}, the expected yields for \mbox{$1.04 \ifb$} are presented.

Since the $Z$+jets, single top and diboson simulations only included photon radiation as QED corrections provided by the PHOTOS package~\cite{photos}
and not at ME level, the predicted yields are only approximations and hence subject to large systematic uncertainties.
The uncertainties were estimated conservatively to 100\% of the yields.
Additionally, all systematic uncertainties as discussed in Ch.~\ref{sec:systematics} were evaluated, in particular the uncertainties
on the detector modelling, but also the uncertainties on the cross sections of the different processes, on the luminosity, and uncertainties due to
limited MC statistics.

\begin{table}[h]
  \center
  \begin{tabular}{|l|r@{}l c r@{}l|r@{}l c r@{}l|}
    \hline
    Process & \multicolumn{5}{c|}{e+jets} & \multicolumn{5}{c|}{$\mu$+jets} \\
    \hline
    $Z$+jets+$\gamma$     &  1 &.3  & $^+_-$ & $^{\emptyplus 2}_{\emptyminus 1}$ & $^{.4\emptyplus}_{.3\emptyminus}$ &  1 &.6  & $^+_-$ & $^{\emptyplus 2}_{\emptyminus 1}$ & $^{.3\emptyplus}_{.6\emptyminus}$ \\
    Single top+$\gamma$ &  0 &.6  & $^+_-$ & $^{\emptyplus 0}_{\emptyminus 0}$ & $^{.7\emptyplus}_{.6\emptyminus}$ &  0 &.2  & $^+_-$ & $^{\emptyplus 0}_{\emptyminus 0}$ & $^{.3\emptyplus}_{.2\emptyminus}$ \\
    Diboson+$\gamma$    &  0 &.16 & $^{+}_{-}$ & $^{\emptyplus 0}_{\emptyminus 0}$ & $^{.34\emptyplus}_{.16\emptyminus}$ &  0 &.04 & $^{+}_{-}$ & $^{\emptyplus 0}_{\emptyminus 0}$ & {$^{.18\emptyplus}_{.04\emptyminus}$} \\
    \hline
  \end{tabular}
  \caption[Expected yields for additional backgrounds with prompt photons]{
    Expected yields for $Z$+jets, single top and diboson production with an additional prompt photon in the final state for \mbox{$1.04 \ifb$}
    from MC simulations.
  }
  \label{tab:Z_st_diboson}
\end{table}

\section{Summary of background processes with prompt photons}
\label{sec:backgroundphoton_combination}

Tab.~\ref{tab:promptphotonbkg} gives a summary of the estimates for background processes with prompt photons in the final state for \mbox{$1.04 \ifb$}
including the total uncertainties.

\begin{table}[h!]
  \center
  \begin{tabular}{|l|r@{}lcr@{}l|r@{}lcr@{}l|}
    \hline
    Process & \multicolumn{5}{c|}{e+jets} & \multicolumn{5}{c|}{$\mu$+jets} \\
    \hline
    Background $\ttg$   &  0 &.8  & $^{+}_{-}$ & $^{\emptyplus 1}_{\emptyminus 0}$ & $^{.1\emptyplus}_{.8\emptyminus}$   &  1 &.3  & $^{+}_{-}$ & $^{\emptyplus 1}_{\emptyminus 1}$ & $^{.9\emptyplus}_{.3\emptyminus}$ \\
    Multijet+$\gamma$   &  1 &.2 & $^{+}_{-}$ & $^{\emptyplus 1}_{\emptyminus 1}$ & $^{.6\emptyplus}_{.2\emptyminus}$ &  0 &.3 & $^{+}_{-}$ & $^{\emptyplus 1}_{\emptyminus 0}$ & $^{.0\emptyplus}_{.3\emptyminus}$ \\
    $W$+jets+$\gamma$     &  1 &.8  & $\pm$ & 0 &.7                        &  3 &.7  & $\pm$ & 1 &.4 \\
    $Z$+jets+$\gamma$     &  1 &.3  & $^+_-$ & $^{\emptyplus 2}_{\emptyminus 1}$ & $^{.4\emptyplus}_{.3\emptyminus}$ &  1 &.6  & $^+_-$ & $^{\emptyplus 2}_{\emptyminus 1}$ & $^{.3\emptyplus}_{.6\emptyminus}$ \\
    Single top+$\gamma$ &  0 &.6  & $^+_-$ & $^{\emptyplus 0}_{\emptyminus 0}$ & $^{.7\emptyplus}_{.6\emptyminus}$ &  0 &.2  & $^+_-$ & $^{\emptyplus 0}_{\emptyminus 0}$ & $^{.3\emptyplus}_{.2\emptyminus}$ \\
    Diboson+$\gamma$    &  0 &.16 & $^{+}_{-}$ & $^{\emptyplus 0}_{\emptyminus 0}$ & $^{.34\emptyplus}_{.16\emptyminus}$ &  0 &.04 & $^{+}_{-}$ & $^{\emptyplus 0}_{\emptyminus 0}$ & {$^{.18\emptyplus}_{.04\emptyminus}$} \\
    \hline
  \end{tabular}
  \caption[Expected yields for background processes with prompt photons]{
    Expected yields for background processes with prompt photons in the final state for \mbox{$1.04 \ifb$}
  }
  \label{tab:promptphotonbkg}
\end{table}

\chapter{Systematic uncertainties}
\label{sec:systematics}

Various effects may lead to systematic biases of the measurement of the $\ttg$ cross section.
The different sources of systematic uncertainties are discussed in the following:
Sec.~\ref{sec:syst_signalmodelling} describes the treatment of systematic uncertainties due to the modelling of the $\ttg$ signal.
Sec.~\ref{sec:syst_backgroundmodelling} discusses systematic uncertainties arising from the modelling of the various background processes.
Effects due to the modelling of the detector are presented in Sec.~\ref{sec:syst_detectormodelling}, and Sec.~\ref{sec:syst_luminosity}
discusses the uncertainty due to the measurement of the integrated luminosity.
The different sources of systematic uncertainties were combined, as described in Sec.~\ref{sec:syst_combination}.

Systematic effects may change the acceptance and efficiency of the $\ttg$ signal and the expectations for the background contributions with
misidentified electrons or prompt photons.
They may also influence the shape of the $\ptcone$ distributions used for the estimation of the background from hadrons misidentified as photons.

For the evaluation of the systematic uncertainties, ensembles of pseudo-data were created according to the expected yields for $\ttg$ and the different
background contributions.
The expected yields for the $\ttg$ processes were taken from the WHIZARD MC simulations (Sec.~\ref{sec:yields}).
The expectations for the background processes with electrons misidentified as photons and with prompt photons were derived in
Ch.~\ref{sec:electronfake} and~\ref{sec:backgroundphotons}, respectively.
For the contribution from hadrons misidentified as photons, the result of the final template fit (Ch.~\ref{sec:results})
was taken as expectation, because this was the most reliable estimate available.

For each systematic uncertainty, 5000 ensembles of pseudo-data were created by fluctuating each bin in the $\ptcone$ distribution according to the
Poissonian uncertainty around the expectation.
The pseudo-data were then fitted with modified efficiencies, background predictions and modified template shapes where applicable.
The mean difference with respect to ensemble tests with the nominal efficiencies, background predictions and template shapes was taken
as a measure of the systematic uncertainty.
The intrinsic statistical uncertainty of the tests with 5000 ensembles was found to be less than 0.4\%, and hence negligible.
Sources of systematic uncertainties which were found to yield an uncertainty smaller than the intrinsic uncertainty from the ensemble test,
were conservatively estimated to be as large as 0.4\%.

All systematic uncertainties were symmetrised by taking the largest systematic variation as a conservative estimate of the upper and lower
uncertainty.

\section{Signal modelling}
\label{sec:syst_signalmodelling}

The choice of the MC generator, the impact of including higher-order QCD effects, as well as the choice of the parton showering model,
the impact of the ISR and FSR modelling, and uncertainties on the choice of the PDFs were evaluated as sources of systematic uncertainties
for the modelling of the $\ttg$ signal.
All of these effects have an impact on the signal acceptance and efficiency only, and not on shape of the $\ptcone$ templates.
Since samples generated with alternative MC generators, a different parton showering and varied settings for ISR and FSR were only available for $\ttbar$
production (Sec.~\ref{sec:backgroundmodelling}) and not for $\ttg$ production, these effects were estimated by comparing the changes in acceptance
and efficiency for $\ttbar$ events for the preselection (Sec.~\ref{sec:preselection}).
The relative change found in $\ttbar$ production was assumed to be an estimate of the relative systematic uncertainty on the $\ttg$ acceptance and
efficiency.
Similarly, the relative uncertainty due to the PDF choice in $\ttbar$ events was taken as an estimate for the relative uncertainty on the $\ttg$ signal.

Additionally, the dependence of the signal acceptance and efficiency on the pile-up conditions was studied.

\paragraph{Monte Carlo generator:}
The acceptances and efficiencies for the preselection from $\ttbar$ samples generated with MC@NLO and POWHEG were compared.
Both generators were interfaced to HERWIG for parton showering.
The relative difference was taken as systematic uncertainty on the choice of the MC generator for $\ttg$ production.

\paragraph{Finite order calculation:}
The acceptances and efficiencies for the preselection from $\ttbar$ samples generated with POWHEG and AcerMC were compared.
While POWHEG is a NLO generator, AcerMC generates $\ttbar$ production in LO.
Both generators were interfaced to HERWIG for parton showering.
The relative difference was taken as systematic uncertainty on the dependence of the $\ttg$ modelling, generated in LO,
on higher-order QCD corrections.

\paragraph{Parton shower:}
The acceptances and efficiencies for the preselection from two $\ttbar$ samples generated with POWHEG were compared, where the HERWIG cluster
model and the PYTHIA Lund string model for parton showering were compared.
The relative difference was taken as systematic uncertainty due to the choice of the parton shower model.

\paragraph{Initial and final state radiation:}
The acceptances and efficiencies for the preselection from different $\ttbar$ samples generated with AcerMC interfaced to PYTHIA for parton
showering were compared, where the parameters which describe the amount of ISR and FSR in PYTHIA were varied in a range comparable to the
Perugia Soft/Hard tune variations~\cite{CSCbook, perugia}.
The largest deviation with respect to the nominal AcerMC sample was taken as systematic uncertainty on the modelling of ISR/FSR
in $\ttg$ production.

\paragraph{Parton density functions:}
The impact of the choice of the set of PDFs was found to be of the order of 2\% in previous cross section measurements in the single lepton
$\ttbar$ decay channel~\cite{ATLAStopXsec_3pb}.
The procedure applied for the evaluation of the systematic uncertainty was based on a reweighting of the events in the MC@NLO $\ttbar$ MC sample
according to different NLO PDF sets as described in~\cite{CSCbook}.
Half of the total envelope of all variations was taken as symmetric systematic uncertainty.

Since this uncertainty is small with respect to the contributions from other sources (Ch.~\ref{sec:results}), it was not evaluated separately
for this analysis.
Twice the uncertainty found in~\cite{ATLAStopXsec_3pb} was taken as a conservative estimate.

\paragraph{Pile-up conditions:}

\begin{figure}[h]
  \begin{center}
    \includegraphics[width=0.49\textwidth]{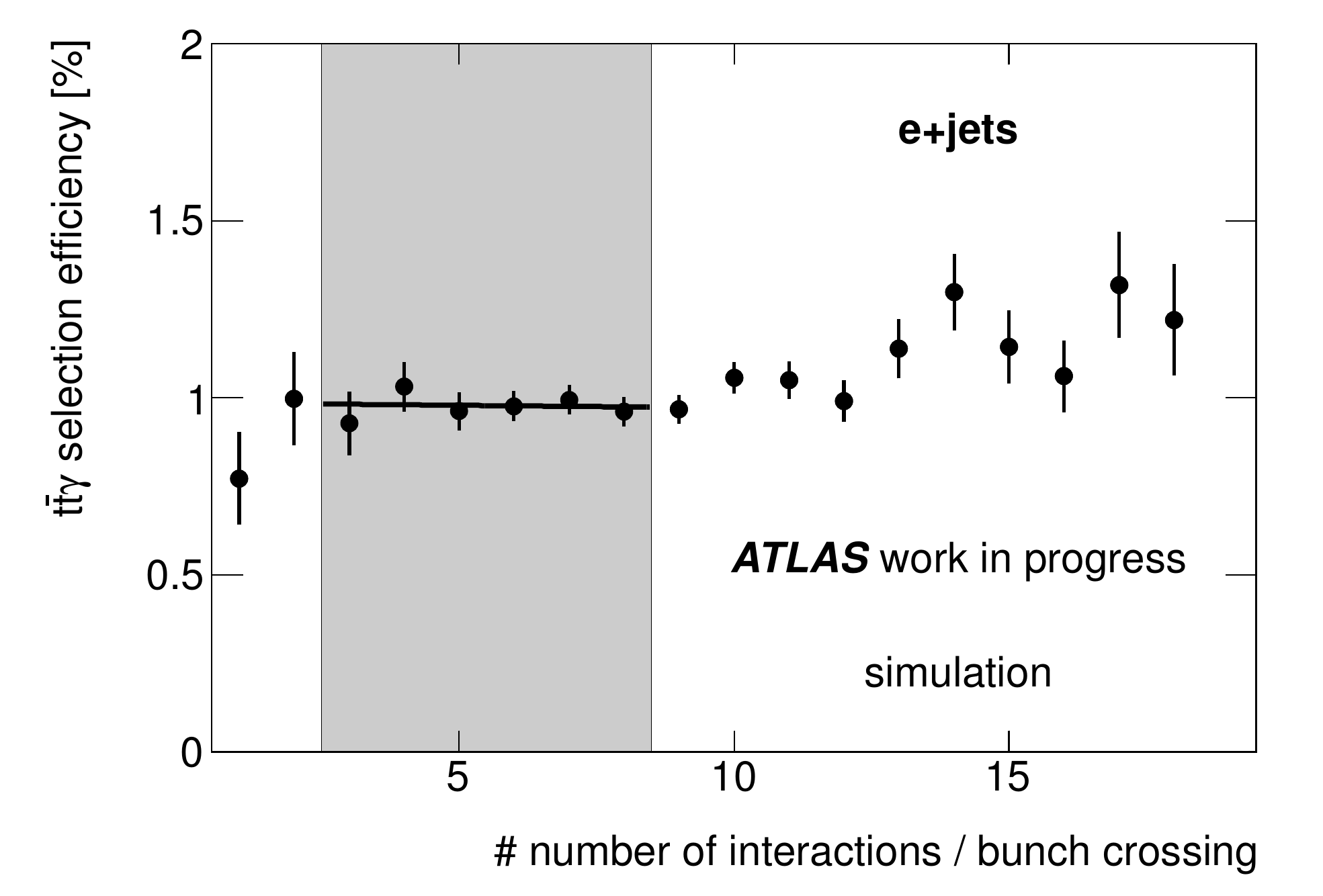}
    \includegraphics[width=0.49\textwidth]{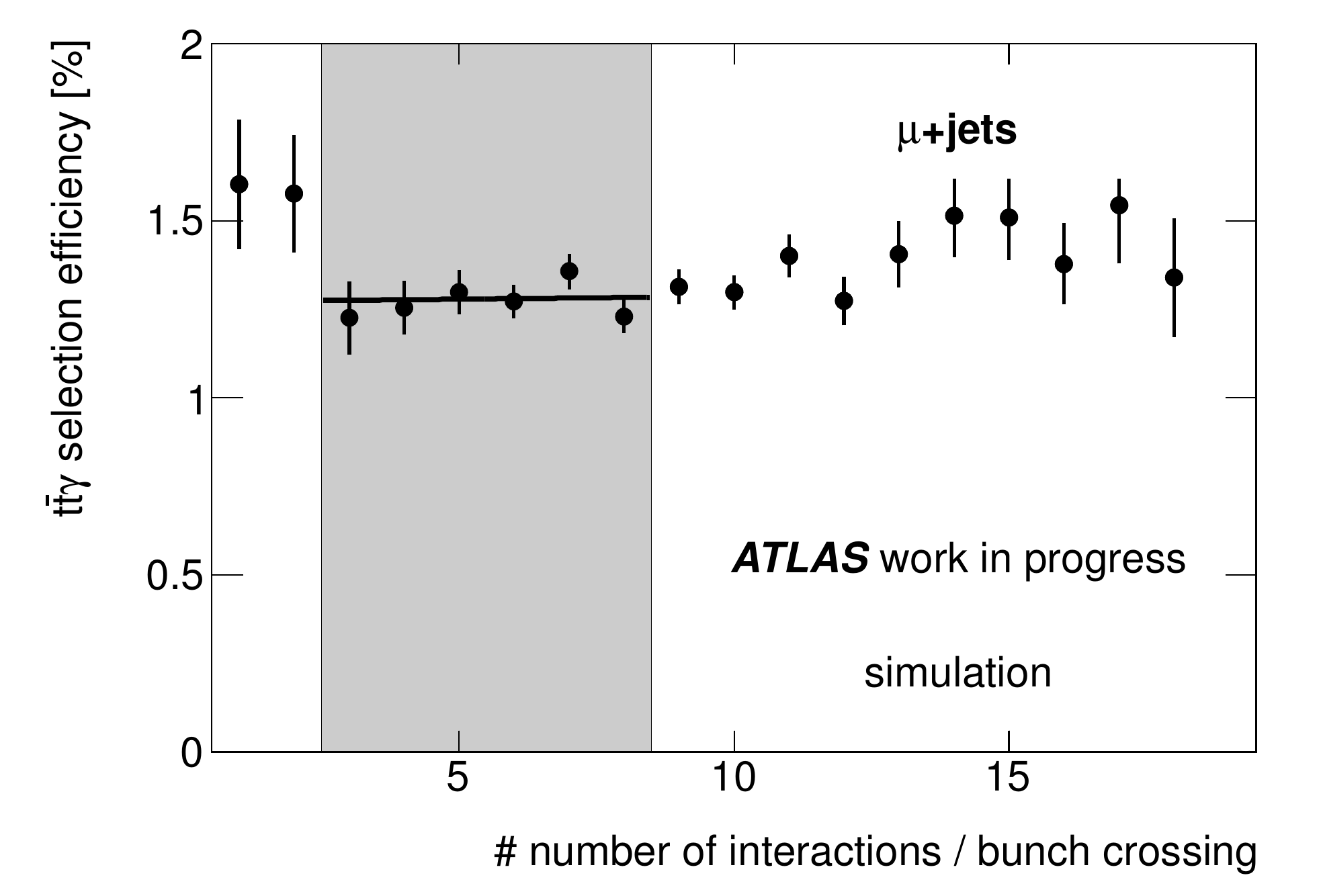}
    \caption[Pile-up dependence of the $\ttg$ selection efficiency]{
      Total selection efficiency of simulated $\ttg$ events as a function of the number of interactions per bunch crossing in the electron
      channel (left) and in the muon channel (right).
      The shaded regions indicate the number of interactions per bunch crossing present in the data analysed in this thesis.
      Additionally, a linear fit to these regions is shown.
    }
    \label{fig:pileup}
  \end{center}
\end{figure}

The stability of the signal modelling with respect to the pile-up conditions was checked by studying a possible dependence of the selection efficiency
of $\ttg$ simulations on the number of interactions per bunch crossing.
This is shown in Fig.~\ref{fig:pileup} for the electron (left) and for the muon channel (right).
The selection efficiency increases slightly with increasing pile-up.
However, as shown in Fig.~\ref{fig:mu}, nearly all data were taken with an average number of interactions between three and eight as indicated
by the grey regions in Fig.~\ref{fig:pileup}.
In this regime, a stable selection efficiency was observed in the simulations, which is illustrated by the linear fits in Fig.~\ref{fig:pileup}.
Also, MC simulations were reweighted to the pile-up conditions present in data, so that pile-up effects were safely ignored as a source
of systematic uncertainty on the signal modelling in this analysis.

\section{Background modelling}
\label{sec:syst_backgroundmodelling}

The number of events from processes with hadrons misidentified as photons was estimated using a fit to the $\ptcone$ distribution of the photon
candidates (Ch.~\ref{sec:strategy}).
Hence, effects which change the shape of either the prompt photon template or the hadron fake template were considered as sources of
systematic uncertainties.

Moreover, uncertainties on the background contributions with electrons misidentified as photons and on the backgrounds with prompt photons were
considered.
Some of the contributing processes were modelled with MC simulations, which are subject to several sources of systematic uncertainty:
the cross sections for the respective processes are only known with a certain precision (Sec.~\ref{sec:backgroundmodelling}) and
the luminosity measurement was subject to uncertainties (Sec.~\ref{sec:syst_luminosity}).
The uncertainties on the cross sections read $^{+\emptynull7\%}_{-10\%}$ for $\ttbar$, $48\%$ for Z+jets, and $5\%$ for diboson production.
For single top production, the uncertainties are as large as $^{+4\%}_{-3\%}$, $\pm 4\%$ and $\pm 7\%$ in the $t$-, $s$- and $Wt$-channels,
cf. Sec.~\ref{sec:backgroundmodelling}.
Additionally, uncertainties in the detector simulation (Sec.~\ref{sec:syst_detectormodelling}) affect all simulated processes.
The luminosity and the detector modelling uncertainties were treated as correlated between all MC samples as described in
Sec.~\ref{sec:syst_detectormodelling} and~\ref{sec:syst_luminosity}.

Additionally, the effect of limited available MC statistics was evaluated for all background contributions estimated from MC simulations.
The upper (lower) limit was estimated by the expectation value of a Poissonian distribution giving a probability smaller than 16\% for
observing less (more) events than estimated in the simulations.

The uncertainties on the cross section predictions were also treated as correlated between the different processes which were estimated from the same
MC sample, as for $Z$+jets+$\gamma$ production and $Z$+jets production with an electron misidentified as a photon.
Template fits were performed with cross section predictions varied within their respective uncertainties and the largest difference
with respect to the nominal template fit was taken as systematic uncertainty on the fit result.

\begin{figure}[p]
  \begin{center}
    \includegraphics[width=0.49\textwidth]{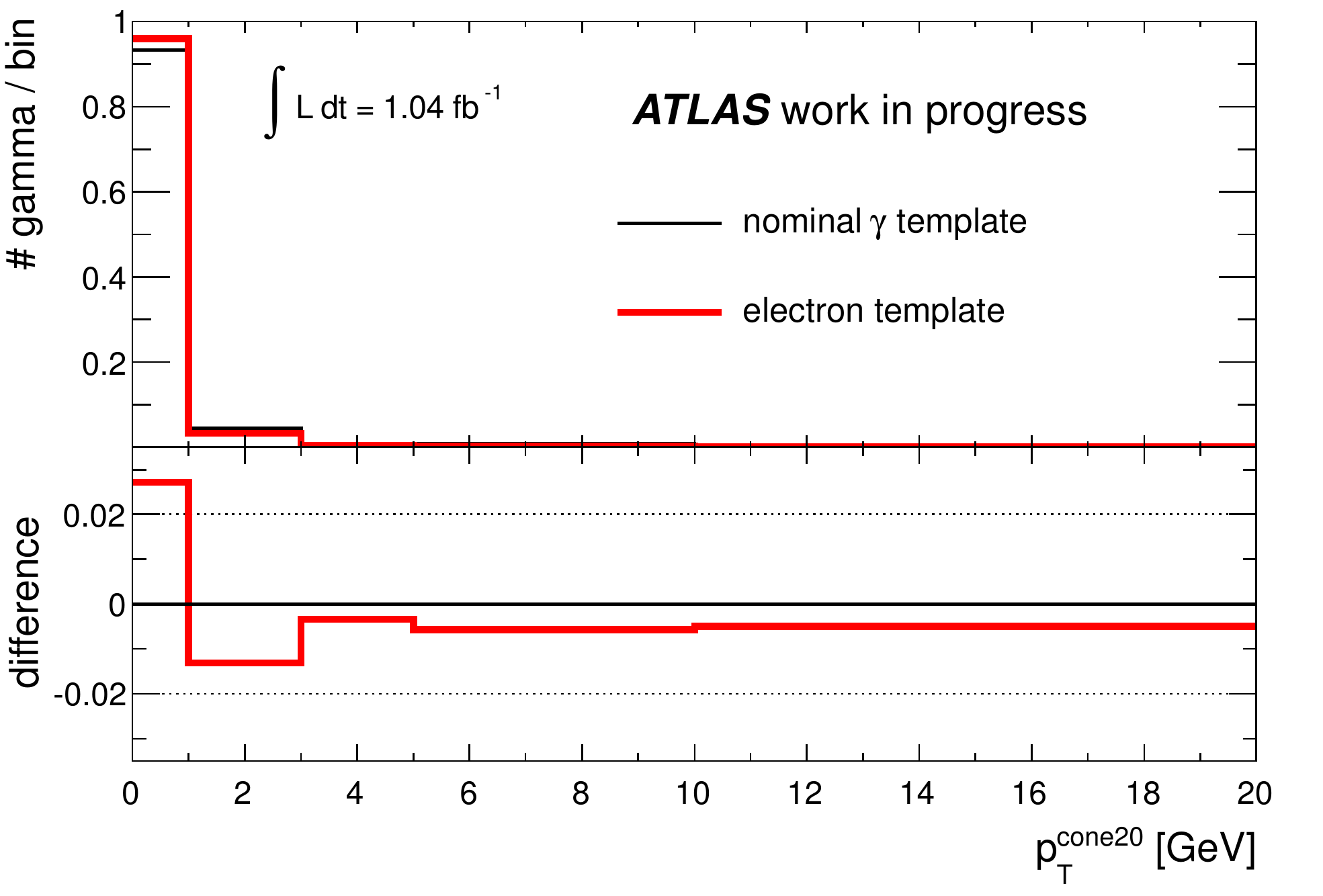} \\
    \includegraphics[width=0.49\textwidth]{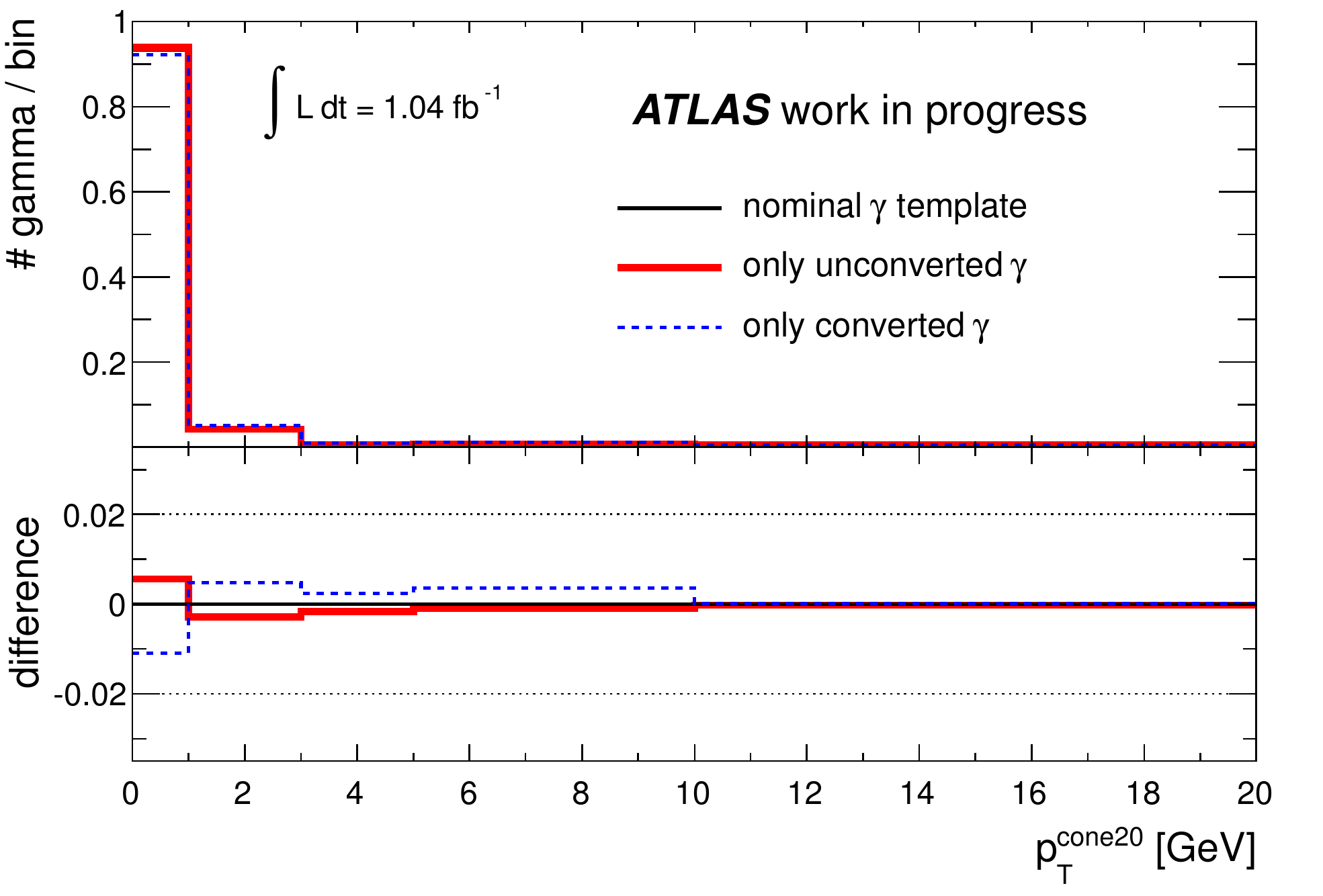}
    \includegraphics[width=0.49\textwidth]{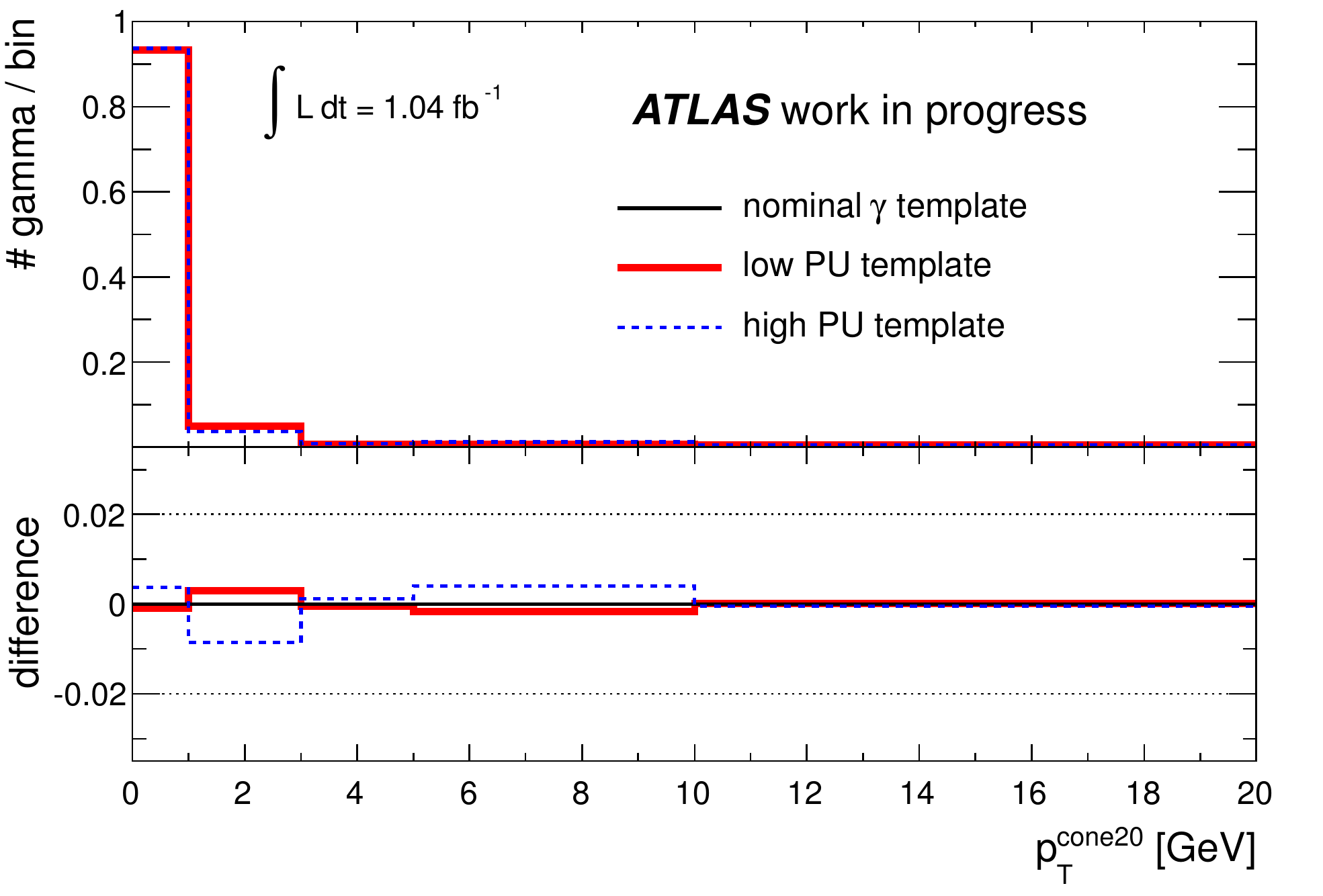}
    \caption[Comparison of the nominal prompt photon $\ptcone$ template with systematically varied templates]{
      Comparison of the nominal prompt photon $\ptcone$ template (thin solid line) with systematically varied templates (thick solid and
      thin dashed lines) due to different effects:
      the difference between photon and electron templates (upper plot), the fraction of unconverted and converted photons (lower left plot), and
      the dependence of the $\ptcone$ distribution on the pile-up conditions (lower right plot).
      Below each plot, the difference between the systematically varied templates and the nominal template is shown.
      In all plots, the last bin includes the overflow bin.
   }
    \label{fig:sigsysttemplates}
    \vspace{0.025\textwidth}
    \includegraphics[width=0.49\textwidth]{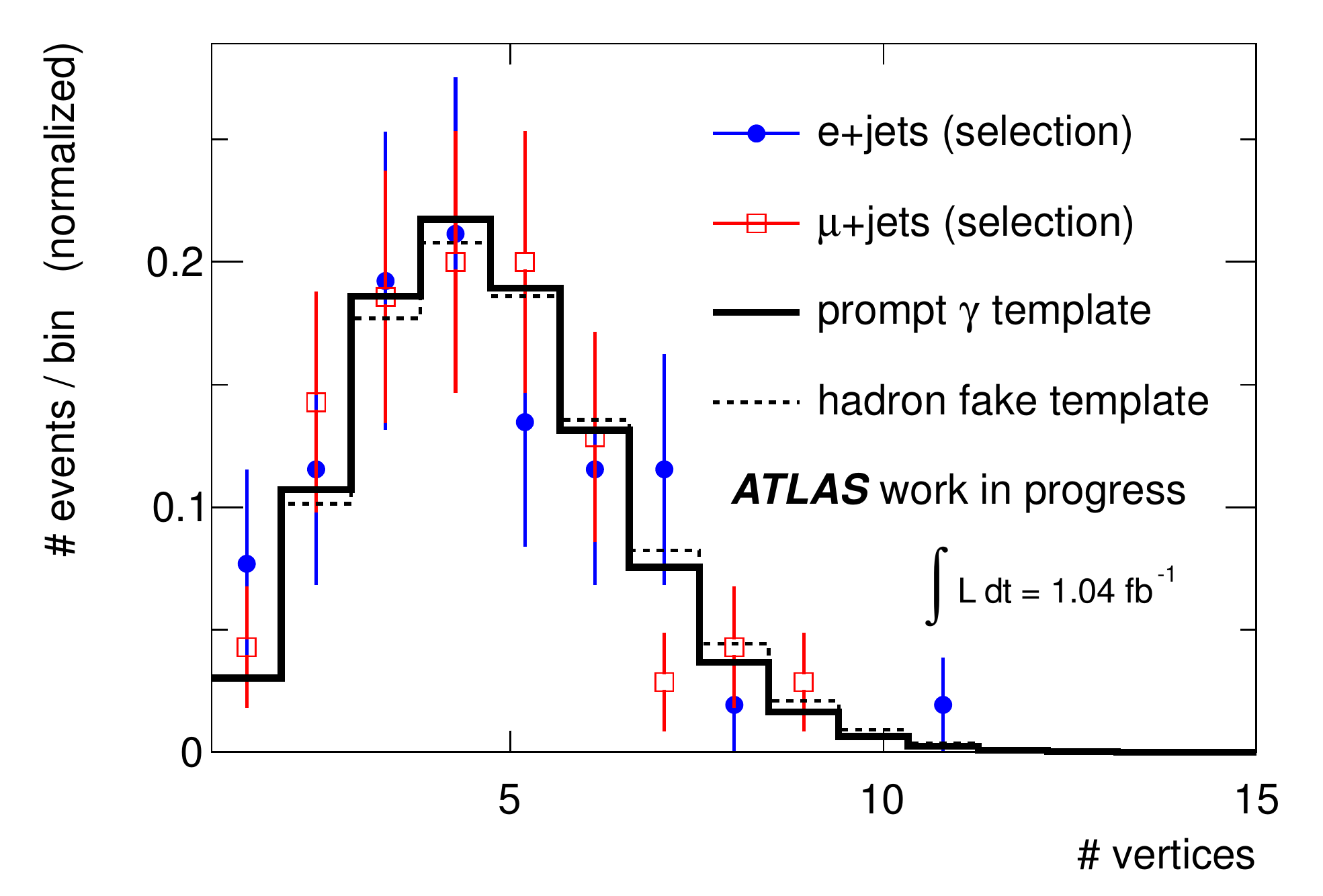}
    \caption[Distribution of the number of primary vertices in different data samples]{
      Normalised distribution of the number of primary vertices of the selected events in the electron channel (solid circles) and in the muon channel
      (open squares).
      Additionally, the distribution for the $\Zee$ sample which was used for the derivation of the electron templates is shown (thick solid line)
      as well as the distribution for the control region used for the derivation of the hadron fake template (thin dashed line).
      The last bin includes the overflow bin.
    }
    \label{fig:nVertices}
  \end{center}
\end{figure}

\paragraph{\boldmath $\ptcone$ \unboldmath template for prompt photons:}
The derivation of the prompt photon template from electron distributions in \Zee events was discussed in Ch.~\ref{sec:photontemplate}.
In particular, a MC correction for the differences between the electron and photon distributions was applied.
The nominal photon template is shown together with the uncorrected electron template in the upper plot in Fig.~\ref{fig:sigsysttemplates}.
The uncertainty due to the MC correction was estimated by comparing template fits with the two different templates.
The difference between the two fits was taken as systematic uncertainty.

The shape of the prompt photon template for unconverted and converted photons only is shown in the lower left plot of Fig.~\ref{fig:sigsysttemplates}
in comparison to the nominal template.
The dependence on the fraction of converted photons is only due to the MC correction applied to the electron templates from \Zee data.
Since this fraction is unknown, the nominal fit was compared to fits with the unconverted- and converted-only templates.
The largest difference with respect to the nominal fit was taken as systematic uncertainty.

The number of primary vertices per event is a measure of the amount of pile-up.
Fig.~\ref{fig:nVertices} shows the distribution of the number of primary vertices for the selected events in the electron and the muon channel.
Additionally, the distribution for the $\Zee$ sample which was used for the derivation of the electron templates is shown
as well as the distribution for the control region (CR) used for the derivation of the hadron fake template as defined in Ch.~\ref{sec:faketemplate}.
The distribution in the \Zee sample is consistent with those in the selected $\ttg$ candidate events in both lepton channels given the
limited statistics, and hence the prompt photon template represents the same pile-up conditions as present in the candidate events.
For the hadron fake CR, a slightly higher average pile-up is observed than in the \Zee sample, which is discussed in the next section treating the
hadron fake template.

Nevertheless, the dependence of the prompt photon template on pile-up was studied and the lower right plot in Fig.~\ref{fig:sigsysttemplates} shows
templates derived for a low pile-up regime with \mbox{1 -- 5} primary vertices per event and a high pile-up regime with \mbox{6 -- 10} primary
vertices.
These two templates with extremely different pile-up conditions were tested in the template fit and the largest difference with respect to the nominal
sample was taken as a very conservative estimate for the systematic uncertainty due to a mismodelling of the pile-up conditions.

\paragraph{\boldmath $\ptcone$ \unboldmath template for hadrons misidentified as photons:}
As discussed in Ch.~\ref{sec:faketemplate}, the template for hadrons which were misidentified as photons showed a dependence on the object's
$\et$ and $\eta$, and the templates were reweighted accordingly.

\begin{figure}[h]
  \begin{center}
    \includegraphics[width=0.49\textwidth]{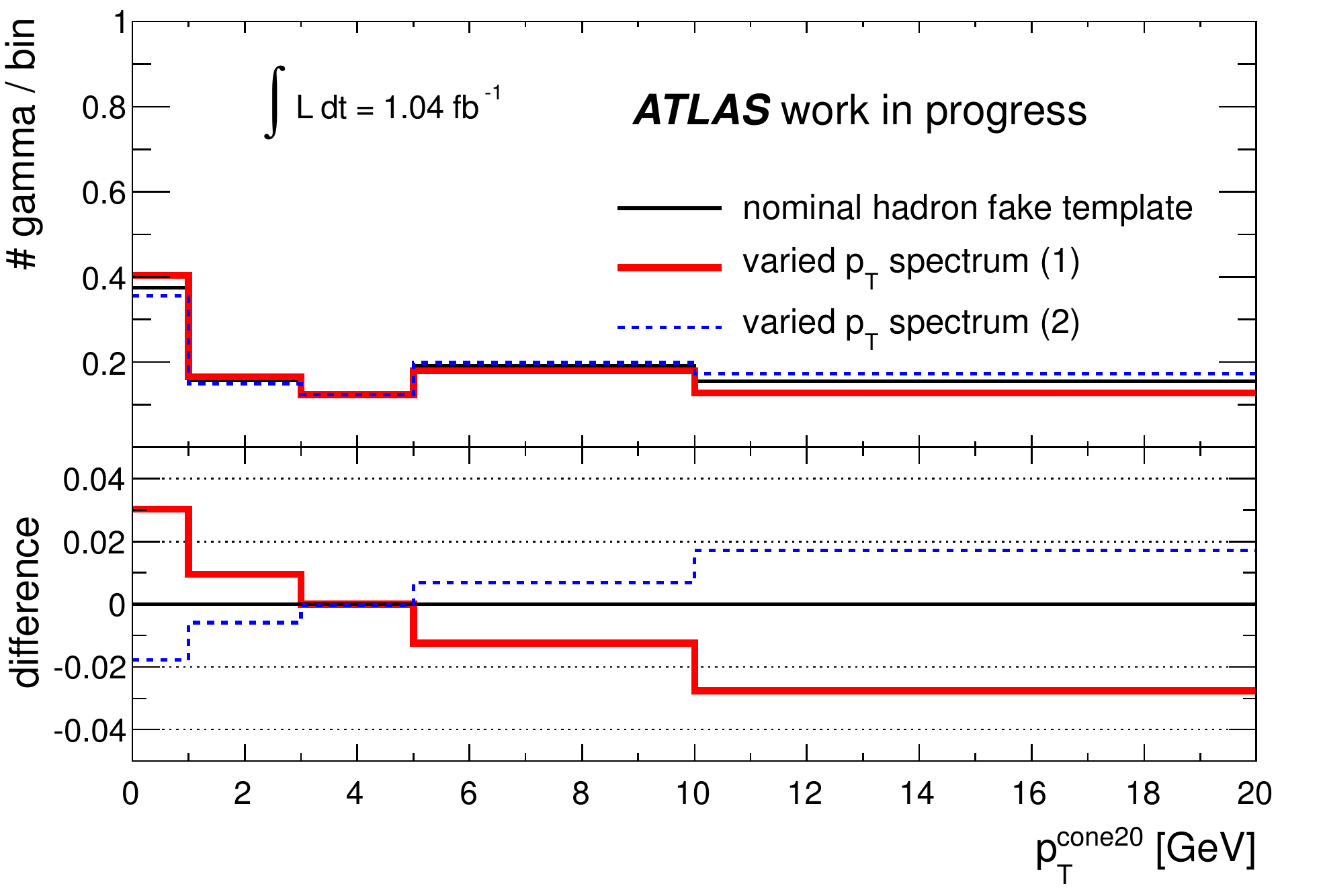}
    \includegraphics[width=0.49\textwidth]{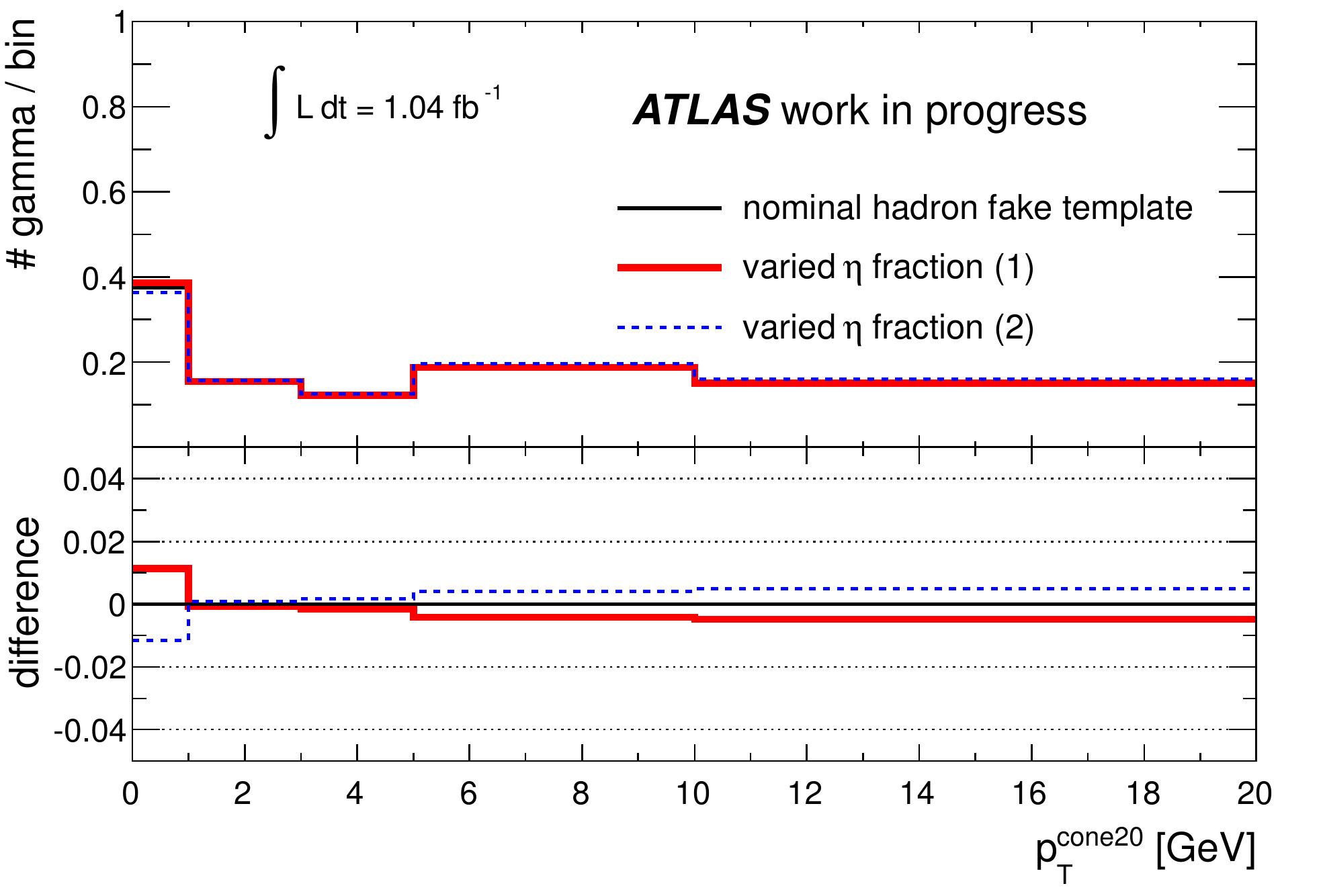} \\
    \includegraphics[width=0.49\textwidth]{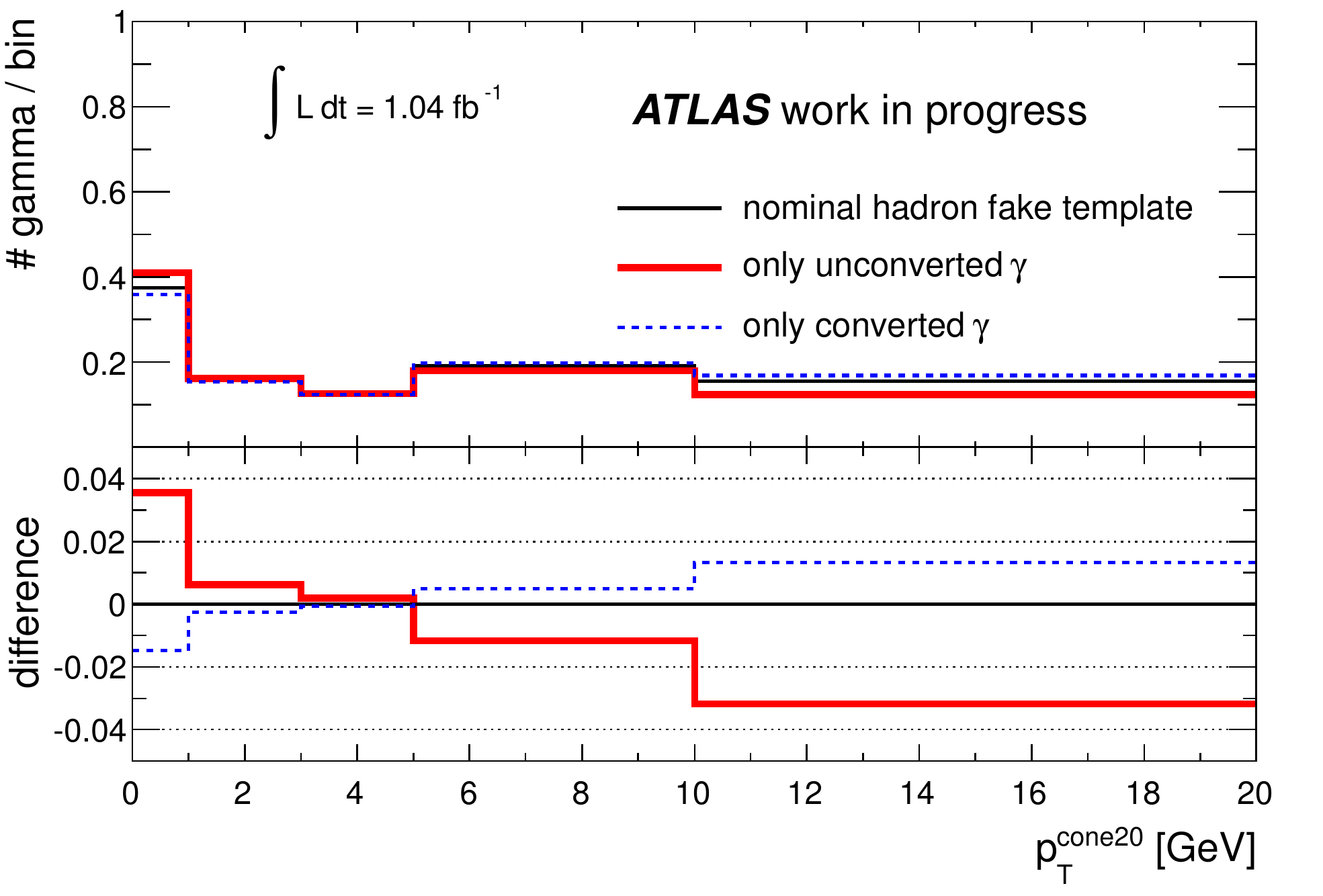}
    \includegraphics[width=0.49\textwidth]{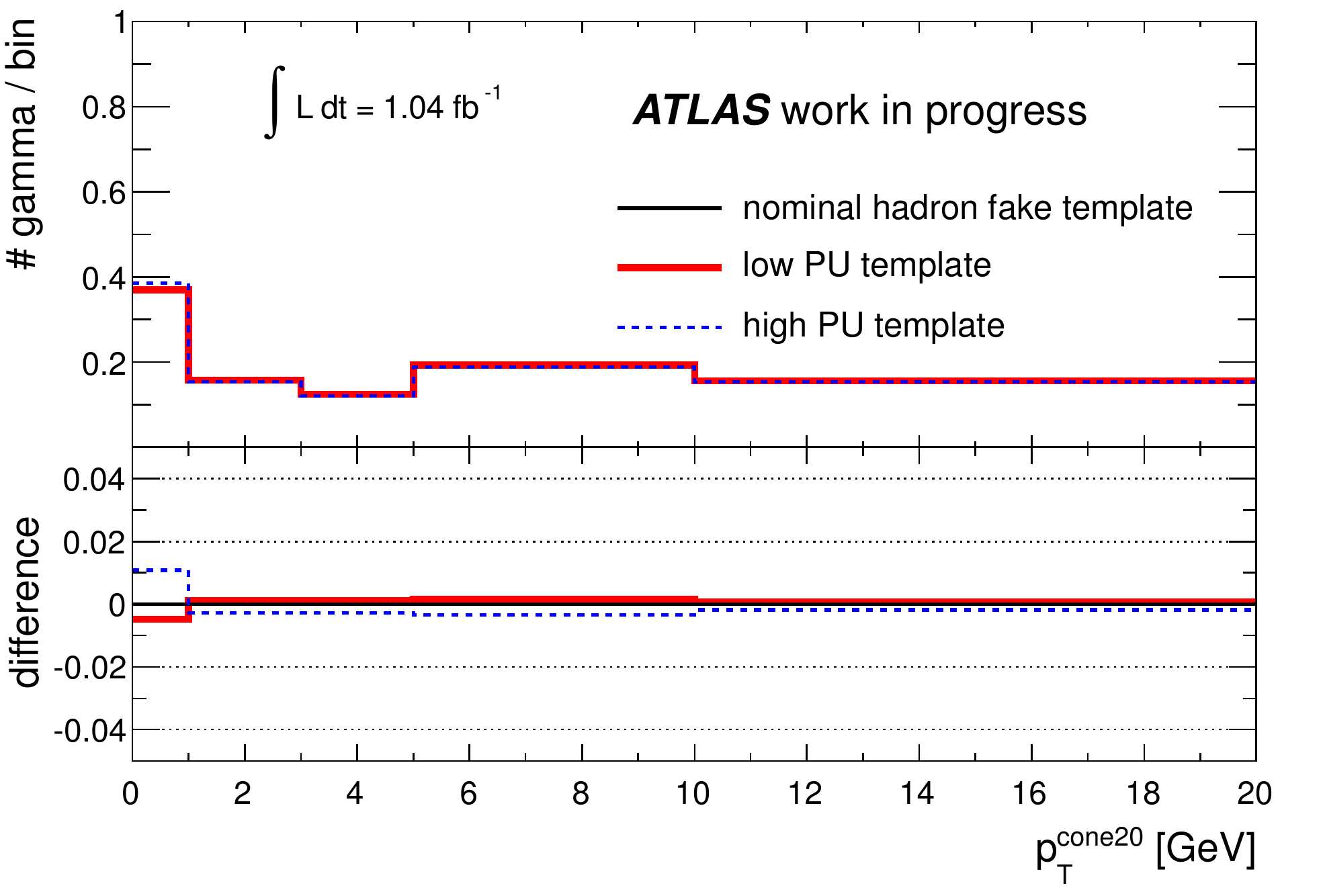}
    \caption[Comparison of the nominal $\ptcone$ template for hadron fakes with systematically varied templates]{
      Comparison of the nominal $\ptcone$ template for hadron fakes (thin solid line) with systematically varied templates (thick solid and
      thin dashed lines) due to different effects:
      the uncertainties on the derivation of the $\et$ spectrum of the hadron fakes (upper left plot) and the amount of hadron fakes with large $|\eta|$
      (upper right plot),
      the fraction of unconverted and converted hadron fakes (lower left plot), and
      the dependence of the $\ptcone$ distribution on the pile-up conditions (lower right plot).
      Below each plot, the difference between the systematically varied templates and the nominal template is shown.
      In all plots, the last bin includes the overflow bin.
    }
    \label{fig:bkgsysttemplates}
  \end{center}
\end{figure}

For the reweighting in $\et$, an exponential fit to the $\et$ distribution in a CR was used, which was subject to significant uncertainties due to
the limited amount of hadron fake candidates in the CR.
The upper left plot in Fig.~\ref{fig:bkgsysttemplates} shows the nominal hadron fake template together with templates which were reweighted with
exponential curves corresponding to variations of the mean lifetime within the uncertainties from the nominal exponential fit.
An additional systematic uncertainty of 25\% was added on the mean lifetime of the exponential due to differences between true hadron fakes and
hadron fake candidates observed in $\ttbar$ simulations (Ch.~\ref{sec:faketemplate}).
The largest difference of the template fits with the two alternative templates with respect to the nominal fit
was taken as systematic uncertainty.

The reweighting in $\eta$ was done in two regions \mbox{$0 \leq |\eta| < 1.81$} and \mbox{$1.81 \leq |\eta| < 2.37$}, where the fraction
of high-$|\eta|$ hadron fakes was also estimated from the CR.
The upper right plot in Fig.~\ref{fig:bkgsysttemplates} shows the nominal hadron fake template together with templates which were reweighted 
according to varied fractions of high-$|\eta|$ hadron fakes within the uncertainties derived in Ch.~\ref{sec:faketemplate}.
The largest difference of the template fits with the two alternative templates with respect to the nominal fit
was taken as systematic uncertainty.

As for the prompt photon templates, the fraction of unconverted and converted photons from misidentified hadrons was unknown.
The lower left plot in Fig.~\ref{fig:bkgsysttemplates} shows the templates for unconverted and converted hadron fakes only
in comparison to the nominal template.
The nominal fit was compared to fits with the unconverted- and converted-only templates, and
the largest difference with respect to the nominal fit was taken as systematic uncertainty.

The average number of primary vertices is slightly larger for the events in the CR used for the derivation of the hadron fake template compared
to the \Zee samples from which the prompt photon template was derived (Fig.~\ref{fig:nVertices}).
This is expected, since with increasing pile-up, the number of events with at least one photon which is faked by a hadron from fragmentation may increase
because of the increasing number of jets, while the number of $Z$ events is not expected to increase significantly.

Although the pile-up conditions are in good agreement with those present in the $\ttg$ candidate events in both lepton channels,
the dependence of the hadron fake template on the amount of pile-up was studied.
The lower right plot in Fig.~\ref{fig:bkgsysttemplates} shows templates derived for a low pile-up regime with \mbox{1 -- 5} primary vertices per
event and a high pile-up regime with \mbox{6 -- 10} primary vertices.
These two templates with extremely different pile-up conditions were tested in the template fit and the largest difference with respect to the nominal
sample was taken as a very conservative estimate for the systematic uncertainty due to a mismodelling of the pile-up conditions.

\paragraph{Estimate of background contributions with electrons misidentified as photons:}
The contribution from processes with electrons which were misidentified as photons was derived in Ch.~\ref{sec:electronfake}.
In particular, scale factors (SFs) for the misidentification rate were derived in order to correct MC simulations.
The resulting yields were presented in Sec.~\ref{sec:egammaapplication} together with systematic uncertainties.
%due to the uncertainty on the SFs,
%the limited MC statistics, the uncertainty on the cross sections of the processes involved, the luminosity uncertainty, and due to the detector
%modelling.

The SFs for the electron-to-photon misidentification rate were varied within their uncertainties.
Moreover, the background estimate was varied within the uncertainty due to the limited MC statistics.
In both cases, the largest difference with respect to the nominal template fit result was taken as systematic uncertainty.

Uncertainties due to detector modelling, the predictions for the cross sections of the different processes, and due to the luminosity measurement,
were treated separately in order to account for the correlations between the processes estimated from MC simulations, as stated above.

\paragraph{Estimate of background contributions with prompt photons:}
The contribution from background processes with prompt photons was discussed in Ch.~\ref{sec:backgroundphotons}.
The systematic uncertainties on the predictions were found to be sizable:
they are at least as large as 100\% for processes which were believed to be only modelled with limited precision in MC simulations.
This holds for $\ttg$ production outside of the signal phase space as well as for backgrounds from $Z$+jets+$\gamma$, single top+$\gamma$ and
diboson+$\gamma$ production.
The limited MC statistics was taken into account in the uncertainties.

The estimates for $W$+jets+$\gamma$ production and for multijet processes with prompt photons were derived with data-driven techniques
involving template fits to the $\ptcone$ distribution of the photon candidates in the respective CRs.
Both estimates feature large uncertainties from the limited statistics in the CRs.
The sources of systematic uncertainties on the prompt photon and hadron fake templates as described above were evaluated also in these template fits.
Additionally, the $W$+jets+$\gamma$ estimate was found to be subject to uncertainties from the extrapolation to the signal region.
The multijet+$\gamma$ estimate features large uncertainties due to the amount of fake leptons.

In order to estimate the systematic uncertainty from backgrounds with prompt photons, the estimate of each contributing process was varied within its
uncertainty in the template fit.
The largest deviation with respect to the nominal fit result was taken as systematic uncertainty in each case.

As stated above, uncertainties on the cross sections, the luminosity, and the detector modelling were treated separately in order to account
for correlations between the processes estimated from MC simulations.

\section{Detector modelling}
\label{sec:syst_detectormodelling}

The MC simulations used in this analysis included a detailed simulation of the ATLAS detector.
The modelling of the detector is subject to systematic uncertainties which translate into uncertainties on the description of the different physics
objects:
uncertainties on the electron, muon, jet and photon modelling were evaluated as well as uncertainties on the description of the $\met$ measurement
and the $b$-tagging performance.
An additional systematic uncertainty was evaluated for the modelling of the broken optical links in the LAr calorimeter (Ch.~\ref{sec:data}).

Systematic uncertainties on the detector modelling do not influence the shape of the $\ptcone$ templates, but only change the respective acceptances
and efficiencies of the signal and background processes.
They were evaluated for all processes estimated from MC simulations, including the $\ttg$ signal as well as the background MC simulations.
The variations were treated consistently between all processes in order to account for the correlations between the samples due to the
different systematic effects.

\paragraph{Electron modelling:}
MC simulations were corrected by SFs for the mismodelling of electron trigger, reconstruction and identification efficiencies, which were derived from
\Zee and \mbox{$W \to e\nu$} data with tag-and-probe methods~\cite{electronperformance} (Sec.~\ref{sec:electron}).
The SFs were varied within their respective uncertainties and the largest difference obtained with respect to the nominal template fit
was considered as systematic uncertainty.

The electron energy scale was corrected in data based on measurements in \Zee events~\cite{electronperformance}.
In order to evaluate the associated systematic uncertainty, the electron scale in MC was varied within the relative scale uncertainty.
The electron energy resolution was also measured in \Zee events~\cite{electronperformance}, and
the resolution in MC simulations was corrected to match the resolution observed in data.
The systematic uncertainty was estimated by comparing template fits with larger and smaller energy resolutions according to the uncertainty
on the resolution measurement.

The largest deviation with respect to the nominal template fit was taken as systematic uncertainty for the energy scale as well as for the
energy resolution.

\paragraph{Muon modelling:}
As for electrons, also for muons, SFs for the mismodelling of the trigger, reconstruction and identification efficiencies were applied to MC
simulations~\cite{muonPerf}
(Sec.~\ref{sec:muon}).
The SFs were derived with tag-and-probe methods from $Z \to \mu^+\mu^-$ events.
They were varied within their respective uncertainties and the largest difference obtained with respect to the nominal fit
was considered as systematic uncertainty.
Since trigger matching was not applied in data in the muon channel, an additional systematic uncertainty was considered, which was estimated to be
as large as 2.2\% for events with zero or one $b$-tags, and 1.5\% for events with at least two $b$-tags.

The muon momentum scale and resolution were corrected in MC simulations based on measurements in $Z \to \mu^+\mu^-$ events~\cite{muonResolution}.
In order to evaluate the associated systematic uncertainties, the template fit was performed with varied
scales and resolutions within their respective uncertainties.
For both effects, the largest deviation with respect to the nominal template fit was taken as systematic uncertainty.

\paragraph{Jet modelling:}
The jet energy scale (JES) was derived using data from test beams and LHC collisions, and from simulations~\cite{jetperf} (Sec.~\ref{sec:jet}).
The total JES uncertainty is of the order of 4\% (6\%) for low-$\pt$ jets in the range \mbox{$0.3 \leq |\eta| < 0.8$} (\mbox{$2.1 \leq |\eta| < 2.8$}) and
decreases to roughly 2\% (2.5\%) for jets with \mbox{$60 \GeV < \pt < 800 \GeV$} (Fig.~\ref{fig:jetperformance}).
Additional uncertainties were added due to the different flavour composition and the higher amount of close-by jets present in $\ttbar$ events
with respect to the dijet topologies used for the derivation of the JES.
In order to account for a possible pile-up dependence, an uncertainty was considered of up to 5\% in the central and up to 7\% in the forward
region (\mbox{$|\eta| > 2.1$}).
Jets originating from $b$-quarks may have a slightly different energy scale than light quark and gluon jets.
This was accounted for by an additional uncertainty on the $b$-JES, as large as 2.5\% for a $\pt$ of \mbox{$20 \GeV$}, and decreasing to 0.0076\% for a $\pt$ of
\mbox{$600 \GeV$}.
The uncertainties on the JES, its pile-up dependence and the JES for $b$-jets were varied within their respective uncertainties and for each effect,
the largest difference with respect to the nominal fit was considered as systematic uncertainty.

The jet energy resolution (JER) was measured in dijet events~\cite{JER} and was found to agree with MC simulations within 10\%.
In order to account for a possible bias due to the JER modelling, the resolution was worsened in the simulations according to the
uncertainty of the JER measurement.
The difference with respect to the nominal fit was symmetrised and considered as systematic uncertainty.

The jet reconstruction efficiency was measured by comparing jets built from calorimeter clusters to jets built from tracks~\cite{jetperf}.
Data and simulations were found to be in agreement within 2\%.
The associated systematic uncertainty was estimated by randomly disregarding jets in MC simulations based on the uncertainty on the reconstruction
efficiency.
The difference of the template fit with respect to the nominal fit was taken as systematic uncertainty.

\paragraph{Photon modelling:}
The photon identification efficiency was measured by shifting the MC shower shapes so that they matched the shower shapes in
data~\cite{diphoton, promptphoton} (Sec.~\ref{sec:photon}).
The associated relative systematic uncertainties were taken from the diphoton cross section measurement~\cite{diphoton}.
The evaluation of the uncertainties included the intrinsic precision of the method, the choice of the photon candidate sample, the knowledge of the
material in front of the calorimeter, the admixture of fragmentation photons and the classification of unconverted and converted photon candidates.
The efficiency was varied within its uncertainty and the largest difference with respect to the nominal result was taken as systematic uncertainty.

Differences in reconstruction efficiencies between data and simulations were found to be smaller than 1\% for electrons with uncertainties
smaller than 0.7\% for \mbox{$|\eta| < 2.37$}.
Since photons were reconstructed within the same algorithm (Sec.~\ref{sec:photon}),
differences in the photon reconstruction efficiency between data and simulations can be expected to be of the same order of magnitude.
The uncertainties due to the photon identification efficiency are significantly larger, and the additional uncertainty due to the mismodelling
of the reconstruction efficiencies was believed to be covered by the uncertainty on the identification efficiencies.

The photon energy scale in data was corrected based on the electron scale measurement in \Zee events~\cite{electronperformance}.
In order to evaluate the associated systematic uncertainty, the scale in MC simulations was varied within the relative scale uncertainty.
The electron energy resolution calibration was used to correct also the photon energy resolution in MC simulations.
The systematic uncertainty was estimated by comparing template fits with larger and smaller energy resolutions according to the uncertainty
on the resolution measurement.
The largest deviation with respect to the nominal template fit was taken as systematic uncertainty due to the energy scale as well as due to the
energy resolution, respectively.

\paragraph{\boldmath $\met$ \unboldmath modelling:}
All systematic variations of the energy scales and resolutions of leptons, jets and photons were propagated consistently into the calculation
of the missing transverse energy.
Additional systematic uncertainties were due to a possible mismodelling of the soft jet and cell out terms in the $\met$, cf. Sec.~\ref{sec:MET}
and Eq.~(\ref{eq:met}).
Moreover, a 10\% uncertainty was added in order to cover a possible pile-up dependence of the $\met$ measurement not modelled in the MC simulations.
All of these sources of systematic uncertainties were treated as correlated.

For each effect, the $\met$ measurement was varied within the respective uncertainties and the largest difference with respect to the
nominal template fit was considered as systematic uncertainty.

\paragraph{Modelling of the \boldmath $b$\unboldmath-tagging performance:}
The $b$-tagging efficiencies and mistag rates were measured in data~\cite{btagcalibration, system8} (Sec.~\ref{sec:btagging}), and SFs for the
correction of the MC simulations were derived.
The uncertainties on the SFs vary between 0.06 and 0.15 for the $b$-tagging efficiencies and between 0.11 and 0.22 for the mistag rates.
Template fits with varied efficiencies according to the uncertainties on the SFs were compared to the nominal fit result, and the largest deviation
was taken as systematic uncertainty.

\paragraph{Modelling of LAr readout issues:}
A large fraction of the data analysed featured a broken optical link in the LAr barrel calorimeter (Ch.~\ref{sec:data}).
In data, events with a jet with a $\pt$ of at least \mbox{$20 \GeV$} which was closer to the affected calorimeter region were disregarded.
In simulations, events with such a jet were reweighted according to the fraction of the integrated luminosity which featured the broken optical link.
The systematic uncertainty on this procedure was evaluated by varying the $\pt$-threshold in MC simulations by $\pm \, 4 \GeV$.
Template fits with these variations were compared to the nominal fit result, and the largest deviation was taken as systematic uncertainty.

\section{Luminosity measurement}
\label{sec:syst_luminosity}

The luminosity of the LHC was measured with van der Meer scans~\cite{vdM}, and
the uncertainty on the luminosity measurement was estimated to 3.7\%~\cite{lumipaper,lumiconf}.
It was considered as common systematic uncertainty to all signal and background processes modelled with MC simulations, and
was treated as correlated between all of these contributions.

\section{Combination of systematic uncertainties}
\label{sec:syst_combination}

The systematic uncertainties due to signal, background and detector modelling were assumed to be uncorrelated and were hence summed in quadrature.
This holds also for the uncertainty on the luminosity measurement.
The only exception is the combination of the uncertainty on the pile-up dependence of the JES and the $\met$ uncertainties,
which were treated fully correlated.

\chapter{Results}
\label{sec:results}

In this chapter, the results of the final template fit to the $\ptcone$ distribution of the photon candidates are presented.
The results were translated into a $\ttg$ cross section estimate.
Moreover, the significance of the measured $\ttg$ signal was estimated.
At the end of this chapter, the evolution of the expected significance with the integrated luminosity is estimated and discussed.

\begin{figure}[h]
\begin{center}
\includegraphics[width=0.49\textwidth]{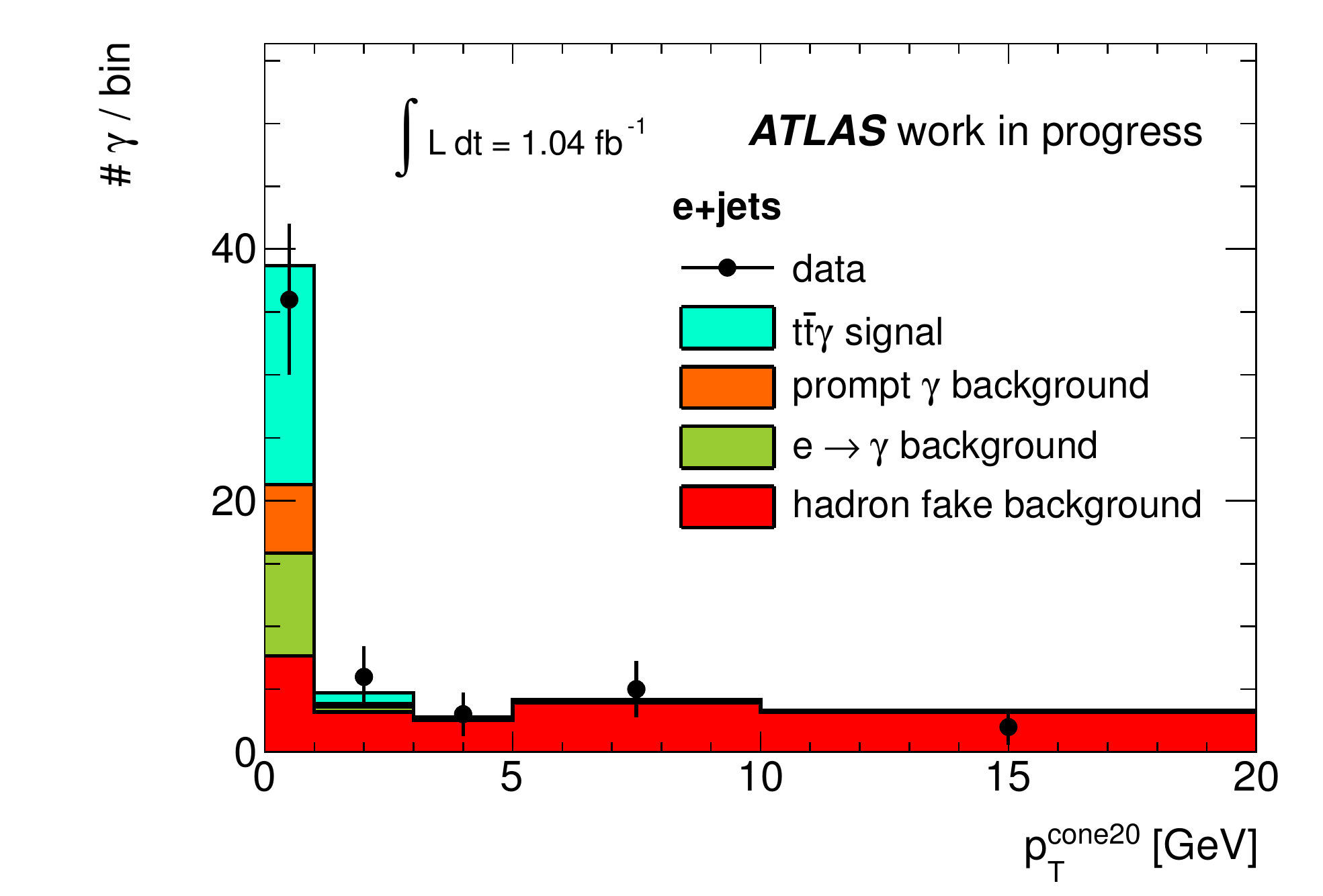}
\includegraphics[width=0.49\textwidth]{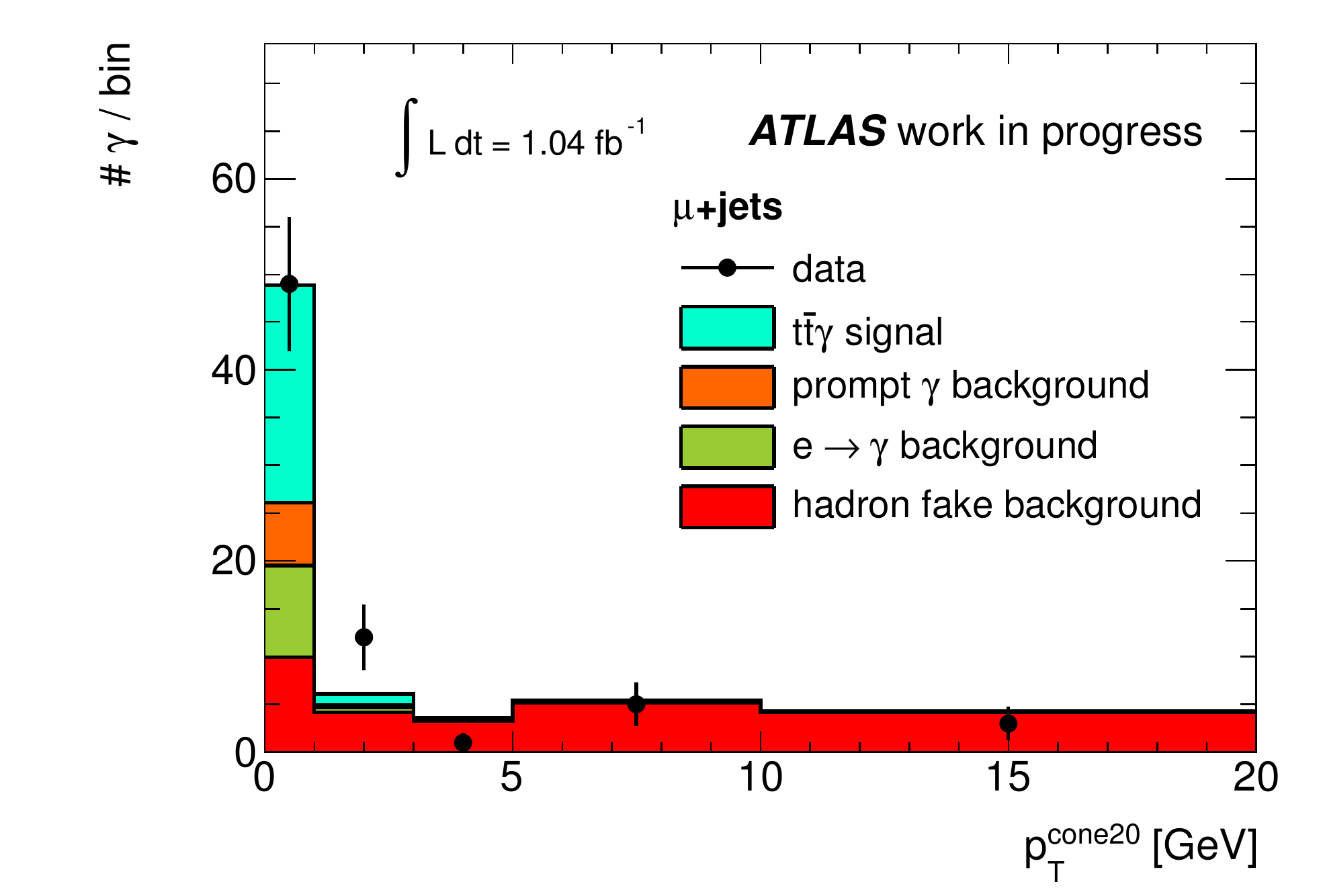}
\caption[Result of the combined template fit]{
  Result of the combined template fit to the photon $\ptcone$ distribution in both lepton channels.
  The left plot shows the distributions in the electron channel, the right plot shows the distributions in the muon channel.
  In both plots, the last bin includes the overflow bin.
}
\label{fig:combinedfit}
\end{center}
\end{figure}

In order to estimate the amount of $\ttg$ signal events and of background events with hadrons misidentified as photons, the template fit described in
Ch.~\ref{sec:strategy} was performed including the background estimates for processes with prompt photons and electrons misidentified as photons
(Ch.~\ref{sec:electronfake} and~\ref{sec:backgroundphotons}).
Fig.~\ref{fig:combinedfit} shows the result of the combined fit in both lepton channels together:
the left plot shows the resulting distributions in the electron channel, and the right plot shows the distributions in the muon channel.

\begin{table}[h]
  \center
  \begin{tabular}{|l|r@{}l c r@{}l|}
    \hline
    Contribution & \multicolumn{5}{c|}{Fit result / yield} \\
    \hline
    $\ttg$ signal                                                   & 43&.0 & $^+_-$ & $^{\emptyplus 10}_{\emptyminus 11}$ & $^{.8\emptyplus}_{.8\emptyminus}$ \\
    Background with prompt photons (e+jets)                         &  5&.9 & & & \\
    Background with prompt photons ($\mu$+jets)                     &  7&.1 & & & \\
    Background with electrons misidentified as photons (e+jets)     &  8&.7 & & & \\
    Background with electrons misidentified as photons ($\mu$+jets) & 10&.3 & & & \\
    Background with hadrons misidentified as photons (e+jets)       & 20&.5 & $^+_-$ & $^{\emptyplus 6}_{\emptyminus 5}$ & $^{.5\emptyplus}_{.5\emptyminus}$ \\
    Background with hadrons misidentified as photons ($\mu$+jets)   & 26&.6 & $^+_-$ & $^{\emptyplus 8}_{\emptyminus 6}$ & $^{.0\emptyplus}_{.7\emptyminus}$ \\
    \hline
    \hline
    Data (e+jets and $\mu$+jets)                                    &122& & & & \\
    \hline
  \end{tabular}
  \caption[Result of the combined template fit]{
    Result of the combined template fit showing the different contributions together with the statistical uncertainty for the contributions which were
    varied in the fit.
    The amount of observed candidate events in data is also shown.
  }
  \label{tab:fitresults}
\end{table}

The signal and the different background contributions as depicted in Fig.~\ref{fig:combinedfit}, are shown in an overview in Tab.~\ref{tab:fitresults}.
The fit yielded \mbox{$43 \, ^{+10.8}_{-11.8}$} $\ttg$ events out of the 122 observed candidate events.
The uncertainties quoted in the table are the statistical uncertainties from the fit only.
While the background contributions with prompt photons and electrons misidentified as photons were fixed in the fit, for the signal contribution and
the contributions with hadrons misidentified as photons, the smallest interval containing 68\% of the marginalised probability density function is
quoted.

\begin{figure}[h]
\begin{center}
\includegraphics[width=0.49\textwidth]{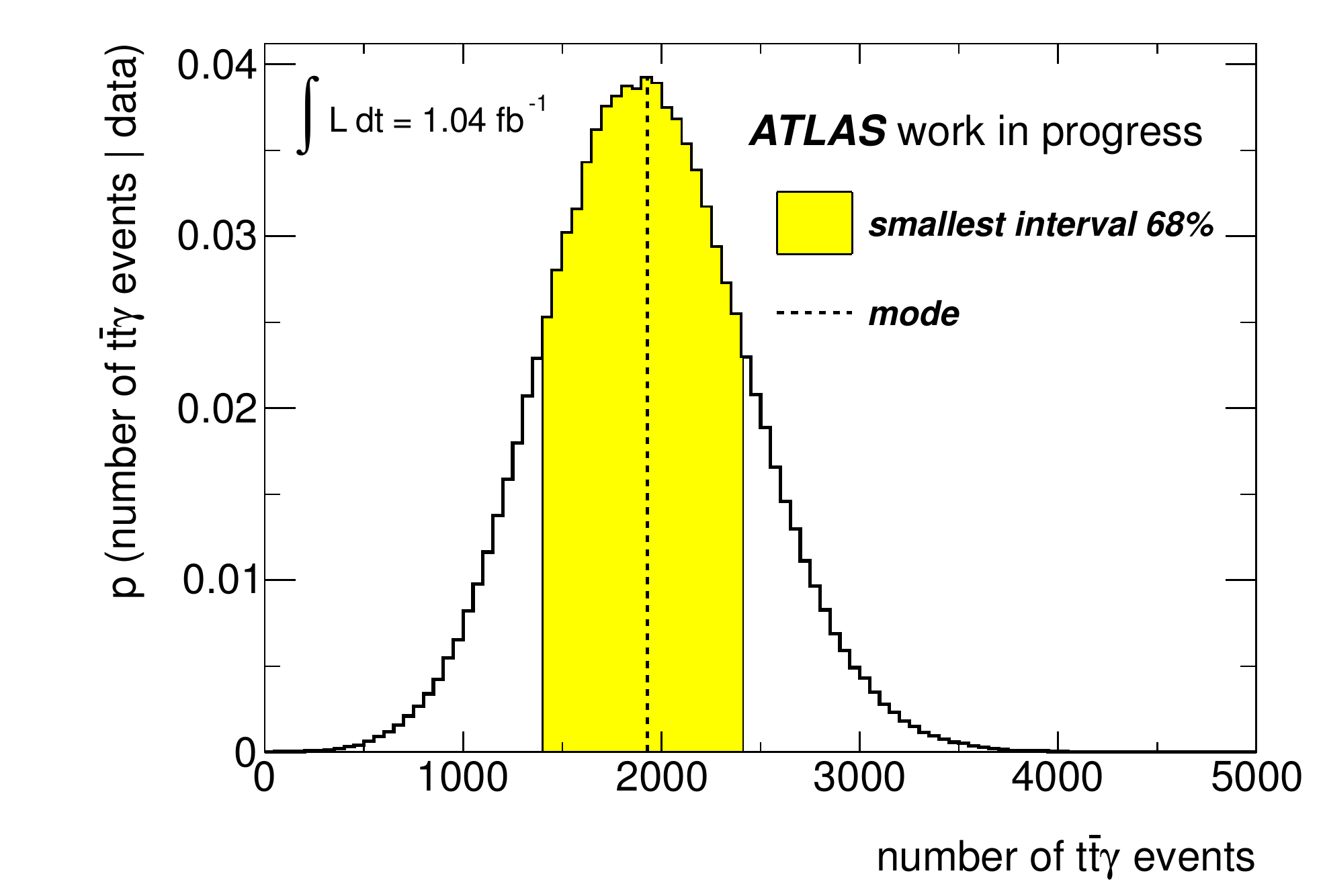}
\caption[Posterior $\ttg$ probability density for the combined template fit]{
  Posterior probability density for the number of $\ttg$ events for the combined template fit.
  The dashed line shows the global mode of the fit.
  The marked region around the mode represents the smallest interval containing 68\% of the distribution.
}
\label{fig:posteriorpdf}
\end{center}
\end{figure}

In the combined fit, the efficiency with respect to the whole $\ttg$ signal phase space was accounted for, as described in
Sec.~\ref{sec:strategy}.
The fit yielded \mbox{$1930 \, ^{+480}_{-530}$} $\ttg$ events in the whole signal phase space before object and event selection including true photon
transverse momenta down to \mbox{$8 \GeV$}.
The uncertainty was again extracted from the smallest interval containing 68\% of the marginalised pdf, which is shown in Fig.~\ref{fig:posteriorpdf}.
The dashed line shows the global mode of the fit.

\begin{figure}[h]
\begin{center}
\includegraphics[width=0.49\textwidth]{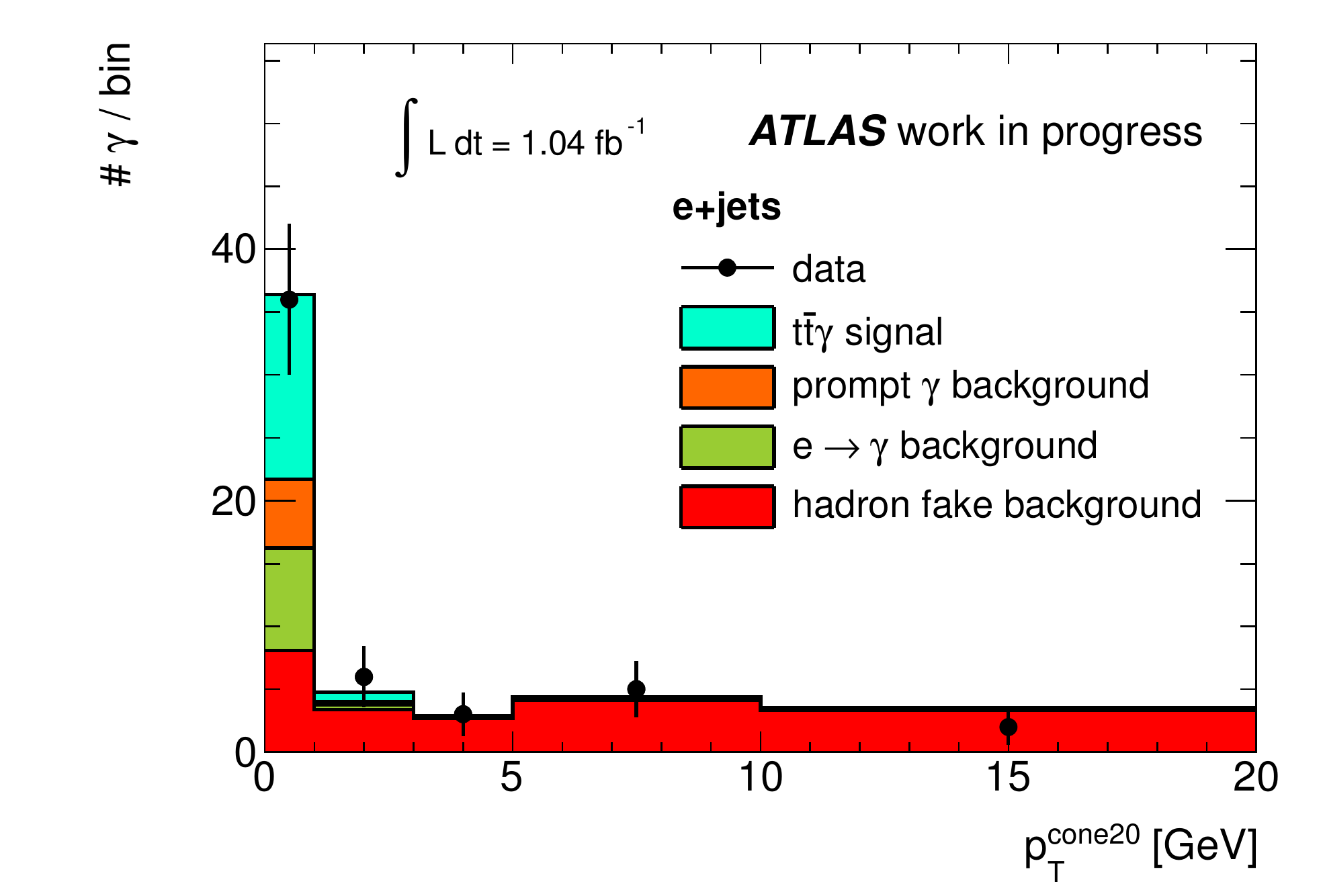}
\includegraphics[width=0.49\textwidth]{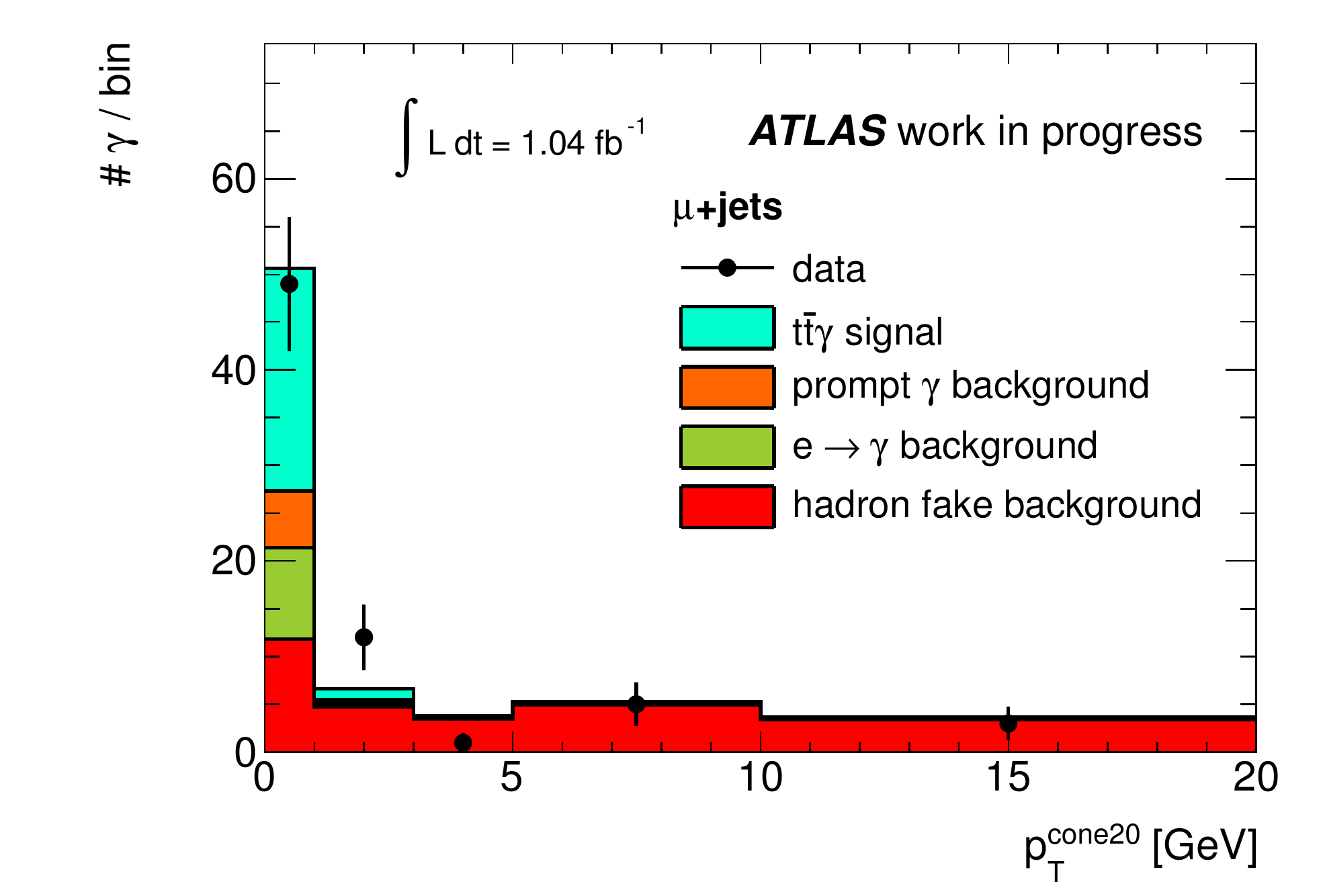}
\caption[Result of the two separate template fits]{
  Result of the two separate template fits to the photon $\ptcone$ distribution in the electron channel (left) and in the muon channel (right).
  In both plots, the last bin includes the overflow bin.
}
\label{fig:singlefits}
\end{center}
\end{figure}

As a cross-check, the template fit was performed in the electron channel and muon channel separately.
The results of these fits are shown in Fig.~\ref{fig:singlefits} for the two channels.
They yielded \mbox{$1640 \, ^{+710}_{-770}$} and \mbox{$1990 \, ^{+700}_{-720}$} $\ttg$ events in the whole signal phase space, respectively.
The fits in the single channels are hence consistent with each other within statistical uncertainties.

\renewcommand\arraystretch{1.15} % default : 1.0
\begin{table}[p]
  \center
  \footnotesize
  \begin{tabular}{|l|c r@{}l||l|c r@{}l|}
    \hline
    Source & \multicolumn{3}{c||}{Uncertainty} & Source & \multicolumn{3}{c|}{Uncertainty} \\
    \hline
    MC generator                              & $\pm$ & 4&.8 \% &
    \multirow{4}{*}{Signal modelling} & \multirow{4}{*}{$\pm$} & \multirow{4}{*}{11} & \multirow{4}{*}{.1 \%} \\
    Finite order calculation                  & $\pm$ & 3&.2 \% & & & & \\
    Showering                                 & $\pm$ & 3&.1 \% & & & & \\
    Initial / final state radiation           & $\pm$ & 8&.0 \% & & & & \\
    Parton density functions                  & $\pm$ & 4&.0 \% & & & & \\
    \hline
    MC correction (photon template)           & $\pm$ & 6&.4 \% &
    \multirow{7}{*}{Template shapes} & \multirow{7}{*}{$\pm$} & \multirow{7}{*}{12} & \multirow{7}{*}{.9 \%} \\
    Converted photons (photon template)       & $\pm$ & 2&.6 \% & & & & \\
    Pile-up (photon template)                 & $\pm$ & 0&.5 \% & & & & \\
    $\eta$-reweighting (hadron fake template) & $\pm$ & 2&.5 \% & & & & \\
    $\et$-reweighting (hadron fake template)  & $\pm$ & 7&.4 \% & & & & \\
    Converted photons (hadron fake template)  & $\pm$ & 7&.3 \% & & & & \\
    Pile-up (hadron fake template)            & $\pm$ & 2&.0 \% & & & & \\
    \hline
    Electron trigger efficiency               & $\pm$ & 1&.0 \% &
    \multirow{5}{*}{Electron modelling} & \multirow{5}{*}{$\pm$} & \multirow{5}{*}{5} & \multirow{5}{*}{.3 \%} \\
    Electron reconstruction efficiency        & $\pm$ & 1&.4 \% & & & & \\
    Electron identification efficiency        & $\pm$ & 4&.5 \% & & & & \\
    Electron energy scale                     & $\pm$ & 1&.5 \% & & & & \\
    Electron energy resolution                & $\pm$ & 1&.7 \% & & & & \\
    \hline
    Muon trigger efficiency                   & $\pm$ & 3&.2 \% &
    \multirow{5}{*}{Muon modelling} & \multirow{5}{*}{$\pm$} & \multirow{5}{*}{3} & \multirow{5}{*}{.4 \%} \\
    Muon reconstruction efficiency            & $\pm$ & 0&.5 \% & & & & \\
    Muon identification efficiency            & $\pm$ & 0&.4 \% & & & & \\
    Muon momentum scale                       & $\pm$ & 0&.4 \% & & & & \\
    Muon momentum resolution                  & $\pm$ & 0&.7 \% & & & & \\
    \hline
    Jet energy scale                          & $\pm$ &18&.4 \% &
    \multirow{6}{*}{Jet and $\met$ modelling} & \multirow{6}{*}{$\pm$} & \multirow{6}{*}{30} & \multirow{6}{*}{.2 \%} \\
    $b$-jet energy scale                      & $\pm$ & 2&.8 \% & & & & \\
    Jet energy scale and $\met$ (pile-up)     & $\pm$ &23&.3 \% & & & & \\
    Jet energy resolution                     & $\pm$ & 4&.2 \% & & & & \\
    Jet reconstruction efficiency             & $\pm$ & 0&.4 \% & & & & \\
    $\met$ (cell-out contribution)            & $\pm$ & 1&.9 \% & & & & \\
    \hline
    Photon identification efficiency          & $\pm$ & 4&.8 \% &
    \multirow{3}{*}{Photon modelling} & \multirow{3}{*}{$\pm$} & \multirow{3}{*}{5} & \multirow{3}{*}{.5 \%} \\
    Photon energy scale                       & $\pm$ & 1&.4 \% & & & & \\
    Photon energy resolution                  & $\pm$ & 2&.2 \% & & & & \\
    \hline
    $b$-tagging performance                   & $\pm$ &12&.5 \% & $b$-tagging & $\pm$ &12&.5 \% \\
    \hline
    Modelling of LAr readout issues           & $\pm$ & 2&.4 \% & LAr readout issues & $\pm$ & 2&.4 \% \\
    \hline
    $\feg$ scale factor                       & $\pm$ & 4&.4 \% & $\feg$ scale factor & $\pm$ & 4&.4 \% \\
    \hline
    Background $\ttg$ estimate                & $\pm$ & 7&.1 \% &
    \multirow{7}{*}{Prompt photon background} & \multirow{7}{*}{$\pm$} & \multirow{7}{*}{10} & \multirow{7}{*}{.9 \%} \\
    $Z$+jets+$\gamma$ estimate                & $\pm$ & 5&.9 \% & & & & \\
    Single top+$\gamma$ estimate              & $\pm$ & 2&.1 \% & & & & \\
    Diboson+$\gamma$ estimate                 & $\pm$ & 1&.0 \% & & & & \\
    Multijet+$\gamma$ (fake leptons)          & $\pm$ & 1&.1 \% & & & & \\
    Multijet+$\gamma$ (prompt photons)        & $\pm$ & 5&.0 \% & & & & \\
    $W$+jets+$\gamma$ estimate                & $\pm$ & 1&.2 \% & & & & \\
    \hline
    $\ttg$ $k$-factor and MC statistics       & $\pm$ & 0&.6 \% &
    \multirow{4}{*}{Background cross sections}& \multirow{5}{*}{$\pm$} & \multirow{5}{*}{10} & \multirow{5}{*}{.2 \%} \\
    $\ttbar$ cross section and MC statistics  & $\pm$ & 4&.4 \% &
    \multirow{4}{*}{and MC statistics}        & & & \\
    $Z$+jets cross section and MC statistics  & $\pm$ & 9&.1 \% & & & & \\
    Single top cross section and MC statistics& $\pm$ & 0&.4 \% & & & & \\
    Diboson cross section and MC statistics   & $\pm$ & 0&.6 \% & & & & \\
    \hline
    \hline
    Total systematic & \multicolumn{7}{c|}{$\pm$ 40.9 \%} \\
    \hline
    Luminosity & \multicolumn{7}{c|}{$\pm$ \emptynull 6.1 \%} \\
    \hline
    Statistical & \multicolumn{7}{c|}{$^{+25.1 \%}_{-27.5 \%}$} \\
    \hline
    Total & \multicolumn{7}{c|}{$^{+48.3 \%}_{-49.6 \%}$} \\
    \hline
  \end{tabular}
  \normalsize
  \caption[Statistical and systematic uncertainties on the number of $\ttg$ events]{
    Statistical and systematic uncertainties from the combined fit.
  }
  \label{tab:systematics}
\end{table}
\renewcommand\arraystretch{1.25} % default : 1.0

Systematic uncertainties were evaluated as described in Ch.~\ref{sec:systematics}.
An overview of the systematic uncertainties is presented in Tab.~\ref{tab:systematics} in terms of fraction of the measured number of $\ttg$ events.
The different sources of systematic uncertainties were summarised in several categories as indicated in the table.

The largest systematic uncertainty is due to the jet and $\met$ modelling (30\%), which is dominated by the nominal uncertainty on the jet
energy scale (JES), and the additional JES uncertainty due to a possible pile-up dependence of the JES.
Both sources of uncertainty change the acceptance of the $\ttg$ signal by roughly $\pm 10\%$.
The acceptance changes for the background processes are of the same order, or even larger, as for example for $Z$+jets events, which feature
a softer jet $\pt$ spectrum, i.e. more jets close to the $\pt$ threshold.
The resulting uncertainties on the cross section
due to the JES and the JES pile-up dependence are of the order of 20\%, which can be understood as follows:
a smaller (larger) background prediction results in a larger (smaller) number of estimated signal events.
Moreover, the change of the $\ttg$ acceptance means that the measured amount of signal corresponds to a smaller (larger) phase space and therefore
is subject to a larger (smaller) acceptance correction.
The acceptance variations of the signal and the background modelling
hence have a positive correlation and lead to the observed large systematic uncertainties on the cross section of the order of 20\%.

In contrast, the modelling of electrons, muons and photons results in only moderate systematic uncertainties (3.4~--~5.5\%).
The largest contributions are the electron and photon identification efficiencies and the muon trigger efficiency, where the latter was increased
to cover effects due to trigger matching (Sec.~\ref{sec:syst_detectormodelling}).
The uncertainty due to the modelling of the $b$-tagging performance was found to be sizable (13\%).
Uncertainties due to the modelling of the readout issues in the LAr calorimeter and due to the systematic uncertainty on the SF for the electron-to-photon
misidentification rate were found to have only moderate contributions (2.4\% and 4.4\%, respectively).

The change of the $\ttg$ acceptance only is of the order of 5\% for both the variations of the photon identification efficiency
and of the $b$-tagging performance.
In the case of the $b$-tagging performance, the background estimate also changes significantly, which leads to an uncertainty on the
cross section of the order of 10\% due to the positive correlation as discussed for the JES uncertainty.
For the photon identification efficiency, however, the background estimate only varies by a small amount:
the main background estimated from MC simulations are events with electrons misidentified as photons.
Although the estimate for these background contributions is expected to increase (decrease) for a larger (smaller) photon identification SF,
it does not vary significantly, because the electron-to-photon misidentification rate $\feg$ changes in the opposite direction.
This is due to the fact that the number of events in the \Zeg sample in MC simulations appears in the denominator in Eq.~(\ref{eq:SF}).
The uncertainty on the cross section is therefore only of the order of 5\%, and hence smaller than the uncertainty due to the $b$-tagging modelling.

Uncertainties due to the background modelling arise from the shape of the templates (13\%), the estimates of the background contributions
with prompt photons (11\%) and the uncertainties due to the predicted cross sections and limited MC statistics for various background processes
(10\%).
The uncertainty due to the template shapes was found to be dominated by the MC correction applied to the electron templates from $\Zee$ data and
the fractions of unconverted and converted photon candidates used for the hadron fake template,
as well as the $\et$ spectrum of the hadron fakes.
The first two uncertainties were estimated conservatively:
the uncertainty on the MC correction was estimated by comparing the result with the correction to the result without any correction.
For the fractions of unconverted and converted photon candidates, the nominal fit was compared to fits in which it was assumed that all hadron fakes
were either unconverted or converted, respectively.
The uncertainty due to the hadron fake $\et$ spectrum, in turn, is dominated by the low statistics in the control region in which the spectrum
was estimated (Ch.~\ref{sec:faketemplate}).

The dominant contributions to the uncertainty from the prompt photon background were found to be
due to the data-driven multijet+$\gamma$ background and the estimate of the contribution from $Z$+jets+$\gamma$ events, but also due to the
estimate of the amount of background $\ttg$ events.
All of these estimates feature conservative uncertainties larger than 100\% (Sec.~\ref{sec:ttgbkg},~\ref{sec:QCDgamma} and~\ref{sec:restgamma}).

The uncertainties on the $\ttbar$ and $Z$+jets cross sections are the dominant uncertainties among the predictions of cross sections for the background
processes as well as the limited statistics of the corresponding MC samples:
dileptonic $\ttbar$ events with an electron misidentified as a photon result in a large background contribution (Sec.~\ref{sec:egammaapplication}).
For the $Z$+jets process, the same simulated sample was used to estimate the contributions from electrons misidentified as photons as well as from
$Z$+jets+$\gamma$ events.
Although the resulting background yields are not very large, the uncertainty on the $Z$+jets cross section still has a non-negligible effect,
because it is only known to 48\%.
A very small uncertainty was found to be due to the uncertainty of the $k$-factor for $\ttg$ production, which originates
from a small corrections from $\ttg$ events in the control region used for the estimate of the $W$+jets+$\gamma$ background.

Uncertainties due to the modelling of the $\ttg$ signal add up to 11\%, where the largest contribution is from the modelling of the initial
and final state radiation.
All systematic uncertainties add up to a total systematic uncertainty of 41\%.

From the measured number of $\ttg$ events, the $\ttg$ cross section in the single lepton and dilepton channels was calculated using the
integrated luminosity of \mbox{$1.04 \ifb$} of the data set analysed.
The uncertainty on the $\ttg$ cross section due to the luminosity measurement was found to be as large as 6.1\%.
It is larger than the intrinsic uncertainty on the luminosity measurement because of the positive correlation between the effects on the
background estimates and on the number of $\ttg$ signal events.
The measurement of the $\ttg$ cross section times branching ratio (BR) into the single lepton and dilepton final state reads:
\begin{equation*}
  \sigma_{t\bar{t}\gamma} \cdot \mathrm{BR} = 1.9 \pm 0.5 \, \mathrm{(stat.)} \pm 0.8 \, \mathrm{(syst.)} \pm 0.1 \, \mathrm{(lumi.)} \pb \, .
\end{equation*}
This result is compatible with the SM expectation (Sec.~\ref{sec:signalmodelling}) of \mbox{$2.1 \pm 0.4 \pb$} within uncertainties.

\begin{figure}[h]
\begin{center}
\includegraphics[width=0.49\textwidth]{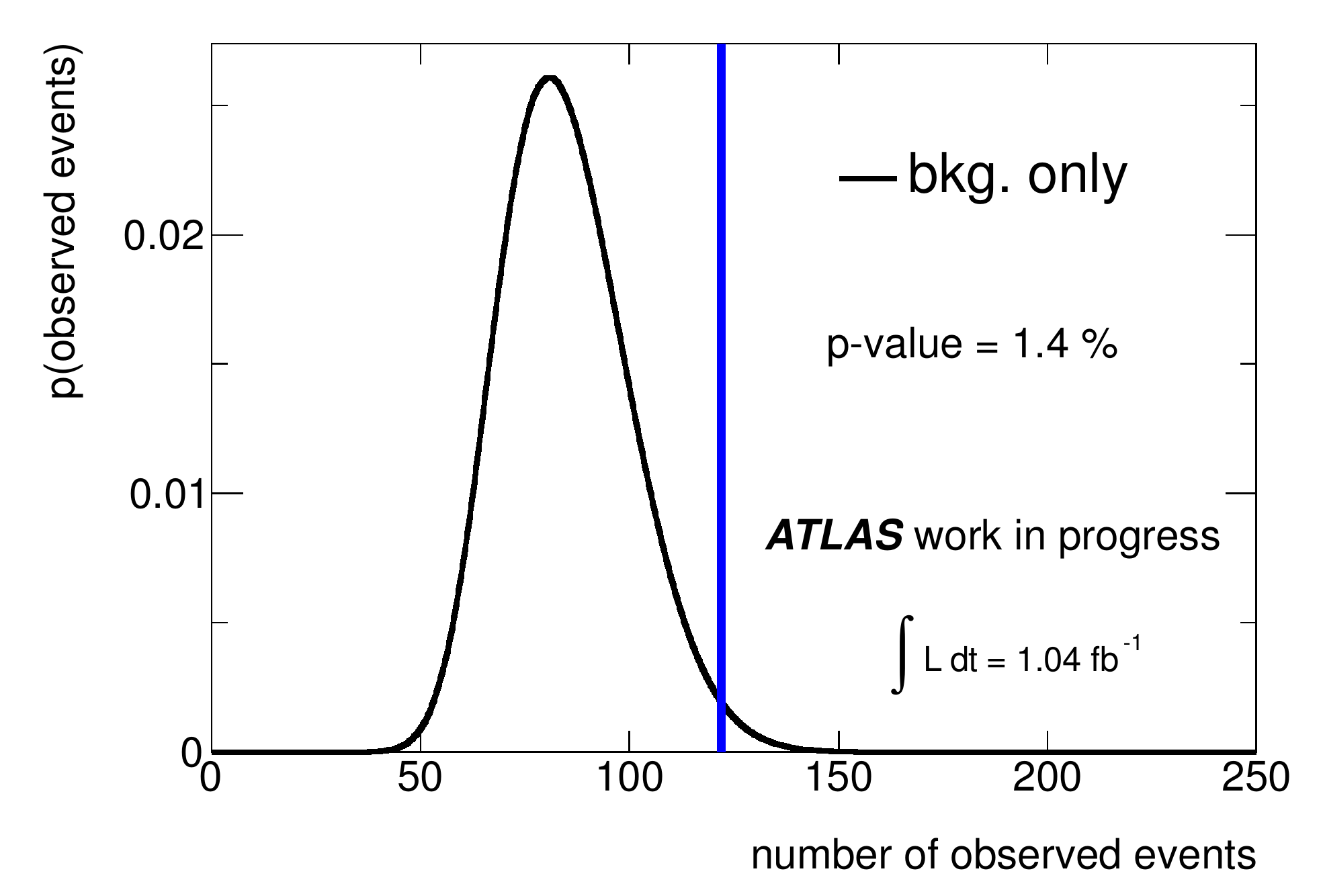}
\includegraphics[width=0.49\textwidth]{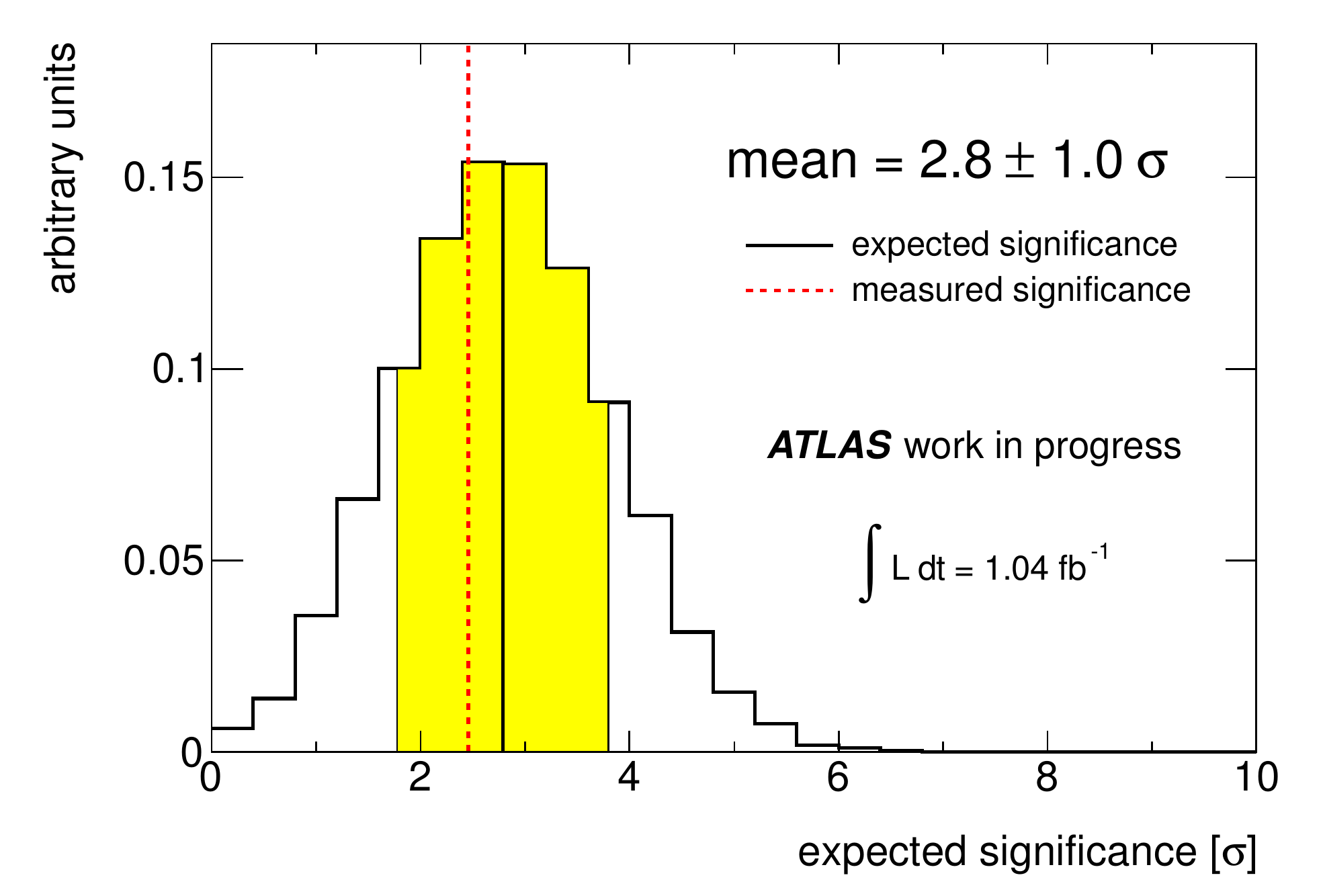}
\caption[Estimation of the significance]{
  Estimation of the significance:
  the left plot shows the probability distribution of the observed number of events for the background-only hypothesis given the statistical
  and systematic uncertainties.
  The p-value for observing at least 122 events is 1.4\%, which corresponds to a significance of $2.5 \, \sigma$.
  The right plot shows the distribution for the expected significance.
  The solid line shows the mean of the distribution at $2.8 \, \sigma$.
  The root mean square is $1.0 \, \sigma$, as indicated by the marked region around the mean.
  The dashed line shows the measured significance.
}
\label{fig:significance}
\end{center}
\end{figure}

In order to estimate the significance of the measured $\ttg$ signal, the probability of the background giving rise to at least 122 observed candidate
events was estimated.
The total background yield in both lepton channels together reads
\mbox{$79 \, ^{+10}_{-\emptynull 9} \, \mathrm{(stat.)} \pm 13 \, \mathrm{(syst.+lumi.)} = 79 \, ^{+17}_{-16}$}.
The total uncertainty was interpreted as the standard deviation of a Gaussian pdf.
The total background pdf was constructed by a convolution of this Gaussian pdf with a Poissonian pdf to allow for statistical fluctuations.
This pdf is shown in the left plot of Fig.~\ref{fig:significance}.
The p-value for the background to produce at least 122 observed events is 1.4\%, which corresponds to a significance of $2.5 \, \sigma$.

The right plot in Fig.~\ref{fig:significance} shows the expected significance given the background estimate as well as the $\ttg$ expectation from
the SM calculation, which reads \mbox{$49 \pm 7 \, \mathrm{(stat.)} \pm 10 \, \mathrm{(syst.)}$}.
The systematic uncertainty is due to the uncertainty on the $\ttg$ $k$-factor.
The mean of the distribution is $2.8 \, \sigma$.
However, the root mean square is as large as $1.0 \, \sigma$ given the statistical and systematic uncertainties on the predicted number of $\ttg$ events.
The observed significance of $2.5 \, \sigma$ is hence consistent with the expected significance within this uncertainty.

\subsubsection{Evolution of the expected significance with increasing integrated luminosity}

\begin{figure}[h]
\begin{center}
\includegraphics[width=0.6\textwidth]{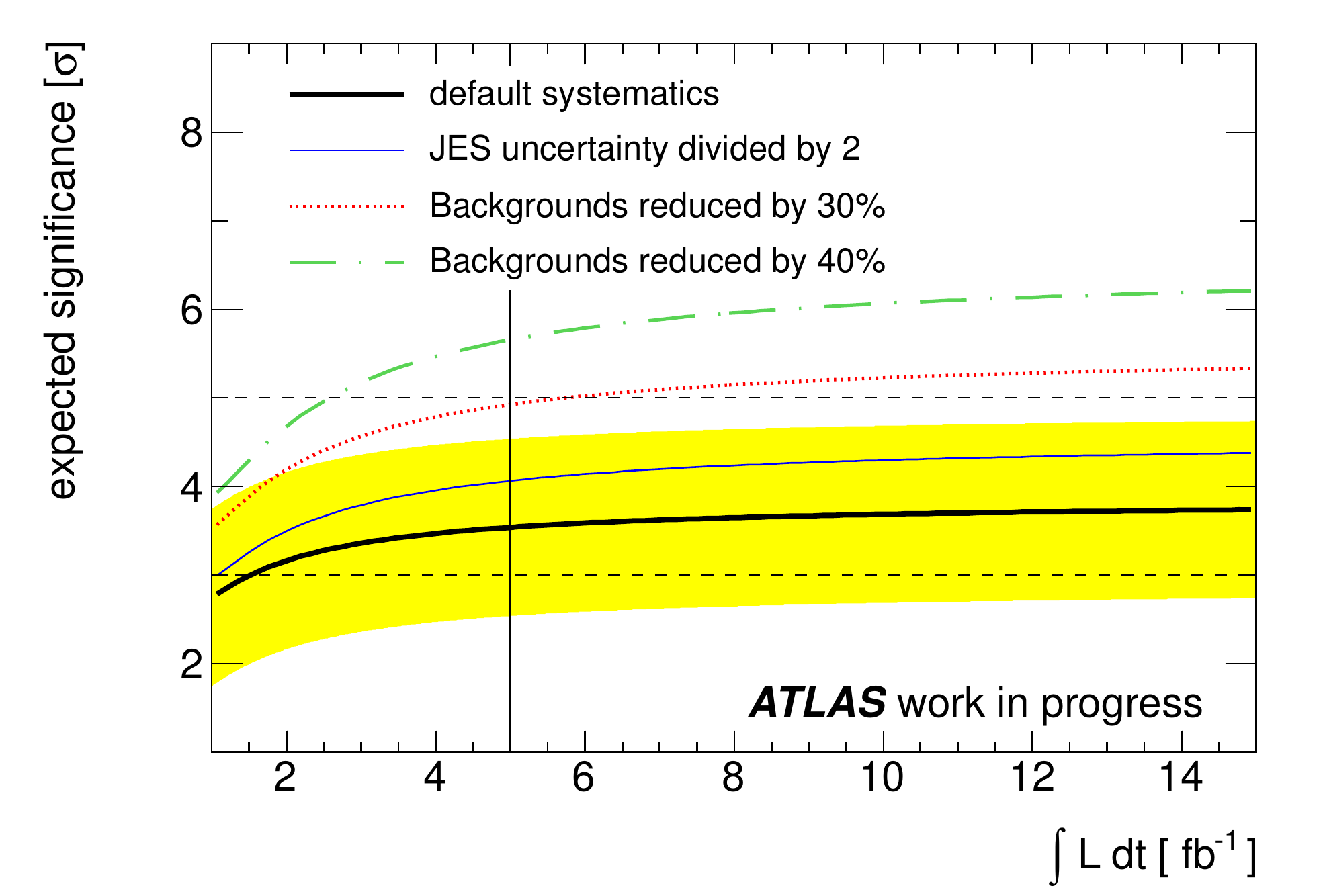}
\caption[Extrapolation of the expected significance]{
  Evolution of the expected significance with the integrated luminosities for four different scenarios:
  systematic uncertainties as in this analysis (thick solid line), uncertainties on the jet energy scale reduced by a factor of two
  (thin solid line), and with all background contributions reduced by a factor of 30\% and 40\% (dotted and dashed-dotted lines).
  The uncertainty band is only shown for the first scenario.
  It corresponds to the root mean square of the distribution of the expected significance.

  The horizontal dashed lines indicate the $3 \, \sigma$ threshold for an evidence of $\ttg$ production
  and the $5 \, \sigma$ threshold for an observation.
  The vertical solid line indicates the maximum available integrated luminosity at \mbox{$\sqrt{s} = 7 \TeV$} of $5 \ifb$.
}
\label{fig:sigma_extrapolation}
\end{center}
\end{figure}

It is interesting to investigate how the expected significance of the $\ttg$ signal will evolve when data sets increase in size.
Fig.~\ref{fig:sigma_extrapolation} shows the evolution of the expected significance for integrated luminosities between $1$ and \mbox{$15 \ifb$}.
Four different scenarios are shown:
the thick solid line represents the systematic uncertainties as presented in this analysis.
The thin solid line shows the significance for a scenario where the uncertainty due to the JES modelling was reduced by a factor of two.
This is a likely scenario for the near future corresponding to a JES uncertainty of the order of 1\%.
The dotted and dashed-dotted lines show the significance for scenarios where the contributions from background processes were reduced by
30\% and 40\%, respectively and the relative systematic uncertainty on the background estimations was assumed to be the same as in this analysis.
In Fig.~\ref{fig:sigma_extrapolation}, the uncertainty band is shown for the first scenario only.
The uncertainty corresponds to the root mean square of the distribution of the expected significance (see right plot in Fig.~\ref{fig:significance}),
which is large and mainly due to the uncertainty on the expected $\ttg$ cross section.

The third and fourth scenarios assume a significant reduction of the background contributions, while
the main background processes feature hadrons or electrons misidentified as photons.
Background contributions from hadrons misidentified as photons could be further reduced by tightening the requirements on the photon shower shapes
or by combining several shower shapes in a multivariate discriminant exploiting their correlations.
The background contribution from electrons misidentified as photons could be reduced by considering only unconverted photon candidates,
because a large fraction of the misidentified electrons results in converted photon candidates.
This requirement would reduce the efficiency for photons from $\ttg$ production
by roughly 50\%, but would clearly enhance the signal-over-background ratio.

Generally, the expected significance increases with an increasing integrated luminosity.
At \mbox{$5 \ifb$}, which is the size of the whole data set available at \mbox{$7 \TeV$},
the measurement is largely dominated by systematic uncertainties.
A significance of more than $3 \, \sigma$, corresponding to an evidence of the $\ttg$ signal, can be expected for systematic uncertainties as presented
in this thesis.
Deviations from the predicted $k$-factor for $\ttg$ production may however lead to significantly smaller or larger values
of the significance.
From a decreased JES uncertainties, an increase of the significance of roughly $0.5 \, \sigma$ can be expected.

A significance of $5 \, \sigma$, corresponding to an observation of
the $\ttg$ signal, will only be possible if the contribution from background processes can be significantly reduced.
With a reduction of \mbox{30 -- 40\%}, a $5 \, \sigma$ observation is in reach with \mbox{$5 \ifb$}
if the relative systematic uncertainties remain of the same order and the signal efficiency is not strongly reduced.

%Another possibility to reduce the impact of systematic uncertainties is the measurement of the ratio of $\ttg$ production with respect to
%$\ttbar$ production.
%The ratio is expected to be less sensitive to systematic uncertainties, because many systematic effects are expected to affect $\ttg$ and $\ttbar$ events
%in a similar way and therefore cancel out to a large extent in the ratio.
%
%Additionally, a measurement in dileptonic $\ttg$ decays is promising with larger data sets.
%Although the branching ratio is reduced with respect to the single lepton channel, the background contributions from $W$+jets+$\gamma$ and
%multijet+$\gamma$ events are strongly suppressed due to the second lepton required in the final state.
%More important, however, is the expected reduction of events with electrons misidentified as photons.
%While in this analysis, a significant background contribution is due to events with two leptons in the final state, such as dileptonic $\ttbar$ and
%$Z$+jets events, these processes would be suppressed in the dilepton $\ttg$ decay channel,
%because processes with three leptons in the final state are rare.
%
Data sets larger than \mbox{$5 \ifb$} will only be available at \mbox{$\sqrt{s} = 8 \TeV$} or at larger energies.
The cross sections for the different processes change with respect to \mbox{$7 \TeV$}, but to first order the extrapolations shown in
Fig.~\ref{fig:sigma_extrapolation} are expected to hold also for \mbox{$8 \TeV$}.
At \mbox{$8 \TeV$}, the detector modelling will need to be studied, and a reduction of systematic uncertainties will only be possible after thorough
studies of the performance of the different objects.
The increasing pile-up may will be a particular issue.

\chapter{Summary, conclusion and outlook}
\label{sec:summary}

Top quark pair production with an additional photon in the final state ($\ttg$) is a process which provides direct sensitivity to the strength and
structure of the top quark's electromagnetic coupling.
Since the top quark decays before producing bound states, $\ttg$ production is a unique process for the direct
measurement of the electromagnetic coupling of a quark and hence an important test of the Standard Model.
Additionally, $\ttg$ production is sensitive to modifications of the $t\gamma$-vertex beyond its Standard Model structure.

In this thesis, the first measurement of the $\ttg$ cross section at the LHC was presented using \mbox{$1.04 \ifb$} of data taken with the ATLAS
detector in proton-proton collisions at a centre-of-mass energy of \mbox{$\sqrt{s} = 7 \TeV$}.
It is hence an important milestone towards tests of the electromagnetic coupling of the top quark.
The treatment of the various background contributions to $\ttg$ production was a particular challenge, and a
strategy was set up which strongly reduced the dependence on Monte Carlo simulations.
\newline

The measurement was performed in the single electron and single muon channels and was hence subject to background contributions from
$W$ and $Z$ boson production with additional jets ($W$+jets and $Z$+jets),
single top, multijet and diboson production, where all of these processes featured an additional photon in the final state.
The main background contributions, however, were due to processes where a hadron from jet fragmentation or an electron were misidentified as a photon.

Since the misidentification of hadrons and electrons as photons is known to be poorly modelled in Monte Carlo simulations,
a strategy which minimised the dependence on simulations was set up for the estimation of these main background contributions:
the contribution from hadrons misidentified as photons was estimated from data with a template fit to the photon isolation distribution.
The probability for electrons to be misidentified as photons was measured in \Zee events in data.

Data-driven techniques were developed for the estimation of the background processes from $W$+jets and multijet production with an additional
real photon in the final state.
These contributions were estimated in control regions in data enhanced with the respective background process and then extrapolated to the signal region.
The remaining background contributions were found to be small and were estimated using Monte Carlo simulations.
\newline

A total of 122 $\ttg$ candidate events were identified in the data set and a combined measurement in the single electron and single muon channel
yielded a cross section of
\begin{equation*}
  \sigma_{t\bar{t}\gamma} \cdot \mathrm{BR} = 1.9 \pm 0.5 \, \mathrm{(stat.)} \pm 0.8 \, \mathrm{(syst.)} \pm 0.1 \, \mathrm{(lumi.)} \pb \, ,
\end{equation*}
where BR is the branching ratio into the single lepton and dilepton $\ttg$ decay channels.
This measurement is in agreement with the expectation from Standard Model calculations.
The statistical significance of the $\ttg$ signal was estimated to $2.5 \, \sigma$ in agreement with the expected significance.

The dominant systematic uncertainty was found to be due to the modelling of jets, in particular due to the measurement of the jet energy scale.
Other large systematic uncertainties were found to be the modelling of the $b$-tagging performance, the shape of the photon isolation templates, and
the modelling of initial and final state radiation.
\newline

Based on the techniques established in this thesis, a $3 \, \sigma$ evidence for $\ttg$ production can be expected with the \mbox{$5 \ifb$} of data
taken at \mbox{$\sqrt{s} = 7 \TeV$}.
If background contributions can further be reduced, a $5 \, \sigma$ observation is in reach.
Moreover, a measurement of the ratio of the $\ttg$ and $\ttbar$ production cross sections may reduce the impact of systematic uncertainties
due to the detector modelling.
With larger data sets, measurements of differential $\ttg$ cross sections may enhance the sensitivity to the $t\gamma$-vertex.
$\ttg$ analyses at the LHC in the next years will hence provide further milestones towards a measurement of the strength and the structure of the
electromagnetic coupling of the top quark and will therefore contribute to our picture of the heaviest elementary particle known to date.

\appendix

\chapter{Additional plots for Chapter~\ref{sec:photontemplate}}
\label{sec:app_extrapolation}

\begin{figure}[h]
  \begin{center}
    \includegraphics[width=0.395\textwidth]{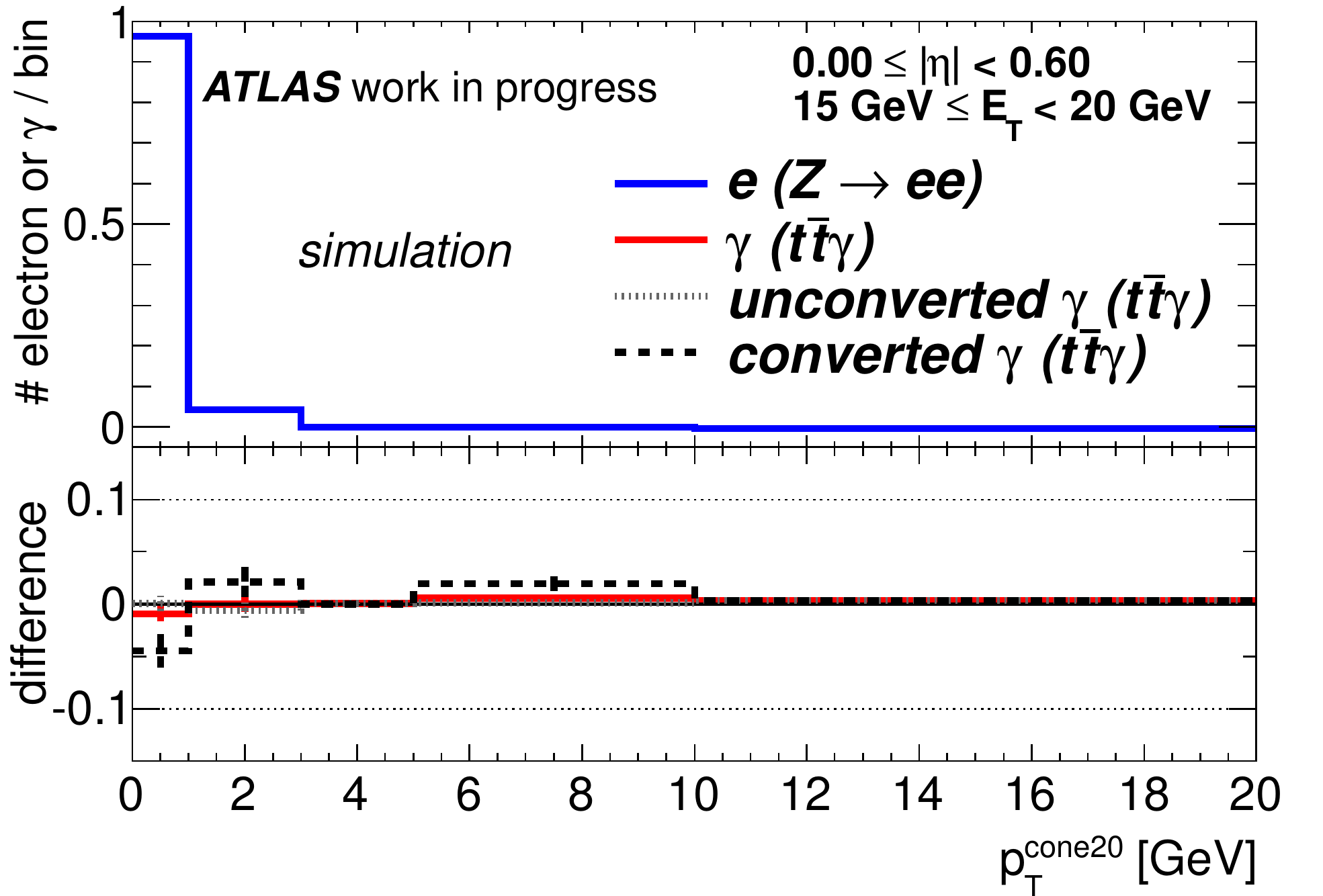}
    \includegraphics[width=0.395\textwidth]{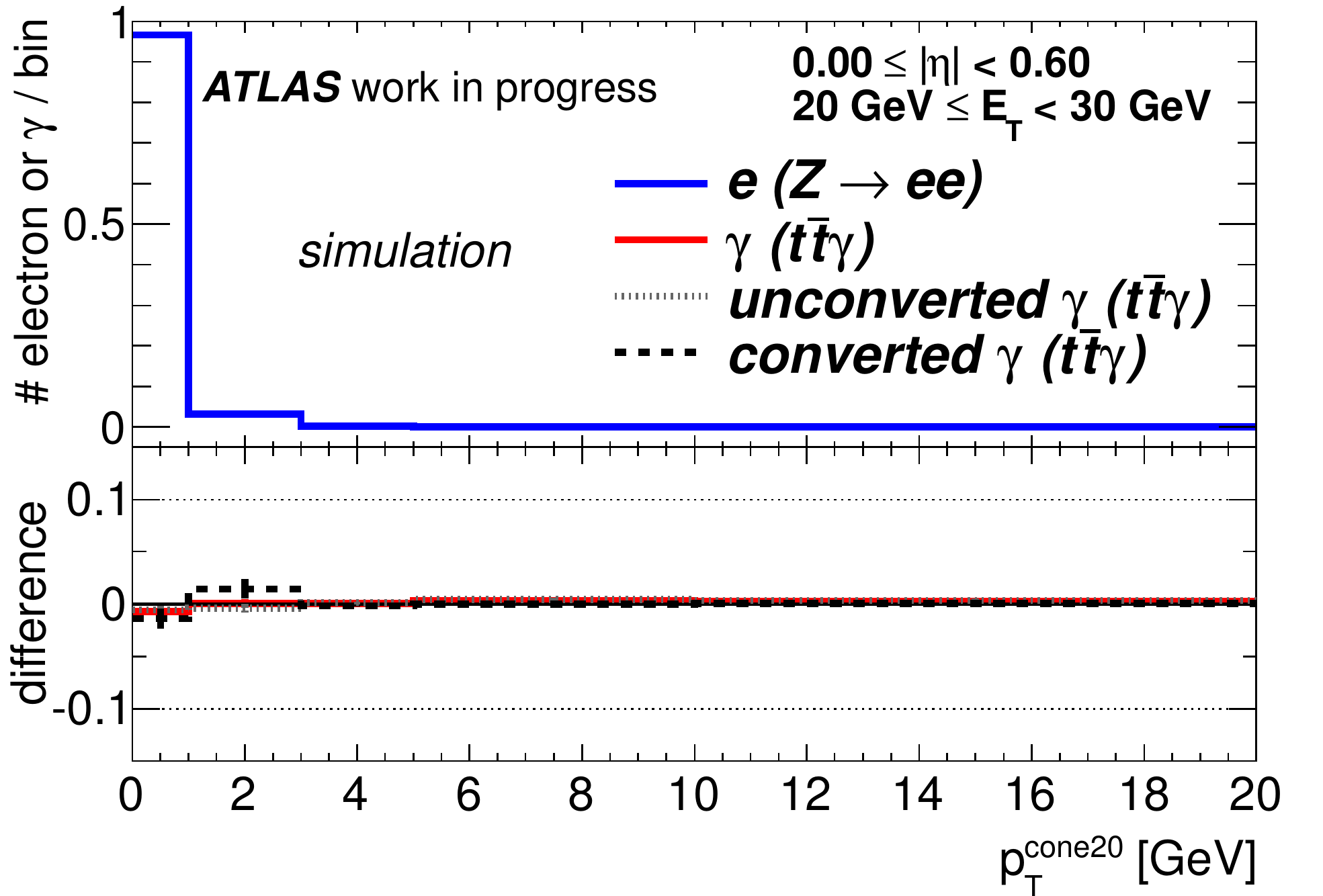}
    \includegraphics[width=0.395\textwidth]{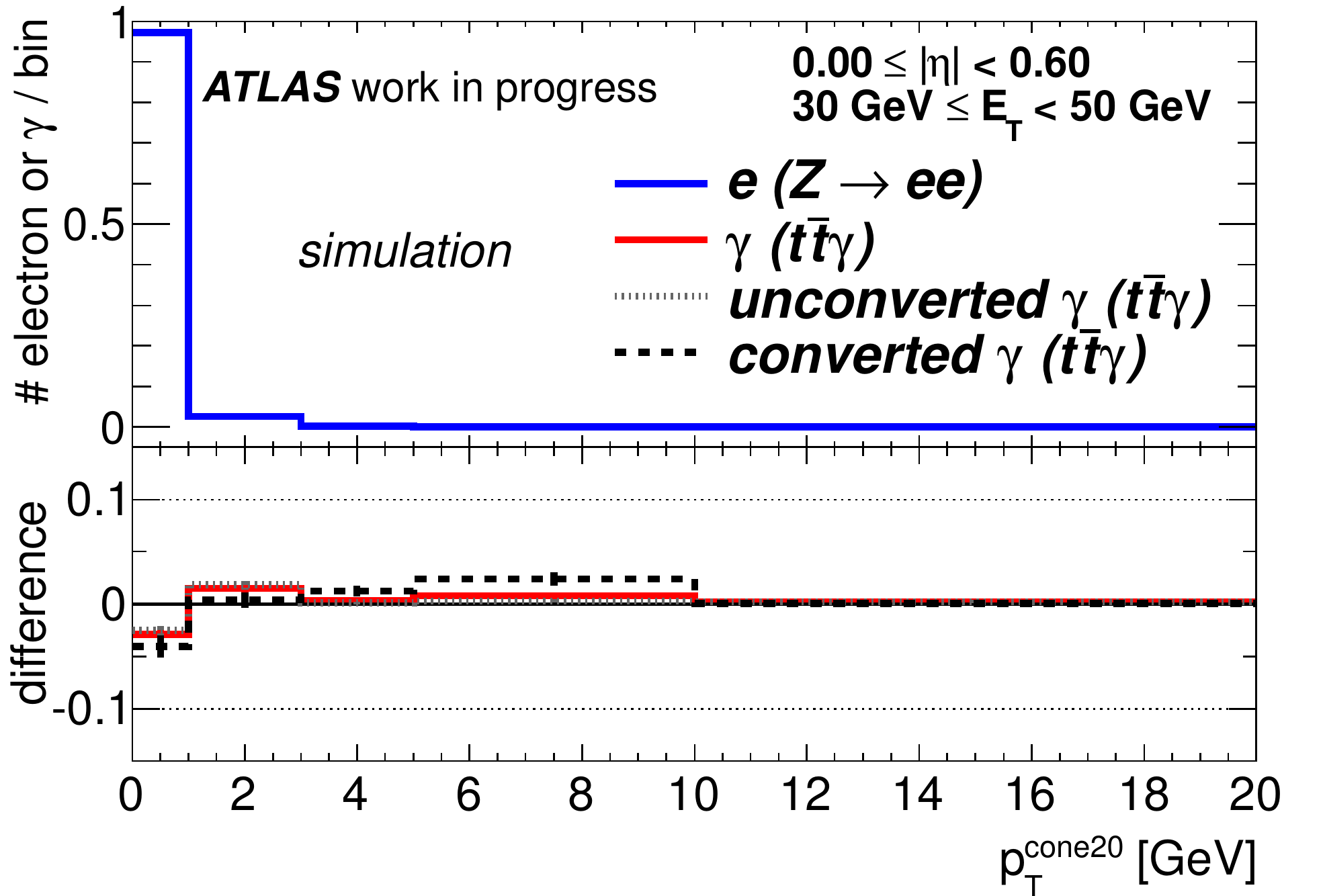}
    \includegraphics[width=0.395\textwidth]{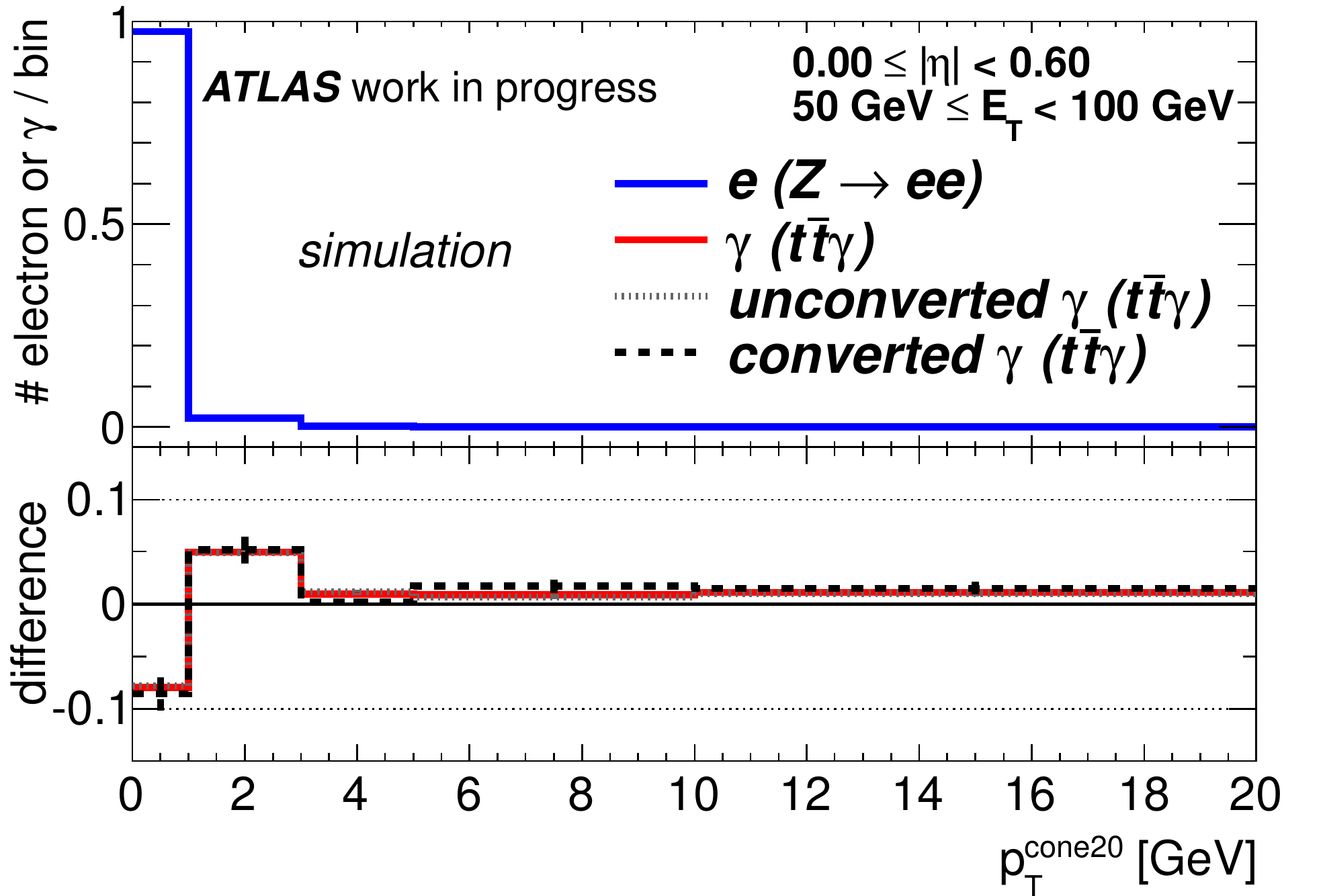}
    \includegraphics[width=0.395\textwidth]{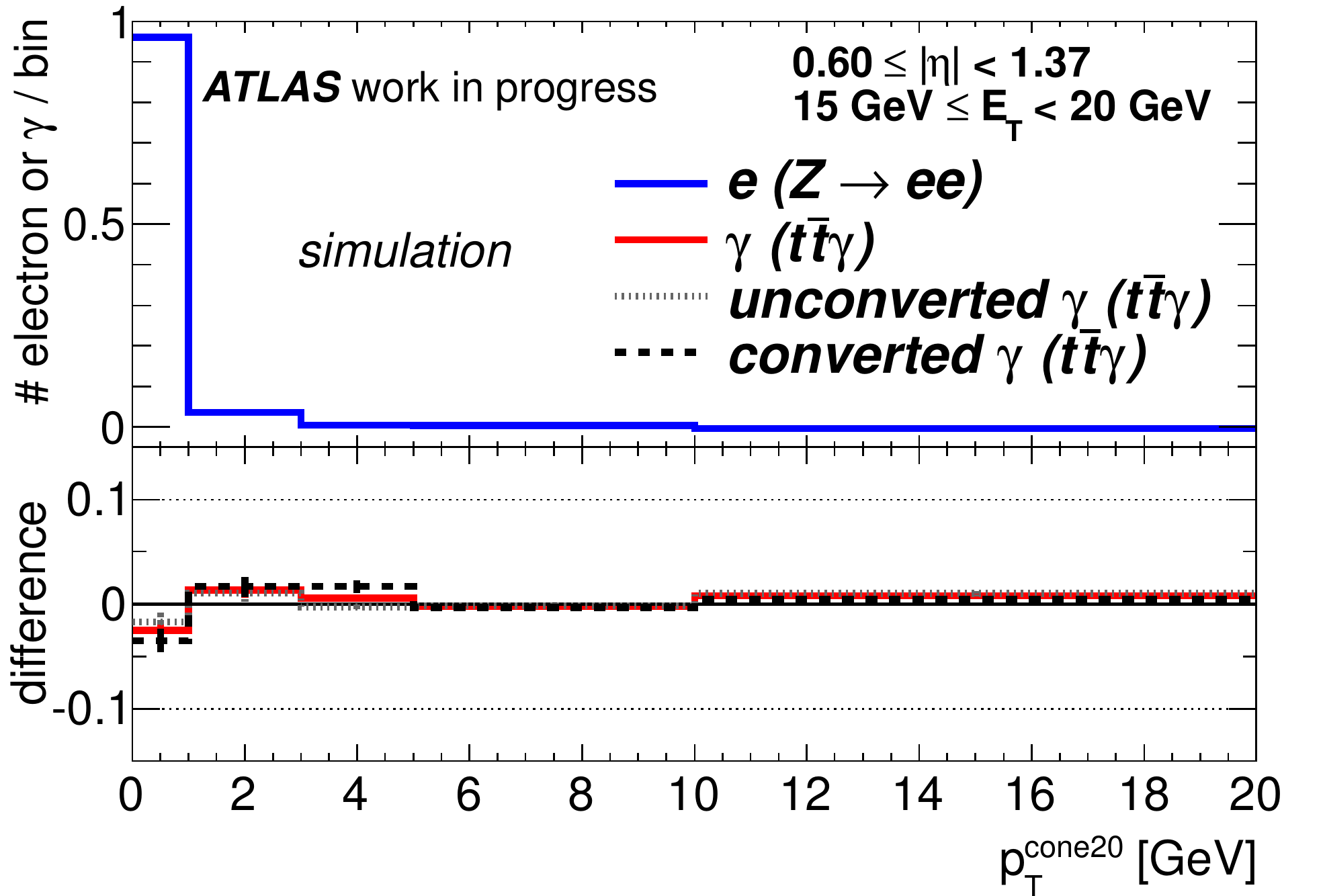}
    \includegraphics[width=0.395\textwidth]{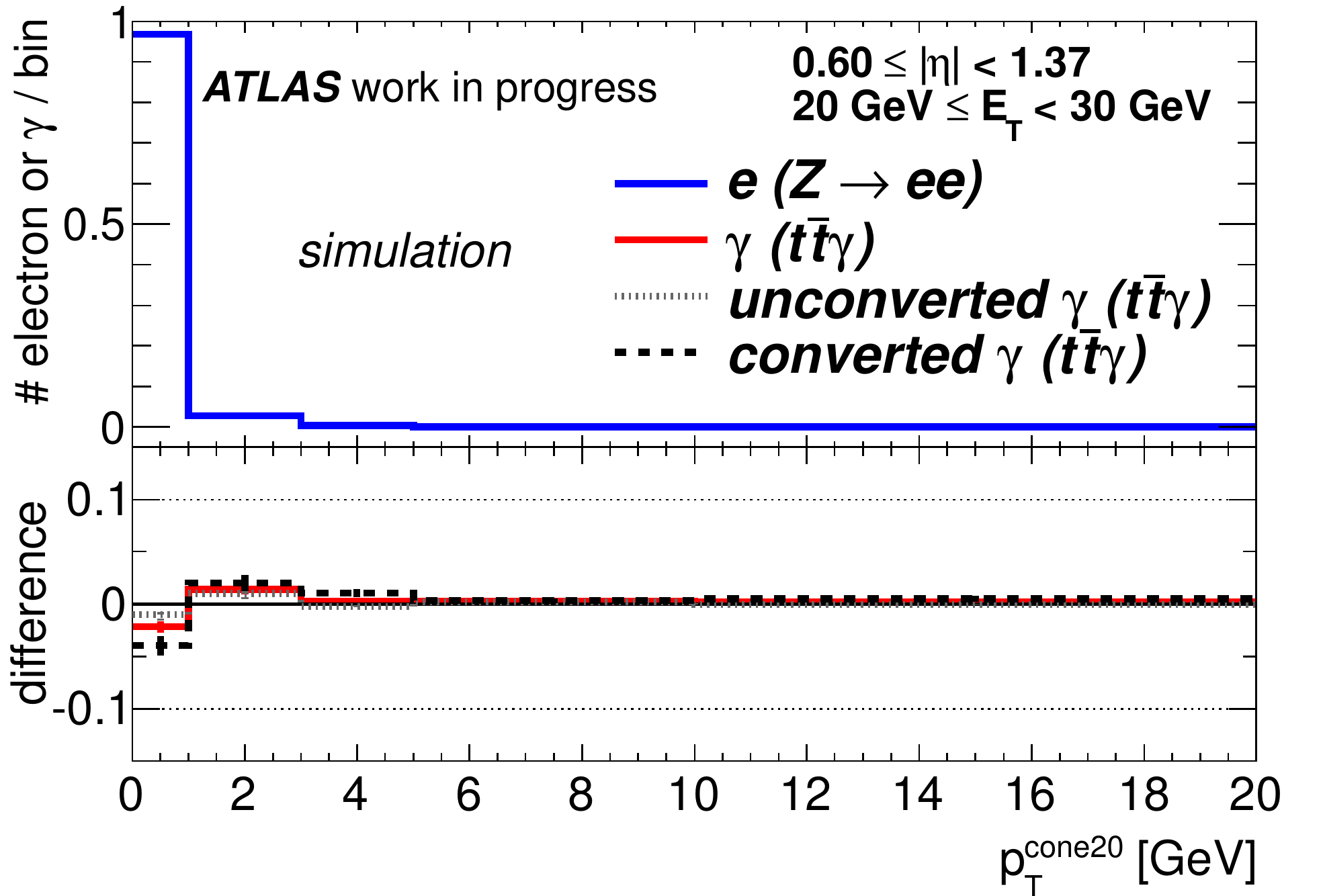}
    \caption[$\ptcone$ distributions for electrons and photons from simulation, $\Delta R(l,\gamma)$ cut (1)]{
      $\ptcone$ distributions for electrons from simulated \mbox{$Z \to e^+e^-$} decays (upper part of each plot)
      in different bins of $\et$ for \mbox{$0 \leq |\eta| < 0.60$} (four upper plots) and \mbox{$0.60 \leq |\eta| < 1.37$} (two lower plots)
      normalised to unity.
      The lower part of each plot shows the difference of the distribution of photons from simulated $\ttg$ events (solid line) with respect
      to the electron distribution.
      Additionally, the distributions for unconverted (dotted grey line) and converted photons (dashed black line) from $\ttg$ simulations are depicted.
      In all plots, the last bin includes the overflow bin.\\
      Photons closer than 0.2 in $\eta$-$\phi$-space to a true lepton were not considered and the agreement with the electron distribution
      is improved with respect to the plots in Fig.~\ref{fig:extrapolation_1} and~\ref{fig:extrapolation_2}.
      These plots complete the example plots shown in Fig.~\ref{fig:extrapolation_3}.
    }
    \label{fig:extrapolation_app_1}
  \end{center}
\end{figure}

\begin{figure}[p]
  \begin{center}
    \includegraphics[width=0.395\textwidth]{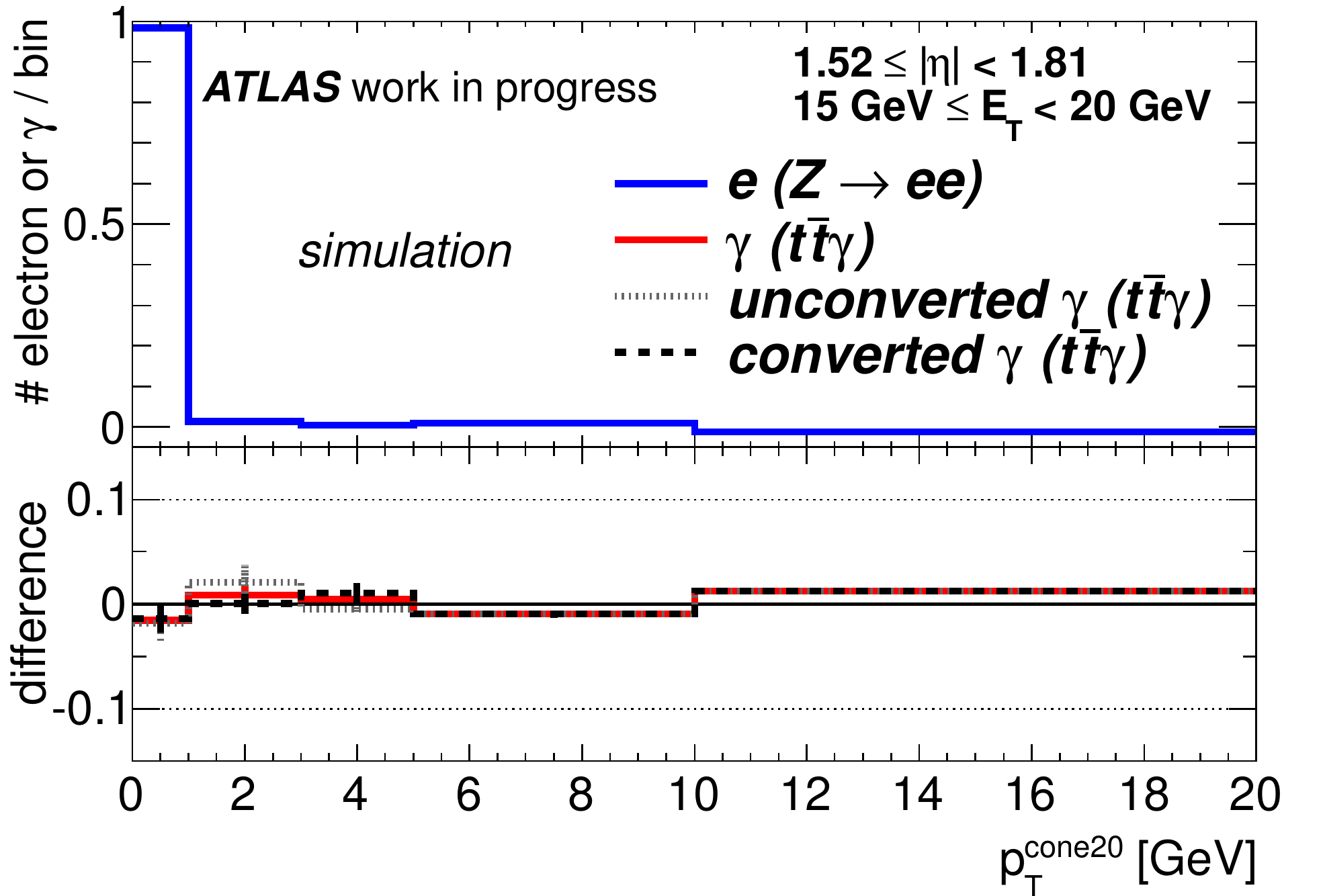}
    \includegraphics[width=0.395\textwidth]{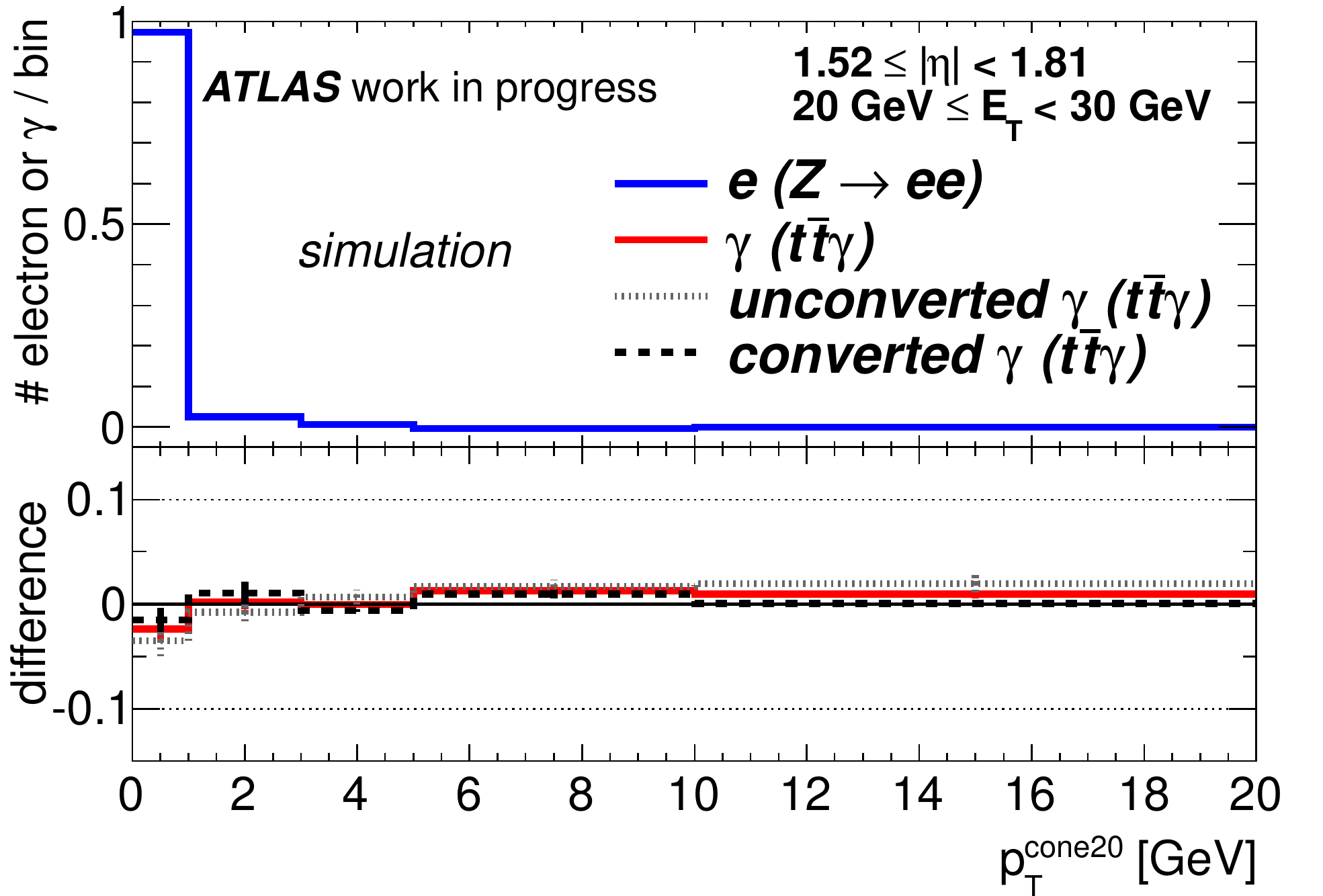}
    \includegraphics[width=0.395\textwidth]{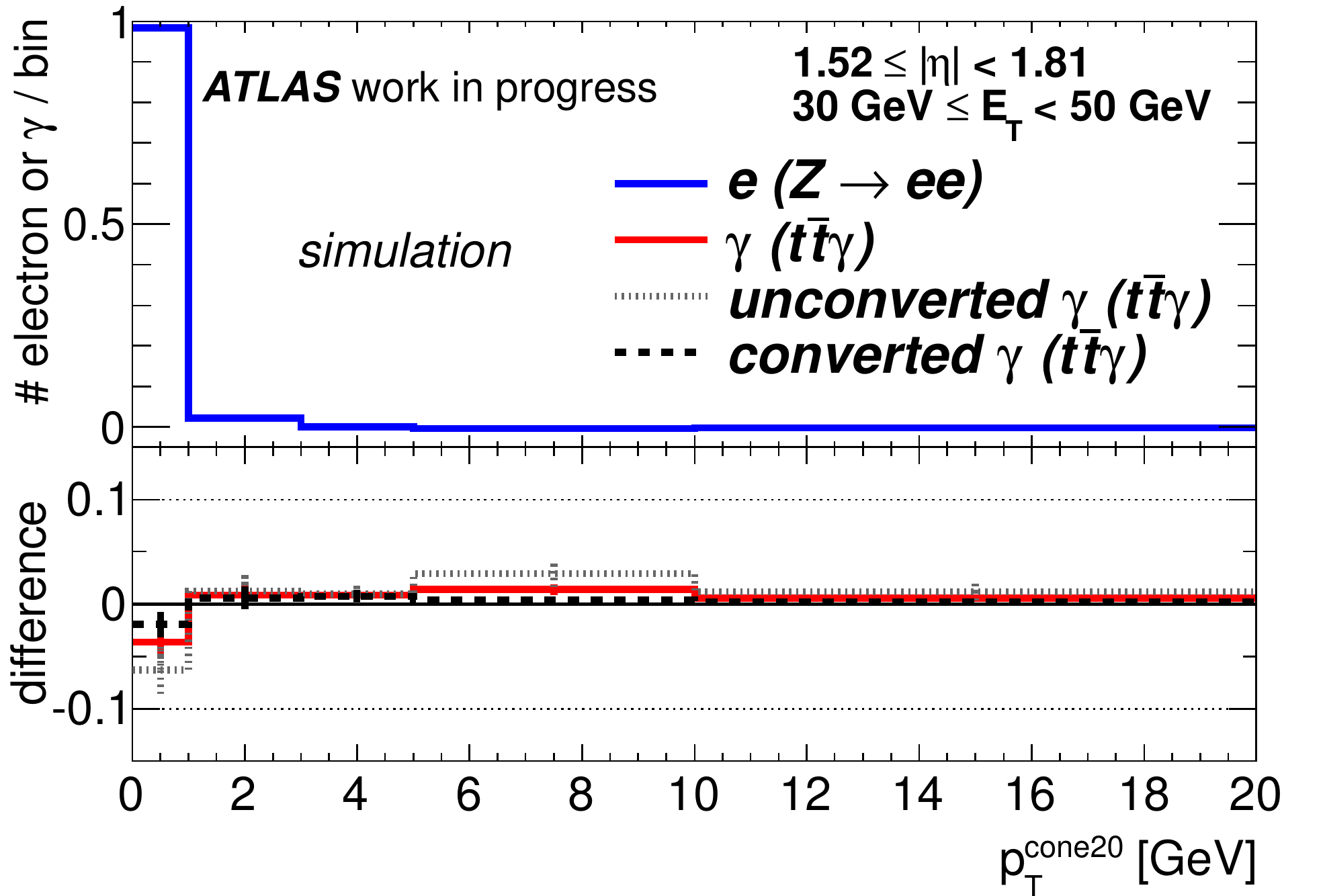}
    \includegraphics[width=0.395\textwidth]{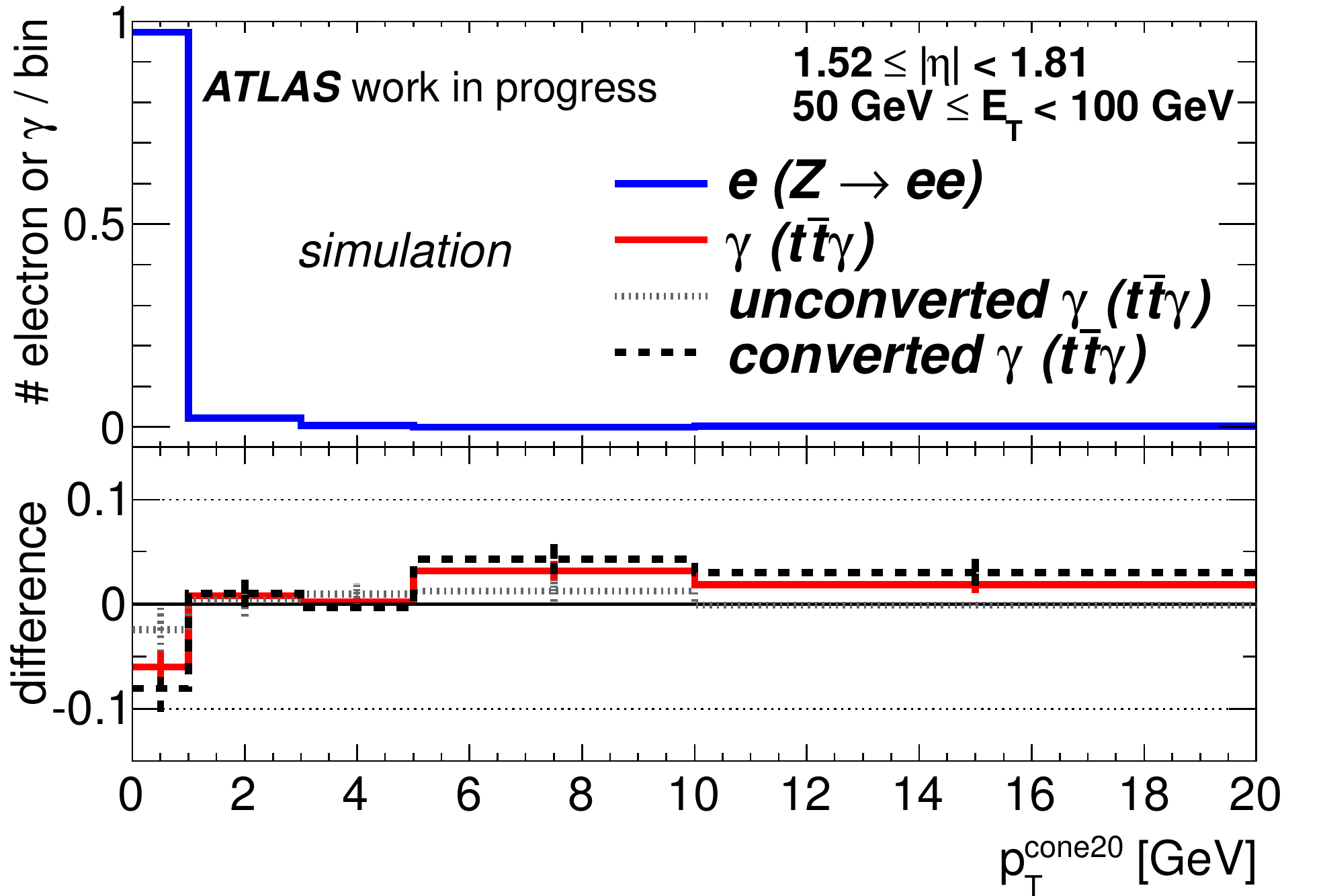}
    \includegraphics[width=0.395\textwidth]{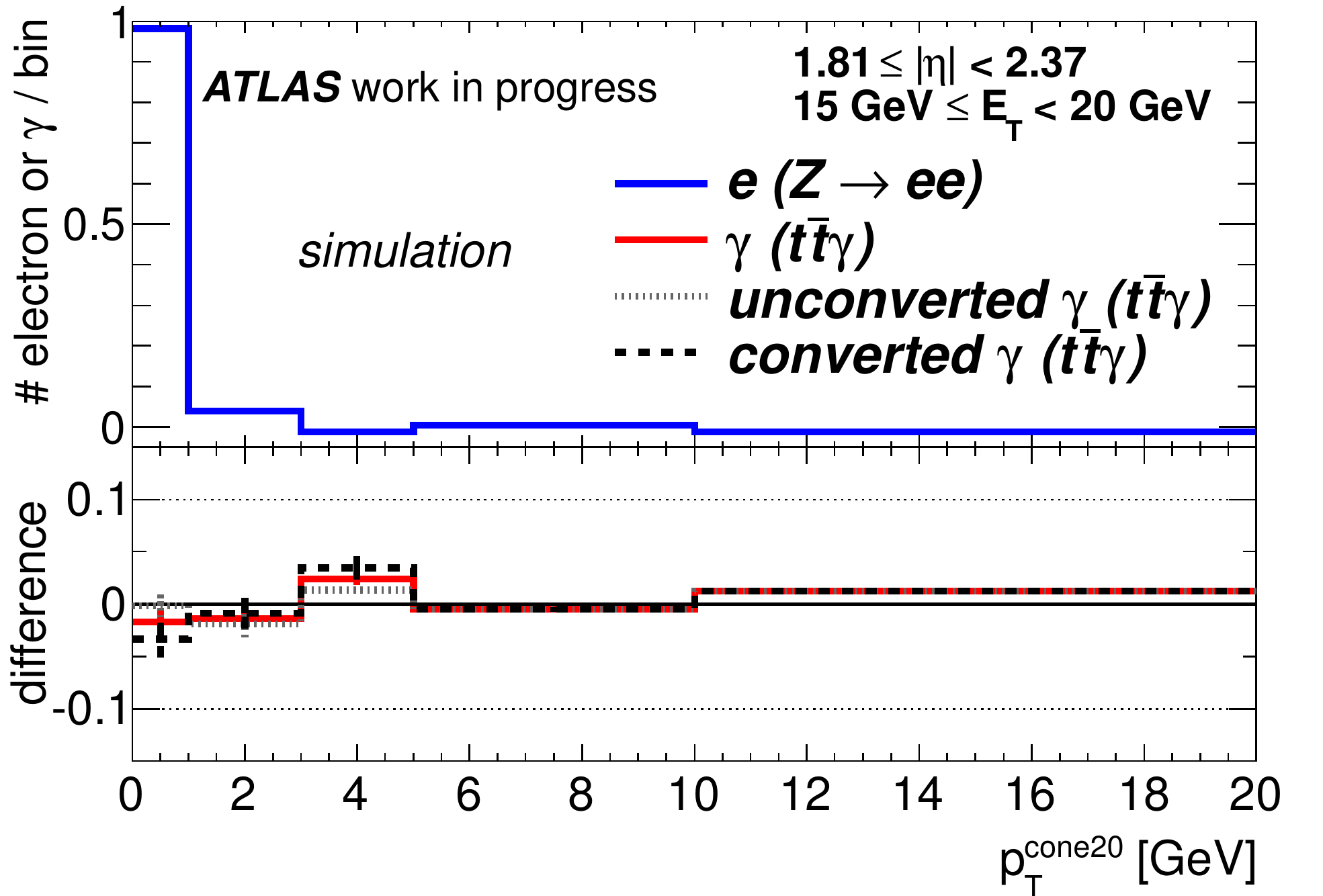}
    \includegraphics[width=0.395\textwidth]{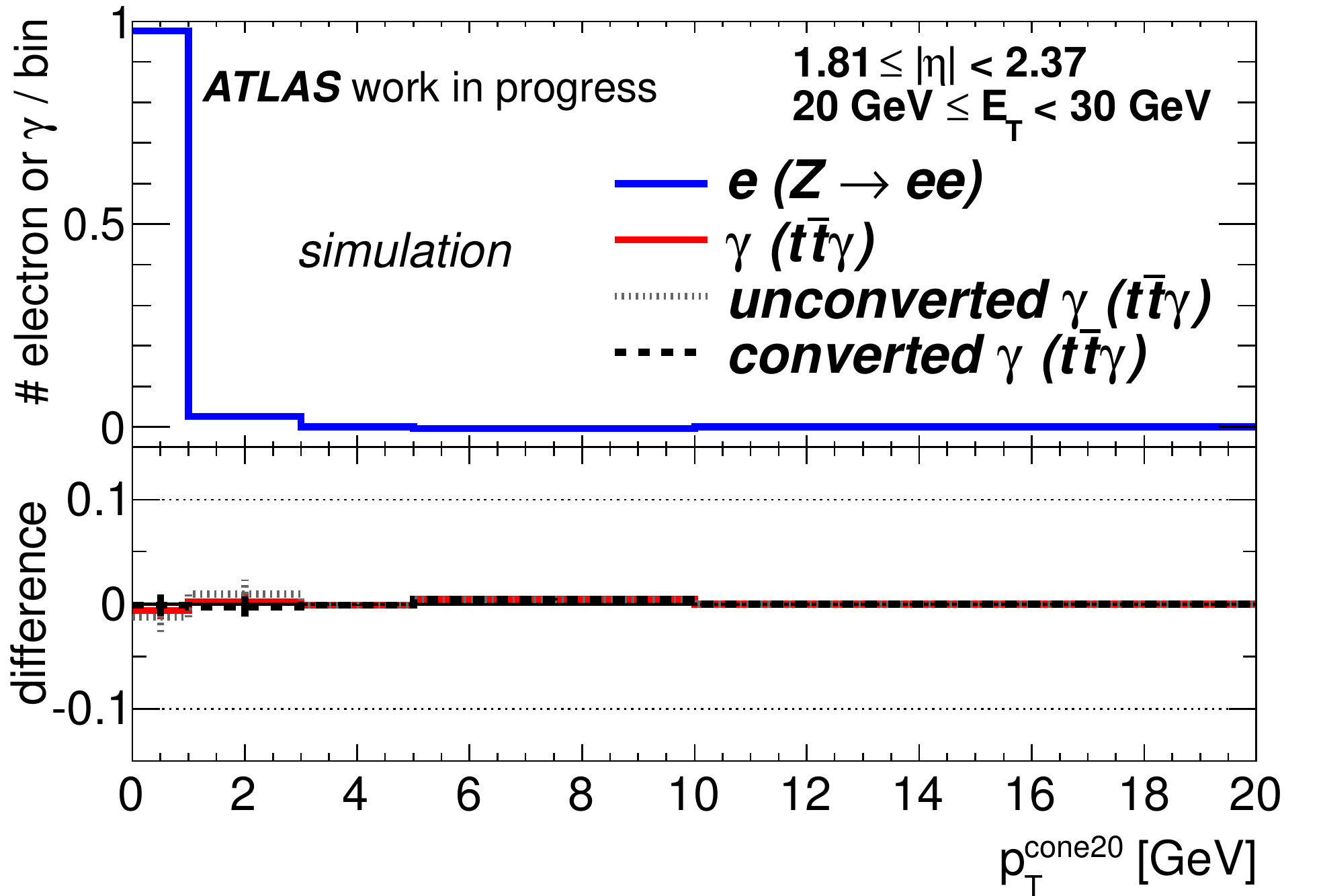}
    \includegraphics[width=0.395\textwidth]{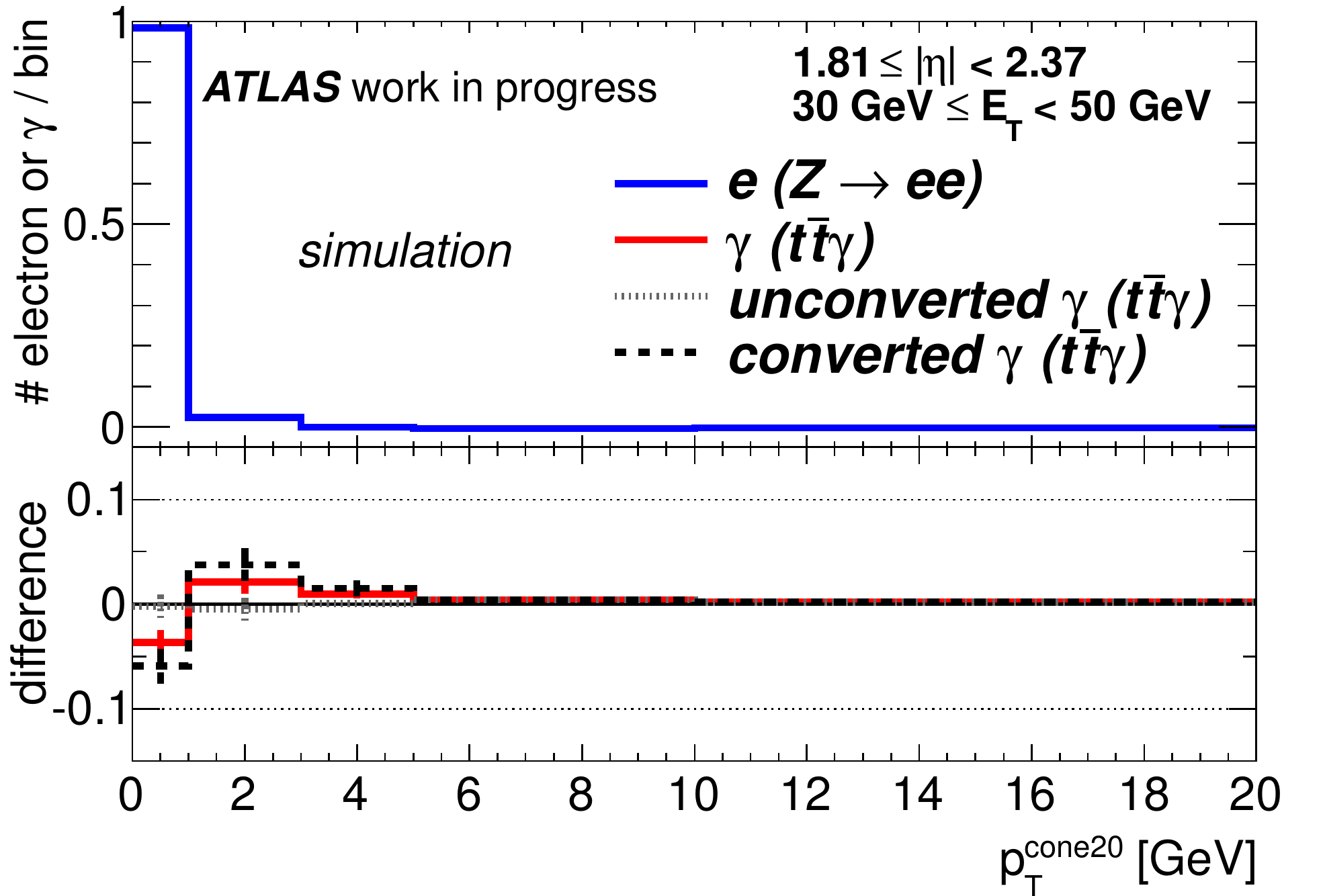}
    \includegraphics[width=0.395\textwidth]{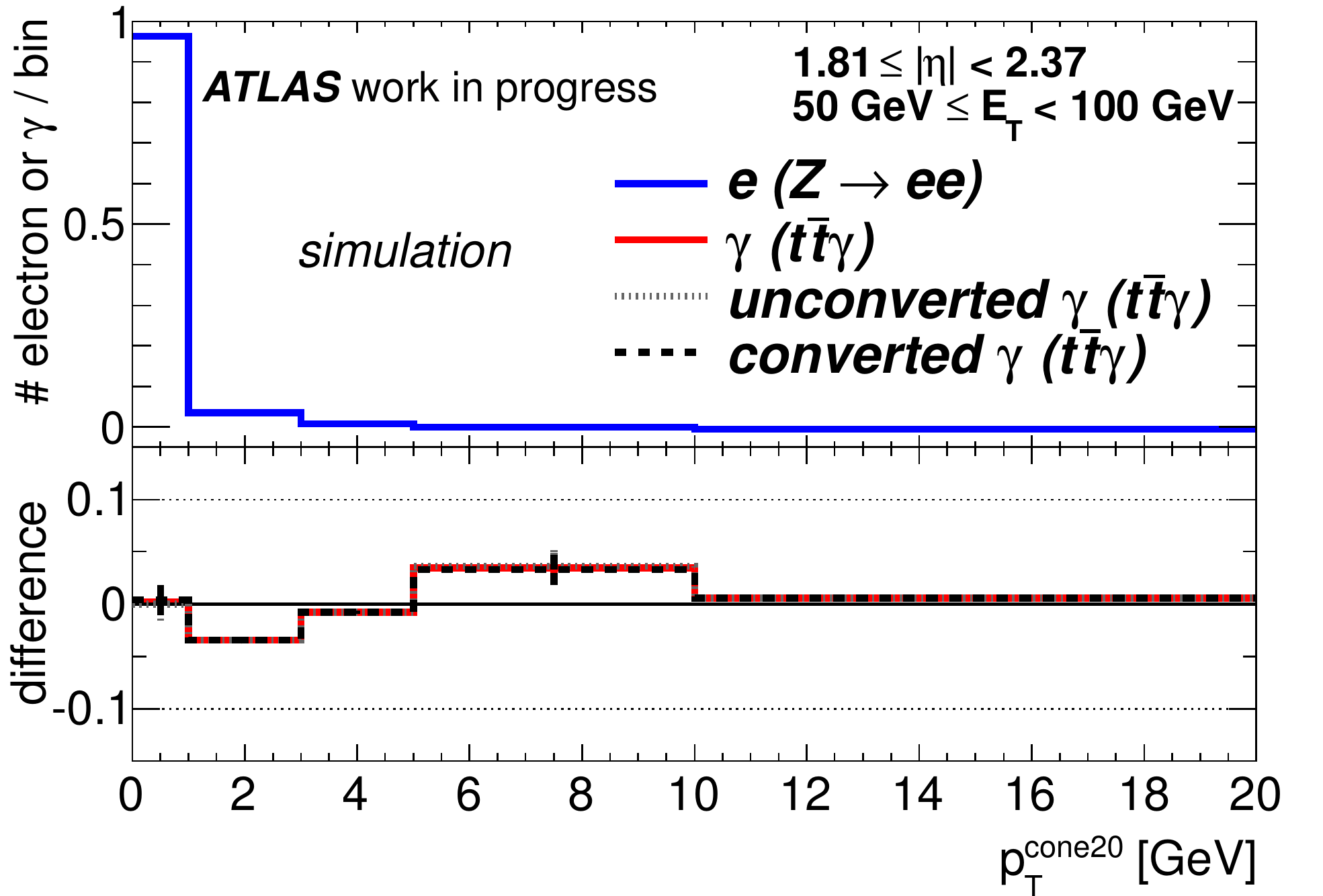}
    \caption[$\ptcone$ distributions for electrons and photons from simulation, $\Delta R(l,\gamma)$ cut (2)]{
      $\ptcone$ distributions for electrons from simulated \mbox{$Z \to e^+e^-$} decays (upper part of each plot)
      in different bins of $\et$ for \mbox{$1.52 \leq |\eta| < 1.81$} (four upper plots) and \mbox{$1.81 \leq |\eta| < 2.37$} (four lower plots)
      normalised to unity.
      The lower part of each plot shows the difference of the distribution of photons from simulated $\ttg$ events (solid line) with respect
      to the electron distribution.
      Additionally, the distributions for unconverted (dotted grey line) and converted photons (dashed black line) from $\ttg$ simulations are depicted.
      In all plots, the last bin includes the overflow bin.\\
      Photons closer than 0.2 in $\eta$-$\phi$-space to a true lepton were not considered and the agreement with the electron distribution
      is improved with respect to the plots in Fig.~\ref{fig:extrapolation_1} and~\ref{fig:extrapolation_2}.
      These plots complete the example plots shown in Fig.~\ref{fig:extrapolation_3}.
    }
    \label{fig:extrapolation_app_2}
  \end{center}
\end{figure}

\clearpage

\phantomsection
\addcontentsline{toc}{chapter}{Bibliography}
\bibliographystyle{bib/thesis}
\bibliography{bib/thesis}

\clearpage

\listoffigures
\addcontentsline{toc}{chapter}{List of Figures}
\phantomsection

\clearpage

\listoftables
\phantomsection
\addcontentsline{toc}{chapter}{List of Tables}

\clearpage

\chapter*{Acknowledgements}
\addcontentsline{toc}{chapter}{Acknowledgements} 

First of all, I want to express my sincere gratitude to Prof.~Dr.~Arnulf Quadt, my thesis adviser, who was at the origin of my enthusiasm
for particle physics in M\"unchen back in 2006.
I want to thank him for the opportunity to join him and his group of motivated top quark physicists in G\"ottingen,
for his trust in a $\ttg$ measurement with early ATLAS data, and for his constant support in physics and non-physics matters.
Moreover, I am particularly grateful for the opportunities to stay, work and extend my knowledge at LAL and at CERN for extended
periods during my time as a PhD student.
I want to thank Jun.-Prof.~Dr.~Lucia Masetti that she kindly agreed to be the second referee of this thesis.

I am very grateful to Kevin Kr\"oninger for his excellent supervision and lots of efficient advice.
Not only was he open for discussions at any time, but also could nothing make him lose his calm.
I want to sincerely thank Lisa Shabalina for the excellent supervision on top physics and Pixel matters during my time at CERN, and for her good
mood regardless of conference stress.
Thanks to J\"org Meyer for tips and help on computing issues and especially for straightforward solutions when time was short.
Special thanks to Kevin, Lisa, Andrea and J\"org for their attentive reading of the drafts of this thesis and their valuable comments.

I want to thank the ATLAS group at LAL, who cordially welcomed me in Orsay, and my colleagues from LAPP for the fruitful collaboration on the
diphoton analysis.
I want to express my special gratitude to Marumi Kado, from whom I learnt so much about photons, for a highly efficient and enjoyable time at LAL.
I also want to thank the group of Prof.~Dr.~Ivor Fleck in Siegen for the great common effort for the HCP conference note.

Thanks to Ms Christa Wohlfahrt, Ms Bernadette Tyson, Ms Heidi Afshar, Heike Ahrens and Ms Gabriela Herbold for constant
and constructive help with administrative issues.
Special thanks to Lucie Hamdi for her patience with creatively filled travel forms.

I want to thank all my colleagues and friends at the II. Institute of Physics for the excellent atmosphere at and off work.
Special thanks to Andrea, Anna, Boris, Daniel, Hans, Olaf and Stefan for distraction in form of football discussions, professional and
office basketball, good movies and series, beer and sports.

Thanks to Andi, Andrea, George, Jarka, Johannes, Jonas, Katharina, Morten and Steffi for a great time at pont de Gremaz with kitchen discussions,
ice-cream meetings, barbecues and hikes to the Reculet.
Thanks to Marilla, Elina and the AOV for being my musical home in G\"ottingen and for a lot of superb rehearsals and concerts.

I am very grateful to my parents for their constant support -- not only, but also -- during my PhD time, and
to Martin and Birgitta for valuable advice at any stage of my PhD thesis.
Last but not least, I want to say thanks to my sister Marion and her husband Simon for support, distraction, conversations and eggs deluxe wherever and
whenever needed.

\clearpage
\selectlanguage{ngerman}

\chapter*{Lebenslauf}
%\chapter*{Lebenslauf f\"ur Johannes Erdmann}

Am 11.10.1982 wurde ich als Sohn des Oberstudienrates Christian Martin Erdmann und der Instrumentalp\"adagogin Katharina Erdmann in Bonn geboren.
Nach dem Besuch der Grundschule besuchte ich das Evangelische Gymnasium in Meinerzhagen, wo ich im Juni 2002 das Abitur erhielt.

Anschlie{\ss}end leistete ich meinen Zivildienst in der evangelischen Tagungsst\"atte \glqq haus nordhelle{\grqq} in Valbert.
Zum Wintersemester 2003/04 nahm ich das Physikstudium an der Ludwig-Maximilians-Universit\"at in M\"unchen auf.
Im August 2005 erhielt ich dort das Vordiplom.
Von September 2005 bis Mai 2006 studierte ich mit einem Erasmus-Stipendium an der Universit\'e de Paris-Sud (Paris XI) in Orsay.

Im Anschlu{\ss} an den Aufenthalt in Frankreich arbeitete ich f\"ur zwei Monate am Max-Planck-Institut f\"ur Physik in M\"unchen bei
Prof.~Dr.~Arnulf Quadt (jetzt Universit\"at G\"ottingen) an der Entwicklung eines kinematischen $\chi^2$-Fitters zur
Rekonstruktion semileptonischer Topquarkpaar-Ereignisse am ATLAS-Experiment.

Nach dem Ablegen der Diplompr\"ufungen begann ich meine Diplomarbeit, die ich am Max-Planck-Institut f\"ur Physik zur Analyse der hadronischen
Kalibration des ATLAS-Endkappen-kalorimeters mit Teststrahl-Daten anfertigte.
Die Gutachter der Arbeit waren Prof.~Dr.~Siegfried Bethke und Prof.~Dr.~Dorothee Schaile.
Im Dezember 2008 erhielt ich das Diplom der Ludwig-Maximilians-Universit\"at M\"unchen mit dem Vermerk \glqq mit Auszeichnung bestanden\grqq.

Im Januar 2009 nahm ich das Promotionsstudium an der Georg-August-Universit\"at G\"ottingen auf.
Von Mai bis August 2010 arbeitete ich am Laboratoire de l'Acc\'el\'erateur Lin\'eaire in Orsay bei Paris an der Messung des inklusiven
Diphoton-Wirkungsquerschnittes am ATLAS-Experiment.
Im Anschluss an den Aufenthalt in Orsay war ich ein Jahr am CERN, wo ich auch Aufgaben f\"ur den Betrieb des ATLAS-Detektors \"ubernahm.
Nach meiner R\"uckkehr nach G\"ottingen wurden die Ergebnisse der $\ttg$-Analyse im November 2011 von der ATLAS-Kollaboration als
\textit{conference note} ver\"offentlicht.

Ich bin deutscher Staatsangeh\"origer.

\vspace{0.1\textheight}
G\"ottingen, den 25.04.2012

\selectlanguage{english}

\end{document}